\newif\ifshort
\makeatletter
\def\solutionfile{figure-short}\edef\solutionfile{\expandafter\strip@prefix\meaning\solutionfile}
\edef\JobName{\jobname}
\ifx\JobName\solutionfile
\shorttrue
\else
\shortfalse
\fi
\newif\ifelsevier

\ifelsevier
\documentclass[twoside,a4paper,amsthm]{elsart}
\usepackage{yjsco}
\else
\documentclass[twoside,a4paper]{amsart}
\fi
\usepackage[pdfauthor={Vladimir V. Kisil},%
  pdftitle={An Extension of Moebius--Lie Geometry
    with Conformal Ensembles of Cycles
    and Its Implementation in a GiNaC Library},%
  pdfsubject={Lie geometry},%
  backref=page,%
  pagebackref=true, %
  pdfkeywords={GiNaC, cycle, CAS},%
  breaklinks=true,%
  bookmarks=true,%
]{hyperref}

\PassOptionsToPackage{british}{babel}
\ifshort
\providecommand{\nwfilename}[1]{}
\providecommand{\nwbegindocs}[1]{}
\providecommand{\nwenddocs}[1]{}
\providecommand{\nwdocspar}{}
\providecommand{\nwendquote}{}
\providecommand{\Tt}{\texttt}
\providecommand{\Rm}{\textrm}
\providecommand{\nwnewline}{}
\else
\usepackage{noweb}
\makeatletter
\let\tab=&
\def\idxexample#1{\nwix@id@uses#1}
\fi
\usepackage{animate}

\usepackage{listings}
\ifelsevier
\noweboptions{scriptsizecode}
\providecommand{\email}[2]{\ead{#2}}
\providecommand{\urladdr}[1]{}
\providecommand{\subjclass}[2][2010]{\par\emph{AMS subject classification (#1)}: #2}
\usepackage{amsfonts,amsmath}
\usepackage[author-year,initials,nobysame,short-journals]{amsrefs}
\else
  \ifshort
  \else
  \fi
  \renewcommand{\baselinestretch}{1}
  \newtheorem{thm}{Theorem}
  \theoremstyle{definition}
  \newtheorem{example}[thm]{Example}
  \newtheorem{defn}[thm]{Definition}
  \theoremstyle{remark}
  \newtheorem{rem}[thm]{Remark}
  \PII{\relax}
  \copyrightinfo{}{\relax}
  \newenvironment{frontmatter}{}{\maketitle}
  \usepackage[initials,nobysame,short-journals]{amsrefs}
\fi

\ifelsevier
\BibSpecAlias{incollection}{inproceedings}
\BibSpec{article}{%
    +{}  {\hspace*{-.5em}\PrintAuthors}                {author}
    +{,} { \PrintDateB}                 {date}
    +{.} { }                            {title}
    +{.} { }                            {part}
    +{:} { }                            {subtitle}
    +{.} { \PrintContributions}         {contribution}
    +{.} { \PrintPartials}              {partial}
    +{.} { \emph}                       {journal}
    +{,} { \textbf}                            {volume}
    +{}  { \parenthesize}               {number}
    +{:} {}                             {pages}
    +{,} { }                            {status}
    +{.} { \PrintTranslation}           {translation}
    +{.} { Reprinted in \PrintReprint}  {reprint}
    +{.} { }                            {note}
    +{.} {}                             {transition}
}

\BibSpec{partial}{%
    +{}  {\hspace*{-.5em}}                             {part}
    +{,} { \PrintDateB}                 {date}
    +{:} { }                            {subtitle}
    +{.} { \PrintContributions}         {contribution}
    +{.} { \emph}                       {journal}
    +{,} { \textbf}                            {volume}
    +{}  { \parenthesize}               {number}
    +{:} {}                             {pages}
}

\BibSpec{book}{%
    +{}  {\hspace*{-.5em}\PrintPrimary}                {transition}
    +{,} { \PrintDateB}                 {date}
    +{.} { \emph}                       {title}
    +{.} { }                            {part}
    +{:} { \emph}                       {subtitle}
    +{.} { }                            {series}
    +{,} { \voltext}                    {volume}
    +{.} { Edited by \PrintNameList}    {editor}
    +{.} { Translated by \PrintNameList}{translator}
    +{.} { \PrintContributions}         {contribution}
    +{.} { }                            {publisher}
    +{.} { }                            {organization}
    +{,} { }                            {address}
    +{,} { \PrintEdition}               {edition}
    +{.} { }                            {note}
    +{.} {}                             {transition}
    +{.} { \PrintTranslation}           {translation}
    +{.} { Reprinted in \PrintReprint}  {reprint}
    +{.} {}                             {transition}
}

\BibSpec{collection.article}{%
    +{}  {\hspace*{-.5em}\PrintAuthors}                {author}
    +{,} { \PrintDateB}                 {date}
    +{.} { }                            {title}
    +{.} { }                            {part}
    +{:} { }                            {subtitle}
    +{.} { \PrintContributions}         {contribution}
    +{.} { \PrintConference}            {conference}
    +{.} { \PrintBook}                  {book}
    +{.} { In \PrintEditorsA}              {editor}
    +{.} { \emph}                         {booktitle}
    +{,} { pages~}                      {pages}
    +{,} { }                            {publisher}
    +{,} { }                            {organization}
    +{,} { }                            {address}
    +{.} { \PrintTranslation}           {translation}
    +{.} { Reprinted in \PrintReprint}  {reprint}
    +{.} { }                            {note}
    +{.} {}                             {transition}
}

 \BibSpec{inproceedings}{%
     +{}  {\hspace*{-.5em}\PrintAuthors}                {author}
     +{,} { \PrintDateB}                 {date}
     +{.} { }                            {title}
     +{.} { }                            {part}
     +{:} { }                            {subtitle}
     +{.} { \PrintContributions}         {contribution}
     +{.} { \PrintConference}            {conference}
     +{.} { \PrintBook}                  {book}
     +{.} { In \PrintEditorsA}              {editor}
     +{.} {  \emph}                         {booktitle}
     +{,} { pages~}                      {pages}
     +{,} { }                            {publisher}
     +{,} { }                            {organization}
     +{,} { }                            {address}
     +{.} { \PrintTranslation}           {translation}
     +{.} { Reprinted in \PrintReprint}  {reprint}
     +{.} { }                            {note}
     +{.} {}                             {transition}
 }

\BibSpec{conference}{%
    +{}  {\hspace*{-.5em}}                        {title}
    +{}  {\PrintConferenceDetails} {transition}
}

\BibSpec{innerbook}{%
    +{.} { \emph}                       {title}
    +{.} { }                            {part}
    +{:} { \emph}                       {subtitle}
    +{.} { }                            {series}
    +{,} { \voltext}                    {volume}
    +{.} { Edited by \PrintNameList}    {editor}
    +{.} { Translated by \PrintNameList}{translator}
    +{.} { \PrintContributions}         {contribution}
    +{.} { }                            {publisher}
    +{.} { }                            {organization}
    +{,} { }                            {address}
    +{,} { \PrintEdition}               {edition}
    +{,} { \PrintDateB}                 {date}
    +{.} { }                            {note}
    +{.} {}                             {transition}
}

\BibSpec{report}{%
    +{}  {\hspace*{-.5em}\PrintPrimary}                {transition}
    +{,} { \PrintDateB}                 {date}
    +{.} { \emph}                       {title}
    +{.} { }                            {part}
    +{:} { \emph}                       {subtitle}
    +{.} { \PrintContributions}         {contribution}
    +{.} { Technical Report }           {number}
    +{,} { }                            {series}
    +{.} { }                            {organization}
    +{,} { }                            {address}
    +{.} { \PrintTranslation}           {translation}
    +{.} { Reprinted in \PrintReprint}  {reprint}
    +{.} { }                            {note}
    +{.} {}                             {transition}
}

\BibSpec{thesis}{%
    +{}  {\hspace*{-.5em}\PrintAuthors}                {author}
    +{,} { \PrintDateB}                 {date}
    +{,} { \emph}                       {title}
    +{:} { \emph}                       {subtitle}
    +{.} { \PrintThesisType}            {type}
    +{.} { }                            {organization}
    +{,} { }                            {address}
    +{.} { \PrintTranslation}           {translation}
    +{.} { Reprinted in \PrintReprint}  {reprint}
    +{.} { }                            {note}
    +{.} {}                             {transition}
}
\else
\BibSpecAlias{incollection}{inproceedings}
\BibSpec{article}{%
    +{}  {\PrintAuthors}                {author}
    +{.} { }                            {title}
    +{.} { }                            {part}
    +{:} { }                            {subtitle}
    +{.} { \PrintContributions}         {contribution}
    +{.} { \PrintPartials}              {partial}
    +{.} { \emph}                       {journal}
    +{,} { \textbf}                            {volume}
    +{}  { \parenthesize}               {number}
    +{:} {}                             {pages}
    +{,} { \PrintDateB}                 {date}
    +{,} { }                            {status}
    +{.} { \PrintTranslation}           {translation}
    +{.} { Reprinted in \PrintReprint}  {reprint}
    +{.} { }                            {note}
    +{.} {}                             {transition}
}

\BibSpec{partial}{%
    +{}  {}                             {part}
    +{:} { }                            {subtitle}
    +{.} { \PrintContributions}         {contribution}
    +{.} { \emph}                       {journal}
    +{,} { \textbf}                            {volume}
    +{}  { \parenthesize}               {number}
    +{:} {}                             {pages}
    +{,} { \PrintDateB}                 {date}
}

\BibSpec{book}{%
    +{}  {\PrintPrimary}                {transition}
    +{.} { \emph}                       {title}
    +{.} { }                            {part}
    +{:} { \emph}                       {subtitle}
    +{.} { }                            {series}
    +{,} { \voltext}                    {volume}
    +{.} { Edited by \PrintNameList}    {editor}
    +{.} { Translated by \PrintNameList}{translator}
    +{.} { \PrintContributions}         {contribution}
    +{.} { }                            {publisher}
    +{.} { }                            {organization}
    +{,} { }                            {address}
    +{,} { \PrintEdition}               {edition}
    +{,} { \PrintDateB}                 {date}
    +{.} { }                            {note}
    +{.} {}                             {transition}
    +{.} { \PrintTranslation}           {translation}
    +{.} { Reprinted in \PrintReprint}  {reprint}
    +{.} {}                             {transition}
}

\BibSpec{collection.article}{%
    +{}  {\PrintAuthors}                {author}
    +{.} { }                            {title}
    +{.} { }                            {part}
    +{:} { }                            {subtitle}
    +{.} { \PrintContributions}         {contribution}
    +{.} { \PrintConference}            {conference}
    +{.} { \PrintBook}                  {book}
    +{.} { In \PrintEditorsA}              {editor}
    +{.} { \emph}                         {booktitle}
    +{,} { pages~}                      {pages}
    +{,} { }                            {publisher}
    +{,} { }                            {organization}
    +{,} { }                            {address}
    +{,} { \PrintDateB}                 {date}
    +{.} { \PrintTranslation}           {translation}
    +{.} { Reprinted in \PrintReprint}  {reprint}
    +{.} { }                            {note}
    +{.} {}                             {transition}
}

 \BibSpec{inproceedings}{%
     +{}  {\PrintAuthors}                {author}
     +{.} { }                            {title}
     +{.} { }                            {part}
     +{:} { }                            {subtitle}
     +{.} { \PrintContributions}         {contribution}
     +{.} { \PrintConference}            {conference}
     +{.} { \PrintBook}                  {book}
     +{.} { In \PrintEditorsA}              {editor}
     +{.} {  \emph}                         {booktitle}
     +{,} { pages~}                      {pages}
     +{,} { }                            {publisher}
     +{,} { }                            {organization}
     +{,} { }                            {address}
     +{,} { \PrintDateB}                 {date}
     +{.} { \PrintTranslation}           {translation}
     +{.} { Reprinted in \PrintReprint}  {reprint}
     +{.} { }                            {note}
     +{.} {}                             {transition}
 }

\BibSpec{conference}{%
    +{}  {}                        {title}
    +{}  {\PrintConferenceDetails} {transition}
}

\BibSpec{innerbook}{%
    +{.} { \emph}                       {title}
    +{.} { }                            {part}
    +{:} { \emph}                       {subtitle}
    +{.} { }                            {series}
    +{,} { \voltext}                    {volume}
    +{.} { Edited by \PrintNameList}    {editor}
    +{.} { Translated by \PrintNameList}{translator}
    +{.} { \PrintContributions}         {contribution}
    +{.} { }                            {publisher}
    +{.} { }                            {organization}
    +{,} { }                            {address}
    +{,} { \PrintEdition}               {edition}
    +{,} { \PrintDateB}                 {date}
    +{.} { }                            {note}
    +{.} {}                             {transition}
}

\BibSpec{report}{%
    +{}  {\PrintPrimary}                {transition}
    +{.} { \emph}                       {title}
    +{.} { }                            {part}
    +{:} { \emph}                       {subtitle}
    +{.} { \PrintContributions}         {contribution}
    +{.} { Technical Report }           {number}
    +{,} { }                            {series}
    +{.} { }                            {organization}
    +{,} { }                            {address}
    +{,} { \PrintDateB}                 {date}
    +{.} { \PrintTranslation}           {translation}
    +{.} { Reprinted in \PrintReprint}  {reprint}
    +{.} { }                            {note}
    +{.} {}                             {transition}
}

\BibSpec{thesis}{%
    +{}  {\PrintAuthors}                {author}
    +{,} { \emph}                       {title}
    +{:} { \emph}                       {subtitle}
    +{.} { \PrintThesisType}            {type}
    +{.} { }                            {organization}
    +{,} { }                            {address}
    +{,} { \PrintDateB}                 {date}
    +{.} { \PrintTranslation}           {translation}
    +{.} { Reprinted in \PrintReprint}  {reprint}
    +{.} { }                            {note}
    +{.} {}                             {transition}
}
\fi
\DeclareFontFamily{OT1}{cyr}{}
\DeclareFontShape{OT1}{cyr}{m}{n}
   {  <5> <6> <7> <8> <9> gen * wncyr
      <10> <10.95> <12> <14.4> <17.28> <20.74> <24.88> wncyr10}{}
\DeclareFontShape{OT1}{cyr}{m}{it}
    {
       <5> <6> <7> <8> <9> gen * wncyi
      <10> <10.95> <12> <14.4> <17.28> <20.74> <24.88>wncyi10
      }{}
\DeclareFontShape{OT1}{cyr}{m}{ss}
    {
       <5> <6> <7> <8> wncyss8
       <9> wncy9
      <10> <10.95> <12> <14.4> <17.28> <20.74> <24.88>wncyss10
      }{}
\DeclareFontShape{OT1}{cyr}{m}{sc}
    {
       <5> <6> <7> <8> <9> <10> <10.95> <12> <14.4> <17.28> <20.74> <24.88>wncysc10
      }{}
\DeclareFontShape{OT1}{cyr}{bx}{n}
   {
       <5> <6> <7> <8> <9> gen * wncyb
      <10> <10.95> <12> <14.4> <17.28> <20.74> <24.88>wncyb10
      }{}
\DeclareTextFontCommand{\textcyr}{\fontfamily{cyr}\selectfont}

\usepackage{graphicx}

\makeatother
\providecommand{\comment}[1]{}
\providecommand{\Space}[3][]{\ensuremath{\mathbb{#2}^{#3}_{#1}{}}}
\providecommand{\Cliff}[2][\comment]{{\ensuremath{%
\mathcal{C}\kern-0.18em\ell(#1,#2)}}}
\providecommand{\norm}[2][\relax]{\left\|#2\right\|\ifx#1\relax\else_{#1}\fi}
\providecommand{\modulus}[2][\relax]{\left| #2 \right|\ifx#1\relax\else_{#1}\fi}

\providecommand{\scalar}[3][\relax]{\left\langle #2,#3
        \right\rangle\ifx#1\relax\else_{#1}\fi}
\providecommand{\nscalar}[3][\relax]{\left[ #2,#3
        \right]\ifx#1\relax\else_{#1}\fi}
\providecommand{\wiki}[2]{\href{http://en.wikipedia.org/wiki/#1}{#2}}

\DeclareMathOperator{\arccosh}{arccosh}

\newcommand*\vtick{\kern -.1em\textsc{\char13}}

\def\nwlbrace{\textbf{\texttt{\char123}}}
\def\nwrbrace{\textbf{\texttt{\char125}}}
\newcommand{\CPP}{\textsf{C++}}
\newcommand{\CPPeleven}{\textsf{C++11}}
\newcommand{\Python}{\textsf{Python}}
\newcommand{\NoWEB}{\texttt{noweb}}

\providecommand{\GiNaC}{\textsf{GiNaC}}

\providecommand{\cycle}[3][]{{#1 C^{#2}_{#3}}}

\providecommand{\Asymptote}{\texttt{Asymptote}}

\newif\iftth

\providecommand{\tr}{\mathop{\mathrm{tr}}}

\providecommand{\clifford}[2][]{\ifcase #1 #2\or \tilde{#2} \or \breve{#2} \fi}
\hypersetup{colorlinks=true,bookmarks=true}

\begin{document}
\def\LA{\begingroup\maybehbox\bgroup\setupmodname\Rm$\langle$}\def\RA{$\rangle$\egroup\endgroup}\providecommand{\MM}{\kern.5pt\raisebox{.4ex}{\begin{math}\scriptscriptstyle-\kern-1pt-\end{math}}\kern.5pt}\providecommand{\PP}{\kern.5pt\raisebox{.4ex}{\begin{math}\scriptscriptstyle+\kern-1pt+\end{math}}\kern.5pt}\def\commopen{/\begin{math}\ast\,\end{math}}\def\commclose{\,\begin{math}\ast\end{math}\kern-.5pt/}\def\begcomm{\begingroup\maybehbox\bgroup\setupmodname}\def\endcomm{\egroup\endgroup}\nwfilename{figure.nw}\nwbegindocs{1}\nwdocspar
\ifshort
\else

\nwenddocs{}\nwbegindocs{2}%
\nwenddocs{}\nwbegindocs{3}%
\nwenddocs{}\nwbegindocs{4}%
\nwenddocs{}\nwbegindocs{5}%
\nwenddocs{}\nwbegindocs{6}%
\nwenddocs{}\nwbegindocs{7}%
\nwenddocs{}\nwbegindocs{8}%
\nwenddocs{}\nwbegindocs{9}%
\nwenddocs{}\nwbegindocs{10}%
\nwenddocs{}\nwbegindocs{11}%
\nwenddocs{}\nwbegindocs{12}%
\nwenddocs{}\nwbegindocs{13}%
\nwenddocs{}\nwbegindocs{14}%
\nwenddocs{}\nwbegindocs{15}%
\nwenddocs{}\nwbegindocs{16}%
\nwenddocs{}\nwbegindocs{17}%
\nwenddocs{}\nwbegindocs{18}%
\nwenddocs{}\nwbegindocs{19}%
\nwenddocs{}\nwbegindocs{20}%
\nwenddocs{}\nwbegindocs{21}%
\nwenddocs{}\nwbegindocs{22}%
\fi

\begin{frontmatter}

\title[Extension of Lie Geometry: Ensembles and their
  Implementation]{An Extension of M\"obius--Lie Geometry\\
    with Conformal Ensembles of Cycles\\
    and Its Implementation in a \GiNaC\ Library}


\author{Vladimir V. Kisil}
\address{School of Mathematics, University of Leeds, Leeds LS2 9JT, England}
\email{\href{mailto:kisilv@maths.leeds.ac.uk}{kisilv@maths.leeds.ac.uk}}
\urladdr{\href{http://www.maths.leeds.ac.uk/~kisilv/}{http://www.maths.leeds.ac.uk/\~{}kisilv/}}

\date{\today\ (v3.1)}

\begin{abstract}
  We propose to consider ensembles of cycles (quadrics), which are
  interconnected through conformal-invariant geometric relations
  (e.g. ``to be orthogonal'', ``to be tangent'', etc.), as new objects
  in an extended M\"obius--Lie geometry.  It was recently
  demonstrated in several related papers, that such ensembles of
  cycles naturally parameterise many other conformally-invariant
  objects, e.g. loxodromes or continued fractions.

  The paper describes a method, which reduces a collection of
  conformally in\-vari\-ant geometric relations to a system of linear
  equations, which may be accompanied by one fixed quadratic
  relation. To show its usefulness, the method is implemented as a {\CPP}
  library. It operates with numeric and symbolic data of cycles in
  spaces of arbitrary dimensionality and metrics with any
  signatures. Numeric calculations can be done in exact or approximate
  arithmetic. In the two- and three-dimensional cases illustrations
  and animations can be produced. An interactive {\Python} wrapper of the
  library is provided as well.
\end{abstract}


\end{frontmatter}

\ifelsevier
\pagestyle{plain}
\fi

\nwenddocs{}\nwbegindocs{23}\nwdocspar
\ifshort
\else
\tableofcontents

\listoffigures
\fi

\nwenddocs{}\nwbegindocs{24}\nwdocspar
\section{Introduction}
\label{sec:introduction}

\nwenddocs{}\nwbegindocs{25}\href{https://en.wikipedia.org/wiki/Lie_sphere_geometry}{Lie sphere
  geometry}~\citelist{ \cite{Cecil08a} \cite{Benz07a}*{Ch.~3}} in the
simplest planar setup unifies circles, lines and points---all together
called \emph{cycles} in this setup.  Symmetries of Lie spheres
geometry include (but are not limited to) fractional linear
transformations (FLT) of the form:
\begin{equation}
  \label{eq:flt-defn}
  \begin{pmatrix}
    a&b\\c&d
  \end{pmatrix}:\  x \mapsto
  \frac{ax+b}{cx+d}\,, \qquad \text{where }
  \det\begin{pmatrix}
    a&b\\c&d
  \end{pmatrix}\neq 0.
\end{equation}
Following other sources, e.g. ~\cite{Simon11a}*{\S~9.2}, we call
\eqref{eq:flt-defn} by FLT and reserve the name ``M\"obius maps'' for
the subgroup of FLT which fixes a particular cycle. For example, on
the complex plane FLT are generated by elements of
\(\mathrm{SL}_2(\Space{C}{})\) and M\"obius maps fixing the real line
are produced by
\(\mathrm{SL}_2(\Space{R}{})\)~\cite{Kisil12a}*{Ch.~1}.

There is a natural set of FLT-invariant geometric relations between
cycles (to be orthogonal, to be tangent, etc.) and the restriction of
Lie sphere geometry to invariants of FLT is called \emph{M\"obius--Lie
  geometry}.  Thus, an ensemble of cycles, structured by a set of such
relations, will be mapped by FLT to another ensemble with the same
structure.

It was shown recently that ensembles of cycles with certain
FLT-invariant relations provide helpful parametrisations of new
objects, e.g. points of the Poincar\'e extended space~\cite{Kisil15a},
loxodromes~\cite{KisilReid18a} or continued
fractions~\cites{BeardonShort14a,Kisil14a}, see
Example~\ref{ex:ensamble-math} below for further details. Thus, we
propose \emph{to extend M\"obius--Lie geometry and consider ensembles
  of cycles as its new objects}, cf. formal
Defn.~\ref{de:extended-Lie-Moebius}. Naturally, ``old''
objects---cycles---are represented by simplest one-element ensembles
without any relation. This paper provides conceptual foundations of
such extension and demonstrates its practical implementation as a
{\CPP} library {\Tt{}\Rm{}{\bf{}figure}\nwendquote}\footnote{All described software is licensed
  under GNU GPLv3~\cite{GNUGPL}.}. Interestingly, the development of
this library shaped the general approach, which leads to specific
realisations in~\cites{Kisil15a,Kisil14a,KisilReid18a}.

\nwenddocs{}\nwbegindocs{26}More specifically, the library {\Tt{}\Rm{}{\bf{}figure}\nwendquote} manipulates ensembles of
cycles (quadrics) interrelated by certain FLT-invariant geometric
conditions.  The code is build on top of the previous library
{\Tt{}\Rm{}{\bf{}cycle}\nwendquote}~\cites{Kisil05b,Kisil12a,Kisil06a}, which manipulates
individual cycles within the \GiNaC~\cite{GiNaC} computer algebra
system. Thinking an ensemble as a graph, one can say that the library
{\Tt{}\Rm{}{\bf{}cycle}\nwendquote} deals with individual vertices (cycles), while {\Tt{}\Rm{}{\bf{}figure}\nwendquote}
considers edges (relations between pairs of cycles) and the whole
graph. Intuitively, an interaction with the library {\Tt{}\Rm{}{\bf{}figure}\nwendquote} reminds
compass-and-straightedge constructions, where new lines or circles are
added to a drawing one-by-one through relations to already presented
objects (the line through two points, the intersection point or the
circle with given centre and a point). See
Example~\ref{ex:touch-centres-collinear} of such interactive
construction from the {\Python} wrapper, which provides an analytic
proof of a simple geometric statement.

It is important that both libraries are capable to work in spaces of
any dimensionality and metrics with an arbitrary signatures:
Euclidean, Minkowski and even degenerate. Parameters of objects can be
symbolic or numeric, the latter admit calculations with exact or
approximate arithmetic.  Drawing routines work with any (elliptic,
parabolic or hyperbolic) metric in two dimensions and the euclidean
metric in three dimensions.

The mathematical formalism employed in the library {\Tt{}\Rm{}{\bf{}cycle}\nwendquote} is based
on Clifford algebras, which are intimately connected to fundamental
geometrical and physical objects
\cites{HestenesSobczyk84a,Hestenes15a}. Thus, it is not surprising
that Clifford algebras have been already used in various geometric
algorithms for a long time, for example see
\cites{Hildenbrand13a,Vince08a,DorstDoranLasenby02a} and further
references there. Our package deals with cycles through
Fillmore--Springer--Cnops construction (FSCc) which also has a long
history, see~\citelist{\cite{Schwerdtfeger79a}*{\S~1.1}
  \cite{Cnops02a}*{\S~4.1}
  \cite{FillmoreSpringer90a} \cite{Kirillov06}*{\S~4.2}
  \cite{Kisil05a} \cite{Kisil12a}*{\S~4.2}} and
section~\ref{sec:lie-spheres-geometry} below. Compared to a plain
analytical treatment~\citelist{\cite{Pedoe95a}*{Ch.~2}
  \cite{Benz07a}*{Ch.~3}}, FSCc is much more efficient and
conceptually coherent in dealing with FLT-invariant properties of
cycles. Correspondingly, the computer code based on FSCc is easy to
write and maintain.

The paper outline is as follows. In Section~\ref{sec:math-backgr} we
sketch the mathematical theory (M\"obius--Lie geometry) covered by the
package of the previous library {\Tt{}\Rm{}{\bf{}cycle}\nwendquote}~\cite{Kisil05b} and the
present library {\Tt{}\Rm{}{\bf{}figure}\nwendquote}. We expose the subject with some
references to its history since this can facilitate further
development.

Sec.~\ref{sec:conn-quadr-cycl} describes the principal mathematical
tool used by the library {\Tt{}\Rm{}{\bf{}figure}\nwendquote}.  It allows to reduce a collection
of various linear and quadratic equations (expressing geometrical
relations like orthogonality and tangency) to a set of linear
equations and \emph{at most one} quadratic
relation~\eqref{eq:det-normalisation-cond}. Notably, the quadratic
relation is the same in all cases, which greatly simplifies its
handling. This approach is the cornerstone of the library
effectiveness both in symbolic and numerical computations.  In
Sec.~\ref{sec:figures-as-families} we present several examples of
ensembles, which were already used in mathematical
theories~\cites{Kisil15a,Kisil14a,KisilReid18a}, then we describe how
ensembles are encoded in the present library {\Tt{}\Rm{}{\bf{}figure}\nwendquote} through the
functional programming framework.

Sec.~\ref{sec:mathematical-results} outlines several typical usages of
the package. An example of a new statement discovered and demonstrated
by the package is given in Thm.~\ref{th:nine-points}.  In
Sec.~\ref{sec:do} we list of some further tasks, which will extend
capacities and usability of the package.

\ifshort
All coding-related material is enclosed as appendices in the full
documentation on the project page~\cite{Kisil05b}. They contain:
\begin{enumerate}
\item Numerous examples of the library usage
  starting from the very simple ones.
\item  A systematic list of callable
  methods.
\item Actual code of the library.
\end{enumerate}
Sec.~\ref{sec:math-backgr}, Example~\ref{ex:touch-centres-collinear}
below or the above-mentioned first two appendices of the full
documentation can serve as an entry point for a reader with respective
preferences and background.  \else All coding-related material is
enclosed as appendices.  App.~\ref{sec:examples} contains examples of
the library usage starting from the very simple ones. A systematic
list of callable methods is given in
Apps~\ref{sec:publ-meth-figure}--\ref{sec:addtional-utilities}. Any of
Sec.~\ref{sec:math-backgr} or
Apps~\ref{sec:examples}--\ref{sec:publ-meth-figure} can serve as an
entry point for a reader with respective preferences and
background. Actual code of the library is collected in
Apps~\ref{sec:figure-header-file}--\ref{sec:impl-class}.  \fi

\nwenddocs{}\nwbegindocs{27}\nwdocspar
\section{M\"obius--Lie Geometry and the {\Tt{}\Rm{}{\bf{}cycle}\nwendquote} Library}
\label{sec:math-backgr}
We briefly outline mathematical formalism of the extend M\"obius--Lie
geometry, which is implemented in the present package. We do not aim
to present the complete theory here, instead we provide a minimal
description with a sufficient amount of references to further
sources. The hierarchical structure of the theory naturally splits the
package into two components: the routines handling individual cycles
(the library {\Tt{}\Rm{}{\bf{}cycle}\nwendquote} briefly reviewed in this section), which were
already introduced elsewhere~\cite{Kisil05b}, and the new component
implemented in this work, which handles families of interrelated
cycles (the library {\Tt{}\Rm{}{\bf{}figure}\nwendquote} introduced in the next section).

\nwenddocs{}\nwbegindocs{28}\nwdocspar
\subsection{M\"obius--Lie geometry and FSC construction}
\label{sec:lie-spheres-geometry}
M\"obius--Lie geometry in \(\Space{R}{n}\) starts from an observation
that points can be treated as spheres of zero radius and planes are
the limiting case of spheres with radii diverging to
infinity. Oriented spheres, planes and points are called together
\emph{cycles}\index{cycle}. Then, the second crucial step is to treat
cycles not as subsets of \(\Space{R}{n}\) but rather as points of some
projective space of higher dimensionality,
see~\citelist{\cite{Benz08a}*{Ch.~3} \cite{Cecil08a} \cite{Pedoe95a}
  \cite{Schwerdtfeger79a}}.

To distinguish two spaces we will call \(\Space{R}{n}\) as the \emph{point
  space}%
\index{point!space}%
\index{space!point} and the higher dimension space, where cycles are
represented by points---the \emph{cycle space}%
\index{cycle!space}%
\index{space!cycle}. Next important observation is that geometrical
relations between cycles as subsets of the point space can be
expressed in term of some indefinite metric on the cycle
space. Therefore, if an indefinite metric shall be considered anyway,
there is no reason to be limited to spheres in Euclidean space \(\Space{R}{n}\)
only. The same approach shall be adopted for quadrics in
spaces \(\Space{R}{pqr}\) of an arbitrary signature
\(p+q+r=n\), including \(r\) nilpotent elements,
cf.~\eqref{eq:clifford-defn} below.

A useful addition to M\"obius--Lie geometry is provided by the
Fillmore--Springer--Cnops construction
(FSCc)~\citelist{\cite{Schwerdtfeger79a}*{\S~1.1}
  \cite{Cnops02a}*{\S~4.1} \cite{Porteous95}*{\S~18}
  \cite{FillmoreSpringer90a} \cite{Kirillov06}*{\S~4.2}
  \cite{Kisil05a} \cite{Kisil12a}*{\S~4.2}}. It is a correspondence
between the cycles (as points of the cycle space) and certain
\(2\times 2\)-matrices defined in~\eqref{eq:spheres-Rn} below. The
main advantages of FSCc are:
\begin{enumerate}
\item The correspondence between cycles and matrices respects the
  projective structure of the cycle space.
\item The correspondence is FLT covariant.
\item The indefinite metric on the cycle space can be expressed
  through natural operations on the respective matrices.
\end{enumerate}
The last observation is that for restricted groups of M\"obius
transformations the metric of the cycle space may not be completely
determined by the metric of the point space,
see~\citelist{\cite{Kisil06a} \cite{Kisil05a}
  \cite{Kisil12a}*{\S~4.2}} for an example in two-dimensional space.

FSCc is useful in consideration of the Poincar\'e extension of
M\"obius maps~\cite{Kisil15a}, loxodromes~\cite{KisilReid18a} and
continued fractions~\cites{Kisil14a}. In theoretical physics FSCc
nicely describes conformal compactifications of various space-time
models~\citelist{\cite{HerranzSantander02b} \cite{Kisil06b}
  \cite{Kisil12a}*{\S~8.1}}.  Regretfully, FSCc have not yet
propagated back to the most fundamental case of complex numbers,
cf.~\cite{Simon11a}*{\S~9.2} or somewhat cumbersome techniques used
in~\cite{Benz07a}*{Ch.~3}. Interestingly, even the founding fathers
were not always strict followers of their own techniques,
see~\cite{FillmoreSpringer00a}.

We turn now to the explicit definitions.

\subsection{Clifford algebras, FLT transformations, and Cycles}
\label{sec:cliff-algebr-mobi}
We describe here the mathematics behind the the first library called
{\Tt{}\Rm{}{\bf{}cycle}\nwendquote}, which implements fundamental geometrical relations between
quadrics in the space \(\Space{R}{pqr}\) with the dimensionality
\(n=p+q+r\) and metric
\(x_1^2+\ldots+x_p^2-x_{p+1}^2-\ldots-x_{p+q}^2\). A version
simplified for complex numbers only can be found
in~\cites{Kisil15a,KisilReid18a,Kisil14a}.

The Clifford algebra \(\Cliff{p,q,r}\) is the associative unital algebra over
\(\Space{R}{}\) generated by the elements \(e_1\),\ldots,\(e_n\)
satisfying the following relation:
\begin{equation}
\label{eq:clifford-defn}
  e_i e_j =- e_je_i\,, \quad \text{ and } \quad e_i^2=\left\{
    \begin{array}{ll}
      -1,&\text{ if } 1\leq i\leq p;\\
      1,&\text{ if } p+1\leq i\leq p+q;\\
      0,&\text{ if } p+q+1\leq i\leq p+q+r.
    \end{array}
  \right.
\end{equation}
It is common~\cites{DelSomSou92,Cnops02a,Porteous95,%
  HestenesSobczyk84a,Hestenes15a} to consider mainly Clifford algebras
\(\Cliff{n}=\Cliff{n,0,0}\) of the Euclidean space or the algebra
\(\Cliff{p,q}=\Cliff{p,q,0}\) of the pseudo-Euclidean (Minkowski)
spaces. However, Clifford algebras \(\Cliff{p,q,r}\), \(r>0\) with
nilpotent generators \(e_i^2=0\) correspond to interesting
geometry~\cites{Kisil12a,Kisil05a,Yaglom79,Mustafa17a} and
physics~\cites{GromovKuratov06a,Gromov10a,Gromov12a,%
  Kisil12c,Kisil09e,Kisil17a} as well. Yet, the geometry with
idempotent units in spaces with dimensionality \(n>2\) is still not
sufficiently elaborated.

An element of
\(\Cliff{p,q,r}\) having the form \(x=x_1e_1+\ldots+x_ne_n\) can be
associated with the vector \((x_1,\ldots,x_n)\in\Space{R}{pqr}\).  The
\emph{reversion} \(a\mapsto a^*\) in
\(\Cliff{p,q,r}\)~\cite{Cnops02a}*{(1.19(ii))} is defined on vectors by
\(x^*=x\) and extended to other elements by the relation
\((ab)^*=b^*a^*\). Similarly the \emph{conjugation} is defined on
vectors by \(\bar{x}=-x\) and the relation
\(\overline{ab}=\bar{b}\bar{a}\). We also use the notation
\(\modulus{a}^2=a\bar{a}\) for any product \(a\) of vectors.
An important observation is that any non-zero \(x\in\Space{R}{n00}\) has a
multiplicative inverse: \(x^{-1}=\frac{\bar{x}}{\modulus{x}^2}\).
For a \(2\times 2\)-matrix \(M=
 \begin{pmatrix}
   a&b\\c&d
 \end{pmatrix}\) with Clifford entries we define,
 cf.~\cite{Cnops02a}*{(4.7)}
\begin{equation}
  \label{eq:matrix-bar-star}
  \bar{M}=
\begin{pmatrix}
  d^*&-b^*\\-c^*&a^*
\end{pmatrix}\qquad \text{ and } \qquad
M^*=\begin{pmatrix}
  \bar{d} &\bar{b}\\\bar{c}&\bar{a}
\end{pmatrix}.
\end{equation}
Then \(M\bar{M}=\delta I\) for the \emph{pseudodeterminant} \(\delta:=ad^*-bc^*\) .


Quadrics in \(\Space{R}{pq}\)---which we continue to
call cycles---can be associated to
\(2\times 2\) matrices through the FSC
construction~\citelist{\cite{FillmoreSpringer90a}
  \cite{Cnops02a}*{(4.12)} \cite{Kisil12a}*{\S~4.4}}:
\begin{equation}
  \label{eq:spheres-Rn}
  k\bar{x}x-l\bar{x}-x\bar{l}+m=0 \quad \leftrightarrow \quad
  \cycle{}{}=
  \begin{pmatrix}
    l & m\\
    k & \bar{l}
  \end{pmatrix},
\end{equation}
where \(k, m\in\Space{R}{}\) and \(l\in\Space{R}{pq}\).  For brevity
we also encode a cycle by its coefficients \((k,l,m)\).  A
justification of~\eqref{eq:spheres-Rn} is provided by the identity:
\begin{displaymath}
  \begin{pmatrix}
    1&\bar{x}
  \end{pmatrix}
  \begin{pmatrix}
    l & m\\
    k & \bar{l}
  \end{pmatrix}
  \begin{pmatrix}
    {x}\\1
  \end{pmatrix}=
  kx\bar{x}-l\bar{x}-x\bar{l}+m,\quad \text{ since } \bar{x}=-x \text{
    for } x\in\Space{R}{pq}.
\end{displaymath}
The identification is also FLT-covariant in the sense that the
transformation~\eqref{eq:flt-defn} associated with the matrix \(M=
\begin{pmatrix}
  a&b\\c&d
\end{pmatrix}
\) sends a cycle
\(\cycle{}{}\) to the cycle \(M\cycle{}{}M^{*}\)~\cite{Cnops02a}*{(4.16)}.
We define the FLT-invariant inner product of cycles
\(\cycle{}{1}\) and \(\cycle{}{2}\) by the
identity
\begin{align}
  \label{eq:cycle-product}
  \scalar{\cycle{}{1}}{\cycle[]{}{2}}&=\Re
\tr(\cycle{}{1}\cycle{}{2})\,,
\end{align}
where \(\Re\) denotes the scalar part of a Clifford number. This
definition in term of matrices immediately implies that the inner
product is FLT-invariant. The explicit expression in terms of
 components of cycles \(\cycle{}{1}=(k_1,l_1,m_1)\) and
\(\cycle{}{2}=(k_2,l_2,m_2)\) is also useful sometimes:
\begin{align}
  \label{eq:cycle-product-expl}
  \scalar{\cycle{}{1}}{\cycle[]{}{2}}&=l_1 l_2+ \bar{l}_1 \bar{l}_2+m_1k_2+m_2k_1\,.
\end{align}
As usual, the relation \(\scalar{\cycle{}{1}}{\cycle[]{}{2}}=0\) is
called the \emph{orthogonality} of cycles \(\cycle{}{1}\) and
\(\cycle{}{2}\). In most cases it corresponds to orthogonality of
quadrics in the point space. More generally, most of
FLT-invariant relations between quadrics may be expressed in
terms FLT-invariant inner product~\eqref{eq:cycle-product}. For
the full description of methods on individual cycles, which are
implemented in the library {\Tt{}\Rm{}{\bf{}cycle}\nwendquote}, see the respective
documentation~\cite{Kisil05b}.

\begin{rem}
  Since cycles are elements of the projective space, the following
  \emph{normalised cycle product}:
  \begin{equation}
    \label{eq:norm-cycle-prod}
    \nscalar{C_1}{C_2}:=\frac{\scalar{C_1}{C_2}}{\sqrt{\scalar{C_1}{C_1}
        \scalar{C_2}{C_2}}}
  \end{equation}
  is more meaningful than the cycle product~\eqref{eq:cycle-product}
  itself. Note that, \(\nscalar{C_1}{C_2}\) is defined only if neither
  \(C_1\) nor \(C_2\) is a zero-radius cycle (i.e. a point). Also, the
  normalised cycle product is \(\mathrm{GL}_2(\Space{C}{})\)-invariant
  in comparison to \(\mathrm{SL}_2(\Space{C}{})\)-invariance
  of~\eqref{eq:cycle-product}.
\end{rem}

We finish this brief review of the library {\Tt{}\Rm{}{\bf{}cycle}\nwendquote} by pointing to
its light version written in \textsf{Asymptote}
language~\cite{Asymptote} and distributed together with the
paper~\cite{KisilReid18a}. Although the light version mostly inherited
API of the library {\Tt{}\Rm{}{\bf{}cycle}\nwendquote},  there are some significant limitations caused
by the absence of {\GiNaC} support:
\begin{enumerate}
\item there is no symbolic computations of any sort;
\item the light version works in two dimensions only;
\item only elliptic metrics in the point and cycle spaces are supported.
\end{enumerate}
On the other hand, being integrated with  \textsf{Asymptote} the light
version simplifies production of illustrations, which are its main target.

\section{Ensembles of Interrelated Cycles and the {\Tt{}\Rm{}{\bf{}figure}\nwendquote} Library}
\label{sec:library-figure}

The library {\Tt{}\Rm{}{\bf{}figure}\nwendquote} has an ability to store and resolve the system of
geometric relations between cycles. We explain below some mathematical
foundations, which greatly simplify this task.

\subsection{Connecting quadrics and cycles}
\label{sec:conn-quadr-cycl}
We need a vocabulary, which translates geometric properties of
quadrics on the point space to corresponding relations in the cycle
space. The key ingredient is the cycle
product~\eqref{eq:cycle-product}--\eqref{eq:cycle-product-expl}, which
is linear in each cycles\vtick\  parameters. However, certain conditions,
e.g. tangency of cycles, involve polynomials of cycle products and
thus are non-linear.  For a successful algorithmic implementation, the
following observation is important: \emph{all non-linear conditions
  below can be linearised if the additional quadratic condition of
  normalisation type is imposed}:
\begin{equation}
  \label{eq:det-normalisation-cond}
  \scalar{\cycle{}{}}{\cycle{}{}}=\pm1.
\end{equation}
This observation in the context of the Apollonius problem was already
made in~\cite{FillmoreSpringer00a}. Conceptually the present work has
a lot in common with the above mentioned paper of Fillmore and
Springer, however a reader need to be warned that our implementation is
totally different (and, interestingly, is more closer to another
paper~\cite{FillmoreSpringer90a} of Fillmore and Springer).
\begin{rem}
  Interestingly, the method of order reduction for algebraic equations is
  conceptually similar to the method of order reduction of
  differential equations used to build a geometric dynamics of quantum
  states in~\cite{AlmalkiKisil18a}.
\end{rem}

Here is the list of relations between cycles implemented in the
current version of the library {\Tt{}\Rm{}{\bf{}figure}\nwendquote}.
\begin{enumerate}
\item  \label{item:quadric-flat}
  A quadric is flat (i.e. is a hyperplane), that is, its equation
  is linear. Then, either of two equivalent conditions can be used:
  \begin{enumerate}
  \item \(k\) component of the cycle vector is zero;
  \item the cycle is orthogonal
    \(\scalar{\cycle{}{1}}{\cycle[]{}{\infty}}=0\) to the ``zero-radius cycle at
    infinity'' \(\cycle[]{}{\infty}=(0,0,1)\).
  \end{enumerate}
\item \label{it:lobachevski-line}
  A quadric on the plane represents a line in Lobachevsky-type
  geometry if it is orthogonal
  \(\scalar{\cycle{}{1}}{\cycle[]{}{\Space{R}{}}}=0\)  to the real line cycle
  \(\cycle{}{\Space{R}{}}\). A similar condition is meaningful in
  higher dimensions as well.
\item \label{it:point-zero-radius}
  A quadric \(\cycle{}{}\) represents a point, that is, it has zero
  radius at given metric of the point space. Then, the determinant of
  the corresponding FSC matrix is zero or, equivalently, the cycle is
  self-orthogonal (isotropic):
  \(\scalar{\cycle{}{}}{\cycle[]{}{}}=0\). Naturally, such a cycle
  cannot be normalised to the form~\eqref{eq:det-normalisation-cond}.
\item Two quadrics are orthogonal in the point space
  \(\Space{R}{pq}\). Then, the matrices representing cycles are
  orthogonal in the sense of the inner
  product~\eqref{eq:cycle-product}.
\item Two cycles \(\cycle{}{}\) and \(\cycle[\tilde]{}{}\)
  are tangent. Then we have the following quadratic condition:
  \begin{equation}
    \label{eq:tangent-condition-defn}
    \scalar{\cycle{}{}}{\cycle[\tilde]{}{}}^2
    =  \scalar{\cycle{}{}}{\cycle{}{}}
    \scalar{\cycle[\tilde]{}{}}{\cycle[\tilde]{}{}}
    \qquad \left(\text{ or }
    \nscalar{\cycle{}{}}{\cycle[\tilde]{}{}}=\pm 1\right).
  \end{equation}
  With the assumption, that the cycle \(\cycle{}{}\) is normalised by
  the condition~\eqref{eq:det-normalisation-cond}, we may re-state
  this condition in the relation, which is linear to components of the cycle
  \(\cycle{}{}\):
  \begin{equation}
    \label{eq:tangent-condition-linear}
    \scalar{\cycle{}{}}{\cycle[\tilde]{}{}}
    = \pm \sqrt{\scalar{\cycle[\tilde]{}{}}{\cycle[\tilde]{}{}}}.
  \end{equation}
  Different signs here represent internal and outer touch.
\item Inversive distance \(\theta\) of two (non-isotropic) cycles is
  defined by the formula:
  \begin{equation}
    \label{eq:inversive-distance}
    \scalar{\cycle{}{}}{\cycle[\tilde]{}{}}
    = \theta \sqrt{
    \scalar{\cycle{}{}}{\cycle{}{}}}
  \sqrt{\scalar{\cycle[\tilde]{}{}}{\cycle[\tilde]{}{}}}
  \end{equation}
  In particular, the above discussed orthogonality corresponds to
  \(\theta=0\) and the tangency to \(\theta=\pm1\). For intersecting
  spheres \(\theta\) provides the cosine of the intersecting
  angle. For other metrics, the geometric interpretation of inversive
  distance shall be modified accordingly.

  If we are looking for a cycle \({\cycle{}{}}\) with a given
  inversive distance \(\theta\) to a given cycle
  \({\cycle[\tilde]{}{}}\), then the
  normalisation~\eqref{eq:det-normalisation-cond} again turns the
  defining relation~\eqref{eq:inversive-distance} into a linear with
  respect to parameters of the unknown cycle  \({\cycle{}{}}\).
\item A generalisation of Steiner power \(d\) of two cycles is defined
  as, cf.~\cite{FillmoreSpringer00a}*{\S~1.1}:
  \begin{equation}
    \label{eq:steiner-power}
    d=    \scalar{\cycle{}{}}{\cycle[\tilde]{}{}}
    + \sqrt{\scalar{\cycle{}{}}{\cycle{}{}}}
  \sqrt{\scalar{\cycle[\tilde]{}{}}{\cycle[\tilde]{}{}}},
  \end{equation}
  where both cycles \(\cycle{}{}\) and \(\cycle[\tilde]{}{}\) are
  \(k\)-normalised, that is the coefficient in front the quadratic
  term in~\eqref{eq:spheres-Rn} is \(1\). Geometrically, the
  generalised Steiner power for spheres provides the square of
  tangential distance. However, this relation is again non-linear for
  the cycle \(\cycle{}{}\).

  If we replace \(\cycle{}{}\) by the cycle
  \(\cycle{}{1}=\frac{1}{\sqrt{\scalar{\cycle{}{}}{\cycle{}{}}}}\cycle{}{}\)
  satisfying~\eqref{eq:det-normalisation-cond}, the
  identity~\eqref{eq:steiner-power} becomes:
  \begin{equation}
    \label{eq:steiner-power-linear}
    d\cdot k=    \scalar{\cycle{}{1}}{\cycle[\tilde]{}{}}
    +   \sqrt{\scalar{\cycle[\tilde]{}{}}{\cycle[\tilde]{}{}}},
  \end{equation}
  where \(k=\frac{1}{\sqrt{\scalar{\cycle{}{}}{\cycle{}{}}}}\) is the
  coefficient in front of the quadratic term of \(\cycle{}{1}\). The
  last identity is linear in terms of the coefficients of
  \(\cycle{}{1}\).
\end{enumerate}
Summing up: if an unknown cycle is connected to already given cycles
by any combination of the above relations, then all conditions can be
expressed as \emph{a system of linear equations for coefficients of the
unknown cycle and at most one quadratic
equation~\eqref{eq:det-normalisation-cond}}.

\nwenddocs{}\nwbegindocs{29}\nwdocspar
\subsection{Figures as families of cycles---functional approach}
\label{sec:figures-as-families}

We start from some examples of ensembles of cycles, which conveniently
describe FLT-invariant families of objects.

\begin{example}
  \label{ex:ensamble-math}
  \begin{enumerate}
  \item \label{it:poincare-extension}
    The Poincar\'e extension of M\"obius transformations from the
  real line to the upper half-plane of complex numbers is described by
  a triple of cycles \(\{\cycle{}{1}, \cycle{}{2}, \cycle{}{3}\}\)
  such that:
  \begin{enumerate}
  \item    \(\cycle{}{1}\) and \(\cycle{}{2}\) are orthogonal  to
    the real line;
  \item \(\scalar{\cycle{}{1}}{\cycle{}{2}}^2\leq
    \scalar{\cycle{}{1}}{\cycle{}{1}}
    \scalar{\cycle{}{2}}{\cycle{}{2}}\);
  \item \(\cycle{}{3}\) is orthogonal to any cycle in the triple
    including itself.
  \end{enumerate}
  A modification~\cites{Kisil14a} with ensembles of four cycles
  describes an extension from the real line to the upper half-plane of
  complex, dual or double numbers.  The construction can be
  generalised to arbitrary dimensions~\cite{Beardon95}.
\item\label{it:param-loxodromes}
  A parametrisation of loxodromes is provided by a triple of
  cycles \(\{\cycle{}{1}, \cycle{}{2}, \cycle{}{3}\}\) such
  that, cf.~\cite{KisilReid18a} and Fig.~\ref{fig:equiv-param-loxodr}: 
  \begin{enumerate}
  \item \(\cycle{}{1}\) is orthogonal to \(\cycle{}{2}\) and  \(\cycle{}{3}\);
  \item \(\scalar{\cycle{}{2}}{\cycle{}{3}}^2\geq
    \scalar{\cycle{}{2}}{\cycle{}{2}}
    \scalar{\cycle{}{3}}{\cycle{}{3}}\).
  \end{enumerate}
  Then, main invariant properties of M\"obius--Lie geometry,
  e.g. tangency of loxodromes, can be expressed in terms of this
  parametrisation~\cite{KisilReid18a}. 
\item A continued fraction is described by an infinite ensemble of
  cycles \((\cycle{}{k})\) such that~\cite{BeardonShort14a}:
  \begin{enumerate}
  \item All \(\cycle{}{k}\) are touching the real line (i.e. are
    \emph{horocycles});
  \item \((\cycle{}{1})\) is a horizontal line passing through
    \((0,1)\);
  \item \(\cycle{}{k+1}\) is tangent to \(\cycle{}{k}\) for all \(k>1\).
  \end{enumerate}
  This setup was extended in \cites{Kisil14a} to several similar
  ensembles. The key analytic properties of continued
  fractions---their convergence---can be linked to asymptotic
  behaviour of such an infinite ensemble~\cite{BeardonShort14a}.
\item A remarkable relation exists between discrete integrable systems
  and M\"obius geometry of finite configurations of
  cycles~\cites{BobenkoSchief18a,%
    KonopelchenkoSchief02a,KonopelchenkoSchief02b,%
    KonopelchenkoSchief05a,SchiefKonopelchenko09a}.  It comes from
  ``reciprocal force diagrams'' used in 19th-century statics, starting
  with J.C.~Maxwell. It is demonstrated in that the geometric
  compatibility of reciprocal figures corresponds to the algebraic
  compatibility of linear systems defining these configurations. On
  the other hand, the algebraic compatibility of linear systems lies
  in the basis of integrable systems. In
  particular~\cites{KonopelchenkoSchief02a,KonopelchenkoSchief02b},
  important integrability conditions encapsulate nothing but a
  fundamental theorem of ancient Greek geometry.
\item \label{it:wave-envelope}
  An important example of an infinite ensemble is provided by the
  representation of an arbitrary wave as the envelope of a continuous
  family of spherical waves. A finite subset of spheres can be used as
  an approximation to the infinite family. Then, discrete snapshots of
  time evolution of sphere wave packets represent a FLT-covariant
  ensemble of cycles~\cite{Bateman55a}. Further physical applications
  of FLT-invariant ensembles may be looked at~\cite{Kastrup08a}.
\end{enumerate}
\end{example}

One can easily note that the above parametrisations of some objects by
ensembles of cycles are not necessary unique. Naturally, two ensembles
parametrising  the same object are again connected by
FLT-invariant conditions. We presented only one example
here, cf.~\cite{KisilReid18a}.
\begin{figure}[htbp]
  \centering
    \animategraphics[controls=true,width=.9\textwidth,poster=first]{50}{_loxodromes}{1}{200}
    \caption[Equivalent parametrisation of a loxodrome]{Animated
      graphics of equivalent three-cycle parametrisations of a
      loxodrome. The green cycle is \(\cycle{}{1}\), two red circles are
      \(\cycle{}{2}\) and \(\cycle{}{3}\).}
  \label{fig:equiv-param-loxodr}
\end{figure}
\begin{example}
  Two non-degenerate triples \(\{\cycle{}{1},\cycle{}{2},\cycle{}{3}\}\) and
  \(\{\cycle[\tilde]{}{1},\cycle[\tilde]{}{2},\cycle[\tilde]{}{3}\}\) parameterise the same loxodrome as
  in Ex.~\ref{ex:ensamble-math}\ref{it:param-loxodromes} if and only if all the following
  conditions are satisfied:
  \begin{enumerate}
  \item \label{item:same-pencil}
    Pairs \(\{\cycle{}{2},\cycle{}{3}\}\) and \(\{\cycle[\tilde]{}{2},\cycle[\tilde]{}{3}\}\)  span the same
    hyperbolic pencil. That is cycles \(\cycle[\tilde]{}{2}\) and \(\cycle[\tilde]{}{3}\) are linear
    combinations of \(\cycle{}{2}\) and \(\cycle{}{3}\) and vise versa.
  \item \label{item:same-lambda}
    Pairs \(\{\cycle{}{2},\cycle{}{3}\}\) and \(\{\cycle[\tilde]{}{2},\cycle[\tilde]{}{3}\}\) have the same
    normalised cycle product~\eqref{eq:norm-cycle-prod}:
    \begin{equation}
      \label{eq:equal-lambdas}
      \nscalar {\cycle{}{2}}{\cycle{}{3}}=\nscalar {\cycle[\tilde]{}{2}}{\cycle[\tilde]{}{3}}.
    \end{equation}
  \item \label{item:ellipt-hyperb-ident}
    The elliptic-hyperbolic identity holds:
    \begin{equation}
      \label{eq:ellipt-hyperb-equat}
      \frac{\arccosh\nscalar {\cycle{}{j}}{\cycle[\tilde]{}{j}}}{\arccosh\nscalar{\cycle{}{2}}{\cycle{}{3}}}
      \equiv
      \frac{1}{2\pi}\arccos\nscalar {\cycle{}{1}}{\cycle[\tilde]{}{1}} \pmod{1}\,,
    \end{equation}
    where \(j\) is either \(2\) or \(3\).
  \end{enumerate}
  Various triples of cycles parametrising the same loxodrome are
  animated on Fig.~\ref{fig:equiv-param-loxodr}.
\end{example}
The respective equivalence relation for parametrisation of Poincar\'e
extension from Ex.~\ref{ex:ensamble-math}\ref{it:poincare-extension} is provided
in~\cite{Kisil15a}*{Prop.~12}.  These examples suggest that one can
expand the subject and applicability of M\"obius--Lie geometry through
the following formal definition.
\begin{defn}
  \label{de:extended-Lie-Moebius}
  Let \(X\) be a set, \(R \subset X\times X\) be an oriented graph
  on \(X\) 
  and \(f\) be a function on \(R\) with values in FLT-invariant
  relations from \S~\ref{sec:conn-quadr-cycl}. Then
  \emph{\((R,f)\)-ensemble} is a collection of cycles
  \(\{C_j\}_{j\in X}\) such that
  \begin{displaymath}
    C_i \text { and } C_j \text{ are in the relation } f(i,j) \text {
      for all } (i,j)\in R.
  \end{displaymath}
  For a fixed FLT-invariant equivalence 
  relations \(\sim\) on the set \(\mathcal{E}\) of all \((R,f)\)-ensembles,
  \emph{the extended M\"obius--Lie geometry} studies properties of cosets
  \(\mathcal{E}/\sim\).
\end{defn}
This definition can be suitably modified for
\begin{enumerate}
\item ensembles with relations of more then two cycles; and/or
\item ensembles parametrised by continuous sets \(X\), cf. wave
  envelopes in Ex.~\ref{ex:ensamble-math}\ref{it:wave-envelope}.
\end{enumerate}

The above extension was developed along with the realisation
the library {\Tt{}\Rm{}{\bf{}figure}\nwendquote} within the \emph{functional programming}
framework. More specifically, an object from the {\Tt{}\Rm{}{\bf{}class} \ {\bf{}figure}\nwendquote} stores
defining relations, which link new cycles to the previously introduced
ones. This also may be treated as classical geometric
compass-and-straightedge constructions, where new lines or circles are
drawn through already existing elements. If requested, an explicit
evaluation of cycles\vtick\ parameters from this data may be
attempted.

To avoid ``chicken or the egg'' dilemma all cycles are stored in a
hierarchical structure of generations, numbered by integers. The basic
principles are:
\begin{enumerate}
\item Any explicitly defined cycle (i.e., a cycle which is not related to any
  previously known cycle) is placed into generation-0;
\item Any new cycle defined by relations to \emph{previous} cycles
  from generations \(k_1\), \(k_2\), \ldots, \(k_n\) is placed to the
  generation \(k\) calculated as:
  \begin{equation}
    \label{eq:generation-calculation}
    k=\max(k_1,k_2,\ldots,k_n)+1 .
  \end{equation}
  This rule does not forbid a cycle to have a relation to itself,
  e.g. isotropy (self-orthogonality) condition
  \(\scalar{\cycle{}{}}{\cycle{}{}}=0\), which specifies point-like
  cycles, cf. relation~\ref{it:point-zero-radius} in~\S~\ref{sec:conn-quadr-cycl}.  In fact, this is the only
  allowed type of relations to cycles in the same (not even speaking
  about younger) generations.
\end{enumerate}
There are the following alterations of the above rules:
\begin{enumerate}
\item From the beginning, every figure has two pre-defined cycles: the
  real line (hyperplane) \(\cycle{}{\Space{R}{}}\), and the zero radius cycle at infinity
  \(\cycle{}{\infty}=(0,0,1)\). These cycles are required for
  relations~\ref{item:quadric-flat} and~\ref{it:lobachevski-line} from
  the previous subsection. As predefined cycles, \(\cycle{}{\Space{R}{}}\) and
  \(\cycle{}{\infty}\) are placed in negative-numbered
  generations defined by the macros {\Tt{}\Rm{}{\it{}REAL\_LINE\_GEN}\nwendquote} and
  {\Tt{}\Rm{}{\it{}INFINITY\_GEN}\nwendquote}.
\item If a point is added to generation-0 of a figure, then it is
  represented by a zero-radius cycle with its centre at the given
  point. Particular parameter of such cycle dependent on the used
  metric, thus this cycle is not considered as explicitly
  defined. Thereafter, the cycle shall have some parents at a
  negative-numbered generation defined by the macro {\Tt{}\Rm{}{\it{}GHOST\_GEN}\nwendquote}.
\end{enumerate}
A figure can be in two different modes: {\Tt{}\Rm{}{\it{}freeze}\nwendquote} or {\Tt{}\Rm{}{\it{}unfreeze}\nwendquote},
the second is default. In the {\Tt{}\Rm{}{\it{}unfreeze}\nwendquote} mode an addition of a new
cycle by its relation prompts an evaluation of its parameters. If the
evaluation was successful then the obtained parameters are stored and
will be used in further calculations for all children of the cycle. Since many
relations (see the previous Subsection) are connected to quadratic
equation~\eqref{eq:det-normalisation-cond}, the solutions may come in
pairs. Furthermore, if the number or nature of conditions is not
sufficient to define the cycle uniquely (up to natural quadratic
multiplicity), then the cycle will depend on a number of free
(symbolic) variable.

There is a macro-like tool, which is called {\Tt{}\Rm{}{\bf{}subfigure}\nwendquote}. Such a
{\Tt{}\Rm{}{\bf{}subfigure}\nwendquote} is a {\Tt{}\Rm{}{\bf{}figure}\nwendquote} itself, such that its inner hierarchy of
generations and relations is not visible from the current
{\Tt{}\Rm{}{\bf{}figure}\nwendquote}. Instead, some cycles (of any generations) of the current
{\Tt{}\Rm{}{\bf{}figure}\nwendquote} are used as predefined cycles of generation-0 of
{\Tt{}\Rm{}{\bf{}subfigure}\nwendquote}. Then only one dependent cycle of {\Tt{}\Rm{}{\bf{}subfigure}\nwendquote}, which
is known as result, is returned back to the current {\Tt{}\Rm{}{\bf{}figure}\nwendquote}. The
generation of the result is calculated from generations of input
cycles by the same formula~\eqref{eq:generation-calculation}.

There is a possibility to test certain conditions (``are two cycles
orthogonal?'') or measure certain quantities (``what is their
intersection angle?'') for already defined cycles. In particular, such
methods can be used to prove geometrical statements according to the
Cartesian programme, that is replacing the synthetic geometry by
purely algebraic manipulations.
\begin{example}
  \label{ex:touch-centres-collinear}
  As an elementary demonstration, let us prove that if a cycle {\Tt{}\Rm{}{\it{}r}\nwendquote}
  is orthogonal to a circle {\Tt{}\Rm{}{\it{}a}\nwendquote} at the point {\Tt{}\Rm{}{\it{}C}\nwendquote} of its contact with a
  tangent line {\Tt{}\Rm{}{\it{}l}\nwendquote}, then {\Tt{}\Rm{}{\it{}r}\nwendquote} is also orthogonal to the line
  {\Tt{}\Rm{}{\it{}l}\nwendquote}. To simplify setup we assume that {\Tt{}\Rm{}{\it{}a}\nwendquote} is the unit
  circle. Here is the {\Python} code:
\begin{lstlisting}
F=figure()
a=F.add_cycle(cycle2D(1,[0,0],-1),"a")
l=symbol("l")
C=symbol("C")
F.add_cycle_rel([is_tangent_i(a),is_orthogonal(F.get_infinity()),only_reals(l)],l)
F.add_cycle_rel([is_orthogonal(C),is_orthogonal(a),is_orthogonal(l),only_reals(C)],C)
r=F.add_cycle_rel([is_orthogonal(C),is_orthogonal(a)],"r")
Res=F.check_rel(l,r,"cycle_orthogonal")
for i in range(len(Res)):
    print "Tangent and radius are orthogonal: %s" %\
    bool(Res[i].subs(pow(cos(wild(0)),2)==1-pow(sin(wild(0)),2)).normal())
\end{lstlisting}
The first line creates an empty figure {\Tt{}\Rm{}{\it{}F}\nwendquote} with the default
euclidean metric. The next line explicitly uses parameters
\((1,0,0,-1)\) of {\Tt{}\Rm{}{\it{}a}\nwendquote} to add it to {\Tt{}\Rm{}{\it{}F}\nwendquote}. Lines~3--4 define symbols
{\Tt{}\Rm{}{\it{}l}\nwendquote} and {\Tt{}\Rm{}{\it{}C}\nwendquote}, which are needed because cycles with these labels are
defined in lines~5--6 through some relations to themselves and the
cycle {\Tt{}\Rm{}{\it{}a}\nwendquote}. In both cases we want to have cycles with real
coefficients only and {\Tt{}\Rm{}{\it{}C}\nwendquote} is additionally self-orthogonal (i.e. is a
zero-radius). Also, {\Tt{}\Rm{}{\it{}l}\nwendquote} is orthogonal to infinity (i.e. is a line)
and {\Tt{}\Rm{}{\it{}C}\nwendquote} is orthogonal to {\Tt{}\Rm{}{\it{}a}\nwendquote} and {\Tt{}\Rm{}{\it{}l}\nwendquote} (i.e. is their common
point). The tangency condition for {\Tt{}\Rm{}{\it{}l}\nwendquote} and self-orthogonality of
{\Tt{}\Rm{}{\it{}C}\nwendquote} are both quadratic relations.  The former has two solutions each
depending on one real parameter, thus line {\Tt{}\Rm{}{\it{}l}\nwendquote} has two
instances. Correspondingly, the point of contact {\Tt{}\Rm{}{\it{}C}\nwendquote} and the
orthogonal cycle {\Tt{}\Rm{}{\it{}r}\nwendquote} through {\Tt{}\Rm{}{\it{}C}\nwendquote} (defined in line~7) each have two
instances as well. Finally, lines~8--11 verify that every instance of
{\Tt{}\Rm{}{\it{}l}\nwendquote} is orthogonal to the respective instance of {\Tt{}\Rm{}{\it{}r}\nwendquote}, this is
assisted by the trigonometric substitution \(\cos^2(*)=1-\sin^2(*)\)
used for parameters of {\Tt{}\Rm{}{\it{}l}\nwendquote} in line~11.  The output predictably is:
\begin{verbatim}
Tangent and circle r are orthogonal: True
Tangent and circle r are orthogonal: True
\end{verbatim}
\end{example}
An original statement proved by the library {\Tt{}\Rm{}{\bf{}figure}\nwendquote} for the first
time will be considered in the next Section.

\nwenddocs{}\nwbegindocs{30}\nwdocspar
\section{Mathematical Usage of the Library}
\label{sec:mathematical-results}

The developed library {\Tt{}\Rm{}{\bf{}figure}\nwendquote} has several different uses:
\begin{itemize}
\item It is easy to produce high-quality illustrations, which are
  fully-accurate in mathematical sence. The user is not responsible
  for evaluation of cycles\vtick\ parameters, all computations are
  done by the library as soon as the figure is defined in terms of few
  geometrical relations. This is especially helpful for complicated
  images which may contain thousands of interrelated cycles. See
  Escher-like Fig.~\ref{fig:action-modular-group} which shows images
  of two circles under the modular group
  action~\cite{StewartTall02a}*{\S~14.4}\ifshort\else,
  cf.~\ref{sec:an-illustr-modul}\fi.
\begin{figure}[htbp]
  \centering
  \includegraphics[,width=.9\textwidth]{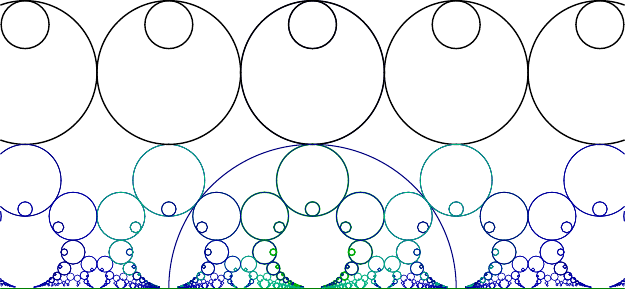}
  \caption{Action of the modular group on the upper half-plane.}
  \label{fig:action-modular-group}
\end{figure}
\begin{figure}[htbp]
  \centering
  \includegraphics[scale=.5]{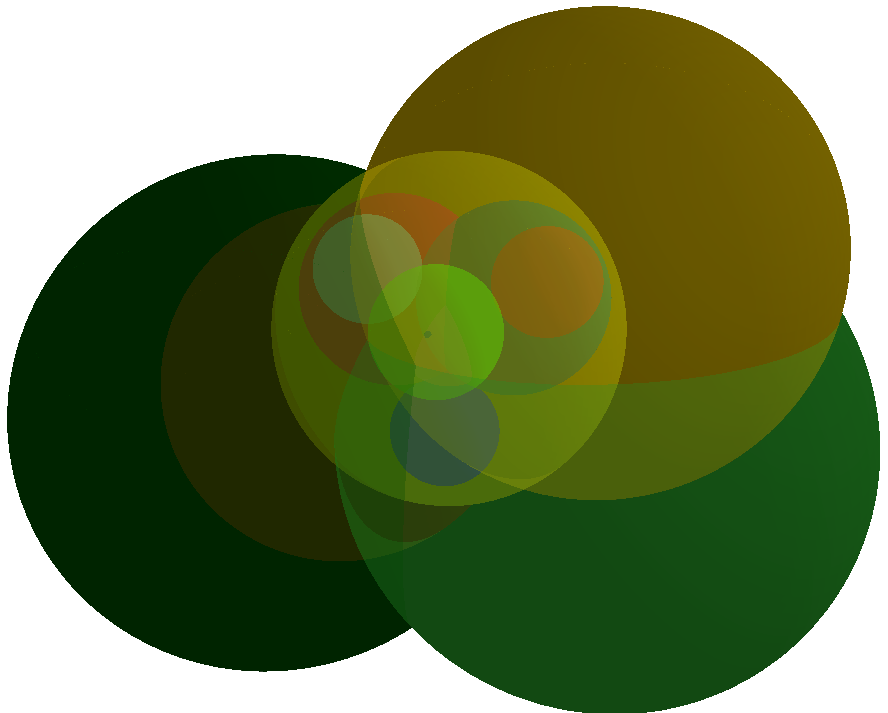}
  \caption{An example of Apollonius problem in three dimensions.}
  \label{fig:apollonius-3D}
\end{figure}
\item The package can be used for computer experiments in M\"obius--Lie
  geometry. There is a possibility to create an arrangement of cycles
  depending on one or several parameters. Then, for particular values
  of those parameters certain conditions, e.g. concurrency of cycles,
  may be numerically tested or graphically visualised. It is possible
  to create animations with gradual change of the parameters, which
  are especially convenient for illustrations, see
  Fig.~\ref{fig:nine-points-anim} and~\cite{Kisil16a}.
\item Since the library is based on the \GiNaC\ system, which provides a
  symbolic computation engine, there is a possibility to make fully
  automatic proofs of various statements in M\"obius--Lie geometry.  Usage
  of computer-supported proofs in geometry is already an established
  practice~\cites{Kisil12a,Pech07a} and it is naturally to expect its further
  rapid growth.
\item Last but not least, the combination of classical beauty of Lie
  sphere geometry and modern computer technologies is a useful
  pedagogical tool to widen interest in mathematics through visual and
  hands-on experience.
\end{itemize}

Computer experiments are especially valuable for Lie geometry of
indefinite or nilpotent metrics since our intuition is not elaborated
there in contrast to the Euclidean
space~\cites{Kisil07a,Kisil06a,Kisil05a}. Some advances in
the two-dimensional space were achieved
recently~\cites{Mustafa17a,Kisil12a}, however further developments in
higher dimensions are still awaiting their researchers.

\begin{figure}[htbp]
  \centering
  \ifelsevier
  \makebox[0pt][l]{(a)}\includegraphics[scale=.6]{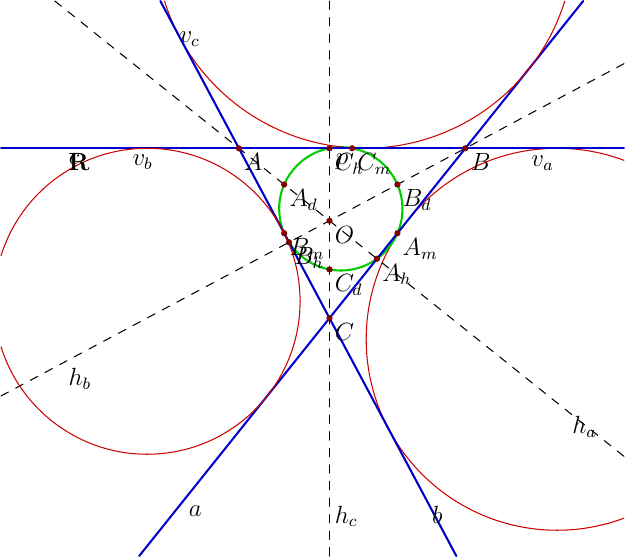}\hfill
  \makebox[0pt][l]{(b)}\includegraphics[scale=.6]{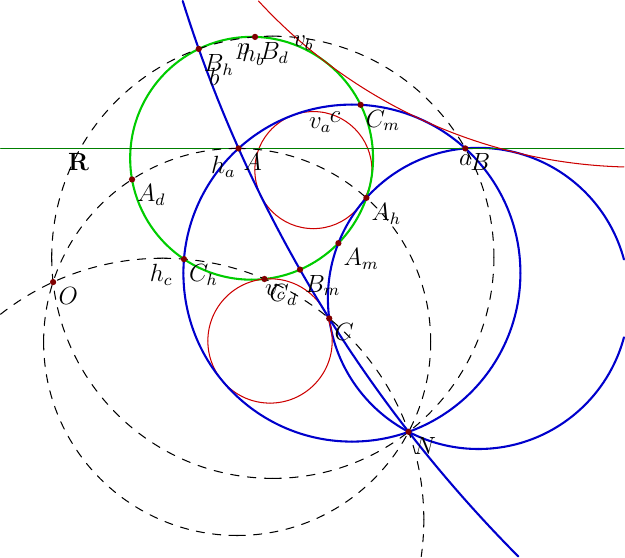}\\
  \makebox[0pt][l]{(c)}\includegraphics[scale=.6]{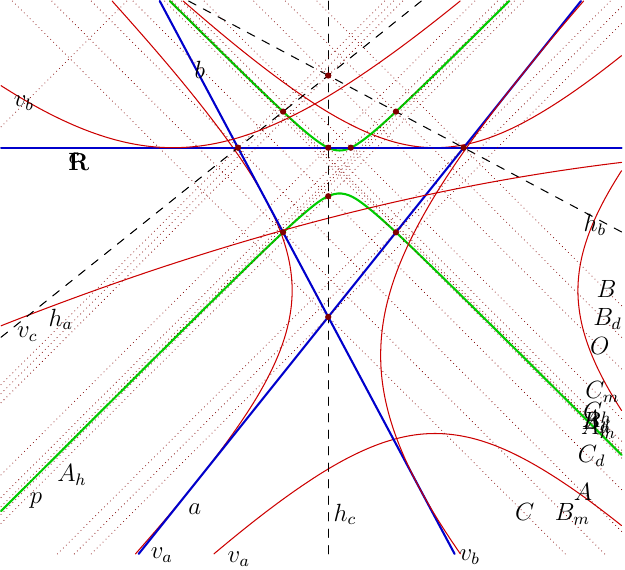}\hfill
  \makebox[0pt][l]{(d)}\includegraphics[scale=.6]{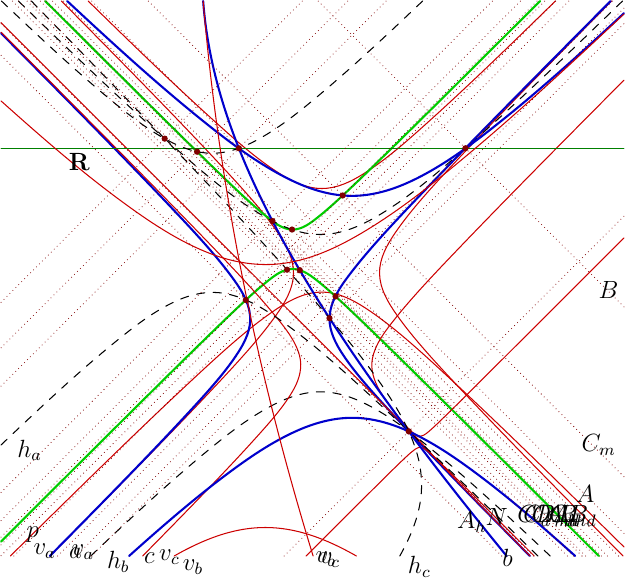}
  \else
  \ifshort
  \makebox[0pt][l]{(a)}\includegraphics[scale=.58]{nine-points-thm-plain.pdf}\hfill
  \makebox[0pt][l]{(b)}\includegraphics[scale=.58]{nine-points-thm.pdf}\\[1em]
  \makebox[0pt][l]{(c)}\includegraphics[scale=.58]{nine-points-thm-plain-hyp.pdf}\hfill
  \makebox[0pt][l]{(d)}\includegraphics[scale=.58]{nine-points-thm-hyp.pdf}
  \else
  \makebox[0pt][l]{(a)}\includegraphics[scale=.8]{nine-points-thm-plain.pdf}\hfill
  \makebox[0pt][l]{(b)}\includegraphics[scale=.8]{nine-points-thm.pdf}\\
  \makebox[0pt][l]{(c)}\includegraphics[scale=.8]{nine-points-thm-plain-hyp.pdf}\hfill
  \makebox[0pt][l]{(d)}\includegraphics[scale=.8]{nine-points-thm-hyp.pdf}
  \fi
  \fi
  \caption[The illustration of the conformal nine-points theorem]
  {The illustration of the conformal nine-points theorem. The
    left column is the statement for a triangle with straight sides
    (the point {\Tt{}\Rm{}{\it{}N}\nwendquote} is at infinity), the right column is its
    conformal version (the point {\Tt{}\Rm{}{\it{}N}\nwendquote} is at the finite part). The
    first row show the elliptic point space, the second row---the
    hyperbolic point space. Thus, the top-left picture shows the
    traditional theorem, three other pictures---its different modifications.}
  \label{fig:illustr-conf-nine}
\end{figure}

As a non-trivial example of automated proof accomplished by the {\Tt{}\Rm{}{\bf{}figure}\nwendquote}
library for the first time, we present a FLT-invariant version of the
classical nine-point theorem~\citelist{\cite{Pedoe95a}*{\S~I.1}
  \cite{CoxeterGreitzer}*{\S~1.8}},
cf. Fig.~\ref{fig:illustr-conf-nine}(a):
\begin{thm}[Nine-point cycle]
  \label{th:nine-points}
  Let \(ABC\) be an arbitrary triangle with the orthocenter (the
  points of intersection of three altitudes) \(H\), then
  the following nine points belongs to the same cycle, which may be a
  circle or a hyperbola:
  \begin{enumerate}
    \item Foots of three altitudes, that is points of pair-wise
      intersections \(AB\) and \(CH\), \(AC\) and \(BH\), \(BC\) and \(AH\).
    \item Midpoints of sides \(AB\), \(BC\) and \(CA\).
    \item Midpoints of intervals \(AH\), \(BH\) and \(CH\).
  \end{enumerate}
\end{thm}
There are many further interesting properties,
e.g. nine-point circle is externally tangent to that triangle three
excircles and internally tangent to its incircle as it seen from
Fig.~\ref{fig:illustr-conf-nine}(a).

To adopt the statement for cycles geometry we need to find a
FLT-invariant meaning of the midpoint \(A_m\) of an interval \(BC\),
because the equality of distances \(BA_m\) and \(A_mC\) is not
FLT-invariant. The definition in cycles geometry can be done by either
of the following equivalent relations:
\begin{itemize}
\item The midpoint \(A_m\) of an interval \(BC\) is defined by the
  cross-ratio \(\frac{BA_m}{CA_m} : \frac{BI}{CI}=1\), where \(I\) is
  the point at infinity.
\item We construct the midpoint \(A_m\) of an interval \(BC\) as the
  intersection of the interval and the line orthogonal to \(BC\) and
  to the cycle, which uses \(BC\) as its diameter. The latter
  condition means that the cycle passes both points \(B\) and \(C\)
  and is orthogonal to the line \(BC\).
\end{itemize}
Both procedures are meaningful if we replace the point at infinity
\(I\) by an arbitrary fixed point \(N\) of the plane. In the second
case all lines will be replaced by cycles passing through \(N\), for
example the line through \(B\) and \(C\) shall be replaced by a cycle
through \(B\), \(C\) and \(N\). If we similarly replace ``lines'' by
``cycles passing through \(N\)'' in Thm.~\ref{th:nine-points} it turns
into a valid FLT-invariant version,
cf. Fig.~\ref{fig:illustr-conf-nine}(b). Some additional properties,
e.g. the tangency of the nine-points circle to the ex-/in-circles, are
preserved in the new version as well.  Furthermore, we can illustrate
the connection between two versions of the theorem by an animation,
where the infinity is transformed to a finite point \(N\) by a
continuous one-parameter group of FLT,
see. Fig.~\ref{fig:nine-points-anim} and further examples
at~\cite{Kisil16a}.

It is natural to test the nine-point theorem in the hyperbolic and the
parabolic spaces. Fortunately, it is very easy under the given
implementation: we only need to change the defining metric of the
point space, this can be done for an already defined figure\ifshort\else,
see~\ref{sec:example:-nine-points}\fi. The corresponding figures
Fig.~\ref{fig:illustr-conf-nine}(c) and~(d) suggest that the
hyperbolic version of the theorem is still true in the plain and even
FLT-invariant forms. We shall clarify that the hyperbolic version
of the Thm.~\ref{th:nine-points} specialises the nine-point conic of a
complete quadrilateral \cites{CerinGianella06a,DeVilliers06a}: in
addition to the existence of this conic, our theorem specifies its
type for this particular arrangement as equilateral hyperbola with the
vertical axis of symmetry.

\begin{figure}[htbp]
  \centering
  \animategraphics[controls=true,width=.9\textwidth]{50}{_nine-points-anim}{}{}
  \caption[Animated transition between the classical and conformal
  nine-point theorems]{Animated transition between the classical and
    conformal versions of the nine-point theorem. Use control buttons to activate
    it. You may need \textsf{Adobe Acrobat Reader} for this feature.}
  \label{fig:nine-points-anim}
\end{figure}

The computational power of the package is sufficient not only to hint
that the new theorem is true but also to make a complete proof. To
this end we define an ensemble of cycles with exactly same
interrelations, but populate the generation-0 with points \(A\), \(B\)
and \(C\) with symbolic coordinates, that is, objects of the \GiNaC\
{\Tt{}\Rm{}{\bf{}class}\ {\bf{}realsymbol}\nwendquote}. Thus, the entire figure defined from them will
be completely general. Then, we may define the hyperbola passing
through three bases of altitudes and check by the symbolic
computations that this hyperbola passes another six ``midpoints'' as
well\ifshort\else, see~\ref{sec:prov-theor-symb}
\fi.

In the parabolic space the nine-point Thm.~\ref{th:nine-points} is not
preserved in this manner. It is already
observed~\cites{Kisil12a,Kisil05a,%
  Kisil15a,Kisil07a,Kisil09e,Kisil11a,Mustafa17a,BarrettBolt10a}, that
the degeneracy of parabolic metric in the point space requires certain
revision of traditional definitions. The parabolic variation of
nine-point theorem may prompt some further considerations as well. An
expanded discussion of various aspects of the nine-point construction
shall be the subject of a separate paper.

\section{To Do List}
\label{sec:do}

The library is still under active development. Along with continuous
bug fixing there is an intention to extend both functionality and
usability. Here are several nearest tasks planned so far:
\begin{itemize}
\item Expand class {\Tt{}\Rm{}{\bf{}subfigure}\nwendquote} in way suitable for encoding
  loxodromes and other objects of an  extended M\"obius--Lie
  geometry~\cites{KisilReid18a,Kisil15a}.
\item Add non-point transformations, extending the package to Lie
  sphere geometry.
\item  Add a method which will apply a FLT
  to the entire figure.
\item Provide an effective parametrisation of solutions of a single
  quadratics condition.
\item Expand drawing facilities in three dimensions to hyperboloids
  and paraboloids.
\item Create  Graphical User Interface which will open the library
  to users without programming skills.
\item Investigate cloud computing options which can free a user from
  the burden of installation.
\end{itemize}
Being an open-source project the library is open for contributions and
suggestions of other developers and users.

\nwenddocs{}\nwbegindocs{31}\nwdocspar
\section*{Acknowledgement}
\label{sec:acknowledgement}

I am grateful to Prof.~Jay~P.~Fillmore for stimulating discussion,
which enriched the library {\Tt{}\Rm{}{\bf{}figure}\nwendquote}. Cameron Kumar wrote
\href{https://sourceforge.net/projects/cycle3dvis.moebinv.p/}{cycle3D-visualiser}.

\nwenddocs{}\nwbegindocs{32}\nwdocspar
\bibliography{arare,aclifford,abbrevmr,akisil,ageometry,algebra,analyse,aphysics}

\nwenddocs{}\nwbegindocs{33}\nwdocspar
\ifshort
\else
\newpage
\appendix

\nwenddocs{}\nwbegindocs{34}\nwdocspar
\textheight 26.5cm
\textwidth 18cm
\oddsidemargin -.9cm
\evensidemargin -.9cm
\topmargin -1.5cm
\renewcommand{\baselinestretch}{1}

\nwenddocs{}\nwbegindocs{35}\nwdocspar
\section{Examples of Usage}
\label{sec:examples}
This section presents several examples, which may be used for quick
start. We begin with very elementary one, but almost all aspects of
the library usage will be illustrated by the end of this section. See
the beginning of Section~\ref{sec:publ-meth-figure} for installation
advise. The collection of these programmes is also serving as a test
suit for the library.

\nwenddocs{}\nwbegindocs{36}\nwdocspar
\nwenddocs{}\nwbegincode{37}\sublabel{NWppJ6t-1fWRR8-1}\nwmargintag{{\nwtagstyle{}\subpageref{NWppJ6t-1fWRR8-1}}}\moddef{separating chunk~{\nwtagstyle{}\subpageref{NWppJ6t-1fWRR8-1}}}\endmoddef\Rm{}\nwstartdeflinemarkup\nwprevnextdefs{\relax}{NWppJ6t-1fWRR8-2}\nwenddeflinemarkup

\nwalsodefined{\\{NWppJ6t-1fWRR8-2}\\{NWppJ6t-1fWRR8-3}\\{NWppJ6t-1fWRR8-4}\\{NWppJ6t-1fWRR8-5}\\{NWppJ6t-1fWRR8-6}\\{NWppJ6t-1fWRR8-7}\\{NWppJ6t-1fWRR8-8}\\{NWppJ6t-1fWRR8-9}}\nwnotused{separating chunk}\nwendcode{}\nwbegindocs{38}\nwdocspar
\subsection{Hello, Cycle!}
\label{sec:hello-cycle}

This is a minimalist example showing how to obtain a simple drawing of
cycles in non-Euclidean geometry. Of course, we are starting from the
library header file.
\nwenddocs{}\nwbegincode{39}\sublabel{NWppJ6t-3rkk58-1}\nwmargintag{{\nwtagstyle{}\subpageref{NWppJ6t-3rkk58-1}}}\moddef{hello-cycle.cpp~{\nwtagstyle{}\subpageref{NWppJ6t-3rkk58-1}}}\endmoddef\Rm{}\nwstartdeflinemarkup\nwprevnextdefs{\relax}{NWppJ6t-3rkk58-2}\nwenddeflinemarkup
\LA{}license~{\nwtagstyle{}\subpageref{NWppJ6t-ZXuKx-1}}\RA{}
{\bf{}\char35{}include}{\tt{} "figure.h"}
\LA{}using all namespaces~{\nwtagstyle{}\subpageref{NWppJ6t-3uHj2c-1}}\RA{}
{\bf{}int} {\it{}main}(){\nwlbrace}\nwindexdefn{\nwixident{main}}{main}{NWppJ6t-3rkk58-1}

\nwalsodefined{\\{NWppJ6t-3rkk58-2}\\{NWppJ6t-3rkk58-3}\\{NWppJ6t-3rkk58-4}\\{NWppJ6t-3rkk58-5}}\nwnotused{hello-cycle.cpp}\nwidentdefs{\\{{\nwixident{main}}{main}}}\nwidentuses{\\{{\nwixident{figure}}{figure}}}\nwindexuse{\nwixident{figure}}{figure}{NWppJ6t-3rkk58-1}\nwendcode{}\nwbegindocs{40}To save keystrokes, we use the following {\Tt{}\Rm{}{\bf{}namespace}\nwendquote}s.
\nwenddocs{}\nwbegincode{41}\sublabel{NWppJ6t-3uHj2c-1}\nwmargintag{{\nwtagstyle{}\subpageref{NWppJ6t-3uHj2c-1}}}\moddef{using all namespaces~{\nwtagstyle{}\subpageref{NWppJ6t-3uHj2c-1}}}\endmoddef\Rm{}\nwstartdeflinemarkup\nwusesondefline{\\{NWppJ6t-3rkk58-1}\\{NWppJ6t-1kKRnF-1}\\{NWppJ6t-2SYluR-1}\\{NWppJ6t-2t2vfS-1}\\{NWppJ6t-30Tc7D-1}\\{NWppJ6t-2zWvnC-1}\\{NWppJ6t-RDDRz-1}\\{NWppJ6t-2pgnFW-1}}\nwenddeflinemarkup
{\bf{}using} {\bf{}namespace} {\it{}std};
{\bf{}using} {\bf{}namespace} {\it{}GiNaC};
{\bf{}using} {\bf{}namespace} {\it{}MoebInv};
\nwindexdefn{\nwixident{MoebInv}}{MoebInv}{NWppJ6t-3uHj2c-1}\eatline
\nwused{\\{NWppJ6t-3rkk58-1}\\{NWppJ6t-1kKRnF-1}\\{NWppJ6t-2SYluR-1}\\{NWppJ6t-2t2vfS-1}\\{NWppJ6t-30Tc7D-1}\\{NWppJ6t-2zWvnC-1}\\{NWppJ6t-RDDRz-1}\\{NWppJ6t-2pgnFW-1}}\nwidentdefs{\\{{\nwixident{MoebInv}}{MoebInv}}}\nwendcode{}\nwbegindocs{42}\nwdocspar
\nwenddocs{}\nwbegindocs{43}We declare the figure {\Tt{}\Rm{}{\it{}F}\nwendquote} which will be constructed with the
default elliptic metric in two dimensions.
\nwenddocs{}\nwbegincode{44}\sublabel{NWppJ6t-3rkk58-2}\nwmargintag{{\nwtagstyle{}\subpageref{NWppJ6t-3rkk58-2}}}\moddef{hello-cycle.cpp~{\nwtagstyle{}\subpageref{NWppJ6t-3rkk58-1}}}\plusendmoddef\Rm{}\nwstartdeflinemarkup\nwprevnextdefs{NWppJ6t-3rkk58-1}{NWppJ6t-3rkk58-3}\nwenddeflinemarkup
    {\bf{}figure} {\it{}F};
\nwindexdefn{\nwixident{figure}}{figure}{NWppJ6t-3rkk58-2}\eatline
\nwidentdefs{\\{{\nwixident{figure}}{figure}}}\nwendcode{}\nwbegindocs{45}\nwdocspar
\nwenddocs{}\nwbegindocs{46}Next we define a couple of points {\Tt{}\Rm{}{\it{}A}\nwendquote} and {\Tt{}\Rm{}{\it{}B}\nwendquote}. Every point is
added to {\Tt{}\Rm{}{\it{}F}\nwendquote} by giving its explicit coordinates as a {\Tt{}\Rm{}{\bf{}lst}\nwendquote} and a
string, which will be used to label the point. The returned value is a
\GiNaC\ expression of {\Tt{}\Rm{}{\bf{}symbol}\nwendquote} class, which will be used as a key
of the respective point. All points are added to the zero generation.
\nwenddocs{}\nwbegincode{47}\sublabel{NWppJ6t-3rkk58-3}\nwmargintag{{\nwtagstyle{}\subpageref{NWppJ6t-3rkk58-3}}}\moddef{hello-cycle.cpp~{\nwtagstyle{}\subpageref{NWppJ6t-3rkk58-1}}}\plusendmoddef\Rm{}\nwstartdeflinemarkup\nwprevnextdefs{NWppJ6t-3rkk58-2}{NWppJ6t-3rkk58-4}\nwenddeflinemarkup
    {\bf{}ex} {\it{}A}={\it{}F}.{\it{}add\_point}({\bf{}lst}{\nwlbrace}-1,.5{\nwrbrace},{\tt{}"A"});
    {\bf{}ex} {\it{}B}={\it{}F}.{\it{}add\_point}({\bf{}lst}{\nwlbrace}1,1.5{\nwrbrace},{\tt{}"B"});
\nwindexdefn{\nwixident{add{\_}point}}{add:unpoint}{NWppJ6t-3rkk58-3}\eatline
\nwidentdefs{\\{{\nwixident{add{\_}point}}{add:unpoint}}}\nwidentuses{\\{{\nwixident{ex}}{ex}}}\nwindexuse{\nwixident{ex}}{ex}{NWppJ6t-3rkk58-3}\nwendcode{}\nwbegindocs{48}\nwdocspar
\nwenddocs{}\nwbegindocs{49}Now we add a ``line'' in the Lobachevsky half-plane. It passes both
points {\Tt{}\Rm{}{\it{}A}\nwendquote} and {\Tt{}\Rm{}{\it{}B}\nwendquote} and is orthogonal to the real line. The real
line and the point at infinity were automatically added to {\Tt{}\Rm{}{\it{}F}\nwendquote} at
its initialisation. The real line is accessible as
{\Tt{}\Rm{}{\it{}F}.{\it{}get\_real\_line}()\nwendquote} method in {\Tt{}\Rm{}{\bf{}figure}\nwendquote} class. A cycle
passes a point if it is orthogonal to the cycle defined by this
point. Thus, the line is defined through a list of three
orthogonalities~\citelist{\cite{Kisil06a} \cite{Kisil12a}*{Defn.~6.1}}
(other pre-defined relations between cycles are listed in
Section~\ref{sec:publ-meth-cycl}). We also supply a string to label
this cycle. The returned valued is a {\Tt{}\Rm{}{\bf{}symbol}\nwendquote}, which is a key for this
cycle.
\nwenddocs{}\nwbegincode{50}\sublabel{NWppJ6t-3rkk58-4}\nwmargintag{{\nwtagstyle{}\subpageref{NWppJ6t-3rkk58-4}}}\moddef{hello-cycle.cpp~{\nwtagstyle{}\subpageref{NWppJ6t-3rkk58-1}}}\plusendmoddef\Rm{}\nwstartdeflinemarkup\nwprevnextdefs{NWppJ6t-3rkk58-3}{NWppJ6t-3rkk58-5}\nwenddeflinemarkup
    {\bf{}ex} {\it{}a}={\it{}F}.{\it{}add\_cycle\_rel}({\bf{}lst}{\nwlbrace}{\it{}is\_orthogonal}({\it{}A}),{\it{}is\_orthogonal}({\it{}B}),{\it{}is\_orthogonal}({\it{}F}.{\it{}get\_real\_line}()){\nwrbrace},{\tt{}"a"});
\nwindexdefn{\nwixident{add{\_}cycle{\_}rel}}{add:uncycle:unrel}{NWppJ6t-3rkk58-4}\nwindexdefn{\nwixident{get{\_}real{\_}line}}{get:unreal:unline}{NWppJ6t-3rkk58-4}\eatline
\nwidentdefs{\\{{\nwixident{add{\_}cycle{\_}rel}}{add:uncycle:unrel}}\\{{\nwixident{get{\_}real{\_}line}}{get:unreal:unline}}}\nwidentuses{\\{{\nwixident{ex}}{ex}}\\{{\nwixident{is{\_}orthogonal}}{is:unorthogonal}}}\nwindexuse{\nwixident{ex}}{ex}{NWppJ6t-3rkk58-4}\nwindexuse{\nwixident{is{\_}orthogonal}}{is:unorthogonal}{NWppJ6t-3rkk58-4}\nwendcode{}\nwbegindocs{51}\nwdocspar
\nwenddocs{}\nwbegindocs{52}Now, we draw our figure to a file. Its format (e.g. EPS, PDF, PNG,
etc.) is determined by your default \Asymptote settings. This can be
overwritten if a format is explicitly requested, see examples below.
The output is shown at Figure~\ref{fig:lobachevky-line}.
\nwenddocs{}\nwbegincode{53}\sublabel{NWppJ6t-3rkk58-5}\nwmargintag{{\nwtagstyle{}\subpageref{NWppJ6t-3rkk58-5}}}\moddef{hello-cycle.cpp~{\nwtagstyle{}\subpageref{NWppJ6t-3rkk58-1}}}\plusendmoddef\Rm{}\nwstartdeflinemarkup\nwprevnextdefs{NWppJ6t-3rkk58-4}{\relax}\nwenddeflinemarkup
    {\it{}F}.{\it{}asy\_write}(300,-3,3,-3,3,{\tt{}"lobachevsky-line"});
    {\bf{}return} 0;
{\nwrbrace}
\nwindexdefn{\nwixident{asy{\_}write}}{asy:unwrite}{NWppJ6t-3rkk58-5}\eatline
\nwidentdefs{\\{{\nwixident{asy{\_}write}}{asy:unwrite}}}\nwendcode{}\nwbegindocs{54}\nwdocspar
\nwenddocs{}\nwbegindocs{55}\nwdocspar
\begin{figure}[htbp]
  \centering
  \includegraphics[scale=.8]{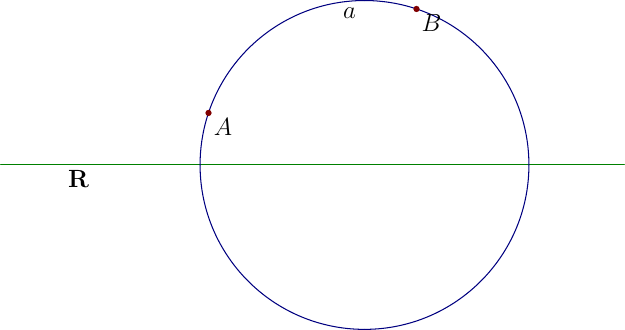}
  \caption{Lobachevky line.}
  \label{fig:lobachevky-line}
\end{figure}
\nwenddocs{}\nwbegindocs{56}\nwdocspar
\nwenddocs{}\nwbegincode{57}\sublabel{NWppJ6t-1fWRR8-2}\nwmargintag{{\nwtagstyle{}\subpageref{NWppJ6t-1fWRR8-2}}}\moddef{separating chunk~{\nwtagstyle{}\subpageref{NWppJ6t-1fWRR8-1}}}\plusendmoddef\Rm{}\nwstartdeflinemarkup\nwprevnextdefs{NWppJ6t-1fWRR8-1}{NWppJ6t-1fWRR8-3}\nwenddeflinemarkup

\nwendcode{}\nwbegindocs{58}\nwdocspar
\subsection{Animated cycle}
\label{sec:animated-cycle}

We use the similar construction to make an animation.
\nwenddocs{}\nwbegincode{59}\sublabel{NWppJ6t-1kKRnF-1}\nwmargintag{{\nwtagstyle{}\subpageref{NWppJ6t-1kKRnF-1}}}\moddef{hello-cycle-anim.cpp~{\nwtagstyle{}\subpageref{NWppJ6t-1kKRnF-1}}}\endmoddef\Rm{}\nwstartdeflinemarkup\nwprevnextdefs{\relax}{NWppJ6t-1kKRnF-2}\nwenddeflinemarkup
\LA{}license~{\nwtagstyle{}\subpageref{NWppJ6t-ZXuKx-1}}\RA{}
{\bf{}\char35{}include}{\tt{} "figure.h"}
\LA{}using all namespaces~{\nwtagstyle{}\subpageref{NWppJ6t-3uHj2c-1}}\RA{}
{\bf{}int} {\it{}main}(){\nwlbrace}\nwindexdefn{\nwixident{main}}{main}{NWppJ6t-1kKRnF-1}

\nwalsodefined{\\{NWppJ6t-1kKRnF-2}\\{NWppJ6t-1kKRnF-3}\\{NWppJ6t-1kKRnF-4}\\{NWppJ6t-1kKRnF-5}\\{NWppJ6t-1kKRnF-6}\\{NWppJ6t-1kKRnF-7}}\nwnotused{hello-cycle-anim.cpp}\nwidentdefs{\\{{\nwixident{main}}{main}}}\nwidentuses{\\{{\nwixident{figure}}{figure}}}\nwindexuse{\nwixident{figure}}{figure}{NWppJ6t-1kKRnF-1}\nwendcode{}\nwbegindocs{60}It is preferable to {\Tt{}\Rm{}{\it{}freeze}\nwendquote} a figure with a symbolic parameter in
order to avoid useless but expensive symbolic computations. It will be
automatically {\Tt{}\Rm{}{\it{}unfreeze}\nwendquote} by {\Tt{}\Rm{}{\it{}asy\_animate}\nwendquote} method below.
\nwenddocs{}\nwbegincode{61}\sublabel{NWppJ6t-1kKRnF-2}\nwmargintag{{\nwtagstyle{}\subpageref{NWppJ6t-1kKRnF-2}}}\moddef{hello-cycle-anim.cpp~{\nwtagstyle{}\subpageref{NWppJ6t-1kKRnF-1}}}\plusendmoddef\Rm{}\nwstartdeflinemarkup\nwprevnextdefs{NWppJ6t-1kKRnF-1}{NWppJ6t-1kKRnF-3}\nwenddeflinemarkup
    {\bf{}figure} {\it{}F}={\bf{}figure}().{\it{}freeze}();
    {\bf{}symbol} {\it{}t}({\tt{}"t"});
\nwindexdefn{\nwixident{freeze}}{freeze}{NWppJ6t-1kKRnF-2}\nwindexdefn{\nwixident{unfreeze}}{unfreeze}{NWppJ6t-1kKRnF-2}\eatline
\nwidentdefs{\\{{\nwixident{freeze}}{freeze}}\\{{\nwixident{unfreeze}}{unfreeze}}}\nwidentuses{\\{{\nwixident{figure}}{figure}}}\nwindexuse{\nwixident{figure}}{figure}{NWppJ6t-1kKRnF-2}\nwendcode{}\nwbegindocs{62}\nwdocspar
\nwenddocs{}\nwbegindocs{63}This time the point {\Tt{}\Rm{}{\it{}A}\nwendquote} on the figure depends from the above parameter
{\Tt{}\Rm{}{\it{}t}\nwendquote} and the point {\Tt{}\Rm{}{\it{}B}\nwendquote} is fixed as before.
\nwenddocs{}\nwbegincode{64}\sublabel{NWppJ6t-1kKRnF-3}\nwmargintag{{\nwtagstyle{}\subpageref{NWppJ6t-1kKRnF-3}}}\moddef{hello-cycle-anim.cpp~{\nwtagstyle{}\subpageref{NWppJ6t-1kKRnF-1}}}\plusendmoddef\Rm{}\nwstartdeflinemarkup\nwprevnextdefs{NWppJ6t-1kKRnF-2}{NWppJ6t-1kKRnF-4}\nwenddeflinemarkup
    {\bf{}ex} {\it{}A}={\it{}F}.{\it{}add\_point}({\bf{}lst}{\nwlbrace}-1\begin{math}\ast\end{math}{\it{}t},.5\begin{math}\ast\end{math}{\it{}t}+.5{\nwrbrace},{\tt{}"A"});
    {\bf{}ex} {\it{}B}={\it{}F}.{\it{}add\_point}({\bf{}lst}{\nwlbrace}1,1.5{\nwrbrace},{\tt{}"B"});

\nwidentuses{\\{{\nwixident{add{\_}point}}{add:unpoint}}\\{{\nwixident{ex}}{ex}}}\nwindexuse{\nwixident{add{\_}point}}{add:unpoint}{NWppJ6t-1kKRnF-3}\nwindexuse{\nwixident{ex}}{ex}{NWppJ6t-1kKRnF-3}\nwendcode{}\nwbegindocs{65}The Lobachevsky line {\Tt{}\Rm{}{\it{}a}\nwendquote} is defined exactly as in the previous
example but is implicitly (through {\Tt{}\Rm{}{\it{}A}\nwendquote}) depending on {\Tt{}\Rm{}{\it{}t}\nwendquote} now.
\nwenddocs{}\nwbegincode{66}\sublabel{NWppJ6t-1kKRnF-4}\nwmargintag{{\nwtagstyle{}\subpageref{NWppJ6t-1kKRnF-4}}}\moddef{hello-cycle-anim.cpp~{\nwtagstyle{}\subpageref{NWppJ6t-1kKRnF-1}}}\plusendmoddef\Rm{}\nwstartdeflinemarkup\nwprevnextdefs{NWppJ6t-1kKRnF-3}{NWppJ6t-1kKRnF-5}\nwenddeflinemarkup
    {\bf{}ex} {\it{}a}={\it{}F}.{\it{}add\_cycle\_rel}({\bf{}lst}{\nwlbrace}{\it{}is\_orthogonal}({\it{}A}),{\it{}is\_orthogonal}({\it{}B}),{\it{}is\_orthogonal}({\it{}F}.{\it{}get\_real\_line}()){\nwrbrace},{\tt{}"a"});

\nwidentuses{\\{{\nwixident{add{\_}cycle{\_}rel}}{add:uncycle:unrel}}\\{{\nwixident{ex}}{ex}}\\{{\nwixident{get{\_}real{\_}line}}{get:unreal:unline}}\\{{\nwixident{is{\_}orthogonal}}{is:unorthogonal}}}\nwindexuse{\nwixident{add{\_}cycle{\_}rel}}{add:uncycle:unrel}{NWppJ6t-1kKRnF-4}\nwindexuse{\nwixident{ex}}{ex}{NWppJ6t-1kKRnF-4}\nwindexuse{\nwixident{get{\_}real{\_}line}}{get:unreal:unline}{NWppJ6t-1kKRnF-4}\nwindexuse{\nwixident{is{\_}orthogonal}}{is:unorthogonal}{NWppJ6t-1kKRnF-4}\nwendcode{}\nwbegindocs{67}The new straight line {\Tt{}\Rm{}{\it{}b}\nwendquote} is defined as a cycle passing
(orthogonal to) the point at infinity. It is accessible by
{\Tt{}\Rm{}{\it{}get\_infinity}\nwendquote} method.
\nwenddocs{}\nwbegincode{68}\sublabel{NWppJ6t-1kKRnF-5}\nwmargintag{{\nwtagstyle{}\subpageref{NWppJ6t-1kKRnF-5}}}\moddef{hello-cycle-anim.cpp~{\nwtagstyle{}\subpageref{NWppJ6t-1kKRnF-1}}}\plusendmoddef\Rm{}\nwstartdeflinemarkup\nwprevnextdefs{NWppJ6t-1kKRnF-4}{NWppJ6t-1kKRnF-6}\nwenddeflinemarkup
    {\bf{}ex} {\it{}b}={\it{}F}.{\it{}add\_cycle\_rel}({\bf{}lst}{\nwlbrace}{\it{}is\_orthogonal}({\it{}A}),{\it{}is\_orthogonal}({\it{}B}),{\it{}is\_orthogonal}({\it{}F}.{\it{}get\_infinity}()){\nwrbrace},{\tt{}"b"});
\nwindexdefn{\nwixident{get{\_}infinity}}{get:uninfinity}{NWppJ6t-1kKRnF-5}\eatline
\nwidentdefs{\\{{\nwixident{get{\_}infinity}}{get:uninfinity}}}\nwidentuses{\\{{\nwixident{add{\_}cycle{\_}rel}}{add:uncycle:unrel}}\\{{\nwixident{ex}}{ex}}\\{{\nwixident{is{\_}orthogonal}}{is:unorthogonal}}}\nwindexuse{\nwixident{add{\_}cycle{\_}rel}}{add:uncycle:unrel}{NWppJ6t-1kKRnF-5}\nwindexuse{\nwixident{ex}}{ex}{NWppJ6t-1kKRnF-5}\nwindexuse{\nwixident{is{\_}orthogonal}}{is:unorthogonal}{NWppJ6t-1kKRnF-5}\nwendcode{}\nwbegindocs{69}\nwdocspar
\nwenddocs{}\nwbegindocs{70}Now we define the set of values for the parameter {\Tt{}\Rm{}{\it{}t}\nwendquote} which will
be used for substitution into the figure.
\nwenddocs{}\nwbegincode{71}\sublabel{NWppJ6t-1kKRnF-6}\nwmargintag{{\nwtagstyle{}\subpageref{NWppJ6t-1kKRnF-6}}}\moddef{hello-cycle-anim.cpp~{\nwtagstyle{}\subpageref{NWppJ6t-1kKRnF-1}}}\plusendmoddef\Rm{}\nwstartdeflinemarkup\nwprevnextdefs{NWppJ6t-1kKRnF-5}{NWppJ6t-1kKRnF-7}\nwenddeflinemarkup
    {\bf{}lst} {\it{}val};
    {\bf{}for} ({\bf{}int} {\it{}i}=0; {\it{}i}\begin{math}<\end{math}40; \protect\PP{\it{}i})
        {\it{}val}.{\it{}append}({\it{}t}\begin{math}\equiv\end{math}{\bf{}numeric}({\it{}i}+2,30));

\nwidentuses{\\{{\nwixident{numeric}}{numeric}}}\nwindexuse{\nwixident{numeric}}{numeric}{NWppJ6t-1kKRnF-6}\nwendcode{}\nwbegindocs{72}Finally animations in different formats are created similarly to the
static picture from the previous example.
\nwenddocs{}\nwbegincode{73}\sublabel{NWppJ6t-1kKRnF-7}\nwmargintag{{\nwtagstyle{}\subpageref{NWppJ6t-1kKRnF-7}}}\moddef{hello-cycle-anim.cpp~{\nwtagstyle{}\subpageref{NWppJ6t-1kKRnF-1}}}\plusendmoddef\Rm{}\nwstartdeflinemarkup\nwprevnextdefs{NWppJ6t-1kKRnF-6}{\relax}\nwenddeflinemarkup
    {\it{}F}.{\it{}asy\_animate}({\it{}val},500,-2.2,3,-2,2,{\tt{}"lobachevsky-anim"},{\tt{}"mng"});
    {\it{}F}.{\it{}asy\_animate}({\it{}val},300,-2.2,3,-2,2,{\tt{}"lobachevsky-anim"},{\tt{}"pdf"});
    {\bf{}return} 0;
{\nwrbrace}
\nwindexdefn{\nwixident{asy{\_}animate}}{asy:unanimate}{NWppJ6t-1kKRnF-7}\eatline
\nwidentdefs{\\{{\nwixident{asy{\_}animate}}{asy:unanimate}}}\nwendcode{}\nwbegindocs{74}\nwdocspar
\nwenddocs{}\nwbegindocs{75}The second command creates two files: \verb|lobachevsky-anim.pdf|
and \verb|_lobachevsky-anim.pdf| (notice the underscore (\verb|_|) in
front of the file name, which makes the difference). The former is a
stand-alone PDF file containing the desired animation. The latter may
be embedded into another PDF document as shown on
Fig.~\ref{fig:lobachevsky-anim}. To this end the \LaTeX\ file need to
have the command
\begin{verbatim}
\usepackage{animate}
\end{verbatim}
in its preamble. To include the animation we use the command:
\begin{verbatim}
\animategraphics[controls]{50}{_lobachevsky-anim}{}{}
\end{verbatim}
More options can be found in the
\href{rors.ctan.org/macros/latex/contrib/animate/animate.pdf}{documentation
  of \texttt{animate} package}.
Finally, the \LaTeX\ file need to be compiled with the
\texttt{pdf}\LaTeX\ command.
\begin{figure}[htbp]
  \centering
  \animategraphics[controls]{50}{_lobachevsky-anim}{}{}
  \caption[Animated Lobachevsky line]{Animated Lobachevsky line: use
    the control buttons to run the animation. You may need
    \textsf{Adobe Acrobat Reader} for this feature.}
  \label{fig:lobachevsky-anim}
\end{figure}
\nwenddocs{}\nwbegindocs{76}\nwdocspar
\nwenddocs{}\nwbegincode{77}\sublabel{NWppJ6t-1fWRR8-3}\nwmargintag{{\nwtagstyle{}\subpageref{NWppJ6t-1fWRR8-3}}}\moddef{separating chunk~{\nwtagstyle{}\subpageref{NWppJ6t-1fWRR8-1}}}\plusendmoddef\Rm{}\nwstartdeflinemarkup\nwprevnextdefs{NWppJ6t-1fWRR8-2}{NWppJ6t-1fWRR8-4}\nwenddeflinemarkup

\nwendcode{}\nwbegindocs{78}\nwdocspar
\subsection{An illustration of the modular group action}
\label{sec:an-illustr-modul}

The library allows to build figures out of cycles which are obtained
from each other by means of FLT. We are going to
illustrate this by the action of the modular group
\(\mathrm{SL}_2(\mathbb{Z})\) on a single
circle~\cite{StewartTall02a}*{\S~14.4}. We repeatedly apply FLT \(T=
\begin{pmatrix}
  1&1\\0&1
\end{pmatrix}\)
for translations and \(S=
\begin{pmatrix}
  0&-1\\1&0
\end{pmatrix}\) for the inversion in the unit circle.

\nwenddocs{}\nwbegindocs{79}Here is the standard start of a programme with some additional
variables being initialised.
\nwenddocs{}\nwbegincode{80}\sublabel{NWppJ6t-2SYluR-1}\nwmargintag{{\nwtagstyle{}\subpageref{NWppJ6t-2SYluR-1}}}\moddef{modular-group.cpp~{\nwtagstyle{}\subpageref{NWppJ6t-2SYluR-1}}}\endmoddef\Rm{}\nwstartdeflinemarkup\nwprevnextdefs{\relax}{NWppJ6t-2SYluR-2}\nwenddeflinemarkup
\LA{}license~{\nwtagstyle{}\subpageref{NWppJ6t-ZXuKx-1}}\RA{}
{\bf{}\char35{}include}{\tt{} "figure.h"}
\LA{}using all namespaces~{\nwtagstyle{}\subpageref{NWppJ6t-3uHj2c-1}}\RA{}
{\bf{}int} {\it{}main}(){\nwlbrace}\nwindexdefn{\nwixident{main}}{main}{NWppJ6t-2SYluR-1}
    {\bf{}char} {\it{}buffer} [50];
    {\bf{}int} {\it{}steps}=3, {\it{}trans}=15;
    {\bf{}double} {\it{}epsilon}=0.00001; // square of radius for a circle to be ignored
    {\bf{}figure} {\it{}F};

\nwalsodefined{\\{NWppJ6t-2SYluR-2}\\{NWppJ6t-2SYluR-3}\\{NWppJ6t-2SYluR-4}\\{NWppJ6t-2SYluR-5}\\{NWppJ6t-2SYluR-6}\\{NWppJ6t-2SYluR-7}\\{NWppJ6t-2SYluR-8}\\{NWppJ6t-2SYluR-9}\\{NWppJ6t-2SYluR-A}\\{NWppJ6t-2SYluR-B}}\nwnotused{modular-group.cpp}\nwidentdefs{\\{{\nwixident{main}}{main}}}\nwidentuses{\\{{\nwixident{epsilon}}{epsilon}}\\{{\nwixident{figure}}{figure}}}\nwindexuse{\nwixident{epsilon}}{epsilon}{NWppJ6t-2SYluR-1}\nwindexuse{\nwixident{figure}}{figure}{NWppJ6t-2SYluR-1}\nwendcode{}\nwbegindocs{81}We will use the metric associated to the figure, it can be extracted
by {\Tt{}\Rm{}{\it{}get\_point\_metric}\nwendquote} method.
\nwenddocs{}\nwbegincode{82}\sublabel{NWppJ6t-2SYluR-2}\nwmargintag{{\nwtagstyle{}\subpageref{NWppJ6t-2SYluR-2}}}\moddef{modular-group.cpp~{\nwtagstyle{}\subpageref{NWppJ6t-2SYluR-1}}}\plusendmoddef\Rm{}\nwstartdeflinemarkup\nwprevnextdefs{NWppJ6t-2SYluR-1}{NWppJ6t-2SYluR-3}\nwenddeflinemarkup
    {\bf{}ex} {\it{}e}={\it{}F}.{\it{}get\_point\_metric}();

\nwindexdefn{\nwixident{get{\_}point{\_}metric}}{get:unpoint:unmetric}{NWppJ6t-2SYluR-2}\eatline
\nwidentdefs{\\{{\nwixident{get{\_}point{\_}metric}}{get:unpoint:unmetric}}}\nwidentuses{\\{{\nwixident{ex}}{ex}}}\nwindexuse{\nwixident{ex}}{ex}{NWppJ6t-2SYluR-2}\nwendcode{}\nwbegindocs{83}\nwdocspar
\nwenddocs{}\nwbegindocs{84}Firstly, we add to the figure an initial cycle and, then, add new
generations of its shifts and reflections.
\nwenddocs{}\nwbegincode{85}\sublabel{NWppJ6t-2SYluR-3}\nwmargintag{{\nwtagstyle{}\subpageref{NWppJ6t-2SYluR-3}}}\moddef{modular-group.cpp~{\nwtagstyle{}\subpageref{NWppJ6t-2SYluR-1}}}\plusendmoddef\Rm{}\nwstartdeflinemarkup\nwprevnextdefs{NWppJ6t-2SYluR-2}{NWppJ6t-2SYluR-4}\nwenddeflinemarkup
    {\bf{}ex} {\it{}a}={\it{}F}.{\it{}add\_cycle}({\bf{}cycle2D}({\bf{}lst}{\nwlbrace}0,{\bf{}numeric}(3,2){\nwrbrace},{\it{}e},{\bf{}numeric}(1,4)),{\tt{}"a"});
    {\bf{}ex} {\it{}c}={\it{}F}.{\it{}add\_cycle}({\bf{}cycle2D}({\bf{}lst}{\nwlbrace}0,{\bf{}numeric}(11,6){\nwrbrace},{\it{}e},{\bf{}numeric}(1,36)),{\tt{}"c"});
    {\bf{}for} ({\bf{}int} {\it{}i}=0; {\it{}i}\begin{math}<\end{math}{\it{}steps};\protect\PP{\it{}i}) {\nwlbrace}

\nwidentuses{\\{{\nwixident{add{\_}cycle}}{add:uncycle}}\\{{\nwixident{ex}}{ex}}\\{{\nwixident{numeric}}{numeric}}}\nwindexuse{\nwixident{add{\_}cycle}}{add:uncycle}{NWppJ6t-2SYluR-3}\nwindexuse{\nwixident{ex}}{ex}{NWppJ6t-2SYluR-3}\nwindexuse{\nwixident{numeric}}{numeric}{NWppJ6t-2SYluR-3}\nwendcode{}\nwbegindocs{86}We want to shift all cycles in the previous
generation. Their key are grasped by {\Tt{}\Rm{}{\it{}get\_all\_keys}\nwendquote} method.
\nwenddocs{}\nwbegincode{87}\sublabel{NWppJ6t-2SYluR-4}\nwmargintag{{\nwtagstyle{}\subpageref{NWppJ6t-2SYluR-4}}}\moddef{modular-group.cpp~{\nwtagstyle{}\subpageref{NWppJ6t-2SYluR-1}}}\plusendmoddef\Rm{}\nwstartdeflinemarkup\nwprevnextdefs{NWppJ6t-2SYluR-3}{NWppJ6t-2SYluR-5}\nwenddeflinemarkup
        {\bf{}lst} {\it{}L}={\it{}ex\_to}\begin{math}<\end{math}{\bf{}lst}\begin{math}>\end{math}({\it{}F}.{\it{}get\_all\_keys}(2\begin{math}\ast\end{math}{\it{}i},2\begin{math}\ast\end{math}{\it{}i}));
        {\bf{}if} ({\it{}L}.{\it{}nops}() \begin{math}\equiv\end{math} 0) {\nwlbrace}
            {\it{}cout} \begin{math}\ll\end{math} {\tt{}"Terminate on iteration "} \begin{math}\ll\end{math} {\it{}i} \begin{math}\ll\end{math} {\it{}endl};
            {\bf{}break};
        {\nwrbrace}
\nwindexdefn{\nwixident{get{\_}all{\_}keys}}{get:unall:unkeys}{NWppJ6t-2SYluR-4}\eatline
\nwidentdefs{\\{{\nwixident{get{\_}all{\_}keys}}{get:unall:unkeys}}}\nwidentuses{\\{{\nwixident{nops}}{nops}}}\nwindexuse{\nwixident{nops}}{nops}{NWppJ6t-2SYluR-4}\nwendcode{}\nwbegindocs{88}\nwdocspar
\nwenddocs{}\nwbegindocs{89}Each cycle with the collected key is shifted horizontally by an
integer \(t\) in range [\(-\){\Tt{}\Rm{}{\it{}trans}\nwendquote},{\Tt{}\Rm{}{\it{}trans}\nwendquote}{}]. This done by
{\Tt{}\Rm{}{\it{}moebius\_transform}\nwendquote} relations and it is our responsibility to
produce proper Clifford-valued entries to the matrix,
see~\cite{Kisil05a}*{\S~2.1} for an advise.
\nwenddocs{}\nwbegincode{90}\sublabel{NWppJ6t-2SYluR-5}\nwmargintag{{\nwtagstyle{}\subpageref{NWppJ6t-2SYluR-5}}}\moddef{modular-group.cpp~{\nwtagstyle{}\subpageref{NWppJ6t-2SYluR-1}}}\plusendmoddef\Rm{}\nwstartdeflinemarkup\nwprevnextdefs{NWppJ6t-2SYluR-4}{NWppJ6t-2SYluR-6}\nwenddeflinemarkup
        {\bf{}for} ({\bf{}const} {\bf{}auto}& {\it{}k}: {\it{}L}) {\nwlbrace}
            {\bf{}lst} {\it{}L1}={\it{}ex\_to}\begin{math}<\end{math}{\bf{}lst}\begin{math}>\end{math}({\it{}F}.{\it{}get\_cycle}({\it{}k}));
            {\bf{}for} ({\bf{}auto} {\it{}x}: {\it{}L1}) {\nwlbrace}
                {\bf{}for} ({\bf{}int} {\it{}t}=-{\it{}trans}; {\it{}t}\begin{math}\leq\end{math}{\it{}trans};\protect\PP{\it{}t}) {\nwlbrace}
                    {\it{}sprintf} ({\it{}buffer}, {\tt{}"

\nwidentuses{\\{{\nwixident{get{\_}cycle}}{get:uncycle}}\\{{\nwixident{k}}{k}}}\nwindexuse{\nwixident{get{\_}cycle}}{get:uncycle}{NWppJ6t-2SYluR-5}\nwindexuse{\nwixident{k}}{k}{NWppJ6t-2SYluR-5}\nwendcode{}\nwbegindocs{91}We shift initial cycles by zero in order to have their copies in the this generation.
\nwenddocs{}\nwbegincode{92}\sublabel{NWppJ6t-2SYluR-6}\nwmargintag{{\nwtagstyle{}\subpageref{NWppJ6t-2SYluR-6}}}\moddef{modular-group.cpp~{\nwtagstyle{}\subpageref{NWppJ6t-2SYluR-1}}}\plusendmoddef\Rm{}\nwstartdeflinemarkup\nwprevnextdefs{NWppJ6t-2SYluR-5}{NWppJ6t-2SYluR-7}\nwenddeflinemarkup
                    {\bf{}if} (({\it{}t} \begin{math}\neq\end{math}0 \begin{math}\vee\end{math} {\it{}i} \begin{math}\equiv\end{math}0)

\nwendcode{}\nwbegindocs{93}To simplify the picture we are skipping circles whose radii would be
smaller than the threshold.
\nwenddocs{}\nwbegincode{94}\sublabel{NWppJ6t-2SYluR-7}\nwmargintag{{\nwtagstyle{}\subpageref{NWppJ6t-2SYluR-7}}}\moddef{modular-group.cpp~{\nwtagstyle{}\subpageref{NWppJ6t-2SYluR-1}}}\plusendmoddef\Rm{}\nwstartdeflinemarkup\nwprevnextdefs{NWppJ6t-2SYluR-6}{NWppJ6t-2SYluR-8}\nwenddeflinemarkup
                        \begin{math}\wedge\end{math} \begin{math}\neg\end{math} (({\it{}ex\_to}\begin{math}<\end{math}{\bf{}cycle}\begin{math}>\end{math}({\it{}x}).{\it{}det}()-({\it{}pow}({\it{}t},2)-1)\begin{math}\ast\end{math}{\it{}epsilon}).{\it{}evalf}()\begin{math}<\end{math}0)){\nwlbrace}
                        {\bf{}ex} {\it{}b}={\it{}F}.{\it{}add\_cycle\_rel}({\it{}moebius\_transform}({\it{}k},{\bf{}true},
                                                               {\bf{}lst}{\nwlbrace}{\it{}dirac\_ONE}(),{\it{}t}\begin{math}\ast\end{math}{\it{}e}.{\it{}subs}({\it{}e}.{\it{}op}(1).{\it{}op}(0)\begin{math}\equiv\end{math}0),0,{\it{}dirac\_ONE}(){\nwrbrace}),{\it{}buffer});
\nwindexdefn{\nwixident{moebius{\_}transform}}{moebius:untransform}{NWppJ6t-2SYluR-7}\eatline
\nwidentdefs{\\{{\nwixident{moebius{\_}transform}}{moebius:untransform}}}\nwidentuses{\\{{\nwixident{add{\_}cycle{\_}rel}}{add:uncycle:unrel}}\\{{\nwixident{epsilon}}{epsilon}}\\{{\nwixident{evalf}}{evalf}}\\{{\nwixident{ex}}{ex}}\\{{\nwixident{k}}{k}}\\{{\nwixident{op}}{op}}\\{{\nwixident{subs}}{subs}}}\nwindexuse{\nwixident{add{\_}cycle{\_}rel}}{add:uncycle:unrel}{NWppJ6t-2SYluR-7}\nwindexuse{\nwixident{epsilon}}{epsilon}{NWppJ6t-2SYluR-7}\nwindexuse{\nwixident{evalf}}{evalf}{NWppJ6t-2SYluR-7}\nwindexuse{\nwixident{ex}}{ex}{NWppJ6t-2SYluR-7}\nwindexuse{\nwixident{k}}{k}{NWppJ6t-2SYluR-7}\nwindexuse{\nwixident{op}}{op}{NWppJ6t-2SYluR-7}\nwindexuse{\nwixident{subs}}{subs}{NWppJ6t-2SYluR-7}\nwendcode{}\nwbegindocs{95}\nwdocspar
\nwenddocs{}\nwbegindocs{96}We want the colour of a cycle reflect its generation, smaller cycles
also need to be drawn by a finer pen. This can be set for each cycle
by {\Tt{}\Rm{}{\it{}set\_asy\_style}\nwendquote} method.
\nwenddocs{}\nwbegincode{97}\sublabel{NWppJ6t-2SYluR-8}\nwmargintag{{\nwtagstyle{}\subpageref{NWppJ6t-2SYluR-8}}}\moddef{modular-group.cpp~{\nwtagstyle{}\subpageref{NWppJ6t-2SYluR-1}}}\plusendmoddef\Rm{}\nwstartdeflinemarkup\nwprevnextdefs{NWppJ6t-2SYluR-7}{NWppJ6t-2SYluR-9}\nwenddeflinemarkup
                        {\it{}sprintf} ({\it{}buffer}, {\tt{}"rgb(0,0,
                        {\it{}F}.{\it{}set\_asy\_style}({\it{}b},{\it{}buffer});
                    {\nwrbrace}
                {\nwrbrace}
            {\nwrbrace}
        {\nwrbrace}
\nwindexdefn{\nwixident{set{\_}asy{\_}style}}{set:unasy:unstyle}{NWppJ6t-2SYluR-8}\nwindexdefn{\nwixident{rgb}}{rgb}{NWppJ6t-2SYluR-8}\eatline
\nwidentdefs{\\{{\nwixident{rgb}}{rgb}}\\{{\nwixident{set{\_}asy{\_}style}}{set:unasy:unstyle}}}\nwendcode{}\nwbegindocs{98}\nwdocspar
\nwenddocs{}\nwbegindocs{99}Similarly, we collect all key from the previous generation cycles
to make their reflection in the unit circle.
\nwenddocs{}\nwbegincode{100}\sublabel{NWppJ6t-2SYluR-9}\nwmargintag{{\nwtagstyle{}\subpageref{NWppJ6t-2SYluR-9}}}\moddef{modular-group.cpp~{\nwtagstyle{}\subpageref{NWppJ6t-2SYluR-1}}}\plusendmoddef\Rm{}\nwstartdeflinemarkup\nwprevnextdefs{NWppJ6t-2SYluR-8}{NWppJ6t-2SYluR-A}\nwenddeflinemarkup
        {\bf{}if} ({\it{}i}\begin{math}<\end{math}{\it{}steps}-1)
            {\it{}L}={\it{}ex\_to}\begin{math}<\end{math}{\bf{}lst}\begin{math}>\end{math}({\it{}F}.{\it{}get\_all\_keys}(2\begin{math}\ast\end{math}{\it{}i}+1,2\begin{math}\ast\end{math}{\it{}i}+1));
        {\bf{}else}
            {\it{}L}={\bf{}lst}{\nwlbrace}{\nwrbrace};
        {\bf{}for} ({\bf{}const} {\bf{}auto}& {\it{}k}: {\it{}L}) {\nwlbrace}
            {\it{}sprintf} ({\it{}buffer}, {\tt{}"

\nwidentuses{\\{{\nwixident{get{\_}all{\_}keys}}{get:unall:unkeys}}\\{{\nwixident{k}}{k}}}\nwindexuse{\nwixident{get{\_}all{\_}keys}}{get:unall:unkeys}{NWppJ6t-2SYluR-9}\nwindexuse{\nwixident{k}}{k}{NWppJ6t-2SYluR-9}\nwendcode{}\nwbegindocs{101}This tame we keep things simple and are using {\Tt{}\Rm{}{\it{}sl2\_transform}\nwendquote}
relation, all Clifford algebra adjustments are taken by the
library. The drawing style is setup accordingly.
\nwenddocs{}\nwbegincode{102}\sublabel{NWppJ6t-2SYluR-A}\nwmargintag{{\nwtagstyle{}\subpageref{NWppJ6t-2SYluR-A}}}\moddef{modular-group.cpp~{\nwtagstyle{}\subpageref{NWppJ6t-2SYluR-1}}}\plusendmoddef\Rm{}\nwstartdeflinemarkup\nwprevnextdefs{NWppJ6t-2SYluR-9}{NWppJ6t-2SYluR-B}\nwenddeflinemarkup
            {\bf{}ex} {\it{}b}={\it{}F}.{\it{}add\_cycle\_rel}({\it{}sl2\_transform}({\it{}k},{\bf{}true},{\bf{}lst}{\nwlbrace}0,-1,1,0{\nwrbrace}),{\it{}buffer});
            {\it{}sprintf} ({\it{}buffer}, {\tt{}"rgb(0,0.7,
            {\it{}F}.{\it{}set\_asy\_style}({\it{}b},{\it{}buffer});
        {\nwrbrace}
    {\nwrbrace}
\nwindexdefn{\nwixident{sl2{\_}transform}}{sl2:untransform}{NWppJ6t-2SYluR-A}\eatline
\nwidentdefs{\\{{\nwixident{sl2{\_}transform}}{sl2:untransform}}}\nwidentuses{\\{{\nwixident{add{\_}cycle{\_}rel}}{add:uncycle:unrel}}\\{{\nwixident{ex}}{ex}}\\{{\nwixident{k}}{k}}\\{{\nwixident{rgb}}{rgb}}\\{{\nwixident{set{\_}asy{\_}style}}{set:unasy:unstyle}}}\nwindexuse{\nwixident{add{\_}cycle{\_}rel}}{add:uncycle:unrel}{NWppJ6t-2SYluR-A}\nwindexuse{\nwixident{ex}}{ex}{NWppJ6t-2SYluR-A}\nwindexuse{\nwixident{k}}{k}{NWppJ6t-2SYluR-A}\nwindexuse{\nwixident{rgb}}{rgb}{NWppJ6t-2SYluR-A}\nwindexuse{\nwixident{set{\_}asy{\_}style}}{set:unasy:unstyle}{NWppJ6t-2SYluR-A}\nwendcode{}\nwbegindocs{103}\nwdocspar

\nwenddocs{}\nwbegindocs{104}Finally, we draw the picture. This time we do not want cycles label
to appear, thus the last parameter {\Tt{}\Rm{}{\it{}with\_labels}\nwendquote} of {\Tt{}\Rm{}{\it{}asy\_write}\nwendquote} is
{\Tt{}\Rm{}{\bf{}false}\nwendquote}. We also want to reduce the size of \Asymptote\ file and
will not print headers of cycles, thus specifying
{\Tt{}\Rm{}{\it{}with\_header}={\bf{}true}\nwendquote}. The remaining parameters are explicitly assigned
their default values.
\nwenddocs{}\nwbegincode{105}\sublabel{NWppJ6t-2SYluR-B}\nwmargintag{{\nwtagstyle{}\subpageref{NWppJ6t-2SYluR-B}}}\moddef{modular-group.cpp~{\nwtagstyle{}\subpageref{NWppJ6t-2SYluR-1}}}\plusendmoddef\Rm{}\nwstartdeflinemarkup\nwprevnextdefs{NWppJ6t-2SYluR-A}{\relax}\nwenddeflinemarkup
    {\bf{}ex} {\it{}u}={\it{}F}.{\it{}add\_cycle}({\bf{}cycle2D}({\bf{}lst}{\nwlbrace}0,0{\nwrbrace},{\it{}e},{\bf{}numeric}(1)),{\tt{}"u"});
    {\it{}F}.{\it{}asy\_write}(300,-2.17,2.17,0,2,{\tt{}"modular-group"},{\tt{}"pdf"},{\it{}default\_asy},{\it{}default\_label},{\bf{}true},{\bf{}false},0,{\tt{}"rgb(0,.9,0)+4pt"},{\bf{}true},{\bf{}false});
    {\bf{}return} 0;
{\nwrbrace}
\nwindexdefn{\nwixident{asy{\_}write}}{asy:unwrite}{NWppJ6t-2SYluR-B}\eatline
\nwidentdefs{\\{{\nwixident{asy{\_}write}}{asy:unwrite}}}\nwidentuses{\\{{\nwixident{add{\_}cycle}}{add:uncycle}}\\{{\nwixident{ex}}{ex}}\\{{\nwixident{numeric}}{numeric}}\\{{\nwixident{rgb}}{rgb}}}\nwindexuse{\nwixident{add{\_}cycle}}{add:uncycle}{NWppJ6t-2SYluR-B}\nwindexuse{\nwixident{ex}}{ex}{NWppJ6t-2SYluR-B}\nwindexuse{\nwixident{numeric}}{numeric}{NWppJ6t-2SYluR-B}\nwindexuse{\nwixident{rgb}}{rgb}{NWppJ6t-2SYluR-B}\nwendcode{}\nwbegindocs{106}\nwdocspar
\nwenddocs{}\nwbegindocs{107}\nwdocspar
\nwenddocs{}\nwbegincode{108}\sublabel{NWppJ6t-1fWRR8-4}\nwmargintag{{\nwtagstyle{}\subpageref{NWppJ6t-1fWRR8-4}}}\moddef{separating chunk~{\nwtagstyle{}\subpageref{NWppJ6t-1fWRR8-1}}}\plusendmoddef\Rm{}\nwstartdeflinemarkup\nwprevnextdefs{NWppJ6t-1fWRR8-3}{NWppJ6t-1fWRR8-5}\nwenddeflinemarkup

\nwendcode{}\nwbegindocs{109}\nwdocspar
\subsection{Simple analysitcal demonstration  }
\label{sec:simple-analys-demons}

The following example essentially repeats the code from
Example~\ref{ex:touch-centres-collinear}. It will be better to start
from a simpler case before we will consider more advanced usage in the
next subsection. Also this example checks how cycle solver is handling
cycles with free parameters if relations do not determine it uniquely.

The first line creates an empty figure {\Tt{}\Rm{}{\it{}F}\nwendquote} with the default
euclidean metric.
\nwenddocs{}\nwbegincode{110}\sublabel{NWppJ6t-2t2vfS-1}\nwmargintag{{\nwtagstyle{}\subpageref{NWppJ6t-2t2vfS-1}}}\moddef{figure-ortho-anlytic-proof.cpp~{\nwtagstyle{}\subpageref{NWppJ6t-2t2vfS-1}}}\endmoddef\Rm{}\nwstartdeflinemarkup\nwprevnextdefs{\relax}{NWppJ6t-2t2vfS-2}\nwenddeflinemarkup
\LA{}license~{\nwtagstyle{}\subpageref{NWppJ6t-ZXuKx-1}}\RA{}
{\bf{}\char35{}include}{\tt{} "figure.h"}
\LA{}using all namespaces~{\nwtagstyle{}\subpageref{NWppJ6t-3uHj2c-1}}\RA{}
{\bf{}int} {\it{}main}(){\nwlbrace}\nwindexdefn{\nwixident{main}}{main}{NWppJ6t-2t2vfS-1}
    {\bf{}figure} {\it{}F}={\bf{}figure}();

\nwalsodefined{\\{NWppJ6t-2t2vfS-2}\\{NWppJ6t-2t2vfS-3}\\{NWppJ6t-2t2vfS-4}\\{NWppJ6t-2t2vfS-5}\\{NWppJ6t-2t2vfS-6}\\{NWppJ6t-2t2vfS-7}\\{NWppJ6t-2t2vfS-8}\\{NWppJ6t-2t2vfS-9}\\{NWppJ6t-2t2vfS-A}\\{NWppJ6t-2t2vfS-B}\\{NWppJ6t-2t2vfS-C}\\{NWppJ6t-2t2vfS-D}}\nwnotused{figure-ortho-anlytic-proof.cpp}\nwidentdefs{\\{{\nwixident{main}}{main}}}\nwidentuses{\\{{\nwixident{figure}}{figure}}}\nwindexuse{\nwixident{figure}}{figure}{NWppJ6t-2t2vfS-1}\nwendcode{}\nwbegindocs{111}The next line explicitly uses parameters \((1,0,0,-1)\) of {\Tt{}\Rm{}{\it{}a}\nwendquote} to add it to {\Tt{}\Rm{}{\it{}F}\nwendquote}.
\nwenddocs{}\nwbegincode{112}\sublabel{NWppJ6t-2t2vfS-2}\nwmargintag{{\nwtagstyle{}\subpageref{NWppJ6t-2t2vfS-2}}}\moddef{figure-ortho-anlytic-proof.cpp~{\nwtagstyle{}\subpageref{NWppJ6t-2t2vfS-1}}}\plusendmoddef\Rm{}\nwstartdeflinemarkup\nwprevnextdefs{NWppJ6t-2t2vfS-1}{NWppJ6t-2t2vfS-3}\nwenddeflinemarkup
    {\bf{}ex} {\it{}a}={\it{}F}.{\it{}add\_cycle}({\bf{}cycle2D}(1,{\bf{}lst}{\nwlbrace}0,0{\nwrbrace},-1),{\tt{}"a"});

\nwidentuses{\\{{\nwixident{add{\_}cycle}}{add:uncycle}}\\{{\nwixident{ex}}{ex}}}\nwindexuse{\nwixident{add{\_}cycle}}{add:uncycle}{NWppJ6t-2t2vfS-2}\nwindexuse{\nwixident{ex}}{ex}{NWppJ6t-2t2vfS-2}\nwendcode{}\nwbegindocs{113}Next lines define symbols {\Tt{}\Rm{}{\it{}l}\nwendquote} and {\Tt{}\Rm{}{\it{}C}\nwendquote}, which are needed because
cycles with these labels are defined in next lines through some
relations to themselves and the cycle {\Tt{}\Rm{}{\it{}a}\nwendquote}.
\nwenddocs{}\nwbegincode{114}\sublabel{NWppJ6t-2t2vfS-3}\nwmargintag{{\nwtagstyle{}\subpageref{NWppJ6t-2t2vfS-3}}}\moddef{figure-ortho-anlytic-proof.cpp~{\nwtagstyle{}\subpageref{NWppJ6t-2t2vfS-1}}}\plusendmoddef\Rm{}\nwstartdeflinemarkup\nwprevnextdefs{NWppJ6t-2t2vfS-2}{NWppJ6t-2t2vfS-4}\nwenddeflinemarkup
    {\bf{}ex} {\it{}l}={\bf{}symbol}({\tt{}"l"});
    {\bf{}ex} {\it{}C}={\bf{}symbol}({\tt{}"C"});

\nwidentuses{\\{{\nwixident{ex}}{ex}}\\{{\nwixident{l}}{l}}}\nwindexuse{\nwixident{ex}}{ex}{NWppJ6t-2t2vfS-3}\nwindexuse{\nwixident{l}}{l}{NWppJ6t-2t2vfS-3}\nwendcode{}\nwbegindocs{115}In both cases we want to have cycles with real
coefficients only and {\Tt{}\Rm{}{\it{}C}\nwendquote} is additionally self-orthogonal (i.e. is a
zero-radius). Also, {\Tt{}\Rm{}{\it{}l}\nwendquote} is orthogonal to infinity (i.e. is a line)
and {\Tt{}\Rm{}{\it{}C}\nwendquote} is orthogonal to {\Tt{}\Rm{}{\it{}a}\nwendquote} and {\Tt{}\Rm{}{\it{}l}\nwendquote} (i.e. is their common
point). The tangency condition for {\Tt{}\Rm{}{\it{}l}\nwendquote} and self-orthogonality of
{\Tt{}\Rm{}{\it{}C}\nwendquote} are both quadratic relations. The former has two solutions each
depending on one real parameter, thus line {\Tt{}\Rm{}{\it{}l}\nwendquote} has two
instances.
\nwenddocs{}\nwbegincode{116}\sublabel{NWppJ6t-2t2vfS-4}\nwmargintag{{\nwtagstyle{}\subpageref{NWppJ6t-2t2vfS-4}}}\moddef{figure-ortho-anlytic-proof.cpp~{\nwtagstyle{}\subpageref{NWppJ6t-2t2vfS-1}}}\plusendmoddef\Rm{}\nwstartdeflinemarkup\nwprevnextdefs{NWppJ6t-2t2vfS-3}{NWppJ6t-2t2vfS-5}\nwenddeflinemarkup
    {\it{}F}.{\it{}add\_cycle\_rel}({\bf{}lst}{\nwlbrace}{\it{}is\_tangent\_i}({\it{}a}),{\it{}is\_orthogonal}({\it{}F}.{\it{}get\_infinity}()),{\it{}only\_reals}({\it{}l}){\nwrbrace},{\it{}l});

\nwidentuses{\\{{\nwixident{add{\_}cycle{\_}rel}}{add:uncycle:unrel}}\\{{\nwixident{get{\_}infinity}}{get:uninfinity}}\\{{\nwixident{is{\_}orthogonal}}{is:unorthogonal}}\\{{\nwixident{is{\_}tangent{\_}i}}{is:untangent:uni}}\\{{\nwixident{l}}{l}}\\{{\nwixident{only{\_}reals}}{only:unreals}}}\nwindexuse{\nwixident{add{\_}cycle{\_}rel}}{add:uncycle:unrel}{NWppJ6t-2t2vfS-4}\nwindexuse{\nwixident{get{\_}infinity}}{get:uninfinity}{NWppJ6t-2t2vfS-4}\nwindexuse{\nwixident{is{\_}orthogonal}}{is:unorthogonal}{NWppJ6t-2t2vfS-4}\nwindexuse{\nwixident{is{\_}tangent{\_}i}}{is:untangent:uni}{NWppJ6t-2t2vfS-4}\nwindexuse{\nwixident{l}}{l}{NWppJ6t-2t2vfS-4}\nwindexuse{\nwixident{only{\_}reals}}{only:unreals}{NWppJ6t-2t2vfS-4}\nwendcode{}\nwbegindocs{117}Correspondingly, the point of contact {\Tt{}\Rm{}{\it{}C}\nwendquote}\ldots
\nwenddocs{}\nwbegincode{118}\sublabel{NWppJ6t-2t2vfS-5}\nwmargintag{{\nwtagstyle{}\subpageref{NWppJ6t-2t2vfS-5}}}\moddef{figure-ortho-anlytic-proof.cpp~{\nwtagstyle{}\subpageref{NWppJ6t-2t2vfS-1}}}\plusendmoddef\Rm{}\nwstartdeflinemarkup\nwprevnextdefs{NWppJ6t-2t2vfS-4}{NWppJ6t-2t2vfS-6}\nwenddeflinemarkup
    {\it{}F}.{\it{}add\_cycle\_rel}({\bf{}lst}{\nwlbrace}{\it{}is\_orthogonal}({\it{}C}),{\it{}is\_orthogonal}({\it{}a}),{\it{}is\_orthogonal}({\it{}l}),{\it{}only\_reals}({\it{}C}){\nwrbrace},{\it{}C});

\nwidentuses{\\{{\nwixident{add{\_}cycle{\_}rel}}{add:uncycle:unrel}}\\{{\nwixident{is{\_}orthogonal}}{is:unorthogonal}}\\{{\nwixident{l}}{l}}\\{{\nwixident{only{\_}reals}}{only:unreals}}}\nwindexuse{\nwixident{add{\_}cycle{\_}rel}}{add:uncycle:unrel}{NWppJ6t-2t2vfS-5}\nwindexuse{\nwixident{is{\_}orthogonal}}{is:unorthogonal}{NWppJ6t-2t2vfS-5}\nwindexuse{\nwixident{l}}{l}{NWppJ6t-2t2vfS-5}\nwindexuse{\nwixident{only{\_}reals}}{only:unreals}{NWppJ6t-2t2vfS-5}\nwendcode{}\nwbegindocs{119}\ldots and the orthogonal cycle {\Tt{}\Rm{}{\it{}r}\nwendquote} through {\Tt{}\Rm{}{\it{}C}\nwendquote} (defined in line~7) each have two
instances as well.
\nwenddocs{}\nwbegincode{120}\sublabel{NWppJ6t-2t2vfS-6}\nwmargintag{{\nwtagstyle{}\subpageref{NWppJ6t-2t2vfS-6}}}\moddef{figure-ortho-anlytic-proof.cpp~{\nwtagstyle{}\subpageref{NWppJ6t-2t2vfS-1}}}\plusendmoddef\Rm{}\nwstartdeflinemarkup\nwprevnextdefs{NWppJ6t-2t2vfS-5}{NWppJ6t-2t2vfS-7}\nwenddeflinemarkup
    {\bf{}ex} {\it{}r}={\it{}F}.{\it{}add\_cycle\_rel}({\bf{}lst}{\nwlbrace}{\it{}is\_orthogonal}({\it{}C}),{\it{}is\_orthogonal}({\it{}a}){\nwrbrace},{\tt{}"r"});

\nwidentuses{\\{{\nwixident{add{\_}cycle{\_}rel}}{add:uncycle:unrel}}\\{{\nwixident{ex}}{ex}}\\{{\nwixident{is{\_}orthogonal}}{is:unorthogonal}}}\nwindexuse{\nwixident{add{\_}cycle{\_}rel}}{add:uncycle:unrel}{NWppJ6t-2t2vfS-6}\nwindexuse{\nwixident{ex}}{ex}{NWppJ6t-2t2vfS-6}\nwindexuse{\nwixident{is{\_}orthogonal}}{is:unorthogonal}{NWppJ6t-2t2vfS-6}\nwendcode{}\nwbegindocs{121} Finally, we  verify that every instance of {\Tt{}\Rm{}{\it{}l}\nwendquote} is orthogonal to
the respective instance of {\Tt{}\Rm{}{\it{}r}\nwendquote}.
\nwenddocs{}\nwbegincode{122}\sublabel{NWppJ6t-2t2vfS-7}\nwmargintag{{\nwtagstyle{}\subpageref{NWppJ6t-2t2vfS-7}}}\moddef{figure-ortho-anlytic-proof.cpp~{\nwtagstyle{}\subpageref{NWppJ6t-2t2vfS-1}}}\plusendmoddef\Rm{}\nwstartdeflinemarkup\nwprevnextdefs{NWppJ6t-2t2vfS-6}{NWppJ6t-2t2vfS-8}\nwenddeflinemarkup
    {\bf{}ex} {\it{}Res}={\it{}F}.{\it{}check\_rel}({\it{}l}, {\it{}r}, {\it{}cycle\_orthogonal});
\nwindexdefn{\nwixident{check{\_}rel}}{check:unrel}{NWppJ6t-2t2vfS-7}\eatline
\nwidentdefs{\\{{\nwixident{check{\_}rel}}{check:unrel}}}\nwidentuses{\\{{\nwixident{cycle{\_}orthogonal}}{cycle:unorthogonal}}\\{{\nwixident{ex}}{ex}}\\{{\nwixident{l}}{l}}}\nwindexuse{\nwixident{cycle{\_}orthogonal}}{cycle:unorthogonal}{NWppJ6t-2t2vfS-7}\nwindexuse{\nwixident{ex}}{ex}{NWppJ6t-2t2vfS-7}\nwindexuse{\nwixident{l}}{l}{NWppJ6t-2t2vfS-7}\nwendcode{}\nwbegindocs{123}\nwdocspar
\nwenddocs{}\nwbegindocs{124}This is assisted by the trigonometric substitution \(\cos^2(*)=1-\sin^2(*)\)
used for parameters of {\Tt{}\Rm{}{\it{}l}\nwendquote}.
\nwenddocs{}\nwbegincode{125}\sublabel{NWppJ6t-2t2vfS-8}\nwmargintag{{\nwtagstyle{}\subpageref{NWppJ6t-2t2vfS-8}}}\moddef{figure-ortho-anlytic-proof.cpp~{\nwtagstyle{}\subpageref{NWppJ6t-2t2vfS-1}}}\plusendmoddef\Rm{}\nwstartdeflinemarkup\nwprevnextdefs{NWppJ6t-2t2vfS-7}{NWppJ6t-2t2vfS-9}\nwenddeflinemarkup
    {\bf{}for} ({\it{}size\_t} {\it{}i}=0; {\it{}i}\begin{math}<\end{math} {\it{}Res}.{\it{}nops}(); \protect\PP{\it{}i}) {\nwlbrace}
        {\it{}cout} \begin{math}\ll\end{math} {\tt{}"Tangent and radius are orthogonal: "} \begin{math}\ll\end{math} {\it{}boolalpha}
             \begin{math}\ll\end{math} {\bf{}bool}({\it{}ex\_to}\begin{math}<\end{math}{\bf{}relational}\begin{math}>\end{math}({\it{}Res}.{\it{}op}({\it{}i}).{\it{}subs}({\it{}pow}({\it{}cos}({\it{}wild}(0)),2)\begin{math}\equiv\end{math}1-{\it{}pow}({\it{}sin}({\it{}wild}(0)),2)).{\it{}normal}()))
             \begin{math}\ll\end{math} {\it{}endl};
    {\nwrbrace}

\nwidentuses{\\{{\nwixident{nops}}{nops}}\\{{\nwixident{op}}{op}}\\{{\nwixident{subs}}{subs}}}\nwindexuse{\nwixident{nops}}{nops}{NWppJ6t-2t2vfS-8}\nwindexuse{\nwixident{op}}{op}{NWppJ6t-2t2vfS-8}\nwindexuse{\nwixident{subs}}{subs}{NWppJ6t-2t2vfS-8}\nwendcode{}\nwbegindocs{126}The output predictably is:
\begin{verbatim}
Tangent and circle r are orthogonal: true
Tangent and circle r are orthogonal: true
\end{verbatim}

\nwenddocs{}\nwbegindocs{127}An additional check. We add a point \((1,0)\) on {\Tt{}\Rm{}{\it{}c}\nwendquote}\ldots
\nwenddocs{}\nwbegincode{128}\sublabel{NWppJ6t-2t2vfS-9}\nwmargintag{{\nwtagstyle{}\subpageref{NWppJ6t-2t2vfS-9}}}\moddef{figure-ortho-anlytic-proof.cpp~{\nwtagstyle{}\subpageref{NWppJ6t-2t2vfS-1}}}\plusendmoddef\Rm{}\nwstartdeflinemarkup\nwprevnextdefs{NWppJ6t-2t2vfS-8}{NWppJ6t-2t2vfS-A}\nwenddeflinemarkup
    {\bf{}ex} {\it{}B}={\it{}F}.{\it{}add\_cycle}({\bf{}cycle2D}({\bf{}lst}{\nwlbrace}1,0{\nwrbrace}),{\tt{}"B"});

\nwidentuses{\\{{\nwixident{add{\_}cycle}}{add:uncycle}}\\{{\nwixident{ex}}{ex}}}\nwindexuse{\nwixident{add{\_}cycle}}{add:uncycle}{NWppJ6t-2t2vfS-9}\nwindexuse{\nwixident{ex}}{ex}{NWppJ6t-2t2vfS-9}\nwendcode{}\nwbegindocs{129}\ldots and a generic cycle touching to {\Tt{}\Rm{}{\it{}c}\nwendquote} at {\Tt{}\Rm{}{\it{}B}\nwendquote}.
\nwenddocs{}\nwbegincode{130}\sublabel{NWppJ6t-2t2vfS-A}\nwmargintag{{\nwtagstyle{}\subpageref{NWppJ6t-2t2vfS-A}}}\moddef{figure-ortho-anlytic-proof.cpp~{\nwtagstyle{}\subpageref{NWppJ6t-2t2vfS-1}}}\plusendmoddef\Rm{}\nwstartdeflinemarkup\nwprevnextdefs{NWppJ6t-2t2vfS-9}{NWppJ6t-2t2vfS-B}\nwenddeflinemarkup
    {\bf{}ex} {\it{}b}={\bf{}symbol}({\tt{}"b"});
    {\it{}F}.{\it{}add\_cycle\_rel}({\bf{}lst}{\nwlbrace}{\it{}is\_tangent}({\it{}a}),{\it{}is\_orthogonal}({\it{}B}),{\it{}only\_reals}({\it{}b}){\nwrbrace}, {\it{}b});

\nwidentuses{\\{{\nwixident{add{\_}cycle{\_}rel}}{add:uncycle:unrel}}\\{{\nwixident{ex}}{ex}}\\{{\nwixident{is{\_}orthogonal}}{is:unorthogonal}}\\{{\nwixident{is{\_}tangent}}{is:untangent}}\\{{\nwixident{only{\_}reals}}{only:unreals}}}\nwindexuse{\nwixident{add{\_}cycle{\_}rel}}{add:uncycle:unrel}{NWppJ6t-2t2vfS-A}\nwindexuse{\nwixident{ex}}{ex}{NWppJ6t-2t2vfS-A}\nwindexuse{\nwixident{is{\_}orthogonal}}{is:unorthogonal}{NWppJ6t-2t2vfS-A}\nwindexuse{\nwixident{is{\_}tangent}}{is:untangent}{NWppJ6t-2t2vfS-A}\nwindexuse{\nwixident{only{\_}reals}}{only:unreals}{NWppJ6t-2t2vfS-A}\nwendcode{}\nwbegindocs{131}Add zero-radius cycles at the centres of {\Tt{}\Rm{}{\it{}a}\nwendquote} and {\Tt{}\Rm{}{\it{}b}\nwendquote}\ldots
\nwenddocs{}\nwbegincode{132}\sublabel{NWppJ6t-2t2vfS-B}\nwmargintag{{\nwtagstyle{}\subpageref{NWppJ6t-2t2vfS-B}}}\moddef{figure-ortho-anlytic-proof.cpp~{\nwtagstyle{}\subpageref{NWppJ6t-2t2vfS-1}}}\plusendmoddef\Rm{}\nwstartdeflinemarkup\nwprevnextdefs{NWppJ6t-2t2vfS-A}{NWppJ6t-2t2vfS-C}\nwenddeflinemarkup
    {\bf{}ex} {\it{}Ca}={\it{}F}.{\it{}add\_cycle}({\bf{}cycle2D}({\it{}ex\_to}\begin{math}<\end{math}{\bf{}lst}\begin{math}>\end{math}({\it{}ex\_to}\begin{math}<\end{math}{\bf{}cycle2D}\begin{math}>\end{math}({\it{}F}.{\it{}get\_cycle}({\it{}a}).{\it{}op}(0)).{\it{}center}())),{\tt{}"Ca"});
    {\bf{}ex} {\it{}Cb}={\it{}F}.{\it{}add\_cycle}({\bf{}cycle2D}({\it{}ex\_to}\begin{math}<\end{math}{\bf{}lst}\begin{math}>\end{math}({\it{}ex\_to}\begin{math}<\end{math}{\bf{}cycle2D}\begin{math}>\end{math}({\it{}F}.{\it{}get\_cycle}({\it{}b}).{\it{}op}(0)).{\it{}center}())),{\tt{}"Cb"});

\nwidentuses{\\{{\nwixident{add{\_}cycle}}{add:uncycle}}\\{{\nwixident{ex}}{ex}}\\{{\nwixident{get{\_}cycle}}{get:uncycle}}\\{{\nwixident{op}}{op}}}\nwindexuse{\nwixident{add{\_}cycle}}{add:uncycle}{NWppJ6t-2t2vfS-B}\nwindexuse{\nwixident{ex}}{ex}{NWppJ6t-2t2vfS-B}\nwindexuse{\nwixident{get{\_}cycle}}{get:uncycle}{NWppJ6t-2t2vfS-B}\nwindexuse{\nwixident{op}}{op}{NWppJ6t-2t2vfS-B}\nwendcode{}\nwbegindocs{133}\ldots and then a cycle passing two centres and the contact point.
\nwenddocs{}\nwbegincode{134}\sublabel{NWppJ6t-2t2vfS-C}\nwmargintag{{\nwtagstyle{}\subpageref{NWppJ6t-2t2vfS-C}}}\moddef{figure-ortho-anlytic-proof.cpp~{\nwtagstyle{}\subpageref{NWppJ6t-2t2vfS-1}}}\plusendmoddef\Rm{}\nwstartdeflinemarkup\nwprevnextdefs{NWppJ6t-2t2vfS-B}{NWppJ6t-2t2vfS-D}\nwenddeflinemarkup
    {\bf{}ex} {\it{}d}={\it{}F}.{\it{}add\_cycle\_rel}({\bf{}lst}{\nwlbrace}{\it{}is\_orthogonal}({\it{}B}), {\it{}is\_orthogonal}({\it{}Ca}), {\it{}is\_orthogonal}({\it{}Cb}){\nwrbrace},{\tt{}"d"});

\nwidentuses{\\{{\nwixident{add{\_}cycle{\_}rel}}{add:uncycle:unrel}}\\{{\nwixident{ex}}{ex}}\\{{\nwixident{is{\_}orthogonal}}{is:unorthogonal}}}\nwindexuse{\nwixident{add{\_}cycle{\_}rel}}{add:uncycle:unrel}{NWppJ6t-2t2vfS-C}\nwindexuse{\nwixident{ex}}{ex}{NWppJ6t-2t2vfS-C}\nwindexuse{\nwixident{is{\_}orthogonal}}{is:unorthogonal}{NWppJ6t-2t2vfS-C}\nwendcode{}\nwbegindocs{135}Finally check that the cycle {\Tt{}\Rm{}{\it{}d}\nwendquote} is a line (passes the infinity).
\nwenddocs{}\nwbegincode{136}\sublabel{NWppJ6t-2t2vfS-D}\nwmargintag{{\nwtagstyle{}\subpageref{NWppJ6t-2t2vfS-D}}}\moddef{figure-ortho-anlytic-proof.cpp~{\nwtagstyle{}\subpageref{NWppJ6t-2t2vfS-1}}}\plusendmoddef\Rm{}\nwstartdeflinemarkup\nwprevnextdefs{NWppJ6t-2t2vfS-C}{\relax}\nwenddeflinemarkup
    {\it{}Res} = {\it{}F}.{\it{}check\_rel}({\it{}d}, {\it{}F}.{\it{}get\_infinity}(), {\it{}cycle\_orthogonal});
    {\bf{}for} ({\it{}size\_t} {\it{}i}=0; {\it{}i}\begin{math}<\end{math} {\it{}Res}.{\it{}nops}(); \protect\PP{\it{}i})
        {\it{}cout} \begin{math}\ll\end{math} {\tt{}"Centres and the contact point are collinear: "}
             \begin{math}\ll\end{math} {\bf{}bool}({\it{}ex\_to}\begin{math}<\end{math}{\bf{}relational}\begin{math}>\end{math}({\it{}Res}.{\it{}op}({\it{}i})))
             \begin{math}\ll\end{math} {\it{}endl};
{\nwrbrace}

\nwidentuses{\\{{\nwixident{check{\_}rel}}{check:unrel}}\\{{\nwixident{cycle{\_}orthogonal}}{cycle:unorthogonal}}\\{{\nwixident{get{\_}infinity}}{get:uninfinity}}\\{{\nwixident{nops}}{nops}}\\{{\nwixident{op}}{op}}}\nwindexuse{\nwixident{check{\_}rel}}{check:unrel}{NWppJ6t-2t2vfS-D}\nwindexuse{\nwixident{cycle{\_}orthogonal}}{cycle:unorthogonal}{NWppJ6t-2t2vfS-D}\nwindexuse{\nwixident{get{\_}infinity}}{get:uninfinity}{NWppJ6t-2t2vfS-D}\nwindexuse{\nwixident{nops}}{nops}{NWppJ6t-2t2vfS-D}\nwindexuse{\nwixident{op}}{op}{NWppJ6t-2t2vfS-D}\nwendcode{}\nwbegindocs{137}The output, as expected, is:
\begin{verbatim}
Centres and the contact point are collinear: true
\end{verbatim}
\nwenddocs{}\nwbegindocs{138}\nwdocspar
\nwenddocs{}\nwbegincode{139}\sublabel{NWppJ6t-1fWRR8-5}\nwmargintag{{\nwtagstyle{}\subpageref{NWppJ6t-1fWRR8-5}}}\moddef{separating chunk~{\nwtagstyle{}\subpageref{NWppJ6t-1fWRR8-1}}}\plusendmoddef\Rm{}\nwstartdeflinemarkup\nwprevnextdefs{NWppJ6t-1fWRR8-4}{NWppJ6t-1fWRR8-6}\nwenddeflinemarkup

\nwendcode{}\nwbegindocs{140}\nwdocspar
\subsection{The nine-points theorem---conformal version}
\label{sec:example:-nine-points}

Here we present further usage of the library by an aesthetically
attractive example, see Section~\ref{sec:mathematical-results}.

\nwenddocs{}\nwbegindocs{141}The start of our file is minimalistic, we definitely need to
include the header of {\Tt{}\Rm{}{\bf{}figure}\nwendquote} library.
\nwenddocs{}\nwbegincode{142}\sublabel{NWppJ6t-30Tc7D-1}\nwmargintag{{\nwtagstyle{}\subpageref{NWppJ6t-30Tc7D-1}}}\moddef{nine-points-thm.cpp~{\nwtagstyle{}\subpageref{NWppJ6t-30Tc7D-1}}}\endmoddef\Rm{}\nwstartdeflinemarkup\nwusesondefline{\\{NWppJ6t-1p0Y9w-1}}\nwprevnextdefs{\relax}{NWppJ6t-30Tc7D-2}\nwenddeflinemarkup
\LA{}license~{\nwtagstyle{}\subpageref{NWppJ6t-ZXuKx-1}}\RA{}
{\bf{}\char35{}include}{\tt{} "figure.h"}
\LA{}using all namespaces~{\nwtagstyle{}\subpageref{NWppJ6t-3uHj2c-1}}\RA{}
{\bf{}int} {\it{}main}(){\nwlbrace}\nwindexdefn{\nwixident{main}}{main}{NWppJ6t-30Tc7D-1}
    \LA{}initial data for drawing~{\nwtagstyle{}\subpageref{NWppJ6t-17vXT1-1}}\RA{}
    \LA{}build medioscribed cycle~{\nwtagstyle{}\subpageref{NWppJ6t-2LTMm-1}}\RA{}

\nwalsodefined{\\{NWppJ6t-30Tc7D-2}\\{NWppJ6t-30Tc7D-3}\\{NWppJ6t-30Tc7D-4}\\{NWppJ6t-30Tc7D-5}\\{NWppJ6t-30Tc7D-6}\\{NWppJ6t-30Tc7D-7}\\{NWppJ6t-30Tc7D-8}\\{NWppJ6t-30Tc7D-9}\\{NWppJ6t-30Tc7D-A}\\{NWppJ6t-30Tc7D-B}\\{NWppJ6t-30Tc7D-C}}\nwused{\\{NWppJ6t-1p0Y9w-1}}\nwidentdefs{\\{{\nwixident{main}}{main}}}\nwidentuses{\\{{\nwixident{figure}}{figure}}}\nwindexuse{\nwixident{figure}}{figure}{NWppJ6t-30Tc7D-1}\nwendcode{}\nwbegindocs{143}We define exact coordinates of points which will be used for our picture.
\nwenddocs{}\nwbegincode{144}\sublabel{NWppJ6t-17vXT1-1}\nwmargintag{{\nwtagstyle{}\subpageref{NWppJ6t-17vXT1-1}}}\moddef{initial data for drawing~{\nwtagstyle{}\subpageref{NWppJ6t-17vXT1-1}}}\endmoddef\Rm{}\nwstartdeflinemarkup\nwusesondefline{\\{NWppJ6t-30Tc7D-1}}\nwenddeflinemarkup
    {\bf{}numeric} {\it{}x1}(-10,10), {\it{}y1}(0,1), {\it{}x2}(10,10), {\it{}y2}(0,1), {\it{}x3}(-1,5), {\it{}y3}(-3,2), {\it{}x4}(1,2), {\it{}y4}(-5,2);
    {\bf{}int} {\it{}sign}=-1;
\nwindexdefn{\nwixident{numeric}}{numeric}{NWppJ6t-17vXT1-1}\eatline
\nwused{\\{NWppJ6t-30Tc7D-1}}\nwidentdefs{\\{{\nwixident{numeric}}{numeric}}}\nwendcode{}\nwbegindocs{145}\nwdocspar
\nwenddocs{}\nwbegindocs{146}We declare the figure {\Tt{}\Rm{}{\it{}F}\nwendquote} which will be constructed.
\nwenddocs{}\nwbegincode{147}\sublabel{NWppJ6t-2LTMm-1}\nwmargintag{{\nwtagstyle{}\subpageref{NWppJ6t-2LTMm-1}}}\moddef{build medioscribed cycle~{\nwtagstyle{}\subpageref{NWppJ6t-2LTMm-1}}}\endmoddef\Rm{}\nwstartdeflinemarkup\nwusesondefline{\\{NWppJ6t-30Tc7D-1}\\{NWppJ6t-2zWvnC-1}}\nwprevnextdefs{\relax}{NWppJ6t-2LTMm-2}\nwenddeflinemarkup
    {\bf{}figure} {\it{}F}({\bf{}lst}{\nwlbrace}-1,{\it{}sign}{\nwrbrace});
\nwindexdefn{\nwixident{figure}}{figure}{NWppJ6t-2LTMm-1}\eatline
\nwalsodefined{\\{NWppJ6t-2LTMm-2}\\{NWppJ6t-2LTMm-3}\\{NWppJ6t-2LTMm-4}\\{NWppJ6t-2LTMm-5}\\{NWppJ6t-2LTMm-6}\\{NWppJ6t-2LTMm-7}\\{NWppJ6t-2LTMm-8}\\{NWppJ6t-2LTMm-9}\\{NWppJ6t-2LTMm-A}\\{NWppJ6t-2LTMm-B}\\{NWppJ6t-2LTMm-C}\\{NWppJ6t-2LTMm-D}\\{NWppJ6t-2LTMm-E}\\{NWppJ6t-2LTMm-F}}\nwused{\\{NWppJ6t-30Tc7D-1}\\{NWppJ6t-2zWvnC-1}}\nwidentdefs{\\{{\nwixident{figure}}{figure}}}\nwendcode{}\nwbegindocs{148}\nwdocspar
\nwenddocs{}\nwbegindocs{149}We will need several ``midpoints'' in our constructions, the
corresponding figure {\Tt{}\Rm{}{\it{}midpoint\_constructor}\nwendquote} is readily available
from the library.
\nwenddocs{}\nwbegincode{150}\sublabel{NWppJ6t-2LTMm-2}\nwmargintag{{\nwtagstyle{}\subpageref{NWppJ6t-2LTMm-2}}}\moddef{build medioscribed cycle~{\nwtagstyle{}\subpageref{NWppJ6t-2LTMm-1}}}\plusendmoddef\Rm{}\nwstartdeflinemarkup\nwusesondefline{\\{NWppJ6t-30Tc7D-1}\\{NWppJ6t-2zWvnC-1}}\nwprevnextdefs{NWppJ6t-2LTMm-1}{NWppJ6t-2LTMm-3}\nwenddeflinemarkup
    {\bf{}figure} {\it{}SF}={\it{}ex\_to}\begin{math}<\end{math}{\bf{}figure}\begin{math}>\end{math}({\it{}midpoint\_constructor}());

\nwused{\\{NWppJ6t-30Tc7D-1}\\{NWppJ6t-2zWvnC-1}}\nwidentuses{\\{{\nwixident{figure}}{figure}}\\{{\nwixident{midpoint{\_}constructor}}{midpoint:unconstructor}}}\nwindexuse{\nwixident{figure}}{figure}{NWppJ6t-2LTMm-2}\nwindexuse{\nwixident{midpoint{\_}constructor}}{midpoint:unconstructor}{NWppJ6t-2LTMm-2}\nwendcode{}\nwbegindocs{151}Next we define vertices of the ``triangle'' {\Tt{}\Rm{}{\it{}A}\nwendquote}, {\Tt{}\Rm{}{\it{}B}\nwendquote}, {\Tt{}\Rm{}{\it{}C}\nwendquote} and
the point {\Tt{}\Rm{}{\it{}N}\nwendquote} which will be an image if infinity. Every point is
added to {\Tt{}\Rm{}{\it{}F}\nwendquote} by giving its explicit coordinates and a string, which
will used to label it. The returned value is a \GiNaC\ expression of
{\Tt{}\Rm{}{\bf{}symbol}\nwendquote} class which will be used as the key of a respective
point. All points are added to the zero generation.
\nwenddocs{}\nwbegincode{152}\sublabel{NWppJ6t-2LTMm-3}\nwmargintag{{\nwtagstyle{}\subpageref{NWppJ6t-2LTMm-3}}}\moddef{build medioscribed cycle~{\nwtagstyle{}\subpageref{NWppJ6t-2LTMm-1}}}\plusendmoddef\Rm{}\nwstartdeflinemarkup\nwusesondefline{\\{NWppJ6t-30Tc7D-1}\\{NWppJ6t-2zWvnC-1}}\nwprevnextdefs{NWppJ6t-2LTMm-2}{NWppJ6t-2LTMm-4}\nwenddeflinemarkup
    {\bf{}ex} {\it{}A}={\it{}F}.{\it{}add\_point}({\bf{}lst}{\nwlbrace}{\it{}x1},{\it{}y1}{\nwrbrace},{\tt{}"A"});
    {\bf{}ex} {\it{}B}={\it{}F}.{\it{}add\_point}({\bf{}lst}{\nwlbrace}{\it{}x2}, {\it{}y2}{\nwrbrace},{\tt{}"B"});
    {\bf{}ex} {\it{}C}={\it{}F}.{\it{}add\_point}({\bf{}lst}{\nwlbrace}{\it{}x3},{\it{}y3}{\nwrbrace},{\tt{}"C"});
\nwindexdefn{\nwixident{add{\_}point}}{add:unpoint}{NWppJ6t-2LTMm-3}\eatline
\nwused{\\{NWppJ6t-30Tc7D-1}\\{NWppJ6t-2zWvnC-1}}\nwidentdefs{\\{{\nwixident{add{\_}point}}{add:unpoint}}}\nwidentuses{\\{{\nwixident{ex}}{ex}}}\nwindexuse{\nwixident{ex}}{ex}{NWppJ6t-2LTMm-3}\nwendcode{}\nwbegindocs{153}\nwdocspar
\nwenddocs{}\nwbegindocs{154}There is the special point in the construction, which play the role
of infinity. We first put this as cycle at infinity to make picture simple.
\nwenddocs{}\nwbegincode{155}\sublabel{NWppJ6t-2LTMm-4}\nwmargintag{{\nwtagstyle{}\subpageref{NWppJ6t-2LTMm-4}}}\moddef{build medioscribed cycle~{\nwtagstyle{}\subpageref{NWppJ6t-2LTMm-1}}}\plusendmoddef\Rm{}\nwstartdeflinemarkup\nwusesondefline{\\{NWppJ6t-30Tc7D-1}\\{NWppJ6t-2zWvnC-1}}\nwprevnextdefs{NWppJ6t-2LTMm-3}{NWppJ6t-2LTMm-5}\nwenddeflinemarkup
    {\bf{}ex} {\it{}N}={\it{}F}.{\it{}add\_cycle}({\bf{}cycle\_data}(0,{\bf{}lst}{\nwlbrace}0,0{\nwrbrace},1),{\tt{}"N"});
\nwindexdefn{\nwixident{add{\_}cycle}}{add:uncycle}{NWppJ6t-2LTMm-4}\nwindexdefn{\nwixident{cycle{\_}data}}{cycle:undata}{NWppJ6t-2LTMm-4}\eatline
\nwused{\\{NWppJ6t-30Tc7D-1}\\{NWppJ6t-2zWvnC-1}}\nwidentdefs{\\{{\nwixident{add{\_}cycle}}{add:uncycle}}\\{{\nwixident{cycle{\_}data}}{cycle:undata}}}\nwidentuses{\\{{\nwixident{ex}}{ex}}}\nwindexuse{\nwixident{ex}}{ex}{NWppJ6t-2LTMm-4}\nwendcode{}\nwbegindocs{156}\nwdocspar
\nwenddocs{}\nwbegindocs{157}This is an alternative selection of point with {\Tt{}\Rm{}{\it{}N}\nwendquote} being at the
centre of the triangle.
\nwenddocs{}\nwbegincode{158}\sublabel{NWppJ6t-2LTMm-5}\nwmargintag{{\nwtagstyle{}\subpageref{NWppJ6t-2LTMm-5}}}\moddef{build medioscribed cycle~{\nwtagstyle{}\subpageref{NWppJ6t-2LTMm-1}}}\plusendmoddef\Rm{}\nwstartdeflinemarkup\nwusesondefline{\\{NWppJ6t-30Tc7D-1}\\{NWppJ6t-2zWvnC-1}}\nwprevnextdefs{NWppJ6t-2LTMm-4}{NWppJ6t-2LTMm-6}\nwenddeflinemarkup
    //Fully symmetric data
    // ex A=F.add\_point(lst{\nwlbrace}-numeric(10,10{\nwrbrace},numeric(0,1)),"A")
    // ex B=F.add\_point(lst{\nwlbrace}numeric(10,10{\nwrbrace},numeric(0,1)),"B")
    // ex C=F.add\_point(lst{\nwlbrace}numeric(0,4{\nwrbrace},-numeric(1732050807,1000000000)),"C")
    // ex N=F.add\_point(lst{\nwlbrace}numeric(0,4{\nwrbrace},-numeric(577350269,1000000000)),"N")

\nwused{\\{NWppJ6t-30Tc7D-1}\\{NWppJ6t-2zWvnC-1}}\nwidentuses{\\{{\nwixident{add{\_}point}}{add:unpoint}}\\{{\nwixident{ex}}{ex}}\\{{\nwixident{numeric}}{numeric}}}\nwindexuse{\nwixident{add{\_}point}}{add:unpoint}{NWppJ6t-2LTMm-5}\nwindexuse{\nwixident{ex}}{ex}{NWppJ6t-2LTMm-5}\nwindexuse{\nwixident{numeric}}{numeric}{NWppJ6t-2LTMm-5}\nwendcode{}\nwbegindocs{159}Now we add ``sides'' of the triangle, that is cycles passing two
vertices and {\Tt{}\Rm{}{\it{}N}\nwendquote}. A cycle passes a point if it is orthogonal to the
cycle defined by this point. Thus, each side is defined through a list
of three orthogonalities~\citelist{\cite{Kisil06a}
  \cite{Kisil12a}*{Defn.~6.1}}. We also supply a string to label this
side. The returned valued is a {\Tt{}\Rm{}{\bf{}symbol}\nwendquote} which is a key for this cycle.
\nwenddocs{}\nwbegincode{160}\sublabel{NWppJ6t-2LTMm-6}\nwmargintag{{\nwtagstyle{}\subpageref{NWppJ6t-2LTMm-6}}}\moddef{build medioscribed cycle~{\nwtagstyle{}\subpageref{NWppJ6t-2LTMm-1}}}\plusendmoddef\Rm{}\nwstartdeflinemarkup\nwusesondefline{\\{NWppJ6t-30Tc7D-1}\\{NWppJ6t-2zWvnC-1}}\nwprevnextdefs{NWppJ6t-2LTMm-5}{NWppJ6t-2LTMm-7}\nwenddeflinemarkup
    {\bf{}ex} {\it{}a}={\it{}F}.{\it{}add\_cycle\_rel}({\bf{}lst}{\nwlbrace}{\it{}is\_orthogonal}({\it{}B}),{\it{}is\_orthogonal}({\it{}C}),{\it{}is\_orthogonal}({\it{}N}){\nwrbrace},{\tt{}"a"});
    {\bf{}ex} {\it{}b}={\it{}F}.{\it{}add\_cycle\_rel}({\bf{}lst}{\nwlbrace}{\it{}is\_orthogonal}({\it{}A}),{\it{}is\_orthogonal}({\it{}C}),{\it{}is\_orthogonal}({\it{}N}){\nwrbrace},{\tt{}"b"});
    {\bf{}ex} {\it{}c}={\it{}F}.{\it{}add\_cycle\_rel}({\bf{}lst}{\nwlbrace}{\it{}is\_orthogonal}({\it{}A}),{\it{}is\_orthogonal}({\it{}B}),{\it{}is\_orthogonal}({\it{}N}){\nwrbrace},{\tt{}"c"});
\nwindexdefn{\nwixident{add{\_}cycle{\_}rel}}{add:uncycle:unrel}{NWppJ6t-2LTMm-6}\nwindexdefn{\nwixident{is{\_}orthogonal}}{is:unorthogonal}{NWppJ6t-2LTMm-6}\eatline
\nwused{\\{NWppJ6t-30Tc7D-1}\\{NWppJ6t-2zWvnC-1}}\nwidentdefs{\\{{\nwixident{add{\_}cycle{\_}rel}}{add:uncycle:unrel}}\\{{\nwixident{is{\_}orthogonal}}{is:unorthogonal}}}\nwidentuses{\\{{\nwixident{ex}}{ex}}}\nwindexuse{\nwixident{ex}}{ex}{NWppJ6t-2LTMm-6}\nwendcode{}\nwbegindocs{161}\nwdocspar
\nwenddocs{}\nwbegindocs{162}We define the custom \Asymptote~\cite{Asymptote} drawing style for
sides of the triangle: the dark blue ({\Tt{}\Rm{}{\it{}rgb}\nwendquote} colour (0,0,0.8)) and line thickness 1pt.
\nwenddocs{}\nwbegincode{163}\sublabel{NWppJ6t-2LTMm-7}\nwmargintag{{\nwtagstyle{}\subpageref{NWppJ6t-2LTMm-7}}}\moddef{build medioscribed cycle~{\nwtagstyle{}\subpageref{NWppJ6t-2LTMm-1}}}\plusendmoddef\Rm{}\nwstartdeflinemarkup\nwusesondefline{\\{NWppJ6t-30Tc7D-1}\\{NWppJ6t-2zWvnC-1}}\nwprevnextdefs{NWppJ6t-2LTMm-6}{NWppJ6t-2LTMm-8}\nwenddeflinemarkup
    {\it{}F}.{\it{}set\_asy\_style}({\it{}a},{\tt{}"rgb(0,0,.8)+1"});
    {\it{}F}.{\it{}set\_asy\_style}({\it{}b},{\tt{}"rgb(0,0,.8)+1"});
    {\it{}F}.{\it{}set\_asy\_style}({\it{}c},{\tt{}"rgb(0,0,.8)+1"});
\nwindexdefn{\nwixident{set{\_}asy{\_}style}}{set:unasy:unstyle}{NWppJ6t-2LTMm-7}\nwindexdefn{\nwixident{rgb}}{rgb}{NWppJ6t-2LTMm-7}\eatline
\nwused{\\{NWppJ6t-30Tc7D-1}\\{NWppJ6t-2zWvnC-1}}\nwidentdefs{\\{{\nwixident{rgb}}{rgb}}\\{{\nwixident{set{\_}asy{\_}style}}{set:unasy:unstyle}}}\nwendcode{}\nwbegindocs{164}\nwdocspar
\nwenddocs{}\nwbegindocs{165}Now we drop ``altitudes'' in our triangle, that is again
provided through three orthogonality relations. They will be draw as
dashed lines.
\nwenddocs{}\nwbegincode{166}\sublabel{NWppJ6t-2LTMm-8}\nwmargintag{{\nwtagstyle{}\subpageref{NWppJ6t-2LTMm-8}}}\moddef{build medioscribed cycle~{\nwtagstyle{}\subpageref{NWppJ6t-2LTMm-1}}}\plusendmoddef\Rm{}\nwstartdeflinemarkup\nwusesondefline{\\{NWppJ6t-30Tc7D-1}\\{NWppJ6t-2zWvnC-1}}\nwprevnextdefs{NWppJ6t-2LTMm-7}{NWppJ6t-2LTMm-9}\nwenddeflinemarkup
       {\bf{}ex} {\it{}ha}={\it{}F}.{\it{}add\_cycle\_rel}({\bf{}lst}{\nwlbrace}{\it{}is\_orthogonal}({\it{}A}),{\it{}is\_orthogonal}({\it{}N}),{\it{}is\_orthogonal}({\it{}a}){\nwrbrace},{\tt{}"h\_a"});
    {\it{}F}.{\it{}set\_asy\_style}({\it{}ha},{\tt{}"dashed"});
    {\bf{}ex} {\it{}hb}={\it{}F}.{\it{}add\_cycle\_rel}({\bf{}lst}{\nwlbrace}{\it{}is\_orthogonal}({\it{}B}),{\it{}is\_orthogonal}({\it{}N}),{\it{}is\_orthogonal}({\it{}b}){\nwrbrace},{\tt{}"h\_b"});
    {\it{}F}.{\it{}set\_asy\_style}({\it{}hb},{\tt{}"dashed"});
    {\bf{}ex} {\it{}hc}={\it{}F}.{\it{}add\_cycle\_rel}({\bf{}lst}{\nwlbrace}{\it{}is\_orthogonal}({\it{}C}),{\it{}is\_orthogonal}({\it{}N}),{\it{}is\_orthogonal}({\it{}c}){\nwrbrace},{\tt{}"h\_c"});
    {\it{}F}.{\it{}set\_asy\_style}({\it{}hc},{\tt{}"dashed"});

\nwused{\\{NWppJ6t-30Tc7D-1}\\{NWppJ6t-2zWvnC-1}}\nwidentuses{\\{{\nwixident{add{\_}cycle{\_}rel}}{add:uncycle:unrel}}\\{{\nwixident{ex}}{ex}}\\{{\nwixident{is{\_}orthogonal}}{is:unorthogonal}}\\{{\nwixident{set{\_}asy{\_}style}}{set:unasy:unstyle}}}\nwindexuse{\nwixident{add{\_}cycle{\_}rel}}{add:uncycle:unrel}{NWppJ6t-2LTMm-8}\nwindexuse{\nwixident{ex}}{ex}{NWppJ6t-2LTMm-8}\nwindexuse{\nwixident{is{\_}orthogonal}}{is:unorthogonal}{NWppJ6t-2LTMm-8}\nwindexuse{\nwixident{set{\_}asy{\_}style}}{set:unasy:unstyle}{NWppJ6t-2LTMm-8}\nwendcode{}\nwbegindocs{167}We need the base of altitude {\Tt{}\Rm{}{\it{}ha}\nwendquote}, which is the intersection points
of the side {\Tt{}\Rm{}{\it{}a}\nwendquote}  and respective altitude {\Tt{}\Rm{}{\it{}ha}\nwendquote}. A point can be can
be characterised as a cycle which is orthogonal to
itself~\citelist{\cite{Kisil12a}*{Defn.~5.13} \cite{Kisil06a}}. To
define a relation of a cycle to itself we first need to define a
symbol {\Tt{}\Rm{}{\it{}A1}\nwendquote} and then add a cycle with the key {\Tt{}\Rm{}{\it{}A1}\nwendquote} and the
relation {\Tt{}\Rm{}{\it{}is\_orthogonal}\nwendquote} to {\Tt{}\Rm{}{\it{}A1}\nwendquote}. Finally, there are two such
points: the base of altitude and {\Tt{}\Rm{}{\it{}N}\nwendquote}. To exclude the second one we
add the relation {\Tt{}\Rm{}{\it{}is\_adifferent}\nwendquote} (``almost different'') to {\Tt{}\Rm{}{\it{}N}\nwendquote}.
\nwenddocs{}\nwbegincode{168}\sublabel{NWppJ6t-2LTMm-9}\nwmargintag{{\nwtagstyle{}\subpageref{NWppJ6t-2LTMm-9}}}\moddef{build medioscribed cycle~{\nwtagstyle{}\subpageref{NWppJ6t-2LTMm-1}}}\plusendmoddef\Rm{}\nwstartdeflinemarkup\nwusesondefline{\\{NWppJ6t-30Tc7D-1}\\{NWppJ6t-2zWvnC-1}}\nwprevnextdefs{NWppJ6t-2LTMm-8}{NWppJ6t-2LTMm-A}\nwenddeflinemarkup
    {\bf{}ex} {\it{}A1}={\bf{}symbol}({\tt{}"A\_h"});
 {\it{}F}.{\it{}add\_cycle\_rel}({\bf{}lst}{\nwlbrace}{\it{}is\_orthogonal}({\it{}a}),{\it{}is\_orthogonal}({\it{}ha}),{\it{}is\_orthogonal}({\it{}A1}),{\it{}is\_adifferent}({\it{}N}){\nwrbrace},{\it{}A1});
\nwindexdefn{\nwixident{is{\_}adifferent}}{is:unadifferent}{NWppJ6t-2LTMm-9}\eatline
\nwused{\\{NWppJ6t-30Tc7D-1}\\{NWppJ6t-2zWvnC-1}}\nwidentdefs{\\{{\nwixident{is{\_}adifferent}}{is:unadifferent}}}\nwidentuses{\\{{\nwixident{add{\_}cycle{\_}rel}}{add:uncycle:unrel}}\\{{\nwixident{ex}}{ex}}\\{{\nwixident{is{\_}orthogonal}}{is:unorthogonal}}}\nwindexuse{\nwixident{add{\_}cycle{\_}rel}}{add:uncycle:unrel}{NWppJ6t-2LTMm-9}\nwindexuse{\nwixident{ex}}{ex}{NWppJ6t-2LTMm-9}\nwindexuse{\nwixident{is{\_}orthogonal}}{is:unorthogonal}{NWppJ6t-2LTMm-9}\nwendcode{}\nwbegindocs{169}\nwdocspar
\nwenddocs{}\nwbegindocs{170}Two other bases of altitude are defined in a similar manner.
\nwenddocs{}\nwbegincode{171}\sublabel{NWppJ6t-2LTMm-A}\nwmargintag{{\nwtagstyle{}\subpageref{NWppJ6t-2LTMm-A}}}\moddef{build medioscribed cycle~{\nwtagstyle{}\subpageref{NWppJ6t-2LTMm-1}}}\plusendmoddef\Rm{}\nwstartdeflinemarkup\nwusesondefline{\\{NWppJ6t-30Tc7D-1}\\{NWppJ6t-2zWvnC-1}}\nwprevnextdefs{NWppJ6t-2LTMm-9}{NWppJ6t-2LTMm-B}\nwenddeflinemarkup
    {\bf{}ex} {\it{}B1}={\bf{}symbol}({\tt{}"B\_h"});
 {\it{}F}.{\it{}add\_cycle\_rel}({\bf{}lst}{\nwlbrace}{\it{}is\_orthogonal}({\it{}b}),{\it{}is\_orthogonal}({\it{}hb}),{\it{}is\_adifferent}({\it{}N}),{\it{}is\_orthogonal}({\it{}B1}){\nwrbrace},{\it{}B1});
    {\bf{}ex} {\it{}C1}={\bf{}symbol}({\tt{}"C\_h"});
    {\it{}F}.{\it{}add\_cycle\_rel}({\bf{}lst}{\nwlbrace}{\it{}is\_adifferent}({\it{}N}),{\it{}is\_orthogonal}({\it{}c}),{\it{}is\_orthogonal}({\it{}hc}),{\it{}is\_orthogonal}({\it{}C1}){\nwrbrace},{\it{}C1});

\nwused{\\{NWppJ6t-30Tc7D-1}\\{NWppJ6t-2zWvnC-1}}\nwidentuses{\\{{\nwixident{add{\_}cycle{\_}rel}}{add:uncycle:unrel}}\\{{\nwixident{ex}}{ex}}\\{{\nwixident{is{\_}adifferent}}{is:unadifferent}}\\{{\nwixident{is{\_}orthogonal}}{is:unorthogonal}}}\nwindexuse{\nwixident{add{\_}cycle{\_}rel}}{add:uncycle:unrel}{NWppJ6t-2LTMm-A}\nwindexuse{\nwixident{ex}}{ex}{NWppJ6t-2LTMm-A}\nwindexuse{\nwixident{is{\_}adifferent}}{is:unadifferent}{NWppJ6t-2LTMm-A}\nwindexuse{\nwixident{is{\_}orthogonal}}{is:unorthogonal}{NWppJ6t-2LTMm-A}\nwendcode{}\nwbegindocs{172}We add the cycle passing all three bases of altitudes.
\nwenddocs{}\nwbegincode{173}\sublabel{NWppJ6t-2LTMm-B}\nwmargintag{{\nwtagstyle{}\subpageref{NWppJ6t-2LTMm-B}}}\moddef{build medioscribed cycle~{\nwtagstyle{}\subpageref{NWppJ6t-2LTMm-1}}}\plusendmoddef\Rm{}\nwstartdeflinemarkup\nwusesondefline{\\{NWppJ6t-30Tc7D-1}\\{NWppJ6t-2zWvnC-1}}\nwprevnextdefs{NWppJ6t-2LTMm-A}{NWppJ6t-2LTMm-C}\nwenddeflinemarkup
    {\bf{}ex} {\it{}p}={\it{}F}.{\it{}add\_cycle\_rel}({\bf{}lst}{\nwlbrace}{\it{}is\_orthogonal}({\it{}A1}),{\it{}is\_orthogonal}({\it{}B1}),{\it{}is\_orthogonal}({\it{}C1}){\nwrbrace},{\tt{}"p"});
    {\it{}F}.{\it{}set\_asy\_style}({\it{}p},{\tt{}"rgb(0,.8,0)+1"});

\nwused{\\{NWppJ6t-30Tc7D-1}\\{NWppJ6t-2zWvnC-1}}\nwidentuses{\\{{\nwixident{add{\_}cycle{\_}rel}}{add:uncycle:unrel}}\\{{\nwixident{ex}}{ex}}\\{{\nwixident{is{\_}orthogonal}}{is:unorthogonal}}\\{{\nwixident{rgb}}{rgb}}\\{{\nwixident{set{\_}asy{\_}style}}{set:unasy:unstyle}}}\nwindexuse{\nwixident{add{\_}cycle{\_}rel}}{add:uncycle:unrel}{NWppJ6t-2LTMm-B}\nwindexuse{\nwixident{ex}}{ex}{NWppJ6t-2LTMm-B}\nwindexuse{\nwixident{is{\_}orthogonal}}{is:unorthogonal}{NWppJ6t-2LTMm-B}\nwindexuse{\nwixident{rgb}}{rgb}{NWppJ6t-2LTMm-B}\nwindexuse{\nwixident{set{\_}asy{\_}style}}{set:unasy:unstyle}{NWppJ6t-2LTMm-B}\nwendcode{}\nwbegindocs{174}We build ``midpoint'' of the arc of {\Tt{}\Rm{}{\it{}a}\nwendquote} between {\Tt{}\Rm{}{\it{}B}\nwendquote} and
{\Tt{}\Rm{}{\it{}C}\nwendquote}. To this end we use subfigure {\Tt{}\Rm{}{\it{}SF}\nwendquote} and supply the list of
parameters {\Tt{}\Rm{}{\it{}B}\nwendquote}, {\Tt{}\Rm{}{\it{}C}\nwendquote} and {\Tt{}\Rm{}{\it{}N}\nwendquote} (``infinity'') which are required
by {\Tt{}\Rm{}{\it{}SF}\nwendquote}.
\nwenddocs{}\nwbegincode{175}\sublabel{NWppJ6t-2LTMm-C}\nwmargintag{{\nwtagstyle{}\subpageref{NWppJ6t-2LTMm-C}}}\moddef{build medioscribed cycle~{\nwtagstyle{}\subpageref{NWppJ6t-2LTMm-1}}}\plusendmoddef\Rm{}\nwstartdeflinemarkup\nwusesondefline{\\{NWppJ6t-30Tc7D-1}\\{NWppJ6t-2zWvnC-1}}\nwprevnextdefs{NWppJ6t-2LTMm-B}{NWppJ6t-2LTMm-D}\nwenddeflinemarkup
    {\bf{}ex} {\it{}A2}={\it{}F}.{\it{}add\_subfigure}({\it{}SF},{\bf{}lst}{\nwlbrace}{\it{}B},{\it{}C},{\it{}N}{\nwrbrace},{\tt{}"A\_m"});
\nwindexdefn{\nwixident{add{\_}subfigure}}{add:unsubfigure}{NWppJ6t-2LTMm-C}\eatline
\nwused{\\{NWppJ6t-30Tc7D-1}\\{NWppJ6t-2zWvnC-1}}\nwidentdefs{\\{{\nwixident{add{\_}subfigure}}{add:unsubfigure}}}\nwidentuses{\\{{\nwixident{ex}}{ex}}}\nwindexuse{\nwixident{ex}}{ex}{NWppJ6t-2LTMm-C}\nwendcode{}\nwbegindocs{176}\nwdocspar
\nwenddocs{}\nwbegindocs{177}Similarly we build other two ``midpoints'', they all will belong to
the cycle {\Tt{}\Rm{}{\it{}p}\nwendquote}.
\nwenddocs{}\nwbegincode{178}\sublabel{NWppJ6t-2LTMm-D}\nwmargintag{{\nwtagstyle{}\subpageref{NWppJ6t-2LTMm-D}}}\moddef{build medioscribed cycle~{\nwtagstyle{}\subpageref{NWppJ6t-2LTMm-1}}}\plusendmoddef\Rm{}\nwstartdeflinemarkup\nwusesondefline{\\{NWppJ6t-30Tc7D-1}\\{NWppJ6t-2zWvnC-1}}\nwprevnextdefs{NWppJ6t-2LTMm-C}{NWppJ6t-2LTMm-E}\nwenddeflinemarkup
    {\bf{}ex} {\it{}B2}={\it{}F}.{\it{}add\_subfigure}({\it{}SF},{\bf{}lst}{\nwlbrace}{\it{}C},{\it{}A},{\it{}N}{\nwrbrace},{\tt{}"B\_m"});
    {\bf{}ex} {\it{}C2}={\it{}F}.{\it{}add\_subfigure}({\it{}SF},{\bf{}lst}{\nwlbrace}{\it{}A},{\it{}B},{\it{}N}{\nwrbrace},{\tt{}"C\_m"});

\nwused{\\{NWppJ6t-30Tc7D-1}\\{NWppJ6t-2zWvnC-1}}\nwidentuses{\\{{\nwixident{add{\_}subfigure}}{add:unsubfigure}}\\{{\nwixident{ex}}{ex}}}\nwindexuse{\nwixident{add{\_}subfigure}}{add:unsubfigure}{NWppJ6t-2LTMm-D}\nwindexuse{\nwixident{ex}}{ex}{NWppJ6t-2LTMm-D}\nwendcode{}\nwbegindocs{179}{\Tt{}\Rm{}{\it{}O}\nwendquote} is the intersection point of altitudes {\Tt{}\Rm{}{\it{}ha}\nwendquote} and {\Tt{}\Rm{}{\it{}hb}\nwendquote},
again it is defined as a cycle with key {\Tt{}\Rm{}{\it{}O}\nwendquote} orthogonal to itself.
\nwenddocs{}\nwbegincode{180}\sublabel{NWppJ6t-2LTMm-E}\nwmargintag{{\nwtagstyle{}\subpageref{NWppJ6t-2LTMm-E}}}\moddef{build medioscribed cycle~{\nwtagstyle{}\subpageref{NWppJ6t-2LTMm-1}}}\plusendmoddef\Rm{}\nwstartdeflinemarkup\nwusesondefline{\\{NWppJ6t-30Tc7D-1}\\{NWppJ6t-2zWvnC-1}}\nwprevnextdefs{NWppJ6t-2LTMm-D}{NWppJ6t-2LTMm-F}\nwenddeflinemarkup
    {\bf{}ex} {\it{}O}={\bf{}symbol}({\tt{}"O"});
 {\it{}F}.{\it{}add\_cycle\_rel}({\bf{}lst}{\nwlbrace}{\it{}is\_orthogonal}({\it{}ha}),{\it{}is\_orthogonal}({\it{}hb}),{\it{}is\_orthogonal}({\it{}O}),{\it{}is\_adifferent}({\it{}N}){\nwrbrace},{\it{}O});

\nwused{\\{NWppJ6t-30Tc7D-1}\\{NWppJ6t-2zWvnC-1}}\nwidentuses{\\{{\nwixident{add{\_}cycle{\_}rel}}{add:uncycle:unrel}}\\{{\nwixident{ex}}{ex}}\\{{\nwixident{is{\_}adifferent}}{is:unadifferent}}\\{{\nwixident{is{\_}orthogonal}}{is:unorthogonal}}}\nwindexuse{\nwixident{add{\_}cycle{\_}rel}}{add:uncycle:unrel}{NWppJ6t-2LTMm-E}\nwindexuse{\nwixident{ex}}{ex}{NWppJ6t-2LTMm-E}\nwindexuse{\nwixident{is{\_}adifferent}}{is:unadifferent}{NWppJ6t-2LTMm-E}\nwindexuse{\nwixident{is{\_}orthogonal}}{is:unorthogonal}{NWppJ6t-2LTMm-E}\nwendcode{}\nwbegindocs{181}We build three more ``midpoints'' which belong to {\Tt{}\Rm{}{\it{}p}\nwendquote} as well.
\nwenddocs{}\nwbegincode{182}\sublabel{NWppJ6t-2LTMm-F}\nwmargintag{{\nwtagstyle{}\subpageref{NWppJ6t-2LTMm-F}}}\moddef{build medioscribed cycle~{\nwtagstyle{}\subpageref{NWppJ6t-2LTMm-1}}}\plusendmoddef\Rm{}\nwstartdeflinemarkup\nwusesondefline{\\{NWppJ6t-30Tc7D-1}\\{NWppJ6t-2zWvnC-1}}\nwprevnextdefs{NWppJ6t-2LTMm-E}{\relax}\nwenddeflinemarkup
    {\bf{}ex} {\it{}A3}={\it{}F}.{\it{}add\_subfigure}({\it{}SF},{\bf{}lst}{\nwlbrace}{\it{}O},{\it{}A},{\it{}N}{\nwrbrace},{\tt{}"A\_d"});
    {\bf{}ex} {\it{}B3}={\it{}F}.{\it{}add\_subfigure}({\it{}SF},{\bf{}lst}{\nwlbrace}{\it{}B},{\it{}O},{\it{}N}{\nwrbrace},{\tt{}"B\_d"});
    {\bf{}ex} {\it{}C3}={\it{}F}.{\it{}add\_subfigure}({\it{}SF},{\bf{}lst}{\nwlbrace}{\it{}C},{\it{}O},{\it{}N}{\nwrbrace},{\tt{}"C\_d"});

    \LA{}check the theorem~{\nwtagstyle{}\subpageref{NWppJ6t-4d7QoM-1}}\RA{}

\nwused{\\{NWppJ6t-30Tc7D-1}\\{NWppJ6t-2zWvnC-1}}\nwidentuses{\\{{\nwixident{add{\_}subfigure}}{add:unsubfigure}}\\{{\nwixident{ex}}{ex}}}\nwindexuse{\nwixident{add{\_}subfigure}}{add:unsubfigure}{NWppJ6t-2LTMm-F}\nwindexuse{\nwixident{ex}}{ex}{NWppJ6t-2LTMm-F}\nwendcode{}\nwbegindocs{183}Now we want to check that the six additional points all belong to
the build cycle {\Tt{}\Rm{}{\it{}p}\nwendquote}. The list of pre-defined conditions which may be
checked is listed in Section~\ref{sec:check-relat-betw}.
\nwenddocs{}\nwbegincode{184}\sublabel{NWppJ6t-4d7QoM-1}\nwmargintag{{\nwtagstyle{}\subpageref{NWppJ6t-4d7QoM-1}}}\moddef{check the theorem~{\nwtagstyle{}\subpageref{NWppJ6t-4d7QoM-1}}}\endmoddef\Rm{}\nwstartdeflinemarkup\nwusesondefline{\\{NWppJ6t-2LTMm-F}\\{NWppJ6t-30Tc7D-5}\\{NWppJ6t-30Tc7D-6}\\{NWppJ6t-30Tc7D-7}\\{NWppJ6t-30Tc7D-8}}\nwenddeflinemarkup
    {\it{}cout} \begin{math}\ll\end{math} {\tt{}"Midpoint BC belongs to the cycle: "} \begin{math}\ll\end{math}  {\it{}F}.{\it{}check\_rel}({\it{}p},{\it{}A2},{\it{}cycle\_orthogonal}) \begin{math}\ll\end{math} {\it{}endl};
    {\it{}cout} \begin{math}\ll\end{math} {\tt{}"Midpoint AC belongs to the cycle: "} \begin{math}\ll\end{math}  {\it{}F}.{\it{}check\_rel}({\it{}p},{\it{}B2},{\it{}cycle\_orthogonal}) \begin{math}\ll\end{math} {\it{}endl};
    {\it{}cout} \begin{math}\ll\end{math} {\tt{}"Midpoint AB belongs to the cycle: "} \begin{math}\ll\end{math}  {\it{}F}.{\it{}check\_rel}({\it{}p},{\it{}C2},{\it{}cycle\_orthogonal}) \begin{math}\ll\end{math} {\it{}endl};
    {\it{}cout} \begin{math}\ll\end{math} {\tt{}"Midpoint OA belongs to the cycle: "} \begin{math}\ll\end{math} {\it{}F}.{\it{}check\_rel}({\it{}p},{\it{}A3},{\it{}cycle\_orthogonal}) \begin{math}\ll\end{math} {\it{}endl};
    {\it{}cout} \begin{math}\ll\end{math} {\tt{}"Midpoint OB belongs to the cycle: "} \begin{math}\ll\end{math}  {\it{}F}.{\it{}check\_rel}({\it{}p},{\it{}B3},{\it{}cycle\_orthogonal}) \begin{math}\ll\end{math} {\it{}endl};
    {\it{}cout} \begin{math}\ll\end{math} {\tt{}"Midpoint OC belongs to the cycle: "} \begin{math}\ll\end{math}  {\it{}F}.{\it{}check\_rel}({\it{}p},{\it{}C3},{\it{}cycle\_orthogonal}) \begin{math}\ll\end{math} {\it{}endl};
\nwindexdefn{\nwixident{check{\_}rel}}{check:unrel}{NWppJ6t-4d7QoM-1}\eatline
\nwused{\\{NWppJ6t-2LTMm-F}\\{NWppJ6t-30Tc7D-5}\\{NWppJ6t-30Tc7D-6}\\{NWppJ6t-30Tc7D-7}\\{NWppJ6t-30Tc7D-8}}\nwidentdefs{\\{{\nwixident{check{\_}rel}}{check:unrel}}}\nwidentuses{\\{{\nwixident{cycle{\_}orthogonal}}{cycle:unorthogonal}}}\nwindexuse{\nwixident{cycle{\_}orthogonal}}{cycle:unorthogonal}{NWppJ6t-4d7QoM-1}\nwendcode{}\nwbegindocs{185}\nwdocspar
\nwenddocs{}\nwbegindocs{186}We inscribe the cycle {\Tt{}\Rm{}{\it{}va}\nwendquote} into the triangle through the relation
{\Tt{}\Rm{}{\it{}is\_tangent\_i}\nwendquote} (that is ``tangent from inside'') and {\Tt{}\Rm{}{\it{}is\_tangent\_}\nwendquote}
(that is ``tangent from outside'')to sides of the triangle. We also
provide custom \Asymptote\ drawing style: dar red colour and line
thickness 0.5pt.
\nwenddocs{}\nwbegincode{187}\sublabel{NWppJ6t-30Tc7D-2}\nwmargintag{{\nwtagstyle{}\subpageref{NWppJ6t-30Tc7D-2}}}\moddef{nine-points-thm.cpp~{\nwtagstyle{}\subpageref{NWppJ6t-30Tc7D-1}}}\plusendmoddef\Rm{}\nwstartdeflinemarkup\nwusesondefline{\\{NWppJ6t-1p0Y9w-1}}\nwprevnextdefs{NWppJ6t-30Tc7D-1}{NWppJ6t-30Tc7D-3}\nwenddeflinemarkup
    {\bf{}ex} {\it{}va}={\it{}F}.{\it{}add\_cycle\_rel}({\bf{}lst}{\nwlbrace}{\it{}is\_tangent\_o}({\it{}a}),{\it{}is\_tangent\_i}({\it{}b}),{\it{}is\_tangent\_i}({\it{}c}){\nwrbrace},{\tt{}"v\_a"});
    {\it{}F}.{\it{}set\_asy\_style}({\it{}va},{\tt{}"rgb(0.8,0,0)+.5"});
\nwindexdefn{\nwixident{is{\_}tangent{\_}o}}{is:untangent:uno}{NWppJ6t-30Tc7D-2}\nwindexdefn{\nwixident{is{\_}tangent{\_}i}}{is:untangent:uni}{NWppJ6t-30Tc7D-2}\eatline
\nwused{\\{NWppJ6t-1p0Y9w-1}}\nwidentdefs{\\{{\nwixident{is{\_}tangent{\_}i}}{is:untangent:uni}}\\{{\nwixident{is{\_}tangent{\_}o}}{is:untangent:uno}}}\nwidentuses{\\{{\nwixident{add{\_}cycle{\_}rel}}{add:uncycle:unrel}}\\{{\nwixident{ex}}{ex}}\\{{\nwixident{rgb}}{rgb}}\\{{\nwixident{set{\_}asy{\_}style}}{set:unasy:unstyle}}}\nwindexuse{\nwixident{add{\_}cycle{\_}rel}}{add:uncycle:unrel}{NWppJ6t-30Tc7D-2}\nwindexuse{\nwixident{ex}}{ex}{NWppJ6t-30Tc7D-2}\nwindexuse{\nwixident{rgb}}{rgb}{NWppJ6t-30Tc7D-2}\nwindexuse{\nwixident{set{\_}asy{\_}style}}{set:unasy:unstyle}{NWppJ6t-30Tc7D-2}\nwendcode{}\nwbegindocs{188}\nwdocspar
\nwenddocs{}\nwbegindocs{189}Similarly we define two other tangent cycles: touching two sides
from inside and  the third from outside (the relation
{\Tt{}\Rm{}{\it{}is\_tangent\_o}\nwendquote}). We also define custom \Asymptote\ styles for the
new cycles.
\nwenddocs{}\nwbegincode{190}\sublabel{NWppJ6t-30Tc7D-3}\nwmargintag{{\nwtagstyle{}\subpageref{NWppJ6t-30Tc7D-3}}}\moddef{nine-points-thm.cpp~{\nwtagstyle{}\subpageref{NWppJ6t-30Tc7D-1}}}\plusendmoddef\Rm{}\nwstartdeflinemarkup\nwusesondefline{\\{NWppJ6t-1p0Y9w-1}}\nwprevnextdefs{NWppJ6t-30Tc7D-2}{NWppJ6t-30Tc7D-4}\nwenddeflinemarkup
    {\bf{}ex} {\it{}vb}={\it{}F}.{\it{}add\_cycle\_rel}({\bf{}lst}{\nwlbrace}{\it{}is\_tangent\_i}({\it{}a}),{\it{}is\_tangent\_o}({\it{}b}),{\it{}is\_tangent\_i}({\it{}c}){\nwrbrace},{\tt{}"v\_b"});
    {\it{}F}.{\it{}set\_asy\_style}({\it{}vb},{\tt{}"rgb(0.8,0,0)+.5"});
    {\bf{}ex} {\it{}vc}={\it{}F}.{\it{}add\_cycle\_rel}({\bf{}lst}{\nwlbrace}{\it{}is\_tangent\_i}({\it{}a}),{\it{}is\_tangent\_i}({\it{}b}),{\it{}is\_tangent\_o}({\it{}c}){\nwrbrace},{\tt{}"v\_c"});
    {\it{}F}.{\it{}set\_asy\_style}({\it{}vc},{\tt{}"rgb(0.8,0,0)+.5"});
    \LA{}check that cycles are tangent~{\nwtagstyle{}\subpageref{NWppJ6t-2XKMIQ-1}}\RA{}

\nwused{\\{NWppJ6t-1p0Y9w-1}}\nwidentuses{\\{{\nwixident{add{\_}cycle{\_}rel}}{add:uncycle:unrel}}\\{{\nwixident{ex}}{ex}}\\{{\nwixident{is{\_}tangent{\_}i}}{is:untangent:uni}}\\{{\nwixident{is{\_}tangent{\_}o}}{is:untangent:uno}}\\{{\nwixident{rgb}}{rgb}}\\{{\nwixident{set{\_}asy{\_}style}}{set:unasy:unstyle}}}\nwindexuse{\nwixident{add{\_}cycle{\_}rel}}{add:uncycle:unrel}{NWppJ6t-30Tc7D-3}\nwindexuse{\nwixident{ex}}{ex}{NWppJ6t-30Tc7D-3}\nwindexuse{\nwixident{is{\_}tangent{\_}i}}{is:untangent:uni}{NWppJ6t-30Tc7D-3}\nwindexuse{\nwixident{is{\_}tangent{\_}o}}{is:untangent:uno}{NWppJ6t-30Tc7D-3}\nwindexuse{\nwixident{rgb}}{rgb}{NWppJ6t-30Tc7D-3}\nwindexuse{\nwixident{set{\_}asy{\_}style}}{set:unasy:unstyle}{NWppJ6t-30Tc7D-3}\nwendcode{}\nwbegindocs{191}We also want to check the touching property between cycles:
\nwenddocs{}\nwbegincode{192}\sublabel{NWppJ6t-2XKMIQ-1}\nwmargintag{{\nwtagstyle{}\subpageref{NWppJ6t-2XKMIQ-1}}}\moddef{check that cycles are tangent~{\nwtagstyle{}\subpageref{NWppJ6t-2XKMIQ-1}}}\endmoddef\Rm{}\nwstartdeflinemarkup\nwusesondefline{\\{NWppJ6t-30Tc7D-3}\\{NWppJ6t-30Tc7D-5}\\{NWppJ6t-30Tc7D-6}\\{NWppJ6t-30Tc7D-7}\\{NWppJ6t-30Tc7D-8}}\nwenddeflinemarkup
    {\it{}cout} \begin{math}\ll\end{math} {\tt{}"p and va are tangent: "} \begin{math}\ll\end{math}  {\it{}F}.{\it{}check\_rel}({\it{}p},{\it{}va},{\it{}check\_tangent}).{\it{}evalf}() \begin{math}\ll\end{math} {\it{}endl};
    {\it{}cout} \begin{math}\ll\end{math} {\tt{}"p and vb are tangent: "} \begin{math}\ll\end{math}  {\it{}F}.{\it{}check\_rel}({\it{}p},{\it{}vb},{\it{}check\_tangent}).{\it{}evalf}() \begin{math}\ll\end{math} {\it{}endl};
    {\it{}cout} \begin{math}\ll\end{math} {\tt{}"p and vc are tangent: "} \begin{math}\ll\end{math}  {\it{}F}.{\it{}check\_rel}({\it{}p},{\it{}vc},{\it{}check\_tangent}).{\it{}evalf}() \begin{math}\ll\end{math} {\it{}endl};

\nwused{\\{NWppJ6t-30Tc7D-3}\\{NWppJ6t-30Tc7D-5}\\{NWppJ6t-30Tc7D-6}\\{NWppJ6t-30Tc7D-7}\\{NWppJ6t-30Tc7D-8}}\nwidentuses{\\{{\nwixident{check{\_}rel}}{check:unrel}}\\{{\nwixident{check{\_}tangent}}{check:untangent}}\\{{\nwixident{evalf}}{evalf}}}\nwindexuse{\nwixident{check{\_}rel}}{check:unrel}{NWppJ6t-2XKMIQ-1}\nwindexuse{\nwixident{check{\_}tangent}}{check:untangent}{NWppJ6t-2XKMIQ-1}\nwindexuse{\nwixident{evalf}}{evalf}{NWppJ6t-2XKMIQ-1}\nwendcode{}\nwbegindocs{193}Now, we draw our figure to the PDF and PNG files, it is shown at
Figure~\ref{fig:illustr-conf-nine}.
\nwenddocs{}\nwbegincode{194}\sublabel{NWppJ6t-30Tc7D-4}\nwmargintag{{\nwtagstyle{}\subpageref{NWppJ6t-30Tc7D-4}}}\moddef{nine-points-thm.cpp~{\nwtagstyle{}\subpageref{NWppJ6t-30Tc7D-1}}}\plusendmoddef\Rm{}\nwstartdeflinemarkup\nwusesondefline{\\{NWppJ6t-1p0Y9w-1}}\nwprevnextdefs{NWppJ6t-30Tc7D-3}{NWppJ6t-30Tc7D-5}\nwenddeflinemarkup
    {\it{}F}.{\it{}asy\_write}(300,-3.1,2.4,-3.6,1.3,{\tt{}"nine-points-thm-plain"}, {\tt{}"pdf"});
    {\it{}F}.{\it{}asy\_write}(600,-3.1,2.4,-3.6,1.3,{\tt{}"nine-points-thm-plain"}, {\tt{}"png"});
\nwindexdefn{\nwixident{asy{\_}write}}{asy:unwrite}{NWppJ6t-30Tc7D-4}\eatline
\nwused{\\{NWppJ6t-1p0Y9w-1}}\nwidentdefs{\\{{\nwixident{asy{\_}write}}{asy:unwrite}}}\nwendcode{}\nwbegindocs{195}\nwdocspar
\nwenddocs{}\nwbegindocs{196}We also can modify a cycle at zero level by {\Tt{}\Rm{}{\it{}move\_point}\nwendquote}. This
time we restore the initial value of {\Tt{}\Rm{}{\it{}N}\nwendquote} as a debug check: this is a
transition from a pre-defined {\Tt{}\Rm{}{\bf{}cycle}\nwendquote} given above to a point
(which is a calculated object due to the internal representation).
\nwenddocs{}\nwbegincode{197}\sublabel{NWppJ6t-30Tc7D-5}\nwmargintag{{\nwtagstyle{}\subpageref{NWppJ6t-30Tc7D-5}}}\moddef{nine-points-thm.cpp~{\nwtagstyle{}\subpageref{NWppJ6t-30Tc7D-1}}}\plusendmoddef\Rm{}\nwstartdeflinemarkup\nwusesondefline{\\{NWppJ6t-1p0Y9w-1}}\nwprevnextdefs{NWppJ6t-30Tc7D-4}{NWppJ6t-30Tc7D-6}\nwenddeflinemarkup
    {\it{}F}.{\it{}move\_point}({\it{}N},{\bf{}lst}{\nwlbrace}{\bf{}numeric}(1,2),-{\bf{}numeric}(5,2){\nwrbrace});
    {\it{}cerr} \begin{math}\ll\end{math} {\it{}F} \begin{math}\ll\end{math} {\it{}endl};
    {\it{}F}.{\it{}asy\_draw}({\it{}cout},{\it{}cerr},{\tt{}""},-3.1,2.4,-3.6,1.3);

    {\it{}F}.{\it{}asy\_write}(300,-3.1,2.4,-3.6,1.3,{\tt{}"nine-points-thm"}, {\tt{}"pdf"});
    {\it{}F}.{\it{}asy\_write}(600,-3.1,2.4,-3.6,1.3,{\tt{}"nine-points-thm"}, {\tt{}"png"});
    \LA{}check the theorem~{\nwtagstyle{}\subpageref{NWppJ6t-4d7QoM-1}}\RA{}
    \LA{}check that cycles are tangent~{\nwtagstyle{}\subpageref{NWppJ6t-2XKMIQ-1}}\RA{}
\nwindexdefn{\nwixident{move{\_}point}}{move:unpoint}{NWppJ6t-30Tc7D-5}\eatline
\nwused{\\{NWppJ6t-1p0Y9w-1}}\nwidentdefs{\\{{\nwixident{move{\_}point}}{move:unpoint}}}\nwidentuses{\\{{\nwixident{asy{\_}draw}}{asy:undraw}}\\{{\nwixident{asy{\_}write}}{asy:unwrite}}\\{{\nwixident{numeric}}{numeric}}}\nwindexuse{\nwixident{asy{\_}draw}}{asy:undraw}{NWppJ6t-30Tc7D-5}\nwindexuse{\nwixident{asy{\_}write}}{asy:unwrite}{NWppJ6t-30Tc7D-5}\nwindexuse{\nwixident{numeric}}{numeric}{NWppJ6t-30Tc7D-5}\nwendcode{}\nwbegindocs{198}\nwdocspar
\nwenddocs{}\nwbegindocs{199}And now we use {\Tt{}\Rm{}{\it{}move\_point}\nwendquote} to change coordinates of the point
(without a change of its type).
\nwenddocs{}\nwbegincode{200}\sublabel{NWppJ6t-30Tc7D-6}\nwmargintag{{\nwtagstyle{}\subpageref{NWppJ6t-30Tc7D-6}}}\moddef{nine-points-thm.cpp~{\nwtagstyle{}\subpageref{NWppJ6t-30Tc7D-1}}}\plusendmoddef\Rm{}\nwstartdeflinemarkup\nwusesondefline{\\{NWppJ6t-1p0Y9w-1}}\nwprevnextdefs{NWppJ6t-30Tc7D-5}{NWppJ6t-30Tc7D-7}\nwenddeflinemarkup
    {\it{}F}.{\it{}move\_point}({\it{}N},{\bf{}lst}{\nwlbrace}{\bf{}numeric}(4,2),-{\bf{}numeric}(5,2){\nwrbrace});
    {\it{}F}.{\it{}asy\_write}(300,-3.1,2.4,-3.6,1.3,{\tt{}"nine-points-thm2"});
    \LA{}check the theorem~{\nwtagstyle{}\subpageref{NWppJ6t-4d7QoM-1}}\RA{}
    \LA{}check that cycles are tangent~{\nwtagstyle{}\subpageref{NWppJ6t-2XKMIQ-1}}\RA{}

\nwused{\\{NWppJ6t-1p0Y9w-1}}\nwidentuses{\\{{\nwixident{asy{\_}write}}{asy:unwrite}}\\{{\nwixident{move{\_}point}}{move:unpoint}}\\{{\nwixident{numeric}}{numeric}}}\nwindexuse{\nwixident{asy{\_}write}}{asy:unwrite}{NWppJ6t-30Tc7D-6}\nwindexuse{\nwixident{move{\_}point}}{move:unpoint}{NWppJ6t-30Tc7D-6}\nwindexuse{\nwixident{numeric}}{numeric}{NWppJ6t-30Tc7D-6}\nwendcode{}\nwbegindocs{201}Then, we move the cycle {\Tt{}\Rm{}{\it{}N}\nwendquote} to represent the point at infinity {\Tt{}\Rm{}(0,{\bf{}lst}{\nwlbrace}0,0{\nwrbrace},1)\nwendquote},
thus the drawing becomes the classical Nine Points Theorem in
Euclidean geometry.
\nwenddocs{}\nwbegincode{202}\sublabel{NWppJ6t-30Tc7D-7}\nwmargintag{{\nwtagstyle{}\subpageref{NWppJ6t-30Tc7D-7}}}\moddef{nine-points-thm.cpp~{\nwtagstyle{}\subpageref{NWppJ6t-30Tc7D-1}}}\plusendmoddef\Rm{}\nwstartdeflinemarkup\nwusesondefline{\\{NWppJ6t-1p0Y9w-1}}\nwprevnextdefs{NWppJ6t-30Tc7D-6}{NWppJ6t-30Tc7D-8}\nwenddeflinemarkup
    {\it{}F}.{\it{}move\_cycle}({\it{}N}, {\bf{}cycle\_data}(0,{\bf{}lst}{\nwlbrace}0,0{\nwrbrace},1));
    {\it{}F}.{\it{}asy\_write}(300,-3.1,2.4,-3.6,1.3,{\tt{}"nine-points-thm1"});
    \LA{}check the theorem~{\nwtagstyle{}\subpageref{NWppJ6t-4d7QoM-1}}\RA{}
    \LA{}check that cycles are tangent~{\nwtagstyle{}\subpageref{NWppJ6t-2XKMIQ-1}}\RA{}
\nwindexdefn{\nwixident{move{\_}cycle}}{move:uncycle}{NWppJ6t-30Tc7D-7}\nwindexdefn{\nwixident{cycle{\_}data}}{cycle:undata}{NWppJ6t-30Tc7D-7}\eatline
\nwused{\\{NWppJ6t-1p0Y9w-1}}\nwidentdefs{\\{{\nwixident{cycle{\_}data}}{cycle:undata}}\\{{\nwixident{move{\_}cycle}}{move:uncycle}}}\nwidentuses{\\{{\nwixident{asy{\_}write}}{asy:unwrite}}}\nwindexuse{\nwixident{asy{\_}write}}{asy:unwrite}{NWppJ6t-30Tc7D-7}\nwendcode{}\nwbegindocs{203}\nwdocspar
\nwenddocs{}\nwbegindocs{204}We can draw the same figures in the hyperbolic metric as well. The
checks show that the nine-point theorem is still valid!
\nwenddocs{}\nwbegincode{205}\sublabel{NWppJ6t-30Tc7D-8}\nwmargintag{{\nwtagstyle{}\subpageref{NWppJ6t-30Tc7D-8}}}\moddef{nine-points-thm.cpp~{\nwtagstyle{}\subpageref{NWppJ6t-30Tc7D-1}}}\plusendmoddef\Rm{}\nwstartdeflinemarkup\nwusesondefline{\\{NWppJ6t-1p0Y9w-1}}\nwprevnextdefs{NWppJ6t-30Tc7D-7}{NWppJ6t-30Tc7D-9}\nwenddeflinemarkup
    {\it{}F}.{\it{}move\_cycle}({\it{}N}, {\bf{}cycle\_data}(0,{\bf{}lst}{\nwlbrace}0,0{\nwrbrace},1));
    {\it{}F}.{\it{}set\_metric}({\it{}diag\_matrix}({\bf{}lst}{\nwlbrace}-1,1{\nwrbrace}));
    {\it{}F}.{\it{}asy\_write}(300,-3.1,2.4,-3.6,1.3,{\tt{}"nine-points-thm-plain-hyp"});
    \LA{}check the theorem~{\nwtagstyle{}\subpageref{NWppJ6t-4d7QoM-1}}\RA{}
    \LA{}check that cycles are tangent~{\nwtagstyle{}\subpageref{NWppJ6t-2XKMIQ-1}}\RA{}
    {\it{}F}.{\it{}move\_point}({\it{}N},{\bf{}lst}{\nwlbrace}{\bf{}numeric}(1,2),-{\bf{}numeric}(5,2){\nwrbrace});
    {\it{}F}.{\it{}asy\_write}(300,-3.1,2.4,-3.6,1.3,{\tt{}"nine-points-thm-hyp"}, {\tt{}"pdf"});
    {\it{}F}.{\it{}asy\_write}(600,-3.1,2.4,-3.6,1.3,{\tt{}"nine-points-thm-hyp"},{\tt{}"png"});
    \LA{}check the theorem~{\nwtagstyle{}\subpageref{NWppJ6t-4d7QoM-1}}\RA{}
    \LA{}check that cycles are tangent~{\nwtagstyle{}\subpageref{NWppJ6t-2XKMIQ-1}}\RA{}
    //F.set\_metric(diag\_matrix(lst{\nwlbrace}-1,0{\nwrbrace}));
    //F.asy\_write(300,-3.1,2.4,-3.6,1.3,"nine-points-thm-par", "pdf");

\nwused{\\{NWppJ6t-1p0Y9w-1}}\nwidentuses{\\{{\nwixident{asy{\_}write}}{asy:unwrite}}\\{{\nwixident{cycle{\_}data}}{cycle:undata}}\\{{\nwixident{move{\_}cycle}}{move:uncycle}}\\{{\nwixident{move{\_}point}}{move:unpoint}}\\{{\nwixident{numeric}}{numeric}}\\{{\nwixident{set{\_}metric}}{set:unmetric}}}\nwindexuse{\nwixident{asy{\_}write}}{asy:unwrite}{NWppJ6t-30Tc7D-8}\nwindexuse{\nwixident{cycle{\_}data}}{cycle:undata}{NWppJ6t-30Tc7D-8}\nwindexuse{\nwixident{move{\_}cycle}}{move:uncycle}{NWppJ6t-30Tc7D-8}\nwindexuse{\nwixident{move{\_}point}}{move:unpoint}{NWppJ6t-30Tc7D-8}\nwindexuse{\nwixident{numeric}}{numeric}{NWppJ6t-30Tc7D-8}\nwindexuse{\nwixident{set{\_}metric}}{set:unmetric}{NWppJ6t-30Tc7D-8}\nwendcode{}\nwbegindocs{206}Finally, we produce an animation, which illustrate the transition
from the traditional nine-point theorem to its conformal version. To
this end we return to the elliptic metric and freeze the figure. This
can be time-consuming and may be not performed by default.
\nwenddocs{}\nwbegincode{207}\sublabel{NWppJ6t-30Tc7D-9}\nwmargintag{{\nwtagstyle{}\subpageref{NWppJ6t-30Tc7D-9}}}\moddef{nine-points-thm.cpp~{\nwtagstyle{}\subpageref{NWppJ6t-30Tc7D-1}}}\plusendmoddef\Rm{}\nwstartdeflinemarkup\nwusesondefline{\\{NWppJ6t-1p0Y9w-1}}\nwprevnextdefs{NWppJ6t-30Tc7D-8}{NWppJ6t-30Tc7D-A}\nwenddeflinemarkup
     {\bf{}if} ({\bf{}true}) {\nwlbrace}
        {\it{}F}.{\it{}set\_metric}({\it{}diag\_matrix}({\bf{}lst}{\nwlbrace}-1,-1{\nwrbrace}));
        {\it{}F}.{\it{}freeze}();

\nwused{\\{NWppJ6t-1p0Y9w-1}}\nwidentuses{\\{{\nwixident{freeze}}{freeze}}\\{{\nwixident{set{\_}metric}}{set:unmetric}}}\nwindexuse{\nwixident{freeze}}{freeze}{NWppJ6t-30Tc7D-9}\nwindexuse{\nwixident{set{\_}metric}}{set:unmetric}{NWppJ6t-30Tc7D-9}\nwendcode{}\nwbegindocs{208}We define a symbolic parameter {\Tt{}\Rm{}{\it{}t}\nwendquote} and make the point {\Tt{}\Rm{}{\it{}N}\nwendquote} depends on it.
\nwenddocs{}\nwbegincode{209}\sublabel{NWppJ6t-30Tc7D-A}\nwmargintag{{\nwtagstyle{}\subpageref{NWppJ6t-30Tc7D-A}}}\moddef{nine-points-thm.cpp~{\nwtagstyle{}\subpageref{NWppJ6t-30Tc7D-1}}}\plusendmoddef\Rm{}\nwstartdeflinemarkup\nwusesondefline{\\{NWppJ6t-1p0Y9w-1}}\nwprevnextdefs{NWppJ6t-30Tc7D-9}{NWppJ6t-30Tc7D-B}\nwenddeflinemarkup
        {\bf{}realsymbol} {\it{}t}({\tt{}"t"});
        {\it{}F}.{\it{}move\_point}({\it{}N},{\bf{}lst}{\nwlbrace}(1.0+{\it{}t})\begin{math}\div\end{math}2.0,-(5.0+{\it{}t})\begin{math}\div\end{math}2.0{\nwrbrace});

\nwused{\\{NWppJ6t-1p0Y9w-1}}\nwidentuses{\\{{\nwixident{move{\_}point}}{move:unpoint}}\\{{\nwixident{realsymbol}}{realsymbol}}}\nwindexuse{\nwixident{move{\_}point}}{move:unpoint}{NWppJ6t-30Tc7D-A}\nwindexuse{\nwixident{realsymbol}}{realsymbol}{NWppJ6t-30Tc7D-A}\nwendcode{}\nwbegindocs{210}Then, the range of values {\Tt{}\Rm{}{\it{}val}\nwendquote} for the parameter {\Tt{}\Rm{}{\it{}t}\nwendquote} and then
produce an animation based on these values. The resulting animation is
presented on the Fig.~\ref{fig:nine-points-anim}.
\nwenddocs{}\nwbegincode{211}\sublabel{NWppJ6t-30Tc7D-B}\nwmargintag{{\nwtagstyle{}\subpageref{NWppJ6t-30Tc7D-B}}}\moddef{nine-points-thm.cpp~{\nwtagstyle{}\subpageref{NWppJ6t-30Tc7D-1}}}\plusendmoddef\Rm{}\nwstartdeflinemarkup\nwusesondefline{\\{NWppJ6t-1p0Y9w-1}}\nwprevnextdefs{NWppJ6t-30Tc7D-A}{NWppJ6t-30Tc7D-C}\nwenddeflinemarkup
        {\bf{}lst} {\it{}val};
         {\bf{}int} {\it{}num}=50;
        {\bf{}for} ({\bf{}int} {\it{}i}=0;{\it{}i}\begin{math}\leq\end{math}{\it{}num};\protect\PP{\it{}i})
            {\it{}val}.{\it{}append}({\it{}t}\begin{math}\equiv\end{math}{\it{}exp}({\it{}pow}(2.2\begin{math}\ast\end{math}({\it{}num}-{\it{}i})\begin{math}\div\end{math}{\it{}num},2.2))-1.0);
        {\it{}F}.{\it{}asy\_animate}({\it{}val},300,-3.1,2.4,-3.6,1.3,{\tt{}"nine-points-anim"}, {\tt{}"pdf"});
    {\nwrbrace}

\nwused{\\{NWppJ6t-1p0Y9w-1}}\nwidentuses{\\{{\nwixident{asy{\_}animate}}{asy:unanimate}}}\nwindexuse{\nwixident{asy{\_}animate}}{asy:unanimate}{NWppJ6t-30Tc7D-B}\nwendcode{}\nwbegindocs{212} We produce an illustration of {\Tt{}\Rm{}{\it{}SF}\nwendquote} in the canonical position. Everything is done now.
\nwenddocs{}\nwbegincode{213}\sublabel{NWppJ6t-30Tc7D-C}\nwmargintag{{\nwtagstyle{}\subpageref{NWppJ6t-30Tc7D-C}}}\moddef{nine-points-thm.cpp~{\nwtagstyle{}\subpageref{NWppJ6t-30Tc7D-1}}}\plusendmoddef\Rm{}\nwstartdeflinemarkup\nwusesondefline{\\{NWppJ6t-1p0Y9w-1}}\nwprevnextdefs{NWppJ6t-30Tc7D-B}{\relax}\nwenddeflinemarkup
    {\bf{}return} 0;
{\nwrbrace}
\nwused{\\{NWppJ6t-1p0Y9w-1}}\nwendcode{}\nwbegindocs{214}\nwdocspar
\nwenddocs{}\nwbegincode{215}\sublabel{NWppJ6t-1fWRR8-6}\nwmargintag{{\nwtagstyle{}\subpageref{NWppJ6t-1fWRR8-6}}}\moddef{separating chunk~{\nwtagstyle{}\subpageref{NWppJ6t-1fWRR8-1}}}\plusendmoddef\Rm{}\nwstartdeflinemarkup\nwprevnextdefs{NWppJ6t-1fWRR8-5}{NWppJ6t-1fWRR8-7}\nwenddeflinemarkup

\nwendcode{}\nwbegindocs{216}\nwdocspar
\nwenddocs{}\nwbegincode{217}\sublabel{NWppJ6t-1p0Y9w-1}\nwmargintag{{\nwtagstyle{}\subpageref{NWppJ6t-1p0Y9w-1}}}\moddef{*~{\nwtagstyle{}\subpageref{NWppJ6t-1p0Y9w-1}}}\endmoddef\Rm{}\nwstartdeflinemarkup\nwenddeflinemarkup
\LA{}nine-points-thm.cpp~{\nwtagstyle{}\subpageref{NWppJ6t-30Tc7D-1}}\RA{}

\nwnotused{*}\nwendcode{}\nwbegindocs{218}\nwdocspar
\subsection{Proving the theorem: Symbolic computations}
\label{sec:prov-theor-symb}

\nwenddocs{}\nwbegindocs{219}\nwdocspar
\nwenddocs{}\nwbegincode{220}\sublabel{NWppJ6t-2zWvnC-1}\nwmargintag{{\nwtagstyle{}\subpageref{NWppJ6t-2zWvnC-1}}}\moddef{nine-points-thm-symb.cpp~{\nwtagstyle{}\subpageref{NWppJ6t-2zWvnC-1}}}\endmoddef\Rm{}\nwstartdeflinemarkup\nwprevnextdefs{\relax}{NWppJ6t-2zWvnC-2}\nwenddeflinemarkup
\LA{}license~{\nwtagstyle{}\subpageref{NWppJ6t-ZXuKx-1}}\RA{}
{\bf{}\char35{}include}{\tt{} "figure.h"}
\LA{}using all namespaces~{\nwtagstyle{}\subpageref{NWppJ6t-3uHj2c-1}}\RA{}
{\bf{}int} {\it{}main}(){\nwlbrace}\nwindexdefn{\nwixident{main}}{main}{NWppJ6t-2zWvnC-1}
    {\it{}cout} \begin{math}\ll\end{math} {\tt{}"Prooving the theorem, this shall take a long time..."}
         \begin{math}\ll\end{math} {\it{}endl};
    \LA{}initial data for proof~{\nwtagstyle{}\subpageref{NWppJ6t-2y20ZP-1}}\RA{}
    \LA{}build medioscribed cycle~{\nwtagstyle{}\subpageref{NWppJ6t-2LTMm-1}}\RA{}

\nwalsodefined{\\{NWppJ6t-2zWvnC-2}}\nwnotused{nine-points-thm-symb.cpp}\nwidentdefs{\\{{\nwixident{main}}{main}}}\nwidentuses{\\{{\nwixident{figure}}{figure}}}\nwindexuse{\nwixident{figure}}{figure}{NWppJ6t-2zWvnC-1}\nwendcode{}\nwbegindocs{221}We define variables from {\Tt{}\Rm{}{\bf{}realsymbol}\nwendquote} class to be used in symbolic computations.
\nwenddocs{}\nwbegincode{222}\sublabel{NWppJ6t-2y20ZP-1}\nwmargintag{{\nwtagstyle{}\subpageref{NWppJ6t-2y20ZP-1}}}\moddef{initial data for proof~{\nwtagstyle{}\subpageref{NWppJ6t-2y20ZP-1}}}\endmoddef\Rm{}\nwstartdeflinemarkup\nwusesondefline{\\{NWppJ6t-2zWvnC-1}}\nwprevnextdefs{\relax}{NWppJ6t-2y20ZP-2}\nwenddeflinemarkup
    {\bf{}realsymbol} {\it{}x1}({\tt{}"x1"}), {\it{}y1}({\tt{}"y1"}), {\it{}x2}({\tt{}"x2"}), {\it{}y2}({\tt{}"y2"}), {\it{}x3}({\tt{}"x3"}), {\it{}y3}({\tt{}"y3"}), {\it{}x4}({\tt{}"x4"}), {\it{}y4}({\tt{}"y4"});
\nwindexdefn{\nwixident{realsymbol}}{realsymbol}{NWppJ6t-2y20ZP-1}\eatline
\nwalsodefined{\\{NWppJ6t-2y20ZP-2}}\nwused{\\{NWppJ6t-2zWvnC-1}}\nwidentdefs{\\{{\nwixident{realsymbol}}{realsymbol}}}\nwendcode{}\nwbegindocs{223}\nwdocspar
\nwenddocs{}\nwbegindocs{224}We also define the sign for the hyperbolic metric. The proof will
work in the elliptic (conformal Euclidean) space as well, however we
have synthetic poofs in this case. Symbolic computations in the
hyperbolic space are mathematically sufficient for demonstration, but
Figure~\ref{fig:illustr-conf-nine} from the previous subsection is
physiologically more convincing on the individual level. A synthetic
proof for hyperbolic space would be interesting to obtain as well.
\nwenddocs{}\nwbegincode{225}\sublabel{NWppJ6t-2y20ZP-2}\nwmargintag{{\nwtagstyle{}\subpageref{NWppJ6t-2y20ZP-2}}}\moddef{initial data for proof~{\nwtagstyle{}\subpageref{NWppJ6t-2y20ZP-1}}}\plusendmoddef\Rm{}\nwstartdeflinemarkup\nwusesondefline{\\{NWppJ6t-2zWvnC-1}}\nwprevnextdefs{NWppJ6t-2y20ZP-1}{\relax}\nwenddeflinemarkup
    {\bf{}int} {\it{}sign}=1;

\nwused{\\{NWppJ6t-2zWvnC-1}}\nwendcode{}\nwbegindocs{226}We got the output, which make a full demonstration that the theorem
holds in the hyperbolic space as well:
\begin{verbatim}
Midpoint BC belongs to the cycle {0==0}
Midpoint AC belongs to the cycle {0==0}
Midpoint AB belongs to the cycle {0==0}
Midpoint OA belongs to the cycle {0==0}
Midpoint OB belongs to the cycle {0==0}
Midpoint OC belongs to the cycle {0==0}
\end{verbatim}
But be prepared, that that will take a long time (about 6 hours of CPU time of my slow PC).
\nwenddocs{}\nwbegincode{227}\sublabel{NWppJ6t-2zWvnC-2}\nwmargintag{{\nwtagstyle{}\subpageref{NWppJ6t-2zWvnC-2}}}\moddef{nine-points-thm-symb.cpp~{\nwtagstyle{}\subpageref{NWppJ6t-2zWvnC-1}}}\plusendmoddef\Rm{}\nwstartdeflinemarkup\nwprevnextdefs{NWppJ6t-2zWvnC-1}{\relax}\nwenddeflinemarkup
    {\bf{}return} 0;
{\nwrbrace}
\nwendcode{}\nwbegindocs{228}\nwdocspar
\nwenddocs{}\nwbegincode{229}\sublabel{NWppJ6t-1fWRR8-7}\nwmargintag{{\nwtagstyle{}\subpageref{NWppJ6t-1fWRR8-7}}}\moddef{separating chunk~{\nwtagstyle{}\subpageref{NWppJ6t-1fWRR8-1}}}\plusendmoddef\Rm{}\nwstartdeflinemarkup\nwprevnextdefs{NWppJ6t-1fWRR8-6}{NWppJ6t-1fWRR8-8}\nwenddeflinemarkup

\nwendcode{}\nwbegindocs{230}\nwdocspar
\subsection{Numerical relations}
\label{sec:numerical-relations}

\nwenddocs{}\nwbegindocs{231}To illustrate the usage of relations with numerical parameters we are
solving the following problem
from~\cite{FillmoreSpringer00a}*{Problem~A}:
\begin{quote}
  Find the cycles having (all three conditions):
  \begin{itemize}
  \item tangential distance \(7\) from the circle
    \begin{displaymath}
      (u-7)^2+(v-1)^2=2^2;
    \end{displaymath}
  \item angle \(\arccos \frac{4}{5}\) with the circle
    \begin{displaymath}
      (u-5)^2+(v-3)^2=5^2;
    \end{displaymath}
  \item centres lying on the line
    \begin{displaymath}
      \frac{5}{13} u + \frac{12}{13} v=0.
    \end{displaymath}
  \end{itemize}
\end{quote}
The statement of the problem uses orientation of cycles. Geometrically
it is given by the inward or outward direction of the normal. In our
library the orientation represented by the direction of the vector in
the projective space, it changes to the opposite if the vector is
multiplied by \(-1\).

\nwenddocs{}\nwbegindocs{232}The start of of our code is similar to the previous one.
\nwenddocs{}\nwbegincode{233}\sublabel{NWppJ6t-RDDRz-1}\nwmargintag{{\nwtagstyle{}\subpageref{NWppJ6t-RDDRz-1}}}\moddef{fillmore-springer-example.cpp~{\nwtagstyle{}\subpageref{NWppJ6t-RDDRz-1}}}\endmoddef\Rm{}\nwstartdeflinemarkup\nwprevnextdefs{\relax}{NWppJ6t-RDDRz-2}\nwenddeflinemarkup
\LA{}license~{\nwtagstyle{}\subpageref{NWppJ6t-ZXuKx-1}}\RA{}
{\bf{}\char35{}include}{\tt{} "figure.h"}
\LA{}using all namespaces~{\nwtagstyle{}\subpageref{NWppJ6t-3uHj2c-1}}\RA{}
{\bf{}int} {\it{}main}(){\nwlbrace}\nwindexdefn{\nwixident{main}}{main}{NWppJ6t-RDDRz-1}
    {\bf{}ex} {\it{}sign}=-{\bf{}numeric}(1);
    {\bf{}varidx} {\it{}nu}({\bf{}symbol}({\tt{}"nu"}, {\tt{}"{\char92}{\char92}nu"}), 2);
    {\bf{}ex} {\it{}e} = {\it{}clifford\_unit}({\it{}nu}, {\it{}diag\_matrix}({\bf{}lst}{\nwlbrace}-{\bf{}numeric}(1),{\it{}sign}{\nwrbrace}));
    {\bf{}figure} {\it{}F}({\it{}e});

\nwalsodefined{\\{NWppJ6t-RDDRz-2}\\{NWppJ6t-RDDRz-3}\\{NWppJ6t-RDDRz-4}\\{NWppJ6t-RDDRz-5}\\{NWppJ6t-RDDRz-6}\\{NWppJ6t-RDDRz-7}\\{NWppJ6t-RDDRz-8}}\nwnotused{fillmore-springer-example.cpp}\nwidentdefs{\\{{\nwixident{main}}{main}}}\nwidentuses{\\{{\nwixident{ex}}{ex}}\\{{\nwixident{figure}}{figure}}\\{{\nwixident{numeric}}{numeric}}}\nwindexuse{\nwixident{ex}}{ex}{NWppJ6t-RDDRz-1}\nwindexuse{\nwixident{figure}}{figure}{NWppJ6t-RDDRz-1}\nwindexuse{\nwixident{numeric}}{numeric}{NWppJ6t-RDDRz-1}\nwendcode{}\nwbegindocs{234}Now we define three circles given in the problem statement above.
\nwenddocs{}\nwbegincode{235}\sublabel{NWppJ6t-RDDRz-2}\nwmargintag{{\nwtagstyle{}\subpageref{NWppJ6t-RDDRz-2}}}\moddef{fillmore-springer-example.cpp~{\nwtagstyle{}\subpageref{NWppJ6t-RDDRz-1}}}\plusendmoddef\Rm{}\nwstartdeflinemarkup\nwprevnextdefs{NWppJ6t-RDDRz-1}{NWppJ6t-RDDRz-3}\nwenddeflinemarkup
    {\bf{}ex} {\it{}A}={\it{}F}.{\it{}add\_cycle}({\bf{}cycle}({\bf{}lst}{\nwlbrace}{\bf{}numeric}(7),{\bf{}numeric}(1){\nwrbrace},{\it{}e},{\bf{}numeric}(4)),{\tt{}"A"});
    {\bf{}ex} {\it{}B}={\it{}F}.{\it{}add\_cycle}({\bf{}cycle}({\bf{}lst}{\nwlbrace}{\bf{}numeric}(5),{\bf{}numeric}(3){\nwrbrace},{\it{}e},{\bf{}numeric}(25)),{\tt{}"B"});
    {\bf{}ex} {\it{}C}={\it{}F}.{\it{}add\_cycle}({\bf{}cycle}({\bf{}numeric}(0),{\bf{}lst}{\nwlbrace}{\bf{}numeric}(5,13),{\bf{}numeric}(12,13){\nwrbrace},0,{\it{}e}),{\tt{}"C"});

\nwidentuses{\\{{\nwixident{add{\_}cycle}}{add:uncycle}}\\{{\nwixident{ex}}{ex}}\\{{\nwixident{numeric}}{numeric}}}\nwindexuse{\nwixident{add{\_}cycle}}{add:uncycle}{NWppJ6t-RDDRz-2}\nwindexuse{\nwixident{ex}}{ex}{NWppJ6t-RDDRz-2}\nwindexuse{\nwixident{numeric}}{numeric}{NWppJ6t-RDDRz-2}\nwendcode{}\nwbegindocs{236}All given data will be drawn in black inc.
\nwenddocs{}\nwbegincode{237}\sublabel{NWppJ6t-RDDRz-3}\nwmargintag{{\nwtagstyle{}\subpageref{NWppJ6t-RDDRz-3}}}\moddef{fillmore-springer-example.cpp~{\nwtagstyle{}\subpageref{NWppJ6t-RDDRz-1}}}\plusendmoddef\Rm{}\nwstartdeflinemarkup\nwprevnextdefs{NWppJ6t-RDDRz-2}{NWppJ6t-RDDRz-4}\nwenddeflinemarkup
    {\it{}F}.{\it{}set\_asy\_style}({\it{}A},{\tt{}"rgb(0,0,0)"});
    {\it{}F}.{\it{}set\_asy\_style}({\it{}B},{\tt{}"rgb(0,0,0)"});
    {\it{}F}.{\it{}set\_asy\_style}({\it{}C},{\tt{}"rgb(0,0,0)"});

\nwidentuses{\\{{\nwixident{rgb}}{rgb}}\\{{\nwixident{set{\_}asy{\_}style}}{set:unasy:unstyle}}}\nwindexuse{\nwixident{rgb}}{rgb}{NWppJ6t-RDDRz-3}\nwindexuse{\nwixident{set{\_}asy{\_}style}}{set:unasy:unstyle}{NWppJ6t-RDDRz-3}\nwendcode{}\nwbegindocs{238}The solution {\Tt{}\Rm{}{\it{}D}\nwendquote} is a circle defined by the three above
conditions. The solution will be drawn in red.
\nwenddocs{}\nwbegincode{239}\sublabel{NWppJ6t-RDDRz-4}\nwmargintag{{\nwtagstyle{}\subpageref{NWppJ6t-RDDRz-4}}}\moddef{fillmore-springer-example.cpp~{\nwtagstyle{}\subpageref{NWppJ6t-RDDRz-1}}}\plusendmoddef\Rm{}\nwstartdeflinemarkup\nwprevnextdefs{NWppJ6t-RDDRz-3}{NWppJ6t-RDDRz-5}\nwenddeflinemarkup
    {\bf{}realsymbol} {\it{}D}({\tt{}"D"}), {\it{}T}({\tt{}"T"});
 {\it{}F}.{\it{}add\_cycle\_rel}({\bf{}lst}{\nwlbrace}{\it{}tangential\_distance}({\it{}A},{\bf{}true},{\bf{}numeric}(7)), // The tangential distance to {\it{}A} is  7
                             {\it{}make\_angle}({\it{}B},{\bf{}true},{\bf{}numeric}(4,5)), // The angle with {\it{}B} is  arccos(4/5)
                             {\it{}is\_orthogonal}({\it{}C}), // If the centre is on {\it{}C}, then {\it{}C} and {\it{}D} are orthogonal
             {\it{}is\_real\_cycle}({\it{}D}){\nwrbrace}, {\it{}D}); // We require {\it{}D} be a real circle, as there are two imaginary solutions as well
    {\it{}F}.{\it{}set\_asy\_style}({\it{}D},{\tt{}"rgb(0.7,0,0)"});
\nwindexdefn{\nwixident{tangential{\_}distance}}{tangential:undistance}{NWppJ6t-RDDRz-4}\nwindexdefn{\nwixident{make{\_}angle}}{make:unangle}{NWppJ6t-RDDRz-4}\nwindexdefn{\nwixident{is{\_}real{\_}cycle}}{is:unreal:uncycle}{NWppJ6t-RDDRz-4}\eatline
\nwidentdefs{\\{{\nwixident{is{\_}real{\_}cycle}}{is:unreal:uncycle}}\\{{\nwixident{make{\_}angle}}{make:unangle}}\\{{\nwixident{tangential{\_}distance}}{tangential:undistance}}}\nwidentuses{\\{{\nwixident{add{\_}cycle{\_}rel}}{add:uncycle:unrel}}\\{{\nwixident{is{\_}orthogonal}}{is:unorthogonal}}\\{{\nwixident{numeric}}{numeric}}\\{{\nwixident{realsymbol}}{realsymbol}}\\{{\nwixident{rgb}}{rgb}}\\{{\nwixident{set{\_}asy{\_}style}}{set:unasy:unstyle}}}\nwindexuse{\nwixident{add{\_}cycle{\_}rel}}{add:uncycle:unrel}{NWppJ6t-RDDRz-4}\nwindexuse{\nwixident{is{\_}orthogonal}}{is:unorthogonal}{NWppJ6t-RDDRz-4}\nwindexuse{\nwixident{numeric}}{numeric}{NWppJ6t-RDDRz-4}\nwindexuse{\nwixident{realsymbol}}{realsymbol}{NWppJ6t-RDDRz-4}\nwindexuse{\nwixident{rgb}}{rgb}{NWppJ6t-RDDRz-4}\nwindexuse{\nwixident{set{\_}asy{\_}style}}{set:unasy:unstyle}{NWppJ6t-RDDRz-4}\nwendcode{}\nwbegindocs{240}\nwdocspar
\nwenddocs{}\nwbegindocs{241}The output tells parameters of two solutions:
\begin{verbatim}
  Solutions: {(-1.0, [0,0]~D, 0.99999999999999999734),
(-0.0069444444444444444444, [-0.089285714285714285705,0.037202380952380952383]~D, -1.0)}
\end{verbatim}
Here, as in {\Tt{}\Rm{}{\bf{}cycle}\nwendquote} library, the set of four numbers \((k,[l,n],m)\)
represent the circle equation \(k(u^2+v^2)-2lu-2nv+m=0\).
\nwenddocs{}\nwbegincode{242}\sublabel{NWppJ6t-RDDRz-5}\nwmargintag{{\nwtagstyle{}\subpageref{NWppJ6t-RDDRz-5}}}\moddef{fillmore-springer-example.cpp~{\nwtagstyle{}\subpageref{NWppJ6t-RDDRz-1}}}\plusendmoddef\Rm{}\nwstartdeflinemarkup\nwprevnextdefs{NWppJ6t-RDDRz-4}{NWppJ6t-RDDRz-6}\nwenddeflinemarkup
    {\it{}cout} \begin{math}\ll\end{math} {\tt{}"Solutions: "}\begin{math}\ll\end{math} {\it{}F}.{\it{}get\_cycle}({\it{}D}).{\it{}evalf}() \begin{math}\ll\end{math} {\it{}endl};

\nwidentuses{\\{{\nwixident{evalf}}{evalf}}\\{{\nwixident{get{\_}cycle}}{get:uncycle}}}\nwindexuse{\nwixident{evalf}}{evalf}{NWppJ6t-RDDRz-5}\nwindexuse{\nwixident{get{\_}cycle}}{get:uncycle}{NWppJ6t-RDDRz-5}\nwendcode{}\nwbegindocs{243}To visualise the tangential distances we may add the joint tangent
lines to the figure. Some solutions are lines with imaginary
coefficients, to avoid them we use {\Tt{}\Rm{}{\it{}only\_reals}\nwendquote} condition. The
tangents will be drawn in blue inc.
\nwenddocs{}\nwbegincode{244}\sublabel{NWppJ6t-RDDRz-6}\nwmargintag{{\nwtagstyle{}\subpageref{NWppJ6t-RDDRz-6}}}\moddef{fillmore-springer-example.cpp~{\nwtagstyle{}\subpageref{NWppJ6t-RDDRz-1}}}\plusendmoddef\Rm{}\nwstartdeflinemarkup\nwprevnextdefs{NWppJ6t-RDDRz-5}{NWppJ6t-RDDRz-7}\nwenddeflinemarkup
    {\it{}F}.{\it{}add\_cycle\_rel}({\bf{}lst}{\nwlbrace}{\it{}is\_tangent\_i}({\it{}D}),{\it{}is\_tangent\_i}({\it{}A}),{\it{}is\_orthogonal}({\it{}F}.{\it{}get\_infinity}()),{\it{}only\_reals}({\it{}T}){\nwrbrace},{\it{}T});
    {\it{}F}.{\it{}set\_asy\_style}({\it{}T},{\tt{}"rgb(0,0,0.7)"});
\nwindexdefn{\nwixident{only{\_}reals}}{only:unreals}{NWppJ6t-RDDRz-6}\eatline
\nwidentdefs{\\{{\nwixident{only{\_}reals}}{only:unreals}}}\nwidentuses{\\{{\nwixident{add{\_}cycle{\_}rel}}{add:uncycle:unrel}}\\{{\nwixident{get{\_}infinity}}{get:uninfinity}}\\{{\nwixident{is{\_}orthogonal}}{is:unorthogonal}}\\{{\nwixident{is{\_}tangent{\_}i}}{is:untangent:uni}}\\{{\nwixident{rgb}}{rgb}}\\{{\nwixident{set{\_}asy{\_}style}}{set:unasy:unstyle}}}\nwindexuse{\nwixident{add{\_}cycle{\_}rel}}{add:uncycle:unrel}{NWppJ6t-RDDRz-6}\nwindexuse{\nwixident{get{\_}infinity}}{get:uninfinity}{NWppJ6t-RDDRz-6}\nwindexuse{\nwixident{is{\_}orthogonal}}{is:unorthogonal}{NWppJ6t-RDDRz-6}\nwindexuse{\nwixident{is{\_}tangent{\_}i}}{is:untangent:uni}{NWppJ6t-RDDRz-6}\nwindexuse{\nwixident{rgb}}{rgb}{NWppJ6t-RDDRz-6}\nwindexuse{\nwixident{set{\_}asy{\_}style}}{set:unasy:unstyle}{NWppJ6t-RDDRz-6}\nwendcode{}\nwbegindocs{245}\nwdocspar
\nwenddocs{}\nwbegindocs{246}Finally we draw the picture, see
Fig.~\ref{fig:illustr-fillm-spring}, which shall be compared
with~\cite{FillmoreSpringer00a}*{Fig.~1}.
\nwenddocs{}\nwbegincode{247}\sublabel{NWppJ6t-RDDRz-7}\nwmargintag{{\nwtagstyle{}\subpageref{NWppJ6t-RDDRz-7}}}\moddef{fillmore-springer-example.cpp~{\nwtagstyle{}\subpageref{NWppJ6t-RDDRz-1}}}\plusendmoddef\Rm{}\nwstartdeflinemarkup\nwprevnextdefs{NWppJ6t-RDDRz-6}{NWppJ6t-RDDRz-8}\nwenddeflinemarkup
    {\it{}F}.{\it{}asy\_write}(400,-4,20,-12.5,9,{\tt{}"fillmore-springer-example"});

\nwidentuses{\\{{\nwixident{asy{\_}write}}{asy:unwrite}}}\nwindexuse{\nwixident{asy{\_}write}}{asy:unwrite}{NWppJ6t-RDDRz-7}\nwendcode{}\nwbegindocs{248}Out of curiosity we may want to know that is square of tangents
intervals which are separate circles {\Tt{}\Rm{}{\it{}A}\nwendquote}, {\Tt{}\Rm{}{\it{}D}\nwendquote}. The output is:
\begin{verbatim}
Sq. cross tangent distance: {41.000000000000000003,-7.571428571428571435}
\end{verbatim}
Thus one solution does have such tangents with length \(\sqrt{41}\),
and for the second solution such tangents are imaginary since the
square is negative. This happens because one solution {\Tt{}\Rm{}{\it{}D}\nwendquote} intersects {\Tt{}\Rm{}{\it{}A}\nwendquote}.
\nwenddocs{}\nwbegincode{249}\sublabel{NWppJ6t-RDDRz-8}\nwmargintag{{\nwtagstyle{}\subpageref{NWppJ6t-RDDRz-8}}}\moddef{fillmore-springer-example.cpp~{\nwtagstyle{}\subpageref{NWppJ6t-RDDRz-1}}}\plusendmoddef\Rm{}\nwstartdeflinemarkup\nwprevnextdefs{NWppJ6t-RDDRz-7}{\relax}\nwenddeflinemarkup
    {\it{}cout} \begin{math}\ll\end{math} {\tt{}"Sq. cross tangent distance: "} \begin{math}\ll\end{math} {\it{}F}.{\it{}measure}({\it{}D},{\it{}A},{\it{}sq\_cross\_t\_distance\_is}).{\it{}evalf}() \begin{math}\ll\end{math} {\it{}endl};
   {\bf{}return} 0;
{\nwrbrace}
\nwindexdefn{\nwixident{measure}}{measure}{NWppJ6t-RDDRz-8}\nwindexdefn{\nwixident{sq{\_}cross{\_}t{\_}distance{\_}is}}{sq:uncross:unt:undistance:unis}{NWppJ6t-RDDRz-8}\eatline
\nwidentdefs{\\{{\nwixident{measure}}{measure}}\\{{\nwixident{sq{\_}cross{\_}t{\_}distance{\_}is}}{sq:uncross:unt:undistance:unis}}}\nwidentuses{\\{{\nwixident{evalf}}{evalf}}}\nwindexuse{\nwixident{evalf}}{evalf}{NWppJ6t-RDDRz-8}\nwendcode{}\nwbegindocs{250}\nwdocspar
\nwenddocs{}\nwbegindocs{251}\nwdocspar
\begin{figure}[htbp]
  \centering
  \includegraphics[scale=.7]{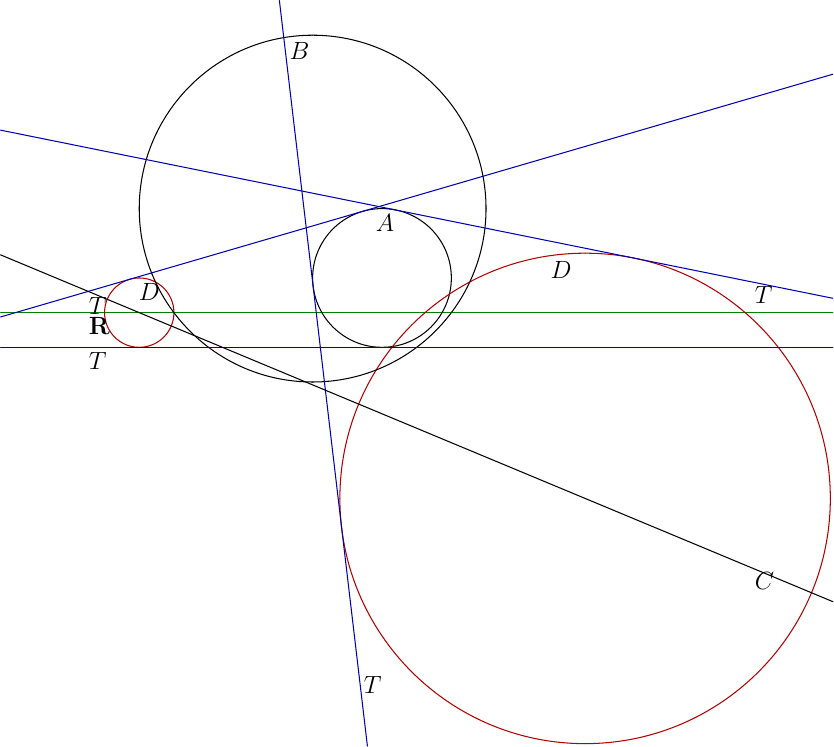}
  \caption[The illustration to Fillmore--Springer example]{The
    illustration to Fillmore--Springer example, which may be compared
    with~\cite{FillmoreSpringer00a}*{Fig.~1}.}
  \label{fig:illustr-fillm-spring}
\end{figure}
\nwenddocs{}\nwbegindocs{252}\nwdocspar
\nwenddocs{}\nwbegincode{253}\sublabel{NWppJ6t-1fWRR8-8}\nwmargintag{{\nwtagstyle{}\subpageref{NWppJ6t-1fWRR8-8}}}\moddef{separating chunk~{\nwtagstyle{}\subpageref{NWppJ6t-1fWRR8-1}}}\plusendmoddef\Rm{}\nwstartdeflinemarkup\nwprevnextdefs{NWppJ6t-1fWRR8-7}{NWppJ6t-1fWRR8-9}\nwenddeflinemarkup

\nwendcode{}\nwbegindocs{254}\nwdocspar
\subsection{Three-dimensional examples}
\label{sec:an-example-calc}

\nwenddocs{}\nwbegindocs{255}The most of the library functionality (except graphical methods) is
literally preserved for quadrics in arbitrary dimensions. We
demonstrate this on the following stereometric problem of
\href{https://en.wikipedia.org/wiki/Problem_of_Apollonius}{Apollonius
  type}, cf.~\cite{FillmoreSpringer00a}*{\S~8}. Let four spheres of
equal radii \(R\) 
have centres at four points \((1,1,1)\), \((-1,-1,1)\), \((-1,1,-1)\),
\((1,-1,-1)\). These points are vertices of a regular
\href{https://en.wikipedia.org/wiki/Platonic_solid}{tetrahedron} and
are every other vertices of a cube with the diagonal \(2\sqrt{3}\).

There are two obvious spheres with the centre at the
origin \((0,0,0)\) touching all four given spheres, they have radii
\(R+\sqrt{3}\) and \(\sqrt{3}-R\). Are there any others?

\nwenddocs{}\nwbegindocs{256}We start from the standard initiation and define the metric of three
dimensional Euclidean space.
\nwenddocs{}\nwbegincode{257}\sublabel{NWppJ6t-4Yd8hY-1}\nwmargintag{{\nwtagstyle{}\subpageref{NWppJ6t-4Yd8hY-1}}}\moddef{3D-figure-example.cpp~{\nwtagstyle{}\subpageref{NWppJ6t-4Yd8hY-1}}}\endmoddef\Rm{}\nwstartdeflinemarkup\nwprevnextdefs{\relax}{NWppJ6t-4Yd8hY-2}\nwenddeflinemarkup
\LA{}3D-figure-example-common~{\nwtagstyle{}\subpageref{NWppJ6t-2pgnFW-1}}\RA{}

\nwalsodefined{\\{NWppJ6t-4Yd8hY-2}\\{NWppJ6t-4Yd8hY-3}}\nwnotused{3D-figure-example.cpp}\nwendcode{}\nwbegindocs{258}The following two chunks are shared with the next example.
\nwenddocs{}\nwbegincode{259}\sublabel{NWppJ6t-2pgnFW-1}\nwmargintag{{\nwtagstyle{}\subpageref{NWppJ6t-2pgnFW-1}}}\moddef{3D-figure-example-common~{\nwtagstyle{}\subpageref{NWppJ6t-2pgnFW-1}}}\endmoddef\Rm{}\nwstartdeflinemarkup\nwusesondefline{\\{NWppJ6t-4Yd8hY-1}\\{NWppJ6t-u6tXb-1}}\nwprevnextdefs{\relax}{NWppJ6t-2pgnFW-2}\nwenddeflinemarkup
\LA{}license~{\nwtagstyle{}\subpageref{NWppJ6t-ZXuKx-1}}\RA{}
{\bf{}\char35{}include}{\tt{} "figure.h"}
\LA{}using all namespaces~{\nwtagstyle{}\subpageref{NWppJ6t-3uHj2c-1}}\RA{}
{\bf{}int} {\it{}main}(){\nwlbrace}\nwindexdefn{\nwixident{main}}{main}{NWppJ6t-2pgnFW-1}
    {\bf{}ex} {\it{}e3D} = {\it{}clifford\_unit}({\bf{}varidx}({\bf{}symbol}({\tt{}"lam"}), 3), {\it{}diag\_matrix}({\bf{}lst}{\nwlbrace}-1,-1,-1{\nwrbrace})); // Metric for 3D space
    {\it{}possymbol} {\it{}R}({\tt{}"R"});
    {\bf{}figure} {\it{}F}({\it{}e3D});

\nwalsodefined{\\{NWppJ6t-2pgnFW-2}}\nwused{\\{NWppJ6t-4Yd8hY-1}\\{NWppJ6t-u6tXb-1}}\nwidentdefs{\\{{\nwixident{main}}{main}}}\nwidentuses{\\{{\nwixident{ex}}{ex}}\\{{\nwixident{figure}}{figure}}}\nwindexuse{\nwixident{ex}}{ex}{NWppJ6t-2pgnFW-1}\nwindexuse{\nwixident{figure}}{figure}{NWppJ6t-2pgnFW-1}\nwendcode{}\nwbegindocs{260}Then we put four given spheres to the figure. They are defined by
their centres and square of radii.
\nwenddocs{}\nwbegincode{261}\sublabel{NWppJ6t-2pgnFW-2}\nwmargintag{{\nwtagstyle{}\subpageref{NWppJ6t-2pgnFW-2}}}\moddef{3D-figure-example-common~{\nwtagstyle{}\subpageref{NWppJ6t-2pgnFW-1}}}\plusendmoddef\Rm{}\nwstartdeflinemarkup\nwusesondefline{\\{NWppJ6t-4Yd8hY-1}\\{NWppJ6t-u6tXb-1}}\nwprevnextdefs{NWppJ6t-2pgnFW-1}{\relax}\nwenddeflinemarkup
    {\commopen} Numerical radii {\commclose}
    \begin{math}\div\end{math}\begin{math}\ast\end{math}  {\bf{}ex} {\it{}P1}={\it{}F}.{\it{}add\_cycle}({\bf{}cycle}({\bf{}lst}{\nwlbrace}1,1,1{\nwrbrace}, {\it{}e3D}, {\bf{}numeric}(3,4)), {\tt{}"P1"});
    {\bf{}ex} {\it{}P2}={\it{}F}.{\it{}add\_cycle}({\bf{}cycle}({\bf{}lst}{\nwlbrace}-1,-1,1{\nwrbrace}, {\it{}e3D}, {\bf{}numeric}(3,4)), {\tt{}"P2"});
    {\bf{}ex} {\it{}P3}={\it{}F}.{\it{}add\_cycle}({\bf{}cycle}({\bf{}lst}{\nwlbrace}1,-1,-1{\nwrbrace}, {\it{}e3D}, {\bf{}numeric}(3,4)), {\tt{}"P3"});
    {\bf{}ex} {\it{}P4}={\it{}F}.{\it{}add\_cycle}({\bf{}cycle}({\bf{}lst}{\nwlbrace}-1,1,-1{\nwrbrace}, {\it{}e3D}, {\bf{}numeric}(3,4)), {\tt{}"P4"});
    \begin{math}\ast\end{math}\begin{math}\div\end{math}
    {\bf{}ex} {\it{}P1}={\it{}F}.{\it{}add\_cycle}({\bf{}cycle}({\bf{}lst}{\nwlbrace}1,1,1{\nwrbrace}, {\it{}e3D}, {\it{}pow}({\it{}R},2)), {\tt{}"P1"});
    {\bf{}ex} {\it{}P2}={\it{}F}.{\it{}add\_cycle}({\bf{}cycle}({\bf{}lst}{\nwlbrace}-1,-1,1{\nwrbrace}, {\it{}e3D}, {\it{}pow}({\it{}R},2)), {\tt{}"P2"});
    {\bf{}ex} {\it{}P3}={\it{}F}.{\it{}add\_cycle}({\bf{}cycle}({\bf{}lst}{\nwlbrace}1,-1,-1{\nwrbrace}, {\it{}e3D}, {\it{}pow}({\it{}R},2)), {\tt{}"P3"});
    {\bf{}ex} {\it{}P4}={\it{}F}.{\it{}add\_cycle}({\bf{}cycle}({\bf{}lst}{\nwlbrace}-1,1,-1{\nwrbrace}, {\it{}e3D}, {\it{}pow}({\it{}R},2)), {\tt{}"P4"});

\nwused{\\{NWppJ6t-4Yd8hY-1}\\{NWppJ6t-u6tXb-1}}\nwidentuses{\\{{\nwixident{add{\_}cycle}}{add:uncycle}}\\{{\nwixident{ex}}{ex}}\\{{\nwixident{numeric}}{numeric}}}\nwindexuse{\nwixident{add{\_}cycle}}{add:uncycle}{NWppJ6t-2pgnFW-2}\nwindexuse{\nwixident{ex}}{ex}{NWppJ6t-2pgnFW-2}\nwindexuse{\nwixident{numeric}}{numeric}{NWppJ6t-2pgnFW-2}\nwendcode{}\nwbegindocs{262}Then we introduce the unknown cycle by the four tangency
conditions to given spheres. We also put two conditions to rule out
non-geometric solutions: {\Tt{}\Rm{}{\it{}is\_real\_cycle}\nwendquote} checks that the radius is
real, {\Tt{}\Rm{}{\it{}only\_reals}\nwendquote} requires that all coefficients are real.
\nwenddocs{}\nwbegincode{263}\sublabel{NWppJ6t-4Yd8hY-2}\nwmargintag{{\nwtagstyle{}\subpageref{NWppJ6t-4Yd8hY-2}}}\moddef{3D-figure-example.cpp~{\nwtagstyle{}\subpageref{NWppJ6t-4Yd8hY-1}}}\plusendmoddef\Rm{}\nwstartdeflinemarkup\nwprevnextdefs{NWppJ6t-4Yd8hY-1}{NWppJ6t-4Yd8hY-3}\nwenddeflinemarkup
    {\bf{}realsymbol} {\it{}N3}({\tt{}"N3"});
    {\it{}F}.{\it{}add\_cycle\_rel}({\bf{}lst}{\nwlbrace}{\it{}is\_tangent}({\it{}P1}), {\it{}is\_tangent}({\it{}P2}), {\it{}is\_tangent}({\it{}P3}),{\it{}is\_tangent}({\it{}P4})
                        {\commopen} Tests below forbid all spheres with symbolic parameters {\commclose}
                        //, only\_reals(N3), is\_real\_cycle(N3)
             {\nwrbrace}, {\it{}N3});
\nwindexdefn{\nwixident{is{\_}tangent}}{is:untangent}{NWppJ6t-4Yd8hY-2}\nwindexdefn{\nwixident{is{\_}real{\_}cycle}}{is:unreal:uncycle}{NWppJ6t-4Yd8hY-2}\nwindexdefn{\nwixident{only{\_}reals}}{only:unreals}{NWppJ6t-4Yd8hY-2}\eatline
\nwidentdefs{\\{{\nwixident{is{\_}real{\_}cycle}}{is:unreal:uncycle}}\\{{\nwixident{is{\_}tangent}}{is:untangent}}\\{{\nwixident{only{\_}reals}}{only:unreals}}}\nwidentuses{\\{{\nwixident{add{\_}cycle{\_}rel}}{add:uncycle:unrel}}\\{{\nwixident{realsymbol}}{realsymbol}}}\nwindexuse{\nwixident{add{\_}cycle{\_}rel}}{add:uncycle:unrel}{NWppJ6t-4Yd8hY-2}\nwindexuse{\nwixident{realsymbol}}{realsymbol}{NWppJ6t-4Yd8hY-2}\nwendcode{}\nwbegindocs{264}\nwdocspar
\nwenddocs{}\nwbegindocs{265}Then we output the solutions and their radii.
\nwenddocs{}\nwbegincode{266}\sublabel{NWppJ6t-4Yd8hY-3}\nwmargintag{{\nwtagstyle{}\subpageref{NWppJ6t-4Yd8hY-3}}}\moddef{3D-figure-example.cpp~{\nwtagstyle{}\subpageref{NWppJ6t-4Yd8hY-1}}}\plusendmoddef\Rm{}\nwstartdeflinemarkup\nwprevnextdefs{NWppJ6t-4Yd8hY-2}{\relax}\nwenddeflinemarkup
    {\bf{}lst} {\it{}L}={\it{}ex\_to}\begin{math}<\end{math}{\bf{}lst}\begin{math}>\end{math}({\it{}F}.{\it{}get\_cycle}({\it{}N3}));
    {\it{}cout} \begin{math}\ll\end{math} {\it{}L}.{\it{}nops}() \begin{math}\ll\end{math} {\tt{}" spheres found"} \begin{math}\ll\end{math} {\it{}endl};
        {\bf{}for} ({\bf{}auto} {\it{}x}: {\it{}L})
        {\it{}cout} \begin{math}\ll\end{math} {\tt{}"Sphere: "} \begin{math}\ll\end{math} {\it{}ex\_to}\begin{math}<\end{math}{\bf{}cycle}\begin{math}>\end{math}({\it{}x}).{\it{}normal}()
         \begin{math}\ll\end{math} {\tt{}", radius sq: "} \begin{math}\ll\end{math} ({\it{}ex\_to}\begin{math}<\end{math}{\bf{}cycle}\begin{math}>\end{math}({\it{}x}).{\it{}det}()).{\it{}normal}()
         \begin{math}\ll\end{math}{\it{}endl};
    {\bf{}return} 0;
{\nwrbrace}

\nwidentuses{\\{{\nwixident{get{\_}cycle}}{get:uncycle}}\\{{\nwixident{nops}}{nops}}}\nwindexuse{\nwixident{get{\_}cycle}}{get:uncycle}{NWppJ6t-4Yd8hY-3}\nwindexuse{\nwixident{nops}}{nops}{NWppJ6t-4Yd8hY-3}\nwendcode{}\nwbegindocs{267}For the numerical value \(R=\frac{\sqrt{3}}{2}\), the program found
\(16\) different solutions which satisfy to {\Tt{}\Rm{}{\it{}is\_real\_cycle}\nwendquote} and
{\Tt{}\Rm{}{\it{}only\_reals}\nwendquote} conditions. If we omit these conditions then additional
\(16\) imaginary spheres will be producing (\(32\) in total).

For the symbolic radii \(R\) again \(32\) different spheres are
found. The condition {\Tt{}\Rm{}{\it{}only\_reals}\nwendquote} leaves only two obvious spheres,
discussed at the beginning of the subsection. This happens because for
some value of \(R\) coefficient of other spheres may turn to be
complex. Finally, if we use the condition {\Tt{}\Rm{}{\it{}is\_real\_cycle}\nwendquote}, then no
sphere passes it---the square of its radius may become negative for
some \(R\).

\nwenddocs{}\nwbegindocs{268}For visualisation we can partially re-use the previous code.
\nwenddocs{}\nwbegincode{269}\sublabel{NWppJ6t-u6tXb-1}\nwmargintag{{\nwtagstyle{}\subpageref{NWppJ6t-u6tXb-1}}}\moddef{3D-figure-visualise.cpp~{\nwtagstyle{}\subpageref{NWppJ6t-u6tXb-1}}}\endmoddef\Rm{}\nwstartdeflinemarkup\nwprevnextdefs{\relax}{NWppJ6t-u6tXb-2}\nwenddeflinemarkup
\LA{}3D-figure-example-common~{\nwtagstyle{}\subpageref{NWppJ6t-2pgnFW-1}}\RA{}

\nwalsodefined{\\{NWppJ6t-u6tXb-2}\\{NWppJ6t-u6tXb-3}}\nwnotused{3D-figure-visualise.cpp}\nwendcode{}\nwbegindocs{270}To simplify the structure we eliminate spheres which are different
only up to the rotational symmetry of the initial set-up. To this end
we explicitly specify inner or outer tangency for different spheres.
\nwenddocs{}\nwbegincode{271}\sublabel{NWppJ6t-u6tXb-2}\nwmargintag{{\nwtagstyle{}\subpageref{NWppJ6t-u6tXb-2}}}\moddef{3D-figure-visualise.cpp~{\nwtagstyle{}\subpageref{NWppJ6t-u6tXb-1}}}\plusendmoddef\Rm{}\nwstartdeflinemarkup\nwprevnextdefs{NWppJ6t-u6tXb-1}{NWppJ6t-u6tXb-3}\nwenddeflinemarkup
    {\bf{}realsymbol} {\it{}N0}({\tt{}"N0"}), {\it{}N1}({\tt{}"N1"}), {\it{}N2}({\tt{}"N2"}), {\it{}N3}({\tt{}"N3"}),  {\it{}N4}({\tt{}"N4"});
    {\it{}F}.{\it{}add\_cycle\_rel}({\bf{}lst}{\nwlbrace}{\it{}is\_tangent\_o}({\it{}P1}), {\it{}is\_tangent\_o}({\it{}P2}), {\it{}is\_tangent\_o}({\it{}P3}),{\it{}is\_tangent\_o}({\it{}P4}),
                {\it{}is\_real\_cycle}({\it{}N0}), {\it{}only\_reals}({\it{}N0}){\nwrbrace}, {\it{}N0});
    {\it{}F}.{\it{}add\_cycle\_rel}({\bf{}lst}{\nwlbrace}{\it{}is\_tangent\_o}({\it{}P1}), {\it{}is\_tangent\_o}({\it{}P2}), {\it{}is\_tangent\_o}({\it{}P3}),{\it{}is\_tangent\_i}({\it{}P4}),
                {\it{}is\_real\_cycle}({\it{}N1}), {\it{}only\_reals}({\it{}N1}){\nwrbrace}, {\it{}N1});
    {\it{}F}.{\it{}add\_cycle\_rel}({\bf{}lst}{\nwlbrace}{\it{}is\_tangent\_o}({\it{}P1}), {\it{}is\_tangent\_o}({\it{}P2}), {\it{}is\_tangent\_i}({\it{}P3}),{\it{}is\_tangent\_i}({\it{}P4}),
                {\it{}is\_real\_cycle}({\it{}N2}), {\it{}only\_reals}({\it{}N2}){\nwrbrace}, {\it{}N2});
    {\it{}F}.{\it{}add\_cycle\_rel}({\bf{}lst}{\nwlbrace}{\it{}is\_tangent\_o}({\it{}P1}), {\it{}is\_tangent\_i}({\it{}P2}), {\it{}is\_tangent\_i}({\it{}P3}),{\it{}is\_tangent\_i}({\it{}P4}),
                {\it{}is\_real\_cycle}({\it{}N3}), {\it{}only\_reals}({\it{}N3}){\nwrbrace}, {\it{}N3});

\nwidentuses{\\{{\nwixident{add{\_}cycle{\_}rel}}{add:uncycle:unrel}}\\{{\nwixident{is{\_}real{\_}cycle}}{is:unreal:uncycle}}\\{{\nwixident{is{\_}tangent{\_}i}}{is:untangent:uni}}\\{{\nwixident{is{\_}tangent{\_}o}}{is:untangent:uno}}\\{{\nwixident{only{\_}reals}}{only:unreals}}\\{{\nwixident{realsymbol}}{realsymbol}}}\nwindexuse{\nwixident{add{\_}cycle{\_}rel}}{add:uncycle:unrel}{NWppJ6t-u6tXb-2}\nwindexuse{\nwixident{is{\_}real{\_}cycle}}{is:unreal:uncycle}{NWppJ6t-u6tXb-2}\nwindexuse{\nwixident{is{\_}tangent{\_}i}}{is:untangent:uni}{NWppJ6t-u6tXb-2}\nwindexuse{\nwixident{is{\_}tangent{\_}o}}{is:untangent:uno}{NWppJ6t-u6tXb-2}\nwindexuse{\nwixident{only{\_}reals}}{only:unreals}{NWppJ6t-u6tXb-2}\nwindexuse{\nwixident{realsymbol}}{realsymbol}{NWppJ6t-u6tXb-2}\nwendcode{}\nwbegindocs{272}Now we save the arrangement for the numerical value \(R^2=\frac{3}{4}\).
\nwenddocs{}\nwbegincode{273}\sublabel{NWppJ6t-u6tXb-3}\nwmargintag{{\nwtagstyle{}\subpageref{NWppJ6t-u6tXb-3}}}\moddef{3D-figure-visualise.cpp~{\nwtagstyle{}\subpageref{NWppJ6t-u6tXb-1}}}\plusendmoddef\Rm{}\nwstartdeflinemarkup\nwprevnextdefs{NWppJ6t-u6tXb-2}{\relax}\nwenddeflinemarkup
    {\it{}F}.{\it{}subs}({\it{}R}\begin{math}\equiv\end{math}{\it{}sqrt}({\it{}ex\_to}\begin{math}<\end{math}{\bf{}numeric}\begin{math}>\end{math}(3))\begin{math}\div\end{math}2).{\it{}arrangement\_write}({\tt{}"appolonius"});
    {\bf{}return} 0;
{\nwrbrace}
\nwindexdefn{\nwixident{arrangement{\_}write}}{arrangement:unwrite}{NWppJ6t-u6tXb-3}\eatline
\nwidentdefs{\\{{\nwixident{arrangement{\_}write}}{arrangement:unwrite}}}\nwidentuses{\\{{\nwixident{numeric}}{numeric}}\\{{\nwixident{subs}}{subs}}}\nwindexuse{\nwixident{numeric}}{numeric}{NWppJ6t-u6tXb-3}\nwindexuse{\nwixident{subs}}{subs}{NWppJ6t-u6tXb-3}\nwendcode{}\nwbegindocs{274}\nwdocspar
\nwenddocs{}\nwbegindocs{275}Now this arrangement can be visualised by loading the file
\texttt{appolonius.txt} into the helper programme
\texttt{cycle3D-visualiser}. A screenshot of such visualisation is
shown on Fig.~\ref{fig:apollonius-3D}.

\nwenddocs{}\nwbegindocs{276}\nwdocspar
\section{Public Methods in the {\Tt{}\Rm{}{\bf{}figure}\nwendquote} class}
\label{sec:publ-meth-figure}

This section lists all methods, which may be of interest to an
end-user. An advanced user may find further advise in
Appendix~\ref{sec:figure-header-file}, which outlines the library
header file. Methods presented here are grouped by their purpose.

The source (interleaved with documentation in a \NoWEB\ file) can be
found at \href{http://moebinv.sourceforge.net/}{SourceForge
  project page}~\cite{Kisil14b} as Git tree.  The code is written
using \NoWEB\ \href{http://en.wikipedia.org/wiki/Literate_programming}{literate
  programming} tool~\cite{NoWEB}.  The code uses some
{\CPPeleven} features, e.g. regeps and {\Tt{}\Rm{}{\it{}std}::{\it{}function}\nwendquote}. Drawing procedures
delegate to \Asymptote~\cite{Asymptote}.

The stable realises and
full documentation are in
\href{https://sourceforge.net/projects/moebinv/files/}{Files section
  of the project}. A release archive contain all {\CPP} files extracted
from the \NoWEB\ source, thus only the standard {\CPP} compiler is
necessary to use them.

Furthermore, a live CD with the compiled library, examples and all
required tools is distributed as an ISO image. You may find a link to
the ISO image at the start of this Web page:
\begin{quote}
\url{http://www.maths.leeds.ac.uk/~kisilv/courses/using_sw.html}
\end{quote}
It also explains how to use the live CD image either to boot your
computer or inside a Virtual Machine.

\nwenddocs{}\nwbegindocs{277}\nwdocspar
\subsection{Creation and re-setting of {\Tt{}\Rm{}{\bf{}figure}\nwendquote}, changing {\Tt{}\Rm{}{\it{}metric}\nwendquote}}

Here are methods to initialise {\Tt{}\Rm{}{\bf{}figure}\nwendquote} and manipulate its basic property.

\label{sec:creation-re-setting}

\nwenddocs{}\nwbegindocs{278}This is the simplest constructor of an initial figure with the
(point) metric {\Tt{}\Rm{}{\it{}Mp}\nwendquote}. By default, any figure contains the {\Tt{}\Rm{}{\it{}real\_line}\nwendquote} and
{\Tt{}\Rm{}{\it{}infinity}\nwendquote}. Parameter {\Tt{}\Rm{}{\it{}M}\nwendquote}, may be same as for definition of
{\Tt{}\Rm{}{\it{}clifford\_unit}\nwendquote} in \GiNaC, that is,  be
represented by a square {\Tt{}\Rm{}{\bf{}matrix}\nwendquote}, {\Tt{}\Rm{}{\bf{}clifford}\nwendquote} or
{\Tt{}\Rm{}{\bf{}indexed}\nwendquote} class object.  If the metric {\Tt{}\Rm{}{\it{}Mp}\nwendquote} is not provided, then the
default elliptic metric in two dimensions is used, it is given by the matrix
\(\begin{pmatrix}
  -1&0\\0&-1
\end{pmatrix}\).

An advanced user may wish to specify a different metric for point
and cycle spaces, see~\cite{Kisil12a}*{\S~4.2} for the discussion. By
default, if the metric in the point space is \(\begin{pmatrix}
  -1&0\\0&\sigma
\end{pmatrix}\) then the metric of cycle space is:
\begin{equation}
  \label{eq:int-heaviside-function}
\begin{pmatrix}
  -1&0\\0&-\chi(-\sigma)
\end{pmatrix},
\qquad \text { where } \quad
  \chi(t)=\left\{
    \begin{array}{ll}
      1,& t\geq 0;\\
      -1,& t<0.
    \end{array}\right.
\end{equation}
is the \emph{\wiki{Heaviside_step_function}{Heaviside function}}%
\index{function!Heaviside}%
\index{Heaviside!function} \(\chi(\sigma)\)
In other word, by default for elliptic and parabolic point space the
cycle space has the same metric and for the parabolic point space the
cycle space is elliptic. If a user want a different combination then
the following constructor need to be used, see also {\Tt{}\Rm{}{\it{}set\_metric}()\nwendquote} below
\nwenddocs{}\nwbegincode{279}\sublabel{NWppJ6t-AWRj3-1}\nwmargintag{{\nwtagstyle{}\subpageref{NWppJ6t-AWRj3-1}}}\moddef{public methods in figure class~{\nwtagstyle{}\subpageref{NWppJ6t-AWRj3-1}}}\endmoddef\Rm{}\nwstartdeflinemarkup\nwusesondefline{\\{NWppJ6t-7vmg0-4}}\nwprevnextdefs{\relax}{NWppJ6t-AWRj3-2}\nwenddeflinemarkup
{\bf{}figure}({\bf{}const} {\bf{}ex} & {\it{}Mp}, {\bf{}const} {\bf{}ex} & {\it{}Mc}=0);
\nwindexdefn{\nwixident{figure}}{figure}{NWppJ6t-AWRj3-1}\eatline
\nwalsodefined{\\{NWppJ6t-AWRj3-2}\\{NWppJ6t-AWRj3-3}\\{NWppJ6t-AWRj3-4}\\{NWppJ6t-AWRj3-5}\\{NWppJ6t-AWRj3-6}\\{NWppJ6t-AWRj3-7}\\{NWppJ6t-AWRj3-8}\\{NWppJ6t-AWRj3-9}\\{NWppJ6t-AWRj3-A}\\{NWppJ6t-AWRj3-B}\\{NWppJ6t-AWRj3-C}\\{NWppJ6t-AWRj3-D}\\{NWppJ6t-AWRj3-E}\\{NWppJ6t-AWRj3-F}\\{NWppJ6t-AWRj3-G}\\{NWppJ6t-AWRj3-H}\\{NWppJ6t-AWRj3-I}\\{NWppJ6t-AWRj3-J}\\{NWppJ6t-AWRj3-K}\\{NWppJ6t-AWRj3-L}\\{NWppJ6t-AWRj3-M}\\{NWppJ6t-AWRj3-N}\\{NWppJ6t-AWRj3-O}\\{NWppJ6t-AWRj3-P}\\{NWppJ6t-AWRj3-Q}\\{NWppJ6t-AWRj3-R}\\{NWppJ6t-AWRj3-S}\\{NWppJ6t-AWRj3-T}\\{NWppJ6t-AWRj3-U}\\{NWppJ6t-AWRj3-V}\\{NWppJ6t-AWRj3-W}\\{NWppJ6t-AWRj3-X}\\{NWppJ6t-AWRj3-Y}}\nwused{\\{NWppJ6t-7vmg0-4}}\nwidentdefs{\\{{\nwixident{figure}}{figure}}}\nwidentuses{\\{{\nwixident{ex}}{ex}}}\nwindexuse{\nwixident{ex}}{ex}{NWppJ6t-AWRj3-1}\nwendcode{}\nwbegindocs{280}\nwdocspar
\nwenddocs{}\nwbegindocs{281}The metrics set in the above constructor can be changed at any stage,
and all cycles will be re-calculated in the figure accordingly. The
parameter {\Tt{}\Rm{}{\it{}Mp}\nwendquote} can be the same type of object as in the first
constructor {\Tt{}\Rm{}{\bf{}figure}({\bf{}const} {\bf{}ex} \& )\nwendquote}.  The first form change the point
space metric and derive respective cycle space metric as described
above. In the second form both metrics are provided explicitly.
\nwenddocs{}\nwbegincode{282}\sublabel{NWppJ6t-AWRj3-2}\nwmargintag{{\nwtagstyle{}\subpageref{NWppJ6t-AWRj3-2}}}\moddef{public methods in figure class~{\nwtagstyle{}\subpageref{NWppJ6t-AWRj3-1}}}\plusendmoddef\Rm{}\nwstartdeflinemarkup\nwusesondefline{\\{NWppJ6t-7vmg0-4}}\nwprevnextdefs{NWppJ6t-AWRj3-1}{NWppJ6t-AWRj3-3}\nwenddeflinemarkup
{\bf{}void} {\it{}set\_metric}({\bf{}const} {\bf{}ex} & {\it{}Mp}, {\bf{}const} {\bf{}ex} & {\it{}Mc}=0);
\nwindexdefn{\nwixident{set{\_}metric}}{set:unmetric}{NWppJ6t-AWRj3-2}\eatline
\nwused{\\{NWppJ6t-7vmg0-4}}\nwidentdefs{\\{{\nwixident{set{\_}metric}}{set:unmetric}}}\nwidentuses{\\{{\nwixident{ex}}{ex}}}\nwindexuse{\nwixident{ex}}{ex}{NWppJ6t-AWRj3-2}\nwendcode{}\nwbegindocs{283}\nwdocspar
\nwenddocs{}\nwbegindocs{284}This constructor can be used to create a figure with the pre-defined
collection {\Tt{}\Rm{}{\it{}N}\nwendquote} of cycles.
\nwenddocs{}\nwbegincode{285}\sublabel{NWppJ6t-AWRj3-3}\nwmargintag{{\nwtagstyle{}\subpageref{NWppJ6t-AWRj3-3}}}\moddef{public methods in figure class~{\nwtagstyle{}\subpageref{NWppJ6t-AWRj3-1}}}\plusendmoddef\Rm{}\nwstartdeflinemarkup\nwusesondefline{\\{NWppJ6t-7vmg0-4}}\nwprevnextdefs{NWppJ6t-AWRj3-2}{NWppJ6t-AWRj3-4}\nwenddeflinemarkup
{\bf{}figure}({\bf{}const} {\bf{}ex} & {\it{}Mp}, {\bf{}const} {\bf{}ex} & {\it{}Mc}, {\bf{}const} {\it{}exhashmap}\begin{math}<\end{math}{\bf{}cycle\_node}\begin{math}>\end{math} & {\it{}N});
\nwindexdefn{\nwixident{figure}}{figure}{NWppJ6t-AWRj3-3}\eatline
\nwused{\\{NWppJ6t-7vmg0-4}}\nwidentdefs{\\{{\nwixident{figure}}{figure}}}\nwidentuses{\\{{\nwixident{cycle{\_}node}}{cycle:unnode}}\\{{\nwixident{ex}}{ex}}}\nwindexuse{\nwixident{cycle{\_}node}}{cycle:unnode}{NWppJ6t-AWRj3-3}\nwindexuse{\nwixident{ex}}{ex}{NWppJ6t-AWRj3-3}\nwendcode{}\nwbegindocs{286}\nwdocspar
\nwenddocs{}\nwbegindocs{287} Remove all {\Tt{}\Rm{}{\bf{}cycle\_node}\nwendquote} from the figure.  Only the
{\Tt{}\Rm{}{\it{}point\_metric}\nwendquote}, {\Tt{}\Rm{}{\it{}cycle\_metric}\nwendquote}, {\Tt{}\Rm{}{\it{}real\_line}\nwendquote} and
{\Tt{}\Rm{}{\it{}infinity}\nwendquote} are preserved.
\nwenddocs{}\nwbegincode{288}\sublabel{NWppJ6t-AWRj3-4}\nwmargintag{{\nwtagstyle{}\subpageref{NWppJ6t-AWRj3-4}}}\moddef{public methods in figure class~{\nwtagstyle{}\subpageref{NWppJ6t-AWRj3-1}}}\plusendmoddef\Rm{}\nwstartdeflinemarkup\nwusesondefline{\\{NWppJ6t-7vmg0-4}}\nwprevnextdefs{NWppJ6t-AWRj3-3}{NWppJ6t-AWRj3-5}\nwenddeflinemarkup
{\bf{}void} {\it{}reset\_figure}();
\nwindexdefn{\nwixident{reset{\_}figure}}{reset:unfigure}{NWppJ6t-AWRj3-4}\eatline
\nwused{\\{NWppJ6t-7vmg0-4}}\nwidentdefs{\\{{\nwixident{reset{\_}figure}}{reset:unfigure}}}\nwendcode{}\nwbegindocs{289}\nwdocspar
\nwenddocs{}\nwbegindocs{290}\nwdocspar
\subsection{Adding elements to figure}
\label{sec:adding-elem-figure}

\nwenddocs{}\nwbegindocs{291}This method add points to the figure. A point is represented
as cycles with radius \(0\) (with respect to the cycle metric) and
coordinates \(x=(x_1, \ldots,x_n)\) of their centre (represented by a
{\Tt{}\Rm{}{\bf{}lst}\nwendquote} of the suitable length). The procedure returns a
symbol, which can be used later to refer this point. In the first form
parameters {\Tt{}\Rm{}{\it{}name}\nwendquote} and (optional) {\Tt{}\Rm{}{\it{}TeXname}\nwendquote} provide respective
string to name this new symbol. In the second form the whole symbol
{\Tt{}\Rm{}{\it{}key}\nwendquote} is provided (and it will be returned by the procedure).
\nwenddocs{}\nwbegincode{292}\sublabel{NWppJ6t-AWRj3-5}\nwmargintag{{\nwtagstyle{}\subpageref{NWppJ6t-AWRj3-5}}}\moddef{public methods in figure class~{\nwtagstyle{}\subpageref{NWppJ6t-AWRj3-1}}}\plusendmoddef\Rm{}\nwstartdeflinemarkup\nwusesondefline{\\{NWppJ6t-7vmg0-4}}\nwprevnextdefs{NWppJ6t-AWRj3-4}{NWppJ6t-AWRj3-6}\nwenddeflinemarkup
{\bf{}ex} {\it{}add\_point}({\bf{}const} {\bf{}ex} & {\it{}x}, {\it{}string} {\it{}name}, {\it{}string} {\it{}TeXname}={\tt{}""});
{\bf{}ex} {\it{}add\_point}({\bf{}const} {\bf{}ex} & {\it{}x}, {\bf{}const} {\bf{}ex} & {\it{}key});
\nwindexdefn{\nwixident{add{\_}point}}{add:unpoint}{NWppJ6t-AWRj3-5}\nwindexdefn{\nwixident{name}}{name}{NWppJ6t-AWRj3-5}\nwindexdefn{\nwixident{TeXname}}{TeXname}{NWppJ6t-AWRj3-5}\nwindexdefn{\nwixident{key}}{key}{NWppJ6t-AWRj3-5}\eatline
\nwused{\\{NWppJ6t-7vmg0-4}}\nwidentdefs{\\{{\nwixident{add{\_}point}}{add:unpoint}}\\{{\nwixident{key}}{key}}\\{{\nwixident{name}}{name}}\\{{\nwixident{TeXname}}{TeXname}}}\nwidentuses{\\{{\nwixident{ex}}{ex}}}\nwindexuse{\nwixident{ex}}{ex}{NWppJ6t-AWRj3-5}\nwendcode{}\nwbegindocs{293}\nwdocspar
\nwenddocs{}\nwbegindocs{294}This method add a cycle at zero generation without parents. The
returned value and parameters {\Tt{}\Rm{}{\it{}name}\nwendquote}, {\Tt{}\Rm{}{\it{}TeXname}\nwendquote} and {\Tt{}\Rm{}{\it{}key}\nwendquote} are as
in the previous methods {\Tt{}\Rm{}{\it{}add\_point}()\nwendquote}.
\nwenddocs{}\nwbegincode{295}\sublabel{NWppJ6t-AWRj3-6}\nwmargintag{{\nwtagstyle{}\subpageref{NWppJ6t-AWRj3-6}}}\moddef{public methods in figure class~{\nwtagstyle{}\subpageref{NWppJ6t-AWRj3-1}}}\plusendmoddef\Rm{}\nwstartdeflinemarkup\nwusesondefline{\\{NWppJ6t-7vmg0-4}}\nwprevnextdefs{NWppJ6t-AWRj3-5}{NWppJ6t-AWRj3-7}\nwenddeflinemarkup
{\bf{}ex} {\it{}add\_cycle}({\bf{}const} {\bf{}ex} & {\it{}C}, {\it{}string} {\it{}name}, {\it{}string} {\it{}TeXname}={\tt{}""});
{\bf{}ex} {\it{}add\_cycle}({\bf{}const} {\bf{}ex} & {\it{}C}, {\bf{}const} {\bf{}ex} & {\it{}key});
\nwindexdefn{\nwixident{add{\_}cycle}}{add:uncycle}{NWppJ6t-AWRj3-6}\eatline
\nwused{\\{NWppJ6t-7vmg0-4}}\nwidentdefs{\\{{\nwixident{add{\_}cycle}}{add:uncycle}}}\nwidentuses{\\{{\nwixident{ex}}{ex}}\\{{\nwixident{key}}{key}}\\{{\nwixident{name}}{name}}\\{{\nwixident{TeXname}}{TeXname}}}\nwindexuse{\nwixident{ex}}{ex}{NWppJ6t-AWRj3-6}\nwindexuse{\nwixident{key}}{key}{NWppJ6t-AWRj3-6}\nwindexuse{\nwixident{name}}{name}{NWppJ6t-AWRj3-6}\nwindexuse{\nwixident{TeXname}}{TeXname}{NWppJ6t-AWRj3-6}\nwendcode{}\nwbegindocs{296}\nwdocspar
\nwenddocs{}\nwbegindocs{297}Add a new node by a set {\Tt{}\Rm{}{\it{}rel}\nwendquote} of relations. The
returned value and parameters {\Tt{}\Rm{}{\it{}name}\nwendquote}, {\Tt{}\Rm{}{\it{}TeXname}\nwendquote} and {\Tt{}\Rm{}{\it{}key}\nwendquote} are as
in methods {\Tt{}\Rm{}{\it{}add\_point}()\nwendquote}.
\nwenddocs{}\nwbegincode{298}\sublabel{NWppJ6t-AWRj3-7}\nwmargintag{{\nwtagstyle{}\subpageref{NWppJ6t-AWRj3-7}}}\moddef{public methods in figure class~{\nwtagstyle{}\subpageref{NWppJ6t-AWRj3-1}}}\plusendmoddef\Rm{}\nwstartdeflinemarkup\nwusesondefline{\\{NWppJ6t-7vmg0-4}}\nwprevnextdefs{NWppJ6t-AWRj3-6}{NWppJ6t-AWRj3-8}\nwenddeflinemarkup
{\bf{}ex} {\it{}add\_cycle\_rel}({\bf{}const} {\bf{}lst} & {\it{}rel}, {\it{}string} {\it{}name}, {\it{}string} {\it{}TeXname}={\tt{}""});
{\bf{}ex} {\it{}add\_cycle\_rel}({\bf{}const} {\bf{}lst} & {\it{}rel}, {\bf{}const} {\bf{}ex} & {\it{}key});
{\bf{}ex} {\it{}add\_cycle\_rel}({\bf{}const} {\bf{}ex} & {\it{}rel}, {\it{}string} {\it{}name}, {\it{}string} {\it{}TeXname}={\tt{}""});
{\bf{}ex} {\it{}add\_cycle\_rel}({\bf{}const} {\bf{}ex} & {\it{}rel}, {\bf{}const} {\bf{}ex} & {\it{}key});
\nwindexdefn{\nwixident{add{\_}cycle{\_}rel}}{add:uncycle:unrel}{NWppJ6t-AWRj3-7}\eatline
\nwused{\\{NWppJ6t-7vmg0-4}}\nwidentdefs{\\{{\nwixident{add{\_}cycle{\_}rel}}{add:uncycle:unrel}}}\nwidentuses{\\{{\nwixident{ex}}{ex}}\\{{\nwixident{key}}{key}}\\{{\nwixident{name}}{name}}\\{{\nwixident{TeXname}}{TeXname}}}\nwindexuse{\nwixident{ex}}{ex}{NWppJ6t-AWRj3-7}\nwindexuse{\nwixident{key}}{key}{NWppJ6t-AWRj3-7}\nwindexuse{\nwixident{name}}{name}{NWppJ6t-AWRj3-7}\nwindexuse{\nwixident{TeXname}}{TeXname}{NWppJ6t-AWRj3-7}\nwendcode{}\nwbegindocs{299}\nwdocspar
\nwenddocs{}\nwbegindocs{300}Add a new cycle as the result of certain subfigure {\Tt{}\Rm{}{\it{}F}\nwendquote}. The list
{\Tt{}\Rm{}{\it{}L}\nwendquote} provides nodes from the present figure, which shall be
substituted to the zero generation of {\Tt{}\Rm{}{\it{}F}\nwendquote}. See
{\Tt{}\Rm{}{\it{}midpoint\_constructor}()\nwendquote} for an example, how subfigure shall be
defined, The returned value and parameters {\Tt{}\Rm{}{\it{}name}\nwendquote}, {\Tt{}\Rm{}{\it{}TeXname}\nwendquote} and
{\Tt{}\Rm{}{\it{}key}\nwendquote} are as in methods {\Tt{}\Rm{}{\it{}add\_point}()\nwendquote}.
\nwenddocs{}\nwbegincode{301}\sublabel{NWppJ6t-AWRj3-8}\nwmargintag{{\nwtagstyle{}\subpageref{NWppJ6t-AWRj3-8}}}\moddef{public methods in figure class~{\nwtagstyle{}\subpageref{NWppJ6t-AWRj3-1}}}\plusendmoddef\Rm{}\nwstartdeflinemarkup\nwusesondefline{\\{NWppJ6t-7vmg0-4}}\nwprevnextdefs{NWppJ6t-AWRj3-7}{NWppJ6t-AWRj3-9}\nwenddeflinemarkup
{\bf{}ex} {\it{}add\_subfigure}({\bf{}const} {\bf{}ex} & {\it{}F}, {\bf{}const} {\bf{}lst} & {\it{}L}, {\it{}string} {\it{}name}, {\it{}string} {\it{}TeXname}={\tt{}""});
{\bf{}ex} {\it{}add\_subfigure}({\bf{}const} {\bf{}ex} & {\it{}F}, {\bf{}const} {\bf{}lst} & {\it{}L}, {\bf{}const} {\bf{}ex} & {\it{}key});
\nwindexdefn{\nwixident{add{\_}subfigure}}{add:unsubfigure}{NWppJ6t-AWRj3-8}\eatline
\nwused{\\{NWppJ6t-7vmg0-4}}\nwidentdefs{\\{{\nwixident{add{\_}subfigure}}{add:unsubfigure}}}\nwidentuses{\\{{\nwixident{ex}}{ex}}\\{{\nwixident{key}}{key}}\\{{\nwixident{name}}{name}}\\{{\nwixident{TeXname}}{TeXname}}}\nwindexuse{\nwixident{ex}}{ex}{NWppJ6t-AWRj3-8}\nwindexuse{\nwixident{key}}{key}{NWppJ6t-AWRj3-8}\nwindexuse{\nwixident{name}}{name}{NWppJ6t-AWRj3-8}\nwindexuse{\nwixident{TeXname}}{TeXname}{NWppJ6t-AWRj3-8}\nwendcode{}\nwbegindocs{302}\nwdocspar
\nwenddocs{}\nwbegindocs{303}\nwdocspar
\subsection{Modification, deletion and searches of nodes}
\label{sec:modification-nodes}
\nwenddocs{}\nwbegindocs{304}This method modifies a node created by {\Tt{}\Rm{}{\it{}add\_point}()\nwendquote} by moving the
centre to new coordinates \(x=(x_1, \ldots,x_n)\) (represented by a
{\Tt{}\Rm{}{\bf{}lst}\nwendquote} of the suitable length).
\nwenddocs{}\nwbegincode{305}\sublabel{NWppJ6t-AWRj3-9}\nwmargintag{{\nwtagstyle{}\subpageref{NWppJ6t-AWRj3-9}}}\moddef{public methods in figure class~{\nwtagstyle{}\subpageref{NWppJ6t-AWRj3-1}}}\plusendmoddef\Rm{}\nwstartdeflinemarkup\nwusesondefline{\\{NWppJ6t-7vmg0-4}}\nwprevnextdefs{NWppJ6t-AWRj3-8}{NWppJ6t-AWRj3-A}\nwenddeflinemarkup
{\bf{}void} {\it{}move\_point}({\bf{}const} {\bf{}ex} & {\it{}key}, {\bf{}const} {\bf{}ex} & {\it{}x});
\nwindexdefn{\nwixident{move{\_}point}}{move:unpoint}{NWppJ6t-AWRj3-9}\eatline
\nwused{\\{NWppJ6t-7vmg0-4}}\nwidentdefs{\\{{\nwixident{move{\_}point}}{move:unpoint}}}\nwidentuses{\\{{\nwixident{ex}}{ex}}\\{{\nwixident{key}}{key}}}\nwindexuse{\nwixident{ex}}{ex}{NWppJ6t-AWRj3-9}\nwindexuse{\nwixident{key}}{key}{NWppJ6t-AWRj3-9}\nwendcode{}\nwbegindocs{306}\nwdocspar
\nwenddocs{}\nwbegindocs{307}This method replaced a node referred by {\Tt{}\Rm{}{\it{}key}\nwendquote} with the value of a
cycle {\Tt{}\Rm{}{\it{}C}\nwendquote}. This can be applied to a node without parents only.
\nwenddocs{}\nwbegincode{308}\sublabel{NWppJ6t-AWRj3-A}\nwmargintag{{\nwtagstyle{}\subpageref{NWppJ6t-AWRj3-A}}}\moddef{public methods in figure class~{\nwtagstyle{}\subpageref{NWppJ6t-AWRj3-1}}}\plusendmoddef\Rm{}\nwstartdeflinemarkup\nwusesondefline{\\{NWppJ6t-7vmg0-4}}\nwprevnextdefs{NWppJ6t-AWRj3-9}{NWppJ6t-AWRj3-B}\nwenddeflinemarkup
{\bf{}void} {\it{}move\_cycle}({\bf{}const} {\bf{}ex} & {\it{}key}, {\bf{}const} {\bf{}ex} & {\it{}C});
\nwindexdefn{\nwixident{move{\_}cycle}}{move:uncycle}{NWppJ6t-AWRj3-A}\eatline
\nwused{\\{NWppJ6t-7vmg0-4}}\nwidentdefs{\\{{\nwixident{move{\_}cycle}}{move:uncycle}}}\nwidentuses{\\{{\nwixident{ex}}{ex}}\\{{\nwixident{key}}{key}}}\nwindexuse{\nwixident{ex}}{ex}{NWppJ6t-AWRj3-A}\nwindexuse{\nwixident{key}}{key}{NWppJ6t-AWRj3-A}\nwendcode{}\nwbegindocs{309}\nwdocspar
\nwenddocs{}\nwbegindocs{310}Remove a node given {\Tt{}\Rm{}{\it{}key}\nwendquote} and all its children and grand
children in all generations
\nwenddocs{}\nwbegincode{311}\sublabel{NWppJ6t-AWRj3-B}\nwmargintag{{\nwtagstyle{}\subpageref{NWppJ6t-AWRj3-B}}}\moddef{public methods in figure class~{\nwtagstyle{}\subpageref{NWppJ6t-AWRj3-1}}}\plusendmoddef\Rm{}\nwstartdeflinemarkup\nwusesondefline{\\{NWppJ6t-7vmg0-4}}\nwprevnextdefs{NWppJ6t-AWRj3-A}{NWppJ6t-AWRj3-C}\nwenddeflinemarkup
{\bf{}void} {\it{}remove\_cycle\_node}({\bf{}const} {\bf{}ex} & {\it{}key});
\nwindexdefn{\nwixident{remove{\_}cycle{\_}node}}{remove:uncycle:unnode}{NWppJ6t-AWRj3-B}\eatline
\nwused{\\{NWppJ6t-7vmg0-4}}\nwidentdefs{\\{{\nwixident{remove{\_}cycle{\_}node}}{remove:uncycle:unnode}}}\nwidentuses{\\{{\nwixident{ex}}{ex}}\\{{\nwixident{key}}{key}}}\nwindexuse{\nwixident{ex}}{ex}{NWppJ6t-AWRj3-B}\nwindexuse{\nwixident{key}}{key}{NWppJ6t-AWRj3-B}\nwendcode{}\nwbegindocs{312}\nwdocspar
\nwenddocs{}\nwbegindocs{313}Return the label for {\Tt{}\Rm{}{\bf{}cycle\_node}\nwendquote} with the first matching
name. If the name is not found, the zero expression is returned.
\nwenddocs{}\nwbegincode{314}\sublabel{NWppJ6t-AWRj3-C}\nwmargintag{{\nwtagstyle{}\subpageref{NWppJ6t-AWRj3-C}}}\moddef{public methods in figure class~{\nwtagstyle{}\subpageref{NWppJ6t-AWRj3-1}}}\plusendmoddef\Rm{}\nwstartdeflinemarkup\nwusesondefline{\\{NWppJ6t-7vmg0-4}}\nwprevnextdefs{NWppJ6t-AWRj3-B}{NWppJ6t-AWRj3-D}\nwenddeflinemarkup
{\bf{}ex} {\it{}get\_cycle\_label}({\it{}string} {\it{}name}) {\bf{}const};
\nwindexdefn{\nwixident{get{\_}cycle{\_}label}}{get:uncycle:unlabel}{NWppJ6t-AWRj3-C}\eatline
\nwused{\\{NWppJ6t-7vmg0-4}}\nwidentdefs{\\{{\nwixident{get{\_}cycle{\_}label}}{get:uncycle:unlabel}}}\nwidentuses{\\{{\nwixident{ex}}{ex}}\\{{\nwixident{name}}{name}}}\nwindexuse{\nwixident{ex}}{ex}{NWppJ6t-AWRj3-C}\nwindexuse{\nwixident{name}}{name}{NWppJ6t-AWRj3-C}\nwendcode{}\nwbegindocs{315}\nwdocspar
\nwenddocs{}\nwbegindocs{316}Finally, we provide the method to obtain the {\Tt{}\Rm{}{\bf{}lst}\nwendquote} of keys for all nodes in
generations between {\Tt{}\Rm{}{\it{}mingen}\nwendquote} and {\Tt{}\Rm{}{\it{}maxgen}\nwendquote} inclusively. The default
value {\Tt{}\Rm{}{\it{}GHOST\_GEN}\nwendquote} of {\Tt{}\Rm{}{\it{}maxgen}\nwendquote} removes the check of the upper
bound. Thus, the call of the method with the default values produce
the list of all key except the ghost generation. The order of keys is
from lowest to highest generation.
\nwenddocs{}\nwbegincode{317}\sublabel{NWppJ6t-AWRj3-D}\nwmargintag{{\nwtagstyle{}\subpageref{NWppJ6t-AWRj3-D}}}\moddef{public methods in figure class~{\nwtagstyle{}\subpageref{NWppJ6t-AWRj3-1}}}\plusendmoddef\Rm{}\nwstartdeflinemarkup\nwusesondefline{\\{NWppJ6t-7vmg0-4}}\nwprevnextdefs{NWppJ6t-AWRj3-C}{NWppJ6t-AWRj3-E}\nwenddeflinemarkup
{\bf{}ex} {\it{}get\_all\_keys}({\bf{}const} {\bf{}int} {\it{}mingen} = {\it{}GHOST\_GEN}+1, {\bf{}const} {\bf{}int} {\it{}maxgen} = {\it{}GHOST\_GEN}) {\bf{}const};
\nwindexdefn{\nwixident{get{\_}all{\_}keys}}{get:unall:unkeys}{NWppJ6t-AWRj3-D}\eatline
\nwused{\\{NWppJ6t-7vmg0-4}}\nwidentdefs{\\{{\nwixident{get{\_}all{\_}keys}}{get:unall:unkeys}}}\nwidentuses{\\{{\nwixident{ex}}{ex}}\\{{\nwixident{GHOST{\_}GEN}}{GHOST:unGEN}}}\nwindexuse{\nwixident{ex}}{ex}{NWppJ6t-AWRj3-D}\nwindexuse{\nwixident{GHOST{\_}GEN}}{GHOST:unGEN}{NWppJ6t-AWRj3-D}\nwendcode{}\nwbegindocs{318}\nwdocspar
\nwenddocs{}\nwbegindocs{319}\nwdocspar
\subsection{Check relations and measure parameters}
\label{sec:check-relat-betw}

\nwenddocs{}\nwbegindocs{320}To prove theorems we need to measure ({\Tt{}\Rm{}{\it{}measure}\nwendquote}) some quantities
or to check ({\Tt{}\Rm{}{\it{}check\_rel}\nwendquote}) if two cycles from the figure are
in a certain relation to each other, which were not explicitly defined by the
construction.

\nwenddocs{}\nwbegindocs{321}\nwdocspar
\subsubsection{Checking relations}
\label{sec:checking-relations}

\nwenddocs{}\nwbegindocs{322}A relation which may holds or not may be checked by the following
method. It returns a {\Tt{}\Rm{}{\bf{}lst}\nwendquote} of {\Tt{}\Rm{}{\it{}GiNaC}::{\bf{}relational}\nwendquote}s, which present
the relation between all pairs of cycles in the nodes with {\Tt{}\Rm{}{\it{}key1}\nwendquote}
and {\Tt{}\Rm{}{\it{}key2}\nwendquote}. Typically two cycles are branching in the synchronous way.
Thus it makes sense to compare only respective pairs, this is achieved
with the default value {\Tt{}\Rm{}{\it{}corresponds}={\bf{}true}\nwendquote}.
\nwenddocs{}\nwbegincode{323}\sublabel{NWppJ6t-AWRj3-E}\nwmargintag{{\nwtagstyle{}\subpageref{NWppJ6t-AWRj3-E}}}\moddef{public methods in figure class~{\nwtagstyle{}\subpageref{NWppJ6t-AWRj3-1}}}\plusendmoddef\Rm{}\nwstartdeflinemarkup\nwusesondefline{\\{NWppJ6t-7vmg0-4}}\nwprevnextdefs{NWppJ6t-AWRj3-D}{NWppJ6t-AWRj3-F}\nwenddeflinemarkup
{\bf{}ex} {\it{}check\_rel}({\bf{}const} {\bf{}ex} & {\it{}key1}, {\bf{}const} {\bf{}ex} & {\it{}key2}, {\it{}PCR} {\it{}rel}, {\bf{}bool} {\it{}use\_cycle\_metric}={\bf{}true},
             {\bf{}const} {\bf{}ex} & {\it{}parameter}=0, {\bf{}bool} {\it{}corresponds}={\bf{}true}) {\bf{}const};
\nwindexdefn{\nwixident{check{\_}rel}}{check:unrel}{NWppJ6t-AWRj3-E}\eatline
\nwused{\\{NWppJ6t-7vmg0-4}}\nwidentdefs{\\{{\nwixident{check{\_}rel}}{check:unrel}}}\nwidentuses{\\{{\nwixident{ex}}{ex}}\\{{\nwixident{PCR}}{PCR}}}\nwindexuse{\nwixident{ex}}{ex}{NWppJ6t-AWRj3-E}\nwindexuse{\nwixident{PCR}}{PCR}{NWppJ6t-AWRj3-E}\nwendcode{}\nwbegindocs{324}\nwdocspar
\nwenddocs{}\nwbegindocs{325}The available cycles properties to check are as follows. Most
of these properties are also behind the cycle relations described
in~\ref{sec:publ-meth-cycl}.

\nwenddocs{}\nwbegindocs{326}Orthogonality of cycles given by \cite{Kisil12a}*{\S~6.1}:
\begin{equation}
  \label{eq:orthog-defn}
  \scalar{\cycle{}{}}{\cycle[\tilde]{}{}}=0.
\end{equation}
For circles it coincides with usual orthogonality, for other
situations see  \cite{Kisil12a}*{Ch.~6} for detailed analysis.
\nwenddocs{}\nwbegincode{327}\sublabel{NWppJ6t-iB8QG-1}\nwmargintag{{\nwtagstyle{}\subpageref{NWppJ6t-iB8QG-1}}}\moddef{relations to check~{\nwtagstyle{}\subpageref{NWppJ6t-iB8QG-1}}}\endmoddef\Rm{}\nwstartdeflinemarkup\nwusesondefline{\\{NWppJ6t-1d2GZW-5}}\nwprevnextdefs{\relax}{NWppJ6t-iB8QG-2}\nwenddeflinemarkup
{\bf{}ex} {\it{}cycle\_orthogonal}({\bf{}const} {\bf{}ex} & {\it{}C1}, {\bf{}const} {\bf{}ex} & {\it{}C2}, {\bf{}const} {\bf{}ex} & {\it{}pr}=0);
\nwindexdefn{\nwixident{cycle{\_}orthogonal}}{cycle:unorthogonal}{NWppJ6t-iB8QG-1}\eatline
\nwalsodefined{\\{NWppJ6t-iB8QG-2}\\{NWppJ6t-iB8QG-3}\\{NWppJ6t-iB8QG-4}\\{NWppJ6t-iB8QG-5}\\{NWppJ6t-iB8QG-6}\\{NWppJ6t-iB8QG-7}}\nwused{\\{NWppJ6t-1d2GZW-5}}\nwidentdefs{\\{{\nwixident{cycle{\_}orthogonal}}{cycle:unorthogonal}}}\nwidentuses{\\{{\nwixident{ex}}{ex}}}\nwindexuse{\nwixident{ex}}{ex}{NWppJ6t-iB8QG-1}\nwendcode{}\nwbegindocs{328}\nwdocspar
\nwenddocs{}\nwbegindocs{329}Focal orthogonality of cycles  \cite{Kisil12a}*{\S~6.6}:
\begin{equation}
  \label{eq:focal-orthog-defn}
  \scalar{\cycle[\tilde]{}{}\cycle{}{}\cycle[\tilde]{}{}}{\Space{R}{}}=0.
\end{equation}
\nwenddocs{}\nwbegincode{330}\sublabel{NWppJ6t-iB8QG-2}\nwmargintag{{\nwtagstyle{}\subpageref{NWppJ6t-iB8QG-2}}}\moddef{relations to check~{\nwtagstyle{}\subpageref{NWppJ6t-iB8QG-1}}}\plusendmoddef\Rm{}\nwstartdeflinemarkup\nwusesondefline{\\{NWppJ6t-1d2GZW-5}}\nwprevnextdefs{NWppJ6t-iB8QG-1}{NWppJ6t-iB8QG-3}\nwenddeflinemarkup
{\bf{}ex} {\it{}cycle\_f\_orthogonal}({\bf{}const} {\bf{}ex} & {\it{}C1}, {\bf{}const} {\bf{}ex} & {\it{}C2}, {\bf{}const} {\bf{}ex} & {\it{}pr}=0);
\nwindexdefn{\nwixident{cycle{\_}f{\_}orthogonal}}{cycle:unf:unorthogonal}{NWppJ6t-iB8QG-2}\eatline
\nwused{\\{NWppJ6t-1d2GZW-5}}\nwidentdefs{\\{{\nwixident{cycle{\_}f{\_}orthogonal}}{cycle:unf:unorthogonal}}}\nwidentuses{\\{{\nwixident{ex}}{ex}}}\nwindexuse{\nwixident{ex}}{ex}{NWppJ6t-iB8QG-2}\nwendcode{}\nwbegindocs{331}\nwdocspar
\nwenddocs{}\nwbegindocs{332}Tangent condition between two cycles which shall be used for checks.
This relation is not suitable for construction, use {\Tt{}\Rm{}{\it{}is\_tangent}\nwendquote} and
the likes from Section~\ref{sec:publ-meth-cycl} for this.
\nwenddocs{}\nwbegincode{333}\sublabel{NWppJ6t-iB8QG-3}\nwmargintag{{\nwtagstyle{}\subpageref{NWppJ6t-iB8QG-3}}}\moddef{relations to check~{\nwtagstyle{}\subpageref{NWppJ6t-iB8QG-1}}}\plusendmoddef\Rm{}\nwstartdeflinemarkup\nwusesondefline{\\{NWppJ6t-1d2GZW-5}}\nwprevnextdefs{NWppJ6t-iB8QG-2}{NWppJ6t-iB8QG-4}\nwenddeflinemarkup
{\bf{}ex} {\it{}check\_tangent}({\bf{}const} {\bf{}ex} & {\it{}C1}, {\bf{}const} {\bf{}ex} & {\it{}C2}, {\bf{}const} {\bf{}ex} & {\it{}pr}=0);
\nwindexdefn{\nwixident{check{\_}tangent}}{check:untangent}{NWppJ6t-iB8QG-3}\eatline
\nwused{\\{NWppJ6t-1d2GZW-5}}\nwidentdefs{\\{{\nwixident{check{\_}tangent}}{check:untangent}}}\nwidentuses{\\{{\nwixident{ex}}{ex}}}\nwindexuse{\nwixident{ex}}{ex}{NWppJ6t-iB8QG-3}\nwendcode{}\nwbegindocs{334}\nwdocspar
\nwenddocs{}\nwbegindocs{335}Check two cycles are different.
\nwenddocs{}\nwbegincode{336}\sublabel{NWppJ6t-iB8QG-4}\nwmargintag{{\nwtagstyle{}\subpageref{NWppJ6t-iB8QG-4}}}\moddef{relations to check~{\nwtagstyle{}\subpageref{NWppJ6t-iB8QG-1}}}\plusendmoddef\Rm{}\nwstartdeflinemarkup\nwusesondefline{\\{NWppJ6t-1d2GZW-5}}\nwprevnextdefs{NWppJ6t-iB8QG-3}{NWppJ6t-iB8QG-5}\nwenddeflinemarkup
{\bf{}ex} {\it{}cycle\_different}({\bf{}const} {\bf{}ex} & {\it{}C1}, {\bf{}const} {\bf{}ex} & {\it{}C2}, {\bf{}const} {\bf{}ex} & {\it{}pr}=0);
\nwindexdefn{\nwixident{cycle{\_}different}}{cycle:undifferent}{NWppJ6t-iB8QG-4}\eatline
\nwused{\\{NWppJ6t-1d2GZW-5}}\nwidentdefs{\\{{\nwixident{cycle{\_}different}}{cycle:undifferent}}}\nwidentuses{\\{{\nwixident{ex}}{ex}}}\nwindexuse{\nwixident{ex}}{ex}{NWppJ6t-iB8QG-4}\nwendcode{}\nwbegindocs{337}\nwdocspar
\nwenddocs{}\nwbegindocs{338}Check two cycles are almost different, counting possible rounding errors.
\nwenddocs{}\nwbegincode{339}\sublabel{NWppJ6t-iB8QG-5}\nwmargintag{{\nwtagstyle{}\subpageref{NWppJ6t-iB8QG-5}}}\moddef{relations to check~{\nwtagstyle{}\subpageref{NWppJ6t-iB8QG-1}}}\plusendmoddef\Rm{}\nwstartdeflinemarkup\nwusesondefline{\\{NWppJ6t-1d2GZW-5}}\nwprevnextdefs{NWppJ6t-iB8QG-4}{NWppJ6t-iB8QG-6}\nwenddeflinemarkup
{\bf{}ex} {\it{}cycle\_adifferent}({\bf{}const} {\bf{}ex} & {\it{}C1}, {\bf{}const} {\bf{}ex} & {\it{}C2}, {\bf{}const} {\bf{}ex} & {\it{}pr}=0);
\nwindexdefn{\nwixident{cycle{\_}adifferent}}{cycle:unadifferent}{NWppJ6t-iB8QG-5}\eatline
\nwused{\\{NWppJ6t-1d2GZW-5}}\nwidentdefs{\\{{\nwixident{cycle{\_}adifferent}}{cycle:unadifferent}}}\nwidentuses{\\{{\nwixident{ex}}{ex}}}\nwindexuse{\nwixident{ex}}{ex}{NWppJ6t-iB8QG-5}\nwendcode{}\nwbegindocs{340}\nwdocspar
\nwenddocs{}\nwbegindocs{341}Check that the cycle product with other cycle (or itself) is
non-positive.
\nwenddocs{}\nwbegincode{342}\sublabel{NWppJ6t-iB8QG-6}\nwmargintag{{\nwtagstyle{}\subpageref{NWppJ6t-iB8QG-6}}}\moddef{relations to check~{\nwtagstyle{}\subpageref{NWppJ6t-iB8QG-1}}}\plusendmoddef\Rm{}\nwstartdeflinemarkup\nwusesondefline{\\{NWppJ6t-1d2GZW-5}}\nwprevnextdefs{NWppJ6t-iB8QG-5}{NWppJ6t-iB8QG-7}\nwenddeflinemarkup
{\bf{}ex} {\it{}product\_sign}({\bf{}const} {\bf{}ex} & {\it{}C1}, {\bf{}const} {\bf{}ex} & {\it{}C2}, {\bf{}const} {\bf{}ex} & {\it{}pr}=1);
\nwindexdefn{\nwixident{product{\_}sign}}{product:unsign}{NWppJ6t-iB8QG-6}\eatline
\nwused{\\{NWppJ6t-1d2GZW-5}}\nwidentdefs{\\{{\nwixident{product{\_}sign}}{product:unsign}}}\nwidentuses{\\{{\nwixident{ex}}{ex}}}\nwindexuse{\nwixident{ex}}{ex}{NWppJ6t-iB8QG-6}\nwendcode{}\nwbegindocs{343}\nwdocspar
\nwenddocs{}\nwbegindocs{344}We may want to exclude cycles with imaginary coefficients, this
condition check it.
\nwenddocs{}\nwbegincode{345}\sublabel{NWppJ6t-iB8QG-7}\nwmargintag{{\nwtagstyle{}\subpageref{NWppJ6t-iB8QG-7}}}\moddef{relations to check~{\nwtagstyle{}\subpageref{NWppJ6t-iB8QG-1}}}\plusendmoddef\Rm{}\nwstartdeflinemarkup\nwusesondefline{\\{NWppJ6t-1d2GZW-5}}\nwprevnextdefs{NWppJ6t-iB8QG-6}{\relax}\nwenddeflinemarkup
{\bf{}ex} {\it{}coefficients\_are\_real}({\bf{}const} {\bf{}ex} & {\it{}C1}, {\bf{}const} {\bf{}ex} & {\it{}C2}, {\bf{}const} {\bf{}ex} & {\it{}pr}=1);
\nwindexdefn{\nwixident{coefficients{\_}are{\_}real}}{coefficients:unare:unreal}{NWppJ6t-iB8QG-7}\eatline
\nwused{\\{NWppJ6t-1d2GZW-5}}\nwidentdefs{\\{{\nwixident{coefficients{\_}are{\_}real}}{coefficients:unare:unreal}}}\nwidentuses{\\{{\nwixident{ex}}{ex}}}\nwindexuse{\nwixident{ex}}{ex}{NWppJ6t-iB8QG-7}\nwendcode{}\nwbegindocs{346}\nwdocspar
\nwenddocs{}\nwbegindocs{347}\nwdocspar
\subsubsection{Measuring quantites}
\label{sec:measuring-quantites}

\nwenddocs{}\nwbegindocs{348}A quantity between two cycles may be measured by this
method. Typically two cycles are branching in the synchronous way.
Thus it makes sense to compare only respective pairs, this is achieved
with the default value {\Tt{}\Rm{}{\it{}corresponds}={\bf{}true}\nwendquote}.
\nwenddocs{}\nwbegincode{349}\sublabel{NWppJ6t-AWRj3-F}\nwmargintag{{\nwtagstyle{}\subpageref{NWppJ6t-AWRj3-F}}}\moddef{public methods in figure class~{\nwtagstyle{}\subpageref{NWppJ6t-AWRj3-1}}}\plusendmoddef\Rm{}\nwstartdeflinemarkup\nwusesondefline{\\{NWppJ6t-7vmg0-4}}\nwprevnextdefs{NWppJ6t-AWRj3-E}{NWppJ6t-AWRj3-G}\nwenddeflinemarkup
{\bf{}ex} {\it{}measure}({\bf{}const} {\bf{}ex} & {\it{}key1}, {\bf{}const} {\bf{}ex} & {\it{}key2}, {\it{}PCR} {\it{}rel}, {\bf{}bool} {\it{}use\_cycle\_metric}={\bf{}true},
           {\bf{}const} {\bf{}ex} & {\it{}parameter}=0, {\bf{}bool} {\it{}corresponds}={\bf{}true}) {\bf{}const};
\nwindexdefn{\nwixident{measure}}{measure}{NWppJ6t-AWRj3-F}\eatline
\nwused{\\{NWppJ6t-7vmg0-4}}\nwidentdefs{\\{{\nwixident{measure}}{measure}}}\nwidentuses{\\{{\nwixident{ex}}{ex}}\\{{\nwixident{PCR}}{PCR}}}\nwindexuse{\nwixident{ex}}{ex}{NWppJ6t-AWRj3-F}\nwindexuse{\nwixident{PCR}}{PCR}{NWppJ6t-AWRj3-F}\nwendcode{}\nwbegindocs{350}\nwdocspar

\nwenddocs{}\nwbegindocs{351}\nwdocspar
\subsection{Accessing elements of the figure}
\label{sec:application-function}

\nwenddocs{}\nwbegindocs{352}We can obtain {\Tt{}\Rm{}{\it{}point\_metric}\nwendquote} and {\Tt{}\Rm{}{\it{}cycle\_metric}\nwendquote} form a figure by
the following methods.
\nwenddocs{}\nwbegincode{353}\sublabel{NWppJ6t-AWRj3-G}\nwmargintag{{\nwtagstyle{}\subpageref{NWppJ6t-AWRj3-G}}}\moddef{public methods in figure class~{\nwtagstyle{}\subpageref{NWppJ6t-AWRj3-1}}}\plusendmoddef\Rm{}\nwstartdeflinemarkup\nwusesondefline{\\{NWppJ6t-7vmg0-4}}\nwprevnextdefs{NWppJ6t-AWRj3-F}{NWppJ6t-AWRj3-H}\nwenddeflinemarkup
{\bf{}inline} {\bf{}ex} {\it{}get\_point\_metric}() {\bf{}const} {\nwlbrace} {\bf{}return} {\it{}point\_metric}; {\nwrbrace}
{\bf{}inline} {\bf{}ex} {\it{}get\_cycle\_metric}() {\bf{}const} {\nwlbrace} {\bf{}return} {\it{}cycle\_metric}; {\nwrbrace}
\nwindexdefn{\nwixident{get{\_}point{\_}metric}}{get:unpoint:unmetric}{NWppJ6t-AWRj3-G}\nwindexdefn{\nwixident{get{\_}cycle{\_}metric}}{get:uncycle:unmetric}{NWppJ6t-AWRj3-G}\eatline
\nwused{\\{NWppJ6t-7vmg0-4}}\nwidentdefs{\\{{\nwixident{get{\_}cycle{\_}metric}}{get:uncycle:unmetric}}\\{{\nwixident{get{\_}point{\_}metric}}{get:unpoint:unmetric}}}\nwidentuses{\\{{\nwixident{cycle{\_}metric}}{cycle:unmetric}}\\{{\nwixident{ex}}{ex}}\\{{\nwixident{point{\_}metric}}{point:unmetric}}}\nwindexuse{\nwixident{cycle{\_}metric}}{cycle:unmetric}{NWppJ6t-AWRj3-G}\nwindexuse{\nwixident{ex}}{ex}{NWppJ6t-AWRj3-G}\nwindexuse{\nwixident{point{\_}metric}}{point:unmetric}{NWppJ6t-AWRj3-G}\nwendcode{}\nwbegindocs{354}\nwdocspar
\nwenddocs{}\nwbegindocs{355}Sometimes, we need to check the dimensionality of the figure,
which is essentially the dimensionality of the metric.
\nwenddocs{}\nwbegincode{356}\sublabel{NWppJ6t-AWRj3-H}\nwmargintag{{\nwtagstyle{}\subpageref{NWppJ6t-AWRj3-H}}}\moddef{public methods in figure class~{\nwtagstyle{}\subpageref{NWppJ6t-AWRj3-1}}}\plusendmoddef\Rm{}\nwstartdeflinemarkup\nwusesondefline{\\{NWppJ6t-7vmg0-4}}\nwprevnextdefs{NWppJ6t-AWRj3-G}{NWppJ6t-AWRj3-I}\nwenddeflinemarkup
{\bf{}inline} {\bf{}ex} {\it{}get\_dim}() {\bf{}const} {\nwlbrace} {\bf{}return} {\it{}ex\_to}\begin{math}<\end{math}{\bf{}varidx}\begin{math}>\end{math}({\it{}point\_metric}.{\it{}op}(1)).{\it{}get\_dim}(); {\nwrbrace}
\nwindexdefn{\nwixident{get{\_}dim()}}{get:undim()}{NWppJ6t-AWRj3-H}\eatline
\nwused{\\{NWppJ6t-7vmg0-4}}\nwidentdefs{\\{{\nwixident{get{\_}dim()}}{get:undim()}}}\nwidentuses{\\{{\nwixident{ex}}{ex}}\\{{\nwixident{op}}{op}}\\{{\nwixident{point{\_}metric}}{point:unmetric}}}\nwindexuse{\nwixident{ex}}{ex}{NWppJ6t-AWRj3-H}\nwindexuse{\nwixident{op}}{op}{NWppJ6t-AWRj3-H}\nwindexuse{\nwixident{point{\_}metric}}{point:unmetric}{NWppJ6t-AWRj3-H}\nwendcode{}\nwbegindocs{357}\nwdocspar
\nwenddocs{}\nwbegindocs{358}All {\Tt{}\Rm{}{\bf{}cycle}\nwendquote} associated with a key {\Tt{}\Rm{}{\it{}k}\nwendquote} can be obtained through
the following method. The optional parameter tell which metric to use:
either {\Tt{}\Rm{}{\it{}point\_metric}\nwendquote} or {\Tt{}\Rm{}{\it{}cycle\_metric}\nwendquote}. The method returns a list
of cycles associated to the key {\Tt{}\Rm{}{\it{}k}\nwendquote}.
\nwenddocs{}\nwbegincode{359}\sublabel{NWppJ6t-AWRj3-I}\nwmargintag{{\nwtagstyle{}\subpageref{NWppJ6t-AWRj3-I}}}\moddef{public methods in figure class~{\nwtagstyle{}\subpageref{NWppJ6t-AWRj3-1}}}\plusendmoddef\Rm{}\nwstartdeflinemarkup\nwusesondefline{\\{NWppJ6t-7vmg0-4}}\nwprevnextdefs{NWppJ6t-AWRj3-H}{NWppJ6t-AWRj3-J}\nwenddeflinemarkup
{\bf{}inline} {\bf{}ex} {\it{}get\_cycle}({\bf{}const} {\bf{}ex} & {\it{}k}, {\bf{}bool} {\it{}use\_point\_metric}={\bf{}true}) {\bf{}const} {\nwlbrace}
    {\bf{}return} {\it{}get\_cycle}({\it{}k},{\it{}use\_point\_metric}?{\it{}point\_metric}:{\it{}cycle\_metric});{\nwrbrace}
\nwindexdefn{\nwixident{get{\_}cycle}}{get:uncycle}{NWppJ6t-AWRj3-I}\eatline
\nwused{\\{NWppJ6t-7vmg0-4}}\nwidentdefs{\\{{\nwixident{get{\_}cycle}}{get:uncycle}}}\nwidentuses{\\{{\nwixident{cycle{\_}metric}}{cycle:unmetric}}\\{{\nwixident{ex}}{ex}}\\{{\nwixident{k}}{k}}\\{{\nwixident{point{\_}metric}}{point:unmetric}}}\nwindexuse{\nwixident{cycle{\_}metric}}{cycle:unmetric}{NWppJ6t-AWRj3-I}\nwindexuse{\nwixident{ex}}{ex}{NWppJ6t-AWRj3-I}\nwindexuse{\nwixident{k}}{k}{NWppJ6t-AWRj3-I}\nwindexuse{\nwixident{point{\_}metric}}{point:unmetric}{NWppJ6t-AWRj3-I}\nwendcode{}\nwbegindocs{360}\nwdocspar
\nwenddocs{}\nwbegindocs{361}In fact, we can use a similar method to get {\Tt{}\Rm{}{\bf{}cycle}\nwendquote} with any
permitted expression as a metric.
\nwenddocs{}\nwbegincode{362}\sublabel{NWppJ6t-AWRj3-J}\nwmargintag{{\nwtagstyle{}\subpageref{NWppJ6t-AWRj3-J}}}\moddef{public methods in figure class~{\nwtagstyle{}\subpageref{NWppJ6t-AWRj3-1}}}\plusendmoddef\Rm{}\nwstartdeflinemarkup\nwusesondefline{\\{NWppJ6t-7vmg0-4}}\nwprevnextdefs{NWppJ6t-AWRj3-I}{NWppJ6t-AWRj3-K}\nwenddeflinemarkup
{\bf{}ex} {\it{}get\_cycle}({\bf{}const} {\bf{}ex} & {\it{}k}, {\bf{}const} {\bf{}ex} & {\it{}metric}) {\bf{}const};
\nwindexdefn{\nwixident{get{\_}cycle}}{get:uncycle}{NWppJ6t-AWRj3-J}\eatline
\nwused{\\{NWppJ6t-7vmg0-4}}\nwidentdefs{\\{{\nwixident{get{\_}cycle}}{get:uncycle}}}\nwidentuses{\\{{\nwixident{ex}}{ex}}\\{{\nwixident{k}}{k}}}\nwindexuse{\nwixident{ex}}{ex}{NWppJ6t-AWRj3-J}\nwindexuse{\nwixident{k}}{k}{NWppJ6t-AWRj3-J}\nwendcode{}\nwbegindocs{363}\nwdocspar
\nwenddocs{}\nwbegindocs{364}The generation of the cycle associated to the key {\Tt{}\Rm{}{\it{}k}\nwendquote} is provided
by the method:
\nwenddocs{}\nwbegincode{365}\sublabel{NWppJ6t-AWRj3-K}\nwmargintag{{\nwtagstyle{}\subpageref{NWppJ6t-AWRj3-K}}}\moddef{public methods in figure class~{\nwtagstyle{}\subpageref{NWppJ6t-AWRj3-1}}}\plusendmoddef\Rm{}\nwstartdeflinemarkup\nwusesondefline{\\{NWppJ6t-7vmg0-4}}\nwprevnextdefs{NWppJ6t-AWRj3-J}{NWppJ6t-AWRj3-L}\nwenddeflinemarkup
{\bf{}inline} {\bf{}ex} {\it{}get\_generation}({\bf{}const} {\bf{}ex} & {\it{}k}) {\bf{}const} {\nwlbrace}
    {\bf{}return} {\it{}ex\_to}\begin{math}<\end{math}{\bf{}cycle\_node}\begin{math}>\end{math}({\it{}get\_cycle\_node}({\it{}k})).{\it{}get\_generation}();{\nwrbrace}
\nwindexdefn{\nwixident{get{\_}generation}}{get:ungeneration}{NWppJ6t-AWRj3-K}\eatline
\nwused{\\{NWppJ6t-7vmg0-4}}\nwidentdefs{\\{{\nwixident{get{\_}generation}}{get:ungeneration}}}\nwidentuses{\\{{\nwixident{cycle{\_}node}}{cycle:unnode}}\\{{\nwixident{ex}}{ex}}\\{{\nwixident{get{\_}cycle{\_}node}}{get:uncycle:unnode}}\\{{\nwixident{k}}{k}}}\nwindexuse{\nwixident{cycle{\_}node}}{cycle:unnode}{NWppJ6t-AWRj3-K}\nwindexuse{\nwixident{ex}}{ex}{NWppJ6t-AWRj3-K}\nwindexuse{\nwixident{get{\_}cycle{\_}node}}{get:uncycle:unnode}{NWppJ6t-AWRj3-K}\nwindexuse{\nwixident{k}}{k}{NWppJ6t-AWRj3-K}\nwendcode{}\nwbegindocs{366}\nwdocspar
\nwenddocs{}\nwbegindocs{367}Sometimes we need to apply a function to all {\Tt{}\Rm{}{\bf{}cycle}\nwendquote}s which
compose the {\Tt{}\Rm{}{\bf{}figure}\nwendquote}. Here we define the type for such a function.
\nwenddocs{}\nwbegincode{368}\sublabel{NWppJ6t-1s4LDF-1}\nwmargintag{{\nwtagstyle{}\subpageref{NWppJ6t-1s4LDF-1}}}\moddef{defining types~{\nwtagstyle{}\subpageref{NWppJ6t-1s4LDF-1}}}\endmoddef\Rm{}\nwstartdeflinemarkup\nwusesondefline{\\{NWppJ6t-3NCnUp-3}}\nwprevnextdefs{\relax}{NWppJ6t-1s4LDF-2}\nwenddeflinemarkup
{\bf{}using} {\it{}PEVAL} = {\it{}std}::{\it{}function}\begin{math}<\end{math}{\bf{}ex}({\bf{}const} {\bf{}ex} &, {\bf{}const} {\bf{}ex} &)\begin{math}>\end{math};

\nwalsodefined{\\{NWppJ6t-1s4LDF-2}}\nwused{\\{NWppJ6t-3NCnUp-3}}\nwidentuses{\\{{\nwixident{ex}}{ex}}}\nwindexuse{\nwixident{ex}}{ex}{NWppJ6t-1s4LDF-1}\nwendcode{}\nwbegindocs{369}This is the method to apply a function {\Tt{}\Rm{}{\it{}func}\nwendquote} to all particular
{\Tt{}\Rm{}{\bf{}cycle}\nwendquote}s which compose the {\Tt{}\Rm{}{\bf{}figure}\nwendquote}. It returns a {\Tt{}\Rm{}{\bf{}lst}\nwendquote} of
{\Tt{}\Rm{}{\bf{}lst}\nwendquote}s. Each sub-list has three elements: the returned value of
{\Tt{}\Rm{}{\it{}func}\nwendquote}, the key of the respective {\Tt{}\Rm{}{\bf{}cycle\_node}\nwendquote} and the number of
{\Tt{}\Rm{}{\bf{}cycle}\nwendquote} in the respective node. The parameter {\Tt{}\Rm{}{\it{}use\_cycle\_metric}\nwendquote}
tells which metric shall be used: either cycle space or point space,
see~\cite{Kisil12a}*{\S~4.2}.
\nwenddocs{}\nwbegincode{370}\sublabel{NWppJ6t-AWRj3-L}\nwmargintag{{\nwtagstyle{}\subpageref{NWppJ6t-AWRj3-L}}}\moddef{public methods in figure class~{\nwtagstyle{}\subpageref{NWppJ6t-AWRj3-1}}}\plusendmoddef\Rm{}\nwstartdeflinemarkup\nwusesondefline{\\{NWppJ6t-7vmg0-4}}\nwprevnextdefs{NWppJ6t-AWRj3-K}{NWppJ6t-AWRj3-M}\nwenddeflinemarkup
{\bf{}ex} {\it{}apply}({\it{}PEVAL} {\it{}func}, {\bf{}bool} {\it{}use\_cycle\_metric}={\bf{}true}, {\bf{}const} {\bf{}ex} & {\it{}param} = 0) {\bf{}const};
\nwindexdefn{\nwixident{apply}}{apply}{NWppJ6t-AWRj3-L}\eatline
\nwused{\\{NWppJ6t-7vmg0-4}}\nwidentdefs{\\{{\nwixident{apply}}{apply}}}\nwidentuses{\\{{\nwixident{ex}}{ex}}}\nwindexuse{\nwixident{ex}}{ex}{NWppJ6t-AWRj3-L}\nwendcode{}\nwbegindocs{371}\nwdocspar
\nwenddocs{}\nwbegindocs{372}\nwdocspar
\subsection{Drawing and printing}
\label{sec:drawing-printing}
There is a collections of methods which help to visualise a figure. We
use \Asymptote\ to produce PostScript, PDF, PNG or other files in
two-dimensions and an interactive visualisation tool is available for
three-dimensional figures.

\nwenddocs{}\nwbegindocs{373}The default behaviour of {\Tt{}\Rm{}{\it{}asy\_write}()\nwendquote} is an attempt to display
files produced by {\Asymptote}. User can disable this visualisation.
\nwenddocs{}\nwbegincode{374}\sublabel{NWppJ6t-1lh2X2-1}\nwmargintag{{\nwtagstyle{}\subpageref{NWppJ6t-1lh2X2-1}}}\moddef{additional functions header~{\nwtagstyle{}\subpageref{NWppJ6t-1lh2X2-1}}}\endmoddef\Rm{}\nwstartdeflinemarkup\nwusesondefline{\\{NWppJ6t-3NCnUp-3}}\nwprevnextdefs{\relax}{NWppJ6t-1lh2X2-2}\nwenddeflinemarkup
    {\bf{}void} {\it{}show\_asy\_on}();
    {\bf{}void} {\it{}show\_asy\_off}();
\nwindexdefn{\nwixident{show{\_}asy{\_}on}}{show:unasy:unon}{NWppJ6t-1lh2X2-1}\nwindexdefn{\nwixident{show{\_}asy{\_}off}}{show:unasy:unoff}{NWppJ6t-1lh2X2-1}\eatline
\nwalsodefined{\\{NWppJ6t-1lh2X2-2}\\{NWppJ6t-1lh2X2-3}\\{NWppJ6t-1lh2X2-4}\\{NWppJ6t-1lh2X2-5}}\nwused{\\{NWppJ6t-3NCnUp-3}}\nwidentdefs{\\{{\nwixident{show{\_}asy{\_}off}}{show:unasy:unoff}}\\{{\nwixident{show{\_}asy{\_}on}}{show:unasy:unon}}}\nwendcode{}\nwbegindocs{375}\nwdocspar
\nwenddocs{}\nwbegindocs{376}\nwdocspar
\subsubsection{Two-dimensional graphics and animation}
\label{sec:two-dimens-graph}

\nwenddocs{}\nwbegindocs{377}The next method returns \Asymptote~\cite{Asymptote} string which
draws the entire figure. The drawing is controlled by two {\Tt{}\Rm{}{\it{}style}\nwendquote}
and {\Tt{}\Rm{}{\it{}lstring}\nwendquote}. Initial parameters have the same meaning as in
{\Tt{}\Rm{}{\bf{}cycle2D}::{\it{}asy\_draw}()\nwendquote}. Explicitly, the drawing is done within the rectangle with the
lower left vertex ({\Tt{}\Rm{}{\it{}xmin}\nwendquote}, {\Tt{}\Rm{}{\it{}ymin}\nwendquote}) and upper right  ({\Tt{}\Rm{}{\it{}xmax}\nwendquote},
{\Tt{}\Rm{}{\it{}ymax}\nwendquote}). The style of drawing is controlled by {\Tt{}\Rm{}{\it{}default\_asy}\nwendquote} and
{\Tt{}\Rm{}{\it{}default\_label}\nwendquote}, see {\Tt{}\Rm{}{\it{}asy\_cycle\_color}()\nwendquote} and {\Tt{}\Rm{}{\it{}label\_pos}()\nwendquote} for
ideas. On
complicated figures, see Fig.~\ref{fig:action-modular-group}, we may
not want cycles label to be printed at all, this can be controlled through
{\Tt{}\Rm{}{\it{}with\_labels}\nwendquote} parameter.
By default the {\Tt{}\Rm{}{\it{}real\_line}\nwendquote} is drawn and the comments in the file are
presented, this can be amended through {\Tt{}\Rm{}{\it{}with\_realline}\nwendquote} and
{\Tt{}\Rm{}{\it{}with\_header}\nwendquote} parameters respectively. The default
number of points per arc is reasonable in most cases, however user can
override this with supplying a value to {\Tt{}\Rm{}{\it{}points\_per\_arc}\nwendquote}.
The result is written to the stream {\Tt{}\Rm{}{\it{}ost}\nwendquote}.
\nwenddocs{}\nwbegincode{378}\sublabel{NWppJ6t-AWRj3-M}\nwmargintag{{\nwtagstyle{}\subpageref{NWppJ6t-AWRj3-M}}}\moddef{public methods in figure class~{\nwtagstyle{}\subpageref{NWppJ6t-AWRj3-1}}}\plusendmoddef\Rm{}\nwstartdeflinemarkup\nwusesondefline{\\{NWppJ6t-7vmg0-4}}\nwprevnextdefs{NWppJ6t-AWRj3-L}{NWppJ6t-AWRj3-N}\nwenddeflinemarkup
{\bf{}void} {\it{}asy\_draw}({\it{}ostream} & {\it{}ost} ={\it{}std}::{\it{}cout}, {\it{}ostream} & {\it{}err}={\it{}std}::{\it{}cerr}, {\bf{}const} {\it{}string} {\it{}picture}={\tt{}""},\nwindexdefn{\nwixident{asy{\_}draw}}{asy:undraw}{NWppJ6t-AWRj3-M}
              {\bf{}const} {\bf{}ex} & {\it{}xmin} = -5, {\bf{}const} {\bf{}ex} & {\it{}xmax} = 5,
              {\bf{}const} {\bf{}ex} & {\it{}ymin} = -5, {\bf{}const} {\bf{}ex} & {\it{}ymax} = 5,
              {\it{}asy\_style} {\it{}style}={\it{}default\_asy}, {\it{}label\_string} {\it{}lstring}={\it{}default\_label},
              {\bf{}bool} {\it{}with\_realline}={\bf{}true}, {\bf{}bool} {\it{}with\_header}={\bf{}true},
              {\bf{}int} {\it{}points\_per\_arc} = 0, {\bf{}const} {\it{}string} {\it{}imaginary\_options}={\tt{}"rgb(0,.9,0)+4pt"},
              {\bf{}bool} {\it{}with\_labels}={\bf{}true}) {\bf{}const};
\nwindexdefn{\nwixident{asy{\_}draw}}{asy:undraw}{NWppJ6t-AWRj3-M}\eatline
\nwused{\\{NWppJ6t-7vmg0-4}}\nwidentdefs{\\{{\nwixident{asy{\_}draw}}{asy:undraw}}}\nwidentuses{\\{{\nwixident{asy{\_}style}}{asy:unstyle}}\\{{\nwixident{ex}}{ex}}\\{{\nwixident{label{\_}string}}{label:unstring}}\\{{\nwixident{rgb}}{rgb}}}\nwindexuse{\nwixident{asy{\_}style}}{asy:unstyle}{NWppJ6t-AWRj3-M}\nwindexuse{\nwixident{ex}}{ex}{NWppJ6t-AWRj3-M}\nwindexuse{\nwixident{label{\_}string}}{label:unstring}{NWppJ6t-AWRj3-M}\nwindexuse{\nwixident{rgb}}{rgb}{NWppJ6t-AWRj3-M}\nwendcode{}\nwbegindocs{379}\nwdocspar
\nwenddocs{}\nwbegindocs{380}This method creates a temporary file with \Asymptote\ commands to
draw the figure, then calls the \Asymptote\ to produce the graphic
file, the temporary file is deleted afterwards. The parameters are the
same as above in {\Tt{}\Rm{}{\it{}asy\_draw}()\nwendquote}. The last parameter {\Tt{}\Rm{}{\it{}rm\_asy\_file}\nwendquote} tells if the
\Asymptote\ file shall be removed. User may keep it and fine-tune the
result manually.
\nwenddocs{}\nwbegincode{381}\sublabel{NWppJ6t-AWRj3-N}\nwmargintag{{\nwtagstyle{}\subpageref{NWppJ6t-AWRj3-N}}}\moddef{public methods in figure class~{\nwtagstyle{}\subpageref{NWppJ6t-AWRj3-1}}}\plusendmoddef\Rm{}\nwstartdeflinemarkup\nwusesondefline{\\{NWppJ6t-7vmg0-4}}\nwprevnextdefs{NWppJ6t-AWRj3-M}{NWppJ6t-AWRj3-O}\nwenddeflinemarkup
{\bf{}void} {\it{}asy\_write}({\bf{}int} {\it{}size}=300, {\bf{}const} {\bf{}ex} & {\it{}xmin} = -5, {\bf{}const} {\bf{}ex} & {\it{}xmax} = 5,\nwindexdefn{\nwixident{asy{\_}write}}{asy:unwrite}{NWppJ6t-AWRj3-N}
               {\bf{}const} {\bf{}ex} & {\it{}ymin} = -5, {\bf{}const} {\bf{}ex} & {\it{}ymax} = 5,
               {\it{}string} {\it{}name}={\tt{}"figure-view-tmp"}, {\it{}string} {\it{}format}={\tt{}""},
               {\it{}asy\_style} {\it{}style}={\it{}default\_asy}, {\it{}label\_string} {\it{}lstring}={\it{}default\_label},
               {\bf{}bool} {\it{}with\_realline}={\bf{}true}, {\bf{}bool} {\it{}with\_header}={\bf{}true},
               {\bf{}int} {\it{}points\_per\_arc}=0, {\bf{}const} {\it{}string} {\it{}imaginary\_options}={\tt{}"rgb(0,.9,0)+4pt"},
               {\bf{}bool} {\it{}rm\_asy\_file}={\bf{}true}, {\bf{}bool} {\it{}with\_labels}={\bf{}true}) {\bf{}const};
\nwindexdefn{\nwixident{asy{\_}write}}{asy:unwrite}{NWppJ6t-AWRj3-N}\eatline
\nwused{\\{NWppJ6t-7vmg0-4}}\nwidentdefs{\\{{\nwixident{asy{\_}write}}{asy:unwrite}}}\nwidentuses{\\{{\nwixident{asy{\_}style}}{asy:unstyle}}\\{{\nwixident{ex}}{ex}}\\{{\nwixident{figure}}{figure}}\\{{\nwixident{label{\_}string}}{label:unstring}}\\{{\nwixident{name}}{name}}\\{{\nwixident{rgb}}{rgb}}}\nwindexuse{\nwixident{asy{\_}style}}{asy:unstyle}{NWppJ6t-AWRj3-N}\nwindexuse{\nwixident{ex}}{ex}{NWppJ6t-AWRj3-N}\nwindexuse{\nwixident{figure}}{figure}{NWppJ6t-AWRj3-N}\nwindexuse{\nwixident{label{\_}string}}{label:unstring}{NWppJ6t-AWRj3-N}\nwindexuse{\nwixident{name}}{name}{NWppJ6t-AWRj3-N}\nwindexuse{\nwixident{rgb}}{rgb}{NWppJ6t-AWRj3-N}\nwendcode{}\nwbegindocs{382}\nwdocspar
\nwenddocs{}\nwbegindocs{383}This a method to produce an animation. The figure may depend from
some parameters, for example of {\Tt{}\Rm{}{\bf{}symbol}\nwendquote}
class. The first argument {\Tt{}\Rm{}{\it{}val}\nwendquote} is a {\Tt{}\Rm{}{\bf{}lst}\nwendquote}, which contains
expressions for substitutions into the figure. That is, elements of
{\Tt{}\Rm{}{\it{}val}\nwendquote} can be any expression suitable to use as the first parameter
of {\Tt{}\Rm{}{\it{}susb}\nwendquote} method in \GiNaC. For example, they may be
{\Tt{}\Rm{}{\bf{}relational}\nwendquote}s (e.g. {\Tt{}\Rm{}{\it{}t}\begin{math}\equiv\end{math}1.5\nwendquote}) or {\Tt{}\Rm{}{\bf{}lst}\nwendquote} of {\Tt{}\Rm{}{\bf{}relational}\nwendquote}s
(e.g. {\Tt{}\Rm{}{\bf{}lst}{\nwlbrace}{\it{}t}\begin{math}\equiv\end{math}1.5,{\it{}s}\begin{math}\equiv\end{math}2.1{\nwrbrace}\nwendquote}. The method make the substitution the
each element of {\Tt{}\Rm{}{\bf{}lst}\nwendquote} into the figure and
uses the resulting \Asymptote\ drawings as a sequence of shots for the
animations. The output {\Tt{}\Rm{}{\it{}format}\nwendquote} may be either predefined {\Tt{}\Rm{}{\tt{}"pdf"}\nwendquote},
{\Tt{}\Rm{}{\tt{}"gif"}\nwendquote}, {\Tt{}\Rm{}{\tt{}"mng"}\nwendquote} or {\Tt{}\Rm{}{\tt{}"mp4"}\nwendquote}, or user-specified \Asymptote\ string.

The values of parameters can be put to the animation. The default bottom-left
position is encoded as {\Tt{}\Rm{}{\tt{}"bl"}\nwendquote} for {\Tt{}\Rm{}{\it{}values\_position}\nwendquote}, other
possible positions are {\Tt{}\Rm{}{\tt{}"br"}\nwendquote} (bottom-right),  {\Tt{}\Rm{}{\tt{}"tl"}\nwendquote} (top-left)
and {\Tt{}\Rm{}{\tt{}"tr"}\nwendquote} (top-right). Any other string (e.g. the empty one) will
preven the parameter values from printing.

The rest of parameters have the same meaning as in
{\Tt{}\Rm{}{\it{}asy\_write}()\nwendquote}. See the end of Sect.~\ref{sec:animated-cycle} for
further advise on animation embedded into PDF files.
\nwenddocs{}\nwbegincode{384}\sublabel{NWppJ6t-AWRj3-O}\nwmargintag{{\nwtagstyle{}\subpageref{NWppJ6t-AWRj3-O}}}\moddef{public methods in figure class~{\nwtagstyle{}\subpageref{NWppJ6t-AWRj3-1}}}\plusendmoddef\Rm{}\nwstartdeflinemarkup\nwusesondefline{\\{NWppJ6t-7vmg0-4}}\nwprevnextdefs{NWppJ6t-AWRj3-N}{NWppJ6t-AWRj3-P}\nwenddeflinemarkup
{\bf{}void} {\it{}asy\_animate}({\bf{}const} {\bf{}ex} & {\it{}val},\nwindexdefn{\nwixident{asy{\_}animate}}{asy:unanimate}{NWppJ6t-AWRj3-O}
                 {\bf{}int} {\it{}size}=300, {\bf{}const} {\bf{}ex} & {\it{}xmin} = -5, {\bf{}const} {\bf{}ex} & {\it{}xmax} = 5,
                 {\bf{}const} {\bf{}ex} & {\it{}ymin} = -5, {\bf{}const} {\bf{}ex} & {\it{}ymax} = 5,
                 {\it{}string} {\it{}name}={\tt{}"figure-animatecf-tmp"}, {\it{}string} {\it{}format}={\tt{}"pdf"},
                 {\it{}asy\_style} {\it{}style}={\it{}default\_asy}, {\it{}label\_string} {\it{}lstring}={\it{}default\_label},
                 {\bf{}bool} {\it{}with\_realline}={\bf{}true}, {\bf{}bool} {\it{}with\_header}={\bf{}true},
                 {\bf{}int} {\it{}points\_per\_arc} = 0, {\bf{}const} {\it{}string} {\it{}imaginary\_options}={\tt{}"rgb(0,.9,0)+4pt"},
                 {\bf{}const} {\it{}string} {\it{}values\_position}={\tt{}"bl"}, {\bf{}bool} {\it{}rm\_asy\_file}={\bf{}true},
                 {\bf{}bool} {\it{}with\_labels}={\bf{}true}) {\bf{}const};
\nwindexdefn{\nwixident{asy{\_}animate}}{asy:unanimate}{NWppJ6t-AWRj3-O}\eatline
\nwused{\\{NWppJ6t-7vmg0-4}}\nwidentdefs{\\{{\nwixident{asy{\_}animate}}{asy:unanimate}}}\nwidentuses{\\{{\nwixident{asy{\_}style}}{asy:unstyle}}\\{{\nwixident{ex}}{ex}}\\{{\nwixident{figure}}{figure}}\\{{\nwixident{label{\_}string}}{label:unstring}}\\{{\nwixident{name}}{name}}\\{{\nwixident{rgb}}{rgb}}}\nwindexuse{\nwixident{asy{\_}style}}{asy:unstyle}{NWppJ6t-AWRj3-O}\nwindexuse{\nwixident{ex}}{ex}{NWppJ6t-AWRj3-O}\nwindexuse{\nwixident{figure}}{figure}{NWppJ6t-AWRj3-O}\nwindexuse{\nwixident{label{\_}string}}{label:unstring}{NWppJ6t-AWRj3-O}\nwindexuse{\nwixident{name}}{name}{NWppJ6t-AWRj3-O}\nwindexuse{\nwixident{rgb}}{rgb}{NWppJ6t-AWRj3-O}\nwendcode{}\nwbegindocs{385}\nwdocspar
\nwenddocs{}\nwbegindocs{386}Evaluation of {\Tt{}\Rm{}{\bf{}cycle}\nwendquote} within a figure with symbolic entries may
took a long time. To prevent this we may use {\Tt{}\Rm{}{\it{}freeze}\nwendquote} method, and
then {\Tt{}\Rm{}{\it{}unfreeze}\nwendquote} after numeric substitution is done.
\nwenddocs{}\nwbegincode{387}\sublabel{NWppJ6t-AWRj3-P}\nwmargintag{{\nwtagstyle{}\subpageref{NWppJ6t-AWRj3-P}}}\moddef{public methods in figure class~{\nwtagstyle{}\subpageref{NWppJ6t-AWRj3-1}}}\plusendmoddef\Rm{}\nwstartdeflinemarkup\nwusesondefline{\\{NWppJ6t-7vmg0-4}}\nwprevnextdefs{NWppJ6t-AWRj3-O}{NWppJ6t-AWRj3-Q}\nwenddeflinemarkup
{\bf{}inline} {\bf{}figure} {\it{}freeze}() {\bf{}const} {\nwlbrace}{\it{}setflag}({\it{}status\_flags}::{\it{}expanded}); {\bf{}return} \begin{math}\ast\end{math}{\it{}this};{\nwrbrace}
{\bf{}inline} {\bf{}figure} {\it{}unfreeze}() {\bf{}const} {\nwlbrace}{\it{}clearflag}({\it{}status\_flags}::{\it{}expanded}); {\bf{}return} \begin{math}\ast\end{math}{\it{}this};{\nwrbrace}
\nwindexdefn{\nwixident{freeze}}{freeze}{NWppJ6t-AWRj3-P}\nwindexdefn{\nwixident{unfreeze}}{unfreeze}{NWppJ6t-AWRj3-P}\eatline
\nwused{\\{NWppJ6t-7vmg0-4}}\nwidentdefs{\\{{\nwixident{freeze}}{freeze}}\\{{\nwixident{unfreeze}}{unfreeze}}}\nwidentuses{\\{{\nwixident{figure}}{figure}}}\nwindexuse{\nwixident{figure}}{figure}{NWppJ6t-AWRj3-P}\nwendcode{}\nwbegindocs{388}\nwdocspar
\nwenddocs{}\nwbegindocs{389}To speed-up evaluation of figures we may force float evaluation
instead of exact arithmetic.
\nwenddocs{}\nwbegincode{390}\sublabel{NWppJ6t-AWRj3-Q}\nwmargintag{{\nwtagstyle{}\subpageref{NWppJ6t-AWRj3-Q}}}\moddef{public methods in figure class~{\nwtagstyle{}\subpageref{NWppJ6t-AWRj3-1}}}\plusendmoddef\Rm{}\nwstartdeflinemarkup\nwusesondefline{\\{NWppJ6t-7vmg0-4}}\nwprevnextdefs{NWppJ6t-AWRj3-P}{NWppJ6t-AWRj3-R}\nwenddeflinemarkup
{\bf{}inline} {\bf{}figure} {\it{}set\_float\_eval}() {\nwlbrace}{\it{}float\_evaluation}={\bf{}true}; {\bf{}return} \begin{math}\ast\end{math}{\it{}this};{\nwrbrace}
{\bf{}inline} {\bf{}figure} {\it{}set\_exact\_eval}() {\nwlbrace}{\it{}float\_evaluation}={\bf{}false}; {\bf{}return} \begin{math}\ast\end{math}{\it{}this};{\nwrbrace}
\nwindexdefn{\nwixident{set{\_}float{\_}eval}}{set:unfloat:uneval}{NWppJ6t-AWRj3-Q}\nwindexdefn{\nwixident{set{\_}exact{\_}eval}}{set:unexact:uneval}{NWppJ6t-AWRj3-Q}\eatline
\nwused{\\{NWppJ6t-7vmg0-4}}\nwidentdefs{\\{{\nwixident{set{\_}exact{\_}eval}}{set:unexact:uneval}}\\{{\nwixident{set{\_}float{\_}eval}}{set:unfloat:uneval}}}\nwidentuses{\\{{\nwixident{figure}}{figure}}\\{{\nwixident{float{\_}evaluation}}{float:unevaluation}}}\nwindexuse{\nwixident{figure}}{figure}{NWppJ6t-AWRj3-Q}\nwindexuse{\nwixident{float{\_}evaluation}}{float:unevaluation}{NWppJ6t-AWRj3-Q}\nwendcode{}\nwbegindocs{391}\nwdocspar
\nwenddocs{}\nwbegindocs{392}These methods allow to specify or read an \Asymptote\ drawing style
for a particular node.
\nwenddocs{}\nwbegincode{393}\sublabel{NWppJ6t-AWRj3-R}\nwmargintag{{\nwtagstyle{}\subpageref{NWppJ6t-AWRj3-R}}}\moddef{public methods in figure class~{\nwtagstyle{}\subpageref{NWppJ6t-AWRj3-1}}}\plusendmoddef\Rm{}\nwstartdeflinemarkup\nwusesondefline{\\{NWppJ6t-7vmg0-4}}\nwprevnextdefs{NWppJ6t-AWRj3-Q}{NWppJ6t-AWRj3-S}\nwenddeflinemarkup
{\bf{}inline} {\bf{}void} {\it{}set\_asy\_style}({\bf{}const} {\bf{}ex} & {\it{}key}, {\it{}string} {\it{}opt}) {\nwlbrace}{\it{}nodes}[{\it{}key}].{\it{}set\_asy\_opt}({\it{}opt});{\nwrbrace}
{\bf{}inline} {\it{}string} {\it{}get\_asy\_style}({\bf{}const} {\bf{}ex} & {\it{}key}) {\bf{}const} {\nwlbrace}{\bf{}return} {\it{}ex\_to}\begin{math}<\end{math}{\bf{}cycle\_node}\begin{math}>\end{math}({\it{}get\_cycle\_node}({\it{}k})).{\it{}get\_asy\_opt}();{\nwrbrace}
\nwindexdefn{\nwixident{set{\_}asy{\_}style}}{set:unasy:unstyle}{NWppJ6t-AWRj3-R}\nwindexdefn{\nwixident{get{\_}asy{\_}style}}{get:unasy:unstyle}{NWppJ6t-AWRj3-R}\eatline
\nwused{\\{NWppJ6t-7vmg0-4}}\nwidentdefs{\\{{\nwixident{get{\_}asy{\_}style}}{get:unasy:unstyle}}\\{{\nwixident{set{\_}asy{\_}style}}{set:unasy:unstyle}}}\nwidentuses{\\{{\nwixident{cycle{\_}node}}{cycle:unnode}}\\{{\nwixident{ex}}{ex}}\\{{\nwixident{get{\_}cycle{\_}node}}{get:uncycle:unnode}}\\{{\nwixident{k}}{k}}\\{{\nwixident{key}}{key}}\\{{\nwixident{nodes}}{nodes}}}\nwindexuse{\nwixident{cycle{\_}node}}{cycle:unnode}{NWppJ6t-AWRj3-R}\nwindexuse{\nwixident{ex}}{ex}{NWppJ6t-AWRj3-R}\nwindexuse{\nwixident{get{\_}cycle{\_}node}}{get:uncycle:unnode}{NWppJ6t-AWRj3-R}\nwindexuse{\nwixident{k}}{k}{NWppJ6t-AWRj3-R}\nwindexuse{\nwixident{key}}{key}{NWppJ6t-AWRj3-R}\nwindexuse{\nwixident{nodes}}{nodes}{NWppJ6t-AWRj3-R}\nwendcode{}\nwbegindocs{394}\nwdocspar
\nwenddocs{}\nwbegindocs{395}\nwdocspar
\subsubsection{Three-dimensional visualisation}
\label{sec:three-dimens-visu}

\nwenddocs{}\nwbegindocs{396}In three dimensions a visualisation is possible with the help of an
additional interactive programme \texttt{cycle3D-visualiser}. The
following method produces a text file {\Tt{}\Rm{}{\it{}name}.{\it{}txt}\nwendquote} (the default suffix {\Tt{}\Rm{}{\tt{}".txt"}\nwendquote} is
added to {\Tt{}\Rm{}{\it{}name}\nwendquote} automatically). The file can be
visualised by a helper programme. All cycles in generations starting
from {\Tt{}\Rm{}{\it{}first\_gen}\nwendquote} are represented by their centres, radii,
generations and labels.
\nwenddocs{}\nwbegincode{397}\sublabel{NWppJ6t-AWRj3-S}\nwmargintag{{\nwtagstyle{}\subpageref{NWppJ6t-AWRj3-S}}}\moddef{public methods in figure class~{\nwtagstyle{}\subpageref{NWppJ6t-AWRj3-1}}}\plusendmoddef\Rm{}\nwstartdeflinemarkup\nwusesondefline{\\{NWppJ6t-7vmg0-4}}\nwprevnextdefs{NWppJ6t-AWRj3-R}{NWppJ6t-AWRj3-T}\nwenddeflinemarkup
{\bf{}void} {\it{}arrangement\_write}({\it{}string} {\it{}name}, {\bf{}int} {\it{}first\_gen}=0) {\bf{}const};\nwindexdefn{\nwixident{arrangement{\_}write}}{arrangement:unwrite}{NWppJ6t-AWRj3-S}
\nwindexdefn{\nwixident{arrangement{\_}write}}{arrangement:unwrite}{NWppJ6t-AWRj3-S}\eatline
\nwused{\\{NWppJ6t-7vmg0-4}}\nwidentdefs{\\{{\nwixident{arrangement{\_}write}}{arrangement:unwrite}}}\nwidentuses{\\{{\nwixident{name}}{name}}}\nwindexuse{\nwixident{name}}{name}{NWppJ6t-AWRj3-S}\nwendcode{}\nwbegindocs{398}\nwdocspar
\nwenddocs{}\nwbegindocs{399}The written file {\Tt{}\Rm{}{\it{}filename}\nwendquote} then can be loaded by
texttt{cycle3D-visualiser} either through command line option or file
choosing dialog. See documentations of the helper programme for
available tools. In particular, it is possible to make screenshots
similar to Fig.~\ref{fig:apollonius-3D}.

\nwenddocs{}\nwbegindocs{400}To print a figure {\Tt{}\Rm{}{\it{}F}\nwendquote} (of any dimensionality) as a list of nodes
and relations between them it is enough to direct the figure to the
stream:
\begin{quote}
  {\Tt{}\Rm{} {\it{}cout} \begin{math}\ll\end{math} {\it{}F} \begin{math}\ll\end{math}{\it{}endl}; \nwendquote}
\end{quote}

\nwenddocs{}\nwbegindocs{401}\nwdocspar
\subsection{Saving and openning}
\label{sec:saving-openning}
\nwenddocs{}\nwbegindocs{402}We can write a figure to a file as a \GiNaC\ archive
(\texttt{*.gar} file) named {\Tt{}\Rm{}{\it{}file\_name}\nwendquote} at a node {\Tt{}\Rm{}{\it{}fig\_name}\nwendquote}.
\nwenddocs{}\nwbegincode{403}\sublabel{NWppJ6t-AWRj3-T}\nwmargintag{{\nwtagstyle{}\subpageref{NWppJ6t-AWRj3-T}}}\moddef{public methods in figure class~{\nwtagstyle{}\subpageref{NWppJ6t-AWRj3-1}}}\plusendmoddef\Rm{}\nwstartdeflinemarkup\nwusesondefline{\\{NWppJ6t-7vmg0-4}}\nwprevnextdefs{NWppJ6t-AWRj3-S}{NWppJ6t-AWRj3-U}\nwenddeflinemarkup
{\bf{}void} {\it{}save}({\bf{}const} {\bf{}char}\begin{math}\ast\end{math} {\it{}file\_name}, {\bf{}const} {\bf{}char}\begin{math}\ast\end{math} {\it{}fig\_name}={\tt{}"myfig"}) {\bf{}const};\nwindexdefn{\nwixident{save}}{save}{NWppJ6t-AWRj3-T}
\nwindexdefn{\nwixident{save}}{save}{NWppJ6t-AWRj3-T}\eatline
\nwused{\\{NWppJ6t-7vmg0-4}}\nwidentdefs{\\{{\nwixident{save}}{save}}}\nwendcode{}\nwbegindocs{404}\nwdocspar
\nwenddocs{}\nwbegindocs{405}This constructor reads a figure stored in a \GiNaC\ archive
(\texttt{*.gar} file) named {\Tt{}\Rm{}{\it{}file\_name}\nwendquote} at a node {\Tt{}\Rm{}{\it{}fig\_name}\nwendquote}.
\nwenddocs{}\nwbegincode{406}\sublabel{NWppJ6t-AWRj3-U}\nwmargintag{{\nwtagstyle{}\subpageref{NWppJ6t-AWRj3-U}}}\moddef{public methods in figure class~{\nwtagstyle{}\subpageref{NWppJ6t-AWRj3-1}}}\plusendmoddef\Rm{}\nwstartdeflinemarkup\nwusesondefline{\\{NWppJ6t-7vmg0-4}}\nwprevnextdefs{NWppJ6t-AWRj3-T}{NWppJ6t-AWRj3-V}\nwenddeflinemarkup
{\bf{}figure}({\bf{}const} {\bf{}char}\begin{math}\ast\end{math} {\it{}file\_name}, {\it{}string} {\it{}fig\_name}={\tt{}"myfig"});
\nwindexdefn{\nwixident{figure}}{figure}{NWppJ6t-AWRj3-U}\eatline
\nwused{\\{NWppJ6t-7vmg0-4}}\nwidentdefs{\\{{\nwixident{figure}}{figure}}}\nwendcode{}\nwbegindocs{407}\nwdocspar
\nwenddocs{}\nwbegindocs{408}\nwdocspar
\section{Public methods in {\Tt{}\Rm{}{\bf{}cycle\_relation}\nwendquote}}
\label{sec:publ-meth-cycl}

Nodes within figure are connected by sets of relations. There is some
essential relations pre-defined in the library. Users can define their
own relations as well.

\nwenddocs{}\nwbegindocs{409} The following relations between cycles are predefined.
Orthogonality of cycles given by
\cite{Kisil12a}*{\S~6.1}:
\begin{equation}
  \label{eq:orthog-defn-1}
  \scalar{\cycle{}{}}{\cycle[\tilde]{}{}}=0.
\end{equation}
\nwenddocs{}\nwbegincode{410}\sublabel{NWppJ6t-2iyHmp-1}\nwmargintag{{\nwtagstyle{}\subpageref{NWppJ6t-2iyHmp-1}}}\moddef{predefined cycle relations~{\nwtagstyle{}\subpageref{NWppJ6t-2iyHmp-1}}}\endmoddef\Rm{}\nwstartdeflinemarkup\nwusesondefline{\\{NWppJ6t-1d2GZW-9}}\nwprevnextdefs{\relax}{NWppJ6t-2iyHmp-2}\nwenddeflinemarkup
{\bf{}inline} {\bf{}cycle\_relation} {\it{}is\_orthogonal}({\bf{}const} {\bf{}ex} & {\it{}key}, {\bf{}bool} {\it{}cm}={\bf{}true})
                    {\nwlbrace}{\bf{}return} {\bf{}cycle\_relation}({\it{}key}, {\it{}cycle\_orthogonal}, {\it{}cm});{\nwrbrace}
\nwindexdefn{\nwixident{is{\_}orthogonal}}{is:unorthogonal}{NWppJ6t-2iyHmp-1}\eatline
\nwalsodefined{\\{NWppJ6t-2iyHmp-2}\\{NWppJ6t-2iyHmp-3}\\{NWppJ6t-2iyHmp-4}\\{NWppJ6t-2iyHmp-5}\\{NWppJ6t-2iyHmp-6}\\{NWppJ6t-2iyHmp-7}\\{NWppJ6t-2iyHmp-8}\\{NWppJ6t-2iyHmp-9}\\{NWppJ6t-2iyHmp-A}\\{NWppJ6t-2iyHmp-B}\\{NWppJ6t-2iyHmp-C}\\{NWppJ6t-2iyHmp-D}\\{NWppJ6t-2iyHmp-E}\\{NWppJ6t-2iyHmp-F}}\nwused{\\{NWppJ6t-1d2GZW-9}}\nwidentdefs{\\{{\nwixident{is{\_}orthogonal}}{is:unorthogonal}}}\nwidentuses{\\{{\nwixident{cycle{\_}orthogonal}}{cycle:unorthogonal}}\\{{\nwixident{cycle{\_}relation}}{cycle:unrelation}}\\{{\nwixident{ex}}{ex}}\\{{\nwixident{key}}{key}}}\nwindexuse{\nwixident{cycle{\_}orthogonal}}{cycle:unorthogonal}{NWppJ6t-2iyHmp-1}\nwindexuse{\nwixident{cycle{\_}relation}}{cycle:unrelation}{NWppJ6t-2iyHmp-1}\nwindexuse{\nwixident{ex}}{ex}{NWppJ6t-2iyHmp-1}\nwindexuse{\nwixident{key}}{key}{NWppJ6t-2iyHmp-1}\nwendcode{}\nwbegindocs{411}\nwdocspar
\nwenddocs{}\nwbegindocs{412}Focal orthogonality of cycles~\eqref{eq:focal-orthog-defn}, see \cite{Kisil12a}*{\S~6.6}.
\nwenddocs{}\nwbegincode{413}\sublabel{NWppJ6t-2iyHmp-2}\nwmargintag{{\nwtagstyle{}\subpageref{NWppJ6t-2iyHmp-2}}}\moddef{predefined cycle relations~{\nwtagstyle{}\subpageref{NWppJ6t-2iyHmp-1}}}\plusendmoddef\Rm{}\nwstartdeflinemarkup\nwusesondefline{\\{NWppJ6t-1d2GZW-9}}\nwprevnextdefs{NWppJ6t-2iyHmp-1}{NWppJ6t-2iyHmp-3}\nwenddeflinemarkup
{\bf{}inline} {\bf{}cycle\_relation} {\it{}is\_f\_orthogonal}({\bf{}const} {\bf{}ex} & {\it{}key}, {\bf{}bool} {\it{}cm}={\bf{}true})
                    {\nwlbrace}{\bf{}return} {\bf{}cycle\_relation}({\it{}key}, {\it{}cycle\_f\_orthogonal}, {\it{}cm});{\nwrbrace}
\nwindexdefn{\nwixident{is{\_}f{\_}orthogonal}}{is:unf:unorthogonal}{NWppJ6t-2iyHmp-2}\eatline
\nwused{\\{NWppJ6t-1d2GZW-9}}\nwidentdefs{\\{{\nwixident{is{\_}f{\_}orthogonal}}{is:unf:unorthogonal}}}\nwidentuses{\\{{\nwixident{cycle{\_}f{\_}orthogonal}}{cycle:unf:unorthogonal}}\\{{\nwixident{cycle{\_}relation}}{cycle:unrelation}}\\{{\nwixident{ex}}{ex}}\\{{\nwixident{key}}{key}}}\nwindexuse{\nwixident{cycle{\_}f{\_}orthogonal}}{cycle:unf:unorthogonal}{NWppJ6t-2iyHmp-2}\nwindexuse{\nwixident{cycle{\_}relation}}{cycle:unrelation}{NWppJ6t-2iyHmp-2}\nwindexuse{\nwixident{ex}}{ex}{NWppJ6t-2iyHmp-2}\nwindexuse{\nwixident{key}}{key}{NWppJ6t-2iyHmp-2}\nwendcode{}\nwbegindocs{414}\nwdocspar
\nwenddocs{}\nwbegindocs{415}We may want a cycle to be different from another. For example, if we
look for intersection of two lines we want to exclude the infinity,
where they are intersected anyway. Then, we may add the condition
{\Tt{}\Rm{}{\it{}is\_different}({\it{}F}.{\it{}get\_infinity}())\nwendquote}.
\nwenddocs{}\nwbegincode{416}\sublabel{NWppJ6t-2iyHmp-3}\nwmargintag{{\nwtagstyle{}\subpageref{NWppJ6t-2iyHmp-3}}}\moddef{predefined cycle relations~{\nwtagstyle{}\subpageref{NWppJ6t-2iyHmp-1}}}\plusendmoddef\Rm{}\nwstartdeflinemarkup\nwusesondefline{\\{NWppJ6t-1d2GZW-9}}\nwprevnextdefs{NWppJ6t-2iyHmp-2}{NWppJ6t-2iyHmp-4}\nwenddeflinemarkup
{\bf{}inline} {\bf{}cycle\_relation} {\it{}is\_different}({\bf{}const} {\bf{}ex} & {\it{}key}, {\bf{}bool} {\it{}cm}={\bf{}true})
                    {\nwlbrace}{\bf{}return} {\bf{}cycle\_relation}({\it{}key}, {\it{}cycle\_different}, {\it{}cm});{\nwrbrace}
\nwindexdefn{\nwixident{is{\_}different}}{is:undifferent}{NWppJ6t-2iyHmp-3}\eatline
\nwused{\\{NWppJ6t-1d2GZW-9}}\nwidentdefs{\\{{\nwixident{is{\_}different}}{is:undifferent}}}\nwidentuses{\\{{\nwixident{cycle{\_}different}}{cycle:undifferent}}\\{{\nwixident{cycle{\_}relation}}{cycle:unrelation}}\\{{\nwixident{ex}}{ex}}\\{{\nwixident{key}}{key}}}\nwindexuse{\nwixident{cycle{\_}different}}{cycle:undifferent}{NWppJ6t-2iyHmp-3}\nwindexuse{\nwixident{cycle{\_}relation}}{cycle:unrelation}{NWppJ6t-2iyHmp-3}\nwindexuse{\nwixident{ex}}{ex}{NWppJ6t-2iyHmp-3}\nwindexuse{\nwixident{key}}{key}{NWppJ6t-2iyHmp-3}\nwendcode{}\nwbegindocs{417}\nwdocspar
\nwenddocs{}\nwbegindocs{418}Due to a possible rounding errors we include an approximate version
of {\Tt{}\Rm{}{\it{}is\_different}\nwendquote}.
\nwenddocs{}\nwbegincode{419}\sublabel{NWppJ6t-2iyHmp-4}\nwmargintag{{\nwtagstyle{}\subpageref{NWppJ6t-2iyHmp-4}}}\moddef{predefined cycle relations~{\nwtagstyle{}\subpageref{NWppJ6t-2iyHmp-1}}}\plusendmoddef\Rm{}\nwstartdeflinemarkup\nwusesondefline{\\{NWppJ6t-1d2GZW-9}}\nwprevnextdefs{NWppJ6t-2iyHmp-3}{NWppJ6t-2iyHmp-5}\nwenddeflinemarkup
{\bf{}inline} {\bf{}cycle\_relation} {\it{}is\_adifferent}({\bf{}const} {\bf{}ex} & {\it{}key}, {\bf{}bool} {\it{}cm}={\bf{}true})
                    {\nwlbrace}{\bf{}return} {\bf{}cycle\_relation}({\it{}key}, {\it{}cycle\_adifferent}, {\it{}cm});{\nwrbrace}
\nwindexdefn{\nwixident{is{\_}adifferent}}{is:unadifferent}{NWppJ6t-2iyHmp-4}\eatline
\nwused{\\{NWppJ6t-1d2GZW-9}}\nwidentdefs{\\{{\nwixident{is{\_}adifferent}}{is:unadifferent}}}\nwidentuses{\\{{\nwixident{cycle{\_}adifferent}}{cycle:unadifferent}}\\{{\nwixident{cycle{\_}relation}}{cycle:unrelation}}\\{{\nwixident{ex}}{ex}}\\{{\nwixident{key}}{key}}}\nwindexuse{\nwixident{cycle{\_}adifferent}}{cycle:unadifferent}{NWppJ6t-2iyHmp-4}\nwindexuse{\nwixident{cycle{\_}relation}}{cycle:unrelation}{NWppJ6t-2iyHmp-4}\nwindexuse{\nwixident{ex}}{ex}{NWppJ6t-2iyHmp-4}\nwindexuse{\nwixident{key}}{key}{NWppJ6t-2iyHmp-4}\nwendcode{}\nwbegindocs{420}\nwdocspar
\nwenddocs{}\nwbegindocs{421}This relation check if a cycle is a non-positive vector, for circles
this corresponds to real (non-imaginary) circles. By default we
check this in the point space metric.
\nwenddocs{}\nwbegincode{422}\sublabel{NWppJ6t-2iyHmp-5}\nwmargintag{{\nwtagstyle{}\subpageref{NWppJ6t-2iyHmp-5}}}\moddef{predefined cycle relations~{\nwtagstyle{}\subpageref{NWppJ6t-2iyHmp-1}}}\plusendmoddef\Rm{}\nwstartdeflinemarkup\nwusesondefline{\\{NWppJ6t-1d2GZW-9}}\nwprevnextdefs{NWppJ6t-2iyHmp-4}{NWppJ6t-2iyHmp-6}\nwenddeflinemarkup
{\bf{}inline} {\bf{}cycle\_relation} {\it{}is\_real\_cycle}({\bf{}const} {\bf{}ex} & {\it{}key}, {\bf{}bool} {\it{}cm}={\bf{}false}, {\bf{}const} {\bf{}ex} & {\it{}pr}=1)
 {\nwlbrace}{\bf{}return} {\bf{}cycle\_relation}({\it{}key}, {\it{}product\_sign}, {\it{}cm}, {\it{}pr});{\nwrbrace}
\nwindexdefn{\nwixident{is{\_}real{\_}cycle}}{is:unreal:uncycle}{NWppJ6t-2iyHmp-5}\eatline
\nwused{\\{NWppJ6t-1d2GZW-9}}\nwidentdefs{\\{{\nwixident{is{\_}real{\_}cycle}}{is:unreal:uncycle}}}\nwidentuses{\\{{\nwixident{cycle{\_}relation}}{cycle:unrelation}}\\{{\nwixident{ex}}{ex}}\\{{\nwixident{key}}{key}}\\{{\nwixident{product{\_}sign}}{product:unsign}}}\nwindexuse{\nwixident{cycle{\_}relation}}{cycle:unrelation}{NWppJ6t-2iyHmp-5}\nwindexuse{\nwixident{ex}}{ex}{NWppJ6t-2iyHmp-5}\nwindexuse{\nwixident{key}}{key}{NWppJ6t-2iyHmp-5}\nwindexuse{\nwixident{product{\_}sign}}{product:unsign}{NWppJ6t-2iyHmp-5}\nwendcode{}\nwbegindocs{423}\nwdocspar
\nwenddocs{}\nwbegindocs{424}Effectively this is the same check but with a different name and
other defaults. It may be used that both cycles are or are not
separated by the light cone in the indefinite metric in space of cycles.
\nwenddocs{}\nwbegincode{425}\sublabel{NWppJ6t-2iyHmp-6}\nwmargintag{{\nwtagstyle{}\subpageref{NWppJ6t-2iyHmp-6}}}\moddef{predefined cycle relations~{\nwtagstyle{}\subpageref{NWppJ6t-2iyHmp-1}}}\plusendmoddef\Rm{}\nwstartdeflinemarkup\nwusesondefline{\\{NWppJ6t-1d2GZW-9}}\nwprevnextdefs{NWppJ6t-2iyHmp-5}{NWppJ6t-2iyHmp-7}\nwenddeflinemarkup
{\bf{}inline} {\bf{}cycle\_relation} {\it{}product\_nonpositive}({\bf{}const} {\bf{}ex} & {\it{}key}, {\bf{}bool} {\it{}cm}={\bf{}true}, {\bf{}const} {\bf{}ex} & {\it{}pr}=1)
 {\nwlbrace}{\bf{}return} {\bf{}cycle\_relation}({\it{}key}, {\it{}product\_sign}, {\it{}cm}, {\it{}pr});{\nwrbrace}
\nwindexdefn{\nwixident{product{\_}nonpositive}}{product:unnonpositive}{NWppJ6t-2iyHmp-6}\eatline
\nwused{\\{NWppJ6t-1d2GZW-9}}\nwidentdefs{\\{{\nwixident{product{\_}nonpositive}}{product:unnonpositive}}}\nwidentuses{\\{{\nwixident{cycle{\_}relation}}{cycle:unrelation}}\\{{\nwixident{ex}}{ex}}\\{{\nwixident{key}}{key}}\\{{\nwixident{product{\_}sign}}{product:unsign}}}\nwindexuse{\nwixident{cycle{\_}relation}}{cycle:unrelation}{NWppJ6t-2iyHmp-6}\nwindexuse{\nwixident{ex}}{ex}{NWppJ6t-2iyHmp-6}\nwindexuse{\nwixident{key}}{key}{NWppJ6t-2iyHmp-6}\nwindexuse{\nwixident{product{\_}sign}}{product:unsign}{NWppJ6t-2iyHmp-6}\nwendcode{}\nwbegindocs{426}\nwdocspar
\nwenddocs{}\nwbegindocs{427}We may want to exclude cycles with imaginary coefficients, this
condition check it.
\nwenddocs{}\nwbegincode{428}\sublabel{NWppJ6t-2iyHmp-7}\nwmargintag{{\nwtagstyle{}\subpageref{NWppJ6t-2iyHmp-7}}}\moddef{predefined cycle relations~{\nwtagstyle{}\subpageref{NWppJ6t-2iyHmp-1}}}\plusendmoddef\Rm{}\nwstartdeflinemarkup\nwusesondefline{\\{NWppJ6t-1d2GZW-9}}\nwprevnextdefs{NWppJ6t-2iyHmp-6}{NWppJ6t-2iyHmp-8}\nwenddeflinemarkup
{\bf{}inline} {\bf{}cycle\_relation} {\it{}only\_reals}({\bf{}const} {\bf{}ex} & {\it{}key}, {\bf{}bool} {\it{}cm}={\bf{}true}, {\bf{}const} {\bf{}ex} & {\it{}pr}=0)
 {\nwlbrace}{\bf{}return} {\bf{}cycle\_relation}({\it{}key}, {\it{}coefficients\_are\_real}, {\it{}cm}, {\it{}pr});{\nwrbrace}
\nwindexdefn{\nwixident{only{\_}reals}}{only:unreals}{NWppJ6t-2iyHmp-7}\eatline
\nwused{\\{NWppJ6t-1d2GZW-9}}\nwidentdefs{\\{{\nwixident{only{\_}reals}}{only:unreals}}}\nwidentuses{\\{{\nwixident{coefficients{\_}are{\_}real}}{coefficients:unare:unreal}}\\{{\nwixident{cycle{\_}relation}}{cycle:unrelation}}\\{{\nwixident{ex}}{ex}}\\{{\nwixident{key}}{key}}}\nwindexuse{\nwixident{coefficients{\_}are{\_}real}}{coefficients:unare:unreal}{NWppJ6t-2iyHmp-7}\nwindexuse{\nwixident{cycle{\_}relation}}{cycle:unrelation}{NWppJ6t-2iyHmp-7}\nwindexuse{\nwixident{ex}}{ex}{NWppJ6t-2iyHmp-7}\nwindexuse{\nwixident{key}}{key}{NWppJ6t-2iyHmp-7}\nwendcode{}\nwbegindocs{429}\nwdocspar
\nwenddocs{}\nwbegindocs{430}This is tangency condition which shall be used to find tangent cycles.
\nwenddocs{}\nwbegincode{431}\sublabel{NWppJ6t-2iyHmp-8}\nwmargintag{{\nwtagstyle{}\subpageref{NWppJ6t-2iyHmp-8}}}\moddef{predefined cycle relations~{\nwtagstyle{}\subpageref{NWppJ6t-2iyHmp-1}}}\plusendmoddef\Rm{}\nwstartdeflinemarkup\nwusesondefline{\\{NWppJ6t-1d2GZW-9}}\nwprevnextdefs{NWppJ6t-2iyHmp-7}{NWppJ6t-2iyHmp-9}\nwenddeflinemarkup
{\bf{}inline} {\bf{}cycle\_relation} {\it{}is\_tangent}({\bf{}const} {\bf{}ex} & {\it{}key}, {\bf{}bool} {\it{}cm}={\bf{}true})
                    {\nwlbrace}{\bf{}return} {\bf{}cycle\_relation}({\it{}key}, {\it{}cycle\_tangent}, {\it{}cm});{\nwrbrace}
\nwindexdefn{\nwixident{is{\_}tangent}}{is:untangent}{NWppJ6t-2iyHmp-8}\eatline
\nwused{\\{NWppJ6t-1d2GZW-9}}\nwidentdefs{\\{{\nwixident{is{\_}tangent}}{is:untangent}}}\nwidentuses{\\{{\nwixident{cycle{\_}relation}}{cycle:unrelation}}\\{{\nwixident{cycle{\_}tangent}}{cycle:untangent}}\\{{\nwixident{ex}}{ex}}\\{{\nwixident{key}}{key}}}\nwindexuse{\nwixident{cycle{\_}relation}}{cycle:unrelation}{NWppJ6t-2iyHmp-8}\nwindexuse{\nwixident{cycle{\_}tangent}}{cycle:untangent}{NWppJ6t-2iyHmp-8}\nwindexuse{\nwixident{ex}}{ex}{NWppJ6t-2iyHmp-8}\nwindexuse{\nwixident{key}}{key}{NWppJ6t-2iyHmp-8}\nwendcode{}\nwbegindocs{432}\nwdocspar
\nwenddocs{}\nwbegindocs{433}The split version for inner and outer tangent cycles.
\nwenddocs{}\nwbegincode{434}\sublabel{NWppJ6t-2iyHmp-9}\nwmargintag{{\nwtagstyle{}\subpageref{NWppJ6t-2iyHmp-9}}}\moddef{predefined cycle relations~{\nwtagstyle{}\subpageref{NWppJ6t-2iyHmp-1}}}\plusendmoddef\Rm{}\nwstartdeflinemarkup\nwusesondefline{\\{NWppJ6t-1d2GZW-9}}\nwprevnextdefs{NWppJ6t-2iyHmp-8}{NWppJ6t-2iyHmp-A}\nwenddeflinemarkup
{\bf{}inline} {\bf{}cycle\_relation} {\it{}is\_tangent\_i}({\bf{}const} {\bf{}ex} & {\it{}key}, {\bf{}bool} {\it{}cm}={\bf{}true})
                    {\nwlbrace}{\bf{}return} {\bf{}cycle\_relation}({\it{}key}, {\it{}cycle\_tangent\_i}, {\it{}cm});{\nwrbrace}
{\bf{}inline} {\bf{}cycle\_relation} {\it{}is\_tangent\_o}({\bf{}const} {\bf{}ex} & {\it{}key}, {\bf{}bool} {\it{}cm}={\bf{}true})
                    {\nwlbrace}{\bf{}return} {\bf{}cycle\_relation}({\it{}key}, {\it{}cycle\_tangent\_o}, {\it{}cm});{\nwrbrace}
\nwindexdefn{\nwixident{is{\_}tangent{\_}i}}{is:untangent:uni}{NWppJ6t-2iyHmp-9}\nwindexdefn{\nwixident{is{\_}tangent{\_}o}}{is:untangent:uno}{NWppJ6t-2iyHmp-9}\eatline
\nwused{\\{NWppJ6t-1d2GZW-9}}\nwidentdefs{\\{{\nwixident{is{\_}tangent{\_}i}}{is:untangent:uni}}\\{{\nwixident{is{\_}tangent{\_}o}}{is:untangent:uno}}}\nwidentuses{\\{{\nwixident{cycle{\_}relation}}{cycle:unrelation}}\\{{\nwixident{cycle{\_}tangent{\_}i}}{cycle:untangent:uni}}\\{{\nwixident{cycle{\_}tangent{\_}o}}{cycle:untangent:uno}}\\{{\nwixident{ex}}{ex}}\\{{\nwixident{key}}{key}}}\nwindexuse{\nwixident{cycle{\_}relation}}{cycle:unrelation}{NWppJ6t-2iyHmp-9}\nwindexuse{\nwixident{cycle{\_}tangent{\_}i}}{cycle:untangent:uni}{NWppJ6t-2iyHmp-9}\nwindexuse{\nwixident{cycle{\_}tangent{\_}o}}{cycle:untangent:uno}{NWppJ6t-2iyHmp-9}\nwindexuse{\nwixident{ex}}{ex}{NWppJ6t-2iyHmp-9}\nwindexuse{\nwixident{key}}{key}{NWppJ6t-2iyHmp-9}\nwendcode{}\nwbegindocs{435}\nwdocspar
\nwenddocs{}\nwbegindocs{436}The relation between cycles to ``intersect with
certain angle'' (but the ``intersection'' may be imaginary). If cycles
are intersecting indeed then the value of {\Tt{}\Rm{}{\it{}pr}\nwendquote} is the cosine of the
angle.
\nwenddocs{}\nwbegincode{437}\sublabel{NWppJ6t-2iyHmp-A}\nwmargintag{{\nwtagstyle{}\subpageref{NWppJ6t-2iyHmp-A}}}\moddef{predefined cycle relations~{\nwtagstyle{}\subpageref{NWppJ6t-2iyHmp-1}}}\plusendmoddef\Rm{}\nwstartdeflinemarkup\nwusesondefline{\\{NWppJ6t-1d2GZW-9}}\nwprevnextdefs{NWppJ6t-2iyHmp-9}{NWppJ6t-2iyHmp-B}\nwenddeflinemarkup
{\bf{}inline} {\bf{}cycle\_relation} {\it{}make\_angle}({\bf{}const} {\bf{}ex} & {\it{}key}, {\bf{}bool} {\it{}cm}={\bf{}true}, {\bf{}const} {\bf{}ex} & {\it{}angle}=0)
                    {\nwlbrace}{\bf{}return} {\bf{}cycle\_relation}({\it{}key}, {\it{}cycle\_angle}, {\it{}cm}, {\it{}angle});{\nwrbrace}
\nwindexdefn{\nwixident{make{\_}angle}}{make:unangle}{NWppJ6t-2iyHmp-A}\eatline
\nwused{\\{NWppJ6t-1d2GZW-9}}\nwidentdefs{\\{{\nwixident{make{\_}angle}}{make:unangle}}}\nwidentuses{\\{{\nwixident{cycle{\_}angle}}{cycle:unangle}}\\{{\nwixident{cycle{\_}relation}}{cycle:unrelation}}\\{{\nwixident{ex}}{ex}}\\{{\nwixident{key}}{key}}}\nwindexuse{\nwixident{cycle{\_}angle}}{cycle:unangle}{NWppJ6t-2iyHmp-A}\nwindexuse{\nwixident{cycle{\_}relation}}{cycle:unrelation}{NWppJ6t-2iyHmp-A}\nwindexuse{\nwixident{ex}}{ex}{NWppJ6t-2iyHmp-A}\nwindexuse{\nwixident{key}}{key}{NWppJ6t-2iyHmp-A}\nwendcode{}\nwbegindocs{438}\nwdocspar
\nwenddocs{}\nwbegindocs{439}The next relation defines a generalisation of a Steiner power of a
point for cycles.
\nwenddocs{}\nwbegincode{440}\sublabel{NWppJ6t-2iyHmp-B}\nwmargintag{{\nwtagstyle{}\subpageref{NWppJ6t-2iyHmp-B}}}\moddef{predefined cycle relations~{\nwtagstyle{}\subpageref{NWppJ6t-2iyHmp-1}}}\plusendmoddef\Rm{}\nwstartdeflinemarkup\nwusesondefline{\\{NWppJ6t-1d2GZW-9}}\nwprevnextdefs{NWppJ6t-2iyHmp-A}{NWppJ6t-2iyHmp-C}\nwenddeflinemarkup
{\bf{}inline} {\bf{}cycle\_relation} {\it{}cycle\_power}({\bf{}const} {\bf{}ex} & {\it{}key}, {\bf{}bool} {\it{}cm}={\bf{}true}, {\bf{}const} {\bf{}ex} & {\it{}cpower}=0)
                    {\nwlbrace}{\bf{}return} {\bf{}cycle\_relation}({\it{}key}, {\it{}steiner\_power}, {\it{}cm}, {\it{}cpower});{\nwrbrace}
\nwindexdefn{\nwixident{cycle{\_}power}}{cycle:unpower}{NWppJ6t-2iyHmp-B}\eatline
\nwused{\\{NWppJ6t-1d2GZW-9}}\nwidentdefs{\\{{\nwixident{cycle{\_}power}}{cycle:unpower}}}\nwidentuses{\\{{\nwixident{cycle{\_}relation}}{cycle:unrelation}}\\{{\nwixident{ex}}{ex}}\\{{\nwixident{key}}{key}}\\{{\nwixident{steiner{\_}power}}{steiner:unpower}}}\nwindexuse{\nwixident{cycle{\_}relation}}{cycle:unrelation}{NWppJ6t-2iyHmp-B}\nwindexuse{\nwixident{ex}}{ex}{NWppJ6t-2iyHmp-B}\nwindexuse{\nwixident{key}}{key}{NWppJ6t-2iyHmp-B}\nwindexuse{\nwixident{steiner{\_}power}}{steiner:unpower}{NWppJ6t-2iyHmp-B}\nwendcode{}\nwbegindocs{441}\nwdocspar
\nwenddocs{}\nwbegindocs{442}The next relation defines tangential distance between cycles.
\nwenddocs{}\nwbegincode{443}\sublabel{NWppJ6t-2iyHmp-C}\nwmargintag{{\nwtagstyle{}\subpageref{NWppJ6t-2iyHmp-C}}}\moddef{predefined cycle relations~{\nwtagstyle{}\subpageref{NWppJ6t-2iyHmp-1}}}\plusendmoddef\Rm{}\nwstartdeflinemarkup\nwusesondefline{\\{NWppJ6t-1d2GZW-9}}\nwprevnextdefs{NWppJ6t-2iyHmp-B}{NWppJ6t-2iyHmp-D}\nwenddeflinemarkup
{\bf{}inline} {\bf{}cycle\_relation} {\it{}tangential\_distance}({\bf{}const} {\bf{}ex} & {\it{}key}, {\bf{}bool} {\it{}cm}={\bf{}true}, {\bf{}const} {\bf{}ex} & {\it{}distance}=0)
                    {\nwlbrace}{\bf{}return} {\bf{}cycle\_relation}({\it{}key}, {\it{}steiner\_power}, {\it{}cm}, {\it{}pow}({\it{}distance},2));{\nwrbrace}
\nwindexdefn{\nwixident{tangential{\_}distance}}{tangential:undistance}{NWppJ6t-2iyHmp-C}\eatline
\nwused{\\{NWppJ6t-1d2GZW-9}}\nwidentdefs{\\{{\nwixident{tangential{\_}distance}}{tangential:undistance}}}\nwidentuses{\\{{\nwixident{cycle{\_}relation}}{cycle:unrelation}}\\{{\nwixident{ex}}{ex}}\\{{\nwixident{key}}{key}}\\{{\nwixident{steiner{\_}power}}{steiner:unpower}}}\nwindexuse{\nwixident{cycle{\_}relation}}{cycle:unrelation}{NWppJ6t-2iyHmp-C}\nwindexuse{\nwixident{ex}}{ex}{NWppJ6t-2iyHmp-C}\nwindexuse{\nwixident{key}}{key}{NWppJ6t-2iyHmp-C}\nwindexuse{\nwixident{steiner{\_}power}}{steiner:unpower}{NWppJ6t-2iyHmp-C}\nwendcode{}\nwbegindocs{444}\nwdocspar
\nwenddocs{}\nwbegindocs{445}The next relation defines cross-tangential distance between cycles.
\nwenddocs{}\nwbegincode{446}\sublabel{NWppJ6t-2iyHmp-D}\nwmargintag{{\nwtagstyle{}\subpageref{NWppJ6t-2iyHmp-D}}}\moddef{predefined cycle relations~{\nwtagstyle{}\subpageref{NWppJ6t-2iyHmp-1}}}\plusendmoddef\Rm{}\nwstartdeflinemarkup\nwusesondefline{\\{NWppJ6t-1d2GZW-9}}\nwprevnextdefs{NWppJ6t-2iyHmp-C}{NWppJ6t-2iyHmp-E}\nwenddeflinemarkup
{\bf{}inline} {\bf{}cycle\_relation} {\it{}cross\_t\_distance}({\bf{}const} {\bf{}ex} & {\it{}key}, {\bf{}bool} {\it{}cm}={\bf{}true}, {\bf{}const} {\bf{}ex} & {\it{}distance}=0)
                    {\nwlbrace}{\bf{}return} {\bf{}cycle\_relation}({\it{}key}, {\it{}cycle\_cross\_t\_distance}, {\it{}cm}, {\it{}distance});{\nwrbrace}
\nwindexdefn{\nwixident{cross{\_}t{\_}distance}}{cross:unt:undistance}{NWppJ6t-2iyHmp-D}\eatline
\nwused{\\{NWppJ6t-1d2GZW-9}}\nwidentdefs{\\{{\nwixident{cross{\_}t{\_}distance}}{cross:unt:undistance}}}\nwidentuses{\\{{\nwixident{cycle{\_}cross{\_}t{\_}distance}}{cycle:uncross:unt:undistance}}\\{{\nwixident{cycle{\_}relation}}{cycle:unrelation}}\\{{\nwixident{ex}}{ex}}\\{{\nwixident{key}}{key}}}\nwindexuse{\nwixident{cycle{\_}cross{\_}t{\_}distance}}{cycle:uncross:unt:undistance}{NWppJ6t-2iyHmp-D}\nwindexuse{\nwixident{cycle{\_}relation}}{cycle:unrelation}{NWppJ6t-2iyHmp-D}\nwindexuse{\nwixident{ex}}{ex}{NWppJ6t-2iyHmp-D}\nwindexuse{\nwixident{key}}{key}{NWppJ6t-2iyHmp-D}\nwendcode{}\nwbegindocs{447}\nwdocspar
\nwenddocs{}\nwbegindocs{448}The next relation creates a cycle, which is a FLT of an existing
cycle. The transformation is defined by
a list of four entries which will make a \(2\times 2\) matrix. The
default value corresponds to the identity map. User will need to use a
proper Clifford algebra for the matrix to make this transformation
works. In two dimensions the next method makes a relief.
\nwenddocs{}\nwbegincode{449}\sublabel{NWppJ6t-2iyHmp-E}\nwmargintag{{\nwtagstyle{}\subpageref{NWppJ6t-2iyHmp-E}}}\moddef{predefined cycle relations~{\nwtagstyle{}\subpageref{NWppJ6t-2iyHmp-1}}}\plusendmoddef\Rm{}\nwstartdeflinemarkup\nwusesondefline{\\{NWppJ6t-1d2GZW-9}}\nwprevnextdefs{NWppJ6t-2iyHmp-D}{NWppJ6t-2iyHmp-F}\nwenddeflinemarkup
{\bf{}inline} {\bf{}cycle\_relation} {\it{}moebius\_transform}({\bf{}const} {\bf{}ex} & {\it{}key}, {\bf{}bool} {\it{}cm}={\bf{}true},
                                        {\bf{}const} {\bf{}ex} & {\bf{}matrix}={\bf{}lst}{\nwlbrace}{\bf{}numeric}(1),0,0,{\bf{}numeric}(1){\nwrbrace})
                    {\nwlbrace}{\bf{}return} {\bf{}cycle\_relation}({\it{}key}, {\it{}cycle\_moebius}, {\it{}cm}, {\bf{}matrix});{\nwrbrace}
\nwindexdefn{\nwixident{moebius{\_}transform}}{moebius:untransform}{NWppJ6t-2iyHmp-E}\eatline
\nwused{\\{NWppJ6t-1d2GZW-9}}\nwidentdefs{\\{{\nwixident{moebius{\_}transform}}{moebius:untransform}}}\nwidentuses{\\{{\nwixident{cycle{\_}moebius}}{cycle:unmoebius}}\\{{\nwixident{cycle{\_}relation}}{cycle:unrelation}}\\{{\nwixident{ex}}{ex}}\\{{\nwixident{key}}{key}}\\{{\nwixident{numeric}}{numeric}}}\nwindexuse{\nwixident{cycle{\_}moebius}}{cycle:unmoebius}{NWppJ6t-2iyHmp-E}\nwindexuse{\nwixident{cycle{\_}relation}}{cycle:unrelation}{NWppJ6t-2iyHmp-E}\nwindexuse{\nwixident{ex}}{ex}{NWppJ6t-2iyHmp-E}\nwindexuse{\nwixident{key}}{key}{NWppJ6t-2iyHmp-E}\nwindexuse{\nwixident{numeric}}{numeric}{NWppJ6t-2iyHmp-E}\nwendcode{}\nwbegindocs{450}\nwdocspar
\nwenddocs{}\nwbegindocs{451}This is a simplified variant of the previous transformations for two
dimension figures and transformations with real entries. The
corresponding check will be carried out by the library. Then, the library
will convert it into the proper Clifford valued matrix.
\nwenddocs{}\nwbegincode{452}\sublabel{NWppJ6t-2iyHmp-F}\nwmargintag{{\nwtagstyle{}\subpageref{NWppJ6t-2iyHmp-F}}}\moddef{predefined cycle relations~{\nwtagstyle{}\subpageref{NWppJ6t-2iyHmp-1}}}\plusendmoddef\Rm{}\nwstartdeflinemarkup\nwusesondefline{\\{NWppJ6t-1d2GZW-9}}\nwprevnextdefs{NWppJ6t-2iyHmp-E}{\relax}\nwenddeflinemarkup
 {\bf{}cycle\_relation} {\it{}sl2\_transform}({\bf{}const} {\bf{}ex} & {\it{}key}, {\bf{}bool} {\it{}cm}={\bf{}true},
                              {\bf{}const} {\bf{}ex} & {\bf{}matrix}={\bf{}lst}{\nwlbrace}{\bf{}numeric}(1),0,0,{\bf{}numeric}(1){\nwrbrace});
\nwindexdefn{\nwixident{sl2{\_}transform}}{sl2:untransform}{NWppJ6t-2iyHmp-F}\eatline
\nwused{\\{NWppJ6t-1d2GZW-9}}\nwidentdefs{\\{{\nwixident{sl2{\_}transform}}{sl2:untransform}}}\nwidentuses{\\{{\nwixident{cycle{\_}relation}}{cycle:unrelation}}\\{{\nwixident{ex}}{ex}}\\{{\nwixident{key}}{key}}\\{{\nwixident{numeric}}{numeric}}}\nwindexuse{\nwixident{cycle{\_}relation}}{cycle:unrelation}{NWppJ6t-2iyHmp-F}\nwindexuse{\nwixident{ex}}{ex}{NWppJ6t-2iyHmp-F}\nwindexuse{\nwixident{key}}{key}{NWppJ6t-2iyHmp-F}\nwindexuse{\nwixident{numeric}}{numeric}{NWppJ6t-2iyHmp-F}\nwendcode{}\nwbegindocs{453}\nwdocspar
\nwenddocs{}\nwbegindocs{454}This is a constructor which creates a relation of the type {\Tt{}\Rm{}{\it{}rel}\nwendquote} to
a node labelled by {\Tt{}\Rm{}{\it{}key}\nwendquote}. Boolean {\Tt{}\Rm{}{\it{}cm}\nwendquote} tells either to chose cycle
metric or point metric for the relation. An additional parameter {\Tt{}\Rm{}{\it{}p}\nwendquote}
can be supplied to the relation.
\nwenddocs{}\nwbegincode{455}\sublabel{NWppJ6t-1HOJwC-1}\nwmargintag{{\nwtagstyle{}\subpageref{NWppJ6t-1HOJwC-1}}}\moddef{public methods for cycle relation~{\nwtagstyle{}\subpageref{NWppJ6t-1HOJwC-1}}}\endmoddef\Rm{}\nwstartdeflinemarkup\nwusesondefline{\\{NWppJ6t-1d2GZW-2}}\nwenddeflinemarkup
{\bf{}cycle\_relation}({\bf{}const} {\bf{}ex} & {\it{}key}, {\it{}PCR} {\it{}rel}, {\bf{}bool} {\it{}cm}={\bf{}true}, {\bf{}const} {\bf{}ex} & {\it{}p}=0);
\nwindexdefn{\nwixident{cycle{\_}relation}}{cycle:unrelation}{NWppJ6t-1HOJwC-1}\eatline
\nwused{\\{NWppJ6t-1d2GZW-2}}\nwidentdefs{\\{{\nwixident{cycle{\_}relation}}{cycle:unrelation}}}\nwidentuses{\\{{\nwixident{ex}}{ex}}\\{{\nwixident{key}}{key}}\\{{\nwixident{PCR}}{PCR}}}\nwindexuse{\nwixident{ex}}{ex}{NWppJ6t-1HOJwC-1}\nwindexuse{\nwixident{key}}{key}{NWppJ6t-1HOJwC-1}\nwindexuse{\nwixident{PCR}}{PCR}{NWppJ6t-1HOJwC-1}\nwendcode{}\nwbegindocs{456}\nwdocspar
\nwenddocs{}\nwbegindocs{457}There is also an additional method to define a joint relation to
several parents by insertion of a {\Tt{}\Rm{}{\bf{}subfigure}\nwendquote}, see
{\Tt{}\Rm{}{\it{}midpoint\_constructor}\nwendquote} below.
\nwenddocs{}\nwbegincode{458}\sublabel{NWppJ6t-25h4gT-1}\nwmargintag{{\nwtagstyle{}\subpageref{NWppJ6t-25h4gT-1}}}\moddef{public methods for subfigure~{\nwtagstyle{}\subpageref{NWppJ6t-25h4gT-1}}}\endmoddef\Rm{}\nwstartdeflinemarkup\nwusesondefline{\\{NWppJ6t-4UgdbB-1}}\nwenddeflinemarkup
{\bf{}subfigure}({\bf{}const} {\bf{}ex} & {\it{}F}, {\bf{}const} {\bf{}ex} & {\it{}L});
\nwindexdefn{\nwixident{subfigure}}{subfigure}{NWppJ6t-25h4gT-1}\eatline
\nwused{\\{NWppJ6t-4UgdbB-1}}\nwidentdefs{\\{{\nwixident{subfigure}}{subfigure}}}\nwidentuses{\\{{\nwixident{ex}}{ex}}}\nwindexuse{\nwixident{ex}}{ex}{NWppJ6t-25h4gT-1}\nwendcode{}\nwbegindocs{459}\nwdocspar
\nwenddocs{}\nwbegindocs{460}\nwdocspar
\section{Addtional utilities}
\label{sec:addtional-utilities}

\nwenddocs{}\nwbegindocs{461}Here is a procedure which returns a figure, which can be used to
build a conformal version of the midpoint. The methods require three
points, say {\Tt{}\Rm{}{\it{}v1}\nwendquote}, {\Tt{}\Rm{}{\it{}v2}\nwendquote} and {\Tt{}\Rm{}{\it{}v3}\nwendquote}. If {\Tt{}\Rm{}{\it{}v3}\nwendquote} is infinity, then the
midpoint between {\Tt{}\Rm{}{\it{}v1}\nwendquote} and {\Tt{}\Rm{}{\it{}v2}\nwendquote} can be build using the orthogonality
only. Put a cycle {\Tt{}\Rm{}{\it{}v4}\nwendquote} joining {\Tt{}\Rm{}{\it{}v1}\nwendquote}, {\Tt{}\Rm{}{\it{}v2}\nwendquote} and {\Tt{}\Rm{}{\it{}v3}\nwendquote}. Then
construct a cycle {\Tt{}\Rm{}{\it{}v5}\nwendquote} with the diameter {\Tt{}\Rm{}{\it{}v1}\nwendquote}--{\Tt{}\Rm{}{\it{}v2}\nwendquote}, that is passing
these points and orthogonal to {\Tt{}\Rm{}{\it{}v4}\nwendquote}. Then, put the cycle {\Tt{}\Rm{}{\it{}v6}\nwendquote}
which passes {\Tt{}\Rm{}{\it{}v3}\nwendquote} and is orthogonal to {\Tt{}\Rm{}{\it{}v4}\nwendquote} and {\Tt{}\Rm{}{\it{}v5}\nwendquote}. The
intersection {\Tt{}\Rm{}{\it{}r}\nwendquote} of {\Tt{}\Rm{}{\it{}v6}\nwendquote} and {\Tt{}\Rm{}{\it{}v4}\nwendquote} is the midpoint of {\Tt{}\Rm{}{\it{}v1}\nwendquote}--{\Tt{}\Rm{}{\it{}v2}\nwendquote}.
\nwenddocs{}\nwbegincode{462}\sublabel{NWppJ6t-1lh2X2-2}\nwmargintag{{\nwtagstyle{}\subpageref{NWppJ6t-1lh2X2-2}}}\moddef{additional functions header~{\nwtagstyle{}\subpageref{NWppJ6t-1lh2X2-1}}}\plusendmoddef\Rm{}\nwstartdeflinemarkup\nwusesondefline{\\{NWppJ6t-3NCnUp-3}}\nwprevnextdefs{NWppJ6t-1lh2X2-1}{NWppJ6t-1lh2X2-3}\nwenddeflinemarkup
{\bf{}ex} {\it{}midpoint\_constructor}();
\nwindexdefn{\nwixident{midpoint{\_}constructor}}{midpoint:unconstructor}{NWppJ6t-1lh2X2-2}\eatline
\nwused{\\{NWppJ6t-3NCnUp-3}}\nwidentdefs{\\{{\nwixident{midpoint{\_}constructor}}{midpoint:unconstructor}}}\nwidentuses{\\{{\nwixident{ex}}{ex}}}\nwindexuse{\nwixident{ex}}{ex}{NWppJ6t-1lh2X2-2}\nwendcode{}\nwbegindocs{463}\nwdocspar
\nwenddocs{}\nwbegindocs{464}This utility make pair-wise comparison of cycles in the list {\Tt{}\Rm{}{\it{}L}\nwendquote}
and deletes duplicates.
\nwenddocs{}\nwbegincode{465}\sublabel{NWppJ6t-1lh2X2-3}\nwmargintag{{\nwtagstyle{}\subpageref{NWppJ6t-1lh2X2-3}}}\moddef{additional functions header~{\nwtagstyle{}\subpageref{NWppJ6t-1lh2X2-1}}}\plusendmoddef\Rm{}\nwstartdeflinemarkup\nwusesondefline{\\{NWppJ6t-3NCnUp-3}}\nwprevnextdefs{NWppJ6t-1lh2X2-2}{NWppJ6t-1lh2X2-4}\nwenddeflinemarkup
{\bf{}ex} {\it{}unique\_cycle}({\bf{}const} {\bf{}ex} & {\it{}L});
\nwindexdefn{\nwixident{unique{\_}cycle}}{unique:uncycle}{NWppJ6t-1lh2X2-3}\eatline
\nwused{\\{NWppJ6t-3NCnUp-3}}\nwidentdefs{\\{{\nwixident{unique{\_}cycle}}{unique:uncycle}}}\nwidentuses{\\{{\nwixident{ex}}{ex}}}\nwindexuse{\nwixident{ex}}{ex}{NWppJ6t-1lh2X2-3}\nwendcode{}\nwbegindocs{466}\nwdocspar
\nwenddocs{}\nwbegindocs{467}The debug output may be switched on and switched off by the
following methods.
\nwenddocs{}\nwbegincode{468}\sublabel{NWppJ6t-1lh2X2-4}\nwmargintag{{\nwtagstyle{}\subpageref{NWppJ6t-1lh2X2-4}}}\moddef{additional functions header~{\nwtagstyle{}\subpageref{NWppJ6t-1lh2X2-1}}}\plusendmoddef\Rm{}\nwstartdeflinemarkup\nwusesondefline{\\{NWppJ6t-3NCnUp-3}}\nwprevnextdefs{NWppJ6t-1lh2X2-3}{NWppJ6t-1lh2X2-5}\nwenddeflinemarkup
{\bf{}void} {\it{}figure\_debug\_on}();
{\bf{}void} {\it{}figure\_debug\_off}();
{\bf{}bool} {\it{}figure\_ask\_debug\_status}();
\nwindexdefn{\nwixident{figure{\_}debug{\_}on}}{figure:undebug:unon}{NWppJ6t-1lh2X2-4}\nwindexdefn{\nwixident{figure{\_}debug{\_}off}}{figure:undebug:unoff}{NWppJ6t-1lh2X2-4}\nwindexdefn{\nwixident{figure{\_}ask{\_}debug{\_}status}}{figure:unask:undebug:unstatus}{NWppJ6t-1lh2X2-4}\eatline
\nwused{\\{NWppJ6t-3NCnUp-3}}\nwidentdefs{\\{{\nwixident{figure{\_}ask{\_}debug{\_}status}}{figure:unask:undebug:unstatus}}\\{{\nwixident{figure{\_}debug{\_}off}}{figure:undebug:unoff}}\\{{\nwixident{figure{\_}debug{\_}on}}{figure:undebug:unon}}}\nwendcode{}\nwbegindocs{469}\nwdocspar
\nwenddocs{}\nwbegindocs{470}Solution of several quadratic equations in a sequence rapidly
increases complexity of expression. We try to resolve this by some
trigonometric or hyperbolic substitutions. Those expression in the
turn need to be simplified as well in {\Tt{}\Rm{}{\it{}evaluate\_cycle}()\nwendquote} for the
condition {\Tt{}\Rm{}{\it{}only\_reals}\nwendquote}. Later this variable will be assigned with a
default list of trigonometric substitutions.  User have a
possibility to adjust this list in the run time.
\nwenddocs{}\nwbegincode{471}\sublabel{NWppJ6t-1lh2X2-5}\nwmargintag{{\nwtagstyle{}\subpageref{NWppJ6t-1lh2X2-5}}}\moddef{additional functions header~{\nwtagstyle{}\subpageref{NWppJ6t-1lh2X2-1}}}\plusendmoddef\Rm{}\nwstartdeflinemarkup\nwusesondefline{\\{NWppJ6t-3NCnUp-3}}\nwprevnextdefs{NWppJ6t-1lh2X2-4}{\relax}\nwenddeflinemarkup
{\bf{}extern} {\bf{}const} {\bf{}ex} {\it{}evaluation\_assist};
\nwindexdefn{\nwixident{evaluation{\_}assist}}{evaluation:unassist}{NWppJ6t-1lh2X2-5}\eatline
\nwused{\\{NWppJ6t-3NCnUp-3}}\nwidentdefs{\\{{\nwixident{evaluation{\_}assist}}{evaluation:unassist}}}\nwidentuses{\\{{\nwixident{ex}}{ex}}}\nwindexuse{\nwixident{ex}}{ex}{NWppJ6t-1lh2X2-5}\nwendcode{}\nwbegindocs{472}\nwdocspar
\nwenddocs{}\nwbegindocs{473}Definition of the simplification rule.
\nwenddocs{}\nwbegincode{474}\sublabel{NWppJ6t-A2bag-1}\nwmargintag{{\nwtagstyle{}\subpageref{NWppJ6t-A2bag-1}}}\moddef{figure library variables and constants~{\nwtagstyle{}\subpageref{NWppJ6t-A2bag-1}}}\endmoddef\Rm{}\nwstartdeflinemarkup\nwusesondefline{\\{NWppJ6t-32jW6L-1}}\nwprevnextdefs{\relax}{NWppJ6t-A2bag-2}\nwenddeflinemarkup
{\bf{}const} {\bf{}ex} {\it{}evaluation\_assist} = {\bf{}lst}{\nwlbrace}{\it{}power}({\it{}cos}({\it{}wild}(0)),2)\begin{math}\equiv\end{math}1-{\it{}power}({\it{}sin}({\it{}wild}(0)),2),\nwindexdefn{\nwixident{ex}}{ex}{NWppJ6t-A2bag-1}
        {\it{}power}({\it{}cosh}({\it{}wild}(1)),2)\begin{math}\equiv\end{math}1+{\it{}power}({\it{}sinh}({\it{}wild}(1)),2){\nwrbrace};
\nwindexdefn{\nwixident{evaluation{\_}assist}}{evaluation:unassist}{NWppJ6t-A2bag-1}\eatline
\nwalsodefined{\\{NWppJ6t-A2bag-2}\\{NWppJ6t-A2bag-3}\\{NWppJ6t-A2bag-4}\\{NWppJ6t-A2bag-5}}\nwused{\\{NWppJ6t-32jW6L-1}}\nwidentdefs{\\{{\nwixident{evaluation{\_}assist}}{evaluation:unassist}}\\{{\nwixident{ex}}{ex}}}\nwendcode{}\nwbegindocs{475}\nwdocspar
\nwenddocs{}\nwbegindocs{476}\nwdocspar
\section{Figure Library Header File}
\label{sec:figure-header-file}

\nwenddocs{}\nwbegindocs{477}Here is the header file of the library. Initially, an end-user does not need to
know its structure much beyond the material presented in
Sections~\ref{sec:publ-meth-figure}--\ref{sec:publ-meth-cycl} and
illustrated in Section~\ref{sec:examples}. Here is some further topics
which can be of interest as well:
\begin{itemize}
\item An intermediate end-user may wish to define his own
  {\Tt{}\Rm{}{\bf{}subfigure}\nwendquote}s, see {\Tt{}\Rm{}{\it{}midpoint\_constructor}\nwendquote} for a sample and
  Subsect.~\ref{sec:subf-class-decl}.
\item Furthermore, an advanced end-user may wish to define some
  additional {\Tt{}\Rm{}{\bf{}cycle\_relation}\nwendquote} to supplement already presented in
  Section~\ref{sec:publ-meth-cycl}, in this case only knowledge of
  {\Tt{}\Rm{}{\bf{}cycle\_relation}\nwendquote} class is required, see
  Subsect.~\ref{sec:cycl-real-class-decl}.
\item To adjust automatically created \Asymptote\ graphics user may
  want to adjust the default styles, see Subsect.~\ref{sec:asympt-cust}.
\end{itemize}

\nwenddocs{}\nwbegindocs{478}\nwdocspar
\nwenddocs{}\nwbegincode{479}\sublabel{NWppJ6t-3NCnUp-1}\nwmargintag{{\nwtagstyle{}\subpageref{NWppJ6t-3NCnUp-1}}}\moddef{figure.h~{\nwtagstyle{}\subpageref{NWppJ6t-3NCnUp-1}}}\endmoddef\Rm{}\nwstartdeflinemarkup\nwprevnextdefs{\relax}{NWppJ6t-3NCnUp-2}\nwenddeflinemarkup
\LA{}license~{\nwtagstyle{}\subpageref{NWppJ6t-ZXuKx-1}}\RA{}
{\bf{}\char35{}ifndef}{\tt{} \_\_\_\_figure\_\_}
{\bf{}\char35{}define}{\tt{} \_\_\_\_figure\_\_}\nwindexdefn{\nwixident{{\_}{\_}{\_}{\_}figure{\_}{\_}}}{:un:un:un:unfigure:un:un}{NWppJ6t-3NCnUp-1}

\nwalsodefined{\\{NWppJ6t-3NCnUp-2}\\{NWppJ6t-3NCnUp-3}}\nwnotused{figure.h}\nwidentdefs{\\{{\nwixident{{\_}{\_}{\_}{\_}figure{\_}{\_}}}{:un:un:un:unfigure:un:un}}}\nwendcode{}\nwbegindocs{480}Some libraries we are using.
\nwenddocs{}\nwbegincode{481}\sublabel{NWppJ6t-3NCnUp-2}\nwmargintag{{\nwtagstyle{}\subpageref{NWppJ6t-3NCnUp-2}}}\moddef{figure.h~{\nwtagstyle{}\subpageref{NWppJ6t-3NCnUp-1}}}\plusendmoddef\Rm{}\nwstartdeflinemarkup\nwprevnextdefs{NWppJ6t-3NCnUp-1}{NWppJ6t-3NCnUp-3}\nwenddeflinemarkup
{\bf{}\char35{}include}{\tt{} \begin{math}<\end{math}iostream\begin{math}>\end{math}}
{\bf{}\char35{}include}{\tt{} \begin{math}<\end{math}cstdlib\begin{math}>\end{math}}
{\bf{}\char35{}include}{\tt{} \begin{math}<\end{math}cstdio\begin{math}>\end{math}}
{\bf{}\char35{}include}{\tt{} \begin{math}<\end{math}fstream\begin{math}>\end{math}}
{\bf{}\char35{}include}{\tt{} \begin{math}<\end{math}regex\begin{math}>\end{math}}
{\bf{}\char35{}include}{\tt{} "cycle.h"}

{\bf{}namespace} {\it{}MoebInv} {\nwlbrace}
{\bf{}using} {\bf{}namespace} {\it{}std};
{\bf{}using} {\bf{}namespace} {\it{}GiNaC};
\nwindexdefn{\nwixident{MoebInv}}{MoebInv}{NWppJ6t-3NCnUp-2}\eatline
\nwidentdefs{\\{{\nwixident{MoebInv}}{MoebInv}}}\nwendcode{}\nwbegindocs{482}\nwdocspar
\nwenddocs{}\nwbegindocs{483}The overview of the header file.
\nwenddocs{}\nwbegincode{484}\sublabel{NWppJ6t-3NCnUp-3}\nwmargintag{{\nwtagstyle{}\subpageref{NWppJ6t-3NCnUp-3}}}\moddef{figure.h~{\nwtagstyle{}\subpageref{NWppJ6t-3NCnUp-1}}}\plusendmoddef\Rm{}\nwstartdeflinemarkup\nwprevnextdefs{NWppJ6t-3NCnUp-2}{\relax}\nwenddeflinemarkup
\LA{}figure define~{\nwtagstyle{}\subpageref{NWppJ6t-3ONz84-1}}\RA{}
\LA{}defining types~{\nwtagstyle{}\subpageref{NWppJ6t-1s4LDF-1}}\RA{}
\LA{}cycle data header~{\nwtagstyle{}\subpageref{NWppJ6t-8XFRe-1}}\RA{}
\LA{}cycle node header~{\nwtagstyle{}\subpageref{NWppJ6t-2uztPH-1}}\RA{}
\LA{}cycle relations~{\nwtagstyle{}\subpageref{NWppJ6t-1d2GZW-1}}\RA{}
\LA{}asy styles~{\nwtagstyle{}\subpageref{NWppJ6t-43RSeU-1}}\RA{}
\LA{}figure header~{\nwtagstyle{}\subpageref{NWppJ6t-7vmg0-1}}\RA{}
\LA{}subfigure header~{\nwtagstyle{}\subpageref{NWppJ6t-4UgdbB-1}}\RA{}
\LA{}additional functions header~{\nwtagstyle{}\subpageref{NWppJ6t-1lh2X2-1}}\RA{}
{\nwrbrace} // namespace MoebInv
{\bf{}\char35{}endif}{\tt{} /* defined(\_\_\_\_figure\_\_) */}

\nwidentuses{\\{{\nwixident{{\_}{\_}{\_}{\_}figure{\_}{\_}}}{:un:un:un:unfigure:un:un}}\\{{\nwixident{MoebInv}}{MoebInv}}}\nwindexuse{\nwixident{{\_}{\_}{\_}{\_}figure{\_}{\_}}}{:un:un:un:unfigure:un:un}{NWppJ6t-3NCnUp-3}\nwindexuse{\nwixident{MoebInv}}{MoebInv}{NWppJ6t-3NCnUp-3}\nwendcode{}\nwbegindocs{485}We use negative numbered generations to save the reference objects.
\nwenddocs{}\nwbegincode{486}\sublabel{NWppJ6t-3ONz84-1}\nwmargintag{{\nwtagstyle{}\subpageref{NWppJ6t-3ONz84-1}}}\moddef{figure define~{\nwtagstyle{}\subpageref{NWppJ6t-3ONz84-1}}}\endmoddef\Rm{}\nwstartdeflinemarkup\nwusesondefline{\\{NWppJ6t-3NCnUp-3}}\nwenddeflinemarkup
{\bf{}\char35{}define}{\tt{} REAL\_LINE\_GEN -1}\nwindexdefn{\nwixident{REAL{\_}LINE{\_}GEN}}{REAL:unLINE:unGEN}{NWppJ6t-3ONz84-1}
{\bf{}\char35{}define}{\tt{} INFINITY\_GEN -2}\nwindexdefn{\nwixident{INFINITY{\_}GEN}}{INFINITY:unGEN}{NWppJ6t-3ONz84-1}
{\bf{}\char35{}define}{\tt{} GHOST\_GEN -3}\nwindexdefn{\nwixident{GHOST{\_}GEN}}{GHOST:unGEN}{NWppJ6t-3ONz84-1}
\nwindexdefn{\nwixident{REAL{\_}LINE{\_}GEN}}{REAL:unLINE:unGEN}{NWppJ6t-3ONz84-1}\nwindexdefn{\nwixident{INFINITY{\_}GEN}}{INFINITY:unGEN}{NWppJ6t-3ONz84-1}\nwindexdefn{\nwixident{GHOST{\_}GEN}}{GHOST:unGEN}{NWppJ6t-3ONz84-1}\eatline
\nwused{\\{NWppJ6t-3NCnUp-3}}\nwidentdefs{\\{{\nwixident{GHOST{\_}GEN}}{GHOST:unGEN}}\\{{\nwixident{INFINITY{\_}GEN}}{INFINITY:unGEN}}\\{{\nwixident{REAL{\_}LINE{\_}GEN}}{REAL:unLINE:unGEN}}}\nwendcode{}\nwbegindocs{487}\nwdocspar
\nwenddocs{}\nwbegindocs{488}\nwdocspar
\subsection{{\Tt{}\Rm{}{\bf{}cycle\_data}\nwendquote} class declaration}
\label{sec:cycl-class-decl}

\nwenddocs{}\nwbegindocs{489}The class to store explicit data of an individual {\Tt{}\Rm{}{\bf{}cycle}\nwendquote}. An
end-user does not need normally to know about it.
\nwenddocs{}\nwbegincode{490}\sublabel{NWppJ6t-8XFRe-1}\nwmargintag{{\nwtagstyle{}\subpageref{NWppJ6t-8XFRe-1}}}\moddef{cycle data header~{\nwtagstyle{}\subpageref{NWppJ6t-8XFRe-1}}}\endmoddef\Rm{}\nwstartdeflinemarkup\nwusesondefline{\\{NWppJ6t-3NCnUp-3}}\nwprevnextdefs{\relax}{NWppJ6t-8XFRe-2}\nwenddeflinemarkup
{\bf{}class} {\bf{}cycle\_data} : {\bf{}public} {\bf{}basic}
{\nwlbrace}
{\it{}GINAC\_DECLARE\_REGISTERED\_CLASS}({\bf{}cycle\_data}, {\bf{}basic})
\nwindexdefn{\nwixident{cycle{\_}data}}{cycle:undata}{NWppJ6t-8XFRe-1}\eatline
\nwalsodefined{\\{NWppJ6t-8XFRe-2}\\{NWppJ6t-8XFRe-3}}\nwused{\\{NWppJ6t-3NCnUp-3}}\nwidentdefs{\\{{\nwixident{cycle{\_}data}}{cycle:undata}}}\nwendcode{}\nwbegindocs{491}\nwdocspar
\nwenddocs{}\nwbegindocs{492}The parameters of the stored {\Tt{}\Rm{}{\bf{}cycle}\nwendquote}.
\nwenddocs{}\nwbegincode{493}\sublabel{NWppJ6t-8XFRe-2}\nwmargintag{{\nwtagstyle{}\subpageref{NWppJ6t-8XFRe-2}}}\moddef{cycle data header~{\nwtagstyle{}\subpageref{NWppJ6t-8XFRe-1}}}\plusendmoddef\Rm{}\nwstartdeflinemarkup\nwusesondefline{\\{NWppJ6t-3NCnUp-3}}\nwprevnextdefs{NWppJ6t-8XFRe-1}{NWppJ6t-8XFRe-3}\nwenddeflinemarkup
{\bf{}protected}:
    {\bf{}ex} {\it{}k},
    {\it{}l},
    {\it{}m};

\nwused{\\{NWppJ6t-3NCnUp-3}}\nwidentuses{\\{{\nwixident{ex}}{ex}}\\{{\nwixident{k}}{k}}\\{{\nwixident{l}}{l}}\\{{\nwixident{m}}{m}}}\nwindexuse{\nwixident{ex}}{ex}{NWppJ6t-8XFRe-2}\nwindexuse{\nwixident{k}}{k}{NWppJ6t-8XFRe-2}\nwindexuse{\nwixident{l}}{l}{NWppJ6t-8XFRe-2}\nwindexuse{\nwixident{m}}{m}{NWppJ6t-8XFRe-2}\nwendcode{}\nwbegindocs{494}Public methods in the class. However, an end-user does not normally
need them.
\nwenddocs{}\nwbegincode{495}\sublabel{NWppJ6t-8XFRe-3}\nwmargintag{{\nwtagstyle{}\subpageref{NWppJ6t-8XFRe-3}}}\moddef{cycle data header~{\nwtagstyle{}\subpageref{NWppJ6t-8XFRe-1}}}\plusendmoddef\Rm{}\nwstartdeflinemarkup\nwusesondefline{\\{NWppJ6t-3NCnUp-3}}\nwprevnextdefs{NWppJ6t-8XFRe-2}{\relax}\nwenddeflinemarkup
{\bf{}public}:
    {\bf{}cycle\_data}({\bf{}const} {\bf{}ex} & {\it{}C});
    {\bf{}cycle\_data}({\bf{}const} {\bf{}ex} & {\it{}k1}, {\bf{}const} {\bf{}ex} {\it{}l1}, {\bf{}const} {\bf{}ex} &{\it{}m1}, {\bf{}bool} {\it{}normalize}={\bf{}false});
    {\bf{}ex} {\it{}get\_cycle}({\bf{}const} {\bf{}ex} & {\it{}metr}) {\bf{}const};
    {\bf{}inline} {\it{}size\_t} {\it{}nops}() {\bf{}const} {\nwlbrace} {\bf{}return} 3; {\nwrbrace}
    {\bf{}ex} {\it{}op}({\it{}size\_t} {\it{}i}) {\bf{}const};
    {\bf{}ex} & {\it{}let\_op}({\it{}size\_t} {\it{}i});
    {\bf{}inline} {\bf{}ex} {\it{}get\_k}() {\bf{}const} {\nwlbrace} {\bf{}return} {\it{}k}; {\nwrbrace}
    {\bf{}inline} {\bf{}ex} {\it{}get\_l}() {\bf{}const} {\nwlbrace} {\bf{}return} {\it{}l}; {\nwrbrace}
    {\bf{}inline} {\bf{}ex} {\it{}get\_l}({\it{}size\_t} {\it{}i}) {\bf{}const} {\nwlbrace} {\bf{}return} {\it{}l}.{\it{}op}(0).{\it{}op}({\it{}i}); {\nwrbrace}
    {\bf{}inline} {\bf{}ex} {\it{}get\_m}() {\bf{}const} {\nwlbrace} {\bf{}return} {\it{}m};{\nwrbrace}
    {\bf{}inline} {\bf{}long} {\bf{}unsigned} {\bf{}int} {\it{}get\_dim}() {\bf{}const} {\nwlbrace} {\bf{}return} {\it{}l}.{\it{}op}(0).{\it{}nops}(); {\nwrbrace}
    {\bf{}void} {\it{}do\_print}({\bf{}const} {\it{}print\_dflt} & {\it{}con}, {\bf{}unsigned} {\it{}level}) {\bf{}const};
    {\bf{}void} {\it{}do\_print\_double}({\bf{}const} {\it{}print\_dflt} & {\it{}con}, {\bf{}unsigned} {\it{}level}) {\bf{}const};
    {\bf{}void} {\it{}archive}({\it{}archive\_node} &{\it{}n}) {\bf{}const};
    {\bf{}inline} {\bf{}ex} {\it{}normalize}() {\bf{}const} {\nwlbrace}{\bf{}return} {\bf{}cycle\_data}({\it{}k},{\it{}l},{\it{}m},{\bf{}true});{\nwrbrace}
    {\bf{}ex} {\it{}num\_normalize}() {\bf{}const};
    {\bf{}void} {\it{}read\_archive}({\bf{}const} {\it{}archive\_node} &{\it{}n}, {\bf{}lst} &{\it{}sym\_lst});
    {\bf{}bool} {\it{}is\_equal}({\bf{}const} {\bf{}basic} & {\it{}other}, {\bf{}bool} {\it{}projectively}) {\bf{}const};
    {\bf{}bool} {\it{}is\_almost\_equal}({\bf{}const} {\bf{}basic} & {\it{}other}, {\bf{}bool} {\it{}projectively}) {\bf{}const};
    {\bf{}cycle\_data} {\it{}subs}({\bf{}const} {\bf{}ex} & {\it{}e}, {\bf{}unsigned} {\it{}options}=0) {\bf{}const};
    {\bf{}ex} {\it{}subs}({\bf{}const} {\it{}exmap} & {\it{}em}, {\bf{}unsigned} {\it{}options}=0) {\bf{}const};
    {\bf{}inline} {\bf{}bool} {\it{}has}({\bf{}const} {\bf{}ex} & {\it{}x}) {\bf{}const} {\nwlbrace} {\bf{}return} ({\it{}k}.{\it{}has}({\it{}x}) \begin{math}\vee\end{math}  {\it{}l}.{\it{}has}({\it{}x}) \begin{math}\vee\end{math}  {\it{}m}.{\it{}has}({\it{}x})); {\nwrbrace}

{\bf{}protected}:
    {\it{}return\_type\_t} {\it{}return\_type\_tinfo}() {\bf{}const};
{\nwrbrace};
{\it{}GINAC\_DECLARE\_UNARCHIVER}({\bf{}cycle\_data});\nwindexdefn{\nwixident{cycle{\_}data}}{cycle:undata}{NWppJ6t-8XFRe-3}

\nwused{\\{NWppJ6t-3NCnUp-3}}\nwidentdefs{\\{{\nwixident{cycle{\_}data}}{cycle:undata}}}\nwidentuses{\\{{\nwixident{archive}}{archive}}\\{{\nwixident{do{\_}print{\_}double}}{do:unprint:undouble}}\\{{\nwixident{ex}}{ex}}\\{{\nwixident{get{\_}cycle}}{get:uncycle}}\\{{\nwixident{get{\_}dim()}}{get:undim()}}\\{{\nwixident{is{\_}almost{\_}equal}}{is:unalmost:unequal}}\\{{\nwixident{k}}{k}}\\{{\nwixident{l}}{l}}\\{{\nwixident{m}}{m}}\\{{\nwixident{nops}}{nops}}\\{{\nwixident{op}}{op}}\\{{\nwixident{read{\_}archive}}{read:unarchive}}\\{{\nwixident{subs}}{subs}}}\nwindexuse{\nwixident{archive}}{archive}{NWppJ6t-8XFRe-3}\nwindexuse{\nwixident{do{\_}print{\_}double}}{do:unprint:undouble}{NWppJ6t-8XFRe-3}\nwindexuse{\nwixident{ex}}{ex}{NWppJ6t-8XFRe-3}\nwindexuse{\nwixident{get{\_}cycle}}{get:uncycle}{NWppJ6t-8XFRe-3}\nwindexuse{\nwixident{get{\_}dim()}}{get:undim()}{NWppJ6t-8XFRe-3}\nwindexuse{\nwixident{is{\_}almost{\_}equal}}{is:unalmost:unequal}{NWppJ6t-8XFRe-3}\nwindexuse{\nwixident{k}}{k}{NWppJ6t-8XFRe-3}\nwindexuse{\nwixident{l}}{l}{NWppJ6t-8XFRe-3}\nwindexuse{\nwixident{m}}{m}{NWppJ6t-8XFRe-3}\nwindexuse{\nwixident{nops}}{nops}{NWppJ6t-8XFRe-3}\nwindexuse{\nwixident{op}}{op}{NWppJ6t-8XFRe-3}\nwindexuse{\nwixident{read{\_}archive}}{read:unarchive}{NWppJ6t-8XFRe-3}\nwindexuse{\nwixident{subs}}{subs}{NWppJ6t-8XFRe-3}\nwendcode{}\nwbegindocs{496}\nwdocspar
\subsection{{\Tt{}\Rm{}{\bf{}cycle\_node}\nwendquote} class declaration}
\label{sec:cycl-node-class-decl}

\nwenddocs{}\nwbegindocs{497}Forward declaration.
\nwenddocs{}\nwbegincode{498}\sublabel{NWppJ6t-2uztPH-1}\nwmargintag{{\nwtagstyle{}\subpageref{NWppJ6t-2uztPH-1}}}\moddef{cycle node header~{\nwtagstyle{}\subpageref{NWppJ6t-2uztPH-1}}}\endmoddef\Rm{}\nwstartdeflinemarkup\nwusesondefline{\\{NWppJ6t-3NCnUp-3}}\nwprevnextdefs{\relax}{NWppJ6t-2uztPH-2}\nwenddeflinemarkup
{\bf{}class} {\bf{}cycle\_relation};

\nwalsodefined{\\{NWppJ6t-2uztPH-2}\\{NWppJ6t-2uztPH-3}\\{NWppJ6t-2uztPH-4}\\{NWppJ6t-2uztPH-5}\\{NWppJ6t-2uztPH-6}\\{NWppJ6t-2uztPH-7}\\{NWppJ6t-2uztPH-8}\\{NWppJ6t-2uztPH-9}\\{NWppJ6t-2uztPH-A}\\{NWppJ6t-2uztPH-B}\\{NWppJ6t-2uztPH-C}\\{NWppJ6t-2uztPH-D}\\{NWppJ6t-2uztPH-E}\\{NWppJ6t-2uztPH-F}\\{NWppJ6t-2uztPH-G}}\nwused{\\{NWppJ6t-3NCnUp-3}}\nwidentuses{\\{{\nwixident{cycle{\_}relation}}{cycle:unrelation}}}\nwindexuse{\nwixident{cycle{\_}relation}}{cycle:unrelation}{NWppJ6t-2uztPH-1}\nwendcode{}\nwbegindocs{499}The class to store nodes containing data of particular cycles and
relations between nodes. An end-user does not need normally to know
about it.
\nwenddocs{}\nwbegincode{500}\sublabel{NWppJ6t-2uztPH-2}\nwmargintag{{\nwtagstyle{}\subpageref{NWppJ6t-2uztPH-2}}}\moddef{cycle node header~{\nwtagstyle{}\subpageref{NWppJ6t-2uztPH-1}}}\plusendmoddef\Rm{}\nwstartdeflinemarkup\nwusesondefline{\\{NWppJ6t-3NCnUp-3}}\nwprevnextdefs{NWppJ6t-2uztPH-1}{NWppJ6t-2uztPH-3}\nwenddeflinemarkup
{\bf{}class} {\bf{}cycle\_node} : {\bf{}public} {\bf{}basic}
{\nwlbrace}
{\it{}GINAC\_DECLARE\_REGISTERED\_CLASS}({\bf{}cycle\_node}, {\bf{}basic})
\nwindexdefn{\nwixident{cycle{\_}node}}{cycle:unnode}{NWppJ6t-2uztPH-2}\eatline
\nwused{\\{NWppJ6t-3NCnUp-3}}\nwidentdefs{\\{{\nwixident{cycle{\_}node}}{cycle:unnode}}}\nwendcode{}\nwbegindocs{501}\nwdocspar
\nwenddocs{}\nwbegindocs{502}Members of the class.
\nwenddocs{}\nwbegincode{503}\sublabel{NWppJ6t-2uztPH-3}\nwmargintag{{\nwtagstyle{}\subpageref{NWppJ6t-2uztPH-3}}}\moddef{cycle node header~{\nwtagstyle{}\subpageref{NWppJ6t-2uztPH-1}}}\plusendmoddef\Rm{}\nwstartdeflinemarkup\nwusesondefline{\\{NWppJ6t-3NCnUp-3}}\nwprevnextdefs{NWppJ6t-2uztPH-2}{NWppJ6t-2uztPH-4}\nwenddeflinemarkup
{\bf{}protected}:
    {\bf{}lst} {\it{}cycles}; // List of cycle data entries
    {\bf{}int} {\it{}generation};
    {\bf{}lst} {\it{}children}; // List of keys to cycle\_nodes
    {\bf{}lst} {\it{}parents}; // List of cycle\_relations or a list containing a single subfigure
    {\it{}string} {\it{}custom\_asy}; // Custom string for Asymptote

\nwused{\\{NWppJ6t-3NCnUp-3}}\nwidentuses{\\{{\nwixident{subfigure}}{subfigure}}}\nwindexuse{\nwixident{subfigure}}{subfigure}{NWppJ6t-2uztPH-3}\nwendcode{}\nwbegindocs{504}Constructors in the class.
\nwenddocs{}\nwbegincode{505}\sublabel{NWppJ6t-2uztPH-4}\nwmargintag{{\nwtagstyle{}\subpageref{NWppJ6t-2uztPH-4}}}\moddef{cycle node header~{\nwtagstyle{}\subpageref{NWppJ6t-2uztPH-1}}}\plusendmoddef\Rm{}\nwstartdeflinemarkup\nwusesondefline{\\{NWppJ6t-3NCnUp-3}}\nwprevnextdefs{NWppJ6t-2uztPH-3}{NWppJ6t-2uztPH-5}\nwenddeflinemarkup
{\bf{}public}:
    {\bf{}cycle\_node}({\bf{}const} {\bf{}ex} & {\it{}C}, {\bf{}int} {\it{}g}=0);
    {\bf{}cycle\_node}({\bf{}const} {\bf{}ex} & {\it{}C}, {\bf{}int} {\it{}g}, {\bf{}const} {\bf{}lst} & {\it{}par});
    {\bf{}cycle\_node}({\bf{}const} {\bf{}ex} & {\it{}C}, {\bf{}int} {\it{}g}, {\bf{}const} {\bf{}lst} & {\it{}par}, {\bf{}const} {\bf{}lst} & {\it{}chil});
    {\bf{}cycle\_node}({\bf{}const} {\bf{}ex} & {\it{}C}, {\bf{}int} {\it{}g}, {\bf{}const} {\bf{}lst} & {\it{}par}, {\bf{}const} {\bf{}lst} & {\it{}chil}, {\it{}string} {\it{}ca});
    {\bf{}cycle\_node} {\it{}subs}({\bf{}const} {\bf{}ex} & {\it{}e}, {\bf{}unsigned} {\it{}options}=0) {\bf{}const};
    {\bf{}void} {\it{}do\_print\_double}({\bf{}const} {\it{}print\_dflt} & {\it{}con}, {\bf{}unsigned} {\it{}level}) {\bf{}const};
    {\bf{}ex} {\it{}subs}({\bf{}const} {\it{}exmap} & {\it{}m}, {\bf{}unsigned} {\it{}options}=0) {\bf{}const};

\nwused{\\{NWppJ6t-3NCnUp-3}}\nwidentuses{\\{{\nwixident{cycle{\_}node}}{cycle:unnode}}\\{{\nwixident{do{\_}print{\_}double}}{do:unprint:undouble}}\\{{\nwixident{ex}}{ex}}\\{{\nwixident{m}}{m}}\\{{\nwixident{subs}}{subs}}}\nwindexuse{\nwixident{cycle{\_}node}}{cycle:unnode}{NWppJ6t-2uztPH-4}\nwindexuse{\nwixident{do{\_}print{\_}double}}{do:unprint:undouble}{NWppJ6t-2uztPH-4}\nwindexuse{\nwixident{ex}}{ex}{NWppJ6t-2uztPH-4}\nwindexuse{\nwixident{m}}{m}{NWppJ6t-2uztPH-4}\nwindexuse{\nwixident{subs}}{subs}{NWppJ6t-2uztPH-4}\nwendcode{}\nwbegindocs{506}Add a chid {\Tt{}\Rm{}{\bf{}cycle\_node}\nwendquote} to the {\Tt{}\Rm{}{\bf{}cycle\_node}\nwendquote}.
\nwenddocs{}\nwbegincode{507}\sublabel{NWppJ6t-2uztPH-5}\nwmargintag{{\nwtagstyle{}\subpageref{NWppJ6t-2uztPH-5}}}\moddef{cycle node header~{\nwtagstyle{}\subpageref{NWppJ6t-2uztPH-1}}}\plusendmoddef\Rm{}\nwstartdeflinemarkup\nwusesondefline{\\{NWppJ6t-3NCnUp-3}}\nwprevnextdefs{NWppJ6t-2uztPH-4}{NWppJ6t-2uztPH-6}\nwenddeflinemarkup
{\bf{}protected}:
    {\bf{}inline} {\bf{}void} {\it{}add\_child}({\bf{}const} {\bf{}ex} & {\it{}c}) {\nwlbrace}{\it{}children}.{\it{}append}({\it{}c});{\nwrbrace}

\nwused{\\{NWppJ6t-3NCnUp-3}}\nwidentuses{\\{{\nwixident{ex}}{ex}}}\nwindexuse{\nwixident{ex}}{ex}{NWppJ6t-2uztPH-5}\nwendcode{}\nwbegindocs{508}Access {\Tt{}\Rm{}{\bf{}cycle}\nwendquote} parameters.
\nwenddocs{}\nwbegincode{509}\sublabel{NWppJ6t-2uztPH-6}\nwmargintag{{\nwtagstyle{}\subpageref{NWppJ6t-2uztPH-6}}}\moddef{cycle node header~{\nwtagstyle{}\subpageref{NWppJ6t-2uztPH-1}}}\plusendmoddef\Rm{}\nwstartdeflinemarkup\nwusesondefline{\\{NWppJ6t-3NCnUp-3}}\nwprevnextdefs{NWppJ6t-2uztPH-5}{NWppJ6t-2uztPH-7}\nwenddeflinemarkup
    {\bf{}inline} {\bf{}ex} {\it{}get\_cycles}() {\bf{}const} {\nwlbrace}{\bf{}return} {\it{}cycles};{\nwrbrace}

\nwused{\\{NWppJ6t-3NCnUp-3}}\nwidentuses{\\{{\nwixident{ex}}{ex}}}\nwindexuse{\nwixident{ex}}{ex}{NWppJ6t-2uztPH-6}\nwendcode{}\nwbegindocs{510}Return the {\Tt{}\Rm{}{\bf{}cycle}\nwendquote} object for every {\Tt{}\Rm{}{\bf{}cycle\_data}\nwendquote} stored in {\Tt{}\Rm{}{\it{}cycles}\nwendquote}.
\nwenddocs{}\nwbegincode{511}\sublabel{NWppJ6t-2uztPH-7}\nwmargintag{{\nwtagstyle{}\subpageref{NWppJ6t-2uztPH-7}}}\moddef{cycle node header~{\nwtagstyle{}\subpageref{NWppJ6t-2uztPH-1}}}\plusendmoddef\Rm{}\nwstartdeflinemarkup\nwusesondefline{\\{NWppJ6t-3NCnUp-3}}\nwprevnextdefs{NWppJ6t-2uztPH-6}{NWppJ6t-2uztPH-8}\nwenddeflinemarkup
    {\bf{}ex} {\it{}get\_cycle}({\bf{}const} {\bf{}ex} & {\it{}metr}) {\bf{}const};
    {\bf{}inline} {\bf{}ex} {\it{}get\_cycle\_data}({\bf{}int} {\it{}i}) {\bf{}const} {\nwlbrace}{\bf{}return} {\it{}cycles}.{\it{}op}({\it{}i});{\nwrbrace}

\nwused{\\{NWppJ6t-3NCnUp-3}}\nwidentuses{\\{{\nwixident{ex}}{ex}}\\{{\nwixident{get{\_}cycle}}{get:uncycle}}\\{{\nwixident{op}}{op}}}\nwindexuse{\nwixident{ex}}{ex}{NWppJ6t-2uztPH-7}\nwindexuse{\nwixident{get{\_}cycle}}{get:uncycle}{NWppJ6t-2uztPH-7}\nwindexuse{\nwixident{op}}{op}{NWppJ6t-2uztPH-7}\nwendcode{}\nwbegindocs{512}Return the generation number.
\nwenddocs{}\nwbegincode{513}\sublabel{NWppJ6t-2uztPH-8}\nwmargintag{{\nwtagstyle{}\subpageref{NWppJ6t-2uztPH-8}}}\moddef{cycle node header~{\nwtagstyle{}\subpageref{NWppJ6t-2uztPH-1}}}\plusendmoddef\Rm{}\nwstartdeflinemarkup\nwusesondefline{\\{NWppJ6t-3NCnUp-3}}\nwprevnextdefs{NWppJ6t-2uztPH-7}{NWppJ6t-2uztPH-9}\nwenddeflinemarkup
    {\bf{}inline} {\bf{}int} {\it{}get\_generation}() {\bf{}const} {\nwlbrace}{\bf{}return} {\it{}generation};{\nwrbrace}

\nwused{\\{NWppJ6t-3NCnUp-3}}\nwidentuses{\\{{\nwixident{get{\_}generation}}{get:ungeneration}}}\nwindexuse{\nwixident{get{\_}generation}}{get:ungeneration}{NWppJ6t-2uztPH-8}\nwendcode{}\nwbegindocs{514}Return the children list
\nwenddocs{}\nwbegincode{515}\sublabel{NWppJ6t-2uztPH-9}\nwmargintag{{\nwtagstyle{}\subpageref{NWppJ6t-2uztPH-9}}}\moddef{cycle node header~{\nwtagstyle{}\subpageref{NWppJ6t-2uztPH-1}}}\plusendmoddef\Rm{}\nwstartdeflinemarkup\nwusesondefline{\\{NWppJ6t-3NCnUp-3}}\nwprevnextdefs{NWppJ6t-2uztPH-8}{NWppJ6t-2uztPH-A}\nwenddeflinemarkup
    {\bf{}inline} {\bf{}lst} {\it{}get\_children}() {\bf{}const} {\nwlbrace}{\bf{}return} {\it{}children};{\nwrbrace}

\nwused{\\{NWppJ6t-3NCnUp-3}}\nwendcode{}\nwbegindocs{516}Replace the current {\Tt{}\Rm{}{\bf{}cycle}\nwendquote} with a new {\Tt{}\Rm{}{\bf{}cycle}\nwendquote}.
\nwenddocs{}\nwbegincode{517}\sublabel{NWppJ6t-2uztPH-A}\nwmargintag{{\nwtagstyle{}\subpageref{NWppJ6t-2uztPH-A}}}\moddef{cycle node header~{\nwtagstyle{}\subpageref{NWppJ6t-2uztPH-1}}}\plusendmoddef\Rm{}\nwstartdeflinemarkup\nwusesondefline{\\{NWppJ6t-3NCnUp-3}}\nwprevnextdefs{NWppJ6t-2uztPH-9}{NWppJ6t-2uztPH-B}\nwenddeflinemarkup
    {\bf{}void} {\it{}set\_cycles}({\bf{}const} {\bf{}ex} & {\it{}C});

\nwused{\\{NWppJ6t-3NCnUp-3}}\nwidentuses{\\{{\nwixident{ex}}{ex}}}\nwindexuse{\nwixident{ex}}{ex}{NWppJ6t-2uztPH-A}\nwendcode{}\nwbegindocs{518}Add one more {\Tt{}\Rm{}{\bf{}cycle}\nwendquote} instance to list of {\Tt{}\Rm{}{\it{}cycles}\nwendquote}.
\nwenddocs{}\nwbegincode{519}\sublabel{NWppJ6t-2uztPH-B}\nwmargintag{{\nwtagstyle{}\subpageref{NWppJ6t-2uztPH-B}}}\moddef{cycle node header~{\nwtagstyle{}\subpageref{NWppJ6t-2uztPH-1}}}\plusendmoddef\Rm{}\nwstartdeflinemarkup\nwusesondefline{\\{NWppJ6t-3NCnUp-3}}\nwprevnextdefs{NWppJ6t-2uztPH-A}{NWppJ6t-2uztPH-C}\nwenddeflinemarkup
    {\bf{}void} {\it{}append\_cycle}({\bf{}const} {\bf{}ex} & {\it{}C});
    {\bf{}void} {\it{}append\_cycle}({\bf{}const} {\bf{}ex} & {\it{}k}, {\bf{}const} {\bf{}ex} & {\it{}l}, {\bf{}const} {\bf{}ex} & {\it{}m});

\nwused{\\{NWppJ6t-3NCnUp-3}}\nwidentuses{\\{{\nwixident{ex}}{ex}}\\{{\nwixident{k}}{k}}\\{{\nwixident{l}}{l}}\\{{\nwixident{m}}{m}}}\nwindexuse{\nwixident{ex}}{ex}{NWppJ6t-2uztPH-B}\nwindexuse{\nwixident{k}}{k}{NWppJ6t-2uztPH-B}\nwindexuse{\nwixident{l}}{l}{NWppJ6t-2uztPH-B}\nwindexuse{\nwixident{m}}{m}{NWppJ6t-2uztPH-B}\nwendcode{}\nwbegindocs{520}Return the parent list.
\nwenddocs{}\nwbegincode{521}\sublabel{NWppJ6t-2uztPH-C}\nwmargintag{{\nwtagstyle{}\subpageref{NWppJ6t-2uztPH-C}}}\moddef{cycle node header~{\nwtagstyle{}\subpageref{NWppJ6t-2uztPH-1}}}\plusendmoddef\Rm{}\nwstartdeflinemarkup\nwusesondefline{\\{NWppJ6t-3NCnUp-3}}\nwprevnextdefs{NWppJ6t-2uztPH-B}{NWppJ6t-2uztPH-D}\nwenddeflinemarkup
    {\bf{}lst} {\it{}get\_parents}() {\bf{}const};

\nwused{\\{NWppJ6t-3NCnUp-3}}\nwendcode{}\nwbegindocs{522}The method returns the list of all keys to parant cycles.
\nwenddocs{}\nwbegincode{523}\sublabel{NWppJ6t-2uztPH-D}\nwmargintag{{\nwtagstyle{}\subpageref{NWppJ6t-2uztPH-D}}}\moddef{cycle node header~{\nwtagstyle{}\subpageref{NWppJ6t-2uztPH-1}}}\plusendmoddef\Rm{}\nwstartdeflinemarkup\nwusesondefline{\\{NWppJ6t-3NCnUp-3}}\nwprevnextdefs{NWppJ6t-2uztPH-C}{NWppJ6t-2uztPH-E}\nwenddeflinemarkup
    {\bf{}lst} {\it{}get\_parent\_keys}() {\bf{}const} ;

\nwused{\\{NWppJ6t-3NCnUp-3}}\nwendcode{}\nwbegindocs{524}Remove a child of the {\Tt{}\Rm{}{\bf{}cycle\_node}\nwendquote}.
\nwenddocs{}\nwbegincode{525}\sublabel{NWppJ6t-2uztPH-E}\nwmargintag{{\nwtagstyle{}\subpageref{NWppJ6t-2uztPH-E}}}\moddef{cycle node header~{\nwtagstyle{}\subpageref{NWppJ6t-2uztPH-1}}}\plusendmoddef\Rm{}\nwstartdeflinemarkup\nwusesondefline{\\{NWppJ6t-3NCnUp-3}}\nwprevnextdefs{NWppJ6t-2uztPH-D}{NWppJ6t-2uztPH-F}\nwenddeflinemarkup
    {\bf{}void} {\it{}remove\_child}({\bf{}const} {\bf{}ex} & {\it{}c});

\nwused{\\{NWppJ6t-3NCnUp-3}}\nwidentuses{\\{{\nwixident{ex}}{ex}}}\nwindexuse{\nwixident{ex}}{ex}{NWppJ6t-2uztPH-E}\nwendcode{}\nwbegindocs{526}Set or read \Asymptote\ option for this particular node.
\nwenddocs{}\nwbegincode{527}\sublabel{NWppJ6t-2uztPH-F}\nwmargintag{{\nwtagstyle{}\subpageref{NWppJ6t-2uztPH-F}}}\moddef{cycle node header~{\nwtagstyle{}\subpageref{NWppJ6t-2uztPH-1}}}\plusendmoddef\Rm{}\nwstartdeflinemarkup\nwusesondefline{\\{NWppJ6t-3NCnUp-3}}\nwprevnextdefs{NWppJ6t-2uztPH-E}{NWppJ6t-2uztPH-G}\nwenddeflinemarkup
    {\bf{}inline} {\bf{}void} {\it{}set\_asy\_opt}({\bf{}const} {\it{}string} {\it{}opt})  {\nwlbrace}{\it{}custom\_asy}={\it{}opt};{\nwrbrace}
    {\bf{}inline} {\it{}string} {\it{}get\_asy\_opt}() {\bf{}const} {\nwlbrace}{\bf{}return} {\it{}custom\_asy};{\nwrbrace}

\nwused{\\{NWppJ6t-3NCnUp-3}}\nwendcode{}\nwbegindocs{528}Service functions including printout the mathematical expression.
\nwenddocs{}\nwbegincode{529}\sublabel{NWppJ6t-2uztPH-G}\nwmargintag{{\nwtagstyle{}\subpageref{NWppJ6t-2uztPH-G}}}\moddef{cycle node header~{\nwtagstyle{}\subpageref{NWppJ6t-2uztPH-1}}}\plusendmoddef\Rm{}\nwstartdeflinemarkup\nwusesondefline{\\{NWppJ6t-3NCnUp-3}}\nwprevnextdefs{NWppJ6t-2uztPH-F}{\relax}\nwenddeflinemarkup
    {\bf{}inline} {\it{}size\_t} {\it{}nops}() {\bf{}const} {\nwlbrace} {\bf{}return} {\it{}cycles}.{\it{}nops}()+{\it{}children}.{\it{}nops}()+{\it{}parents}.{\it{}nops}(); {\nwrbrace}
    {\bf{}ex} {\it{}op}({\it{}size\_t} {\it{}i}) {\bf{}const};
    {\bf{}ex} & {\it{}let\_op}({\it{}size\_t} {\it{}i});
    {\bf{}void} {\it{}do\_print}({\bf{}const} {\it{}print\_dflt} & {\it{}con}, {\bf{}unsigned} {\it{}level}) {\bf{}const};
    {\bf{}void} {\it{}do\_print\_tree}({\bf{}const} {\it{}print\_tree} & {\it{}con}, {\bf{}unsigned} {\it{}level}) {\bf{}const};
{\bf{}protected}:
    {\it{}return\_type\_t} {\it{}return\_type\_tinfo}() {\bf{}const};
    {\bf{}void} {\it{}archive}({\it{}archive\_node} &{\it{}n}) {\bf{}const};
    {\bf{}void} {\it{}read\_archive}({\bf{}const} {\it{}archive\_node} &{\it{}n}, {\bf{}lst} &{\it{}sym\_lst});

{\bf{}friend} {\bf{}class} {\bf{}cycle\_relation};
{\bf{}friend} {\bf{}class} {\bf{}figure};
{\nwrbrace};
{\it{}GINAC\_DECLARE\_UNARCHIVER}({\bf{}cycle\_node});\nwindexdefn{\nwixident{cycle{\_}node}}{cycle:unnode}{NWppJ6t-2uztPH-G}

\nwused{\\{NWppJ6t-3NCnUp-3}}\nwidentdefs{\\{{\nwixident{cycle{\_}node}}{cycle:unnode}}}\nwidentuses{\\{{\nwixident{archive}}{archive}}\\{{\nwixident{cycle{\_}relation}}{cycle:unrelation}}\\{{\nwixident{ex}}{ex}}\\{{\nwixident{figure}}{figure}}\\{{\nwixident{nops}}{nops}}\\{{\nwixident{op}}{op}}\\{{\nwixident{read{\_}archive}}{read:unarchive}}}\nwindexuse{\nwixident{archive}}{archive}{NWppJ6t-2uztPH-G}\nwindexuse{\nwixident{cycle{\_}relation}}{cycle:unrelation}{NWppJ6t-2uztPH-G}\nwindexuse{\nwixident{ex}}{ex}{NWppJ6t-2uztPH-G}\nwindexuse{\nwixident{figure}}{figure}{NWppJ6t-2uztPH-G}\nwindexuse{\nwixident{nops}}{nops}{NWppJ6t-2uztPH-G}\nwindexuse{\nwixident{op}}{op}{NWppJ6t-2uztPH-G}\nwindexuse{\nwixident{read{\_}archive}}{read:unarchive}{NWppJ6t-2uztPH-G}\nwendcode{}\nwbegindocs{530}\nwdocspar
\subsection{{\Tt{}\Rm{}{\bf{}cycle\_relation}\nwendquote} class declaration}
\label{sec:cycl-real-class-decl}

\nwenddocs{}\nwbegindocs{531}First, we define a type to hold cycle relations. That is a pointer
to a functions with two arguments. See the definition of
{\Tt{}\Rm{}{\it{}cycle\_orthogonal}\nwendquote}, {\Tt{}\Rm{}{\it{}cycle\_different}\nwendquote}, for samples.
\nwenddocs{}\nwbegincode{532}\sublabel{NWppJ6t-1s4LDF-2}\nwmargintag{{\nwtagstyle{}\subpageref{NWppJ6t-1s4LDF-2}}}\moddef{defining types~{\nwtagstyle{}\subpageref{NWppJ6t-1s4LDF-1}}}\plusendmoddef\Rm{}\nwstartdeflinemarkup\nwusesondefline{\\{NWppJ6t-3NCnUp-3}}\nwprevnextdefs{NWppJ6t-1s4LDF-1}{\relax}\nwenddeflinemarkup
{\bf{}using} {\it{}PCR} = {\it{}std}::{\it{}function}\begin{math}<\end{math}{\bf{}ex}({\bf{}const} {\bf{}ex} &, {\bf{}const} {\bf{}ex} &, {\bf{}const} {\bf{}ex} &)\begin{math}>\end{math};
\nwindexdefn{\nwixident{PCR}}{PCR}{NWppJ6t-1s4LDF-2}\eatline
\nwused{\\{NWppJ6t-3NCnUp-3}}\nwidentdefs{\\{{\nwixident{PCR}}{PCR}}}\nwidentuses{\\{{\nwixident{ex}}{ex}}}\nwindexuse{\nwixident{ex}}{ex}{NWppJ6t-1s4LDF-2}\nwendcode{}\nwbegindocs{533}\nwdocspar
\nwenddocs{}\nwbegindocs{534}This class describes relations between {\Tt{}\Rm{}{\bf{}cycle\_node}\nwendquote}s. An
advanced end-user may want to add some new relations similar to
already provided in Section~\ref{sec:publ-meth-cycl}. Note however,
that archiving (saving) of user-defined relations cannot be done as
they contain pointers to functions which are not portable.

\nwenddocs{}\nwbegindocs{535}Memebrs of the class.
\nwenddocs{}\nwbegincode{536}\sublabel{NWppJ6t-1d2GZW-1}\nwmargintag{{\nwtagstyle{}\subpageref{NWppJ6t-1d2GZW-1}}}\moddef{cycle relations~{\nwtagstyle{}\subpageref{NWppJ6t-1d2GZW-1}}}\endmoddef\Rm{}\nwstartdeflinemarkup\nwusesondefline{\\{NWppJ6t-3NCnUp-3}}\nwprevnextdefs{\relax}{NWppJ6t-1d2GZW-2}\nwenddeflinemarkup
{\bf{}class} {\bf{}cycle\_relation} : {\bf{}public} {\bf{}basic}
{\nwlbrace}
    {\it{}GINAC\_DECLARE\_REGISTERED\_CLASS}({\bf{}cycle\_relation}, {\bf{}basic})
{\bf{}protected}:
    {\bf{}ex} {\it{}parkey}; // A key to a parent cycle\_node in figure
    {\it{}PCR} {\it{}rel}; // A pointer to function which produces the relation
    {\bf{}ex} {\it{}parameter}; // The value, which is supplied to rel() as the third parameter
    {\bf{}bool} {\it{}use\_cycle\_metric}; // If true uses the cycle space metric, otherwise the point space metric
\nwindexdefn{\nwixident{cycle{\_}relation}}{cycle:unrelation}{NWppJ6t-1d2GZW-1}\eatline
\nwalsodefined{\\{NWppJ6t-1d2GZW-2}\\{NWppJ6t-1d2GZW-3}\\{NWppJ6t-1d2GZW-4}\\{NWppJ6t-1d2GZW-5}\\{NWppJ6t-1d2GZW-6}\\{NWppJ6t-1d2GZW-7}\\{NWppJ6t-1d2GZW-8}\\{NWppJ6t-1d2GZW-9}\\{NWppJ6t-1d2GZW-A}\\{NWppJ6t-1d2GZW-B}}\nwused{\\{NWppJ6t-3NCnUp-3}}\nwidentdefs{\\{{\nwixident{cycle{\_}relation}}{cycle:unrelation}}}\nwidentuses{\\{{\nwixident{cycle{\_}node}}{cycle:unnode}}\\{{\nwixident{ex}}{ex}}\\{{\nwixident{figure}}{figure}}\\{{\nwixident{key}}{key}}\\{{\nwixident{PCR}}{PCR}}}\nwindexuse{\nwixident{cycle{\_}node}}{cycle:unnode}{NWppJ6t-1d2GZW-1}\nwindexuse{\nwixident{ex}}{ex}{NWppJ6t-1d2GZW-1}\nwindexuse{\nwixident{figure}}{figure}{NWppJ6t-1d2GZW-1}\nwindexuse{\nwixident{key}}{key}{NWppJ6t-1d2GZW-1}\nwindexuse{\nwixident{PCR}}{PCR}{NWppJ6t-1d2GZW-1}\nwendcode{}\nwbegindocs{537}\nwdocspar
\nwenddocs{}\nwbegindocs{538}Public methods in the class.
\nwenddocs{}\nwbegincode{539}\sublabel{NWppJ6t-1d2GZW-2}\nwmargintag{{\nwtagstyle{}\subpageref{NWppJ6t-1d2GZW-2}}}\moddef{cycle relations~{\nwtagstyle{}\subpageref{NWppJ6t-1d2GZW-1}}}\plusendmoddef\Rm{}\nwstartdeflinemarkup\nwusesondefline{\\{NWppJ6t-3NCnUp-3}}\nwprevnextdefs{NWppJ6t-1d2GZW-1}{NWppJ6t-1d2GZW-3}\nwenddeflinemarkup
{\bf{}public}:
    \LA{}public methods for cycle relation~{\nwtagstyle{}\subpageref{NWppJ6t-1HOJwC-1}}\RA{}
    {\bf{}inline} {\bf{}ex} {\it{}get\_parkey}() {\bf{}const} {\nwlbrace}{\bf{}return} {\it{}parkey};{\nwrbrace}
    {\bf{}inline} {\it{}PCR} {\it{}get\_PCR}() {\bf{}const} {\nwlbrace}{\bf{}return} {\it{}rel};{\nwrbrace}
    {\bf{}inline} {\bf{}ex} {\it{}get\_parameter}() {\bf{}const} {\nwlbrace}{\bf{}return} {\it{}parameter};{\nwrbrace}
    {\bf{}inline} {\bf{}bool} {\it{}cycle\_metric\_inuse}() {\bf{}const} {\nwlbrace}{\bf{}return} {\it{}use\_cycle\_metric};{\nwrbrace}
    {\bf{}inline} {\bf{}ex} {\it{}subs}({\bf{}const} {\it{}exmap} & {\it{}em}, {\bf{}unsigned} {\it{}options}=0) {\bf{}const}
    {\nwlbrace}{\bf{}return} {\bf{}cycle\_relation}({\it{}parkey}, {\it{}rel}, {\it{}use\_cycle\_metric}, {\it{}parameter}.{\it{}subs}({\it{}em},{\it{}options}));{\nwrbrace}

\nwused{\\{NWppJ6t-3NCnUp-3}}\nwidentuses{\\{{\nwixident{cycle{\_}relation}}{cycle:unrelation}}\\{{\nwixident{ex}}{ex}}\\{{\nwixident{PCR}}{PCR}}\\{{\nwixident{subs}}{subs}}}\nwindexuse{\nwixident{cycle{\_}relation}}{cycle:unrelation}{NWppJ6t-1d2GZW-2}\nwindexuse{\nwixident{ex}}{ex}{NWppJ6t-1d2GZW-2}\nwindexuse{\nwixident{PCR}}{PCR}{NWppJ6t-1d2GZW-2}\nwindexuse{\nwixident{subs}}{subs}{NWppJ6t-1d2GZW-2}\nwendcode{}\nwbegindocs{540}Protected methods in the class.
\nwenddocs{}\nwbegincode{541}\sublabel{NWppJ6t-1d2GZW-3}\nwmargintag{{\nwtagstyle{}\subpageref{NWppJ6t-1d2GZW-3}}}\moddef{cycle relations~{\nwtagstyle{}\subpageref{NWppJ6t-1d2GZW-1}}}\plusendmoddef\Rm{}\nwstartdeflinemarkup\nwusesondefline{\\{NWppJ6t-3NCnUp-3}}\nwprevnextdefs{NWppJ6t-1d2GZW-2}{NWppJ6t-1d2GZW-4}\nwenddeflinemarkup
{\bf{}protected}:
    {\bf{}ex} {\it{}rel\_to\_parent}({\bf{}const} {\bf{}cycle\_data} & {\it{}C1}, {\bf{}const} {\bf{}ex} & {\it{}pmetric}, {\bf{}const} {\bf{}ex} & {\it{}cmetric},
                     {\bf{}const} {\it{}exhashmap}\begin{math}<\end{math}{\bf{}cycle\_node}\begin{math}>\end{math} & {\it{}N}) {\bf{}const};
    {\it{}return\_type\_t} {\it{}return\_type\_tinfo}() {\bf{}const};
    {\bf{}void} {\it{}do\_print}({\bf{}const} {\it{}print\_dflt} & {\it{}con}, {\bf{}unsigned} {\it{}level}) {\bf{}const};
    {\bf{}void} {\it{}do\_print\_tree}({\bf{}const} {\it{}print\_tree} & {\it{}con}, {\bf{}unsigned} {\it{}level}) {\bf{}const};

\nwused{\\{NWppJ6t-3NCnUp-3}}\nwidentuses{\\{{\nwixident{cycle{\_}data}}{cycle:undata}}\\{{\nwixident{cycle{\_}node}}{cycle:unnode}}\\{{\nwixident{ex}}{ex}}}\nwindexuse{\nwixident{cycle{\_}data}}{cycle:undata}{NWppJ6t-1d2GZW-3}\nwindexuse{\nwixident{cycle{\_}node}}{cycle:unnode}{NWppJ6t-1d2GZW-3}\nwindexuse{\nwixident{ex}}{ex}{NWppJ6t-1d2GZW-3}\nwendcode{}\nwbegindocs{542} (un)Archiving of {\Tt{}\Rm{}{\bf{}cycle\_relation}\nwendquote} is not universal. At present it
only can handle relations declared in the header file {\Tt{}\Rm{}{\it{}cycle\_orthogonal}\nwendquote},
{\Tt{}\Rm{}{\it{}cycle\_f\_orthogonal}\nwendquote}, {\Tt{}\Rm{}{\it{}cycle\_adifferent}\nwendquote}, {\Tt{}\Rm{}{\it{}cycle\_different}\nwendquote},
{\Tt{}\Rm{}{\it{}cycle\_tangent}\nwendquote},  {\Tt{}\Rm{}{\it{}cycle\_power}\nwendquote} etc. from
Subsection~\ref{sec:publ-meth-cycl}.
\nwenddocs{}\nwbegincode{543}\sublabel{NWppJ6t-1d2GZW-4}\nwmargintag{{\nwtagstyle{}\subpageref{NWppJ6t-1d2GZW-4}}}\moddef{cycle relations~{\nwtagstyle{}\subpageref{NWppJ6t-1d2GZW-1}}}\plusendmoddef\Rm{}\nwstartdeflinemarkup\nwusesondefline{\\{NWppJ6t-3NCnUp-3}}\nwprevnextdefs{NWppJ6t-1d2GZW-3}{NWppJ6t-1d2GZW-5}\nwenddeflinemarkup
    {\bf{}void} {\it{}archive}({\it{}archive\_node} &{\it{}n}) {\bf{}const};
    {\bf{}void} {\it{}read\_archive}({\bf{}const} {\it{}archive\_node} &{\it{}n}, {\bf{}lst} &{\it{}sym\_lst});

    {\bf{}inline} {\it{}size\_t} {\it{}nops}() {\bf{}const} {\nwlbrace} {\bf{}return} 2; {\nwrbrace}
    {\bf{}ex} {\it{}op}({\it{}size\_t} {\it{}i}) {\bf{}const};
    {\bf{}ex} & {\it{}let\_op}({\it{}size\_t} {\it{}i});

{\bf{}friend} {\bf{}class} {\bf{}cycle\_node};
{\bf{}friend} {\bf{}class} {\bf{}figure};
{\nwrbrace};
{\it{}GINAC\_DECLARE\_UNARCHIVER}({\bf{}cycle\_relation});\nwindexdefn{\nwixident{cycle{\_}relation}}{cycle:unrelation}{NWppJ6t-1d2GZW-4}

\nwused{\\{NWppJ6t-3NCnUp-3}}\nwidentdefs{\\{{\nwixident{cycle{\_}relation}}{cycle:unrelation}}}\nwidentuses{\\{{\nwixident{archive}}{archive}}\\{{\nwixident{cycle{\_}node}}{cycle:unnode}}\\{{\nwixident{ex}}{ex}}\\{{\nwixident{figure}}{figure}}\\{{\nwixident{nops}}{nops}}\\{{\nwixident{op}}{op}}\\{{\nwixident{read{\_}archive}}{read:unarchive}}}\nwindexuse{\nwixident{archive}}{archive}{NWppJ6t-1d2GZW-4}\nwindexuse{\nwixident{cycle{\_}node}}{cycle:unnode}{NWppJ6t-1d2GZW-4}\nwindexuse{\nwixident{ex}}{ex}{NWppJ6t-1d2GZW-4}\nwindexuse{\nwixident{figure}}{figure}{NWppJ6t-1d2GZW-4}\nwindexuse{\nwixident{nops}}{nops}{NWppJ6t-1d2GZW-4}\nwindexuse{\nwixident{op}}{op}{NWppJ6t-1d2GZW-4}\nwindexuse{\nwixident{read{\_}archive}}{read:unarchive}{NWppJ6t-1d2GZW-4}\nwendcode{}\nwbegindocs{544}The following functions are used as {\Tt{}\Rm{}{\it{}PCR}\nwendquote} pointers for corresponding cycle
relations.
\nwenddocs{}\nwbegincode{545}\sublabel{NWppJ6t-1d2GZW-5}\nwmargintag{{\nwtagstyle{}\subpageref{NWppJ6t-1d2GZW-5}}}\moddef{cycle relations~{\nwtagstyle{}\subpageref{NWppJ6t-1d2GZW-1}}}\plusendmoddef\Rm{}\nwstartdeflinemarkup\nwusesondefline{\\{NWppJ6t-3NCnUp-3}}\nwprevnextdefs{NWppJ6t-1d2GZW-4}{NWppJ6t-1d2GZW-6}\nwenddeflinemarkup
\LA{}relations to check~{\nwtagstyle{}\subpageref{NWppJ6t-iB8QG-1}}\RA{}

\nwused{\\{NWppJ6t-3NCnUp-3}}\nwendcode{}\nwbegindocs{546}The following procedures are used to construct relations but are
impractical to check.
\nwenddocs{}\nwbegincode{547}\sublabel{NWppJ6t-1d2GZW-6}\nwmargintag{{\nwtagstyle{}\subpageref{NWppJ6t-1d2GZW-6}}}\moddef{cycle relations~{\nwtagstyle{}\subpageref{NWppJ6t-1d2GZW-1}}}\plusendmoddef\Rm{}\nwstartdeflinemarkup\nwusesondefline{\\{NWppJ6t-3NCnUp-3}}\nwprevnextdefs{NWppJ6t-1d2GZW-5}{NWppJ6t-1d2GZW-7}\nwenddeflinemarkup
{\bf{}ex} {\it{}cycle\_tangent}({\bf{}const} {\bf{}ex} & {\it{}C1}, {\bf{}const} {\bf{}ex} & {\it{}C2}, {\bf{}const} {\bf{}ex} & {\it{}pr}=0);
{\bf{}ex} {\it{}cycle\_tangent\_i}({\bf{}const} {\bf{}ex} & {\it{}C1}, {\bf{}const} {\bf{}ex} & {\it{}C2}, {\bf{}const} {\bf{}ex} & {\it{}pr}=0);
{\bf{}ex} {\it{}cycle\_tangent\_o}({\bf{}const} {\bf{}ex} & {\it{}C1}, {\bf{}const} {\bf{}ex} & {\it{}C2}, {\bf{}const} {\bf{}ex} & {\it{}pr}=0);
{\bf{}ex} {\it{}cycle\_angle}({\bf{}const} {\bf{}ex} & {\it{}C1}, {\bf{}const} {\bf{}ex} & {\it{}C2}, {\bf{}const} {\bf{}ex} & {\it{}pr});
{\bf{}ex} {\it{}steiner\_power}({\bf{}const} {\bf{}ex} & {\it{}C1}, {\bf{}const} {\bf{}ex} & {\it{}C2}, {\bf{}const} {\bf{}ex} & {\it{}pr});
{\bf{}ex} {\it{}cycle\_cross\_t\_distance}({\bf{}const} {\bf{}ex} & {\it{}C1}, {\bf{}const} {\bf{}ex} & {\it{}C2}, {\bf{}const} {\bf{}ex} & {\it{}pr});
\nwindexdefn{\nwixident{cycle{\_}tangent}}{cycle:untangent}{NWppJ6t-1d2GZW-6}\nwindexdefn{\nwixident{cycle{\_}tangent{\_}i}}{cycle:untangent:uni}{NWppJ6t-1d2GZW-6}\nwindexdefn{\nwixident{cycle{\_}tangent{\_}o}}{cycle:untangent:uno}{NWppJ6t-1d2GZW-6}\nwindexdefn{\nwixident{cycle{\_}angle}}{cycle:unangle}{NWppJ6t-1d2GZW-6}\nwindexdefn{\nwixident{steiner{\_}power}}{steiner:unpower}{NWppJ6t-1d2GZW-6}\nwindexdefn{\nwixident{cycle{\_}cross{\_}t{\_}distance}}{cycle:uncross:unt:undistance}{NWppJ6t-1d2GZW-6}\eatline
\nwused{\\{NWppJ6t-3NCnUp-3}}\nwidentdefs{\\{{\nwixident{cycle{\_}angle}}{cycle:unangle}}\\{{\nwixident{cycle{\_}cross{\_}t{\_}distance}}{cycle:uncross:unt:undistance}}\\{{\nwixident{cycle{\_}tangent}}{cycle:untangent}}\\{{\nwixident{cycle{\_}tangent{\_}i}}{cycle:untangent:uni}}\\{{\nwixident{cycle{\_}tangent{\_}o}}{cycle:untangent:uno}}\\{{\nwixident{steiner{\_}power}}{steiner:unpower}}}\nwidentuses{\\{{\nwixident{ex}}{ex}}}\nwindexuse{\nwixident{ex}}{ex}{NWppJ6t-1d2GZW-6}\nwendcode{}\nwbegindocs{548}\nwdocspar
\nwenddocs{}\nwbegindocs{549}Fractional linear transformations.
\nwenddocs{}\nwbegincode{550}\sublabel{NWppJ6t-1d2GZW-7}\nwmargintag{{\nwtagstyle{}\subpageref{NWppJ6t-1d2GZW-7}}}\moddef{cycle relations~{\nwtagstyle{}\subpageref{NWppJ6t-1d2GZW-1}}}\plusendmoddef\Rm{}\nwstartdeflinemarkup\nwusesondefline{\\{NWppJ6t-3NCnUp-3}}\nwprevnextdefs{NWppJ6t-1d2GZW-6}{NWppJ6t-1d2GZW-8}\nwenddeflinemarkup
{\bf{}ex} {\it{}cycle\_moebius}({\bf{}const} {\bf{}ex} & {\it{}C1}, {\bf{}const} {\bf{}ex} & {\it{}C2}, {\bf{}const} {\bf{}ex} & {\it{}pr});
{\bf{}ex} {\it{}cycle\_sl2}({\bf{}const} {\bf{}ex} & {\it{}C1}, {\bf{}const} {\bf{}ex} & {\it{}C2}, {\bf{}const} {\bf{}ex} & {\it{}pr});
\nwindexdefn{\nwixident{cycle{\_}moebius}}{cycle:unmoebius}{NWppJ6t-1d2GZW-7}\nwindexdefn{\nwixident{cycle{\_}sl2}}{cycle:unsl2}{NWppJ6t-1d2GZW-7}\eatline
\nwused{\\{NWppJ6t-3NCnUp-3}}\nwidentdefs{\\{{\nwixident{cycle{\_}moebius}}{cycle:unmoebius}}\\{{\nwixident{cycle{\_}sl2}}{cycle:unsl2}}}\nwidentuses{\\{{\nwixident{ex}}{ex}}}\nwindexuse{\nwixident{ex}}{ex}{NWppJ6t-1d2GZW-7}\nwendcode{}\nwbegindocs{551}\nwdocspar
\nwenddocs{}\nwbegindocs{552}The next functions are used to measure certain quantities between cycles.
\nwenddocs{}\nwbegincode{553}\sublabel{NWppJ6t-1d2GZW-8}\nwmargintag{{\nwtagstyle{}\subpageref{NWppJ6t-1d2GZW-8}}}\moddef{cycle relations~{\nwtagstyle{}\subpageref{NWppJ6t-1d2GZW-1}}}\plusendmoddef\Rm{}\nwstartdeflinemarkup\nwusesondefline{\\{NWppJ6t-3NCnUp-3}}\nwprevnextdefs{NWppJ6t-1d2GZW-7}{NWppJ6t-1d2GZW-9}\nwenddeflinemarkup
{\bf{}ex} {\it{}power\_is}({\bf{}const} {\bf{}ex} & {\it{}C1}, {\bf{}const} {\bf{}ex} & {\it{}C2}, {\bf{}const} {\bf{}ex} & {\it{}pr}=1);
{\bf{}inline} {\bf{}ex} {\it{}sq\_t\_distance\_is}({\bf{}const} {\bf{}ex} & {\it{}C1}, {\bf{}const} {\bf{}ex} & {\it{}C2}, {\bf{}const} {\bf{}ex} & {\it{}pr}=1)
 {\nwlbrace}{\bf{}return} {\it{}power\_is}({\it{}C1},{\it{}C2},1);{\nwrbrace}
{\bf{}inline} {\bf{}ex} {\it{}sq\_cross\_t\_distance\_is}({\bf{}const} {\bf{}ex} & {\it{}C1}, {\bf{}const} {\bf{}ex} & {\it{}C2}, {\bf{}const} {\bf{}ex} & {\it{}pr}=-1)
 {\nwlbrace}{\bf{}return} {\it{}power\_is}({\it{}C1},{\it{}C2},-1);{\nwrbrace}
{\bf{}ex} {\it{}angle\_is}({\bf{}const} {\bf{}ex} & {\it{}C1}, {\bf{}const} {\bf{}ex} & {\it{}C2}, {\bf{}const} {\bf{}ex} & {\it{}pr}=0);
\nwindexdefn{\nwixident{power{\_}is}}{power:unis}{NWppJ6t-1d2GZW-8}\nwindexdefn{\nwixident{sq{\_}t{\_}distance{\_}is}}{sq:unt:undistance:unis}{NWppJ6t-1d2GZW-8}\nwindexdefn{\nwixident{sq{\_}cross{\_}t{\_}distance{\_}is}}{sq:uncross:unt:undistance:unis}{NWppJ6t-1d2GZW-8}\nwindexdefn{\nwixident{angle{\_}is}}{angle:unis}{NWppJ6t-1d2GZW-8}\eatline
\nwused{\\{NWppJ6t-3NCnUp-3}}\nwidentdefs{\\{{\nwixident{angle{\_}is}}{angle:unis}}\\{{\nwixident{power{\_}is}}{power:unis}}\\{{\nwixident{sq{\_}cross{\_}t{\_}distance{\_}is}}{sq:uncross:unt:undistance:unis}}\\{{\nwixident{sq{\_}t{\_}distance{\_}is}}{sq:unt:undistance:unis}}}\nwidentuses{\\{{\nwixident{ex}}{ex}}}\nwindexuse{\nwixident{ex}}{ex}{NWppJ6t-1d2GZW-8}\nwendcode{}\nwbegindocs{554}\nwdocspar
\nwenddocs{}\nwbegindocs{555}We include the list of pre-defined metrics in two dimensions.
\nwenddocs{}\nwbegincode{556}\sublabel{NWppJ6t-1d2GZW-9}\nwmargintag{{\nwtagstyle{}\subpageref{NWppJ6t-1d2GZW-9}}}\moddef{cycle relations~{\nwtagstyle{}\subpageref{NWppJ6t-1d2GZW-1}}}\plusendmoddef\Rm{}\nwstartdeflinemarkup\nwusesondefline{\\{NWppJ6t-3NCnUp-3}}\nwprevnextdefs{NWppJ6t-1d2GZW-8}{NWppJ6t-1d2GZW-A}\nwenddeflinemarkup
\LA{}predefined cycle relations~{\nwtagstyle{}\subpageref{NWppJ6t-2iyHmp-1}}\RA{}

\nwused{\\{NWppJ6t-3NCnUp-3}}\nwendcode{}\nwbegindocs{557}We explicitly define three types of metrics on a plane: elliptic, parabolic, hyperbolic.
\nwenddocs{}\nwbegincode{558}\sublabel{NWppJ6t-1d2GZW-A}\nwmargintag{{\nwtagstyle{}\subpageref{NWppJ6t-1d2GZW-A}}}\moddef{cycle relations~{\nwtagstyle{}\subpageref{NWppJ6t-1d2GZW-1}}}\plusendmoddef\Rm{}\nwstartdeflinemarkup\nwusesondefline{\\{NWppJ6t-3NCnUp-3}}\nwprevnextdefs{NWppJ6t-1d2GZW-9}{NWppJ6t-1d2GZW-B}\nwenddeflinemarkup
{\bf{}extern} {\bf{}const} {\bf{}ex} {\it{}metric\_e}, {\it{}metric\_p}, {\it{}metric\_h};
\nwindexdefn{\nwixident{metric{\_}e}}{metric:une}{NWppJ6t-1d2GZW-A}\nwindexdefn{\nwixident{metric{\_}p}}{metric:unp}{NWppJ6t-1d2GZW-A}\nwindexdefn{\nwixident{metric{\_}h}}{metric:unh}{NWppJ6t-1d2GZW-A}\eatline
\nwused{\\{NWppJ6t-3NCnUp-3}}\nwidentdefs{\\{{\nwixident{metric{\_}e}}{metric:une}}\\{{\nwixident{metric{\_}h}}{metric:unh}}\\{{\nwixident{metric{\_}p}}{metric:unp}}}\nwidentuses{\\{{\nwixident{ex}}{ex}}}\nwindexuse{\nwixident{ex}}{ex}{NWppJ6t-1d2GZW-A}\nwendcode{}\nwbegindocs{559}\nwdocspar
\nwenddocs{}\nwbegindocs{560}The predefined metrics are based on diagonal matrices with different signatures.
\nwenddocs{}\nwbegincode{561}\sublabel{NWppJ6t-A2bag-2}\nwmargintag{{\nwtagstyle{}\subpageref{NWppJ6t-A2bag-2}}}\moddef{figure library variables and constants~{\nwtagstyle{}\subpageref{NWppJ6t-A2bag-1}}}\plusendmoddef\Rm{}\nwstartdeflinemarkup\nwusesondefline{\\{NWppJ6t-32jW6L-1}}\nwprevnextdefs{NWppJ6t-A2bag-1}{NWppJ6t-A2bag-3}\nwenddeflinemarkup
{\bf{}const} {\bf{}ex} {\it{}metric\_e} = {\it{}clifford\_unit}({\bf{}varidx}({\bf{}symbol}({\tt{}"i"}), {\bf{}numeric}(2)), {\bf{}indexed}({\it{}diag\_matrix}({\bf{}lst}{\nwlbrace}-1,-1{\nwrbrace}), {\it{}sy\_symm}(),\nwindexdefn{\nwixident{ex}}{ex}{NWppJ6t-A2bag-2}
                                                                           {\bf{}varidx}({\bf{}symbol}({\tt{}"j"}), {\bf{}numeric}(2)), {\bf{}varidx}({\bf{}symbol}({\tt{}"k"}), {\bf{}numeric}(2))));
{\bf{}const} {\bf{}ex} {\it{}metric\_p} = {\it{}clifford\_unit}({\bf{}varidx}({\bf{}symbol}({\tt{}"i"}), {\bf{}numeric}(2)), {\bf{}indexed}({\it{}diag\_matrix}({\bf{}lst}{\nwlbrace}-1,0{\nwrbrace}), {\it{}sy\_symm}(),\nwindexdefn{\nwixident{ex}}{ex}{NWppJ6t-A2bag-2}
                                                                           {\bf{}varidx}({\bf{}symbol}({\tt{}"j"}), {\bf{}numeric}(2)), {\bf{}varidx}({\bf{}symbol}({\tt{}"k"}), {\bf{}numeric}(2))));
{\bf{}const} {\bf{}ex} {\it{}metric\_h} = {\it{}clifford\_unit}({\bf{}varidx}({\bf{}symbol}({\tt{}"i"}), {\bf{}numeric}(2)), {\bf{}indexed}({\it{}diag\_matrix}({\bf{}lst}{\nwlbrace}-1,1{\nwrbrace}), {\it{}sy\_symm}(),\nwindexdefn{\nwixident{ex}}{ex}{NWppJ6t-A2bag-2}
{\bf{}varidx}({\bf{}symbol}({\tt{}"j"}), {\bf{}numeric}(2)), {\bf{}varidx}({\bf{}symbol}({\tt{}"k"}), {\bf{}numeric}(2))));
\nwindexdefn{\nwixident{metric{\_}e}}{metric:une}{NWppJ6t-A2bag-2}\nwindexdefn{\nwixident{metric{\_}p}}{metric:unp}{NWppJ6t-A2bag-2}\nwindexdefn{\nwixident{metric{\_}h}}{metric:unh}{NWppJ6t-A2bag-2}\eatline
\nwused{\\{NWppJ6t-32jW6L-1}}\nwidentdefs{\\{{\nwixident{ex}}{ex}}\\{{\nwixident{metric{\_}e}}{metric:une}}\\{{\nwixident{metric{\_}h}}{metric:unh}}\\{{\nwixident{metric{\_}p}}{metric:unp}}}\nwidentuses{\\{{\nwixident{k}}{k}}\\{{\nwixident{numeric}}{numeric}}}\nwindexuse{\nwixident{k}}{k}{NWppJ6t-A2bag-2}\nwindexuse{\nwixident{numeric}}{numeric}{NWppJ6t-A2bag-2}\nwendcode{}\nwbegindocs{562}\nwdocspar
\nwenddocs{}\nwbegindocs{563}There is the list of pre-defined metrics in two dimensions cycle
relations. Orthogonality of cycles of three types independent from a
metric stored in the figure.
\nwenddocs{}\nwbegincode{564}\sublabel{NWppJ6t-1d2GZW-B}\nwmargintag{{\nwtagstyle{}\subpageref{NWppJ6t-1d2GZW-B}}}\moddef{cycle relations~{\nwtagstyle{}\subpageref{NWppJ6t-1d2GZW-1}}}\plusendmoddef\Rm{}\nwstartdeflinemarkup\nwusesondefline{\\{NWppJ6t-3NCnUp-3}}\nwprevnextdefs{NWppJ6t-1d2GZW-A}{\relax}\nwenddeflinemarkup
{\bf{}inline} {\bf{}ex} {\it{}cycle\_orthogonal\_e}({\bf{}const} {\bf{}ex} & {\it{}C1}, {\bf{}const} {\bf{}ex} & {\it{}C2}, {\bf{}const} {\bf{}ex} & {\it{}pr}=0) {\nwlbrace}
    {\bf{}return} {\bf{}lst}{\nwlbrace}({\bf{}ex}){\bf{}lst}{\nwlbrace}{\it{}ex\_to}\begin{math}<\end{math}{\bf{}cycle}\begin{math}>\end{math}({\it{}C1}).{\it{}is\_orthogonal}({\it{}ex\_to}\begin{math}<\end{math}{\bf{}cycle}\begin{math}>\end{math}({\it{}C2}), {\it{}metric\_e}){\nwrbrace}{\nwrbrace};{\nwrbrace}

{\bf{}inline} {\bf{}ex} {\it{}cycle\_orthogonal\_p}({\bf{}const} {\bf{}ex} & {\it{}C1}, {\bf{}const} {\bf{}ex} & {\it{}C2}, {\bf{}const} {\bf{}ex} & {\it{}pr}=0) {\nwlbrace}
    {\bf{}return} {\bf{}lst}{\nwlbrace}({\bf{}ex}){\bf{}lst}{\nwlbrace}{\it{}ex\_to}\begin{math}<\end{math}{\bf{}cycle}\begin{math}>\end{math}({\it{}C1}).{\it{}is\_orthogonal}({\it{}ex\_to}\begin{math}<\end{math}{\bf{}cycle}\begin{math}>\end{math}({\it{}C2}), {\it{}metric\_p}){\nwrbrace}{\nwrbrace};{\nwrbrace}

{\bf{}inline} {\bf{}ex} {\it{}cycle\_orthogonal\_h}({\bf{}const} {\bf{}ex} & {\it{}C1}, {\bf{}const} {\bf{}ex} & {\it{}C2}, {\bf{}const} {\bf{}ex} & {\it{}pr}=0) {\nwlbrace}
    {\bf{}return} {\bf{}lst}{\nwlbrace}({\bf{}ex}){\bf{}lst}{\nwlbrace}{\it{}ex\_to}\begin{math}<\end{math}{\bf{}cycle}\begin{math}>\end{math}({\it{}C1}).{\it{}is\_orthogonal}({\it{}ex\_to}\begin{math}<\end{math}{\bf{}cycle}\begin{math}>\end{math}({\it{}C2}), {\it{}metric\_h}){\nwrbrace}{\nwrbrace};{\nwrbrace}
\nwindexdefn{\nwixident{cycle{\_}orthogonal{\_}e}}{cycle:unorthogonal:une}{NWppJ6t-1d2GZW-B}\nwindexdefn{\nwixident{cycle{\_}orthogonal{\_}p}}{cycle:unorthogonal:unp}{NWppJ6t-1d2GZW-B}\nwindexdefn{\nwixident{cycle{\_}orthogonal{\_}h}}{cycle:unorthogonal:unh}{NWppJ6t-1d2GZW-B}\eatline
\nwused{\\{NWppJ6t-3NCnUp-3}}\nwidentdefs{\\{{\nwixident{cycle{\_}orthogonal{\_}e}}{cycle:unorthogonal:une}}\\{{\nwixident{cycle{\_}orthogonal{\_}h}}{cycle:unorthogonal:unh}}\\{{\nwixident{cycle{\_}orthogonal{\_}p}}{cycle:unorthogonal:unp}}}\nwidentuses{\\{{\nwixident{ex}}{ex}}\\{{\nwixident{is{\_}orthogonal}}{is:unorthogonal}}\\{{\nwixident{metric{\_}e}}{metric:une}}\\{{\nwixident{metric{\_}h}}{metric:unh}}\\{{\nwixident{metric{\_}p}}{metric:unp}}}\nwindexuse{\nwixident{ex}}{ex}{NWppJ6t-1d2GZW-B}\nwindexuse{\nwixident{is{\_}orthogonal}}{is:unorthogonal}{NWppJ6t-1d2GZW-B}\nwindexuse{\nwixident{metric{\_}e}}{metric:une}{NWppJ6t-1d2GZW-B}\nwindexuse{\nwixident{metric{\_}h}}{metric:unh}{NWppJ6t-1d2GZW-B}\nwindexuse{\nwixident{metric{\_}p}}{metric:unp}{NWppJ6t-1d2GZW-B}\nwendcode{}\nwbegindocs{565}\nwdocspar
\nwenddocs{}\nwbegindocs{566}\nwdocspar
\subsection{{\Tt{}\Rm{}{\bf{}subfigure}\nwendquote} class declaration}
\label{sec:subf-class-decl}

\nwenddocs{}\nwbegindocs{567}{\Tt{}\Rm{}{\bf{}subfigure}] {\bf{}class} {\it{}allows} {\it{}to} {\it{}encapsulate} {\it{}some} {\it{}common} {\it{}constructions}.\nwdocspar\nwnewline
{\it{}The} {\it{}library} {\it{}provides} {\it{}an} {\it{}important} {\it{}example} [[{\it{}midpoint}\nwendquote}. End-user may
define his own {\Tt{}\Rm{}{\bf{}subfigure}\nwendquote}s, they will not be handled as native
ones, including (un)archiving.

In the essence {\Tt{}\Rm{}{\bf{}subfigure}\nwendquote} is created from a {\Tt{}\Rm{}{\bf{}figure}\nwendquote}, which were
designed to be included in another figures.
\nwenddocs{}\nwbegincode{568}\sublabel{NWppJ6t-4UgdbB-1}\nwmargintag{{\nwtagstyle{}\subpageref{NWppJ6t-4UgdbB-1}}}\moddef{subfigure header~{\nwtagstyle{}\subpageref{NWppJ6t-4UgdbB-1}}}\endmoddef\Rm{}\nwstartdeflinemarkup\nwusesondefline{\\{NWppJ6t-3NCnUp-3}}\nwprevnextdefs{\relax}{NWppJ6t-4UgdbB-2}\nwenddeflinemarkup
{\bf{}class} {\bf{}subfigure} : {\bf{}public} {\bf{}basic}
{\nwlbrace}
    {\it{}GINAC\_DECLARE\_REGISTERED\_CLASS}({\bf{}subfigure}, {\bf{}basic})
{\bf{}protected}:
    {\bf{}ex} {\it{}subf}; // A figure to be inserted
    {\bf{}lst} {\it{}parlist}; // A list of key to a parent cycle\_node in figure
{\bf{}public}:
    \LA{}public methods for subfigure~{\nwtagstyle{}\subpageref{NWppJ6t-25h4gT-1}}\RA{}
    {\bf{}inline} {\bf{}ex} {\it{}subs}({\bf{}const} {\it{}exmap} & {\it{}em}, {\bf{}unsigned} {\it{}options}=0) {\bf{}const};
\nwindexdefn{\nwixident{subfigure}}{subfigure}{NWppJ6t-4UgdbB-1}\eatline
\nwalsodefined{\\{NWppJ6t-4UgdbB-2}\\{NWppJ6t-4UgdbB-3}}\nwused{\\{NWppJ6t-3NCnUp-3}}\nwidentdefs{\\{{\nwixident{subfigure}}{subfigure}}}\nwidentuses{\\{{\nwixident{cycle{\_}node}}{cycle:unnode}}\\{{\nwixident{ex}}{ex}}\\{{\nwixident{figure}}{figure}}\\{{\nwixident{key}}{key}}\\{{\nwixident{subs}}{subs}}}\nwindexuse{\nwixident{cycle{\_}node}}{cycle:unnode}{NWppJ6t-4UgdbB-1}\nwindexuse{\nwixident{ex}}{ex}{NWppJ6t-4UgdbB-1}\nwindexuse{\nwixident{figure}}{figure}{NWppJ6t-4UgdbB-1}\nwindexuse{\nwixident{key}}{key}{NWppJ6t-4UgdbB-1}\nwindexuse{\nwixident{subs}}{subs}{NWppJ6t-4UgdbB-1}\nwendcode{}\nwbegindocs{569}\nwdocspar
\nwenddocs{}\nwbegindocs{570}Some service methods.
\nwenddocs{}\nwbegincode{571}\sublabel{NWppJ6t-4UgdbB-2}\nwmargintag{{\nwtagstyle{}\subpageref{NWppJ6t-4UgdbB-2}}}\moddef{subfigure header~{\nwtagstyle{}\subpageref{NWppJ6t-4UgdbB-1}}}\plusendmoddef\Rm{}\nwstartdeflinemarkup\nwusesondefline{\\{NWppJ6t-3NCnUp-3}}\nwprevnextdefs{NWppJ6t-4UgdbB-1}{NWppJ6t-4UgdbB-3}\nwenddeflinemarkup
{\bf{}protected}:
    {\bf{}inline} {\bf{}ex} {\it{}get\_parlist}() {\bf{}const} {\nwlbrace}{\bf{}return} {\it{}parlist};{\nwrbrace}
    {\bf{}inline} {\bf{}ex} {\it{}get\_subf}() {\bf{}const} {\nwlbrace}{\bf{}return} {\it{}subf};{\nwrbrace}
    {\it{}return\_type\_t} {\it{}return\_type\_tinfo}() {\bf{}const};
    {\bf{}void} {\it{}do\_print}({\bf{}const} {\it{}print\_dflt} & {\it{}con}, {\bf{}unsigned} {\it{}level}) {\bf{}const};
    {\bf{}void} {\it{}do\_print\_tree}({\bf{}const} {\it{}print\_tree} & {\it{}con}, {\bf{}unsigned} {\it{}level}) {\bf{}const};

\nwused{\\{NWppJ6t-3NCnUp-3}}\nwidentuses{\\{{\nwixident{ex}}{ex}}}\nwindexuse{\nwixident{ex}}{ex}{NWppJ6t-4UgdbB-2}\nwendcode{}\nwbegindocs{572} (un)Archiving of {\Tt{}\Rm{}{\bf{}cycle\_relation}\nwendquote} is not universal. At present it
only can handle relations declared in the header file {\Tt{}\Rm{}{\it{}p\_orthogonal}\nwendquote},
{\Tt{}\Rm{}{\it{}p\_f\_orthogonal}\nwendquote}, {\Tt{}\Rm{}{\it{}p\_adifferent}\nwendquote}, {\Tt{}\Rm{}{\it{}p\_different}\nwendquote} and {\Tt{}\Rm{}{\it{}p\_tangent}\nwendquote} etc. from
Subsection~\ref{sec:publ-meth-cycl}.
\nwenddocs{}\nwbegincode{573}\sublabel{NWppJ6t-4UgdbB-3}\nwmargintag{{\nwtagstyle{}\subpageref{NWppJ6t-4UgdbB-3}}}\moddef{subfigure header~{\nwtagstyle{}\subpageref{NWppJ6t-4UgdbB-1}}}\plusendmoddef\Rm{}\nwstartdeflinemarkup\nwusesondefline{\\{NWppJ6t-3NCnUp-3}}\nwprevnextdefs{NWppJ6t-4UgdbB-2}{\relax}\nwenddeflinemarkup
    {\bf{}void} {\it{}archive}({\it{}archive\_node} &{\it{}n}) {\bf{}const};
    {\bf{}void} {\it{}read\_archive}({\bf{}const} {\it{}archive\_node} &{\it{}n}, {\bf{}lst} &{\it{}sym\_lst});

{\bf{}friend} {\bf{}class} {\bf{}cycle\_node};
{\bf{}friend} {\bf{}class} {\bf{}figure};
{\nwrbrace};
{\it{}GINAC\_DECLARE\_UNARCHIVER}({\bf{}subfigure});\nwindexdefn{\nwixident{subfigure}}{subfigure}{NWppJ6t-4UgdbB-3}

\nwused{\\{NWppJ6t-3NCnUp-3}}\nwidentdefs{\\{{\nwixident{subfigure}}{subfigure}}}\nwidentuses{\\{{\nwixident{archive}}{archive}}\\{{\nwixident{cycle{\_}node}}{cycle:unnode}}\\{{\nwixident{figure}}{figure}}\\{{\nwixident{read{\_}archive}}{read:unarchive}}}\nwindexuse{\nwixident{archive}}{archive}{NWppJ6t-4UgdbB-3}\nwindexuse{\nwixident{cycle{\_}node}}{cycle:unnode}{NWppJ6t-4UgdbB-3}\nwindexuse{\nwixident{figure}}{figure}{NWppJ6t-4UgdbB-3}\nwindexuse{\nwixident{read{\_}archive}}{read:unarchive}{NWppJ6t-4UgdbB-3}\nwendcode{}\nwbegindocs{574}\nwdocspar
\subsection{{\Tt{}\Rm{}{\bf{}figure}\nwendquote} class declaration}
\label{sec:figure-class-header}

The essential interface to {\Tt{}\Rm{}{\bf{}figure}\nwendquote} class was already presented in
Section~\ref{sec:publ-meth-figure}, here we keep the less-used
elements. An advanced end-user may be interested in {\Tt{}\Rm{}{\bf{}figure}\nwendquote} class
members given in \S~\ref{sec:members-figure-class}.

\nwenddocs{}\nwbegindocs{575}We define {\Tt{}\Rm{}{\bf{}figure}\nwendquote} class as a children of \GiNaC\ {\Tt{}\Rm{}{\bf{}basic}\nwendquote}.
\nwenddocs{}\nwbegincode{576}\sublabel{NWppJ6t-7vmg0-1}\nwmargintag{{\nwtagstyle{}\subpageref{NWppJ6t-7vmg0-1}}}\moddef{figure header~{\nwtagstyle{}\subpageref{NWppJ6t-7vmg0-1}}}\endmoddef\Rm{}\nwstartdeflinemarkup\nwusesondefline{\\{NWppJ6t-3NCnUp-3}}\nwprevnextdefs{\relax}{NWppJ6t-7vmg0-2}\nwenddeflinemarkup
{\bf{}class} {\bf{}figure} : {\bf{}public} {\bf{}basic}
{\nwlbrace}
{\it{}GINAC\_DECLARE\_REGISTERED\_CLASS}({\bf{}figure}, {\bf{}basic})

\LA{}member of figure class~{\nwtagstyle{}\subpageref{NWppJ6t-4OZeJw-1}}\RA{}
\nwindexdefn{\nwixident{figure}}{figure}{NWppJ6t-7vmg0-1}\eatline
\nwalsodefined{\\{NWppJ6t-7vmg0-2}\\{NWppJ6t-7vmg0-3}\\{NWppJ6t-7vmg0-4}\\{NWppJ6t-7vmg0-5}\\{NWppJ6t-7vmg0-6}\\{NWppJ6t-7vmg0-7}\\{NWppJ6t-7vmg0-8}}\nwused{\\{NWppJ6t-3NCnUp-3}}\nwidentdefs{\\{{\nwixident{figure}}{figure}}}\nwendcode{}\nwbegindocs{577}\nwdocspar
\nwenddocs{}\nwbegindocs{578}The method to update {\Tt{}\Rm{}{\bf{}cycle\_node}\nwendquote} with labelled by the
{\Tt{}\Rm{}{\it{}key}\nwendquote}. Since the list of conditions may branches and  has a variable
length the method runs recursively with {\Tt{}\Rm{}{\it{}level}\nwendquote} parameterising the
depth of nested calls.
\nwenddocs{}\nwbegincode{579}\sublabel{NWppJ6t-7vmg0-2}\nwmargintag{{\nwtagstyle{}\subpageref{NWppJ6t-7vmg0-2}}}\moddef{figure header~{\nwtagstyle{}\subpageref{NWppJ6t-7vmg0-1}}}\plusendmoddef\Rm{}\nwstartdeflinemarkup\nwusesondefline{\\{NWppJ6t-3NCnUp-3}}\nwprevnextdefs{NWppJ6t-7vmg0-1}{NWppJ6t-7vmg0-3}\nwenddeflinemarkup
{\bf{}protected}:
    {\bf{}ex} {\it{}update\_cycle\_node}({\bf{}const} {\bf{}ex} & {\it{}key}, {\bf{}const} {\bf{}lst} & {\it{}eq\_cond}={\bf{}lst}{\nwlbrace}{\nwrbrace},
                           {\bf{}const} {\bf{}lst} & {\it{}neq\_cond}={\bf{}lst}{\nwlbrace}{\nwrbrace}, {\bf{}lst} {\it{}res}={\bf{}lst}{\nwlbrace}{\nwrbrace}, {\it{}size\_t} {\it{}level}=0);
    {\bf{}void} {\it{}set\_cycle}({\bf{}const} {\bf{}ex} & {\it{}key}, {\bf{}const} {\bf{}ex} & {\it{}C});
\nwindexdefn{\nwixident{set{\_}cycle}}{set:uncycle}{NWppJ6t-7vmg0-2}\nwindexdefn{\nwixident{update{\_}cycle{\_}node}}{update:uncycle:unnode}{NWppJ6t-7vmg0-2}\eatline
\nwused{\\{NWppJ6t-3NCnUp-3}}\nwidentdefs{\\{{\nwixident{set{\_}cycle}}{set:uncycle}}\\{{\nwixident{update{\_}cycle{\_}node}}{update:uncycle:unnode}}}\nwidentuses{\\{{\nwixident{ex}}{ex}}\\{{\nwixident{key}}{key}}}\nwindexuse{\nwixident{ex}}{ex}{NWppJ6t-7vmg0-2}\nwindexuse{\nwixident{key}}{key}{NWppJ6t-7vmg0-2}\nwendcode{}\nwbegindocs{580}\nwdocspar
\nwenddocs{}\nwbegindocs{581}Evaluate a cycle through a list of conditions.
\nwenddocs{}\nwbegincode{582}\sublabel{NWppJ6t-7vmg0-3}\nwmargintag{{\nwtagstyle{}\subpageref{NWppJ6t-7vmg0-3}}}\moddef{figure header~{\nwtagstyle{}\subpageref{NWppJ6t-7vmg0-1}}}\plusendmoddef\Rm{}\nwstartdeflinemarkup\nwusesondefline{\\{NWppJ6t-3NCnUp-3}}\nwprevnextdefs{NWppJ6t-7vmg0-2}{NWppJ6t-7vmg0-4}\nwenddeflinemarkup
    {\bf{}ex} {\it{}evaluate\_cycle}({\bf{}const} {\bf{}ex} & {\it{}symbolic}, {\bf{}const} {\bf{}lst} & {\it{}cond}) {\bf{}const};

\nwused{\\{NWppJ6t-3NCnUp-3}}\nwidentuses{\\{{\nwixident{evaluate{\_}cycle}}{evaluate:uncycle}}\\{{\nwixident{ex}}{ex}}}\nwindexuse{\nwixident{evaluate{\_}cycle}}{evaluate:uncycle}{NWppJ6t-7vmg0-3}\nwindexuse{\nwixident{ex}}{ex}{NWppJ6t-7vmg0-3}\nwendcode{}\nwbegindocs{583}We include here methods from Section~\ref{sec:publ-meth-figure},
which are of interest for an end-user.
\nwenddocs{}\nwbegincode{584}\sublabel{NWppJ6t-7vmg0-4}\nwmargintag{{\nwtagstyle{}\subpageref{NWppJ6t-7vmg0-4}}}\moddef{figure header~{\nwtagstyle{}\subpageref{NWppJ6t-7vmg0-1}}}\plusendmoddef\Rm{}\nwstartdeflinemarkup\nwusesondefline{\\{NWppJ6t-3NCnUp-3}}\nwprevnextdefs{NWppJ6t-7vmg0-3}{NWppJ6t-7vmg0-5}\nwenddeflinemarkup
{\bf{}public}:
    \LA{}public methods in figure class~{\nwtagstyle{}\subpageref{NWppJ6t-AWRj3-1}}\RA{}

\nwused{\\{NWppJ6t-3NCnUp-3}}\nwendcode{}\nwbegindocs{585}The following methods are public as well however may be less used.
\nwenddocs{}\nwbegincode{586}\sublabel{NWppJ6t-7vmg0-5}\nwmargintag{{\nwtagstyle{}\subpageref{NWppJ6t-7vmg0-5}}}\moddef{figure header~{\nwtagstyle{}\subpageref{NWppJ6t-7vmg0-1}}}\plusendmoddef\Rm{}\nwstartdeflinemarkup\nwusesondefline{\\{NWppJ6t-3NCnUp-3}}\nwprevnextdefs{NWppJ6t-7vmg0-4}{NWppJ6t-7vmg0-6}\nwenddeflinemarkup
    {\bf{}inline} {\bf{}ex} {\it{}get\_cycle\_node}({\bf{}const} {\bf{}ex} & {\it{}k}) {\bf{}const} {\nwlbrace}{\bf{}return} {\it{}nodes}.{\it{}find}({\it{}k})\begin{math}\rightarrow\end{math}{\it{}second};{\nwrbrace}
    {\bf{}void} {\it{}do\_print\_double}({\bf{}const} {\it{}print\_dflt} & {\it{}con}, {\bf{}unsigned} {\it{}level}) {\bf{}const};
\nwindexdefn{\nwixident{get{\_}cycle{\_}node}}{get:uncycle:unnode}{NWppJ6t-7vmg0-5}\nwindexdefn{\nwixident{do{\_}print{\_}double}}{do:unprint:undouble}{NWppJ6t-7vmg0-5}\eatline
\nwused{\\{NWppJ6t-3NCnUp-3}}\nwidentdefs{\\{{\nwixident{do{\_}print{\_}double}}{do:unprint:undouble}}\\{{\nwixident{get{\_}cycle{\_}node}}{get:uncycle:unnode}}}\nwidentuses{\\{{\nwixident{ex}}{ex}}\\{{\nwixident{k}}{k}}\\{{\nwixident{nodes}}{nodes}}}\nwindexuse{\nwixident{ex}}{ex}{NWppJ6t-7vmg0-5}\nwindexuse{\nwixident{k}}{k}{NWppJ6t-7vmg0-5}\nwindexuse{\nwixident{nodes}}{nodes}{NWppJ6t-7vmg0-5}\nwendcode{}\nwbegindocs{587}\nwdocspar
\nwenddocs{}\nwbegindocs{588}The method returning all nodes.
\nwenddocs{}\nwbegincode{589}\sublabel{NWppJ6t-AWRj3-V}\nwmargintag{{\nwtagstyle{}\subpageref{NWppJ6t-AWRj3-V}}}\moddef{public methods in figure class~{\nwtagstyle{}\subpageref{NWppJ6t-AWRj3-1}}}\plusendmoddef\Rm{}\nwstartdeflinemarkup\nwusesondefline{\\{NWppJ6t-7vmg0-4}}\nwprevnextdefs{NWppJ6t-AWRj3-U}{NWppJ6t-AWRj3-W}\nwenddeflinemarkup
    {\bf{}inline} {\it{}exhashmap}\begin{math}<\end{math}{\bf{}cycle\_node}\begin{math}>\end{math} {\it{}get\_nodes}() {\bf{}const} {\nwlbrace}{\bf{}return} {\it{}nodes};{\nwrbrace}

\nwused{\\{NWppJ6t-7vmg0-4}}\nwidentuses{\\{{\nwixident{cycle{\_}node}}{cycle:unnode}}\\{{\nwixident{nodes}}{nodes}}}\nwindexuse{\nwixident{cycle{\_}node}}{cycle:unnode}{NWppJ6t-AWRj3-V}\nwindexuse{\nwixident{nodes}}{nodes}{NWppJ6t-AWRj3-V}\nwendcode{}\nwbegindocs{590}Sometimes we need access to predefined {\Tt{}\Rm{}{\it{}infinity}\nwendquote} or the
{\Tt{}\Rm{}{\it{}real\_line}\nwendquote}, for example to specify a cycle relation to them.
\nwenddocs{}\nwbegincode{591}\sublabel{NWppJ6t-AWRj3-W}\nwmargintag{{\nwtagstyle{}\subpageref{NWppJ6t-AWRj3-W}}}\moddef{public methods in figure class~{\nwtagstyle{}\subpageref{NWppJ6t-AWRj3-1}}}\plusendmoddef\Rm{}\nwstartdeflinemarkup\nwusesondefline{\\{NWppJ6t-7vmg0-4}}\nwprevnextdefs{NWppJ6t-AWRj3-V}{NWppJ6t-AWRj3-X}\nwenddeflinemarkup
    {\bf{}inline} {\bf{}ex} {\it{}get\_real\_line}() {\bf{}const} {\nwlbrace}{\bf{}return} {\it{}real\_line};{\nwrbrace}
    {\bf{}inline} {\bf{}ex} {\it{}get\_infinity}() {\bf{}const} {\nwlbrace}{\bf{}return} {\it{}infinity};{\nwrbrace}
\nwindexdefn{\nwixident{get{\_}real{\_}line}}{get:unreal:unline}{NWppJ6t-AWRj3-W}\nwindexdefn{\nwixident{get{\_}infinity}}{get:uninfinity}{NWppJ6t-AWRj3-W}\eatline
\nwused{\\{NWppJ6t-7vmg0-4}}\nwidentdefs{\\{{\nwixident{get{\_}infinity}}{get:uninfinity}}\\{{\nwixident{get{\_}real{\_}line}}{get:unreal:unline}}}\nwidentuses{\\{{\nwixident{ex}}{ex}}\\{{\nwixident{infinity}}{infinity}}\\{{\nwixident{real{\_}line}}{real:unline}}}\nwindexuse{\nwixident{ex}}{ex}{NWppJ6t-AWRj3-W}\nwindexuse{\nwixident{infinity}}{infinity}{NWppJ6t-AWRj3-W}\nwindexuse{\nwixident{real{\_}line}}{real:unline}{NWppJ6t-AWRj3-W}\nwendcode{}\nwbegindocs{592}\nwdocspar
\nwenddocs{}\nwbegindocs{593}Return the maximal generation number of cycles in this figure.
\nwenddocs{}\nwbegincode{594}\sublabel{NWppJ6t-AWRj3-X}\nwmargintag{{\nwtagstyle{}\subpageref{NWppJ6t-AWRj3-X}}}\moddef{public methods in figure class~{\nwtagstyle{}\subpageref{NWppJ6t-AWRj3-1}}}\plusendmoddef\Rm{}\nwstartdeflinemarkup\nwusesondefline{\\{NWppJ6t-7vmg0-4}}\nwprevnextdefs{NWppJ6t-AWRj3-W}{NWppJ6t-AWRj3-Y}\nwenddeflinemarkup
    {\bf{}int} {\it{}get\_max\_generation}() {\bf{}const};
\nwindexdefn{\nwixident{get{\_}max{\_}generation}}{get:unmax:ungeneration}{NWppJ6t-AWRj3-X}\eatline
\nwused{\\{NWppJ6t-7vmg0-4}}\nwidentdefs{\\{{\nwixident{get{\_}max{\_}generation}}{get:unmax:ungeneration}}}\nwendcode{}\nwbegindocs{595}\nwdocspar
\nwenddocs{}\nwbegindocs{596} Some standard \GiNaC\ methods which are not very interesting for
end-user, who is working within functional programming set-up.
\nwenddocs{}\nwbegincode{597}\sublabel{NWppJ6t-AWRj3-Y}\nwmargintag{{\nwtagstyle{}\subpageref{NWppJ6t-AWRj3-Y}}}\moddef{public methods in figure class~{\nwtagstyle{}\subpageref{NWppJ6t-AWRj3-1}}}\plusendmoddef\Rm{}\nwstartdeflinemarkup\nwusesondefline{\\{NWppJ6t-7vmg0-4}}\nwprevnextdefs{NWppJ6t-AWRj3-X}{\relax}\nwenddeflinemarkup
    {\bf{}inline} {\it{}size\_t} {\it{}nops}() {\bf{}const} {\nwlbrace}{\bf{}return} 4+{\it{}nodes}.{\it{}size}();{\nwrbrace}
    {\bf{}ex} {\it{}op}({\it{}size\_t} {\it{}i}) {\bf{}const};
    //ex & let\_op(size\_t i);
    {\bf{}ex} {\it{}evalf}({\bf{}int} {\it{}level}=0) {\bf{}const};
    {\bf{}figure} {\it{}subs}({\bf{}const} {\bf{}ex} & {\it{}e}, {\bf{}unsigned} {\it{}options}=0) {\bf{}const};
    {\bf{}ex} {\it{}subs}({\bf{}const} {\it{}exmap} & {\it{}m}, {\bf{}unsigned} {\it{}options}=0) {\bf{}const};
    {\bf{}void} {\it{}archive}({\it{}archive\_node} &{\it{}n}) {\bf{}const};
    {\bf{}void} {\it{}read\_archive}({\bf{}const} {\it{}archive\_node} &{\it{}n}, {\bf{}lst} &{\it{}sym\_lst});
    {\bf{}bool} {\it{}info}({\bf{}unsigned} {\it{}inf}) {\bf{}const};
\nwindexdefn{\nwixident{nops}}{nops}{NWppJ6t-AWRj3-Y}\nwindexdefn{\nwixident{op}}{op}{NWppJ6t-AWRj3-Y}\nwindexdefn{\nwixident{evalf}}{evalf}{NWppJ6t-AWRj3-Y}\nwindexdefn{\nwixident{subs}}{subs}{NWppJ6t-AWRj3-Y}\nwindexdefn{\nwixident{archive}}{archive}{NWppJ6t-AWRj3-Y}\nwindexdefn{\nwixident{read{\_}archive}}{read:unarchive}{NWppJ6t-AWRj3-Y}\nwindexdefn{\nwixident{info}}{info}{NWppJ6t-AWRj3-Y}\eatline
\nwused{\\{NWppJ6t-7vmg0-4}}\nwidentdefs{\\{{\nwixident{archive}}{archive}}\\{{\nwixident{evalf}}{evalf}}\\{{\nwixident{info}}{info}}\\{{\nwixident{nops}}{nops}}\\{{\nwixident{op}}{op}}\\{{\nwixident{read{\_}archive}}{read:unarchive}}\\{{\nwixident{subs}}{subs}}}\nwidentuses{\\{{\nwixident{ex}}{ex}}\\{{\nwixident{figure}}{figure}}\\{{\nwixident{m}}{m}}\\{{\nwixident{nodes}}{nodes}}}\nwindexuse{\nwixident{ex}}{ex}{NWppJ6t-AWRj3-Y}\nwindexuse{\nwixident{figure}}{figure}{NWppJ6t-AWRj3-Y}\nwindexuse{\nwixident{m}}{m}{NWppJ6t-AWRj3-Y}\nwindexuse{\nwixident{nodes}}{nodes}{NWppJ6t-AWRj3-Y}\nwendcode{}\nwbegindocs{598}\nwdocspar
\nwenddocs{}\nwbegindocs{599}Printing and returning the objects list.
\nwenddocs{}\nwbegincode{600}\sublabel{NWppJ6t-7vmg0-6}\nwmargintag{{\nwtagstyle{}\subpageref{NWppJ6t-7vmg0-6}}}\moddef{figure header~{\nwtagstyle{}\subpageref{NWppJ6t-7vmg0-1}}}\plusendmoddef\Rm{}\nwstartdeflinemarkup\nwusesondefline{\\{NWppJ6t-3NCnUp-3}}\nwprevnextdefs{NWppJ6t-7vmg0-5}{NWppJ6t-7vmg0-7}\nwenddeflinemarkup
{\bf{}protected}:
    {\bf{}void} {\it{}do\_print}({\bf{}const} {\it{}print\_dflt} & {\it{}con}, {\bf{}unsigned} {\it{}level}) {\bf{}const};
    {\it{}return\_type\_t} {\it{}return\_type\_tinfo}() {\bf{}const};

\nwused{\\{NWppJ6t-3NCnUp-3}}\nwendcode{}\nwbegindocs{601}Update all cycles (with all children) in the given list.
\nwenddocs{}\nwbegincode{602}\sublabel{NWppJ6t-7vmg0-7}\nwmargintag{{\nwtagstyle{}\subpageref{NWppJ6t-7vmg0-7}}}\moddef{figure header~{\nwtagstyle{}\subpageref{NWppJ6t-7vmg0-1}}}\plusendmoddef\Rm{}\nwstartdeflinemarkup\nwusesondefline{\\{NWppJ6t-3NCnUp-3}}\nwprevnextdefs{NWppJ6t-7vmg0-6}{NWppJ6t-7vmg0-8}\nwenddeflinemarkup
    {\bf{}void} {\it{}update\_node\_lst}({\bf{}const} {\bf{}ex} & {\it{}inlist});
\nwindexdefn{\nwixident{update{\_}node{\_}lst}}{update:unnode:unlst}{NWppJ6t-7vmg0-7}\eatline
\nwused{\\{NWppJ6t-3NCnUp-3}}\nwidentdefs{\\{{\nwixident{update{\_}node{\_}lst}}{update:unnode:unlst}}}\nwidentuses{\\{{\nwixident{ex}}{ex}}}\nwindexuse{\nwixident{ex}}{ex}{NWppJ6t-7vmg0-7}\nwendcode{}\nwbegindocs{603}\nwdocspar
\nwenddocs{}\nwbegindocs{604}Update the entire figure.
\nwenddocs{}\nwbegincode{605}\sublabel{NWppJ6t-7vmg0-8}\nwmargintag{{\nwtagstyle{}\subpageref{NWppJ6t-7vmg0-8}}}\moddef{figure header~{\nwtagstyle{}\subpageref{NWppJ6t-7vmg0-1}}}\plusendmoddef\Rm{}\nwstartdeflinemarkup\nwusesondefline{\\{NWppJ6t-3NCnUp-3}}\nwprevnextdefs{NWppJ6t-7vmg0-7}{\relax}\nwenddeflinemarkup
    {\bf{}figure} {\it{}update\_cycles}();
{\nwrbrace};
{\it{}GINAC\_DECLARE\_UNARCHIVER}({\bf{}figure});\nwindexdefn{\nwixident{figure}}{figure}{NWppJ6t-7vmg0-8}
\nwindexdefn{\nwixident{update{\_}cycles}}{update:uncycles}{NWppJ6t-7vmg0-8}\eatline
\nwused{\\{NWppJ6t-3NCnUp-3}}\nwidentdefs{\\{{\nwixident{figure}}{figure}}\\{{\nwixident{update{\_}cycles}}{update:uncycles}}}\nwendcode{}\nwbegindocs{606}\nwdocspar
\nwenddocs{}\nwbegindocs{607}\nwdocspar
\subsubsection{Members of {\Tt{}\Rm{}{\bf{}figure}\nwendquote} class}
\label{sec:members-figure-class}
A knowledge of {\Tt{}\Rm{}{\bf{}figure}\nwendquote} class members may be useful for advanced users.

\nwenddocs{}\nwbegindocs{608}The real line and infinity are two cycles which are present at any figure.
\nwenddocs{}\nwbegincode{609}\sublabel{NWppJ6t-4OZeJw-1}\nwmargintag{{\nwtagstyle{}\subpageref{NWppJ6t-4OZeJw-1}}}\moddef{member of figure class~{\nwtagstyle{}\subpageref{NWppJ6t-4OZeJw-1}}}\endmoddef\Rm{}\nwstartdeflinemarkup\nwusesondefline{\\{NWppJ6t-7vmg0-1}}\nwprevnextdefs{\relax}{NWppJ6t-4OZeJw-2}\nwenddeflinemarkup
{\bf{}protected}:
 {\bf{}ex} {\it{}real\_line}, // the key for the real line
      {\it{}infinity}; // the key for cycle at infinity
\nwindexdefn{\nwixident{real{\_}line}}{real:unline}{NWppJ6t-4OZeJw-1}\nwindexdefn{\nwixident{infinity}}{infinity}{NWppJ6t-4OZeJw-1}\eatline
\nwalsodefined{\\{NWppJ6t-4OZeJw-2}\\{NWppJ6t-4OZeJw-3}\\{NWppJ6t-4OZeJw-4}\\{NWppJ6t-4OZeJw-5}}\nwused{\\{NWppJ6t-7vmg0-1}}\nwidentdefs{\\{{\nwixident{infinity}}{infinity}}\\{{\nwixident{real{\_}line}}{real:unline}}}\nwidentuses{\\{{\nwixident{ex}}{ex}}\\{{\nwixident{key}}{key}}}\nwindexuse{\nwixident{ex}}{ex}{NWppJ6t-4OZeJw-1}\nwindexuse{\nwixident{key}}{key}{NWppJ6t-4OZeJw-1}\nwendcode{}\nwbegindocs{610}\nwdocspar
\nwenddocs{}\nwbegindocs{611}We define separate metrics for the point and cycle spaces, see~\cite{Kisil12a}*{\S~4.2}.
\nwenddocs{}\nwbegincode{612}\sublabel{NWppJ6t-4OZeJw-2}\nwmargintag{{\nwtagstyle{}\subpageref{NWppJ6t-4OZeJw-2}}}\moddef{member of figure class~{\nwtagstyle{}\subpageref{NWppJ6t-4OZeJw-1}}}\plusendmoddef\Rm{}\nwstartdeflinemarkup\nwusesondefline{\\{NWppJ6t-7vmg0-1}}\nwprevnextdefs{NWppJ6t-4OZeJw-1}{NWppJ6t-4OZeJw-3}\nwenddeflinemarkup
    {\bf{}ex} {\it{}point\_metric}; // The metric of the point space encoded as a clifford\_unit object
    {\bf{}ex} {\it{}cycle\_metric}; // The metric of the cycle space encoded as a clifford\_unit object
\nwindexdefn{\nwixident{point{\_}metric}}{point:unmetric}{NWppJ6t-4OZeJw-2}\nwindexdefn{\nwixident{cycle{\_}metric}}{cycle:unmetric}{NWppJ6t-4OZeJw-2}\eatline
\nwused{\\{NWppJ6t-7vmg0-1}}\nwidentdefs{\\{{\nwixident{cycle{\_}metric}}{cycle:unmetric}}\\{{\nwixident{point{\_}metric}}{point:unmetric}}}\nwidentuses{\\{{\nwixident{ex}}{ex}}}\nwindexuse{\nwixident{ex}}{ex}{NWppJ6t-4OZeJw-2}\nwendcode{}\nwbegindocs{613}\nwdocspar
\nwenddocs{}\nwbegindocs{614}This is the {\Tt{}\Rm{}{\it{}hashmap}\nwendquote} of {\Tt{}\Rm{}{\bf{}cycle\_node}\nwendquote} which encode the relation in the figure.
\nwenddocs{}\nwbegincode{615}\sublabel{NWppJ6t-4OZeJw-3}\nwmargintag{{\nwtagstyle{}\subpageref{NWppJ6t-4OZeJw-3}}}\moddef{member of figure class~{\nwtagstyle{}\subpageref{NWppJ6t-4OZeJw-1}}}\plusendmoddef\Rm{}\nwstartdeflinemarkup\nwusesondefline{\\{NWppJ6t-7vmg0-1}}\nwprevnextdefs{NWppJ6t-4OZeJw-2}{NWppJ6t-4OZeJw-4}\nwenddeflinemarkup
    {\it{}exhashmap}\begin{math}<\end{math}{\bf{}cycle\_node}\begin{math}>\end{math} {\it{}nodes}; // List of cycle\_node, exhashmap\begin{math}<\end{math}cycle\_node\begin{math}>\end{math} object
\nwindexdefn{\nwixident{nodes}}{nodes}{NWppJ6t-4OZeJw-3}\eatline
\nwused{\\{NWppJ6t-7vmg0-1}}\nwidentdefs{\\{{\nwixident{nodes}}{nodes}}}\nwidentuses{\\{{\nwixident{cycle{\_}node}}{cycle:unnode}}}\nwindexuse{\nwixident{cycle{\_}node}}{cycle:unnode}{NWppJ6t-4OZeJw-3}\nwendcode{}\nwbegindocs{616}\nwdocspar
\nwenddocs{}\nwbegindocs{617}The following variable controls either we are doing exact or float
evaluations of cycles parameters.
\nwenddocs{}\nwbegincode{618}\sublabel{NWppJ6t-4OZeJw-4}\nwmargintag{{\nwtagstyle{}\subpageref{NWppJ6t-4OZeJw-4}}}\moddef{member of figure class~{\nwtagstyle{}\subpageref{NWppJ6t-4OZeJw-1}}}\plusendmoddef\Rm{}\nwstartdeflinemarkup\nwusesondefline{\\{NWppJ6t-7vmg0-1}}\nwprevnextdefs{NWppJ6t-4OZeJw-3}{NWppJ6t-4OZeJw-5}\nwenddeflinemarkup
    {\bf{}bool} {\it{}float\_evaluation}={\bf{}false};
\nwindexdefn{\nwixident{float{\_}evaluation}}{float:unevaluation}{NWppJ6t-4OZeJw-4}\eatline
\nwused{\\{NWppJ6t-7vmg0-1}}\nwidentdefs{\\{{\nwixident{float{\_}evaluation}}{float:unevaluation}}}\nwendcode{}\nwbegindocs{619}\nwdocspar
\nwenddocs{}\nwbegindocs{620}These are symbols for internal calculations, they are out of the interest we do not
count them in {\Tt{}\Rm{}{\it{}nops}()\nwendquote} methods.
\nwenddocs{}\nwbegincode{621}\sublabel{NWppJ6t-4OZeJw-5}\nwmargintag{{\nwtagstyle{}\subpageref{NWppJ6t-4OZeJw-5}}}\moddef{member of figure class~{\nwtagstyle{}\subpageref{NWppJ6t-4OZeJw-1}}}\plusendmoddef\Rm{}\nwstartdeflinemarkup\nwusesondefline{\\{NWppJ6t-7vmg0-1}}\nwprevnextdefs{NWppJ6t-4OZeJw-4}{\relax}\nwenddeflinemarkup
    {\bf{}ex} {\it{}k}, {\it{}m}; // realsymbols for symbolic calculations
    {\bf{}lst} {\it{}l};
\nwindexdefn{\nwixident{k}}{k}{NWppJ6t-4OZeJw-5}\nwindexdefn{\nwixident{l}}{l}{NWppJ6t-4OZeJw-5}\nwindexdefn{\nwixident{m}}{m}{NWppJ6t-4OZeJw-5}\eatline
\nwused{\\{NWppJ6t-7vmg0-1}}\nwidentdefs{\\{{\nwixident{k}}{k}}\\{{\nwixident{l}}{l}}\\{{\nwixident{m}}{m}}}\nwidentuses{\\{{\nwixident{ex}}{ex}}}\nwindexuse{\nwixident{ex}}{ex}{NWppJ6t-4OZeJw-5}\nwendcode{}\nwbegindocs{622}\nwdocspar
\nwenddocs{}\nwbegindocs{623}\nwdocspar
\subsection{\Asymptote\ customization}
\label{sec:asympt-cust}
The library provides a possibility to fine-tune \Asymptote\ output. We
provide some default styles, a user may customise them according to
existing needs.

\nwenddocs{}\nwbegindocs{624} We define a type for producing colouring scheme for \Asymptote\ drawing.
\nwenddocs{}\nwbegincode{625}\sublabel{NWppJ6t-43RSeU-1}\nwmargintag{{\nwtagstyle{}\subpageref{NWppJ6t-43RSeU-1}}}\moddef{asy styles~{\nwtagstyle{}\subpageref{NWppJ6t-43RSeU-1}}}\endmoddef\Rm{}\nwstartdeflinemarkup\nwusesondefline{\\{NWppJ6t-3NCnUp-3}}\nwprevnextdefs{\relax}{NWppJ6t-43RSeU-2}\nwenddeflinemarkup
{\bf{}using} {\it{}asy\_style}={\it{}std}::{\it{}function}\begin{math}<\end{math}{\it{}string}({\bf{}const} {\bf{}ex} &, {\bf{}const} {\bf{}ex} &, {\bf{}lst} &)\begin{math}>\end{math};
//typedef string (*asy\_style)(const ex &, const ex &, lst &);
{\bf{}inline} {\it{}string} {\it{}no\_color}({\bf{}const} {\bf{}ex} & {\it{}label}, {\bf{}const} {\bf{}ex} & {\it{}C}, {\bf{}lst} & {\it{}color}) {\nwlbrace}{\it{}color}={\bf{}lst}{\nwlbrace}0,0,0{\nwrbrace}; {\bf{}return} {\tt{}""};{\nwrbrace}
{\it{}string} {\it{}asy\_cycle\_color}({\bf{}const} {\bf{}ex} & {\it{}label}, {\bf{}const} {\bf{}ex} & {\it{}C}, {\bf{}lst} & {\it{}color});
{\bf{}const} {\it{}asy\_style} {\it{}default\_asy}={\it{}asy\_cycle\_color};\nwindexdefn{\nwixident{asy{\_}style}}{asy:unstyle}{NWppJ6t-43RSeU-1}

\nwalsodefined{\\{NWppJ6t-43RSeU-2}}\nwused{\\{NWppJ6t-3NCnUp-3}}\nwidentdefs{\\{{\nwixident{asy{\_}style}}{asy:unstyle}}}\nwidentuses{\\{{\nwixident{asy{\_}cycle{\_}color}}{asy:uncycle:uncolor}}\\{{\nwixident{ex}}{ex}}}\nwindexuse{\nwixident{asy{\_}cycle{\_}color}}{asy:uncycle:uncolor}{NWppJ6t-43RSeU-1}\nwindexuse{\nwixident{ex}}{ex}{NWppJ6t-43RSeU-1}\nwendcode{}\nwbegindocs{626}Similarly we produce a default labelling style.
\nwenddocs{}\nwbegincode{627}\sublabel{NWppJ6t-43RSeU-2}\nwmargintag{{\nwtagstyle{}\subpageref{NWppJ6t-43RSeU-2}}}\moddef{asy styles~{\nwtagstyle{}\subpageref{NWppJ6t-43RSeU-1}}}\plusendmoddef\Rm{}\nwstartdeflinemarkup\nwusesondefline{\\{NWppJ6t-3NCnUp-3}}\nwprevnextdefs{NWppJ6t-43RSeU-1}{\relax}\nwenddeflinemarkup
{\bf{}using} {\it{}label\_string}={\it{}std}::{\it{}function}\begin{math}<\end{math}{\it{}string}({\bf{}const} {\bf{}ex} &, {\bf{}const} {\bf{}ex} &, {\bf{}const} {\it{}string})\begin{math}>\end{math};
{\it{}string} {\it{}label\_pos}({\bf{}const} {\bf{}ex} & {\it{}label}, {\bf{}const} {\bf{}ex} & {\it{}C}, {\bf{}const} {\it{}string} {\it{}draw\_str});
{\bf{}inline} {\it{}string} {\it{}no\_label}({\bf{}const} {\bf{}ex} & {\it{}label}, {\bf{}const} {\bf{}ex} & {\it{}C}, {\bf{}const} {\it{}string} {\it{}draw\_str}) {\nwlbrace}{\bf{}return} {\tt{}""};{\nwrbrace}
{\bf{}const} {\it{}label\_string} {\it{}default\_label}={\it{}label\_pos};\nwindexdefn{\nwixident{label{\_}string}}{label:unstring}{NWppJ6t-43RSeU-2}

\nwused{\\{NWppJ6t-3NCnUp-3}}\nwidentdefs{\\{{\nwixident{label{\_}string}}{label:unstring}}}\nwidentuses{\\{{\nwixident{ex}}{ex}}\\{{\nwixident{label{\_}pos}}{label:unpos}}}\nwindexuse{\nwixident{ex}}{ex}{NWppJ6t-43RSeU-2}\nwindexuse{\nwixident{label{\_}pos}}{label:unpos}{NWppJ6t-43RSeU-2}\nwendcode{}\nwbegindocs{628}\nwdocspar
\section{Implementation of Classes}
\label{sec:impl-class}

\nwenddocs{}\nwbegindocs{629}This is the outline of the code.
\nwenddocs{}\nwbegincode{630}\sublabel{NWppJ6t-32jW6L-1}\nwmargintag{{\nwtagstyle{}\subpageref{NWppJ6t-32jW6L-1}}}\moddef{figure.cpp~{\nwtagstyle{}\subpageref{NWppJ6t-32jW6L-1}}}\endmoddef\Rm{}\nwstartdeflinemarkup\nwenddeflinemarkup
\LA{}license~{\nwtagstyle{}\subpageref{NWppJ6t-ZXuKx-1}}\RA{}
{\bf{}\char35{}include}{\tt{} "figure.h"}

{\bf{}namespace} {\it{}MoebInv} {\nwlbrace}
{\bf{}using} {\bf{}namespace} {\it{}std};
{\bf{}using} {\bf{}namespace} {\it{}GiNaC};
\LA{}figure library variables and constants~{\nwtagstyle{}\subpageref{NWppJ6t-A2bag-1}}\RA{}
\LA{}GiNaC declarations~{\nwtagstyle{}\subpageref{NWppJ6t-3vIsfh-1}}\RA{}
\LA{}auxillary function~{\nwtagstyle{}\subpageref{NWppJ6t-1KYtgD-1}}\RA{}
\LA{}add cycle relations~{\nwtagstyle{}\subpageref{NWppJ6t-3fVAGh-1}}\RA{}
\LA{}cycle data class~{\nwtagstyle{}\subpageref{NWppJ6t-2oFbmT-1}}\RA{}
\LA{}cycle relation class~{\nwtagstyle{}\subpageref{NWppJ6t-3bfNK9-1}}\RA{}
\LA{}subfigure class~{\nwtagstyle{}\subpageref{NWppJ6t-3dX0u0-1}}\RA{}
\LA{}cycle node class~{\nwtagstyle{}\subpageref{NWppJ6t-1HqLYY-1}}\RA{}
\LA{}figure class~{\nwtagstyle{}\subpageref{NWppJ6t-DKmU5-1}}\RA{}
\LA{}addional functions~{\nwtagstyle{}\subpageref{NWppJ6t-3LImg1-1}}\RA{}
{\nwrbrace} // namespace MoebInv

\nwnotused{figure.cpp}\nwidentuses{\\{{\nwixident{figure}}{figure}}\\{{\nwixident{MoebInv}}{MoebInv}}}\nwindexuse{\nwixident{figure}}{figure}{NWppJ6t-32jW6L-1}\nwindexuse{\nwixident{MoebInv}}{MoebInv}{NWppJ6t-32jW6L-1}\nwendcode{}\nwbegindocs{631}\nwdocspar
\nwenddocs{}\nwbegincode{632}\sublabel{NWppJ6t-A2bag-3}\nwmargintag{{\nwtagstyle{}\subpageref{NWppJ6t-A2bag-3}}}\moddef{figure library variables and constants~{\nwtagstyle{}\subpageref{NWppJ6t-A2bag-1}}}\plusendmoddef\Rm{}\nwstartdeflinemarkup\nwusesondefline{\\{NWppJ6t-32jW6L-1}}\nwprevnextdefs{NWppJ6t-A2bag-2}{NWppJ6t-A2bag-4}\nwenddeflinemarkup
{\bf{}unsigned} {\it{}do\_not\_update\_subfigure} = 0x0100;\nwindexdefn{\nwixident{do{\_}not{\_}update{\_}subfigure}}{do:unnot:unupdate:unsubfigure}{NWppJ6t-A2bag-3}

\nwused{\\{NWppJ6t-32jW6L-1}}\nwidentdefs{\\{{\nwixident{do{\_}not{\_}update{\_}subfigure}}{do:unnot:unupdate:unsubfigure}}}\nwendcode{}\nwbegindocs{633}This can de defined {\Tt{}\Rm{}{\bf{}false}\nwendquote} to prevent some diagnostic output to {\Tt{}\Rm{}{\it{}std}::{\it{}cerr}\nwendquote}.
\nwenddocs{}\nwbegincode{634}\sublabel{NWppJ6t-A2bag-4}\nwmargintag{{\nwtagstyle{}\subpageref{NWppJ6t-A2bag-4}}}\moddef{figure library variables and constants~{\nwtagstyle{}\subpageref{NWppJ6t-A2bag-1}}}\plusendmoddef\Rm{}\nwstartdeflinemarkup\nwusesondefline{\\{NWppJ6t-32jW6L-1}}\nwprevnextdefs{NWppJ6t-A2bag-3}{NWppJ6t-A2bag-5}\nwenddeflinemarkup
{\bf{}bool} {\it{}FIGURE\_DEBUG}={\bf{}true};
\nwindexdefn{\nwixident{FIGURE{\_}DEBUG}}{FIGURE:unDEBUG}{NWppJ6t-A2bag-4}\eatline
\nwused{\\{NWppJ6t-32jW6L-1}}\nwidentdefs{\\{{\nwixident{FIGURE{\_}DEBUG}}{FIGURE:unDEBUG}}}\nwendcode{}\nwbegindocs{635}\nwdocspar
\nwenddocs{}\nwbegindocs{636}This can de defined {\Tt{}\Rm{}{\bf{}false}\nwendquote} to prevent some diagnostic output to {\Tt{}\Rm{}{\it{}std}::{\it{}cerr}\nwendquote}.
\nwenddocs{}\nwbegincode{637}\sublabel{NWppJ6t-A2bag-5}\nwmargintag{{\nwtagstyle{}\subpageref{NWppJ6t-A2bag-5}}}\moddef{figure library variables and constants~{\nwtagstyle{}\subpageref{NWppJ6t-A2bag-1}}}\plusendmoddef\Rm{}\nwstartdeflinemarkup\nwusesondefline{\\{NWppJ6t-32jW6L-1}}\nwprevnextdefs{NWppJ6t-A2bag-4}{\relax}\nwenddeflinemarkup
{\bf{}bool} {\it{}show\_asy\_graphics}={\bf{}true};
\nwindexdefn{\nwixident{show{\_}asy{\_}graphics}}{show:unasy:ungraphics}{NWppJ6t-A2bag-5}\eatline
\nwused{\\{NWppJ6t-32jW6L-1}}\nwidentdefs{\\{{\nwixident{show{\_}asy{\_}graphics}}{show:unasy:ungraphics}}}\nwendcode{}\nwbegindocs{638}\nwdocspar
\nwenddocs{}\nwbegindocs{639}We use \GiNaC\ implementation macros for our classes.
\nwenddocs{}\nwbegincode{640}\sublabel{NWppJ6t-3vIsfh-1}\nwmargintag{{\nwtagstyle{}\subpageref{NWppJ6t-3vIsfh-1}}}\moddef{GiNaC declarations~{\nwtagstyle{}\subpageref{NWppJ6t-3vIsfh-1}}}\endmoddef\Rm{}\nwstartdeflinemarkup\nwusesondefline{\\{NWppJ6t-32jW6L-1}}\nwenddeflinemarkup
{\it{}GINAC\_IMPLEMENT\_REGISTERED\_CLASS\_OPT}({\bf{}cycle\_data}, {\bf{}basic},
                                     {\it{}print\_func}\begin{math}<\end{math}{\it{}print\_dflt}\begin{math}>\end{math}(&{\bf{}cycle\_data}::{\it{}do\_print}))

{\it{}GINAC\_IMPLEMENT\_REGISTERED\_CLASS\_OPT}({\bf{}cycle\_relation}, {\bf{}basic},
                                     {\it{}print\_func}\begin{math}<\end{math}{\it{}print\_dflt}\begin{math}>\end{math}(&{\bf{}cycle\_relation}::{\it{}do\_print}).
{\it{}print\_func}\begin{math}<\end{math}{\it{}print\_tree}\begin{math}>\end{math}(&{\bf{}cycle\_relation}::{\it{}do\_print\_tree}))

{\it{}GINAC\_IMPLEMENT\_REGISTERED\_CLASS\_OPT}({\bf{}subfigure}, {\bf{}basic},
                                     {\it{}print\_func}\begin{math}<\end{math}{\it{}print\_dflt}\begin{math}>\end{math}(&{\bf{}subfigure}::{\it{}do\_print}))

{\it{}GINAC\_IMPLEMENT\_REGISTERED\_CLASS\_OPT}({\bf{}cycle\_node}, {\bf{}basic},
                                     {\it{}print\_func}\begin{math}<\end{math}{\it{}print\_dflt}\begin{math}>\end{math}(&{\bf{}cycle\_node}::{\it{}do\_print}).
{\it{}print\_func}\begin{math}<\end{math}{\it{}print\_tree}\begin{math}>\end{math}(&{\bf{}cycle\_relation}::{\it{}do\_print\_tree}))

{\it{}GINAC\_IMPLEMENT\_REGISTERED\_CLASS\_OPT}({\bf{}figure}, {\bf{}basic},
                                     {\it{}print\_func}\begin{math}<\end{math}{\it{}print\_dflt}\begin{math}>\end{math}(&{\bf{}figure}::{\it{}do\_print}))

\nwused{\\{NWppJ6t-32jW6L-1}}\nwidentuses{\\{{\nwixident{cycle{\_}data}}{cycle:undata}}\\{{\nwixident{cycle{\_}node}}{cycle:unnode}}\\{{\nwixident{cycle{\_}relation}}{cycle:unrelation}}\\{{\nwixident{figure}}{figure}}\\{{\nwixident{subfigure}}{subfigure}}}\nwindexuse{\nwixident{cycle{\_}data}}{cycle:undata}{NWppJ6t-3vIsfh-1}\nwindexuse{\nwixident{cycle{\_}node}}{cycle:unnode}{NWppJ6t-3vIsfh-1}\nwindexuse{\nwixident{cycle{\_}relation}}{cycle:unrelation}{NWppJ6t-3vIsfh-1}\nwindexuse{\nwixident{figure}}{figure}{NWppJ6t-3vIsfh-1}\nwindexuse{\nwixident{subfigure}}{subfigure}{NWppJ6t-3vIsfh-1}\nwendcode{}\nwbegindocs{641}Exact solving of quadratic equations is not always practical, thus
we relay on some rounding methods. If the outcome is not good for you
increase the precision with {\Tt{}\Rm{}{\it{}GiNaC}::{\it{}Digits}\nwendquote}.
\nwenddocs{}\nwbegincode{642}\sublabel{NWppJ6t-1KYtgD-1}\nwmargintag{{\nwtagstyle{}\subpageref{NWppJ6t-1KYtgD-1}}}\moddef{auxillary function~{\nwtagstyle{}\subpageref{NWppJ6t-1KYtgD-1}}}\endmoddef\Rm{}\nwstartdeflinemarkup\nwusesondefline{\\{NWppJ6t-32jW6L-1}}\nwprevnextdefs{\relax}{NWppJ6t-1KYtgD-2}\nwenddeflinemarkup
{\bf{}const} {\bf{}ex} {\it{}epsilon}={\it{}GiNaC}::{\it{}pow}(10,-{\it{}Digits}\begin{math}\div\end{math}2);\nwindexdefn{\nwixident{ex}}{ex}{NWppJ6t-1KYtgD-1}
\nwindexdefn{\nwixident{epsilon}}{epsilon}{NWppJ6t-1KYtgD-1}\eatline
\nwalsodefined{\\{NWppJ6t-1KYtgD-2}}\nwused{\\{NWppJ6t-32jW6L-1}}\nwidentdefs{\\{{\nwixident{epsilon}}{epsilon}}\\{{\nwixident{ex}}{ex}}}\nwendcode{}\nwbegindocs{643}\nwdocspar
\nwenddocs{}\nwbegindocs{644}an auxillary function to find small numbers
\nwenddocs{}\nwbegincode{645}\sublabel{NWppJ6t-1KYtgD-2}\nwmargintag{{\nwtagstyle{}\subpageref{NWppJ6t-1KYtgD-2}}}\moddef{auxillary function~{\nwtagstyle{}\subpageref{NWppJ6t-1KYtgD-1}}}\plusendmoddef\Rm{}\nwstartdeflinemarkup\nwusesondefline{\\{NWppJ6t-32jW6L-1}}\nwprevnextdefs{NWppJ6t-1KYtgD-1}{\relax}\nwenddeflinemarkup
{\bf{}bool} {\it{}is\_less\_than\_epsilon}({\bf{}const} {\bf{}ex} & {\it{}x})
{\nwlbrace}
        {\bf{}return} ( {\it{}x}.{\it{}is\_zero}() \begin{math}\vee\end{math} {\it{}abs}({\it{}x}).{\it{}evalf}() \begin{math}<\end{math} {\it{}epsilon} ) ;
{\nwrbrace}
\nwindexdefn{\nwixident{is{\_}less{\_}than{\_}epsilon}}{is:unless:unthan:unepsilon}{NWppJ6t-1KYtgD-2}\eatline
\nwused{\\{NWppJ6t-32jW6L-1}}\nwidentdefs{\\{{\nwixident{is{\_}less{\_}than{\_}epsilon}}{is:unless:unthan:unepsilon}}}\nwidentuses{\\{{\nwixident{epsilon}}{epsilon}}\\{{\nwixident{evalf}}{evalf}}\\{{\nwixident{ex}}{ex}}}\nwindexuse{\nwixident{epsilon}}{epsilon}{NWppJ6t-1KYtgD-2}\nwindexuse{\nwixident{evalf}}{evalf}{NWppJ6t-1KYtgD-2}\nwindexuse{\nwixident{ex}}{ex}{NWppJ6t-1KYtgD-2}\nwendcode{}\nwbegindocs{646}\nwdocspar
\nwenddocs{}\nwbegindocs{647}\nwdocspar

\subsection{Implementation of {\Tt{}\Rm{}{\bf{}cycle\_data}\nwendquote} class}
\label{sec:impl-cycl-class}

\nwenddocs{}\nwbegindocs{648}Constructors
\nwenddocs{}\nwbegincode{649}\sublabel{NWppJ6t-2oFbmT-1}\nwmargintag{{\nwtagstyle{}\subpageref{NWppJ6t-2oFbmT-1}}}\moddef{cycle data class~{\nwtagstyle{}\subpageref{NWppJ6t-2oFbmT-1}}}\endmoddef\Rm{}\nwstartdeflinemarkup\nwusesondefline{\\{NWppJ6t-32jW6L-1}}\nwprevnextdefs{\relax}{NWppJ6t-2oFbmT-2}\nwenddeflinemarkup
{\bf{}cycle\_data}::{\bf{}cycle\_data}() : {\it{}k}(), {\it{}l}(), {\it{}m}()
{\nwlbrace}
    ;
{\nwrbrace}

\nwalsodefined{\\{NWppJ6t-2oFbmT-2}\\{NWppJ6t-2oFbmT-3}\\{NWppJ6t-2oFbmT-4}\\{NWppJ6t-2oFbmT-5}\\{NWppJ6t-2oFbmT-6}\\{NWppJ6t-2oFbmT-7}\\{NWppJ6t-2oFbmT-8}\\{NWppJ6t-2oFbmT-9}\\{NWppJ6t-2oFbmT-A}\\{NWppJ6t-2oFbmT-B}\\{NWppJ6t-2oFbmT-C}\\{NWppJ6t-2oFbmT-D}\\{NWppJ6t-2oFbmT-E}\\{NWppJ6t-2oFbmT-F}\\{NWppJ6t-2oFbmT-G}\\{NWppJ6t-2oFbmT-H}\\{NWppJ6t-2oFbmT-I}\\{NWppJ6t-2oFbmT-J}\\{NWppJ6t-2oFbmT-K}\\{NWppJ6t-2oFbmT-L}\\{NWppJ6t-2oFbmT-M}\\{NWppJ6t-2oFbmT-N}\\{NWppJ6t-2oFbmT-O}}\nwused{\\{NWppJ6t-32jW6L-1}}\nwidentuses{\\{{\nwixident{cycle{\_}data}}{cycle:undata}}\\{{\nwixident{k}}{k}}\\{{\nwixident{l}}{l}}\\{{\nwixident{m}}{m}}}\nwindexuse{\nwixident{cycle{\_}data}}{cycle:undata}{NWppJ6t-2oFbmT-1}\nwindexuse{\nwixident{k}}{k}{NWppJ6t-2oFbmT-1}\nwindexuse{\nwixident{l}}{l}{NWppJ6t-2oFbmT-1}\nwindexuse{\nwixident{m}}{m}{NWppJ6t-2oFbmT-1}\nwendcode{}\nwbegindocs{650}Constructors
\nwenddocs{}\nwbegincode{651}\sublabel{NWppJ6t-2oFbmT-2}\nwmargintag{{\nwtagstyle{}\subpageref{NWppJ6t-2oFbmT-2}}}\moddef{cycle data class~{\nwtagstyle{}\subpageref{NWppJ6t-2oFbmT-1}}}\plusendmoddef\Rm{}\nwstartdeflinemarkup\nwusesondefline{\\{NWppJ6t-32jW6L-1}}\nwprevnextdefs{NWppJ6t-2oFbmT-1}{NWppJ6t-2oFbmT-3}\nwenddeflinemarkup
{\bf{}cycle\_data}::{\bf{}cycle\_data}({\bf{}const} {\bf{}ex} & {\it{}C})
{\nwlbrace}
    {\bf{}if} ({\it{}is\_a}\begin{math}<\end{math}{\bf{}cycle}\begin{math}>\end{math}({\it{}C})) {\nwlbrace}
        {\bf{}cycle} {\it{}C\_new}={\it{}ex\_to}\begin{math}<\end{math}{\bf{}cycle}\begin{math}>\end{math}({\it{}C}).{\it{}normalize}();
        {\it{}k}={\it{}C\_new}.{\it{}get\_k}();
        {\it{}l}={\it{}C\_new}.{\it{}get\_l}();
        {\it{}m}={\it{}C\_new}.{\it{}get\_m}();
    {\nwrbrace} {\bf{}else} {\bf{}if} ({\it{}is\_a}\begin{math}<\end{math}{\bf{}cycle\_data}\begin{math}>\end{math}({\it{}C}))
        \begin{math}\ast\end{math}{\it{}this}={\it{}ex\_to}\begin{math}<\end{math}{\bf{}cycle\_data}\begin{math}>\end{math}({\it{}C});
    {\bf{}else}
        {\bf{}throw}({\it{}std}::{\it{}invalid\_argument}({\tt{}"cycle\_data(): accept only cycle or cycle\_data"}
                                    {\tt{}" as the parameter"}));

{\nwrbrace}

\nwused{\\{NWppJ6t-32jW6L-1}}\nwidentuses{\\{{\nwixident{cycle{\_}data}}{cycle:undata}}\\{{\nwixident{ex}}{ex}}\\{{\nwixident{k}}{k}}\\{{\nwixident{l}}{l}}\\{{\nwixident{m}}{m}}}\nwindexuse{\nwixident{cycle{\_}data}}{cycle:undata}{NWppJ6t-2oFbmT-2}\nwindexuse{\nwixident{ex}}{ex}{NWppJ6t-2oFbmT-2}\nwindexuse{\nwixident{k}}{k}{NWppJ6t-2oFbmT-2}\nwindexuse{\nwixident{l}}{l}{NWppJ6t-2oFbmT-2}\nwindexuse{\nwixident{m}}{m}{NWppJ6t-2oFbmT-2}\nwendcode{}\nwbegindocs{652}Constructors.
\nwenddocs{}\nwbegincode{653}\sublabel{NWppJ6t-2oFbmT-3}\nwmargintag{{\nwtagstyle{}\subpageref{NWppJ6t-2oFbmT-3}}}\moddef{cycle data class~{\nwtagstyle{}\subpageref{NWppJ6t-2oFbmT-1}}}\plusendmoddef\Rm{}\nwstartdeflinemarkup\nwusesondefline{\\{NWppJ6t-32jW6L-1}}\nwprevnextdefs{NWppJ6t-2oFbmT-2}{NWppJ6t-2oFbmT-4}\nwenddeflinemarkup
{\bf{}cycle\_data}::{\bf{}cycle\_data}({\bf{}const} {\bf{}ex} & {\it{}k1}, {\bf{}const} {\bf{}ex} {\it{}l1}, {\bf{}const} {\bf{}ex} &{\it{}m1}, {\bf{}bool} {\it{}normalize}) : {\it{}k}({\it{}k1}), {\it{}l}({\it{}l1}), {\it{}m}({\it{}m1})
{\nwlbrace}
    {\bf{}if} ({\it{}normalize}) {\nwlbrace}
        {\bf{}ex} {\it{}ratio} = 0;
        {\bf{}if} (\begin{math}\neg\end{math}{\it{}k}.{\it{}is\_zero}()) // First non-zero coefficient among k, m, l\_0, l\_1, ... is set to 1
            {\it{}ratio} = {\it{}k};
        {\bf{}else} {\bf{}if} (\begin{math}\neg\end{math}{\it{}m}.{\it{}is\_zero}())
            {\it{}ratio} = {\it{}m};
        {\bf{}else} {\nwlbrace}
            {\bf{}for} ({\bf{}unsigned} {\bf{}int} {\it{}i}=0; {\it{}i}\begin{math}<\end{math}{\it{}get\_dim}(); {\it{}i}\protect\PP)
                {\bf{}if} (\begin{math}\neg\end{math}{\it{}l}.{\it{}subs}({\it{}l}.{\it{}op}(1) \begin{math}\equiv\end{math} {\it{}i}).{\it{}is\_zero}()) {\nwlbrace}
                    {\it{}ratio} = {\it{}l}.{\it{}subs}({\it{}l}.{\it{}op}(1) \begin{math}\equiv\end{math} {\it{}i});
                    {\bf{}break};
                {\nwrbrace}
        {\nwrbrace}

        {\bf{}if} (\begin{math}\neg\end{math}{\it{}ratio}.{\it{}is\_zero}()) {\nwlbrace}
            {\it{}k}={\it{}k}\begin{math}\div\end{math}{\it{}ratio};
            {\it{}l}={\bf{}indexed}(({\it{}l}.{\it{}op}(0)\begin{math}\div\end{math}{\it{}ratio}).{\it{}evalm}().{\it{}normal}(), {\it{}l}.{\it{}op}(1));
            {\it{}m}=({\it{}m}\begin{math}\div\end{math}{\it{}ratio}).{\it{}normal}();
        {\nwrbrace}
    {\nwrbrace}
{\nwrbrace}

\nwused{\\{NWppJ6t-32jW6L-1}}\nwidentuses{\\{{\nwixident{cycle{\_}data}}{cycle:undata}}\\{{\nwixident{ex}}{ex}}\\{{\nwixident{get{\_}dim()}}{get:undim()}}\\{{\nwixident{k}}{k}}\\{{\nwixident{l}}{l}}\\{{\nwixident{m}}{m}}\\{{\nwixident{op}}{op}}\\{{\nwixident{subs}}{subs}}}\nwindexuse{\nwixident{cycle{\_}data}}{cycle:undata}{NWppJ6t-2oFbmT-3}\nwindexuse{\nwixident{ex}}{ex}{NWppJ6t-2oFbmT-3}\nwindexuse{\nwixident{get{\_}dim()}}{get:undim()}{NWppJ6t-2oFbmT-3}\nwindexuse{\nwixident{k}}{k}{NWppJ6t-2oFbmT-3}\nwindexuse{\nwixident{l}}{l}{NWppJ6t-2oFbmT-3}\nwindexuse{\nwixident{m}}{m}{NWppJ6t-2oFbmT-3}\nwindexuse{\nwixident{op}}{op}{NWppJ6t-2oFbmT-3}\nwindexuse{\nwixident{subs}}{subs}{NWppJ6t-2oFbmT-3}\nwendcode{}\nwbegindocs{654}\nwdocspar
\nwenddocs{}\nwbegincode{655}\sublabel{NWppJ6t-2oFbmT-4}\nwmargintag{{\nwtagstyle{}\subpageref{NWppJ6t-2oFbmT-4}}}\moddef{cycle data class~{\nwtagstyle{}\subpageref{NWppJ6t-2oFbmT-1}}}\plusendmoddef\Rm{}\nwstartdeflinemarkup\nwusesondefline{\\{NWppJ6t-32jW6L-1}}\nwprevnextdefs{NWppJ6t-2oFbmT-3}{NWppJ6t-2oFbmT-5}\nwenddeflinemarkup
{\it{}return\_type\_t} {\bf{}cycle\_data}::{\it{}return\_type\_tinfo}() {\bf{}const}
{\nwlbrace}
    {\bf{}return} {\it{}make\_return\_type\_t}\begin{math}<\end{math}{\bf{}cycle\_data}\begin{math}>\end{math}();
{\nwrbrace}

\nwused{\\{NWppJ6t-32jW6L-1}}\nwidentuses{\\{{\nwixident{cycle{\_}data}}{cycle:undata}}}\nwindexuse{\nwixident{cycle{\_}data}}{cycle:undata}{NWppJ6t-2oFbmT-4}\nwendcode{}\nwbegindocs{656}\nwdocspar
\nwenddocs{}\nwbegincode{657}\sublabel{NWppJ6t-2oFbmT-5}\nwmargintag{{\nwtagstyle{}\subpageref{NWppJ6t-2oFbmT-5}}}\moddef{cycle data class~{\nwtagstyle{}\subpageref{NWppJ6t-2oFbmT-1}}}\plusendmoddef\Rm{}\nwstartdeflinemarkup\nwusesondefline{\\{NWppJ6t-32jW6L-1}}\nwprevnextdefs{NWppJ6t-2oFbmT-4}{NWppJ6t-2oFbmT-6}\nwenddeflinemarkup
{\bf{}int} {\bf{}cycle\_data}::{\it{}compare\_same\_type}({\bf{}const} {\bf{}basic} &{\it{}other}) {\bf{}const}\nwindexdefn{\nwixident{cycle{\_}data}}{cycle:undata}{NWppJ6t-2oFbmT-5}
{\nwlbrace}
       {\it{}GINAC\_ASSERT}({\it{}is\_a}\begin{math}<\end{math}{\bf{}cycle\_data}\begin{math}>\end{math}({\it{}other}));
       {\bf{}return} {\it{}inherited}::{\it{}compare\_same\_type}({\it{}other});
{\nwrbrace}

\nwused{\\{NWppJ6t-32jW6L-1}}\nwidentdefs{\\{{\nwixident{cycle{\_}data}}{cycle:undata}}}\nwendcode{}\nwbegindocs{658}Printing the cycle data
\nwenddocs{}\nwbegincode{659}\sublabel{NWppJ6t-2oFbmT-6}\nwmargintag{{\nwtagstyle{}\subpageref{NWppJ6t-2oFbmT-6}}}\moddef{cycle data class~{\nwtagstyle{}\subpageref{NWppJ6t-2oFbmT-1}}}\plusendmoddef\Rm{}\nwstartdeflinemarkup\nwusesondefline{\\{NWppJ6t-32jW6L-1}}\nwprevnextdefs{NWppJ6t-2oFbmT-5}{NWppJ6t-2oFbmT-7}\nwenddeflinemarkup
{\bf{}void} {\bf{}cycle\_data}::{\it{}do\_print}({\bf{}const} {\it{}print\_dflt} & {\it{}con}, {\bf{}unsigned} {\it{}level}) {\bf{}const}\nwindexdefn{\nwixident{cycle{\_}data}}{cycle:undata}{NWppJ6t-2oFbmT-6}
{\nwlbrace}
    {\it{}con}.{\it{}s} \begin{math}\ll\end{math} {\tt{}"("};
    {\it{}k}.{\it{}print}({\it{}con}, {\it{}level});
    {\it{}con}.{\it{}s} \begin{math}\ll\end{math} {\tt{}", "};
    {\it{}l}.{\it{}print}({\it{}con}, {\it{}level});
    {\it{}con}.{\it{}s} \begin{math}\ll\end{math} {\tt{}", "};
    {\it{}m}.{\it{}print}({\it{}con}, {\it{}level});
    {\it{}con}.{\it{}s} \begin{math}\ll\end{math} {\tt{}")"};
{\nwrbrace}

\nwused{\\{NWppJ6t-32jW6L-1}}\nwidentdefs{\\{{\nwixident{cycle{\_}data}}{cycle:undata}}}\nwidentuses{\\{{\nwixident{k}}{k}}\\{{\nwixident{l}}{l}}\\{{\nwixident{m}}{m}}}\nwindexuse{\nwixident{k}}{k}{NWppJ6t-2oFbmT-6}\nwindexuse{\nwixident{l}}{l}{NWppJ6t-2oFbmT-6}\nwindexuse{\nwixident{m}}{m}{NWppJ6t-2oFbmT-6}\nwendcode{}\nwbegindocs{660}Printing the cycle data in the float mode if possible.
\nwenddocs{}\nwbegincode{661}\sublabel{NWppJ6t-2oFbmT-7}\nwmargintag{{\nwtagstyle{}\subpageref{NWppJ6t-2oFbmT-7}}}\moddef{cycle data class~{\nwtagstyle{}\subpageref{NWppJ6t-2oFbmT-1}}}\plusendmoddef\Rm{}\nwstartdeflinemarkup\nwusesondefline{\\{NWppJ6t-32jW6L-1}}\nwprevnextdefs{NWppJ6t-2oFbmT-6}{NWppJ6t-2oFbmT-8}\nwenddeflinemarkup
{\bf{}void} {\bf{}cycle\_data}::{\it{}do\_print\_double}({\bf{}const} {\it{}print\_dflt} & {\it{}con}, {\bf{}unsigned} {\it{}level}) {\bf{}const}\nwindexdefn{\nwixident{cycle{\_}data}}{cycle:undata}{NWppJ6t-2oFbmT-7}
{\nwlbrace}
    {\bf{}if} (\begin{math}\neg\end{math} {\it{}is\_a}\begin{math}<\end{math}{\bf{}numeric}\begin{math}>\end{math}({\it{}get\_dim}())) {\nwlbrace}
        {\it{}do\_print}({\it{}con}, {\it{}level});
    {\nwrbrace} {\bf{}else} {\nwlbrace}

\nwused{\\{NWppJ6t-32jW6L-1}}\nwidentdefs{\\{{\nwixident{cycle{\_}data}}{cycle:undata}}}\nwidentuses{\\{{\nwixident{do{\_}print{\_}double}}{do:unprint:undouble}}\\{{\nwixident{get{\_}dim()}}{get:undim()}}\\{{\nwixident{numeric}}{numeric}}}\nwindexuse{\nwixident{do{\_}print{\_}double}}{do:unprint:undouble}{NWppJ6t-2oFbmT-7}\nwindexuse{\nwixident{get{\_}dim()}}{get:undim()}{NWppJ6t-2oFbmT-7}\nwindexuse{\nwixident{numeric}}{numeric}{NWppJ6t-2oFbmT-7}\nwendcode{}\nwbegindocs{662}Check if conversion to double is possible and accurate.
\nwenddocs{}\nwbegincode{663}\sublabel{NWppJ6t-2oFbmT-8}\nwmargintag{{\nwtagstyle{}\subpageref{NWppJ6t-2oFbmT-8}}}\moddef{cycle data class~{\nwtagstyle{}\subpageref{NWppJ6t-2oFbmT-1}}}\plusendmoddef\Rm{}\nwstartdeflinemarkup\nwusesondefline{\\{NWppJ6t-32jW6L-1}}\nwprevnextdefs{NWppJ6t-2oFbmT-7}{NWppJ6t-2oFbmT-9}\nwenddeflinemarkup
        {\it{}con}.{\it{}s} \begin{math}\ll\end{math} {\tt{}"("};
        {\bf{}if} (({\it{}is\_a}\begin{math}<\end{math}{\bf{}numeric}\begin{math}>\end{math}({\it{}k}) \begin{math}\wedge\end{math} \begin{math}\neg\end{math} {\it{}ex\_to}\begin{math}<\end{math}{\bf{}numeric}\begin{math}>\end{math}({\it{}k}).{\it{}is\_crational}())
            \begin{math}\vee\end{math} {\it{}is\_a}\begin{math}<\end{math}{\bf{}numeric}\begin{math}>\end{math}({\it{}k}.{\it{}evalf}())) {\nwlbrace}
            {\bf{}ex} {\it{}f}={\it{}k}.{\it{}evalf}();
            \LA{}common part of float output~{\nwtagstyle{}\subpageref{NWppJ6t-tXDC3-1}}\RA{}

\nwused{\\{NWppJ6t-32jW6L-1}}\nwidentuses{\\{{\nwixident{evalf}}{evalf}}\\{{\nwixident{ex}}{ex}}\\{{\nwixident{k}}{k}}\\{{\nwixident{numeric}}{numeric}}}\nwindexuse{\nwixident{evalf}}{evalf}{NWppJ6t-2oFbmT-8}\nwindexuse{\nwixident{ex}}{ex}{NWppJ6t-2oFbmT-8}\nwindexuse{\nwixident{k}}{k}{NWppJ6t-2oFbmT-8}\nwindexuse{\nwixident{numeric}}{numeric}{NWppJ6t-2oFbmT-8}\nwendcode{}\nwbegindocs{664}Here is the repeating part
\nwenddocs{}\nwbegincode{665}\sublabel{NWppJ6t-tXDC3-1}\nwmargintag{{\nwtagstyle{}\subpageref{NWppJ6t-tXDC3-1}}}\moddef{common part of float output~{\nwtagstyle{}\subpageref{NWppJ6t-tXDC3-1}}}\endmoddef\Rm{}\nwstartdeflinemarkup\nwusesondefline{\\{NWppJ6t-2oFbmT-8}\\{NWppJ6t-2oFbmT-A}\\{NWppJ6t-2oFbmT-B}}\nwenddeflinemarkup
            {\it{}con}.{\it{}s} \begin{math}\ll\end{math} {\it{}ex\_to}\begin{math}<\end{math}{\bf{}numeric}\begin{math}>\end{math}({\it{}f}).{\it{}to\_double}(); // only real part is converted
            {\bf{}if} (\begin{math}\neg\end{math} {\it{}ex\_to}\begin{math}<\end{math}{\bf{}numeric}\begin{math}>\end{math}({\it{}f}).{\it{}is\_real}()) {\nwlbrace}
                {\bf{}double} {\it{}b}={\it{}ex\_to}\begin{math}<\end{math}{\bf{}numeric}\begin{math}>\end{math}({\it{}f}.{\it{}imag\_part}()).{\it{}to\_double}();
                {\bf{}if} ({\it{}b}\begin{math}>\end{math}0)
                    {\it{}con}.{\it{}s} \begin{math}\ll\end{math} {\tt{}"+"};
                {\it{}con}.{\it{}s} \begin{math}\ll\end{math} {\it{}b} \begin{math}\ll\end{math} {\tt{}"*I"};
            {\nwrbrace}

\nwused{\\{NWppJ6t-2oFbmT-8}\\{NWppJ6t-2oFbmT-A}\\{NWppJ6t-2oFbmT-B}}\nwidentuses{\\{{\nwixident{numeric}}{numeric}}}\nwindexuse{\nwixident{numeric}}{numeric}{NWppJ6t-tXDC3-1}\nwendcode{}\nwbegindocs{666}back to our routine.
\nwenddocs{}\nwbegincode{667}\sublabel{NWppJ6t-2oFbmT-9}\nwmargintag{{\nwtagstyle{}\subpageref{NWppJ6t-2oFbmT-9}}}\moddef{cycle data class~{\nwtagstyle{}\subpageref{NWppJ6t-2oFbmT-1}}}\plusendmoddef\Rm{}\nwstartdeflinemarkup\nwusesondefline{\\{NWppJ6t-32jW6L-1}}\nwprevnextdefs{NWppJ6t-2oFbmT-8}{NWppJ6t-2oFbmT-A}\nwenddeflinemarkup
        {\nwrbrace} {\bf{}else}
            {\it{}k}.{\it{}print}({\it{}con}, {\it{}level});
        {\it{}con}.{\it{}s} \begin{math}\ll\end{math} {\tt{}", [["};

\nwused{\\{NWppJ6t-32jW6L-1}}\nwidentuses{\\{{\nwixident{k}}{k}}}\nwindexuse{\nwixident{k}}{k}{NWppJ6t-2oFbmT-9}\nwendcode{}\nwbegindocs{668}Run through all elements of the {\Tt{}\Rm{}{\it{}l}\nwendquote} vector.
\nwenddocs{}\nwbegincode{669}\sublabel{NWppJ6t-2oFbmT-A}\nwmargintag{{\nwtagstyle{}\subpageref{NWppJ6t-2oFbmT-A}}}\moddef{cycle data class~{\nwtagstyle{}\subpageref{NWppJ6t-2oFbmT-1}}}\plusendmoddef\Rm{}\nwstartdeflinemarkup\nwusesondefline{\\{NWppJ6t-32jW6L-1}}\nwprevnextdefs{NWppJ6t-2oFbmT-9}{NWppJ6t-2oFbmT-B}\nwenddeflinemarkup
        {\bf{}int} {\it{}D}={\it{}ex\_to}\begin{math}<\end{math}{\bf{}numeric}\begin{math}>\end{math}({\it{}get\_dim}()).{\it{}to\_int}();
        {\bf{}for}({\bf{}int} {\it{}i}=0; {\it{}i}\begin{math}<\end{math} {\it{}D}; \protect\PP{\it{}i}) {\nwlbrace}
            {\bf{}if} (({\it{}is\_a}\begin{math}<\end{math}{\bf{}numeric}\begin{math}>\end{math}({\it{}l}.{\it{}op}(0).{\it{}op}({\it{}i})) \begin{math}\wedge\end{math} \begin{math}\neg\end{math} {\it{}ex\_to}\begin{math}<\end{math}{\bf{}numeric}\begin{math}>\end{math}({\it{}l}.{\it{}op}(0).{\it{}op}({\it{}i})).{\it{}is\_crational}())
                \begin{math}\vee\end{math} {\it{}is\_a}\begin{math}<\end{math}{\bf{}numeric}\begin{math}>\end{math}({\it{}l}.{\it{}op}(0).{\it{}op}({\it{}i}).{\it{}evalf}())) {\nwlbrace}
                {\bf{}ex} {\it{}f}={\it{}ex\_to}\begin{math}<\end{math}{\bf{}numeric}\begin{math}>\end{math}({\it{}l}.{\it{}op}(0).{\it{}op}({\it{}i})).{\it{}evalf}();
                \LA{}common part of float output~{\nwtagstyle{}\subpageref{NWppJ6t-tXDC3-1}}\RA{}
            {\nwrbrace} {\bf{}else}
                {\it{}l}.{\it{}op}(0).{\it{}op}({\it{}i}).{\it{}print}({\it{}con}, {\it{}level});
            {\bf{}if} ({\it{}i}\begin{math}<\end{math}{\it{}D}-1)
                {\it{}con}.{\it{}s} \begin{math}\ll\end{math} {\tt{}","};
        {\nwrbrace}
        {\it{}con}.{\it{}s} \begin{math}\ll\end{math} {\tt{}"]]"};
        {\it{}l}.{\it{}op}(1).{\it{}print}({\it{}con}, {\it{}level});

\nwused{\\{NWppJ6t-32jW6L-1}}\nwidentuses{\\{{\nwixident{evalf}}{evalf}}\\{{\nwixident{ex}}{ex}}\\{{\nwixident{get{\_}dim()}}{get:undim()}}\\{{\nwixident{l}}{l}}\\{{\nwixident{numeric}}{numeric}}\\{{\nwixident{op}}{op}}}\nwindexuse{\nwixident{evalf}}{evalf}{NWppJ6t-2oFbmT-A}\nwindexuse{\nwixident{ex}}{ex}{NWppJ6t-2oFbmT-A}\nwindexuse{\nwixident{get{\_}dim()}}{get:undim()}{NWppJ6t-2oFbmT-A}\nwindexuse{\nwixident{l}}{l}{NWppJ6t-2oFbmT-A}\nwindexuse{\nwixident{numeric}}{numeric}{NWppJ6t-2oFbmT-A}\nwindexuse{\nwixident{op}}{op}{NWppJ6t-2oFbmT-A}\nwendcode{}\nwbegindocs{670}Finishing with the {\Tt{}\Rm{}{\it{}m}\nwendquote} part.
\nwenddocs{}\nwbegincode{671}\sublabel{NWppJ6t-2oFbmT-B}\nwmargintag{{\nwtagstyle{}\subpageref{NWppJ6t-2oFbmT-B}}}\moddef{cycle data class~{\nwtagstyle{}\subpageref{NWppJ6t-2oFbmT-1}}}\plusendmoddef\Rm{}\nwstartdeflinemarkup\nwusesondefline{\\{NWppJ6t-32jW6L-1}}\nwprevnextdefs{NWppJ6t-2oFbmT-A}{NWppJ6t-2oFbmT-C}\nwenddeflinemarkup
        {\it{}con}.{\it{}s} \begin{math}\ll\end{math} {\tt{}", "};
        {\bf{}if} (({\it{}is\_a}\begin{math}<\end{math}{\bf{}numeric}\begin{math}>\end{math}({\it{}m}) \begin{math}\wedge\end{math} \begin{math}\neg\end{math} {\it{}ex\_to}\begin{math}<\end{math}{\bf{}numeric}\begin{math}>\end{math}({\it{}m}).{\it{}is\_crational}())
            \begin{math}\vee\end{math} {\it{}is\_a}\begin{math}<\end{math}{\bf{}numeric}\begin{math}>\end{math}({\it{}m}.{\it{}evalf}())) {\nwlbrace}
            {\bf{}ex} {\it{}f}={\it{}m}.{\it{}evalf}();
            \LA{}common part of float output~{\nwtagstyle{}\subpageref{NWppJ6t-tXDC3-1}}\RA{}
        {\nwrbrace} {\bf{}else}
            {\it{}m}.{\it{}print}({\it{}con}, {\it{}level});
        {\it{}con}.{\it{}s} \begin{math}\ll\end{math} {\tt{}")"};
    {\nwrbrace}
{\nwrbrace}

\nwused{\\{NWppJ6t-32jW6L-1}}\nwidentuses{\\{{\nwixident{evalf}}{evalf}}\\{{\nwixident{ex}}{ex}}\\{{\nwixident{m}}{m}}\\{{\nwixident{numeric}}{numeric}}}\nwindexuse{\nwixident{evalf}}{evalf}{NWppJ6t-2oFbmT-B}\nwindexuse{\nwixident{ex}}{ex}{NWppJ6t-2oFbmT-B}\nwindexuse{\nwixident{m}}{m}{NWppJ6t-2oFbmT-B}\nwindexuse{\nwixident{numeric}}{numeric}{NWppJ6t-2oFbmT-B}\nwendcode{}\nwbegindocs{672}\nwdocspar
\nwenddocs{}\nwbegincode{673}\sublabel{NWppJ6t-2oFbmT-C}\nwmargintag{{\nwtagstyle{}\subpageref{NWppJ6t-2oFbmT-C}}}\moddef{cycle data class~{\nwtagstyle{}\subpageref{NWppJ6t-2oFbmT-1}}}\plusendmoddef\Rm{}\nwstartdeflinemarkup\nwusesondefline{\\{NWppJ6t-32jW6L-1}}\nwprevnextdefs{NWppJ6t-2oFbmT-B}{NWppJ6t-2oFbmT-D}\nwenddeflinemarkup
{\bf{}void} {\bf{}cycle\_data}::{\it{}archive}({\it{}archive\_node} &{\it{}n}) {\bf{}const}\nwindexdefn{\nwixident{cycle{\_}data}}{cycle:undata}{NWppJ6t-2oFbmT-C}
{\nwlbrace}
    {\it{}inherited}::{\it{}archive}({\it{}n});
    {\it{}n}.{\it{}add\_ex}({\tt{}"k-val"}, {\it{}k});
    {\it{}n}.{\it{}add\_ex}({\tt{}"l-val"}, {\it{}l});
    {\it{}n}.{\it{}add\_ex}({\tt{}"m-val"}, {\it{}m});
{\nwrbrace}

\nwused{\\{NWppJ6t-32jW6L-1}}\nwidentdefs{\\{{\nwixident{cycle{\_}data}}{cycle:undata}}}\nwidentuses{\\{{\nwixident{archive}}{archive}}\\{{\nwixident{k}}{k}}\\{{\nwixident{l}}{l}}\\{{\nwixident{m}}{m}}}\nwindexuse{\nwixident{archive}}{archive}{NWppJ6t-2oFbmT-C}\nwindexuse{\nwixident{k}}{k}{NWppJ6t-2oFbmT-C}\nwindexuse{\nwixident{l}}{l}{NWppJ6t-2oFbmT-C}\nwindexuse{\nwixident{m}}{m}{NWppJ6t-2oFbmT-C}\nwendcode{}\nwbegindocs{674}\nwdocspar
\nwenddocs{}\nwbegincode{675}\sublabel{NWppJ6t-2oFbmT-D}\nwmargintag{{\nwtagstyle{}\subpageref{NWppJ6t-2oFbmT-D}}}\moddef{cycle data class~{\nwtagstyle{}\subpageref{NWppJ6t-2oFbmT-1}}}\plusendmoddef\Rm{}\nwstartdeflinemarkup\nwusesondefline{\\{NWppJ6t-32jW6L-1}}\nwprevnextdefs{NWppJ6t-2oFbmT-C}{NWppJ6t-2oFbmT-E}\nwenddeflinemarkup
{\bf{}void} {\bf{}cycle\_data}::{\it{}read\_archive}({\bf{}const} {\it{}archive\_node} &{\it{}n}, {\bf{}lst} &{\it{}sym\_lst})\nwindexdefn{\nwixident{cycle{\_}data}}{cycle:undata}{NWppJ6t-2oFbmT-D}
{\nwlbrace}
    {\it{}inherited}::{\it{}read\_archive}({\it{}n}, {\it{}sym\_lst});
    {\it{}n}.{\it{}find\_ex}({\tt{}"k-val"}, {\it{}k}, {\it{}sym\_lst});
    {\it{}n}.{\it{}find\_ex}({\tt{}"l-val"}, {\it{}l}, {\it{}sym\_lst});
    {\it{}n}.{\it{}find\_ex}({\tt{}"m-val"}, {\it{}m}, {\it{}sym\_lst});
{\nwrbrace}

\nwused{\\{NWppJ6t-32jW6L-1}}\nwidentdefs{\\{{\nwixident{cycle{\_}data}}{cycle:undata}}}\nwidentuses{\\{{\nwixident{k}}{k}}\\{{\nwixident{l}}{l}}\\{{\nwixident{m}}{m}}\\{{\nwixident{read{\_}archive}}{read:unarchive}}}\nwindexuse{\nwixident{k}}{k}{NWppJ6t-2oFbmT-D}\nwindexuse{\nwixident{l}}{l}{NWppJ6t-2oFbmT-D}\nwindexuse{\nwixident{m}}{m}{NWppJ6t-2oFbmT-D}\nwindexuse{\nwixident{read{\_}archive}}{read:unarchive}{NWppJ6t-2oFbmT-D}\nwendcode{}\nwbegindocs{676}\nwdocspar
\nwenddocs{}\nwbegincode{677}\sublabel{NWppJ6t-2oFbmT-E}\nwmargintag{{\nwtagstyle{}\subpageref{NWppJ6t-2oFbmT-E}}}\moddef{cycle data class~{\nwtagstyle{}\subpageref{NWppJ6t-2oFbmT-1}}}\plusendmoddef\Rm{}\nwstartdeflinemarkup\nwusesondefline{\\{NWppJ6t-32jW6L-1}}\nwprevnextdefs{NWppJ6t-2oFbmT-D}{NWppJ6t-2oFbmT-F}\nwenddeflinemarkup
{\it{}GINAC\_BIND\_UNARCHIVER}({\bf{}cycle\_data});\nwindexdefn{\nwixident{cycle{\_}data}}{cycle:undata}{NWppJ6t-2oFbmT-E}

\nwused{\\{NWppJ6t-32jW6L-1}}\nwidentdefs{\\{{\nwixident{cycle{\_}data}}{cycle:undata}}}\nwendcode{}\nwbegindocs{678}\nwdocspar
\nwenddocs{}\nwbegincode{679}\sublabel{NWppJ6t-2oFbmT-F}\nwmargintag{{\nwtagstyle{}\subpageref{NWppJ6t-2oFbmT-F}}}\moddef{cycle data class~{\nwtagstyle{}\subpageref{NWppJ6t-2oFbmT-1}}}\plusendmoddef\Rm{}\nwstartdeflinemarkup\nwusesondefline{\\{NWppJ6t-32jW6L-1}}\nwprevnextdefs{NWppJ6t-2oFbmT-E}{NWppJ6t-2oFbmT-G}\nwenddeflinemarkup
{\bf{}ex} {\bf{}cycle\_data}::{\it{}op}({\it{}size\_t} {\it{}i}) {\bf{}const}
{\nwlbrace}
 {\it{}GINAC\_ASSERT}({\it{}i}\begin{math}<\end{math}{\it{}nops}());
    {\bf{}switch}({\it{}i}) {\nwlbrace}
    {\bf{}case} 0:
        {\bf{}return} {\it{}k};
    {\bf{}case} 1:
        {\bf{}return} {\it{}l};
    {\bf{}case} 2:
        {\bf{}return} {\it{}m};
    {\bf{}default}:
        {\bf{}throw}({\it{}std}::{\it{}invalid\_argument}({\tt{}"cycle\_data::op(): requested operand out of the range (3)"}));
    {\nwrbrace}
{\nwrbrace}

\nwused{\\{NWppJ6t-32jW6L-1}}\nwidentuses{\\{{\nwixident{cycle{\_}data}}{cycle:undata}}\\{{\nwixident{ex}}{ex}}\\{{\nwixident{k}}{k}}\\{{\nwixident{l}}{l}}\\{{\nwixident{m}}{m}}\\{{\nwixident{nops}}{nops}}\\{{\nwixident{op}}{op}}}\nwindexuse{\nwixident{cycle{\_}data}}{cycle:undata}{NWppJ6t-2oFbmT-F}\nwindexuse{\nwixident{ex}}{ex}{NWppJ6t-2oFbmT-F}\nwindexuse{\nwixident{k}}{k}{NWppJ6t-2oFbmT-F}\nwindexuse{\nwixident{l}}{l}{NWppJ6t-2oFbmT-F}\nwindexuse{\nwixident{m}}{m}{NWppJ6t-2oFbmT-F}\nwindexuse{\nwixident{nops}}{nops}{NWppJ6t-2oFbmT-F}\nwindexuse{\nwixident{op}}{op}{NWppJ6t-2oFbmT-F}\nwendcode{}\nwbegindocs{680}\nwdocspar
\nwenddocs{}\nwbegincode{681}\sublabel{NWppJ6t-2oFbmT-G}\nwmargintag{{\nwtagstyle{}\subpageref{NWppJ6t-2oFbmT-G}}}\moddef{cycle data class~{\nwtagstyle{}\subpageref{NWppJ6t-2oFbmT-1}}}\plusendmoddef\Rm{}\nwstartdeflinemarkup\nwusesondefline{\\{NWppJ6t-32jW6L-1}}\nwprevnextdefs{NWppJ6t-2oFbmT-F}{NWppJ6t-2oFbmT-H}\nwenddeflinemarkup
{\bf{}ex} & {\bf{}cycle\_data}::{\it{}let\_op}({\it{}size\_t} {\it{}i})
{\nwlbrace}
    {\it{}ensure\_if\_modifiable}();
    {\it{}GINAC\_ASSERT}({\it{}i}\begin{math}<\end{math}{\it{}nops}());
    {\bf{}switch}({\it{}i}) {\nwlbrace}
    {\bf{}case} 0:
        {\bf{}return} {\it{}k};
    {\bf{}case} 1:
        {\bf{}return} {\it{}l};
    {\bf{}case} 2:
        {\bf{}return} {\it{}m};
    {\bf{}default}:
        {\bf{}throw}({\it{}std}::{\it{}invalid\_argument}({\tt{}"cycle\_data::let\_op(): requested operand out of the range (3)"}));
    {\nwrbrace}
{\nwrbrace}

\nwused{\\{NWppJ6t-32jW6L-1}}\nwidentuses{\\{{\nwixident{cycle{\_}data}}{cycle:undata}}\\{{\nwixident{ex}}{ex}}\\{{\nwixident{k}}{k}}\\{{\nwixident{l}}{l}}\\{{\nwixident{m}}{m}}\\{{\nwixident{nops}}{nops}}}\nwindexuse{\nwixident{cycle{\_}data}}{cycle:undata}{NWppJ6t-2oFbmT-G}\nwindexuse{\nwixident{ex}}{ex}{NWppJ6t-2oFbmT-G}\nwindexuse{\nwixident{k}}{k}{NWppJ6t-2oFbmT-G}\nwindexuse{\nwixident{l}}{l}{NWppJ6t-2oFbmT-G}\nwindexuse{\nwixident{m}}{m}{NWppJ6t-2oFbmT-G}\nwindexuse{\nwixident{nops}}{nops}{NWppJ6t-2oFbmT-G}\nwendcode{}\nwbegindocs{682}\nwdocspar
\nwenddocs{}\nwbegincode{683}\sublabel{NWppJ6t-2oFbmT-H}\nwmargintag{{\nwtagstyle{}\subpageref{NWppJ6t-2oFbmT-H}}}\moddef{cycle data class~{\nwtagstyle{}\subpageref{NWppJ6t-2oFbmT-1}}}\plusendmoddef\Rm{}\nwstartdeflinemarkup\nwusesondefline{\\{NWppJ6t-32jW6L-1}}\nwprevnextdefs{NWppJ6t-2oFbmT-G}{NWppJ6t-2oFbmT-I}\nwenddeflinemarkup
{\bf{}ex} {\bf{}cycle\_data}::{\it{}get\_cycle}({\bf{}const} {\bf{}ex} & {\it{}metr}) {\bf{}const}
{\nwlbrace}
    {\bf{}return} {\bf{}cycle}({\it{}k},{\it{}l},{\it{}m},{\it{}metr});
{\nwrbrace}

\nwused{\\{NWppJ6t-32jW6L-1}}\nwidentuses{\\{{\nwixident{cycle{\_}data}}{cycle:undata}}\\{{\nwixident{ex}}{ex}}\\{{\nwixident{get{\_}cycle}}{get:uncycle}}\\{{\nwixident{k}}{k}}\\{{\nwixident{l}}{l}}\\{{\nwixident{m}}{m}}}\nwindexuse{\nwixident{cycle{\_}data}}{cycle:undata}{NWppJ6t-2oFbmT-H}\nwindexuse{\nwixident{ex}}{ex}{NWppJ6t-2oFbmT-H}\nwindexuse{\nwixident{get{\_}cycle}}{get:uncycle}{NWppJ6t-2oFbmT-H}\nwindexuse{\nwixident{k}}{k}{NWppJ6t-2oFbmT-H}\nwindexuse{\nwixident{l}}{l}{NWppJ6t-2oFbmT-H}\nwindexuse{\nwixident{m}}{m}{NWppJ6t-2oFbmT-H}\nwendcode{}\nwbegindocs{684}\nwdocspar
\nwenddocs{}\nwbegincode{685}\sublabel{NWppJ6t-2oFbmT-I}\nwmargintag{{\nwtagstyle{}\subpageref{NWppJ6t-2oFbmT-I}}}\moddef{cycle data class~{\nwtagstyle{}\subpageref{NWppJ6t-2oFbmT-1}}}\plusendmoddef\Rm{}\nwstartdeflinemarkup\nwusesondefline{\\{NWppJ6t-32jW6L-1}}\nwprevnextdefs{NWppJ6t-2oFbmT-H}{NWppJ6t-2oFbmT-J}\nwenddeflinemarkup
{\bf{}bool} {\bf{}cycle\_data}::{\it{}is\_equal}({\bf{}const} {\bf{}basic} & {\it{}other}, {\bf{}bool} {\it{}projectively}) {\bf{}const}
{\nwlbrace}
    {\bf{}if} ({\it{}not} {\it{}is\_a}\begin{math}<\end{math}{\bf{}cycle\_data}\begin{math}>\end{math}({\it{}other}))
        {\bf{}return} {\bf{}false};
    {\bf{}const} {\bf{}cycle\_data} {\it{}o} = {\it{}ex\_to}\begin{math}<\end{math}{\bf{}cycle\_data}\begin{math}>\end{math}({\it{}other});
    {\bf{}ex} {\it{}factor}=0, {\it{}ofactor}=0;

    {\bf{}if} ({\it{}projectively}) {\nwlbrace}
        // Check that coefficients are scalar multiples of other
        {\bf{}if} ({\it{}not} (({\it{}m}\begin{math}\ast\end{math}{\it{}o}.{\it{}get\_k}()-{\it{}o}.{\it{}get\_m}()\begin{math}\ast\end{math}{\it{}k}).{\it{}normal}().{\it{}is\_zero}()))
            {\bf{}return} {\bf{}false};
        // Set up coefficients for proportionality
        {\bf{}if} ({\it{}get\_k}().{\it{}normal}().{\it{}is\_zero}()) {\nwlbrace}
            {\it{}factor}={\it{}get\_m}();
            {\it{}ofactor}={\it{}o}.{\it{}get\_m}();
        {\nwrbrace} {\bf{}else} {\nwlbrace}
            {\it{}factor}={\it{}get\_k}();
            {\it{}ofactor}={\it{}o}.{\it{}get\_k}();
        {\nwrbrace}

    {\nwrbrace} {\bf{}else}
        // Check the exact equality of coefficients
        {\bf{}if} ({\it{}not} (({\it{}get\_k}()-{\it{}o}.{\it{}get\_k}()).{\it{}normal}().{\it{}is\_zero}()
                 \begin{math}\wedge\end{math} ({\it{}get\_m}()-{\it{}o}.{\it{}get\_m}()).{\it{}normal}().{\it{}is\_zero}()))
            {\bf{}return} {\bf{}false};

\nwused{\\{NWppJ6t-32jW6L-1}}\nwidentuses{\\{{\nwixident{cycle{\_}data}}{cycle:undata}}\\{{\nwixident{ex}}{ex}}\\{{\nwixident{k}}{k}}\\{{\nwixident{m}}{m}}}\nwindexuse{\nwixident{cycle{\_}data}}{cycle:undata}{NWppJ6t-2oFbmT-I}\nwindexuse{\nwixident{ex}}{ex}{NWppJ6t-2oFbmT-I}\nwindexuse{\nwixident{k}}{k}{NWppJ6t-2oFbmT-I}\nwindexuse{\nwixident{m}}{m}{NWppJ6t-2oFbmT-I}\nwendcode{}\nwbegindocs{686}Now we iterate through the coefficients of {\Tt{}\Rm{}{\it{}l}\nwendquote}.
\nwenddocs{}\nwbegincode{687}\sublabel{NWppJ6t-2oFbmT-J}\nwmargintag{{\nwtagstyle{}\subpageref{NWppJ6t-2oFbmT-J}}}\moddef{cycle data class~{\nwtagstyle{}\subpageref{NWppJ6t-2oFbmT-1}}}\plusendmoddef\Rm{}\nwstartdeflinemarkup\nwusesondefline{\\{NWppJ6t-32jW6L-1}}\nwprevnextdefs{NWppJ6t-2oFbmT-I}{NWppJ6t-2oFbmT-K}\nwenddeflinemarkup
    {\bf{}for} ({\bf{}unsigned} {\bf{}int} {\it{}i}=0; {\it{}i}\begin{math}<\end{math}{\it{}get\_dim}(); {\it{}i}\protect\PP)
        {\bf{}if} ({\it{}projectively}) {\nwlbrace}
            // search the the first non-zero coefficient
            {\bf{}if} ({\it{}factor}.{\it{}is\_zero}()) {\nwlbrace}
                {\it{}factor}={\it{}get\_l}({\it{}i});
                {\it{}ofactor}={\it{}o}.{\it{}get\_l}({\it{}i});
            {\nwrbrace} {\bf{}else}
                {\bf{}if} (\begin{math}\neg\end{math} ({\it{}get\_l}({\it{}i})\begin{math}\ast\end{math}{\it{}ofactor}-{\it{}o}.{\it{}get\_l}({\it{}i})\begin{math}\ast\end{math}{\it{}factor}).{\it{}normal}().{\it{}is\_zero}())
                    {\bf{}return} {\bf{}false};
        {\nwrbrace} {\bf{}else}
            {\bf{}if} (\begin{math}\neg\end{math} ({\it{}get\_l}({\it{}i})-{\it{}o}.{\it{}get\_l}({\it{}i})).{\it{}normal}().{\it{}is\_zero}())
                {\bf{}return} {\bf{}false};

    {\bf{}return} {\bf{}true};
{\nwrbrace}

\nwused{\\{NWppJ6t-32jW6L-1}}\nwidentuses{\\{{\nwixident{get{\_}dim()}}{get:undim()}}}\nwindexuse{\nwixident{get{\_}dim()}}{get:undim()}{NWppJ6t-2oFbmT-J}\nwendcode{}\nwbegindocs{688}\nwdocspar
\nwenddocs{}\nwbegincode{689}\sublabel{NWppJ6t-2oFbmT-K}\nwmargintag{{\nwtagstyle{}\subpageref{NWppJ6t-2oFbmT-K}}}\moddef{cycle data class~{\nwtagstyle{}\subpageref{NWppJ6t-2oFbmT-1}}}\plusendmoddef\Rm{}\nwstartdeflinemarkup\nwusesondefline{\\{NWppJ6t-32jW6L-1}}\nwprevnextdefs{NWppJ6t-2oFbmT-J}{NWppJ6t-2oFbmT-L}\nwenddeflinemarkup
{\bf{}bool} {\bf{}cycle\_data}::{\it{}is\_almost\_equal}({\bf{}const} {\bf{}basic} & {\it{}other}, {\bf{}bool} {\it{}projectively}) {\bf{}const}
{\nwlbrace}
    {\bf{}if} ({\it{}not} {\it{}is\_a}\begin{math}<\end{math}{\bf{}cycle\_data}\begin{math}>\end{math}({\it{}other}))
        {\bf{}return} {\bf{}false};
    {\bf{}const} {\bf{}cycle\_data} {\it{}o} = {\it{}ex\_to}\begin{math}<\end{math}{\bf{}cycle\_data}\begin{math}>\end{math}({\it{}other});
    {\bf{}ex} {\it{}factor}=0, {\it{}ofactor}=0;

    {\bf{}if} ({\it{}projectively}) {\nwlbrace}
        // Check that coefficients are scalar multiples of other
        {\bf{}if} (\begin{math}\neg\end{math} ({\it{}is\_less\_than\_epsilon}({\it{}m}\begin{math}\ast\end{math}{\it{}o}.{\it{}get\_k}()-{\it{}o}.{\it{}get\_m}()\begin{math}\ast\end{math}{\it{}k})))
            {\bf{}return} {\bf{}false};
        // Set up coefficients for proportionality
        {\bf{}if} ({\it{}is\_less\_than\_epsilon}({\it{}get\_k}())) {\nwlbrace}
            {\it{}factor}={\it{}get\_m}();
            {\it{}ofactor}={\it{}o}.{\it{}get\_m}();
        {\nwrbrace} {\bf{}else} {\nwlbrace}
            {\it{}factor}={\it{}get\_k}();
            {\it{}ofactor}={\it{}o}.{\it{}get\_k}();
        {\nwrbrace}

    {\nwrbrace} {\bf{}else}
        // Check the exact equality of coefficients
        {\bf{}if} (\begin{math}\neg\end{math} ({\it{}is\_less\_than\_epsilon}(({\it{}get\_k}()-{\it{}o}.{\it{}get\_k}()))
                 \begin{math}\wedge\end{math} {\it{}is\_less\_than\_epsilon}({\it{}get\_m}()-{\it{}o}.{\it{}get\_m}())))
            {\bf{}return} {\bf{}false};

\nwused{\\{NWppJ6t-32jW6L-1}}\nwidentuses{\\{{\nwixident{cycle{\_}data}}{cycle:undata}}\\{{\nwixident{ex}}{ex}}\\{{\nwixident{is{\_}almost{\_}equal}}{is:unalmost:unequal}}\\{{\nwixident{is{\_}less{\_}than{\_}epsilon}}{is:unless:unthan:unepsilon}}\\{{\nwixident{k}}{k}}\\{{\nwixident{m}}{m}}}\nwindexuse{\nwixident{cycle{\_}data}}{cycle:undata}{NWppJ6t-2oFbmT-K}\nwindexuse{\nwixident{ex}}{ex}{NWppJ6t-2oFbmT-K}\nwindexuse{\nwixident{is{\_}almost{\_}equal}}{is:unalmost:unequal}{NWppJ6t-2oFbmT-K}\nwindexuse{\nwixident{is{\_}less{\_}than{\_}epsilon}}{is:unless:unthan:unepsilon}{NWppJ6t-2oFbmT-K}\nwindexuse{\nwixident{k}}{k}{NWppJ6t-2oFbmT-K}\nwindexuse{\nwixident{m}}{m}{NWppJ6t-2oFbmT-K}\nwendcode{}\nwbegindocs{690}Now we iterate through the coefficients of {\Tt{}\Rm{}{\it{}l}\nwendquote}.
\nwenddocs{}\nwbegincode{691}\sublabel{NWppJ6t-2oFbmT-L}\nwmargintag{{\nwtagstyle{}\subpageref{NWppJ6t-2oFbmT-L}}}\moddef{cycle data class~{\nwtagstyle{}\subpageref{NWppJ6t-2oFbmT-1}}}\plusendmoddef\Rm{}\nwstartdeflinemarkup\nwusesondefline{\\{NWppJ6t-32jW6L-1}}\nwprevnextdefs{NWppJ6t-2oFbmT-K}{NWppJ6t-2oFbmT-M}\nwenddeflinemarkup
    {\bf{}for} ({\bf{}unsigned} {\bf{}int} {\it{}i}=0; {\it{}i}\begin{math}<\end{math}{\it{}get\_dim}(); {\it{}i}\protect\PP)
        {\bf{}if} ({\it{}projectively}) {\nwlbrace}
            // search the the first non-zero coefficient
            {\bf{}if} ({\it{}factor}.{\it{}is\_zero}()) {\nwlbrace}
                {\it{}factor}={\it{}get\_l}({\it{}i});
                {\it{}ofactor}={\it{}o}.{\it{}get\_l}({\it{}i});
            {\nwrbrace} {\bf{}else}
                {\bf{}if} (\begin{math}\neg\end{math} {\it{}is\_less\_than\_epsilon}({\it{}get\_l}({\it{}i})\begin{math}\ast\end{math}{\it{}ofactor}-{\it{}o}.{\it{}get\_l}({\it{}i})\begin{math}\ast\end{math}{\it{}factor}))
                    {\bf{}return} {\bf{}false};
        {\nwrbrace} {\bf{}else}
            {\bf{}if} (\begin{math}\neg\end{math} {\it{}is\_less\_than\_epsilon}({\it{}get\_l}({\it{}i})-{\it{}o}.{\it{}get\_l}({\it{}i})))
                {\bf{}return} {\bf{}false};

    {\bf{}return} {\bf{}true};
{\nwrbrace}

\nwused{\\{NWppJ6t-32jW6L-1}}\nwidentuses{\\{{\nwixident{get{\_}dim()}}{get:undim()}}\\{{\nwixident{is{\_}less{\_}than{\_}epsilon}}{is:unless:unthan:unepsilon}}}\nwindexuse{\nwixident{get{\_}dim()}}{get:undim()}{NWppJ6t-2oFbmT-L}\nwindexuse{\nwixident{is{\_}less{\_}than{\_}epsilon}}{is:unless:unthan:unepsilon}{NWppJ6t-2oFbmT-L}\nwendcode{}\nwbegindocs{692}\nwdocspar
\nwenddocs{}\nwbegincode{693}\sublabel{NWppJ6t-2oFbmT-M}\nwmargintag{{\nwtagstyle{}\subpageref{NWppJ6t-2oFbmT-M}}}\moddef{cycle data class~{\nwtagstyle{}\subpageref{NWppJ6t-2oFbmT-1}}}\plusendmoddef\Rm{}\nwstartdeflinemarkup\nwusesondefline{\\{NWppJ6t-32jW6L-1}}\nwprevnextdefs{NWppJ6t-2oFbmT-L}{NWppJ6t-2oFbmT-N}\nwenddeflinemarkup
{\bf{}cycle\_data} {\bf{}cycle\_data}::{\it{}subs}({\bf{}const} {\bf{}ex} & {\it{}e}, {\bf{}unsigned} {\it{}options}) {\bf{}const}
{\nwlbrace}
    {\bf{}return} {\bf{}cycle\_data}({\it{}k}.{\it{}subs}({\it{}e},{\it{}options}),{\it{}l}.{\it{}subs}({\it{}e},{\it{}options}),{\it{}m}.{\it{}subs}({\it{}e},{\it{}options}),{\bf{}false});
{\nwrbrace}
\nwused{\\{NWppJ6t-32jW6L-1}}\nwidentuses{\\{{\nwixident{cycle{\_}data}}{cycle:undata}}\\{{\nwixident{ex}}{ex}}\\{{\nwixident{k}}{k}}\\{{\nwixident{l}}{l}}\\{{\nwixident{m}}{m}}\\{{\nwixident{subs}}{subs}}}\nwindexuse{\nwixident{cycle{\_}data}}{cycle:undata}{NWppJ6t-2oFbmT-M}\nwindexuse{\nwixident{ex}}{ex}{NWppJ6t-2oFbmT-M}\nwindexuse{\nwixident{k}}{k}{NWppJ6t-2oFbmT-M}\nwindexuse{\nwixident{l}}{l}{NWppJ6t-2oFbmT-M}\nwindexuse{\nwixident{m}}{m}{NWppJ6t-2oFbmT-M}\nwindexuse{\nwixident{subs}}{subs}{NWppJ6t-2oFbmT-M}\nwendcode{}\nwbegindocs{694}\nwdocspar
\nwenddocs{}\nwbegincode{695}\sublabel{NWppJ6t-2oFbmT-N}\nwmargintag{{\nwtagstyle{}\subpageref{NWppJ6t-2oFbmT-N}}}\moddef{cycle data class~{\nwtagstyle{}\subpageref{NWppJ6t-2oFbmT-1}}}\plusendmoddef\Rm{}\nwstartdeflinemarkup\nwusesondefline{\\{NWppJ6t-32jW6L-1}}\nwprevnextdefs{NWppJ6t-2oFbmT-M}{NWppJ6t-2oFbmT-O}\nwenddeflinemarkup
{\bf{}ex} {\bf{}cycle\_data}::{\it{}subs}({\bf{}const} {\it{}exmap} & {\it{}em}, {\bf{}unsigned} {\it{}options}) {\bf{}const}
{\nwlbrace}
    {\bf{}return} {\bf{}cycle\_data}({\it{}k}.{\it{}subs}({\it{}em},{\it{}options}),{\it{}l}.{\it{}subs}({\it{}em},{\it{}options}),{\it{}m}.{\it{}subs}({\it{}em},{\it{}options}),{\bf{}false});
{\nwrbrace}

\nwused{\\{NWppJ6t-32jW6L-1}}\nwidentuses{\\{{\nwixident{cycle{\_}data}}{cycle:undata}}\\{{\nwixident{ex}}{ex}}\\{{\nwixident{k}}{k}}\\{{\nwixident{l}}{l}}\\{{\nwixident{m}}{m}}\\{{\nwixident{subs}}{subs}}}\nwindexuse{\nwixident{cycle{\_}data}}{cycle:undata}{NWppJ6t-2oFbmT-N}\nwindexuse{\nwixident{ex}}{ex}{NWppJ6t-2oFbmT-N}\nwindexuse{\nwixident{k}}{k}{NWppJ6t-2oFbmT-N}\nwindexuse{\nwixident{l}}{l}{NWppJ6t-2oFbmT-N}\nwindexuse{\nwixident{m}}{m}{NWppJ6t-2oFbmT-N}\nwindexuse{\nwixident{subs}}{subs}{NWppJ6t-2oFbmT-N}\nwendcode{}\nwbegindocs{696}\nwdocspar
\nwenddocs{}\nwbegincode{697}\sublabel{NWppJ6t-2oFbmT-O}\nwmargintag{{\nwtagstyle{}\subpageref{NWppJ6t-2oFbmT-O}}}\moddef{cycle data class~{\nwtagstyle{}\subpageref{NWppJ6t-2oFbmT-1}}}\plusendmoddef\Rm{}\nwstartdeflinemarkup\nwusesondefline{\\{NWppJ6t-32jW6L-1}}\nwprevnextdefs{NWppJ6t-2oFbmT-N}{\relax}\nwenddeflinemarkup
{\bf{}ex} {\bf{}cycle\_data}::{\it{}num\_normalize}() {\bf{}const}
{\nwlbrace}
    {\bf{}if} (\begin{math}\neg\end{math} ({\it{}is\_a}\begin{math}<\end{math}{\bf{}numeric}\begin{math}>\end{math}({\it{}k}) \begin{math}\wedge\end{math} {\it{}is\_a}\begin{math}<\end{math}{\bf{}numeric}\begin{math}>\end{math}({\it{}m})
           \begin{math}\wedge\end{math} {\it{}is\_a}\begin{math}<\end{math}{\bf{}numeric}\begin{math}>\end{math}({\it{}l}.{\it{}op}(0).{\it{}op}(0))  \begin{math}\wedge\end{math} {\it{}is\_a}\begin{math}<\end{math}{\bf{}numeric}\begin{math}>\end{math}({\it{}l}.{\it{}op}(0).{\it{}op}(1))))
        {\bf{}return} {\bf{}cycle\_data}({\it{}k},{\it{}l},{\it{}m},{\bf{}true});

    {\bf{}numeric} {\it{}k1}={\it{}ex\_to}\begin{math}<\end{math}{\bf{}numeric}\begin{math}>\end{math}({\it{}k}),
        {\it{}m1}={\it{}ex\_to}\begin{math}<\end{math}{\bf{}numeric}\begin{math}>\end{math}({\it{}m});
    {\bf{}numeric} {\it{}r}={\it{}max}({\it{}abs}({\it{}k1}),{\it{}abs}({\it{}m1}));
    {\bf{}for} ({\bf{}unsigned} {\bf{}int} {\it{}i}=0; {\it{}i}\begin{math}<\end{math}{\it{}get\_dim}(); \protect\PP{\it{}i})
        {\it{}r}={\it{}max}({\it{}r},{\it{}abs}({\it{}ex\_to}\begin{math}<\end{math}{\bf{}numeric}\begin{math}>\end{math}({\it{}l}.{\it{}op}(0).{\it{}op}({\it{}i}))));

    {\bf{}if} ({\it{}is\_less\_than\_epsilon}({\it{}r}))
        {\bf{}return} {\bf{}cycle\_data}({\it{}k},{\it{}l},{\it{}m},{\bf{}true});
    {\it{}k1}\begin{math}\div\end{math}={\it{}r}; {\it{}k1}=({\it{}is\_less\_than\_epsilon}({\it{}k1})?0:{\it{}k1});
    {\it{}m1}\begin{math}\div\end{math}={\it{}r}; {\it{}m1}=({\it{}is\_less\_than\_epsilon}({\it{}m1})?0:{\it{}m1});
    {\bf{}lst} {\it{}l0};
    {\bf{}for} ({\bf{}unsigned} {\bf{}int} {\it{}i}=0; {\it{}i}\begin{math}<\end{math}{\it{}get\_dim}(); \protect\PP{\it{}i}) {\nwlbrace}
        {\bf{}numeric} {\it{}li}= {\it{}ex\_to}\begin{math}<\end{math}{\bf{}numeric}\begin{math}>\end{math}({\it{}l}.{\it{}op}(0).{\it{}op}({\it{}i}))\begin{math}\div\end{math}{\it{}r};
        {\it{}l0}.{\it{}append}({\it{}is\_less\_than\_epsilon}({\it{}li})?0:{\it{}li});
    {\nwrbrace}

    {\bf{}return} {\bf{}cycle\_data}({\it{}k1},{\bf{}indexed}({\bf{}matrix}(1, {\it{}get\_dim}(), {\it{}l0}), {\it{}l}.{\it{}op}(1)),{\it{}m1});
{\nwrbrace}

\nwused{\\{NWppJ6t-32jW6L-1}}\nwidentuses{\\{{\nwixident{cycle{\_}data}}{cycle:undata}}\\{{\nwixident{ex}}{ex}}\\{{\nwixident{get{\_}dim()}}{get:undim()}}\\{{\nwixident{is{\_}less{\_}than{\_}epsilon}}{is:unless:unthan:unepsilon}}\\{{\nwixident{k}}{k}}\\{{\nwixident{l}}{l}}\\{{\nwixident{m}}{m}}\\{{\nwixident{numeric}}{numeric}}\\{{\nwixident{op}}{op}}}\nwindexuse{\nwixident{cycle{\_}data}}{cycle:undata}{NWppJ6t-2oFbmT-O}\nwindexuse{\nwixident{ex}}{ex}{NWppJ6t-2oFbmT-O}\nwindexuse{\nwixident{get{\_}dim()}}{get:undim()}{NWppJ6t-2oFbmT-O}\nwindexuse{\nwixident{is{\_}less{\_}than{\_}epsilon}}{is:unless:unthan:unepsilon}{NWppJ6t-2oFbmT-O}\nwindexuse{\nwixident{k}}{k}{NWppJ6t-2oFbmT-O}\nwindexuse{\nwixident{l}}{l}{NWppJ6t-2oFbmT-O}\nwindexuse{\nwixident{m}}{m}{NWppJ6t-2oFbmT-O}\nwindexuse{\nwixident{numeric}}{numeric}{NWppJ6t-2oFbmT-O}\nwindexuse{\nwixident{op}}{op}{NWppJ6t-2oFbmT-O}\nwendcode{}\nwbegindocs{698}\nwdocspar
\subsection{Implementation of {\Tt{}\Rm{}{\bf{}cycle\_relation}\nwendquote} class}
\label{sec:impl-cycl}

\nwenddocs{}\nwbegindocs{699}\nwdocspar
\nwenddocs{}\nwbegincode{700}\sublabel{NWppJ6t-3bfNK9-1}\nwmargintag{{\nwtagstyle{}\subpageref{NWppJ6t-3bfNK9-1}}}\moddef{cycle relation class~{\nwtagstyle{}\subpageref{NWppJ6t-3bfNK9-1}}}\endmoddef\Rm{}\nwstartdeflinemarkup\nwusesondefline{\\{NWppJ6t-32jW6L-1}}\nwprevnextdefs{\relax}{NWppJ6t-3bfNK9-2}\nwenddeflinemarkup
{\bf{}cycle\_relation}::{\bf{}cycle\_relation}() : {\it{}parkey}(), {\it{}parameter}()
{\nwlbrace}
  {\it{}rel} = {\it{}cycle\_orthogonal};
  {\it{}use\_cycle\_metric} = {\bf{}true};
{\nwrbrace}

\nwalsodefined{\\{NWppJ6t-3bfNK9-2}\\{NWppJ6t-3bfNK9-3}\\{NWppJ6t-3bfNK9-4}\\{NWppJ6t-3bfNK9-5}\\{NWppJ6t-3bfNK9-6}\\{NWppJ6t-3bfNK9-7}\\{NWppJ6t-3bfNK9-8}\\{NWppJ6t-3bfNK9-9}\\{NWppJ6t-3bfNK9-A}\\{NWppJ6t-3bfNK9-B}\\{NWppJ6t-3bfNK9-C}\\{NWppJ6t-3bfNK9-D}\\{NWppJ6t-3bfNK9-E}}\nwused{\\{NWppJ6t-32jW6L-1}}\nwidentuses{\\{{\nwixident{cycle{\_}orthogonal}}{cycle:unorthogonal}}\\{{\nwixident{cycle{\_}relation}}{cycle:unrelation}}}\nwindexuse{\nwixident{cycle{\_}orthogonal}}{cycle:unorthogonal}{NWppJ6t-3bfNK9-1}\nwindexuse{\nwixident{cycle{\_}relation}}{cycle:unrelation}{NWppJ6t-3bfNK9-1}\nwendcode{}\nwbegindocs{701}\nwdocspar
\nwenddocs{}\nwbegincode{702}\sublabel{NWppJ6t-3bfNK9-2}\nwmargintag{{\nwtagstyle{}\subpageref{NWppJ6t-3bfNK9-2}}}\moddef{cycle relation class~{\nwtagstyle{}\subpageref{NWppJ6t-3bfNK9-1}}}\plusendmoddef\Rm{}\nwstartdeflinemarkup\nwusesondefline{\\{NWppJ6t-32jW6L-1}}\nwprevnextdefs{NWppJ6t-3bfNK9-1}{NWppJ6t-3bfNK9-3}\nwenddeflinemarkup
{\bf{}cycle\_relation}::{\bf{}cycle\_relation}({\bf{}const} {\bf{}ex} & {\it{}k}, {\it{}PCR} {\it{}r}, {\bf{}bool} {\it{}cm}, {\bf{}const} {\bf{}ex} & {\it{}p}) {\nwlbrace}
    {\it{}parkey} = {\it{}k};
    {\it{}rel} = {\it{}r};
    {\it{}use\_cycle\_metric} = {\it{}cm};
    {\it{}parameter}={\it{}p};
{\nwrbrace}

\nwused{\\{NWppJ6t-32jW6L-1}}\nwidentuses{\\{{\nwixident{cycle{\_}relation}}{cycle:unrelation}}\\{{\nwixident{ex}}{ex}}\\{{\nwixident{k}}{k}}\\{{\nwixident{PCR}}{PCR}}}\nwindexuse{\nwixident{cycle{\_}relation}}{cycle:unrelation}{NWppJ6t-3bfNK9-2}\nwindexuse{\nwixident{ex}}{ex}{NWppJ6t-3bfNK9-2}\nwindexuse{\nwixident{k}}{k}{NWppJ6t-3bfNK9-2}\nwindexuse{\nwixident{PCR}}{PCR}{NWppJ6t-3bfNK9-2}\nwendcode{}\nwbegindocs{703}\nwdocspar
\nwenddocs{}\nwbegincode{704}\sublabel{NWppJ6t-3bfNK9-3}\nwmargintag{{\nwtagstyle{}\subpageref{NWppJ6t-3bfNK9-3}}}\moddef{cycle relation class~{\nwtagstyle{}\subpageref{NWppJ6t-3bfNK9-1}}}\plusendmoddef\Rm{}\nwstartdeflinemarkup\nwusesondefline{\\{NWppJ6t-32jW6L-1}}\nwprevnextdefs{NWppJ6t-3bfNK9-2}{NWppJ6t-3bfNK9-4}\nwenddeflinemarkup
{\it{}return\_type\_t} {\bf{}cycle\_relation}::{\it{}return\_type\_tinfo}() {\bf{}const}
{\nwlbrace}
    {\bf{}return} {\it{}make\_return\_type\_t}\begin{math}<\end{math}{\bf{}cycle\_relation}\begin{math}>\end{math}();
{\nwrbrace}

\nwused{\\{NWppJ6t-32jW6L-1}}\nwidentuses{\\{{\nwixident{cycle{\_}relation}}{cycle:unrelation}}}\nwindexuse{\nwixident{cycle{\_}relation}}{cycle:unrelation}{NWppJ6t-3bfNK9-3}\nwendcode{}\nwbegindocs{705}\nwdocspar
\nwenddocs{}\nwbegincode{706}\sublabel{NWppJ6t-3bfNK9-4}\nwmargintag{{\nwtagstyle{}\subpageref{NWppJ6t-3bfNK9-4}}}\moddef{cycle relation class~{\nwtagstyle{}\subpageref{NWppJ6t-3bfNK9-1}}}\plusendmoddef\Rm{}\nwstartdeflinemarkup\nwusesondefline{\\{NWppJ6t-32jW6L-1}}\nwprevnextdefs{NWppJ6t-3bfNK9-3}{NWppJ6t-3bfNK9-5}\nwenddeflinemarkup
{\bf{}int} {\bf{}cycle\_relation}::{\it{}compare\_same\_type}({\bf{}const} {\bf{}basic} &{\it{}other}) {\bf{}const}\nwindexdefn{\nwixident{cycle{\_}relation}}{cycle:unrelation}{NWppJ6t-3bfNK9-4}
{\nwlbrace}
       {\it{}GINAC\_ASSERT}({\it{}is\_a}\begin{math}<\end{math}{\bf{}cycle\_relation}\begin{math}>\end{math}({\it{}other}));
       {\bf{}return} {\it{}inherited}::{\it{}compare\_same\_type}({\it{}other});
       \begin{math}\div\end{math}\begin{math}\ast\end{math}
  {\bf{}const} {\bf{}cycle\_relation} &{\it{}o} = {\bf{}static\_cast}\begin{math}<\end{math}{\bf{}const} {\bf{}cycle\_relation} &\begin{math}>\end{math}({\it{}other});
    {\bf{}if} (({\it{}parkey} \begin{math}\equiv\end{math} {\it{}o}.{\it{}parkey}) \begin{math}\wedge\end{math} (&{\it{}rel} \begin{math}\equiv\end{math} &{\it{}o}.{\it{}rel}))
        {\bf{}return} 0;
    {\bf{}else} {\bf{}if} (({\it{}parkey} \begin{math}<\end{math} {\it{}o}.{\it{}parkey}) \begin{math}\vee\end{math} (&{\it{}rel} \begin{math}<\end{math} &{\it{}o}.{\it{}rel}))
        {\bf{}return} -1;
    {\bf{}else}
    {\bf{}return} 1;\begin{math}\ast\end{math}\begin{math}\div\end{math}
{\nwrbrace}

\nwused{\\{NWppJ6t-32jW6L-1}}\nwidentdefs{\\{{\nwixident{cycle{\_}relation}}{cycle:unrelation}}}\nwendcode{}\nwbegindocs{707}(un)Archiving of {\Tt{}\Rm{}{\bf{}cycle\_relation}\nwendquote} is not universal. At present it
only can handle relations declared in the header file: {\Tt{}\Rm{}{\it{}cycle\_orthogonal}\nwendquote},
{\Tt{}\Rm{}{\it{}cycle\_f\_orthogonal}\nwendquote}, {\Tt{}\Rm{}{\it{}cycle\_adifferent}\nwendquote}, {\Tt{}\Rm{}{\it{}cycle\_different}\nwendquote} and {\Tt{}\Rm{}{\it{}cycle\_tangent}\nwendquote}.
\nwenddocs{}\nwbegincode{708}\sublabel{NWppJ6t-3bfNK9-5}\nwmargintag{{\nwtagstyle{}\subpageref{NWppJ6t-3bfNK9-5}}}\moddef{cycle relation class~{\nwtagstyle{}\subpageref{NWppJ6t-3bfNK9-1}}}\plusendmoddef\Rm{}\nwstartdeflinemarkup\nwusesondefline{\\{NWppJ6t-32jW6L-1}}\nwprevnextdefs{NWppJ6t-3bfNK9-4}{NWppJ6t-3bfNK9-6}\nwenddeflinemarkup
{\bf{}void} {\bf{}cycle\_relation}::{\it{}archive}({\it{}archive\_node} &{\it{}n}) {\bf{}const}\nwindexdefn{\nwixident{cycle{\_}relation}}{cycle:unrelation}{NWppJ6t-3bfNK9-5}
{\nwlbrace}
    {\it{}inherited}::{\it{}archive}({\it{}n});
    {\it{}n}.{\it{}add\_ex}({\tt{}"cr-parkey"}, {\it{}parkey});
    {\it{}n}.{\it{}add\_bool}({\tt{}"use\_cycle\_metric"}, {\it{}use\_cycle\_metric});
    {\it{}n}.{\it{}add\_ex}({\tt{}"parameter"}, {\it{}parameter});
    {\bf{}ex} (\begin{math}\ast\end{math}{\bf{}const}\begin{math}\ast\end{math} {\it{}ptr})({\bf{}const} {\bf{}ex} &, {\bf{}const} {\bf{}ex} &, {\bf{}const} {\bf{}ex} &)
        = {\it{}rel}.{\it{}target}\begin{math}<\end{math}{\bf{}ex}(\begin{math}\ast\end{math})({\bf{}const} {\bf{}ex}&, {\bf{}const} {\bf{}ex} &,{\bf{}const} {\bf{}ex}&)\begin{math}>\end{math}();
    {\bf{}if} ({\it{}ptr} \begin{math}\wedge\end{math} \begin{math}\ast\end{math}{\it{}ptr} \begin{math}\equiv\end{math} {\it{}cycle\_orthogonal})
        {\it{}n}.{\it{}add\_string}({\tt{}"relation"}, {\tt{}"orthogonal"});
    {\bf{}else} {\bf{}if} ({\it{}ptr} \begin{math}\wedge\end{math} \begin{math}\ast\end{math}{\it{}ptr} \begin{math}\equiv\end{math} {\it{}cycle\_f\_orthogonal})
        {\it{}n}.{\it{}add\_string}({\tt{}"relation"}, {\tt{}"f\_orthogonal"});
    {\bf{}else} {\bf{}if} ({\it{}ptr} \begin{math}\wedge\end{math} \begin{math}\ast\end{math}{\it{}ptr} \begin{math}\equiv\end{math} {\it{}cycle\_different})
        {\it{}n}.{\it{}add\_string}({\tt{}"relation"}, {\tt{}"different"});
    {\bf{}else} {\bf{}if} ({\it{}ptr} \begin{math}\wedge\end{math} \begin{math}\ast\end{math}{\it{}ptr} \begin{math}\equiv\end{math} {\it{}cycle\_adifferent})
        {\it{}n}.{\it{}add\_string}({\tt{}"relation"}, {\tt{}"adifferent"});
    {\bf{}else} {\bf{}if} ({\it{}ptr} \begin{math}\wedge\end{math} \begin{math}\ast\end{math}{\it{}ptr} \begin{math}\equiv\end{math} {\it{}cycle\_tangent})
        {\it{}n}.{\it{}add\_string}({\tt{}"relation"}, {\tt{}"tangent"});
    {\bf{}else} {\bf{}if} ({\it{}ptr} \begin{math}\wedge\end{math} \begin{math}\ast\end{math}{\it{}ptr} \begin{math}\equiv\end{math} {\it{}cycle\_tangent\_i})
        {\it{}n}.{\it{}add\_string}({\tt{}"relation"}, {\tt{}"tangent\_i"});
    {\bf{}else} {\bf{}if} ({\it{}ptr} \begin{math}\wedge\end{math} \begin{math}\ast\end{math}{\it{}ptr} \begin{math}\equiv\end{math} {\it{}cycle\_tangent\_o})
        {\it{}n}.{\it{}add\_string}({\tt{}"relation"}, {\tt{}"tangent\_o"});
    {\bf{}else} {\bf{}if} ({\it{}ptr} \begin{math}\wedge\end{math} \begin{math}\ast\end{math}{\it{}ptr} \begin{math}\equiv\end{math} {\it{}cycle\_angle})
        {\it{}n}.{\it{}add\_string}({\tt{}"relation"}, {\tt{}"angle"});
    {\bf{}else} {\bf{}if} ({\it{}ptr} \begin{math}\wedge\end{math} \begin{math}\ast\end{math}{\it{}ptr} \begin{math}\equiv\end{math} {\it{}steiner\_power})
        {\it{}n}.{\it{}add\_string}({\tt{}"relation"}, {\tt{}"steiner\_power"});
    {\bf{}else} {\bf{}if} ({\it{}ptr} \begin{math}\wedge\end{math} \begin{math}\ast\end{math}{\it{}ptr} \begin{math}\equiv\end{math} {\it{}cycle\_cross\_t\_distance})
        {\it{}n}.{\it{}add\_string}({\tt{}"relation"}, {\tt{}"cross\_distance"});
    {\bf{}else} {\bf{}if} ({\it{}ptr} \begin{math}\wedge\end{math} \begin{math}\ast\end{math}{\it{}ptr} \begin{math}\equiv\end{math} {\it{}product\_sign})
        {\it{}n}.{\it{}add\_string}({\tt{}"relation"}, {\tt{}"product\_sign"});
    {\bf{}else} {\bf{}if} ({\it{}ptr} \begin{math}\wedge\end{math} \begin{math}\ast\end{math}{\it{}ptr} \begin{math}\equiv\end{math} {\it{}coefficients\_are\_real})
        {\it{}n}.{\it{}add\_string}({\tt{}"relation"}, {\tt{}"are\_real"});
    {\bf{}else} {\bf{}if} ({\it{}ptr} \begin{math}\wedge\end{math} \begin{math}\ast\end{math}{\it{}ptr} \begin{math}\equiv\end{math} {\it{}cycle\_moebius})
        {\it{}n}.{\it{}add\_string}({\tt{}"relation"}, {\tt{}"moebius"});
    {\bf{}else} {\bf{}if} ({\it{}ptr} \begin{math}\wedge\end{math} \begin{math}\ast\end{math}{\it{}ptr} \begin{math}\equiv\end{math} {\it{}cycle\_sl2})
        {\it{}n}.{\it{}add\_string}({\tt{}"relation"}, {\tt{}"sl2"});
    {\bf{}else}
        {\bf{}throw}({\it{}std}::{\it{}invalid\_argument}({\tt{}"cycle\_relation::archive(): archiving of this relation is not"}
                                    {\tt{}" implemented"}));
{\nwrbrace}

\nwused{\\{NWppJ6t-32jW6L-1}}\nwidentdefs{\\{{\nwixident{cycle{\_}relation}}{cycle:unrelation}}}\nwidentuses{\\{{\nwixident{archive}}{archive}}\\{{\nwixident{coefficients{\_}are{\_}real}}{coefficients:unare:unreal}}\\{{\nwixident{cycle{\_}adifferent}}{cycle:unadifferent}}\\{{\nwixident{cycle{\_}angle}}{cycle:unangle}}\\{{\nwixident{cycle{\_}cross{\_}t{\_}distance}}{cycle:uncross:unt:undistance}}\\{{\nwixident{cycle{\_}different}}{cycle:undifferent}}\\{{\nwixident{cycle{\_}f{\_}orthogonal}}{cycle:unf:unorthogonal}}\\{{\nwixident{cycle{\_}moebius}}{cycle:unmoebius}}\\{{\nwixident{cycle{\_}orthogonal}}{cycle:unorthogonal}}\\{{\nwixident{cycle{\_}sl2}}{cycle:unsl2}}\\{{\nwixident{cycle{\_}tangent}}{cycle:untangent}}\\{{\nwixident{cycle{\_}tangent{\_}i}}{cycle:untangent:uni}}\\{{\nwixident{cycle{\_}tangent{\_}o}}{cycle:untangent:uno}}\\{{\nwixident{ex}}{ex}}\\{{\nwixident{product{\_}sign}}{product:unsign}}\\{{\nwixident{steiner{\_}power}}{steiner:unpower}}}\nwindexuse{\nwixident{archive}}{archive}{NWppJ6t-3bfNK9-5}\nwindexuse{\nwixident{coefficients{\_}are{\_}real}}{coefficients:unare:unreal}{NWppJ6t-3bfNK9-5}\nwindexuse{\nwixident{cycle{\_}adifferent}}{cycle:unadifferent}{NWppJ6t-3bfNK9-5}\nwindexuse{\nwixident{cycle{\_}angle}}{cycle:unangle}{NWppJ6t-3bfNK9-5}\nwindexuse{\nwixident{cycle{\_}cross{\_}t{\_}distance}}{cycle:uncross:unt:undistance}{NWppJ6t-3bfNK9-5}\nwindexuse{\nwixident{cycle{\_}different}}{cycle:undifferent}{NWppJ6t-3bfNK9-5}\nwindexuse{\nwixident{cycle{\_}f{\_}orthogonal}}{cycle:unf:unorthogonal}{NWppJ6t-3bfNK9-5}\nwindexuse{\nwixident{cycle{\_}moebius}}{cycle:unmoebius}{NWppJ6t-3bfNK9-5}\nwindexuse{\nwixident{cycle{\_}orthogonal}}{cycle:unorthogonal}{NWppJ6t-3bfNK9-5}\nwindexuse{\nwixident{cycle{\_}sl2}}{cycle:unsl2}{NWppJ6t-3bfNK9-5}\nwindexuse{\nwixident{cycle{\_}tangent}}{cycle:untangent}{NWppJ6t-3bfNK9-5}\nwindexuse{\nwixident{cycle{\_}tangent{\_}i}}{cycle:untangent:uni}{NWppJ6t-3bfNK9-5}\nwindexuse{\nwixident{cycle{\_}tangent{\_}o}}{cycle:untangent:uno}{NWppJ6t-3bfNK9-5}\nwindexuse{\nwixident{ex}}{ex}{NWppJ6t-3bfNK9-5}\nwindexuse{\nwixident{product{\_}sign}}{product:unsign}{NWppJ6t-3bfNK9-5}\nwindexuse{\nwixident{steiner{\_}power}}{steiner:unpower}{NWppJ6t-3bfNK9-5}\nwendcode{}\nwbegindocs{709}\nwdocspar
\nwenddocs{}\nwbegincode{710}\sublabel{NWppJ6t-3bfNK9-6}\nwmargintag{{\nwtagstyle{}\subpageref{NWppJ6t-3bfNK9-6}}}\moddef{cycle relation class~{\nwtagstyle{}\subpageref{NWppJ6t-3bfNK9-1}}}\plusendmoddef\Rm{}\nwstartdeflinemarkup\nwusesondefline{\\{NWppJ6t-32jW6L-1}}\nwprevnextdefs{NWppJ6t-3bfNK9-5}{NWppJ6t-3bfNK9-7}\nwenddeflinemarkup
{\bf{}void} {\bf{}cycle\_relation}::{\it{}read\_archive}({\bf{}const} {\it{}archive\_node} &{\it{}n}, {\bf{}lst} &{\it{}sym\_lst})\nwindexdefn{\nwixident{cycle{\_}relation}}{cycle:unrelation}{NWppJ6t-3bfNK9-6}
{\nwlbrace}
    {\bf{}ex} {\it{}e};
    {\it{}inherited}::{\it{}read\_archive}({\it{}n}, {\it{}sym\_lst});
    {\it{}n}.{\it{}find\_ex}({\tt{}"cr-parkey"}, {\it{}e}, {\it{}sym\_lst});
    {\bf{}if} ({\it{}is\_a}\begin{math}<\end{math}{\bf{}symbol}\begin{math}>\end{math}({\it{}e}))
        {\it{}parkey}={\it{}e};
    {\bf{}else}
        {\bf{}throw}({\it{}std}::{\it{}invalid\_argument}({\tt{}"cycle\_relation::read\_archive(): read a non-symbol as"}
                                    {\tt{}" a parkey from the archive"}));
    {\it{}n}.{\it{}find\_ex}({\tt{}"parameter"}, {\it{}parameter}, {\it{}sym\_lst});
    {\it{}n}.{\it{}find\_bool}({\tt{}"use\_cycle\_metric"}, {\it{}use\_cycle\_metric});
    {\it{}string} {\it{}relation};
    {\it{}n}.{\it{}find\_string}({\tt{}"relation"}, {\it{}relation});
    {\bf{}if} ({\it{}relation} \begin{math}\equiv\end{math} {\tt{}"orthogonal"})
        {\it{}rel} = {\it{}cycle\_orthogonal};
    {\bf{}else} {\bf{}if} ({\it{}relation} \begin{math}\equiv\end{math} {\tt{}"f\_orthogonal"})
        {\it{}rel} = {\it{}cycle\_f\_orthogonal};
    {\bf{}else} {\bf{}if} ({\it{}relation} \begin{math}\equiv\end{math} {\tt{}"different"})
        {\it{}rel} = {\it{}cycle\_different};
    {\bf{}else} {\bf{}if} ({\it{}relation} \begin{math}\equiv\end{math} {\tt{}"adifferent"})
        {\it{}rel} = {\it{}cycle\_adifferent};
    {\bf{}else} {\bf{}if} ({\it{}relation} \begin{math}\equiv\end{math} {\tt{}"tangent"})
        {\it{}rel} = {\it{}cycle\_tangent};
    {\bf{}else} {\bf{}if} ({\it{}relation} \begin{math}\equiv\end{math} {\tt{}"tangent\_i"})
        {\it{}rel} = {\it{}cycle\_tangent\_i};
    {\bf{}else} {\bf{}if} ({\it{}relation} \begin{math}\equiv\end{math} {\tt{}"tangent\_o"})
        {\it{}rel} = {\it{}cycle\_tangent\_o};
    {\bf{}else} {\bf{}if} ({\it{}relation} \begin{math}\equiv\end{math} {\tt{}"angle"})
        {\it{}rel} = {\it{}cycle\_angle};
    {\bf{}else} {\bf{}if} ({\it{}relation} \begin{math}\equiv\end{math} {\tt{}"steiner\_power"})
        {\it{}rel} = {\it{}steiner\_power};
    {\bf{}else} {\bf{}if} ({\it{}relation} \begin{math}\equiv\end{math} {\tt{}"cross\_distance"})
        {\it{}rel} = {\it{}cycle\_cross\_t\_distance};
    {\bf{}else} {\bf{}if} ({\it{}relation} \begin{math}\equiv\end{math} {\tt{}"product\_sign"})
        {\it{}rel} = {\it{}product\_sign};
    {\bf{}else} {\bf{}if} ({\it{}relation} \begin{math}\equiv\end{math} {\tt{}"are\_real"})
        {\it{}rel} = {\it{}coefficients\_are\_real};
    {\bf{}else} {\bf{}if} ({\it{}relation} \begin{math}\equiv\end{math} {\tt{}"moebius"})
        {\it{}rel} = {\it{}cycle\_moebius};
    {\bf{}else} {\bf{}if} ({\it{}relation} \begin{math}\equiv\end{math} {\tt{}"sl2"})
        {\it{}rel} = {\it{}cycle\_sl2};
    {\bf{}else}
        {\bf{}throw}({\it{}std}::{\it{}invalid\_argument}({\tt{}"cycle\_relation::read\_archive(): archive contains unknown"}
                                    {\tt{}" relation"}));
{\nwrbrace}

\nwused{\\{NWppJ6t-32jW6L-1}}\nwidentdefs{\\{{\nwixident{cycle{\_}relation}}{cycle:unrelation}}}\nwidentuses{\\{{\nwixident{archive}}{archive}}\\{{\nwixident{coefficients{\_}are{\_}real}}{coefficients:unare:unreal}}\\{{\nwixident{cycle{\_}adifferent}}{cycle:unadifferent}}\\{{\nwixident{cycle{\_}angle}}{cycle:unangle}}\\{{\nwixident{cycle{\_}cross{\_}t{\_}distance}}{cycle:uncross:unt:undistance}}\\{{\nwixident{cycle{\_}different}}{cycle:undifferent}}\\{{\nwixident{cycle{\_}f{\_}orthogonal}}{cycle:unf:unorthogonal}}\\{{\nwixident{cycle{\_}moebius}}{cycle:unmoebius}}\\{{\nwixident{cycle{\_}orthogonal}}{cycle:unorthogonal}}\\{{\nwixident{cycle{\_}sl2}}{cycle:unsl2}}\\{{\nwixident{cycle{\_}tangent}}{cycle:untangent}}\\{{\nwixident{cycle{\_}tangent{\_}i}}{cycle:untangent:uni}}\\{{\nwixident{cycle{\_}tangent{\_}o}}{cycle:untangent:uno}}\\{{\nwixident{ex}}{ex}}\\{{\nwixident{product{\_}sign}}{product:unsign}}\\{{\nwixident{read{\_}archive}}{read:unarchive}}\\{{\nwixident{steiner{\_}power}}{steiner:unpower}}}\nwindexuse{\nwixident{archive}}{archive}{NWppJ6t-3bfNK9-6}\nwindexuse{\nwixident{coefficients{\_}are{\_}real}}{coefficients:unare:unreal}{NWppJ6t-3bfNK9-6}\nwindexuse{\nwixident{cycle{\_}adifferent}}{cycle:unadifferent}{NWppJ6t-3bfNK9-6}\nwindexuse{\nwixident{cycle{\_}angle}}{cycle:unangle}{NWppJ6t-3bfNK9-6}\nwindexuse{\nwixident{cycle{\_}cross{\_}t{\_}distance}}{cycle:uncross:unt:undistance}{NWppJ6t-3bfNK9-6}\nwindexuse{\nwixident{cycle{\_}different}}{cycle:undifferent}{NWppJ6t-3bfNK9-6}\nwindexuse{\nwixident{cycle{\_}f{\_}orthogonal}}{cycle:unf:unorthogonal}{NWppJ6t-3bfNK9-6}\nwindexuse{\nwixident{cycle{\_}moebius}}{cycle:unmoebius}{NWppJ6t-3bfNK9-6}\nwindexuse{\nwixident{cycle{\_}orthogonal}}{cycle:unorthogonal}{NWppJ6t-3bfNK9-6}\nwindexuse{\nwixident{cycle{\_}sl2}}{cycle:unsl2}{NWppJ6t-3bfNK9-6}\nwindexuse{\nwixident{cycle{\_}tangent}}{cycle:untangent}{NWppJ6t-3bfNK9-6}\nwindexuse{\nwixident{cycle{\_}tangent{\_}i}}{cycle:untangent:uni}{NWppJ6t-3bfNK9-6}\nwindexuse{\nwixident{cycle{\_}tangent{\_}o}}{cycle:untangent:uno}{NWppJ6t-3bfNK9-6}\nwindexuse{\nwixident{ex}}{ex}{NWppJ6t-3bfNK9-6}\nwindexuse{\nwixident{product{\_}sign}}{product:unsign}{NWppJ6t-3bfNK9-6}\nwindexuse{\nwixident{read{\_}archive}}{read:unarchive}{NWppJ6t-3bfNK9-6}\nwindexuse{\nwixident{steiner{\_}power}}{steiner:unpower}{NWppJ6t-3bfNK9-6}\nwendcode{}\nwbegindocs{711}\nwdocspar
\nwenddocs{}\nwbegincode{712}\sublabel{NWppJ6t-3bfNK9-7}\nwmargintag{{\nwtagstyle{}\subpageref{NWppJ6t-3bfNK9-7}}}\moddef{cycle relation class~{\nwtagstyle{}\subpageref{NWppJ6t-3bfNK9-1}}}\plusendmoddef\Rm{}\nwstartdeflinemarkup\nwusesondefline{\\{NWppJ6t-32jW6L-1}}\nwprevnextdefs{NWppJ6t-3bfNK9-6}{NWppJ6t-3bfNK9-8}\nwenddeflinemarkup
{\it{}GINAC\_BIND\_UNARCHIVER}({\bf{}cycle\_relation});\nwindexdefn{\nwixident{cycle{\_}relation}}{cycle:unrelation}{NWppJ6t-3bfNK9-7}

\nwused{\\{NWppJ6t-32jW6L-1}}\nwidentdefs{\\{{\nwixident{cycle{\_}relation}}{cycle:unrelation}}}\nwendcode{}\nwbegindocs{713}\nwdocspar
\nwenddocs{}\nwbegincode{714}\sublabel{NWppJ6t-3bfNK9-8}\nwmargintag{{\nwtagstyle{}\subpageref{NWppJ6t-3bfNK9-8}}}\moddef{cycle relation class~{\nwtagstyle{}\subpageref{NWppJ6t-3bfNK9-1}}}\plusendmoddef\Rm{}\nwstartdeflinemarkup\nwusesondefline{\\{NWppJ6t-32jW6L-1}}\nwprevnextdefs{NWppJ6t-3bfNK9-7}{NWppJ6t-3bfNK9-9}\nwenddeflinemarkup
{\bf{}ex} {\bf{}cycle\_relation}::{\it{}rel\_to\_parent}({\bf{}const} {\bf{}cycle\_data} & {\it{}C1}, {\bf{}const} {\bf{}ex} & {\it{}pmetric}, {\bf{}const} {\bf{}ex} & {\it{}cmetric},
                                 {\bf{}const} {\it{}exhashmap}\begin{math}<\end{math}{\bf{}cycle\_node}\begin{math}>\end{math} & {\it{}N}) {\bf{}const}
{\nwlbrace}

\nwused{\\{NWppJ6t-32jW6L-1}}\nwidentuses{\\{{\nwixident{cycle{\_}data}}{cycle:undata}}\\{{\nwixident{cycle{\_}node}}{cycle:unnode}}\\{{\nwixident{cycle{\_}relation}}{cycle:unrelation}}\\{{\nwixident{ex}}{ex}}}\nwindexuse{\nwixident{cycle{\_}data}}{cycle:undata}{NWppJ6t-3bfNK9-8}\nwindexuse{\nwixident{cycle{\_}node}}{cycle:unnode}{NWppJ6t-3bfNK9-8}\nwindexuse{\nwixident{cycle{\_}relation}}{cycle:unrelation}{NWppJ6t-3bfNK9-8}\nwindexuse{\nwixident{ex}}{ex}{NWppJ6t-3bfNK9-8}\nwendcode{}\nwbegindocs{715}First we check if the required key exists in the cycles list. If
there is no such key, we return the relation to the calling cycle itself.
\nwenddocs{}\nwbegincode{716}\sublabel{NWppJ6t-3bfNK9-9}\nwmargintag{{\nwtagstyle{}\subpageref{NWppJ6t-3bfNK9-9}}}\moddef{cycle relation class~{\nwtagstyle{}\subpageref{NWppJ6t-3bfNK9-1}}}\plusendmoddef\Rm{}\nwstartdeflinemarkup\nwusesondefline{\\{NWppJ6t-32jW6L-1}}\nwprevnextdefs{NWppJ6t-3bfNK9-8}{NWppJ6t-3bfNK9-A}\nwenddeflinemarkup
    {\it{}exhashmap}\begin{math}<\end{math}{\bf{}cycle\_node}\begin{math}>\end{math}::{\it{}const\_iterator}  {\it{}cnode}={\it{}N}.{\it{}find}({\it{}parkey});

\nwused{\\{NWppJ6t-32jW6L-1}}\nwidentuses{\\{{\nwixident{cycle{\_}node}}{cycle:unnode}}}\nwindexuse{\nwixident{cycle{\_}node}}{cycle:unnode}{NWppJ6t-3bfNK9-9}\nwendcode{}\nwbegindocs{717}Otherwise the list of equations is constructed for the found key.
\nwenddocs{}\nwbegincode{718}\sublabel{NWppJ6t-3bfNK9-A}\nwmargintag{{\nwtagstyle{}\subpageref{NWppJ6t-3bfNK9-A}}}\moddef{cycle relation class~{\nwtagstyle{}\subpageref{NWppJ6t-3bfNK9-1}}}\plusendmoddef\Rm{}\nwstartdeflinemarkup\nwusesondefline{\\{NWppJ6t-32jW6L-1}}\nwprevnextdefs{NWppJ6t-3bfNK9-9}{NWppJ6t-3bfNK9-B}\nwenddeflinemarkup
    {\bf{}lst} {\it{}res},
        {\it{}cycles}={\it{}ex\_to}\begin{math}<\end{math}{\bf{}lst}\begin{math}>\end{math}({\it{}cnode}\begin{math}\rightarrow\end{math}{\it{}second}.{\it{}get\_cycle}({\it{}use\_cycle\_metric}? {\it{}cmetric} : {\it{}pmetric}));

    {\bf{}for} ({\bf{}const} {\bf{}auto}& {\it{}it} : {\it{}cycles}) {\nwlbrace}
        {\bf{}lst} {\it{}calc}={\it{}ex\_to}\begin{math}<\end{math}{\bf{}lst}\begin{math}>\end{math}({\it{}rel}({\it{}C1}.{\it{}get\_cycle}({\it{}use\_cycle\_metric}? {\it{}cmetric} : {\it{}pmetric}), {\it{}ex\_to}\begin{math}<\end{math}{\bf{}cycle}\begin{math}>\end{math}({\it{}it}), {\it{}parameter}));
        {\bf{}for} ({\bf{}const} {\bf{}auto}& {\it{}it1} : {\it{}calc})
            {\it{}res}.{\it{}append}({\it{}it1});
    {\nwrbrace}
    {\bf{}return} {\it{}res};
{\nwrbrace}

\nwused{\\{NWppJ6t-32jW6L-1}}\nwidentuses{\\{{\nwixident{get{\_}cycle}}{get:uncycle}}}\nwindexuse{\nwixident{get{\_}cycle}}{get:uncycle}{NWppJ6t-3bfNK9-A}\nwendcode{}\nwbegindocs{719}\nwdocspar
\nwenddocs{}\nwbegincode{720}\sublabel{NWppJ6t-3bfNK9-B}\nwmargintag{{\nwtagstyle{}\subpageref{NWppJ6t-3bfNK9-B}}}\moddef{cycle relation class~{\nwtagstyle{}\subpageref{NWppJ6t-3bfNK9-1}}}\plusendmoddef\Rm{}\nwstartdeflinemarkup\nwusesondefline{\\{NWppJ6t-32jW6L-1}}\nwprevnextdefs{NWppJ6t-3bfNK9-A}{NWppJ6t-3bfNK9-C}\nwenddeflinemarkup
{\bf{}void} {\bf{}cycle\_relation}::{\it{}do\_print}({\bf{}const} {\it{}print\_dflt} & {\it{}con}, {\bf{}unsigned} {\it{}level}) {\bf{}const}\nwindexdefn{\nwixident{cycle{\_}relation}}{cycle:unrelation}{NWppJ6t-3bfNK9-B}
{\nwlbrace}
    {\it{}con}.{\it{}s} \begin{math}\ll\end{math} {\it{}parkey} \begin{math}\ll\end{math} ({\it{}use\_cycle\_metric}? {\tt{}"|"} : {\tt{}"/"});
    {\bf{}ex} (\begin{math}\ast\end{math}{\bf{}const}\begin{math}\ast\end{math} {\it{}ptr})({\bf{}const} {\bf{}ex} &, {\bf{}const} {\bf{}ex} &, {\bf{}const} {\bf{}ex} &)
        = {\it{}rel}.{\it{}target}\begin{math}<\end{math}{\bf{}ex}(\begin{math}\ast\end{math})({\bf{}const} {\bf{}ex}&, {\bf{}const} {\bf{}ex} &, {\bf{}const} {\bf{}ex} &)\begin{math}>\end{math}();
    {\bf{}if} ({\it{}ptr} \begin{math}\wedge\end{math} \begin{math}\ast\end{math}{\it{}ptr} \begin{math}\equiv\end{math} {\it{}cycle\_orthogonal})
        {\it{}con}.{\it{}s} \begin{math}\ll\end{math} {\tt{}"o"};
    {\bf{}else} {\bf{}if} ({\it{}ptr} \begin{math}\wedge\end{math} \begin{math}\ast\end{math}{\it{}ptr} \begin{math}\equiv\end{math} {\it{}cycle\_f\_orthogonal})
        {\it{}con}.{\it{}s} \begin{math}\ll\end{math} {\tt{}"f"};
    {\bf{}else} {\bf{}if} ({\it{}ptr} \begin{math}\wedge\end{math} \begin{math}\ast\end{math}{\it{}ptr} \begin{math}\equiv\end{math} {\it{}cycle\_different})
        {\it{}con}.{\it{}s} \begin{math}\ll\end{math} {\tt{}"d"};
    {\bf{}else} {\bf{}if} ({\it{}ptr} \begin{math}\wedge\end{math} \begin{math}\ast\end{math}{\it{}ptr} \begin{math}\equiv\end{math} {\it{}cycle\_adifferent})
        {\it{}con}.{\it{}s} \begin{math}\ll\end{math} {\tt{}"ad"};
    {\bf{}else} {\bf{}if} ({\it{}ptr} \begin{math}\wedge\end{math} \begin{math}\ast\end{math}{\it{}ptr} \begin{math}\equiv\end{math} {\it{}cycle\_tangent})
        {\it{}con}.{\it{}s} \begin{math}\ll\end{math} {\tt{}"t"};
    {\bf{}else} {\bf{}if} ({\it{}ptr} \begin{math}\wedge\end{math} \begin{math}\ast\end{math}{\it{}ptr} \begin{math}\equiv\end{math} {\it{}cycle\_tangent\_i})
        {\it{}con}.{\it{}s} \begin{math}\ll\end{math} {\tt{}"ti"};
    {\bf{}else} {\bf{}if} ({\it{}ptr} \begin{math}\wedge\end{math} \begin{math}\ast\end{math}{\it{}ptr} \begin{math}\equiv\end{math} {\it{}cycle\_tangent\_o})
        {\it{}con}.{\it{}s} \begin{math}\ll\end{math} {\tt{}"to"};
    {\bf{}else} {\bf{}if} ({\it{}ptr} \begin{math}\wedge\end{math} \begin{math}\ast\end{math}{\it{}ptr} \begin{math}\equiv\end{math} {\it{}steiner\_power})
        {\it{}con}.{\it{}s} \begin{math}\ll\end{math} {\tt{}"s"};
    {\bf{}else} {\bf{}if} ({\it{}ptr} \begin{math}\wedge\end{math} \begin{math}\ast\end{math}{\it{}ptr} \begin{math}\equiv\end{math} {\it{}cycle\_angle})
        {\it{}con}.{\it{}s} \begin{math}\ll\end{math} {\tt{}"a"};
    {\bf{}else} {\bf{}if} ({\it{}ptr} \begin{math}\wedge\end{math} \begin{math}\ast\end{math}{\it{}ptr} \begin{math}\equiv\end{math} {\it{}cycle\_cross\_t\_distance})
        {\it{}con}.{\it{}s} \begin{math}\ll\end{math} {\tt{}"c"};
    {\bf{}else} {\bf{}if} ({\it{}ptr} \begin{math}\wedge\end{math} \begin{math}\ast\end{math}{\it{}ptr} \begin{math}\equiv\end{math} {\it{}product\_sign})
        {\it{}con}.{\it{}s} \begin{math}\ll\end{math} {\tt{}"p"};
    {\bf{}else} {\bf{}if} ({\it{}ptr} \begin{math}\wedge\end{math} \begin{math}\ast\end{math}{\it{}ptr} \begin{math}\equiv\end{math} {\it{}coefficients\_are\_real})
        {\it{}con}.{\it{}s} \begin{math}\ll\end{math} {\tt{}"r"};
    {\bf{}else} {\bf{}if} ({\it{}ptr} \begin{math}\wedge\end{math} \begin{math}\ast\end{math}{\it{}ptr} \begin{math}\equiv\end{math} {\it{}cycle\_moebius})
        {\it{}con}.{\it{}s} \begin{math}\ll\end{math} {\tt{}"m"};
    {\bf{}else} {\bf{}if} ({\it{}ptr} \begin{math}\wedge\end{math} \begin{math}\ast\end{math}{\it{}ptr} \begin{math}\equiv\end{math} {\it{}cycle\_sl2})
        {\it{}con}.{\it{}s} \begin{math}\ll\end{math} {\tt{}"l"};
    {\bf{}else}
        {\it{}con}.{\it{}s} \begin{math}\ll\end{math} {\tt{}"u"}; // unknown type of relations
    {\bf{}if} (\begin{math}\neg\end{math} {\it{}parameter}.{\it{}is\_zero}())
        {\it{}con}.{\it{}s} \begin{math}\ll\end{math} {\tt{}"["} \begin{math}\ll\end{math} {\it{}parameter} \begin{math}\ll\end{math} {\tt{}"]"};
{\nwrbrace}

\nwused{\\{NWppJ6t-32jW6L-1}}\nwidentdefs{\\{{\nwixident{cycle{\_}relation}}{cycle:unrelation}}}\nwidentuses{\\{{\nwixident{coefficients{\_}are{\_}real}}{coefficients:unare:unreal}}\\{{\nwixident{cycle{\_}adifferent}}{cycle:unadifferent}}\\{{\nwixident{cycle{\_}angle}}{cycle:unangle}}\\{{\nwixident{cycle{\_}cross{\_}t{\_}distance}}{cycle:uncross:unt:undistance}}\\{{\nwixident{cycle{\_}different}}{cycle:undifferent}}\\{{\nwixident{cycle{\_}f{\_}orthogonal}}{cycle:unf:unorthogonal}}\\{{\nwixident{cycle{\_}moebius}}{cycle:unmoebius}}\\{{\nwixident{cycle{\_}orthogonal}}{cycle:unorthogonal}}\\{{\nwixident{cycle{\_}sl2}}{cycle:unsl2}}\\{{\nwixident{cycle{\_}tangent}}{cycle:untangent}}\\{{\nwixident{cycle{\_}tangent{\_}i}}{cycle:untangent:uni}}\\{{\nwixident{cycle{\_}tangent{\_}o}}{cycle:untangent:uno}}\\{{\nwixident{ex}}{ex}}\\{{\nwixident{l}}{l}}\\{{\nwixident{m}}{m}}\\{{\nwixident{product{\_}sign}}{product:unsign}}\\{{\nwixident{steiner{\_}power}}{steiner:unpower}}}\nwindexuse{\nwixident{coefficients{\_}are{\_}real}}{coefficients:unare:unreal}{NWppJ6t-3bfNK9-B}\nwindexuse{\nwixident{cycle{\_}adifferent}}{cycle:unadifferent}{NWppJ6t-3bfNK9-B}\nwindexuse{\nwixident{cycle{\_}angle}}{cycle:unangle}{NWppJ6t-3bfNK9-B}\nwindexuse{\nwixident{cycle{\_}cross{\_}t{\_}distance}}{cycle:uncross:unt:undistance}{NWppJ6t-3bfNK9-B}\nwindexuse{\nwixident{cycle{\_}different}}{cycle:undifferent}{NWppJ6t-3bfNK9-B}\nwindexuse{\nwixident{cycle{\_}f{\_}orthogonal}}{cycle:unf:unorthogonal}{NWppJ6t-3bfNK9-B}\nwindexuse{\nwixident{cycle{\_}moebius}}{cycle:unmoebius}{NWppJ6t-3bfNK9-B}\nwindexuse{\nwixident{cycle{\_}orthogonal}}{cycle:unorthogonal}{NWppJ6t-3bfNK9-B}\nwindexuse{\nwixident{cycle{\_}sl2}}{cycle:unsl2}{NWppJ6t-3bfNK9-B}\nwindexuse{\nwixident{cycle{\_}tangent}}{cycle:untangent}{NWppJ6t-3bfNK9-B}\nwindexuse{\nwixident{cycle{\_}tangent{\_}i}}{cycle:untangent:uni}{NWppJ6t-3bfNK9-B}\nwindexuse{\nwixident{cycle{\_}tangent{\_}o}}{cycle:untangent:uno}{NWppJ6t-3bfNK9-B}\nwindexuse{\nwixident{ex}}{ex}{NWppJ6t-3bfNK9-B}\nwindexuse{\nwixident{l}}{l}{NWppJ6t-3bfNK9-B}\nwindexuse{\nwixident{m}}{m}{NWppJ6t-3bfNK9-B}\nwindexuse{\nwixident{product{\_}sign}}{product:unsign}{NWppJ6t-3bfNK9-B}\nwindexuse{\nwixident{steiner{\_}power}}{steiner:unpower}{NWppJ6t-3bfNK9-B}\nwendcode{}\nwbegindocs{721}\nwdocspar
\nwenddocs{}\nwbegincode{722}\sublabel{NWppJ6t-3bfNK9-C}\nwmargintag{{\nwtagstyle{}\subpageref{NWppJ6t-3bfNK9-C}}}\moddef{cycle relation class~{\nwtagstyle{}\subpageref{NWppJ6t-3bfNK9-1}}}\plusendmoddef\Rm{}\nwstartdeflinemarkup\nwusesondefline{\\{NWppJ6t-32jW6L-1}}\nwprevnextdefs{NWppJ6t-3bfNK9-B}{NWppJ6t-3bfNK9-D}\nwenddeflinemarkup
{\bf{}void} {\bf{}cycle\_relation}::{\it{}do\_print\_tree}({\bf{}const} {\it{}print\_tree} & {\it{}con}, {\bf{}unsigned} {\it{}level}) {\bf{}const}\nwindexdefn{\nwixident{cycle{\_}relation}}{cycle:unrelation}{NWppJ6t-3bfNK9-C}
{\nwlbrace}
    //  inherited::do\_print\_tree(con,level);
    {\it{}parkey}.{\it{}print}({\it{}con},{\it{}level}+{\it{}con}.{\it{}delta\_indent});
    //  con.s \begin{math}<\end{math}\begin{math}<\end{math}  std::string(level+con.delta\_indent, ' ') \begin{math}<\end{math}\begin{math}<\end{math} (int)rel \begin{math}<\end{math}\begin{math}<\end{math}endl;
{\nwrbrace}

\nwused{\\{NWppJ6t-32jW6L-1}}\nwidentdefs{\\{{\nwixident{cycle{\_}relation}}{cycle:unrelation}}}\nwendcode{}\nwbegindocs{723}\nwdocspar
\nwenddocs{}\nwbegincode{724}\sublabel{NWppJ6t-3bfNK9-D}\nwmargintag{{\nwtagstyle{}\subpageref{NWppJ6t-3bfNK9-D}}}\moddef{cycle relation class~{\nwtagstyle{}\subpageref{NWppJ6t-3bfNK9-1}}}\plusendmoddef\Rm{}\nwstartdeflinemarkup\nwusesondefline{\\{NWppJ6t-32jW6L-1}}\nwprevnextdefs{NWppJ6t-3bfNK9-C}{NWppJ6t-3bfNK9-E}\nwenddeflinemarkup
{\bf{}ex} {\bf{}cycle\_relation}::{\it{}op}({\it{}size\_t} {\it{}i}) {\bf{}const}
{\nwlbrace}
 {\it{}GINAC\_ASSERT}({\it{}i}\begin{math}<\end{math}{\it{}nops}());
    {\bf{}switch}({\it{}i}) {\nwlbrace}
    {\bf{}case} 0:
        {\bf{}return} {\it{}parkey};
    {\bf{}case} 1:
        {\bf{}return} {\it{}parameter};
    {\bf{}default}:
        {\bf{}throw}({\it{}std}::{\it{}invalid\_argument}({\tt{}"cycle\_relation::op(): requested operand out of the range (1)"}));
    {\nwrbrace}
{\nwrbrace}

\nwused{\\{NWppJ6t-32jW6L-1}}\nwidentuses{\\{{\nwixident{cycle{\_}relation}}{cycle:unrelation}}\\{{\nwixident{ex}}{ex}}\\{{\nwixident{nops}}{nops}}\\{{\nwixident{op}}{op}}}\nwindexuse{\nwixident{cycle{\_}relation}}{cycle:unrelation}{NWppJ6t-3bfNK9-D}\nwindexuse{\nwixident{ex}}{ex}{NWppJ6t-3bfNK9-D}\nwindexuse{\nwixident{nops}}{nops}{NWppJ6t-3bfNK9-D}\nwindexuse{\nwixident{op}}{op}{NWppJ6t-3bfNK9-D}\nwendcode{}\nwbegindocs{725}\nwdocspar
\nwenddocs{}\nwbegincode{726}\sublabel{NWppJ6t-3bfNK9-E}\nwmargintag{{\nwtagstyle{}\subpageref{NWppJ6t-3bfNK9-E}}}\moddef{cycle relation class~{\nwtagstyle{}\subpageref{NWppJ6t-3bfNK9-1}}}\plusendmoddef\Rm{}\nwstartdeflinemarkup\nwusesondefline{\\{NWppJ6t-32jW6L-1}}\nwprevnextdefs{NWppJ6t-3bfNK9-D}{\relax}\nwenddeflinemarkup
{\bf{}ex} & {\bf{}cycle\_relation}::{\it{}let\_op}({\it{}size\_t} {\it{}i})
{\nwlbrace}
    {\it{}ensure\_if\_modifiable}();
    {\it{}GINAC\_ASSERT}({\it{}i}\begin{math}<\end{math}{\it{}nops}());
    {\bf{}switch}({\it{}i}) {\nwlbrace}
    {\bf{}case} 0:
        {\bf{}return} {\it{}parkey};
    {\bf{}case} 1:
        {\bf{}return} {\it{}parameter};
    {\bf{}default}:
        {\bf{}throw}({\it{}std}::{\it{}invalid\_argument}({\tt{}"cycle\_relation::let\_op(): requested operand out of the range (1)"}));
    {\nwrbrace}
{\nwrbrace}

\nwused{\\{NWppJ6t-32jW6L-1}}\nwidentuses{\\{{\nwixident{cycle{\_}relation}}{cycle:unrelation}}\\{{\nwixident{ex}}{ex}}\\{{\nwixident{nops}}{nops}}}\nwindexuse{\nwixident{cycle{\_}relation}}{cycle:unrelation}{NWppJ6t-3bfNK9-E}\nwindexuse{\nwixident{ex}}{ex}{NWppJ6t-3bfNK9-E}\nwindexuse{\nwixident{nops}}{nops}{NWppJ6t-3bfNK9-E}\nwendcode{}\nwbegindocs{727}\nwdocspar

\subsection{Implementation of {\Tt{}\Rm{}{\bf{}subfigure}\nwendquote} class}
\label{sec:impl-subf-class}

\nwenddocs{}\nwbegindocs{728}\nwdocspar
\nwenddocs{}\nwbegincode{729}\sublabel{NWppJ6t-3dX0u0-1}\nwmargintag{{\nwtagstyle{}\subpageref{NWppJ6t-3dX0u0-1}}}\moddef{subfigure class~{\nwtagstyle{}\subpageref{NWppJ6t-3dX0u0-1}}}\endmoddef\Rm{}\nwstartdeflinemarkup\nwusesondefline{\\{NWppJ6t-32jW6L-1}}\nwprevnextdefs{\relax}{NWppJ6t-3dX0u0-2}\nwenddeflinemarkup
{\bf{}subfigure}::{\bf{}subfigure}() : {\it{}inherited}()
{\nwlbrace}
{\nwrbrace}

\nwalsodefined{\\{NWppJ6t-3dX0u0-2}\\{NWppJ6t-3dX0u0-3}\\{NWppJ6t-3dX0u0-4}\\{NWppJ6t-3dX0u0-5}\\{NWppJ6t-3dX0u0-6}\\{NWppJ6t-3dX0u0-7}\\{NWppJ6t-3dX0u0-8}\\{NWppJ6t-3dX0u0-9}}\nwused{\\{NWppJ6t-32jW6L-1}}\nwidentuses{\\{{\nwixident{subfigure}}{subfigure}}}\nwindexuse{\nwixident{subfigure}}{subfigure}{NWppJ6t-3dX0u0-1}\nwendcode{}\nwbegindocs{730}\nwdocspar
\nwenddocs{}\nwbegincode{731}\sublabel{NWppJ6t-3dX0u0-2}\nwmargintag{{\nwtagstyle{}\subpageref{NWppJ6t-3dX0u0-2}}}\moddef{subfigure class~{\nwtagstyle{}\subpageref{NWppJ6t-3dX0u0-1}}}\plusendmoddef\Rm{}\nwstartdeflinemarkup\nwusesondefline{\\{NWppJ6t-32jW6L-1}}\nwprevnextdefs{NWppJ6t-3dX0u0-1}{NWppJ6t-3dX0u0-3}\nwenddeflinemarkup
{\bf{}subfigure}::{\bf{}subfigure}({\bf{}const} {\bf{}ex} & {\it{}F}, {\bf{}const} {\bf{}ex} & {\it{}l}) {\nwlbrace}
    {\it{}parlist} = {\it{}ex\_to}\begin{math}<\end{math}{\bf{}lst}\begin{math}>\end{math}({\it{}l});
    {\it{}subf} = {\it{}F};
{\nwrbrace}

\nwused{\\{NWppJ6t-32jW6L-1}}\nwidentuses{\\{{\nwixident{ex}}{ex}}\\{{\nwixident{l}}{l}}\\{{\nwixident{subfigure}}{subfigure}}}\nwindexuse{\nwixident{ex}}{ex}{NWppJ6t-3dX0u0-2}\nwindexuse{\nwixident{l}}{l}{NWppJ6t-3dX0u0-2}\nwindexuse{\nwixident{subfigure}}{subfigure}{NWppJ6t-3dX0u0-2}\nwendcode{}\nwbegindocs{732}\nwdocspar
\nwenddocs{}\nwbegincode{733}\sublabel{NWppJ6t-3dX0u0-3}\nwmargintag{{\nwtagstyle{}\subpageref{NWppJ6t-3dX0u0-3}}}\moddef{subfigure class~{\nwtagstyle{}\subpageref{NWppJ6t-3dX0u0-1}}}\plusendmoddef\Rm{}\nwstartdeflinemarkup\nwusesondefline{\\{NWppJ6t-32jW6L-1}}\nwprevnextdefs{NWppJ6t-3dX0u0-2}{NWppJ6t-3dX0u0-4}\nwenddeflinemarkup
{\it{}return\_type\_t} {\bf{}subfigure}::{\it{}return\_type\_tinfo}() {\bf{}const}
{\nwlbrace}
    {\bf{}return} {\it{}make\_return\_type\_t}\begin{math}<\end{math}{\bf{}subfigure}\begin{math}>\end{math}();
{\nwrbrace}

\nwused{\\{NWppJ6t-32jW6L-1}}\nwidentuses{\\{{\nwixident{subfigure}}{subfigure}}}\nwindexuse{\nwixident{subfigure}}{subfigure}{NWppJ6t-3dX0u0-3}\nwendcode{}\nwbegindocs{734}\nwdocspar
\nwenddocs{}\nwbegincode{735}\sublabel{NWppJ6t-3dX0u0-4}\nwmargintag{{\nwtagstyle{}\subpageref{NWppJ6t-3dX0u0-4}}}\moddef{subfigure class~{\nwtagstyle{}\subpageref{NWppJ6t-3dX0u0-1}}}\plusendmoddef\Rm{}\nwstartdeflinemarkup\nwusesondefline{\\{NWppJ6t-32jW6L-1}}\nwprevnextdefs{NWppJ6t-3dX0u0-3}{NWppJ6t-3dX0u0-5}\nwenddeflinemarkup
{\bf{}int} {\bf{}subfigure}::{\it{}compare\_same\_type}({\bf{}const} {\bf{}basic} &{\it{}other}) {\bf{}const}\nwindexdefn{\nwixident{subfigure}}{subfigure}{NWppJ6t-3dX0u0-4}
{\nwlbrace}
       {\it{}GINAC\_ASSERT}({\it{}is\_a}\begin{math}<\end{math}{\bf{}subfigure}\begin{math}>\end{math}({\it{}other}));
       {\bf{}return} {\it{}inherited}::{\it{}compare\_same\_type}({\it{}other});
{\nwrbrace}

\nwused{\\{NWppJ6t-32jW6L-1}}\nwidentdefs{\\{{\nwixident{subfigure}}{subfigure}}}\nwendcode{}\nwbegindocs{736}(un)Archiving of {\Tt{}\Rm{}{\bf{}subfigure}\nwendquote} is not universal. At present it
only can handle relations declared in the header file: {\Tt{}\Rm{}{\it{}cycle\_orthogonal}\nwendquote}
and {\Tt{}\Rm{}{\it{}cycle\_f\_orthogonal}\nwendquote}.

\nwenddocs{}\nwbegincode{737}\sublabel{NWppJ6t-3dX0u0-5}\nwmargintag{{\nwtagstyle{}\subpageref{NWppJ6t-3dX0u0-5}}}\moddef{subfigure class~{\nwtagstyle{}\subpageref{NWppJ6t-3dX0u0-1}}}\plusendmoddef\Rm{}\nwstartdeflinemarkup\nwusesondefline{\\{NWppJ6t-32jW6L-1}}\nwprevnextdefs{NWppJ6t-3dX0u0-4}{NWppJ6t-3dX0u0-6}\nwenddeflinemarkup
{\bf{}void} {\bf{}subfigure}::{\it{}archive}({\it{}archive\_node} &{\it{}n}) {\bf{}const}\nwindexdefn{\nwixident{subfigure}}{subfigure}{NWppJ6t-3dX0u0-5}
{\nwlbrace}
    {\it{}inherited}::{\it{}archive}({\it{}n});
    {\it{}n}.{\it{}add\_ex}({\tt{}"parlist"}, {\it{}ex\_to}\begin{math}<\end{math}{\bf{}lst}\begin{math}>\end{math}({\it{}parlist}));
    {\it{}n}.{\it{}add\_ex}({\tt{}"subf"}, {\it{}ex\_to}\begin{math}<\end{math}{\bf{}figure}\begin{math}>\end{math}({\it{}subf}));
{\nwrbrace}

\nwused{\\{NWppJ6t-32jW6L-1}}\nwidentdefs{\\{{\nwixident{subfigure}}{subfigure}}}\nwidentuses{\\{{\nwixident{archive}}{archive}}\\{{\nwixident{figure}}{figure}}}\nwindexuse{\nwixident{archive}}{archive}{NWppJ6t-3dX0u0-5}\nwindexuse{\nwixident{figure}}{figure}{NWppJ6t-3dX0u0-5}\nwendcode{}\nwbegindocs{738}\nwdocspar
\nwenddocs{}\nwbegincode{739}\sublabel{NWppJ6t-3dX0u0-6}\nwmargintag{{\nwtagstyle{}\subpageref{NWppJ6t-3dX0u0-6}}}\moddef{subfigure class~{\nwtagstyle{}\subpageref{NWppJ6t-3dX0u0-1}}}\plusendmoddef\Rm{}\nwstartdeflinemarkup\nwusesondefline{\\{NWppJ6t-32jW6L-1}}\nwprevnextdefs{NWppJ6t-3dX0u0-5}{NWppJ6t-3dX0u0-7}\nwenddeflinemarkup
{\bf{}void} {\bf{}subfigure}::{\it{}read\_archive}({\bf{}const} {\it{}archive\_node} &{\it{}n}, {\bf{}lst} &{\it{}sym\_lst})\nwindexdefn{\nwixident{subfigure}}{subfigure}{NWppJ6t-3dX0u0-6}
{\nwlbrace}
    {\bf{}ex} {\it{}e};
    {\it{}inherited}::{\it{}read\_archive}({\it{}n}, {\it{}sym\_lst});
    {\it{}n}.{\it{}find\_ex}({\tt{}"parlist"}, {\it{}e}, {\it{}sym\_lst});
    {\bf{}if} ({\it{}is\_a}\begin{math}<\end{math}{\bf{}lst}\begin{math}>\end{math}({\it{}e}))
        {\it{}parlist}={\it{}ex\_to}\begin{math}<\end{math}{\bf{}lst}\begin{math}>\end{math}({\it{}e});
    {\bf{}else}
        {\bf{}throw}({\it{}std}::{\it{}invalid\_argument}({\tt{}"subfigure::read\_archive(): read a non-lst as a parlist from"}
                                    {\tt{}" the archive"}));
    {\it{}n}.{\it{}find\_ex}({\tt{}"subf"}, {\it{}e}, {\it{}sym\_lst});
    {\bf{}if} ({\it{}is\_a}\begin{math}<\end{math}{\bf{}figure}\begin{math}>\end{math}({\it{}e}))
        {\it{}subf}={\it{}ex\_to}\begin{math}<\end{math}{\bf{}figure}\begin{math}>\end{math}({\it{}e});
    {\bf{}else}
        {\bf{}throw}({\it{}std}::{\it{}invalid\_argument}({\tt{}"subfigure::read\_archive(): read a non-figure as a subf from"}
                                    {\tt{}" the archive"}));
{\nwrbrace}

\nwused{\\{NWppJ6t-32jW6L-1}}\nwidentdefs{\\{{\nwixident{subfigure}}{subfigure}}}\nwidentuses{\\{{\nwixident{archive}}{archive}}\\{{\nwixident{ex}}{ex}}\\{{\nwixident{figure}}{figure}}\\{{\nwixident{read{\_}archive}}{read:unarchive}}}\nwindexuse{\nwixident{archive}}{archive}{NWppJ6t-3dX0u0-6}\nwindexuse{\nwixident{ex}}{ex}{NWppJ6t-3dX0u0-6}\nwindexuse{\nwixident{figure}}{figure}{NWppJ6t-3dX0u0-6}\nwindexuse{\nwixident{read{\_}archive}}{read:unarchive}{NWppJ6t-3dX0u0-6}\nwendcode{}\nwbegindocs{740}\nwdocspar
\nwenddocs{}\nwbegincode{741}\sublabel{NWppJ6t-3dX0u0-7}\nwmargintag{{\nwtagstyle{}\subpageref{NWppJ6t-3dX0u0-7}}}\moddef{subfigure class~{\nwtagstyle{}\subpageref{NWppJ6t-3dX0u0-1}}}\plusendmoddef\Rm{}\nwstartdeflinemarkup\nwusesondefline{\\{NWppJ6t-32jW6L-1}}\nwprevnextdefs{NWppJ6t-3dX0u0-6}{NWppJ6t-3dX0u0-8}\nwenddeflinemarkup
{\it{}GINAC\_BIND\_UNARCHIVER}({\bf{}subfigure});\nwindexdefn{\nwixident{subfigure}}{subfigure}{NWppJ6t-3dX0u0-7}

\nwused{\\{NWppJ6t-32jW6L-1}}\nwidentdefs{\\{{\nwixident{subfigure}}{subfigure}}}\nwendcode{}\nwbegindocs{742}\nwdocspar
\nwenddocs{}\nwbegincode{743}\sublabel{NWppJ6t-3dX0u0-8}\nwmargintag{{\nwtagstyle{}\subpageref{NWppJ6t-3dX0u0-8}}}\moddef{subfigure class~{\nwtagstyle{}\subpageref{NWppJ6t-3dX0u0-1}}}\plusendmoddef\Rm{}\nwstartdeflinemarkup\nwusesondefline{\\{NWppJ6t-32jW6L-1}}\nwprevnextdefs{NWppJ6t-3dX0u0-7}{NWppJ6t-3dX0u0-9}\nwenddeflinemarkup
{\bf{}void} {\bf{}subfigure}::{\it{}do\_print}({\bf{}const} {\it{}print\_dflt} & {\it{}con}, {\bf{}unsigned} {\it{}level}) {\bf{}const}\nwindexdefn{\nwixident{subfigure}}{subfigure}{NWppJ6t-3dX0u0-8}
{\nwlbrace}
    {\it{}con}.{\it{}s} \begin{math}\ll\end{math} {\tt{}"subfig( "} ;
        {\it{}parlist}.{\it{}print}({\it{}con}, {\it{}level}+1);
        //        subf.print(con, level+1);
    {\it{}con}.{\it{}s} \begin{math}\ll\end{math} {\tt{}")"} ;
{\nwrbrace}

\nwused{\\{NWppJ6t-32jW6L-1}}\nwidentdefs{\\{{\nwixident{subfigure}}{subfigure}}}\nwendcode{}\nwbegindocs{744}\nwdocspar
\nwenddocs{}\nwbegincode{745}\sublabel{NWppJ6t-3dX0u0-9}\nwmargintag{{\nwtagstyle{}\subpageref{NWppJ6t-3dX0u0-9}}}\moddef{subfigure class~{\nwtagstyle{}\subpageref{NWppJ6t-3dX0u0-1}}}\plusendmoddef\Rm{}\nwstartdeflinemarkup\nwusesondefline{\\{NWppJ6t-32jW6L-1}}\nwprevnextdefs{NWppJ6t-3dX0u0-8}{\relax}\nwenddeflinemarkup
{\bf{}inline} {\bf{}ex} {\bf{}subfigure}::{\it{}subs}({\bf{}const} {\it{}exmap} & {\it{}em}, {\bf{}unsigned} {\it{}options}) {\bf{}const} {\nwlbrace}
    {\bf{}return} {\bf{}subfigure}({\it{}subf}.{\it{}subs}({\it{}em},{\it{}options} \begin{math}\mid\end{math} {\it{}do\_not\_update\_subfigure}), {\it{}parlist});
{\nwrbrace}

\nwused{\\{NWppJ6t-32jW6L-1}}\nwidentuses{\\{{\nwixident{do{\_}not{\_}update{\_}subfigure}}{do:unnot:unupdate:unsubfigure}}\\{{\nwixident{ex}}{ex}}\\{{\nwixident{subfigure}}{subfigure}}\\{{\nwixident{subs}}{subs}}}\nwindexuse{\nwixident{do{\_}not{\_}update{\_}subfigure}}{do:unnot:unupdate:unsubfigure}{NWppJ6t-3dX0u0-9}\nwindexuse{\nwixident{ex}}{ex}{NWppJ6t-3dX0u0-9}\nwindexuse{\nwixident{subfigure}}{subfigure}{NWppJ6t-3dX0u0-9}\nwindexuse{\nwixident{subs}}{subs}{NWppJ6t-3dX0u0-9}\nwendcode{}\nwbegindocs{746}\nwdocspar
\subsection{Implementation of {\Tt{}\Rm{}{\bf{}cycle\_node}\nwendquote} class}
\label{sec:impem-cycl}

\nwenddocs{}\nwbegindocs{747}Default constructor.
\nwenddocs{}\nwbegincode{748}\sublabel{NWppJ6t-1HqLYY-1}\nwmargintag{{\nwtagstyle{}\subpageref{NWppJ6t-1HqLYY-1}}}\moddef{cycle node class~{\nwtagstyle{}\subpageref{NWppJ6t-1HqLYY-1}}}\endmoddef\Rm{}\nwstartdeflinemarkup\nwusesondefline{\\{NWppJ6t-32jW6L-1}}\nwprevnextdefs{\relax}{NWppJ6t-1HqLYY-2}\nwenddeflinemarkup
{\bf{}cycle\_node}::{\bf{}cycle\_node}()
{\nwlbrace}
    {\it{}generation}=0;
{\nwrbrace}

\nwalsodefined{\\{NWppJ6t-1HqLYY-2}\\{NWppJ6t-1HqLYY-3}\\{NWppJ6t-1HqLYY-4}\\{NWppJ6t-1HqLYY-5}\\{NWppJ6t-1HqLYY-6}\\{NWppJ6t-1HqLYY-7}\\{NWppJ6t-1HqLYY-8}\\{NWppJ6t-1HqLYY-9}\\{NWppJ6t-1HqLYY-A}\\{NWppJ6t-1HqLYY-B}\\{NWppJ6t-1HqLYY-C}\\{NWppJ6t-1HqLYY-D}\\{NWppJ6t-1HqLYY-E}\\{NWppJ6t-1HqLYY-F}\\{NWppJ6t-1HqLYY-G}\\{NWppJ6t-1HqLYY-H}\\{NWppJ6t-1HqLYY-I}\\{NWppJ6t-1HqLYY-J}\\{NWppJ6t-1HqLYY-K}\\{NWppJ6t-1HqLYY-L}\\{NWppJ6t-1HqLYY-M}\\{NWppJ6t-1HqLYY-N}\\{NWppJ6t-1HqLYY-O}\\{NWppJ6t-1HqLYY-P}\\{NWppJ6t-1HqLYY-Q}\\{NWppJ6t-1HqLYY-R}\\{NWppJ6t-1HqLYY-S}}\nwused{\\{NWppJ6t-32jW6L-1}}\nwidentuses{\\{{\nwixident{cycle{\_}node}}{cycle:unnode}}}\nwindexuse{\nwixident{cycle{\_}node}}{cycle:unnode}{NWppJ6t-1HqLYY-1}\nwendcode{}\nwbegindocs{749}Create a {\Tt{}\Rm{}{\bf{}cycle\_node}\nwendquote} out of {\Tt{}\Rm{}{\bf{}cycle}\nwendquote} or {\Tt{}\Rm{}{\bf{}cycle\_node}\nwendquote}.
\nwenddocs{}\nwbegincode{750}\sublabel{NWppJ6t-1HqLYY-2}\nwmargintag{{\nwtagstyle{}\subpageref{NWppJ6t-1HqLYY-2}}}\moddef{cycle node class~{\nwtagstyle{}\subpageref{NWppJ6t-1HqLYY-1}}}\plusendmoddef\Rm{}\nwstartdeflinemarkup\nwusesondefline{\\{NWppJ6t-32jW6L-1}}\nwprevnextdefs{NWppJ6t-1HqLYY-1}{NWppJ6t-1HqLYY-3}\nwenddeflinemarkup
{\bf{}cycle\_node}::{\bf{}cycle\_node}({\bf{}const} {\bf{}ex} & {\it{}C}, {\bf{}int} {\it{}g})
{\nwlbrace}
    {\it{}generation}={\it{}g};
    \LA{}set cycle data to the node~{\nwtagstyle{}\subpageref{NWppJ6t-26cGQh-1}}\RA{}
{\nwrbrace}

\nwused{\\{NWppJ6t-32jW6L-1}}\nwidentuses{\\{{\nwixident{cycle{\_}node}}{cycle:unnode}}\\{{\nwixident{ex}}{ex}}}\nwindexuse{\nwixident{cycle{\_}node}}{cycle:unnode}{NWppJ6t-1HqLYY-2}\nwindexuse{\nwixident{ex}}{ex}{NWppJ6t-1HqLYY-2}\nwendcode{}\nwbegindocs{751}We use this check to initialise or change cycle info of the node.
\nwenddocs{}\nwbegincode{752}\sublabel{NWppJ6t-26cGQh-1}\nwmargintag{{\nwtagstyle{}\subpageref{NWppJ6t-26cGQh-1}}}\moddef{set cycle data to the node~{\nwtagstyle{}\subpageref{NWppJ6t-26cGQh-1}}}\endmoddef\Rm{}\nwstartdeflinemarkup\nwusesondefline{\\{NWppJ6t-1HqLYY-2}}\nwenddeflinemarkup
    {\bf{}if} ({\it{}is\_a}\begin{math}<\end{math}{\bf{}cycle\_node}\begin{math}>\end{math}({\it{}C})) {\nwlbrace}
        {\it{}cycles}={\it{}ex\_to}\begin{math}<\end{math}{\bf{}lst}\begin{math}>\end{math}({\it{}ex\_to}\begin{math}<\end{math}{\bf{}cycle\_node}\begin{math}>\end{math}({\it{}C}).{\it{}get\_cycles}());
        {\it{}generation} = {\it{}ex\_to}\begin{math}<\end{math}{\bf{}cycle\_node}\begin{math}>\end{math}({\it{}C}).{\it{}get\_generation}();
        {\it{}children} = {\it{}ex\_to}\begin{math}<\end{math}{\bf{}cycle\_node}\begin{math}>\end{math}({\it{}C}).{\it{}get\_children}();
        {\it{}parents} = {\it{}ex\_to}\begin{math}<\end{math}{\bf{}cycle\_node}\begin{math}>\end{math}({\it{}C}).{\it{}get\_parents}();
    {\nwrbrace} {\bf{}else}
        \LA{}check cycles are valid~{\nwtagstyle{}\subpageref{NWppJ6t-1qiQ4a-1}}\RA{}

\nwused{\\{NWppJ6t-1HqLYY-2}}\nwidentuses{\\{{\nwixident{cycle{\_}node}}{cycle:unnode}}\\{{\nwixident{get{\_}generation}}{get:ungeneration}}}\nwindexuse{\nwixident{cycle{\_}node}}{cycle:unnode}{NWppJ6t-26cGQh-1}\nwindexuse{\nwixident{get{\_}generation}}{get:ungeneration}{NWppJ6t-26cGQh-1}\nwendcode{}\nwbegindocs{753}\nwdocspar
\nwenddocs{}\nwbegincode{754}\sublabel{NWppJ6t-1HqLYY-3}\nwmargintag{{\nwtagstyle{}\subpageref{NWppJ6t-1HqLYY-3}}}\moddef{cycle node class~{\nwtagstyle{}\subpageref{NWppJ6t-1HqLYY-1}}}\plusendmoddef\Rm{}\nwstartdeflinemarkup\nwusesondefline{\\{NWppJ6t-32jW6L-1}}\nwprevnextdefs{NWppJ6t-1HqLYY-2}{NWppJ6t-1HqLYY-4}\nwenddeflinemarkup
{\bf{}cycle\_node}::{\bf{}cycle\_node}({\bf{}const} {\bf{}ex} & {\it{}C}, {\bf{}int} {\it{}g}, {\bf{}const} {\bf{}lst} & {\it{}par})
{\nwlbrace}
    {\it{}generation}={\it{}g};
    \LA{}check cycles are valid~{\nwtagstyle{}\subpageref{NWppJ6t-1qiQ4a-1}}\RA{}
    \LA{}check parents are valid~{\nwtagstyle{}\subpageref{NWppJ6t-2RJcfD-1}}\RA{}
{\nwrbrace}

\nwused{\\{NWppJ6t-32jW6L-1}}\nwidentuses{\\{{\nwixident{cycle{\_}node}}{cycle:unnode}}\\{{\nwixident{ex}}{ex}}}\nwindexuse{\nwixident{cycle{\_}node}}{cycle:unnode}{NWppJ6t-1HqLYY-3}\nwindexuse{\nwixident{ex}}{ex}{NWppJ6t-1HqLYY-3}\nwendcode{}\nwbegindocs{755}\nwdocspar
\nwenddocs{}\nwbegincode{756}\sublabel{NWppJ6t-1qiQ4a-1}\nwmargintag{{\nwtagstyle{}\subpageref{NWppJ6t-1qiQ4a-1}}}\moddef{check cycles are valid~{\nwtagstyle{}\subpageref{NWppJ6t-1qiQ4a-1}}}\endmoddef\Rm{}\nwstartdeflinemarkup\nwusesondefline{\\{NWppJ6t-26cGQh-1}\\{NWppJ6t-1HqLYY-3}\\{NWppJ6t-1HqLYY-4}\\{NWppJ6t-1HqLYY-5}\\{NWppJ6t-1HqLYY-B}}\nwenddeflinemarkup
    {\bf{}if} ({\it{}is\_a}\begin{math}<\end{math}{\bf{}lst}\begin{math}>\end{math}({\it{}C})) {\nwlbrace}
        {\bf{}for} ({\bf{}const} {\bf{}auto}& {\it{}it} : {\it{}ex\_to}\begin{math}<\end{math}{\bf{}lst}\begin{math}>\end{math}({\it{}C}))
            {\bf{}if} ( {\it{}is\_a}\begin{math}<\end{math}{\bf{}cycle\_data}\begin{math}>\end{math}({\it{}it}) \begin{math}\vee\end{math} {\it{}is\_a}\begin{math}<\end{math}{\bf{}cycle}\begin{math}>\end{math}({\it{}it}))
                {\it{}cycles}.{\it{}append}({\bf{}cycle\_data}({\it{}it}));
            {\bf{}else}
                {\bf{}throw}({\it{}std}::{\it{}invalid\_argument}({\tt{}"cycle\_node::cycle\_node(): "}
                                            {\tt{}"the parameter is list of something which is not"}
                                            {\tt{}" cycle\_data"}));
    {\nwrbrace} {\bf{}else} {\bf{}if} ({\it{}is\_a}\begin{math}<\end{math}{\bf{}cycle\_data}\begin{math}>\end{math}({\it{}C})) {\nwlbrace}
        {\it{}cycles} = {\bf{}lst}{\nwlbrace}{\it{}C}{\nwrbrace};
    {\nwrbrace} {\bf{}else} {\bf{}if} ({\it{}is\_a}\begin{math}<\end{math}{\bf{}cycle}\begin{math}>\end{math}({\it{}C})) {\nwlbrace}
        {\it{}cycles}={\bf{}lst}{\nwlbrace}{\bf{}cycle\_data}({\it{}ex\_to}\begin{math}<\end{math}{\bf{}cycle}\begin{math}>\end{math}({\it{}C}).{\it{}get\_k}(), {\it{}ex\_to}\begin{math}<\end{math}{\bf{}cycle}\begin{math}>\end{math}({\it{}C}).{\it{}get\_l}(),
                              {\it{}ex\_to}\begin{math}<\end{math}{\bf{}cycle}\begin{math}>\end{math}({\it{}C}).{\it{}get\_m}()){\nwrbrace};
    {\nwrbrace} {\bf{}else}
        {\bf{}throw}({\it{}std}::{\it{}invalid\_argument}({\tt{}"cycle\_node::cycle\_node(): "}
                                    {\tt{}"the first parameters must be either cycle, cycle\_data,"}
                                    {\tt{}" cycle\_node or list of cycle\_data"}));

\nwused{\\{NWppJ6t-26cGQh-1}\\{NWppJ6t-1HqLYY-3}\\{NWppJ6t-1HqLYY-4}\\{NWppJ6t-1HqLYY-5}\\{NWppJ6t-1HqLYY-B}}\nwidentuses{\\{{\nwixident{cycle{\_}data}}{cycle:undata}}\\{{\nwixident{cycle{\_}node}}{cycle:unnode}}}\nwindexuse{\nwixident{cycle{\_}data}}{cycle:undata}{NWppJ6t-1qiQ4a-1}\nwindexuse{\nwixident{cycle{\_}node}}{cycle:unnode}{NWppJ6t-1qiQ4a-1}\nwendcode{}\nwbegindocs{757}\nwdocspar
\nwenddocs{}\nwbegincode{758}\sublabel{NWppJ6t-2RJcfD-1}\nwmargintag{{\nwtagstyle{}\subpageref{NWppJ6t-2RJcfD-1}}}\moddef{check parents are valid~{\nwtagstyle{}\subpageref{NWppJ6t-2RJcfD-1}}}\endmoddef\Rm{}\nwstartdeflinemarkup\nwusesondefline{\\{NWppJ6t-1HqLYY-3}\\{NWppJ6t-1HqLYY-4}\\{NWppJ6t-1HqLYY-5}}\nwenddeflinemarkup
    {\it{}GINAC\_ASSERT}({\it{}is\_a}\begin{math}<\end{math}{\bf{}lst}\begin{math}>\end{math}({\it{}par}));
    {\it{}parents} = {\it{}ex\_to}\begin{math}<\end{math}{\bf{}lst}\begin{math}>\end{math}({\it{}par});

\nwused{\\{NWppJ6t-1HqLYY-3}\\{NWppJ6t-1HqLYY-4}\\{NWppJ6t-1HqLYY-5}}\nwendcode{}\nwbegindocs{759}\nwdocspar
\nwenddocs{}\nwbegincode{760}\sublabel{NWppJ6t-1HqLYY-4}\nwmargintag{{\nwtagstyle{}\subpageref{NWppJ6t-1HqLYY-4}}}\moddef{cycle node class~{\nwtagstyle{}\subpageref{NWppJ6t-1HqLYY-1}}}\plusendmoddef\Rm{}\nwstartdeflinemarkup\nwusesondefline{\\{NWppJ6t-32jW6L-1}}\nwprevnextdefs{NWppJ6t-1HqLYY-3}{NWppJ6t-1HqLYY-5}\nwenddeflinemarkup
{\bf{}cycle\_node}::{\bf{}cycle\_node}({\bf{}const} {\bf{}ex} & {\it{}C}, {\bf{}int} {\it{}g}, {\bf{}const} {\bf{}lst} & {\it{}par}, {\bf{}const} {\bf{}lst} & {\it{}ch})
{\nwlbrace}
    {\it{}generation}={\it{}g};
    {\it{}children}={\it{}ch};
    {\it{}custom\_asy}={\tt{}""};
    \LA{}check cycles are valid~{\nwtagstyle{}\subpageref{NWppJ6t-1qiQ4a-1}}\RA{}
    \LA{}check parents are valid~{\nwtagstyle{}\subpageref{NWppJ6t-2RJcfD-1}}\RA{}
{\nwrbrace}

\nwused{\\{NWppJ6t-32jW6L-1}}\nwidentuses{\\{{\nwixident{cycle{\_}node}}{cycle:unnode}}\\{{\nwixident{ex}}{ex}}}\nwindexuse{\nwixident{cycle{\_}node}}{cycle:unnode}{NWppJ6t-1HqLYY-4}\nwindexuse{\nwixident{ex}}{ex}{NWppJ6t-1HqLYY-4}\nwendcode{}\nwbegindocs{761}\nwdocspar
\nwenddocs{}\nwbegincode{762}\sublabel{NWppJ6t-1HqLYY-5}\nwmargintag{{\nwtagstyle{}\subpageref{NWppJ6t-1HqLYY-5}}}\moddef{cycle node class~{\nwtagstyle{}\subpageref{NWppJ6t-1HqLYY-1}}}\plusendmoddef\Rm{}\nwstartdeflinemarkup\nwusesondefline{\\{NWppJ6t-32jW6L-1}}\nwprevnextdefs{NWppJ6t-1HqLYY-4}{NWppJ6t-1HqLYY-6}\nwenddeflinemarkup
{\bf{}cycle\_node}::{\bf{}cycle\_node}({\bf{}const} {\bf{}ex} & {\it{}C}, {\bf{}int} {\it{}g}, {\bf{}const} {\bf{}lst} & {\it{}par}, {\bf{}const} {\bf{}lst} & {\it{}ch}, {\it{}string} {\it{}ca})
{\nwlbrace}
    {\it{}generation}={\it{}g};
    {\it{}children}={\it{}ch};
    {\it{}custom\_asy}={\it{}ca};
    \LA{}check cycles are valid~{\nwtagstyle{}\subpageref{NWppJ6t-1qiQ4a-1}}\RA{}
    \LA{}check parents are valid~{\nwtagstyle{}\subpageref{NWppJ6t-2RJcfD-1}}\RA{}
{\nwrbrace}

\nwused{\\{NWppJ6t-32jW6L-1}}\nwidentuses{\\{{\nwixident{cycle{\_}node}}{cycle:unnode}}\\{{\nwixident{ex}}{ex}}}\nwindexuse{\nwixident{cycle{\_}node}}{cycle:unnode}{NWppJ6t-1HqLYY-5}\nwindexuse{\nwixident{ex}}{ex}{NWppJ6t-1HqLYY-5}\nwendcode{}\nwbegindocs{763}\nwdocspar
\nwenddocs{}\nwbegincode{764}\sublabel{NWppJ6t-1HqLYY-6}\nwmargintag{{\nwtagstyle{}\subpageref{NWppJ6t-1HqLYY-6}}}\moddef{cycle node class~{\nwtagstyle{}\subpageref{NWppJ6t-1HqLYY-1}}}\plusendmoddef\Rm{}\nwstartdeflinemarkup\nwusesondefline{\\{NWppJ6t-32jW6L-1}}\nwprevnextdefs{NWppJ6t-1HqLYY-5}{NWppJ6t-1HqLYY-7}\nwenddeflinemarkup
{\it{}return\_type\_t} {\bf{}cycle\_node}::{\it{}return\_type\_tinfo}() {\bf{}const}
{\nwlbrace}
    {\bf{}return} {\it{}make\_return\_type\_t}\begin{math}<\end{math}{\bf{}cycle\_node}\begin{math}>\end{math}();
{\nwrbrace}

\nwused{\\{NWppJ6t-32jW6L-1}}\nwidentuses{\\{{\nwixident{cycle{\_}node}}{cycle:unnode}}}\nwindexuse{\nwixident{cycle{\_}node}}{cycle:unnode}{NWppJ6t-1HqLYY-6}\nwendcode{}\nwbegindocs{765}\nwdocspar
\nwenddocs{}\nwbegincode{766}\sublabel{NWppJ6t-1HqLYY-7}\nwmargintag{{\nwtagstyle{}\subpageref{NWppJ6t-1HqLYY-7}}}\moddef{cycle node class~{\nwtagstyle{}\subpageref{NWppJ6t-1HqLYY-1}}}\plusendmoddef\Rm{}\nwstartdeflinemarkup\nwusesondefline{\\{NWppJ6t-32jW6L-1}}\nwprevnextdefs{NWppJ6t-1HqLYY-6}{NWppJ6t-1HqLYY-8}\nwenddeflinemarkup
{\bf{}ex} {\bf{}cycle\_node}::{\it{}op}({\it{}size\_t} {\it{}i}) {\bf{}const}
{\nwlbrace}
    {\it{}GINAC\_ASSERT}({\it{}i}\begin{math}<\end{math}{\it{}nops}());
    {\it{}size\_t} {\it{}ncyc}={\it{}cycles}.{\it{}nops}(), {\it{}nchil}={\it{}children}.{\it{}nops}(), {\it{}npar}={\it{}parents}.{\it{}nops}();
    {\bf{}if} ( {\it{}i} \begin{math}<\end{math} {\it{}ncyc})
        {\bf{}return} {\it{}cycles}.{\it{}op}({\it{}i});
    {\bf{}else} {\bf{}if} ( {\it{}i} \begin{math}<\end{math} {\it{}ncyc} + {\it{}nchil})
        {\bf{}return} {\it{}children}.{\it{}op}({\it{}i}-{\it{}ncyc});
    {\bf{}else} {\bf{}if} ( {\it{}i} \begin{math}<\end{math} {\it{}ncyc} + {\it{}nchil} + {\it{}npar})
        {\bf{}return} {\it{}parents}.{\it{}op}({\it{}i}-{\it{}ncyc}-{\it{}nchil});
    {\bf{}else}
        {\bf{}throw}({\it{}std}::{\it{}invalid\_argument}({\tt{}"cycle\_node::op(): requested operand out of the range"}));
{\nwrbrace}

\nwused{\\{NWppJ6t-32jW6L-1}}\nwidentuses{\\{{\nwixident{cycle{\_}node}}{cycle:unnode}}\\{{\nwixident{ex}}{ex}}\\{{\nwixident{nops}}{nops}}\\{{\nwixident{op}}{op}}}\nwindexuse{\nwixident{cycle{\_}node}}{cycle:unnode}{NWppJ6t-1HqLYY-7}\nwindexuse{\nwixident{ex}}{ex}{NWppJ6t-1HqLYY-7}\nwindexuse{\nwixident{nops}}{nops}{NWppJ6t-1HqLYY-7}\nwindexuse{\nwixident{op}}{op}{NWppJ6t-1HqLYY-7}\nwendcode{}\nwbegindocs{767}\nwdocspar
\nwenddocs{}\nwbegincode{768}\sublabel{NWppJ6t-1HqLYY-8}\nwmargintag{{\nwtagstyle{}\subpageref{NWppJ6t-1HqLYY-8}}}\moddef{cycle node class~{\nwtagstyle{}\subpageref{NWppJ6t-1HqLYY-1}}}\plusendmoddef\Rm{}\nwstartdeflinemarkup\nwusesondefline{\\{NWppJ6t-32jW6L-1}}\nwprevnextdefs{NWppJ6t-1HqLYY-7}{NWppJ6t-1HqLYY-9}\nwenddeflinemarkup
{\bf{}ex} & {\bf{}cycle\_node}::{\it{}let\_op}({\it{}size\_t} {\it{}i})
{\nwlbrace}
    {\it{}ensure\_if\_modifiable}();
    {\it{}GINAC\_ASSERT}({\it{}i}\begin{math}<\end{math}{\it{}nops}());
    {\it{}size\_t} {\it{}ncyc}={\it{}cycles}.{\it{}nops}(), {\it{}nchil}={\it{}children}.{\it{}nops}(), {\it{}npar}={\it{}parents}.{\it{}nops}();
    {\bf{}if} ( {\it{}i} \begin{math}<\end{math} {\it{}ncyc})
        {\bf{}return} {\it{}cycles}.{\it{}let\_op}({\it{}i});
    {\bf{}else} {\bf{}if} ( {\it{}i} \begin{math}<\end{math} {\it{}ncyc} + {\it{}nchil})
        {\bf{}return} {\it{}children}.{\it{}let\_op}({\it{}i}-{\it{}ncyc});
    {\bf{}else} {\bf{}if} ( {\it{}i} \begin{math}<\end{math} {\it{}ncyc} + {\it{}nchil} + {\it{}npar})
        {\bf{}return} {\it{}parents}.{\it{}let\_op}({\it{}i}-{\it{}ncyc}-{\it{}nchil});
    {\bf{}else}
        {\bf{}throw}({\it{}std}::{\it{}invalid\_argument}({\tt{}"cycle\_node::let\_op(): requested operand out of the range"}));
{\nwrbrace}

\nwused{\\{NWppJ6t-32jW6L-1}}\nwidentuses{\\{{\nwixident{cycle{\_}node}}{cycle:unnode}}\\{{\nwixident{ex}}{ex}}\\{{\nwixident{nops}}{nops}}}\nwindexuse{\nwixident{cycle{\_}node}}{cycle:unnode}{NWppJ6t-1HqLYY-8}\nwindexuse{\nwixident{ex}}{ex}{NWppJ6t-1HqLYY-8}\nwindexuse{\nwixident{nops}}{nops}{NWppJ6t-1HqLYY-8}\nwendcode{}\nwbegindocs{769}\nwdocspar
\nwenddocs{}\nwbegincode{770}\sublabel{NWppJ6t-1HqLYY-9}\nwmargintag{{\nwtagstyle{}\subpageref{NWppJ6t-1HqLYY-9}}}\moddef{cycle node class~{\nwtagstyle{}\subpageref{NWppJ6t-1HqLYY-1}}}\plusendmoddef\Rm{}\nwstartdeflinemarkup\nwusesondefline{\\{NWppJ6t-32jW6L-1}}\nwprevnextdefs{NWppJ6t-1HqLYY-8}{NWppJ6t-1HqLYY-A}\nwenddeflinemarkup
{\bf{}int} {\bf{}cycle\_node}::{\it{}compare\_same\_type}({\bf{}const} {\bf{}basic} &{\it{}other}) {\bf{}const}\nwindexdefn{\nwixident{cycle{\_}node}}{cycle:unnode}{NWppJ6t-1HqLYY-9}
{\nwlbrace}
       {\it{}GINAC\_ASSERT}({\it{}is\_a}\begin{math}<\end{math}{\bf{}cycle\_node}\begin{math}>\end{math}({\it{}other}));
       {\bf{}return} {\it{}inherited}::{\it{}compare\_same\_type}({\it{}other});
{\nwrbrace}

\nwused{\\{NWppJ6t-32jW6L-1}}\nwidentdefs{\\{{\nwixident{cycle{\_}node}}{cycle:unnode}}}\nwendcode{}\nwbegindocs{771}If neither of parameters has multiply values we return a cycle.
\nwenddocs{}\nwbegincode{772}\sublabel{NWppJ6t-1HqLYY-A}\nwmargintag{{\nwtagstyle{}\subpageref{NWppJ6t-1HqLYY-A}}}\moddef{cycle node class~{\nwtagstyle{}\subpageref{NWppJ6t-1HqLYY-1}}}\plusendmoddef\Rm{}\nwstartdeflinemarkup\nwusesondefline{\\{NWppJ6t-32jW6L-1}}\nwprevnextdefs{NWppJ6t-1HqLYY-9}{NWppJ6t-1HqLYY-B}\nwenddeflinemarkup
{\bf{}ex} {\bf{}cycle\_node}::{\it{}get\_cycle}({\bf{}const} {\bf{}ex} & {\it{}metr}) {\bf{}const}
{\nwlbrace}
    {\bf{}lst} {\it{}res};
    {\bf{}for} ({\bf{}const} {\bf{}auto}&  {\it{}it} : {\it{}cycles})
        {\it{}res}.{\it{}append}({\it{}ex\_to}\begin{math}<\end{math}{\bf{}cycle\_data}\begin{math}>\end{math}({\it{}it}).{\it{}get\_cycle}({\it{}metr}));
    {\bf{}return} {\it{}res};
{\nwrbrace}

\nwused{\\{NWppJ6t-32jW6L-1}}\nwidentuses{\\{{\nwixident{cycle{\_}data}}{cycle:undata}}\\{{\nwixident{cycle{\_}node}}{cycle:unnode}}\\{{\nwixident{ex}}{ex}}\\{{\nwixident{get{\_}cycle}}{get:uncycle}}}\nwindexuse{\nwixident{cycle{\_}data}}{cycle:undata}{NWppJ6t-1HqLYY-A}\nwindexuse{\nwixident{cycle{\_}node}}{cycle:unnode}{NWppJ6t-1HqLYY-A}\nwindexuse{\nwixident{ex}}{ex}{NWppJ6t-1HqLYY-A}\nwindexuse{\nwixident{get{\_}cycle}}{get:uncycle}{NWppJ6t-1HqLYY-A}\nwendcode{}\nwbegindocs{773}\nwdocspar
\nwenddocs{}\nwbegincode{774}\sublabel{NWppJ6t-1HqLYY-B}\nwmargintag{{\nwtagstyle{}\subpageref{NWppJ6t-1HqLYY-B}}}\moddef{cycle node class~{\nwtagstyle{}\subpageref{NWppJ6t-1HqLYY-1}}}\plusendmoddef\Rm{}\nwstartdeflinemarkup\nwusesondefline{\\{NWppJ6t-32jW6L-1}}\nwprevnextdefs{NWppJ6t-1HqLYY-A}{NWppJ6t-1HqLYY-C}\nwenddeflinemarkup
{\bf{}void} {\bf{}cycle\_node}::{\it{}set\_cycles}({\bf{}const} {\bf{}ex} & {\it{}C})\nwindexdefn{\nwixident{cycle{\_}node}}{cycle:unnode}{NWppJ6t-1HqLYY-B}
{\nwlbrace}
    {\it{}cycles}.{\it{}remove\_all}();
    \LA{}check cycles are valid~{\nwtagstyle{}\subpageref{NWppJ6t-1qiQ4a-1}}\RA{}
{\nwrbrace}

\nwused{\\{NWppJ6t-32jW6L-1}}\nwidentdefs{\\{{\nwixident{cycle{\_}node}}{cycle:unnode}}}\nwidentuses{\\{{\nwixident{ex}}{ex}}}\nwindexuse{\nwixident{ex}}{ex}{NWppJ6t-1HqLYY-B}\nwendcode{}\nwbegindocs{775}\nwdocspar
\nwenddocs{}\nwbegincode{776}\sublabel{NWppJ6t-1HqLYY-C}\nwmargintag{{\nwtagstyle{}\subpageref{NWppJ6t-1HqLYY-C}}}\moddef{cycle node class~{\nwtagstyle{}\subpageref{NWppJ6t-1HqLYY-1}}}\plusendmoddef\Rm{}\nwstartdeflinemarkup\nwusesondefline{\\{NWppJ6t-32jW6L-1}}\nwprevnextdefs{NWppJ6t-1HqLYY-B}{NWppJ6t-1HqLYY-D}\nwenddeflinemarkup
{\bf{}void} {\bf{}cycle\_node}::{\it{}append\_cycle}({\bf{}const} {\bf{}ex} & {\it{}k}, {\bf{}const} {\bf{}ex} & {\it{}l}, {\bf{}const} {\bf{}ex} & {\it{}m})\nwindexdefn{\nwixident{cycle{\_}node}}{cycle:unnode}{NWppJ6t-1HqLYY-C}
{\nwlbrace}
    {\it{}cycles}.{\it{}append}({\bf{}cycle\_data}({\it{}k},{\it{}l},{\it{}m}));
{\nwrbrace}

\nwused{\\{NWppJ6t-32jW6L-1}}\nwidentdefs{\\{{\nwixident{cycle{\_}node}}{cycle:unnode}}}\nwidentuses{\\{{\nwixident{cycle{\_}data}}{cycle:undata}}\\{{\nwixident{ex}}{ex}}\\{{\nwixident{k}}{k}}\\{{\nwixident{l}}{l}}\\{{\nwixident{m}}{m}}}\nwindexuse{\nwixident{cycle{\_}data}}{cycle:undata}{NWppJ6t-1HqLYY-C}\nwindexuse{\nwixident{ex}}{ex}{NWppJ6t-1HqLYY-C}\nwindexuse{\nwixident{k}}{k}{NWppJ6t-1HqLYY-C}\nwindexuse{\nwixident{l}}{l}{NWppJ6t-1HqLYY-C}\nwindexuse{\nwixident{m}}{m}{NWppJ6t-1HqLYY-C}\nwendcode{}\nwbegindocs{777}\nwdocspar
\nwenddocs{}\nwbegincode{778}\sublabel{NWppJ6t-1HqLYY-D}\nwmargintag{{\nwtagstyle{}\subpageref{NWppJ6t-1HqLYY-D}}}\moddef{cycle node class~{\nwtagstyle{}\subpageref{NWppJ6t-1HqLYY-1}}}\plusendmoddef\Rm{}\nwstartdeflinemarkup\nwusesondefline{\\{NWppJ6t-32jW6L-1}}\nwprevnextdefs{NWppJ6t-1HqLYY-C}{NWppJ6t-1HqLYY-E}\nwenddeflinemarkup
{\bf{}void} {\bf{}cycle\_node}::{\it{}append\_cycle}({\bf{}const} {\bf{}ex} & {\it{}C})\nwindexdefn{\nwixident{cycle{\_}node}}{cycle:unnode}{NWppJ6t-1HqLYY-D}
{\nwlbrace}
    {\bf{}if} ({\it{}is\_a}\begin{math}<\end{math}{\bf{}cycle}\begin{math}>\end{math}({\it{}C}))
        {\it{}cycles}.{\it{}append}({\bf{}cycle\_data}({\it{}ex\_to}\begin{math}<\end{math}{\bf{}cycle}\begin{math}>\end{math}({\it{}C}).{\it{}get\_k}(), {\it{}ex\_to}\begin{math}<\end{math}{\bf{}cycle}\begin{math}>\end{math}({\it{}C}).{\it{}get\_l}(),
                                   {\it{}ex\_to}\begin{math}<\end{math}{\bf{}cycle}\begin{math}>\end{math}({\it{}C}).{\it{}get\_m}()));
    {\bf{}else}    {\bf{}if} ({\it{}is\_a}\begin{math}<\end{math}{\bf{}cycle\_data}\begin{math}>\end{math}({\it{}C}))
        {\it{}cycles}.{\it{}append}({\it{}ex\_to}\begin{math}<\end{math}{\bf{}cycle\_data}\begin{math}>\end{math}({\it{}C}));
    {\bf{}else}
        {\bf{}throw}({\it{}std}::{\it{}invalid\_argument}({\tt{}"cycle\_node::append\_cycle(const ex &): the parameter must be"}
                                    {\tt{}" either cycle or cycle\_data"}));
{\nwrbrace}

\nwused{\\{NWppJ6t-32jW6L-1}}\nwidentdefs{\\{{\nwixident{cycle{\_}node}}{cycle:unnode}}}\nwidentuses{\\{{\nwixident{cycle{\_}data}}{cycle:undata}}\\{{\nwixident{ex}}{ex}}}\nwindexuse{\nwixident{cycle{\_}data}}{cycle:undata}{NWppJ6t-1HqLYY-D}\nwindexuse{\nwixident{ex}}{ex}{NWppJ6t-1HqLYY-D}\nwendcode{}\nwbegindocs{779}Return the list of parents---either {\Tt{}\Rm{}{\it{}cycle\_relations}\nwendquote} or {\Tt{}\Rm{}{\bf{}subfigure}\nwendquote}
\nwenddocs{}\nwbegincode{780}\sublabel{NWppJ6t-1HqLYY-E}\nwmargintag{{\nwtagstyle{}\subpageref{NWppJ6t-1HqLYY-E}}}\moddef{cycle node class~{\nwtagstyle{}\subpageref{NWppJ6t-1HqLYY-1}}}\plusendmoddef\Rm{}\nwstartdeflinemarkup\nwusesondefline{\\{NWppJ6t-32jW6L-1}}\nwprevnextdefs{NWppJ6t-1HqLYY-D}{NWppJ6t-1HqLYY-F}\nwenddeflinemarkup
{\bf{}lst} {\bf{}cycle\_node}::{\it{}get\_parents}() {\bf{}const}
{\nwlbrace}
    {\bf{}return} {\it{}parents};
{\nwrbrace}

\nwused{\\{NWppJ6t-32jW6L-1}}\nwidentuses{\\{{\nwixident{cycle{\_}node}}{cycle:unnode}}}\nwindexuse{\nwixident{cycle{\_}node}}{cycle:unnode}{NWppJ6t-1HqLYY-E}\nwendcode{}\nwbegindocs{781}The method returns the list of all keys to parant cycles.
\nwenddocs{}\nwbegincode{782}\sublabel{NWppJ6t-1HqLYY-F}\nwmargintag{{\nwtagstyle{}\subpageref{NWppJ6t-1HqLYY-F}}}\moddef{cycle node class~{\nwtagstyle{}\subpageref{NWppJ6t-1HqLYY-1}}}\plusendmoddef\Rm{}\nwstartdeflinemarkup\nwusesondefline{\\{NWppJ6t-32jW6L-1}}\nwprevnextdefs{NWppJ6t-1HqLYY-E}{NWppJ6t-1HqLYY-G}\nwenddeflinemarkup
{\bf{}lst} {\bf{}cycle\_node}::{\it{}get\_parent\_keys}() {\bf{}const}
{\nwlbrace}
    {\bf{}lst} {\it{}pkeys};
    {\bf{}if} ( ({\it{}parents}.{\it{}nops}() \begin{math}\equiv\end{math} 1) \begin{math}\wedge\end{math} ({\it{}is\_a}\begin{math}<\end{math}{\bf{}subfigure}\begin{math}>\end{math}({\it{}parents}.{\it{}op}(0)))) {\nwlbrace}
        {\it{}pkeys}={\it{}ex\_to}\begin{math}<\end{math}{\bf{}lst}\begin{math}>\end{math}({\it{}ex\_to}\begin{math}<\end{math}{\bf{}subfigure}\begin{math}>\end{math}({\it{}parents}.{\it{}op}(0)).{\it{}get\_parlist}());
    {\nwrbrace} {\bf{}else} {\nwlbrace}
        {\bf{}for} ({\bf{}const} {\bf{}auto}& {\it{}it} : {\it{}parents})
            {\it{}pkeys}.{\it{}append}({\it{}ex\_to}\begin{math}<\end{math}{\bf{}cycle\_relation}\begin{math}>\end{math}({\it{}it}).{\it{}get\_parkey}());
    {\nwrbrace}
    {\bf{}return} {\it{}pkeys};
{\nwrbrace}

\nwused{\\{NWppJ6t-32jW6L-1}}\nwidentuses{\\{{\nwixident{cycle{\_}node}}{cycle:unnode}}\\{{\nwixident{cycle{\_}relation}}{cycle:unrelation}}\\{{\nwixident{nops}}{nops}}\\{{\nwixident{op}}{op}}\\{{\nwixident{subfigure}}{subfigure}}}\nwindexuse{\nwixident{cycle{\_}node}}{cycle:unnode}{NWppJ6t-1HqLYY-F}\nwindexuse{\nwixident{cycle{\_}relation}}{cycle:unrelation}{NWppJ6t-1HqLYY-F}\nwindexuse{\nwixident{nops}}{nops}{NWppJ6t-1HqLYY-F}\nwindexuse{\nwixident{op}}{op}{NWppJ6t-1HqLYY-F}\nwindexuse{\nwixident{subfigure}}{subfigure}{NWppJ6t-1HqLYY-F}\nwendcode{}\nwbegindocs{783}Printing of a {\Tt{}\Rm{}{\bf{}cycle\_node}\nwendquote} has two almost identical form: accurate
and float.
\nwenddocs{}\nwbegincode{784}\sublabel{NWppJ6t-1HqLYY-G}\nwmargintag{{\nwtagstyle{}\subpageref{NWppJ6t-1HqLYY-G}}}\moddef{cycle node class~{\nwtagstyle{}\subpageref{NWppJ6t-1HqLYY-1}}}\plusendmoddef\Rm{}\nwstartdeflinemarkup\nwusesondefline{\\{NWppJ6t-32jW6L-1}}\nwprevnextdefs{NWppJ6t-1HqLYY-F}{NWppJ6t-1HqLYY-H}\nwenddeflinemarkup
{\bf{}void} {\bf{}cycle\_node}::{\it{}do\_print}({\bf{}const} {\it{}print\_dflt} & {\it{}con}, {\bf{}unsigned} {\it{}level}) {\bf{}const}\nwindexdefn{\nwixident{cycle{\_}node}}{cycle:unnode}{NWppJ6t-1HqLYY-G}
{\nwlbrace}
    \LA{}start to print cycle node~{\nwtagstyle{}\subpageref{NWppJ6t-1o3xH5-1}}\RA{}
        {\it{}ex\_to}\begin{math}<\end{math}{\bf{}cycle\_data}\begin{math}>\end{math}({\it{}it}).{\it{}do\_print}({\it{}con}, {\it{}level});
        \LA{}end to print cycle node~{\nwtagstyle{}\subpageref{NWppJ6t-1tclWh-1}}\RA{}
{\nwrbrace}

\nwused{\\{NWppJ6t-32jW6L-1}}\nwidentdefs{\\{{\nwixident{cycle{\_}node}}{cycle:unnode}}}\nwidentuses{\\{{\nwixident{cycle{\_}data}}{cycle:undata}}}\nwindexuse{\nwixident{cycle{\_}data}}{cycle:undata}{NWppJ6t-1HqLYY-G}\nwendcode{}\nwbegindocs{785}And a similar one for the float printing
\nwenddocs{}\nwbegincode{786}\sublabel{NWppJ6t-1HqLYY-H}\nwmargintag{{\nwtagstyle{}\subpageref{NWppJ6t-1HqLYY-H}}}\moddef{cycle node class~{\nwtagstyle{}\subpageref{NWppJ6t-1HqLYY-1}}}\plusendmoddef\Rm{}\nwstartdeflinemarkup\nwusesondefline{\\{NWppJ6t-32jW6L-1}}\nwprevnextdefs{NWppJ6t-1HqLYY-G}{NWppJ6t-1HqLYY-I}\nwenddeflinemarkup
{\bf{}void} {\bf{}cycle\_node}::{\it{}do\_print\_double}({\bf{}const} {\it{}print\_dflt} & {\it{}con}, {\bf{}unsigned} {\it{}level}) {\bf{}const}\nwindexdefn{\nwixident{cycle{\_}node}}{cycle:unnode}{NWppJ6t-1HqLYY-H}
{\nwlbrace}
    \LA{}start to print cycle node~{\nwtagstyle{}\subpageref{NWppJ6t-1o3xH5-1}}\RA{}
        {\it{}ex\_to}\begin{math}<\end{math}{\bf{}cycle\_data}\begin{math}>\end{math}({\it{}it}).{\it{}do\_print\_double}({\it{}con}, {\it{}level});
        \LA{}end to print cycle node~{\nwtagstyle{}\subpageref{NWppJ6t-1tclWh-1}}\RA{}
{\nwrbrace}

\nwused{\\{NWppJ6t-32jW6L-1}}\nwidentdefs{\\{{\nwixident{cycle{\_}node}}{cycle:unnode}}}\nwidentuses{\\{{\nwixident{cycle{\_}data}}{cycle:undata}}\\{{\nwixident{do{\_}print{\_}double}}{do:unprint:undouble}}}\nwindexuse{\nwixident{cycle{\_}data}}{cycle:undata}{NWppJ6t-1HqLYY-H}\nwindexuse{\nwixident{do{\_}print{\_}double}}{do:unprint:undouble}{NWppJ6t-1HqLYY-H}\nwendcode{}\nwbegindocs{787}We output generation and all children, \ldots
\nwenddocs{}\nwbegincode{788}\sublabel{NWppJ6t-1o3xH5-1}\nwmargintag{{\nwtagstyle{}\subpageref{NWppJ6t-1o3xH5-1}}}\moddef{start to print cycle node~{\nwtagstyle{}\subpageref{NWppJ6t-1o3xH5-1}}}\endmoddef\Rm{}\nwstartdeflinemarkup\nwusesondefline{\\{NWppJ6t-1HqLYY-G}\\{NWppJ6t-1HqLYY-H}}\nwenddeflinemarkup
    {\it{}con}.{\it{}s} \begin{math}\ll\end{math} {\tt{}'{\char123}'};
    {\bf{}for} ({\bf{}const} {\bf{}auto}& {\it{}it} : {\it{}cycles}) {\nwlbrace}

\nwused{\\{NWppJ6t-1HqLYY-G}\\{NWppJ6t-1HqLYY-H}}\nwendcode{}\nwbegindocs{789}\nwdocspar
\nwenddocs{}\nwbegincode{790}\sublabel{NWppJ6t-1tclWh-1}\nwmargintag{{\nwtagstyle{}\subpageref{NWppJ6t-1tclWh-1}}}\moddef{end to print cycle node~{\nwtagstyle{}\subpageref{NWppJ6t-1tclWh-1}}}\endmoddef\Rm{}\nwstartdeflinemarkup\nwusesondefline{\\{NWppJ6t-1HqLYY-G}\\{NWppJ6t-1HqLYY-H}}\nwprevnextdefs{\relax}{NWppJ6t-1tclWh-2}\nwenddeflinemarkup
        {\it{}con}.{\it{}s} \begin{math}\ll\end{math} {\tt{}", "};
    {\nwrbrace}
    {\it{}con}.{\it{}s} \begin{math}\ll\end{math} {\it{}generation} \begin{math}\ll\end{math} {\tt{}'{\char125}'} \begin{math}\ll\end{math} {\tt{}" --> ("};
    // list all children
    {\bf{}for} ({\bf{}lst}::{\it{}const\_iterator} {\it{}it} = {\it{}children}.{\it{}begin}(); {\it{}it} \begin{math}\neq\end{math} {\it{}children}.{\it{}end}();) {\nwlbrace}
        {\it{}con}.{\it{}s} \begin{math}\ll\end{math} (\begin{math}\ast\end{math}{\it{}it});
        \protect\PP{\it{}it};
        {\bf{}if} ({\it{}it} \begin{math}\neq\end{math} {\it{}children}.{\it{}end}())
            {\it{}con}.{\it{}s} \begin{math}\ll\end{math}{\tt{}","};
    {\nwrbrace}

\nwalsodefined{\\{NWppJ6t-1tclWh-2}\\{NWppJ6t-1tclWh-3}}\nwused{\\{NWppJ6t-1HqLYY-G}\\{NWppJ6t-1HqLYY-H}}\nwendcode{}\nwbegindocs{791}\ldots then all parents.
\nwenddocs{}\nwbegincode{792}\sublabel{NWppJ6t-1tclWh-2}\nwmargintag{{\nwtagstyle{}\subpageref{NWppJ6t-1tclWh-2}}}\moddef{end to print cycle node~{\nwtagstyle{}\subpageref{NWppJ6t-1tclWh-1}}}\plusendmoddef\Rm{}\nwstartdeflinemarkup\nwusesondefline{\\{NWppJ6t-1HqLYY-G}\\{NWppJ6t-1HqLYY-H}}\nwprevnextdefs{NWppJ6t-1tclWh-1}{NWppJ6t-1tclWh-3}\nwenddeflinemarkup
    {\it{}con}.{\it{}s} \begin{math}\ll\end{math} {\tt{}");  <-- ("};
    {\bf{}if} ({\it{}generation} \begin{math}>\end{math} 0 \begin{math}\vee\end{math} {\it{}FIGURE\_DEBUG})
        {\bf{}for} ({\bf{}lst}::{\it{}const\_iterator} {\it{}it} = {\it{}parents}.{\it{}begin}(); {\it{}it} \begin{math}\neq\end{math} {\it{}parents}.{\it{}end}();) {\nwlbrace}
            {\bf{}if} ({\it{}is\_a}\begin{math}<\end{math}{\bf{}cycle\_relation}\begin{math}>\end{math}(\begin{math}\ast\end{math}{\it{}it}))
                {\it{}ex\_to}\begin{math}<\end{math}{\bf{}cycle\_relation}\begin{math}>\end{math}(\begin{math}\ast\end{math}{\it{}it}).{\it{}do\_print}({\it{}con},{\it{}level});
            {\bf{}else} {\bf{}if} ({\it{}is\_a}\begin{math}<\end{math}{\bf{}subfigure}\begin{math}>\end{math}(\begin{math}\ast\end{math}{\it{}it}))
                {\it{}ex\_to}\begin{math}<\end{math}{\bf{}subfigure}\begin{math}>\end{math}(\begin{math}\ast\end{math}{\it{}it}).{\it{}do\_print}({\it{}con},{\it{}level});
            \protect\PP{\it{}it};
            {\bf{}if} ({\it{}it} \begin{math}\neq\end{math} {\it{}parents}.{\it{}end}())
                {\it{}con}.{\it{}s} \begin{math}\ll\end{math}{\tt{}","};
        {\nwrbrace}
    {\it{}con}.{\it{}s} \begin{math}\ll\end{math} {\tt{}")"};

\nwused{\\{NWppJ6t-1HqLYY-G}\\{NWppJ6t-1HqLYY-H}}\nwidentuses{\\{{\nwixident{cycle{\_}relation}}{cycle:unrelation}}\\{{\nwixident{FIGURE{\_}DEBUG}}{FIGURE:unDEBUG}}\\{{\nwixident{subfigure}}{subfigure}}}\nwindexuse{\nwixident{cycle{\_}relation}}{cycle:unrelation}{NWppJ6t-1tclWh-2}\nwindexuse{\nwixident{FIGURE{\_}DEBUG}}{FIGURE:unDEBUG}{NWppJ6t-1tclWh-2}\nwindexuse{\nwixident{subfigure}}{subfigure}{NWppJ6t-1tclWh-2}\nwendcode{}\nwbegindocs{793}Finally if the custom \Asymptote\ style is not empty we print it as well.
\nwenddocs{}\nwbegincode{794}\sublabel{NWppJ6t-1tclWh-3}\nwmargintag{{\nwtagstyle{}\subpageref{NWppJ6t-1tclWh-3}}}\moddef{end to print cycle node~{\nwtagstyle{}\subpageref{NWppJ6t-1tclWh-1}}}\plusendmoddef\Rm{}\nwstartdeflinemarkup\nwusesondefline{\\{NWppJ6t-1HqLYY-G}\\{NWppJ6t-1HqLYY-H}}\nwprevnextdefs{NWppJ6t-1tclWh-2}{\relax}\nwenddeflinemarkup
    {\bf{}if} ({\it{}custom\_asy} \begin{math}\neq\end{math} {\tt{}""})
        {\it{}con}.{\it{}s}\begin{math}\ll\end{math} {\tt{}" /"} \begin{math}\ll\end{math} {\it{}custom\_asy} \begin{math}\ll\end{math} {\tt{}"/"};
    {\it{}con}.{\it{}s} \begin{math}\ll\end{math} {\it{}endl};

\nwused{\\{NWppJ6t-1HqLYY-G}\\{NWppJ6t-1HqLYY-H}}\nwendcode{}\nwbegindocs{795}\nwdocspar
\nwenddocs{}\nwbegincode{796}\sublabel{NWppJ6t-1HqLYY-I}\nwmargintag{{\nwtagstyle{}\subpageref{NWppJ6t-1HqLYY-I}}}\moddef{cycle node class~{\nwtagstyle{}\subpageref{NWppJ6t-1HqLYY-1}}}\plusendmoddef\Rm{}\nwstartdeflinemarkup\nwusesondefline{\\{NWppJ6t-32jW6L-1}}\nwprevnextdefs{NWppJ6t-1HqLYY-H}{NWppJ6t-1HqLYY-J}\nwenddeflinemarkup
{\bf{}void} {\bf{}cycle\_node}::{\it{}do\_print\_tree}({\bf{}const} {\it{}print\_tree} & {\it{}con}, {\bf{}unsigned} {\it{}level}) {\bf{}const}\nwindexdefn{\nwixident{cycle{\_}node}}{cycle:unnode}{NWppJ6t-1HqLYY-I}
{\nwlbrace}
    {\bf{}for} ({\bf{}const} {\bf{}auto}& {\it{}it} : {\it{}cycles})
        {\it{}it}.{\it{}print}({\it{}con}, {\it{}level});
    {\it{}con}.{\it{}s} \begin{math}\ll\end{math} {\it{}std}::{\it{}string}({\it{}level}+{\it{}con}.{\it{}delta\_indent}, {\tt{}' '}) \begin{math}\ll\end{math} {\tt{}"generation: "}\begin{math}\ll\end{math} {\it{}generation} \begin{math}\ll\end{math} {\it{}endl};
    {\it{}con}.{\it{}s} \begin{math}\ll\end{math} {\it{}std}::{\it{}string}({\it{}level}+{\it{}con}.{\it{}delta\_indent}, {\tt{}' '}) \begin{math}\ll\end{math} {\tt{}"children"} \begin{math}\ll\end{math}{\it{}endl};
    {\it{}children}.{\it{}print}({\it{}con},{\it{}level}+2\begin{math}\ast\end{math}{\it{}con}.{\it{}delta\_indent});
    {\it{}con}.{\it{}s} \begin{math}\ll\end{math} {\it{}std}::{\it{}string}({\it{}level}+{\it{}con}.{\it{}delta\_indent}, {\tt{}' '}) \begin{math}\ll\end{math} {\tt{}"parents"} \begin{math}\ll\end{math}{\it{}endl};
    {\it{}parents}.{\it{}print}({\it{}con},{\it{}level}+2\begin{math}\ast\end{math}{\it{}con}.{\it{}delta\_indent});
    {\it{}con}.{\it{}s} \begin{math}\ll\end{math} {\it{}std}::{\it{}string}({\it{}level}+{\it{}con}.{\it{}delta\_indent}, {\tt{}' '}) \begin{math}\ll\end{math} {\tt{}"custom\_asy: "} \begin{math}\ll\end{math} {\it{}custom\_asy} \begin{math}\ll\end{math}{\it{}endl};
{\nwrbrace}

\nwused{\\{NWppJ6t-32jW6L-1}}\nwidentdefs{\\{{\nwixident{cycle{\_}node}}{cycle:unnode}}}\nwendcode{}\nwbegindocs{797}\nwdocspar
\nwenddocs{}\nwbegincode{798}\sublabel{NWppJ6t-1HqLYY-J}\nwmargintag{{\nwtagstyle{}\subpageref{NWppJ6t-1HqLYY-J}}}\moddef{cycle node class~{\nwtagstyle{}\subpageref{NWppJ6t-1HqLYY-1}}}\plusendmoddef\Rm{}\nwstartdeflinemarkup\nwusesondefline{\\{NWppJ6t-32jW6L-1}}\nwprevnextdefs{NWppJ6t-1HqLYY-I}{NWppJ6t-1HqLYY-K}\nwenddeflinemarkup
{\bf{}void} {\bf{}cycle\_node}::{\it{}remove\_child}({\bf{}const} {\bf{}ex} & {\it{}other})\nwindexdefn{\nwixident{cycle{\_}node}}{cycle:unnode}{NWppJ6t-1HqLYY-J}
{\nwlbrace}
    {\bf{}lst} {\it{}nchildren};
    {\bf{}for} ({\bf{}const} {\bf{}auto}& {\it{}it} : {\it{}children})
        {\bf{}if} ({\it{}it} \begin{math}\neq\end{math} {\it{}other})
            {\it{}nchildren}.{\it{}append}({\it{}it});
    {\it{}children}={\it{}nchildren};
{\nwrbrace}

\nwused{\\{NWppJ6t-32jW6L-1}}\nwidentdefs{\\{{\nwixident{cycle{\_}node}}{cycle:unnode}}}\nwidentuses{\\{{\nwixident{ex}}{ex}}}\nwindexuse{\nwixident{ex}}{ex}{NWppJ6t-1HqLYY-J}\nwendcode{}\nwbegindocs{799}\nwdocspar
\nwenddocs{}\nwbegincode{800}\sublabel{NWppJ6t-1HqLYY-K}\nwmargintag{{\nwtagstyle{}\subpageref{NWppJ6t-1HqLYY-K}}}\moddef{cycle node class~{\nwtagstyle{}\subpageref{NWppJ6t-1HqLYY-1}}}\plusendmoddef\Rm{}\nwstartdeflinemarkup\nwusesondefline{\\{NWppJ6t-32jW6L-1}}\nwprevnextdefs{NWppJ6t-1HqLYY-J}{NWppJ6t-1HqLYY-L}\nwenddeflinemarkup
{\bf{}cycle\_node} {\bf{}cycle\_node}::{\it{}subs}({\bf{}const} {\bf{}ex} & {\it{}e}, {\bf{}unsigned} {\it{}options}) {\bf{}const}
{\nwlbrace}
    {\it{}exmap} {\it{}em};
    {\bf{}if} ({\it{}e}.{\it{}info}({\it{}info\_flags}::{\it{}list})) {\nwlbrace}
        {\bf{}lst} {\it{}l} = {\it{}ex\_to}\begin{math}<\end{math}{\bf{}lst}\begin{math}>\end{math}({\it{}e});
        {\bf{}for} ({\bf{}const} {\bf{}auto}& {\it{}i} : {\it{}l})
            {\it{}em}.{\it{}insert}({\it{}std}::{\it{}make\_pair}({\it{}i}.{\it{}op}(0), {\it{}i}.{\it{}op}(1)));
    {\nwrbrace} {\bf{}else} {\bf{}if} ({\it{}is\_a}\begin{math}<\end{math}{\bf{}relational}\begin{math}>\end{math}({\it{}e})) {\nwlbrace}
        {\it{}em}.{\it{}insert}({\it{}std}::{\it{}make\_pair}({\it{}e}.{\it{}op}(0), {\it{}e}.{\it{}op}(1)));
    {\nwrbrace} {\bf{}else}
        {\bf{}throw}({\it{}std}::{\it{}invalid\_argument}({\tt{}"cycle::subs(): the parameter should be a relational or a lst"}));

    {\bf{}return} {\it{}ex\_to}\begin{math}<\end{math}{\bf{}cycle\_node}\begin{math}>\end{math}({\it{}subs}({\it{}em}, {\it{}options}));
{\nwrbrace}

\nwused{\\{NWppJ6t-32jW6L-1}}\nwidentuses{\\{{\nwixident{cycle{\_}node}}{cycle:unnode}}\\{{\nwixident{ex}}{ex}}\\{{\nwixident{info}}{info}}\\{{\nwixident{l}}{l}}\\{{\nwixident{op}}{op}}\\{{\nwixident{subs}}{subs}}}\nwindexuse{\nwixident{cycle{\_}node}}{cycle:unnode}{NWppJ6t-1HqLYY-K}\nwindexuse{\nwixident{ex}}{ex}{NWppJ6t-1HqLYY-K}\nwindexuse{\nwixident{info}}{info}{NWppJ6t-1HqLYY-K}\nwindexuse{\nwixident{l}}{l}{NWppJ6t-1HqLYY-K}\nwindexuse{\nwixident{op}}{op}{NWppJ6t-1HqLYY-K}\nwindexuse{\nwixident{subs}}{subs}{NWppJ6t-1HqLYY-K}\nwendcode{}\nwbegindocs{801}\nwdocspar
\nwenddocs{}\nwbegincode{802}\sublabel{NWppJ6t-1HqLYY-L}\nwmargintag{{\nwtagstyle{}\subpageref{NWppJ6t-1HqLYY-L}}}\moddef{cycle node class~{\nwtagstyle{}\subpageref{NWppJ6t-1HqLYY-1}}}\plusendmoddef\Rm{}\nwstartdeflinemarkup\nwusesondefline{\\{NWppJ6t-32jW6L-1}}\nwprevnextdefs{NWppJ6t-1HqLYY-K}{NWppJ6t-1HqLYY-M}\nwenddeflinemarkup
{\bf{}ex} {\bf{}cycle\_node}::{\it{}subs}({\bf{}const} {\it{}exmap} & {\it{}em}, {\bf{}unsigned} {\it{}options}) {\bf{}const}
{\nwlbrace}
    {\bf{}return} {\bf{}cycle\_node}({\it{}cycles}.{\it{}subs}({\it{}em}, {\it{}options}), {\it{}generation}, {\it{}ex\_to}\begin{math}<\end{math}{\bf{}lst}\begin{math}>\end{math}({\it{}parents}.{\it{}subs}({\it{}em}, {\it{}options})), {\it{}children}, {\it{}custom\_asy});
{\nwrbrace}

\nwused{\\{NWppJ6t-32jW6L-1}}\nwidentuses{\\{{\nwixident{cycle{\_}node}}{cycle:unnode}}\\{{\nwixident{ex}}{ex}}\\{{\nwixident{subs}}{subs}}}\nwindexuse{\nwixident{cycle{\_}node}}{cycle:unnode}{NWppJ6t-1HqLYY-L}\nwindexuse{\nwixident{ex}}{ex}{NWppJ6t-1HqLYY-L}\nwindexuse{\nwixident{subs}}{subs}{NWppJ6t-1HqLYY-L}\nwendcode{}\nwbegindocs{803}\nwdocspar
\nwenddocs{}\nwbegincode{804}\sublabel{NWppJ6t-1HqLYY-M}\nwmargintag{{\nwtagstyle{}\subpageref{NWppJ6t-1HqLYY-M}}}\moddef{cycle node class~{\nwtagstyle{}\subpageref{NWppJ6t-1HqLYY-1}}}\plusendmoddef\Rm{}\nwstartdeflinemarkup\nwusesondefline{\\{NWppJ6t-32jW6L-1}}\nwprevnextdefs{NWppJ6t-1HqLYY-L}{NWppJ6t-1HqLYY-N}\nwenddeflinemarkup
{\bf{}void} {\bf{}cycle\_node}::{\it{}archive}({\it{}archive\_node} &{\it{}n}) {\bf{}const}\nwindexdefn{\nwixident{cycle{\_}node}}{cycle:unnode}{NWppJ6t-1HqLYY-M}
{\nwlbrace}
    {\it{}inherited}::{\it{}archive}({\it{}n});
    {\it{}n}.{\it{}add\_ex}({\tt{}"cycles"}, {\it{}cycles});
    {\it{}n}.{\it{}add\_unsigned}({\tt{}"children\_size"}, {\it{}children}.{\it{}nops}());
    {\bf{}if} ({\it{}children}.{\it{}nops}()\begin{math}>\end{math}0)
        {\bf{}for} ({\bf{}const} {\bf{}auto}& {\it{}it} : {\it{}children})
            {\it{}n}.{\it{}add\_ex}({\tt{}"chil"}, {\it{}it});

    {\it{}n}.{\it{}add\_unsigned}({\tt{}"parent\_size"}, {\it{}parents}.{\it{}nops}());
    {\bf{}if} ({\it{}parents}.{\it{}nops}()\begin{math}>\end{math}0) {\nwlbrace}
        {\it{}n}.{\it{}add\_bool}({\tt{}"has\_subfigure"}, {\bf{}false});
        {\bf{}for} ({\bf{}const} {\bf{}auto}& {\it{}it} : {\it{}parents})
            {\it{}n}.{\it{}add\_ex}({\tt{}"par"}, {\it{}ex\_to}\begin{math}<\end{math}{\bf{}cycle\_relation}\begin{math}>\end{math}({\it{}it}));
    {\nwrbrace}

\nwused{\\{NWppJ6t-32jW6L-1}}\nwidentdefs{\\{{\nwixident{cycle{\_}node}}{cycle:unnode}}}\nwidentuses{\\{{\nwixident{archive}}{archive}}\\{{\nwixident{cycle{\_}relation}}{cycle:unrelation}}\\{{\nwixident{nops}}{nops}}}\nwindexuse{\nwixident{archive}}{archive}{NWppJ6t-1HqLYY-M}\nwindexuse{\nwixident{cycle{\_}relation}}{cycle:unrelation}{NWppJ6t-1HqLYY-M}\nwindexuse{\nwixident{nops}}{nops}{NWppJ6t-1HqLYY-M}\nwendcode{}\nwbegindocs{805}storing the generation with its sign.
\nwenddocs{}\nwbegincode{806}\sublabel{NWppJ6t-1HqLYY-N}\nwmargintag{{\nwtagstyle{}\subpageref{NWppJ6t-1HqLYY-N}}}\moddef{cycle node class~{\nwtagstyle{}\subpageref{NWppJ6t-1HqLYY-1}}}\plusendmoddef\Rm{}\nwstartdeflinemarkup\nwusesondefline{\\{NWppJ6t-32jW6L-1}}\nwprevnextdefs{NWppJ6t-1HqLYY-M}{NWppJ6t-1HqLYY-O}\nwenddeflinemarkup
    {\bf{}bool} {\it{}neg\_generation}=({\it{}generation}\begin{math}<\end{math}0);
    {\it{}n}.{\it{}add\_bool}({\tt{}"neg\_generation"}, {\it{}neg\_generation});
    {\bf{}if} ({\it{}neg\_generation})
        {\it{}n}.{\it{}add\_unsigned}({\tt{}"abs\_generation"}, -{\it{}generation});
    {\bf{}else}
        {\it{}n}.{\it{}add\_unsigned}({\tt{}"abs\_generation"}, {\it{}generation});

\nwused{\\{NWppJ6t-32jW6L-1}}\nwendcode{}\nwbegindocs{807}saving the asymptote options
\nwenddocs{}\nwbegincode{808}\sublabel{NWppJ6t-1HqLYY-O}\nwmargintag{{\nwtagstyle{}\subpageref{NWppJ6t-1HqLYY-O}}}\moddef{cycle node class~{\nwtagstyle{}\subpageref{NWppJ6t-1HqLYY-1}}}\plusendmoddef\Rm{}\nwstartdeflinemarkup\nwusesondefline{\\{NWppJ6t-32jW6L-1}}\nwprevnextdefs{NWppJ6t-1HqLYY-N}{NWppJ6t-1HqLYY-P}\nwenddeflinemarkup
    {\it{}n}.{\it{}add\_string}({\tt{}"custom\_asy"}, {\it{}custom\_asy});
{\nwrbrace}

\nwused{\\{NWppJ6t-32jW6L-1}}\nwendcode{}\nwbegindocs{809}\nwdocspar
\nwenddocs{}\nwbegincode{810}\sublabel{NWppJ6t-1HqLYY-P}\nwmargintag{{\nwtagstyle{}\subpageref{NWppJ6t-1HqLYY-P}}}\moddef{cycle node class~{\nwtagstyle{}\subpageref{NWppJ6t-1HqLYY-1}}}\plusendmoddef\Rm{}\nwstartdeflinemarkup\nwusesondefline{\\{NWppJ6t-32jW6L-1}}\nwprevnextdefs{NWppJ6t-1HqLYY-O}{NWppJ6t-1HqLYY-Q}\nwenddeflinemarkup
{\bf{}void} {\bf{}cycle\_node}::{\it{}read\_archive}({\bf{}const} {\it{}archive\_node} &{\it{}n}, {\bf{}lst} &{\it{}sym\_lst})\nwindexdefn{\nwixident{cycle{\_}node}}{cycle:unnode}{NWppJ6t-1HqLYY-P}
{\nwlbrace}
    {\it{}inherited}::{\it{}read\_archive}({\it{}n}, {\it{}sym\_lst});
    {\bf{}ex} {\it{}e};
    {\it{}n}.{\it{}find\_ex}({\tt{}"cycles"}, {\it{}e}, {\it{}sym\_lst});
    {\it{}cycles}={\it{}ex\_to}\begin{math}<\end{math}{\bf{}lst}\begin{math}>\end{math}({\it{}e});
    {\bf{}ex} {\it{}ch}, {\it{}par};
    {\bf{}unsigned} {\bf{}int} {\it{}c\_size};
    {\it{}n}.{\it{}find\_unsigned}({\tt{}"children\_size"}, {\it{}c\_size});

    {\bf{}if} ({\it{}c\_size}\begin{math}>\end{math}0) {\nwlbrace}
        {\it{}archive\_node}::{\it{}archive\_node\_cit} {\it{}first} = {\it{}n}.{\it{}find\_first}({\tt{}"chil"});
        {\it{}archive\_node}::{\it{}archive\_node\_cit} {\it{}last} = {\it{}n}.{\it{}find\_last}({\tt{}"chil"});
        \protect\PP{\it{}last};
        {\bf{}for} ({\it{}archive\_node}::{\it{}archive\_node\_cit} {\it{}i}={\it{}first}; {\it{}i} \begin{math}\neq\end{math} {\it{}last}; \protect\PP{\it{}i}) {\nwlbrace}
            {\bf{}ex} {\it{}e};
            {\it{}n}.{\it{}find\_ex\_by\_loc}({\it{}i}, {\it{}e}, {\it{}sym\_lst});
            {\it{}children}.{\it{}append}({\it{}e});
        {\nwrbrace}
    {\nwrbrace}

    {\bf{}unsigned} {\bf{}int} {\it{}p\_size};
    {\it{}n}.{\it{}find\_unsigned}({\tt{}"parent\_size"}, {\it{}p\_size});

    {\bf{}if} ({\it{}p\_size}\begin{math}>\end{math}0) {\nwlbrace}
        {\it{}archive\_node}::{\it{}archive\_node\_cit} {\it{}first} = {\it{}n}.{\it{}find\_first}({\tt{}"par"});
        {\it{}archive\_node}::{\it{}archive\_node\_cit} {\it{}last} = {\it{}n}.{\it{}find\_last}({\tt{}"par"});
        \protect\PP{\it{}last};
        {\bf{}for} ({\it{}archive\_node}::{\it{}archive\_node\_cit} {\it{}i}={\it{}first}; {\it{}i} \begin{math}\neq\end{math} {\it{}last}; \protect\PP{\it{}i}) {\nwlbrace}
            {\bf{}ex} {\it{}e};
            {\it{}n}.{\it{}find\_ex\_by\_loc}({\it{}i}, {\it{}e}, {\it{}sym\_lst});
            {\it{}parents}.{\it{}append}({\it{}e});
        {\nwrbrace}
    {\nwrbrace}

\nwused{\\{NWppJ6t-32jW6L-1}}\nwidentdefs{\\{{\nwixident{cycle{\_}node}}{cycle:unnode}}}\nwidentuses{\\{{\nwixident{ex}}{ex}}\\{{\nwixident{read{\_}archive}}{read:unarchive}}}\nwindexuse{\nwixident{ex}}{ex}{NWppJ6t-1HqLYY-P}\nwindexuse{\nwixident{read{\_}archive}}{read:unarchive}{NWppJ6t-1HqLYY-P}\nwendcode{}\nwbegindocs{811}restoring the generation with its sign
\nwenddocs{}\nwbegincode{812}\sublabel{NWppJ6t-1HqLYY-Q}\nwmargintag{{\nwtagstyle{}\subpageref{NWppJ6t-1HqLYY-Q}}}\moddef{cycle node class~{\nwtagstyle{}\subpageref{NWppJ6t-1HqLYY-1}}}\plusendmoddef\Rm{}\nwstartdeflinemarkup\nwusesondefline{\\{NWppJ6t-32jW6L-1}}\nwprevnextdefs{NWppJ6t-1HqLYY-P}{NWppJ6t-1HqLYY-R}\nwenddeflinemarkup
    {\bf{}bool} {\it{}neg\_generation};
    {\it{}n}.{\it{}find\_bool}({\tt{}"neg\_generation"}, {\it{}neg\_generation});
    {\bf{}unsigned} {\bf{}int} {\it{}abs\_generation};
    {\it{}n}.{\it{}find\_unsigned}({\tt{}"abs\_generation"}, {\it{}abs\_generation});
    {\bf{}if} ({\it{}neg\_generation})
        {\it{}generation} = -{\it{}abs\_generation};
    {\bf{}else}
        {\it{}generation} = {\it{}abs\_generation};

\nwused{\\{NWppJ6t-32jW6L-1}}\nwendcode{}\nwbegindocs{813}restoring the asymptote options
\nwenddocs{}\nwbegincode{814}\sublabel{NWppJ6t-1HqLYY-R}\nwmargintag{{\nwtagstyle{}\subpageref{NWppJ6t-1HqLYY-R}}}\moddef{cycle node class~{\nwtagstyle{}\subpageref{NWppJ6t-1HqLYY-1}}}\plusendmoddef\Rm{}\nwstartdeflinemarkup\nwusesondefline{\\{NWppJ6t-32jW6L-1}}\nwprevnextdefs{NWppJ6t-1HqLYY-Q}{NWppJ6t-1HqLYY-S}\nwenddeflinemarkup
    {\it{}n}.{\it{}find\_string}({\tt{}"custom\_asy"}, {\it{}custom\_asy});
{\nwrbrace}

\nwused{\\{NWppJ6t-32jW6L-1}}\nwendcode{}\nwbegindocs{815}\nwdocspar
\nwenddocs{}\nwbegincode{816}\sublabel{NWppJ6t-1HqLYY-S}\nwmargintag{{\nwtagstyle{}\subpageref{NWppJ6t-1HqLYY-S}}}\moddef{cycle node class~{\nwtagstyle{}\subpageref{NWppJ6t-1HqLYY-1}}}\plusendmoddef\Rm{}\nwstartdeflinemarkup\nwusesondefline{\\{NWppJ6t-32jW6L-1}}\nwprevnextdefs{NWppJ6t-1HqLYY-R}{\relax}\nwenddeflinemarkup
{\it{}GINAC\_BIND\_UNARCHIVER}({\bf{}cycle\_node});\nwindexdefn{\nwixident{cycle{\_}node}}{cycle:unnode}{NWppJ6t-1HqLYY-S}

\nwused{\\{NWppJ6t-32jW6L-1}}\nwidentdefs{\\{{\nwixident{cycle{\_}node}}{cycle:unnode}}}\nwendcode{}\nwbegindocs{817}\nwdocspar
\subsection{Implementation of {\Tt{}\Rm{}{\bf{}figure}\nwendquote} class}
\label{sec:implem-figure-class}

Since this is the main class of the library, its implementation is
most evolved.

\nwenddocs{}\nwbegindocs{818}\nwdocspar
\subsubsection{{\Tt{}\Rm{}{\bf{}figure}\nwendquote} conctructors}
\label{sec:figure-conctructors}

\nwenddocs{}\nwbegindocs{819}We create a figure with two initial objects: the cycle at infinity
and the real line.
\nwenddocs{}\nwbegincode{820}\sublabel{NWppJ6t-DKmU5-1}\nwmargintag{{\nwtagstyle{}\subpageref{NWppJ6t-DKmU5-1}}}\moddef{figure class~{\nwtagstyle{}\subpageref{NWppJ6t-DKmU5-1}}}\endmoddef\Rm{}\nwstartdeflinemarkup\nwusesondefline{\\{NWppJ6t-32jW6L-1}}\nwprevnextdefs{\relax}{NWppJ6t-DKmU5-2}\nwenddeflinemarkup
{\bf{}figure}::{\bf{}figure}() : {\it{}inherited}(), {\it{}k}({\bf{}realsymbol}({\tt{}"k"})), {\it{}m}({\bf{}realsymbol}({\tt{}"m"})), {\it{}l}()
{\nwlbrace}
    {\it{}l}.{\it{}append}({\bf{}realsymbol}({\tt{}"l0"}));
    {\it{}l}.{\it{}append}({\bf{}realsymbol}({\tt{}"l1"}));
    {\it{}infinity}={\bf{}symbol}({\tt{}"infty"},{\tt{}"{\char92}{\char92}infty"});
    {\it{}real\_line}={\bf{}symbol}({\tt{}"R"},{\tt{}"{\char92}{\char92}mathbf{\char123}R{\char125}"});
    {\it{}point\_metric} = {\it{}clifford\_unit}({\bf{}varidx}({\it{}real\_line}, 2), {\bf{}indexed}(-({\bf{}new} {\it{}tensdelta})\begin{math}\rightarrow\end{math}{\it{}setflag}({\it{}status\_flags}::{\it{}dynallocated}),
                                                               {\it{}sy\_symm}(), {\bf{}varidx}({\bf{}symbol}({\tt{}"i"}), 2), {\bf{}varidx}({\bf{}symbol}({\tt{}"j"}), 2)));
    {\it{}cycle\_metric} = {\it{}clifford\_unit}({\bf{}varidx}({\it{}real\_line}, 2), {\bf{}indexed}(-({\bf{}new} {\it{}tensdelta})\begin{math}\rightarrow\end{math}{\it{}setflag}({\it{}status\_flags}::{\it{}dynallocated}),
                                                               {\it{}sy\_symm}(), {\bf{}varidx}({\bf{}symbol}({\tt{}"ic"}), 2), {\bf{}varidx}({\bf{}symbol}({\tt{}"jc"}), 2)));
    \LA{}set the infinity~{\nwtagstyle{}\subpageref{NWppJ6t-2j51zU-1}}\RA{}
    \LA{}set the real line~{\nwtagstyle{}\subpageref{NWppJ6t-2q1QAq-1}}\RA{}
{\nwrbrace}
\nwindexdefn{\nwixident{figure}}{figure}{NWppJ6t-DKmU5-1}\eatline
\nwalsodefined{\\{NWppJ6t-DKmU5-2}\\{NWppJ6t-DKmU5-3}\\{NWppJ6t-DKmU5-4}\\{NWppJ6t-DKmU5-5}\\{NWppJ6t-DKmU5-6}\\{NWppJ6t-DKmU5-7}\\{NWppJ6t-DKmU5-8}\\{NWppJ6t-DKmU5-9}\\{NWppJ6t-DKmU5-A}\\{NWppJ6t-DKmU5-B}\\{NWppJ6t-DKmU5-C}\\{NWppJ6t-DKmU5-D}\\{NWppJ6t-DKmU5-E}\\{NWppJ6t-DKmU5-F}\\{NWppJ6t-DKmU5-G}\\{NWppJ6t-DKmU5-H}\\{NWppJ6t-DKmU5-I}\\{NWppJ6t-DKmU5-J}\\{NWppJ6t-DKmU5-K}\\{NWppJ6t-DKmU5-L}\\{NWppJ6t-DKmU5-M}\\{NWppJ6t-DKmU5-N}\\{NWppJ6t-DKmU5-O}\\{NWppJ6t-DKmU5-P}\\{NWppJ6t-DKmU5-Q}\\{NWppJ6t-DKmU5-R}\\{NWppJ6t-DKmU5-S}\\{NWppJ6t-DKmU5-T}\\{NWppJ6t-DKmU5-U}\\{NWppJ6t-DKmU5-V}\\{NWppJ6t-DKmU5-W}\\{NWppJ6t-DKmU5-X}\\{NWppJ6t-DKmU5-Y}\\{NWppJ6t-DKmU5-Z}\\{NWppJ6t-DKmU5-a}\\{NWppJ6t-DKmU5-b}\\{NWppJ6t-DKmU5-c}\\{NWppJ6t-DKmU5-d}\\{NWppJ6t-DKmU5-e}\\{NWppJ6t-DKmU5-f}\\{NWppJ6t-DKmU5-g}\\{NWppJ6t-DKmU5-h}\\{NWppJ6t-DKmU5-i}\\{NWppJ6t-DKmU5-j}\\{NWppJ6t-DKmU5-k}\\{NWppJ6t-DKmU5-l}\\{NWppJ6t-DKmU5-m}\\{NWppJ6t-DKmU5-n}\\{NWppJ6t-DKmU5-o}\\{NWppJ6t-DKmU5-p}\\{NWppJ6t-DKmU5-q}\\{NWppJ6t-DKmU5-r}\\{NWppJ6t-DKmU5-s}\\{NWppJ6t-DKmU5-t}\\{NWppJ6t-DKmU5-u}\\{NWppJ6t-DKmU5-v}\\{NWppJ6t-DKmU5-w}\\{NWppJ6t-DKmU5-x}\\{NWppJ6t-DKmU5-y}\\{NWppJ6t-DKmU5-z}\\{NWppJ6t-DKmU5-10}\\{NWppJ6t-DKmU5-11}\\{NWppJ6t-DKmU5-12}\\{NWppJ6t-DKmU5-13}\\{NWppJ6t-DKmU5-14}\\{NWppJ6t-DKmU5-15}\\{NWppJ6t-DKmU5-16}\\{NWppJ6t-DKmU5-17}\\{NWppJ6t-DKmU5-18}\\{NWppJ6t-DKmU5-19}\\{NWppJ6t-DKmU5-1A}\\{NWppJ6t-DKmU5-1B}\\{NWppJ6t-DKmU5-1C}\\{NWppJ6t-DKmU5-1D}\\{NWppJ6t-DKmU5-1E}\\{NWppJ6t-DKmU5-1F}\\{NWppJ6t-DKmU5-1G}\\{NWppJ6t-DKmU5-1H}\\{NWppJ6t-DKmU5-1I}\\{NWppJ6t-DKmU5-1J}\\{NWppJ6t-DKmU5-1K}\\{NWppJ6t-DKmU5-1L}\\{NWppJ6t-DKmU5-1M}\\{NWppJ6t-DKmU5-1N}\\{NWppJ6t-DKmU5-1O}\\{NWppJ6t-DKmU5-1P}\\{NWppJ6t-DKmU5-1Q}\\{NWppJ6t-DKmU5-1R}\\{NWppJ6t-DKmU5-1S}\\{NWppJ6t-DKmU5-1T}\\{NWppJ6t-DKmU5-1U}\\{NWppJ6t-DKmU5-1V}\\{NWppJ6t-DKmU5-1W}\\{NWppJ6t-DKmU5-1X}\\{NWppJ6t-DKmU5-1Y}\\{NWppJ6t-DKmU5-1Z}\\{NWppJ6t-DKmU5-1a}\\{NWppJ6t-DKmU5-1b}\\{NWppJ6t-DKmU5-1c}\\{NWppJ6t-DKmU5-1d}\\{NWppJ6t-DKmU5-1e}\\{NWppJ6t-DKmU5-1f}\\{NWppJ6t-DKmU5-1g}\\{NWppJ6t-DKmU5-1h}\\{NWppJ6t-DKmU5-1i}\\{NWppJ6t-DKmU5-1j}\\{NWppJ6t-DKmU5-1k}\\{NWppJ6t-DKmU5-1l}\\{NWppJ6t-DKmU5-1m}\\{NWppJ6t-DKmU5-1n}\\{NWppJ6t-DKmU5-1o}\\{NWppJ6t-DKmU5-1p}\\{NWppJ6t-DKmU5-1q}\\{NWppJ6t-DKmU5-1r}\\{NWppJ6t-DKmU5-1s}\\{NWppJ6t-DKmU5-1t}\\{NWppJ6t-DKmU5-1u}\\{NWppJ6t-DKmU5-1v}\\{NWppJ6t-DKmU5-1w}\\{NWppJ6t-DKmU5-1x}\\{NWppJ6t-DKmU5-1y}\\{NWppJ6t-DKmU5-1z}\\{NWppJ6t-DKmU5-20}\\{NWppJ6t-DKmU5-21}\\{NWppJ6t-DKmU5-22}\\{NWppJ6t-DKmU5-23}\\{NWppJ6t-DKmU5-24}\\{NWppJ6t-DKmU5-25}\\{NWppJ6t-DKmU5-26}\\{NWppJ6t-DKmU5-27}\\{NWppJ6t-DKmU5-28}\\{NWppJ6t-DKmU5-29}\\{NWppJ6t-DKmU5-2A}\\{NWppJ6t-DKmU5-2B}\\{NWppJ6t-DKmU5-2C}\\{NWppJ6t-DKmU5-2D}\\{NWppJ6t-DKmU5-2E}\\{NWppJ6t-DKmU5-2F}\\{NWppJ6t-DKmU5-2G}\\{NWppJ6t-DKmU5-2H}\\{NWppJ6t-DKmU5-2I}\\{NWppJ6t-DKmU5-2J}\\{NWppJ6t-DKmU5-2K}}\nwused{\\{NWppJ6t-32jW6L-1}}\nwidentdefs{\\{{\nwixident{figure}}{figure}}}\nwidentuses{\\{{\nwixident{cycle{\_}metric}}{cycle:unmetric}}\\{{\nwixident{infinity}}{infinity}}\\{{\nwixident{k}}{k}}\\{{\nwixident{l}}{l}}\\{{\nwixident{m}}{m}}\\{{\nwixident{point{\_}metric}}{point:unmetric}}\\{{\nwixident{real{\_}line}}{real:unline}}\\{{\nwixident{realsymbol}}{realsymbol}}}\nwindexuse{\nwixident{cycle{\_}metric}}{cycle:unmetric}{NWppJ6t-DKmU5-1}\nwindexuse{\nwixident{infinity}}{infinity}{NWppJ6t-DKmU5-1}\nwindexuse{\nwixident{k}}{k}{NWppJ6t-DKmU5-1}\nwindexuse{\nwixident{l}}{l}{NWppJ6t-DKmU5-1}\nwindexuse{\nwixident{m}}{m}{NWppJ6t-DKmU5-1}\nwindexuse{\nwixident{point{\_}metric}}{point:unmetric}{NWppJ6t-DKmU5-1}\nwindexuse{\nwixident{real{\_}line}}{real:unline}{NWppJ6t-DKmU5-1}\nwindexuse{\nwixident{realsymbol}}{realsymbol}{NWppJ6t-DKmU5-1}\nwendcode{}\nwbegindocs{821}\nwdocspar
\nwenddocs{}\nwbegindocs{822}Dimension of the fiigure is taken and the respective vector is created.
\nwenddocs{}\nwbegincode{823}\sublabel{NWppJ6t-2V7UoL-1}\nwmargintag{{\nwtagstyle{}\subpageref{NWppJ6t-2V7UoL-1}}}\moddef{initialise the dimension and vector~{\nwtagstyle{}\subpageref{NWppJ6t-2V7UoL-1}}}\endmoddef\Rm{}\nwstartdeflinemarkup\nwusesondefline{\\{NWppJ6t-2j51zU-1}\\{NWppJ6t-1NAwBA-1}}\nwenddeflinemarkup
   {\bf{}unsigned} {\bf{}int} {\it{}dim}={\it{}ex\_to}\begin{math}<\end{math}{\bf{}numeric}\begin{math}>\end{math}({\it{}get\_dim}()).{\it{}to\_int}();
   {\bf{}lst} {\it{}l0};
   {\bf{}for}({\bf{}unsigned} {\bf{}int} {\it{}i}=0; {\it{}i}\begin{math}<\end{math}{\it{}dim}; \protect\PP{\it{}i})
       {\it{}l0}.{\it{}append}(0);

\nwused{\\{NWppJ6t-2j51zU-1}\\{NWppJ6t-1NAwBA-1}}\nwidentuses{\\{{\nwixident{get{\_}dim()}}{get:undim()}}\\{{\nwixident{numeric}}{numeric}}}\nwindexuse{\nwixident{get{\_}dim()}}{get:undim()}{NWppJ6t-2V7UoL-1}\nwindexuse{\nwixident{numeric}}{numeric}{NWppJ6t-2V7UoL-1}\nwendcode{}\nwbegindocs{824}\nwdocspar
\nwenddocs{}\nwbegincode{825}\sublabel{NWppJ6t-2j51zU-1}\nwmargintag{{\nwtagstyle{}\subpageref{NWppJ6t-2j51zU-1}}}\moddef{set the infinity~{\nwtagstyle{}\subpageref{NWppJ6t-2j51zU-1}}}\endmoddef\Rm{}\nwstartdeflinemarkup\nwusesondefline{\\{NWppJ6t-DKmU5-1}\\{NWppJ6t-1NAwBA-1}\\{NWppJ6t-DKmU5-1V}}\nwenddeflinemarkup
   \LA{}initialise the dimension and vector~{\nwtagstyle{}\subpageref{NWppJ6t-2V7UoL-1}}\RA{}
   {\it{}nodes}[{\it{}infinity}] = {\bf{}cycle\_node}({\bf{}cycle\_data}({\bf{}numeric}(0),{\bf{}indexed}({\bf{}matrix}(1, {\it{}dim}, {\it{}l0}),
                                                              {\bf{}varidx}({\it{}infinity}, {\it{}dim})),{\bf{}numeric}(1)),{\it{}INFINITY\_GEN});

\nwused{\\{NWppJ6t-DKmU5-1}\\{NWppJ6t-1NAwBA-1}\\{NWppJ6t-DKmU5-1V}}\nwidentuses{\\{{\nwixident{cycle{\_}data}}{cycle:undata}}\\{{\nwixident{cycle{\_}node}}{cycle:unnode}}\\{{\nwixident{infinity}}{infinity}}\\{{\nwixident{INFINITY{\_}GEN}}{INFINITY:unGEN}}\\{{\nwixident{nodes}}{nodes}}\\{{\nwixident{numeric}}{numeric}}}\nwindexuse{\nwixident{cycle{\_}data}}{cycle:undata}{NWppJ6t-2j51zU-1}\nwindexuse{\nwixident{cycle{\_}node}}{cycle:unnode}{NWppJ6t-2j51zU-1}\nwindexuse{\nwixident{infinity}}{infinity}{NWppJ6t-2j51zU-1}\nwindexuse{\nwixident{INFINITY{\_}GEN}}{INFINITY:unGEN}{NWppJ6t-2j51zU-1}\nwindexuse{\nwixident{nodes}}{nodes}{NWppJ6t-2j51zU-1}\nwindexuse{\nwixident{numeric}}{numeric}{NWppJ6t-2j51zU-1}\nwendcode{}\nwbegindocs{826}\nwdocspar
\nwenddocs{}\nwbegincode{827}\sublabel{NWppJ6t-2q1QAq-1}\nwmargintag{{\nwtagstyle{}\subpageref{NWppJ6t-2q1QAq-1}}}\moddef{set the real line~{\nwtagstyle{}\subpageref{NWppJ6t-2q1QAq-1}}}\endmoddef\Rm{}\nwstartdeflinemarkup\nwusesondefline{\\{NWppJ6t-DKmU5-1}\\{NWppJ6t-1NAwBA-1}\\{NWppJ6t-DKmU5-1V}}\nwenddeflinemarkup
    {\it{}l0}.{\it{}remove\_last}();
    {\it{}l0}.{\it{}append}(1);
    {\it{}nodes}[{\it{}real\_line}] = {\bf{}cycle\_node}({\bf{}cycle\_data}({\bf{}numeric}(0),{\bf{}indexed}({\bf{}matrix}(1, {\it{}dim}, {\it{}l0}),
                                                                    {\bf{}varidx}({\it{}real\_line}, {\it{}dim})),{\bf{}numeric}(0)),{\it{}REAL\_LINE\_GEN});

\nwused{\\{NWppJ6t-DKmU5-1}\\{NWppJ6t-1NAwBA-1}\\{NWppJ6t-DKmU5-1V}}\nwidentuses{\\{{\nwixident{cycle{\_}data}}{cycle:undata}}\\{{\nwixident{cycle{\_}node}}{cycle:unnode}}\\{{\nwixident{nodes}}{nodes}}\\{{\nwixident{numeric}}{numeric}}\\{{\nwixident{real{\_}line}}{real:unline}}\\{{\nwixident{REAL{\_}LINE{\_}GEN}}{REAL:unLINE:unGEN}}}\nwindexuse{\nwixident{cycle{\_}data}}{cycle:undata}{NWppJ6t-2q1QAq-1}\nwindexuse{\nwixident{cycle{\_}node}}{cycle:unnode}{NWppJ6t-2q1QAq-1}\nwindexuse{\nwixident{nodes}}{nodes}{NWppJ6t-2q1QAq-1}\nwindexuse{\nwixident{numeric}}{numeric}{NWppJ6t-2q1QAq-1}\nwindexuse{\nwixident{real{\_}line}}{real:unline}{NWppJ6t-2q1QAq-1}\nwindexuse{\nwixident{REAL{\_}LINE{\_}GEN}}{REAL:unLINE:unGEN}{NWppJ6t-2q1QAq-1}\nwendcode{}\nwbegindocs{828}This constructor may be called with several different inputs.
\nwenddocs{}\nwbegincode{829}\sublabel{NWppJ6t-DKmU5-2}\nwmargintag{{\nwtagstyle{}\subpageref{NWppJ6t-DKmU5-2}}}\moddef{figure class~{\nwtagstyle{}\subpageref{NWppJ6t-DKmU5-1}}}\plusendmoddef\Rm{}\nwstartdeflinemarkup\nwusesondefline{\\{NWppJ6t-32jW6L-1}}\nwprevnextdefs{NWppJ6t-DKmU5-1}{NWppJ6t-DKmU5-3}\nwenddeflinemarkup
{\bf{}figure}::{\bf{}figure}({\bf{}const} {\bf{}ex} & {\it{}Mp}, {\bf{}const} {\bf{}ex} & {\it{}Mc}) : {\it{}inherited}(), {\it{}k}({\bf{}realsymbol}({\tt{}"k"})), {\it{}m}({\bf{}realsymbol}({\tt{}"m"})), {\it{}l}()
{\nwlbrace}
    {\it{}infinity}={\bf{}symbol}({\tt{}"infty"},{\tt{}"{\char92}{\char92}infty"});
    {\it{}real\_line}={\bf{}symbol}({\tt{}"R"},{\tt{}"{\char92}{\char92}mathbf{\char123}R{\char125}"});
    {\bf{}bool} {\it{}inf\_missing}={\bf{}true}, {\it{}R\_missing}={\bf{}true};
    \LA{}set point metric in figure~{\nwtagstyle{}\subpageref{NWppJ6t-4ALcRV-1}}\RA{}

\nwused{\\{NWppJ6t-32jW6L-1}}\nwidentuses{\\{{\nwixident{ex}}{ex}}\\{{\nwixident{figure}}{figure}}\\{{\nwixident{infinity}}{infinity}}\\{{\nwixident{k}}{k}}\\{{\nwixident{l}}{l}}\\{{\nwixident{m}}{m}}\\{{\nwixident{real{\_}line}}{real:unline}}\\{{\nwixident{realsymbol}}{realsymbol}}}\nwindexuse{\nwixident{ex}}{ex}{NWppJ6t-DKmU5-2}\nwindexuse{\nwixident{figure}}{figure}{NWppJ6t-DKmU5-2}\nwindexuse{\nwixident{infinity}}{infinity}{NWppJ6t-DKmU5-2}\nwindexuse{\nwixident{k}}{k}{NWppJ6t-DKmU5-2}\nwindexuse{\nwixident{l}}{l}{NWppJ6t-DKmU5-2}\nwindexuse{\nwixident{m}}{m}{NWppJ6t-DKmU5-2}\nwindexuse{\nwixident{real{\_}line}}{real:unline}{NWppJ6t-DKmU5-2}\nwindexuse{\nwixident{realsymbol}}{realsymbol}{NWppJ6t-DKmU5-2}\nwendcode{}\nwbegindocs{830}Below are various parameters which can define a metric in the same
way as it used to create a {\Tt{}\Rm{}{\it{}cliffordunit}\nwendquote} object in \GiNaC.
\nwenddocs{}\nwbegincode{831}\sublabel{NWppJ6t-4ALcRV-1}\nwmargintag{{\nwtagstyle{}\subpageref{NWppJ6t-4ALcRV-1}}}\moddef{set point metric in figure~{\nwtagstyle{}\subpageref{NWppJ6t-4ALcRV-1}}}\endmoddef\Rm{}\nwstartdeflinemarkup\nwusesondefline{\\{NWppJ6t-DKmU5-2}\\{NWppJ6t-DKmU5-1Q}}\nwprevnextdefs{\relax}{NWppJ6t-4ALcRV-2}\nwenddeflinemarkup
{\bf{}if} ({\it{}is\_a}\begin{math}<\end{math}{\bf{}clifford}\begin{math}>\end{math}({\it{}Mp})) {\nwlbrace}
    {\it{}point\_metric} = {\it{}clifford\_unit}({\bf{}varidx}({\it{}real\_line},
                                        {\it{}ex\_to}\begin{math}<\end{math}{\bf{}idx}\begin{math}>\end{math}({\it{}ex\_to}\begin{math}<\end{math}{\bf{}clifford}\begin{math}>\end{math}({\it{}Mp}).{\it{}get\_metric}().{\it{}op}(1)).{\it{}get\_dim}()),
                                 {\it{}ex\_to}\begin{math}<\end{math}{\bf{}clifford}\begin{math}>\end{math}({\it{}Mp}).{\it{}get\_metric}());
 {\nwrbrace} {\bf{}else} {\bf{}if} ({\it{}is\_a}\begin{math}<\end{math}{\bf{}matrix}\begin{math}>\end{math}({\it{}Mp})) {\nwlbrace}
    {\bf{}ex} {\it{}D};
    {\bf{}if} ({\it{}ex\_to}\begin{math}<\end{math}{\bf{}matrix}\begin{math}>\end{math}({\it{}Mp}).{\it{}rows}() \begin{math}\equiv\end{math} {\it{}ex\_to}\begin{math}<\end{math}{\bf{}matrix}\begin{math}>\end{math}({\it{}Mp}).{\it{}cols}())
        {\it{}D}={\it{}ex\_to}\begin{math}<\end{math}{\bf{}matrix}\begin{math}>\end{math}({\it{}Mp}).{\it{}rows}();
    {\bf{}else}
        {\bf{}throw}({\it{}std}::{\it{}invalid\_argument}({\tt{}"figure::figure(const ex &, const ex &):"}
                                    {\tt{}" only square matrices are admitted as point metric"}));
    {\it{}point\_metric} = {\it{}clifford\_unit}({\bf{}varidx}({\it{}real\_line}, {\it{}D}), {\bf{}indexed}({\it{}Mp}, {\it{}sy\_symm}(), {\bf{}varidx}({\bf{}symbol}({\tt{}"i"}), {\it{}D}), {\bf{}varidx}({\bf{}symbol}({\tt{}"j"}), {\it{}D})));
 {\nwrbrace} {\bf{}else} {\bf{}if} ({\it{}is\_a}\begin{math}<\end{math}{\bf{}indexed}\begin{math}>\end{math}({\it{}Mp})) {\nwlbrace}
    {\it{}point\_metric} = {\it{}clifford\_unit}({\bf{}varidx}({\it{}real\_line}, {\it{}ex\_to}\begin{math}<\end{math}{\bf{}idx}\begin{math}>\end{math}({\it{}Mp}.{\it{}op}(1)).{\it{}get\_dim}()), {\it{}Mp});

\nwalsodefined{\\{NWppJ6t-4ALcRV-2}}\nwused{\\{NWppJ6t-DKmU5-2}\\{NWppJ6t-DKmU5-1Q}}\nwidentuses{\\{{\nwixident{ex}}{ex}}\\{{\nwixident{figure}}{figure}}\\{{\nwixident{get{\_}dim()}}{get:undim()}}\\{{\nwixident{op}}{op}}\\{{\nwixident{point{\_}metric}}{point:unmetric}}\\{{\nwixident{real{\_}line}}{real:unline}}}\nwindexuse{\nwixident{ex}}{ex}{NWppJ6t-4ALcRV-1}\nwindexuse{\nwixident{figure}}{figure}{NWppJ6t-4ALcRV-1}\nwindexuse{\nwixident{get{\_}dim()}}{get:undim()}{NWppJ6t-4ALcRV-1}\nwindexuse{\nwixident{op}}{op}{NWppJ6t-4ALcRV-1}\nwindexuse{\nwixident{point{\_}metric}}{point:unmetric}{NWppJ6t-4ALcRV-1}\nwindexuse{\nwixident{real{\_}line}}{real:unline}{NWppJ6t-4ALcRV-1}\nwendcode{}\nwbegindocs{832}If a {\Tt{}\Rm{}{\bf{}lst}\nwendquote} is supplied we use as the signature of
metric, entries {\Tt{}\Rm{}{\it{}Mp}\nwendquote} as the diagonal elements of the matrix.
\nwenddocs{}\nwbegincode{833}\sublabel{NWppJ6t-4ALcRV-2}\nwmargintag{{\nwtagstyle{}\subpageref{NWppJ6t-4ALcRV-2}}}\moddef{set point metric in figure~{\nwtagstyle{}\subpageref{NWppJ6t-4ALcRV-1}}}\plusendmoddef\Rm{}\nwstartdeflinemarkup\nwusesondefline{\\{NWppJ6t-DKmU5-2}\\{NWppJ6t-DKmU5-1Q}}\nwprevnextdefs{NWppJ6t-4ALcRV-1}{\relax}\nwenddeflinemarkup
    {\nwrbrace} {\bf{}else} {\bf{}if} ({\it{}is\_a}\begin{math}<\end{math}{\bf{}lst}\begin{math}>\end{math}({\it{}Mp})) {\nwlbrace}
        {\it{}point\_metric}={\it{}clifford\_unit}({\bf{}varidx}({\it{}real\_line}, {\it{}Mp}.{\it{}nops}()), {\bf{}indexed}({\it{}diag\_matrix}({\it{}ex\_to}\begin{math}<\end{math}{\bf{}lst}\begin{math}>\end{math}({\it{}Mp})), {\it{}sy\_symm}(),
                                                                         {\bf{}varidx}({\bf{}symbol}({\tt{}"i"}), {\it{}Mp}.{\it{}nops}()), {\bf{}varidx}({\bf{}symbol}({\tt{}"j"}), {\it{}Mp}.{\it{}nops}())));
    {\nwrbrace}

\nwused{\\{NWppJ6t-DKmU5-2}\\{NWppJ6t-DKmU5-1Q}}\nwidentuses{\\{{\nwixident{nops}}{nops}}\\{{\nwixident{point{\_}metric}}{point:unmetric}}\\{{\nwixident{real{\_}line}}{real:unline}}}\nwindexuse{\nwixident{nops}}{nops}{NWppJ6t-4ALcRV-2}\nwindexuse{\nwixident{point{\_}metric}}{point:unmetric}{NWppJ6t-4ALcRV-2}\nwindexuse{\nwixident{real{\_}line}}{real:unline}{NWppJ6t-4ALcRV-2}\nwendcode{}\nwbegindocs{834}If {\Tt{}\Rm{}{\it{}Mp}\nwendquote} is a figure we effectively copy it.
\nwenddocs{}\nwbegincode{835}\sublabel{NWppJ6t-DKmU5-3}\nwmargintag{{\nwtagstyle{}\subpageref{NWppJ6t-DKmU5-3}}}\moddef{figure class~{\nwtagstyle{}\subpageref{NWppJ6t-DKmU5-1}}}\plusendmoddef\Rm{}\nwstartdeflinemarkup\nwusesondefline{\\{NWppJ6t-32jW6L-1}}\nwprevnextdefs{NWppJ6t-DKmU5-2}{NWppJ6t-DKmU5-4}\nwenddeflinemarkup
    {\bf{}else} {\bf{}if} ({\it{}is\_a}\begin{math}<\end{math}{\bf{}figure}\begin{math}>\end{math}({\it{}Mp})) {\nwlbrace}
        {\it{}point\_metric} = {\it{}ex\_to}\begin{math}<\end{math}{\bf{}figure}\begin{math}>\end{math}({\it{}Mp}).{\it{}get\_point\_metric}();
        {\it{}cycle\_metric} = {\it{}ex\_to}\begin{math}<\end{math}{\bf{}figure}\begin{math}>\end{math}({\it{}Mp}).{\it{}get\_cycle\_metric}();
        {\it{}exhashmap}\begin{math}<\end{math}{\bf{}cycle\_node}\begin{math}>\end{math} {\it{}nnodes} = {\it{}ex\_to}\begin{math}<\end{math}{\bf{}figure}\begin{math}>\end{math}({\it{}Mp}).{\it{}get\_nodes}();
        {\bf{}for} ({\bf{}const} {\bf{}auto}& {\it{}x}: {\it{}nnodes}) {\nwlbrace}
            {\it{}nodes}[{\it{}x}.{\it{}first}]={\it{}x}.{\it{}second};
            \LA{}identify infinity and real line~{\nwtagstyle{}\subpageref{NWppJ6t-1pIH3B-1}}\RA{}
            {\nwrbrace}

\nwused{\\{NWppJ6t-32jW6L-1}}\nwidentuses{\\{{\nwixident{cycle{\_}metric}}{cycle:unmetric}}\\{{\nwixident{cycle{\_}node}}{cycle:unnode}}\\{{\nwixident{figure}}{figure}}\\{{\nwixident{get{\_}cycle{\_}metric}}{get:uncycle:unmetric}}\\{{\nwixident{get{\_}point{\_}metric}}{get:unpoint:unmetric}}\\{{\nwixident{nodes}}{nodes}}\\{{\nwixident{point{\_}metric}}{point:unmetric}}}\nwindexuse{\nwixident{cycle{\_}metric}}{cycle:unmetric}{NWppJ6t-DKmU5-3}\nwindexuse{\nwixident{cycle{\_}node}}{cycle:unnode}{NWppJ6t-DKmU5-3}\nwindexuse{\nwixident{figure}}{figure}{NWppJ6t-DKmU5-3}\nwindexuse{\nwixident{get{\_}cycle{\_}metric}}{get:uncycle:unmetric}{NWppJ6t-DKmU5-3}\nwindexuse{\nwixident{get{\_}point{\_}metric}}{get:unpoint:unmetric}{NWppJ6t-DKmU5-3}\nwindexuse{\nwixident{nodes}}{nodes}{NWppJ6t-DKmU5-3}\nwindexuse{\nwixident{point{\_}metric}}{point:unmetric}{NWppJ6t-DKmU5-3}\nwendcode{}\nwbegindocs{836}We need to set {\Tt{}\Rm{}{\it{}real\_line}\nwendquote} and {\Tt{}\Rm{}{\it{}infinity}\nwendquote} accordingly.
\nwenddocs{}\nwbegincode{837}\sublabel{NWppJ6t-1pIH3B-1}\nwmargintag{{\nwtagstyle{}\subpageref{NWppJ6t-1pIH3B-1}}}\moddef{identify infinity and real line~{\nwtagstyle{}\subpageref{NWppJ6t-1pIH3B-1}}}\endmoddef\Rm{}\nwstartdeflinemarkup\nwusesondefline{\\{NWppJ6t-DKmU5-3}\\{NWppJ6t-DKmU5-A}}\nwenddeflinemarkup
            {\bf{}if} ({\it{}x}.{\it{}second}.{\it{}get\_generation}() \begin{math}\equiv\end{math} {\it{}REAL\_LINE\_GEN}) {\nwlbrace}
                {\it{}real\_line} = {\it{}x}.{\it{}first};
                {\it{}R\_missing}={\bf{}false};
            {\nwrbrace}
            {\bf{}else} {\bf{}if} ({\it{}x}.{\it{}second}.{\it{}get\_generation}() \begin{math}\equiv\end{math} {\it{}INFINITY\_GEN}) {\nwlbrace}
                {\it{}infinity} = {\it{}x}.{\it{}first};
                {\it{}inf\_missing}={\bf{}false};
            {\nwrbrace}

\nwused{\\{NWppJ6t-DKmU5-3}\\{NWppJ6t-DKmU5-A}}\nwidentuses{\\{{\nwixident{get{\_}generation}}{get:ungeneration}}\\{{\nwixident{infinity}}{infinity}}\\{{\nwixident{INFINITY{\_}GEN}}{INFINITY:unGEN}}\\{{\nwixident{real{\_}line}}{real:unline}}\\{{\nwixident{REAL{\_}LINE{\_}GEN}}{REAL:unLINE:unGEN}}}\nwindexuse{\nwixident{get{\_}generation}}{get:ungeneration}{NWppJ6t-1pIH3B-1}\nwindexuse{\nwixident{infinity}}{infinity}{NWppJ6t-1pIH3B-1}\nwindexuse{\nwixident{INFINITY{\_}GEN}}{INFINITY:unGEN}{NWppJ6t-1pIH3B-1}\nwindexuse{\nwixident{real{\_}line}}{real:unline}{NWppJ6t-1pIH3B-1}\nwindexuse{\nwixident{REAL{\_}LINE{\_}GEN}}{REAL:unLINE:unGEN}{NWppJ6t-1pIH3B-1}\nwendcode{}\nwbegindocs{838}For an unknown type parameter we throw an exception.
\nwenddocs{}\nwbegincode{839}\sublabel{NWppJ6t-DKmU5-4}\nwmargintag{{\nwtagstyle{}\subpageref{NWppJ6t-DKmU5-4}}}\moddef{figure class~{\nwtagstyle{}\subpageref{NWppJ6t-DKmU5-1}}}\plusendmoddef\Rm{}\nwstartdeflinemarkup\nwusesondefline{\\{NWppJ6t-32jW6L-1}}\nwprevnextdefs{NWppJ6t-DKmU5-3}{NWppJ6t-DKmU5-5}\nwenddeflinemarkup
    {\nwrbrace} {\bf{}else}
            {\bf{}throw}({\it{}std}::{\it{}invalid\_argument}({\tt{}"figure::figure(const ex &, const ex &):"}
                                        {\tt{}" the first parameter shall be a figure, a lst, "}
                                        {\tt{}" a metric (can be either tensor, matrix,"}
                                        {\tt{}" Clifford unit or indexed by two indices) "}));
    \LA{}set cycle metric in figure~{\nwtagstyle{}\subpageref{NWppJ6t-1wPO4G-1}}\RA{}

\nwused{\\{NWppJ6t-32jW6L-1}}\nwidentuses{\\{{\nwixident{ex}}{ex}}\\{{\nwixident{figure}}{figure}}}\nwindexuse{\nwixident{ex}}{ex}{NWppJ6t-DKmU5-4}\nwindexuse{\nwixident{figure}}{figure}{NWppJ6t-DKmU5-4}\nwendcode{}\nwbegindocs{840}If a metric is not supplied or is zero then we clone the point space
metric by the rule defined in equation~\eqref{eq:int-heaviside-function}.
\nwenddocs{}\nwbegincode{841}\sublabel{NWppJ6t-1wPO4G-1}\nwmargintag{{\nwtagstyle{}\subpageref{NWppJ6t-1wPO4G-1}}}\moddef{set cycle metric in figure~{\nwtagstyle{}\subpageref{NWppJ6t-1wPO4G-1}}}\endmoddef\Rm{}\nwstartdeflinemarkup\nwusesondefline{\\{NWppJ6t-DKmU5-4}\\{NWppJ6t-DKmU5-1Q}}\nwprevnextdefs{\relax}{NWppJ6t-1wPO4G-2}\nwenddeflinemarkup
    {\bf{}if} ({\it{}Mc}.{\it{}is\_zero}()) {\nwlbrace}
        {\bf{}ex} {\it{}D}={\it{}get\_dim}();
        {\bf{}if} ({\it{}is\_a}\begin{math}<\end{math}{\bf{}numeric}\begin{math}>\end{math}({\it{}D})) {\nwlbrace}
            {\bf{}lst} {\it{}l0};
            {\bf{}for}({\bf{}int} {\it{}i}=0; {\it{}i}\begin{math}<\end{math} {\it{}ex\_to}\begin{math}<\end{math}{\bf{}numeric}\begin{math}>\end{math}({\it{}D}).{\it{}to\_int}(); \protect\PP{\it{}i})
                {\it{}l0}.{\it{}append}(-{\it{}jump\_fnct}(-{\it{}ex\_to}\begin{math}<\end{math}{\bf{}clifford}\begin{math}>\end{math}({\it{}point\_metric}).{\it{}get\_metric}({\bf{}idx}({\it{}i},{\it{}D}),{\bf{}idx}({\it{}i},{\it{}D}))));
            {\it{}cycle\_metric} = {\it{}clifford\_unit}({\bf{}varidx}({\it{}real\_line}, {\it{}D}), {\bf{}indexed}({\it{}diag\_matrix}({\it{}l0}), {\it{}sy\_symm}(),
                                                                       {\bf{}varidx}({\bf{}symbol}({\tt{}"ic"}), {\it{}D}), {\bf{}varidx}({\bf{}symbol}({\tt{}"jc"}), {\it{}D})));

\nwalsodefined{\\{NWppJ6t-1wPO4G-2}\\{NWppJ6t-1wPO4G-3}\\{NWppJ6t-1wPO4G-4}}\nwused{\\{NWppJ6t-DKmU5-4}\\{NWppJ6t-DKmU5-1Q}}\nwidentuses{\\{{\nwixident{cycle{\_}metric}}{cycle:unmetric}}\\{{\nwixident{ex}}{ex}}\\{{\nwixident{get{\_}dim()}}{get:undim()}}\\{{\nwixident{numeric}}{numeric}}\\{{\nwixident{point{\_}metric}}{point:unmetric}}\\{{\nwixident{real{\_}line}}{real:unline}}}\nwindexuse{\nwixident{cycle{\_}metric}}{cycle:unmetric}{NWppJ6t-1wPO4G-1}\nwindexuse{\nwixident{ex}}{ex}{NWppJ6t-1wPO4G-1}\nwindexuse{\nwixident{get{\_}dim()}}{get:undim()}{NWppJ6t-1wPO4G-1}\nwindexuse{\nwixident{numeric}}{numeric}{NWppJ6t-1wPO4G-1}\nwindexuse{\nwixident{point{\_}metric}}{point:unmetric}{NWppJ6t-1wPO4G-1}\nwindexuse{\nwixident{real{\_}line}}{real:unline}{NWppJ6t-1wPO4G-1}\nwendcode{}\nwbegindocs{842}If dimensionality is not integer, then the point metric is copied.
\nwenddocs{}\nwbegincode{843}\sublabel{NWppJ6t-1wPO4G-2}\nwmargintag{{\nwtagstyle{}\subpageref{NWppJ6t-1wPO4G-2}}}\moddef{set cycle metric in figure~{\nwtagstyle{}\subpageref{NWppJ6t-1wPO4G-1}}}\plusendmoddef\Rm{}\nwstartdeflinemarkup\nwusesondefline{\\{NWppJ6t-DKmU5-4}\\{NWppJ6t-DKmU5-1Q}}\nwprevnextdefs{NWppJ6t-1wPO4G-1}{NWppJ6t-1wPO4G-3}\nwenddeflinemarkup
        {\nwrbrace} {\bf{}else}
            {\it{}cycle\_metric} = {\it{}clifford\_unit}({\bf{}varidx}({\it{}real\_line}, {\it{}D}), {\bf{}indexed}({\it{}point\_metric}.{\it{}op}(0), {\it{}sy\_symm}(),
                                                                       {\bf{}varidx}({\bf{}symbol}({\tt{}"ic"}), {\it{}D}), {\bf{}varidx}({\bf{}symbol}({\tt{}"jc"}), {\it{}D})));

\nwused{\\{NWppJ6t-DKmU5-4}\\{NWppJ6t-DKmU5-1Q}}\nwidentuses{\\{{\nwixident{cycle{\_}metric}}{cycle:unmetric}}\\{{\nwixident{op}}{op}}\\{{\nwixident{point{\_}metric}}{point:unmetric}}\\{{\nwixident{real{\_}line}}{real:unline}}}\nwindexuse{\nwixident{cycle{\_}metric}}{cycle:unmetric}{NWppJ6t-1wPO4G-2}\nwindexuse{\nwixident{op}}{op}{NWppJ6t-1wPO4G-2}\nwindexuse{\nwixident{point{\_}metric}}{point:unmetric}{NWppJ6t-1wPO4G-2}\nwindexuse{\nwixident{real{\_}line}}{real:unline}{NWppJ6t-1wPO4G-2}\nwendcode{}\nwbegindocs{844}If the metric is supplied, we repeat the same procedure to set-up
the metric of the cycle space as was done for point space.
\nwenddocs{}\nwbegincode{845}\sublabel{NWppJ6t-1wPO4G-3}\nwmargintag{{\nwtagstyle{}\subpageref{NWppJ6t-1wPO4G-3}}}\moddef{set cycle metric in figure~{\nwtagstyle{}\subpageref{NWppJ6t-1wPO4G-1}}}\plusendmoddef\Rm{}\nwstartdeflinemarkup\nwusesondefline{\\{NWppJ6t-DKmU5-4}\\{NWppJ6t-DKmU5-1Q}}\nwprevnextdefs{NWppJ6t-1wPO4G-2}{NWppJ6t-1wPO4G-4}\nwenddeflinemarkup
    {\nwrbrace} {\bf{}else} {\bf{}if} ({\it{}is\_a}\begin{math}<\end{math}{\bf{}clifford}\begin{math}>\end{math}({\it{}Mc})) {\nwlbrace}
        {\it{}cycle\_metric} = {\it{}clifford\_unit}({\bf{}varidx}({\it{}real\_line},
                                            {\it{}ex\_to}\begin{math}<\end{math}{\bf{}idx}\begin{math}>\end{math}({\it{}ex\_to}\begin{math}<\end{math}{\bf{}clifford}\begin{math}>\end{math}({\it{}Mc}).{\it{}get\_metric}().{\it{}op}(1)).{\it{}get\_dim}()),
                                     {\it{}ex\_to}\begin{math}<\end{math}{\bf{}clifford}\begin{math}>\end{math}({\it{}Mc}).{\it{}get\_metric}());
    {\nwrbrace} {\bf{}else} {\bf{}if} ({\it{}is\_a}\begin{math}<\end{math}{\bf{}matrix}\begin{math}>\end{math}({\it{}Mc})) {\nwlbrace}
        {\bf{}if} ({\it{}ex\_to}\begin{math}<\end{math}{\bf{}matrix}\begin{math}>\end{math}({\it{}Mp}).{\it{}rows}() \begin{math}\neq\end{math} {\it{}ex\_to}\begin{math}<\end{math}{\bf{}matrix}\begin{math}>\end{math}({\it{}Mp}).{\it{}cols}())
            {\bf{}throw}({\it{}std}::{\it{}invalid\_argument}({\tt{}"figure::figure(const ex &, const ex &):"}
                                        {\tt{}" only square matrices are admitted as cycle metric"}));

        {\it{}cycle\_metric} = {\it{}clifford\_unit}({\bf{}varidx}({\it{}real\_line}, {\it{}get\_dim}()), {\bf{}indexed}({\it{}Mc}, {\it{}sy\_symm}(), {\bf{}varidx}({\bf{}symbol}({\tt{}"ic"}),
                                                                                                 {\it{}get\_dim}()), {\bf{}varidx}({\bf{}symbol}({\tt{}"jc"}), {\it{}get\_dim}())));

\nwused{\\{NWppJ6t-DKmU5-4}\\{NWppJ6t-DKmU5-1Q}}\nwidentuses{\\{{\nwixident{cycle{\_}metric}}{cycle:unmetric}}\\{{\nwixident{ex}}{ex}}\\{{\nwixident{figure}}{figure}}\\{{\nwixident{get{\_}dim()}}{get:undim()}}\\{{\nwixident{op}}{op}}\\{{\nwixident{real{\_}line}}{real:unline}}}\nwindexuse{\nwixident{cycle{\_}metric}}{cycle:unmetric}{NWppJ6t-1wPO4G-3}\nwindexuse{\nwixident{ex}}{ex}{NWppJ6t-1wPO4G-3}\nwindexuse{\nwixident{figure}}{figure}{NWppJ6t-1wPO4G-3}\nwindexuse{\nwixident{get{\_}dim()}}{get:undim()}{NWppJ6t-1wPO4G-3}\nwindexuse{\nwixident{op}}{op}{NWppJ6t-1wPO4G-3}\nwindexuse{\nwixident{real{\_}line}}{real:unline}{NWppJ6t-1wPO4G-3}\nwendcode{}\nwbegindocs{846}Other types of metric.
\nwenddocs{}\nwbegincode{847}\sublabel{NWppJ6t-1wPO4G-4}\nwmargintag{{\nwtagstyle{}\subpageref{NWppJ6t-1wPO4G-4}}}\moddef{set cycle metric in figure~{\nwtagstyle{}\subpageref{NWppJ6t-1wPO4G-1}}}\plusendmoddef\Rm{}\nwstartdeflinemarkup\nwusesondefline{\\{NWppJ6t-DKmU5-4}\\{NWppJ6t-DKmU5-1Q}}\nwprevnextdefs{NWppJ6t-1wPO4G-3}{\relax}\nwenddeflinemarkup
    {\nwrbrace} {\bf{}else} {\bf{}if} ({\it{}is\_a}\begin{math}<\end{math}{\bf{}indexed}\begin{math}>\end{math}({\it{}Mc})) {\nwlbrace}
        {\it{}cycle\_metric} = {\it{}clifford\_unit}({\bf{}varidx}({\it{}real\_line}, {\it{}ex\_to}\begin{math}<\end{math}{\bf{}idx}\begin{math}>\end{math}({\it{}Mc}.{\it{}op}(1)).{\it{}get\_dim}()), {\it{}Mc});
    {\nwrbrace} {\bf{}else} {\bf{}if} ({\it{}is\_a}\begin{math}<\end{math}{\bf{}lst}\begin{math}>\end{math}({\it{}Mc})) {\nwlbrace}
        {\it{}cycle\_metric}={\it{}clifford\_unit}({\bf{}varidx}({\it{}real\_line}, {\it{}Mc}.{\it{}nops}()), {\bf{}indexed}({\it{}diag\_matrix}({\it{}ex\_to}\begin{math}<\end{math}{\bf{}lst}\begin{math}>\end{math}({\it{}Mc})), {\it{}sy\_symm}(),
                            {\bf{}varidx}({\bf{}symbol}({\tt{}"ic"}), {\it{}Mc}.{\it{}nops}()), {\bf{}varidx}({\bf{}symbol}({\tt{}"jc"}), {\it{}Mc}.{\it{}nops}())));
    {\nwrbrace}

\nwused{\\{NWppJ6t-DKmU5-4}\\{NWppJ6t-DKmU5-1Q}}\nwidentuses{\\{{\nwixident{cycle{\_}metric}}{cycle:unmetric}}\\{{\nwixident{get{\_}dim()}}{get:undim()}}\\{{\nwixident{nops}}{nops}}\\{{\nwixident{op}}{op}}\\{{\nwixident{real{\_}line}}{real:unline}}}\nwindexuse{\nwixident{cycle{\_}metric}}{cycle:unmetric}{NWppJ6t-1wPO4G-4}\nwindexuse{\nwixident{get{\_}dim()}}{get:undim()}{NWppJ6t-1wPO4G-4}\nwindexuse{\nwixident{nops}}{nops}{NWppJ6t-1wPO4G-4}\nwindexuse{\nwixident{op}}{op}{NWppJ6t-1wPO4G-4}\nwindexuse{\nwixident{real{\_}line}}{real:unline}{NWppJ6t-1wPO4G-4}\nwendcode{}\nwbegindocs{848}The error message
\nwenddocs{}\nwbegincode{849}\sublabel{NWppJ6t-DKmU5-5}\nwmargintag{{\nwtagstyle{}\subpageref{NWppJ6t-DKmU5-5}}}\moddef{figure class~{\nwtagstyle{}\subpageref{NWppJ6t-DKmU5-1}}}\plusendmoddef\Rm{}\nwstartdeflinemarkup\nwusesondefline{\\{NWppJ6t-32jW6L-1}}\nwprevnextdefs{NWppJ6t-DKmU5-4}{NWppJ6t-DKmU5-6}\nwenddeflinemarkup
    {\bf{}else}
        {\bf{}throw}({\it{}std}::{\it{}invalid\_argument}({\tt{}"figure::figure(const ex &, const ex &):"}
                                    {\tt{}" the second parameter"}
                                    {\tt{}" shall be omitted, equal to zero "}
                                    {\tt{}" or be a lst, a metric (can be either tensor, matrix,"}
                                    {\tt{}" Clifford unit or indexed by two indices)"}));

\nwused{\\{NWppJ6t-32jW6L-1}}\nwidentuses{\\{{\nwixident{ex}}{ex}}\\{{\nwixident{figure}}{figure}}}\nwindexuse{\nwixident{ex}}{ex}{NWppJ6t-DKmU5-5}\nwindexuse{\nwixident{figure}}{figure}{NWppJ6t-DKmU5-5}\nwendcode{}\nwbegindocs{850}Finally we check that point and cycle metrics have the same dimensionality.
\nwenddocs{}\nwbegincode{851}\sublabel{NWppJ6t-DKmU5-6}\nwmargintag{{\nwtagstyle{}\subpageref{NWppJ6t-DKmU5-6}}}\moddef{figure class~{\nwtagstyle{}\subpageref{NWppJ6t-DKmU5-1}}}\plusendmoddef\Rm{}\nwstartdeflinemarkup\nwusesondefline{\\{NWppJ6t-32jW6L-1}}\nwprevnextdefs{NWppJ6t-DKmU5-5}{NWppJ6t-DKmU5-7}\nwenddeflinemarkup
    {\bf{}if} (\begin{math}\neg\end{math} ({\it{}get\_dim}()-{\it{}ex\_to}\begin{math}<\end{math}{\bf{}idx}\begin{math}>\end{math}({\it{}cycle\_metric}.{\it{}op}(1)).{\it{}get\_dim}()).{\it{}is\_zero}())
        {\bf{}throw}({\it{}std}::{\it{}invalid\_argument}({\tt{}"figure::figure(const ex &, const ex &):"}
                                    {\tt{}"the point and cycle metrics shall have "}
                                    {\tt{}"the same dimensions"}));

\nwused{\\{NWppJ6t-32jW6L-1}}\nwidentuses{\\{{\nwixident{cycle{\_}metric}}{cycle:unmetric}}\\{{\nwixident{ex}}{ex}}\\{{\nwixident{figure}}{figure}}\\{{\nwixident{get{\_}dim()}}{get:undim()}}\\{{\nwixident{op}}{op}}}\nwindexuse{\nwixident{cycle{\_}metric}}{cycle:unmetric}{NWppJ6t-DKmU5-6}\nwindexuse{\nwixident{ex}}{ex}{NWppJ6t-DKmU5-6}\nwindexuse{\nwixident{figure}}{figure}{NWppJ6t-DKmU5-6}\nwindexuse{\nwixident{get{\_}dim()}}{get:undim()}{NWppJ6t-DKmU5-6}\nwindexuse{\nwixident{op}}{op}{NWppJ6t-DKmU5-6}\nwendcode{}\nwbegindocs{852}We also check that {\Tt{}\Rm{}{\it{}point\_metric}\nwendquote} and {\Tt{}\Rm{}{\it{}cycle\_metric}\nwendquote} has the
same dimensionality.
\nwenddocs{}\nwbegincode{853}\sublabel{NWppJ6t-DKmU5-7}\nwmargintag{{\nwtagstyle{}\subpageref{NWppJ6t-DKmU5-7}}}\moddef{figure class~{\nwtagstyle{}\subpageref{NWppJ6t-DKmU5-1}}}\plusendmoddef\Rm{}\nwstartdeflinemarkup\nwusesondefline{\\{NWppJ6t-32jW6L-1}}\nwprevnextdefs{NWppJ6t-DKmU5-6}{NWppJ6t-DKmU5-8}\nwenddeflinemarkup
    \LA{}check dimensionalities point and cycle metrics~{\nwtagstyle{}\subpageref{NWppJ6t-11clwc-1}}\RA{}
    \LA{}add symbols to match dimensionality~{\nwtagstyle{}\subpageref{NWppJ6t-3QR99o-1}}\RA{}

\nwused{\\{NWppJ6t-32jW6L-1}}\nwendcode{}\nwbegindocs{854}\nwdocspar
\nwenddocs{}\nwbegincode{855}\sublabel{NWppJ6t-11clwc-1}\nwmargintag{{\nwtagstyle{}\subpageref{NWppJ6t-11clwc-1}}}\moddef{check dimensionalities point and cycle metrics~{\nwtagstyle{}\subpageref{NWppJ6t-11clwc-1}}}\endmoddef\Rm{}\nwstartdeflinemarkup\nwusesondefline{\\{NWppJ6t-DKmU5-7}\\{NWppJ6t-DKmU5-1Q}}\nwenddeflinemarkup
    {\bf{}if} (\begin{math}\neg\end{math}({\it{}get\_dim}()-{\it{}ex\_to}\begin{math}<\end{math}{\bf{}varidx}\begin{math}>\end{math}({\it{}cycle\_metric}.{\it{}op}(1)).{\it{}get\_dim}()).{\it{}is\_zero}())
        {\bf{}throw}({\it{}std}::{\it{}invalid\_argument}({\tt{}"Metrics for point and cycle spaces have"}
                                    {\tt{}" different dimensionalities!"}));

\nwused{\\{NWppJ6t-DKmU5-7}\\{NWppJ6t-DKmU5-1Q}}\nwidentuses{\\{{\nwixident{cycle{\_}metric}}{cycle:unmetric}}\\{{\nwixident{get{\_}dim()}}{get:undim()}}\\{{\nwixident{op}}{op}}}\nwindexuse{\nwixident{cycle{\_}metric}}{cycle:unmetric}{NWppJ6t-11clwc-1}\nwindexuse{\nwixident{get{\_}dim()}}{get:undim()}{NWppJ6t-11clwc-1}\nwindexuse{\nwixident{op}}{op}{NWppJ6t-11clwc-1}\nwendcode{}\nwbegindocs{856}We produce enough symbols to match dimensionality.
\nwenddocs{}\nwbegincode{857}\sublabel{NWppJ6t-3QR99o-1}\nwmargintag{{\nwtagstyle{}\subpageref{NWppJ6t-3QR99o-1}}}\moddef{add symbols to match dimensionality~{\nwtagstyle{}\subpageref{NWppJ6t-3QR99o-1}}}\endmoddef\Rm{}\nwstartdeflinemarkup\nwusesondefline{\\{NWppJ6t-DKmU5-7}\\{NWppJ6t-DKmU5-A}}\nwenddeflinemarkup
    {\bf{}int} {\it{}D};
    {\bf{}if} ({\it{}is\_a}\begin{math}<\end{math}{\bf{}numeric}\begin{math}>\end{math}({\it{}get\_dim}())) {\nwlbrace}
        {\it{}D}={\it{}ex\_to}\begin{math}<\end{math}{\bf{}numeric}\begin{math}>\end{math}({\it{}get\_dim}()).{\it{}to\_int}();
        {\bf{}char} {\it{}name}[6];
        {\bf{}for}({\bf{}int} {\it{}i}=0; {\it{}i}\begin{math}<\end{math}{\it{}D}; \protect\PP{\it{}i}) {\nwlbrace}
            {\it{}sprintf}({\it{}name}, {\tt{}"l
            {\it{}l}.{\it{}append}({\bf{}realsymbol}({\it{}name}));
        {\nwrbrace}
    {\nwrbrace}

\nwused{\\{NWppJ6t-DKmU5-7}\\{NWppJ6t-DKmU5-A}}\nwidentuses{\\{{\nwixident{get{\_}dim()}}{get:undim()}}\\{{\nwixident{l}}{l}}\\{{\nwixident{name}}{name}}\\{{\nwixident{numeric}}{numeric}}\\{{\nwixident{realsymbol}}{realsymbol}}}\nwindexuse{\nwixident{get{\_}dim()}}{get:undim()}{NWppJ6t-3QR99o-1}\nwindexuse{\nwixident{l}}{l}{NWppJ6t-3QR99o-1}\nwindexuse{\nwixident{name}}{name}{NWppJ6t-3QR99o-1}\nwindexuse{\nwixident{numeric}}{numeric}{NWppJ6t-3QR99o-1}\nwindexuse{\nwixident{realsymbol}}{realsymbol}{NWppJ6t-3QR99o-1}\nwendcode{}\nwbegindocs{858}Finally, we set-up two elements which present at any figure: the
real line and infinity.
\nwenddocs{}\nwbegincode{859}\sublabel{NWppJ6t-DKmU5-8}\nwmargintag{{\nwtagstyle{}\subpageref{NWppJ6t-DKmU5-8}}}\moddef{figure class~{\nwtagstyle{}\subpageref{NWppJ6t-DKmU5-1}}}\plusendmoddef\Rm{}\nwstartdeflinemarkup\nwusesondefline{\\{NWppJ6t-32jW6L-1}}\nwprevnextdefs{NWppJ6t-DKmU5-7}{NWppJ6t-DKmU5-9}\nwenddeflinemarkup
    \LA{}setup real line and infinity~{\nwtagstyle{}\subpageref{NWppJ6t-1NAwBA-1}}\RA{}
{\nwrbrace}

\nwused{\\{NWppJ6t-32jW6L-1}}\nwendcode{}\nwbegindocs{860}Finally, we supply nodes for the real line and the cycle at infinity.
\nwenddocs{}\nwbegincode{861}\sublabel{NWppJ6t-1NAwBA-1}\nwmargintag{{\nwtagstyle{}\subpageref{NWppJ6t-1NAwBA-1}}}\moddef{setup real line and infinity~{\nwtagstyle{}\subpageref{NWppJ6t-1NAwBA-1}}}\endmoddef\Rm{}\nwstartdeflinemarkup\nwusesondefline{\\{NWppJ6t-DKmU5-8}\\{NWppJ6t-DKmU5-A}}\nwenddeflinemarkup
    {\bf{}if} ({\it{}inf\_missing}) {\nwlbrace}
        \LA{}set the infinity~{\nwtagstyle{}\subpageref{NWppJ6t-2j51zU-1}}\RA{}
    {\nwrbrace}
    {\bf{}if} ({\it{}R\_missing}) {\nwlbrace}
        \LA{}initialise the dimension and vector~{\nwtagstyle{}\subpageref{NWppJ6t-2V7UoL-1}}\RA{}
        \LA{}set the real line~{\nwtagstyle{}\subpageref{NWppJ6t-2q1QAq-1}}\RA{}
    {\nwrbrace}

\nwused{\\{NWppJ6t-DKmU5-8}\\{NWppJ6t-DKmU5-A}}\nwendcode{}\nwbegindocs{862}\nwdocspar
\nwenddocs{}\nwbegincode{863}\sublabel{NWppJ6t-DKmU5-9}\nwmargintag{{\nwtagstyle{}\subpageref{NWppJ6t-DKmU5-9}}}\moddef{figure class~{\nwtagstyle{}\subpageref{NWppJ6t-DKmU5-1}}}\plusendmoddef\Rm{}\nwstartdeflinemarkup\nwusesondefline{\\{NWppJ6t-32jW6L-1}}\nwprevnextdefs{NWppJ6t-DKmU5-8}{NWppJ6t-DKmU5-A}\nwenddeflinemarkup
{\bf{}figure}::{\bf{}figure}({\bf{}const} {\bf{}ex} & {\it{}Mp}, {\bf{}const} {\bf{}ex} & {\it{}Mc}, {\bf{}const} {\it{}exhashmap}\begin{math}<\end{math}{\bf{}cycle\_node}\begin{math}>\end{math} & {\it{}N}):
              {\it{}inherited}(), {\it{}k}({\bf{}realsymbol}({\tt{}"k"})), {\it{}m}({\bf{}realsymbol}({\tt{}"m"})), {\it{}l}()
{\nwlbrace}
    {\it{}infinity}={\bf{}symbol}({\tt{}"infty"},{\tt{}"{\char92}{\char92}infty"});
    {\it{}real\_line}={\bf{}symbol}({\tt{}"R"},{\tt{}"{\char92}{\char92}mathbf{\char123}R{\char125}"});
    {\bf{}bool} {\it{}inf\_missing}={\bf{}true}, {\it{}R\_missing}={\bf{}true};
    {\bf{}if} ({\it{}is\_a}\begin{math}<\end{math}{\bf{}clifford}\begin{math}>\end{math}({\it{}Mp}) \begin{math}\wedge\end{math} {\it{}is\_a}\begin{math}<\end{math}{\bf{}clifford}\begin{math}>\end{math}({\it{}Mc})) {\nwlbrace}
        {\it{}point\_metric} = {\it{}Mp};
        {\it{}cycle\_metric} = {\it{}Mc};
    {\nwrbrace} {\bf{}else}
        {\bf{}throw}({\it{}std}::{\it{}invalid\_argument}({\tt{}"figure::figure(const ex &, const ex &, exhashmap<cycle\_node>):"}
                                    {\tt{}" the point\_metric and cycle\_metric should be clifford\_unit. "}));

\nwused{\\{NWppJ6t-32jW6L-1}}\nwidentuses{\\{{\nwixident{cycle{\_}metric}}{cycle:unmetric}}\\{{\nwixident{cycle{\_}node}}{cycle:unnode}}\\{{\nwixident{ex}}{ex}}\\{{\nwixident{figure}}{figure}}\\{{\nwixident{infinity}}{infinity}}\\{{\nwixident{k}}{k}}\\{{\nwixident{l}}{l}}\\{{\nwixident{m}}{m}}\\{{\nwixident{point{\_}metric}}{point:unmetric}}\\{{\nwixident{real{\_}line}}{real:unline}}\\{{\nwixident{realsymbol}}{realsymbol}}}\nwindexuse{\nwixident{cycle{\_}metric}}{cycle:unmetric}{NWppJ6t-DKmU5-9}\nwindexuse{\nwixident{cycle{\_}node}}{cycle:unnode}{NWppJ6t-DKmU5-9}\nwindexuse{\nwixident{ex}}{ex}{NWppJ6t-DKmU5-9}\nwindexuse{\nwixident{figure}}{figure}{NWppJ6t-DKmU5-9}\nwindexuse{\nwixident{infinity}}{infinity}{NWppJ6t-DKmU5-9}\nwindexuse{\nwixident{k}}{k}{NWppJ6t-DKmU5-9}\nwindexuse{\nwixident{l}}{l}{NWppJ6t-DKmU5-9}\nwindexuse{\nwixident{m}}{m}{NWppJ6t-DKmU5-9}\nwindexuse{\nwixident{point{\_}metric}}{point:unmetric}{NWppJ6t-DKmU5-9}\nwindexuse{\nwixident{real{\_}line}}{real:unline}{NWppJ6t-DKmU5-9}\nwindexuse{\nwixident{realsymbol}}{realsymbol}{NWppJ6t-DKmU5-9}\nwendcode{}\nwbegindocs{864}We coming nodes of cycle to the new figure.
\nwenddocs{}\nwbegincode{865}\sublabel{NWppJ6t-DKmU5-A}\nwmargintag{{\nwtagstyle{}\subpageref{NWppJ6t-DKmU5-A}}}\moddef{figure class~{\nwtagstyle{}\subpageref{NWppJ6t-DKmU5-1}}}\plusendmoddef\Rm{}\nwstartdeflinemarkup\nwusesondefline{\\{NWppJ6t-32jW6L-1}}\nwprevnextdefs{NWppJ6t-DKmU5-9}{NWppJ6t-DKmU5-B}\nwenddeflinemarkup
    {\bf{}for} ({\bf{}const} {\bf{}auto}& {\it{}x}: {\it{}N}) {\nwlbrace}
        {\it{}nodes}[{\it{}x}.{\it{}first}]={\it{}x}.{\it{}second};
        \LA{}identify infinity and real line~{\nwtagstyle{}\subpageref{NWppJ6t-1pIH3B-1}}\RA{}
    {\nwrbrace}
    \LA{}add symbols to match dimensionality~{\nwtagstyle{}\subpageref{NWppJ6t-3QR99o-1}}\RA{}
    \LA{}setup real line and infinity~{\nwtagstyle{}\subpageref{NWppJ6t-1NAwBA-1}}\RA{}
{\nwrbrace}

\nwused{\\{NWppJ6t-32jW6L-1}}\nwidentuses{\\{{\nwixident{nodes}}{nodes}}}\nwindexuse{\nwixident{nodes}}{nodes}{NWppJ6t-DKmU5-A}\nwendcode{}\nwbegindocs{866}This constructor reads a figure from a file given by name.
\nwenddocs{}\nwbegincode{867}\sublabel{NWppJ6t-DKmU5-B}\nwmargintag{{\nwtagstyle{}\subpageref{NWppJ6t-DKmU5-B}}}\moddef{figure class~{\nwtagstyle{}\subpageref{NWppJ6t-DKmU5-1}}}\plusendmoddef\Rm{}\nwstartdeflinemarkup\nwusesondefline{\\{NWppJ6t-32jW6L-1}}\nwprevnextdefs{NWppJ6t-DKmU5-A}{NWppJ6t-DKmU5-C}\nwenddeflinemarkup
{\bf{}figure}::{\bf{}figure}({\bf{}const} {\bf{}char}\begin{math}\ast\end{math} {\it{}file\_name}, {\it{}string} {\it{}fig\_name}) : {\it{}inherited}(), {\it{}k}({\bf{}realsymbol}({\tt{}"k"})), {\it{}m}({\bf{}realsymbol}({\tt{}"m"})), {\it{}l}()
{\nwlbrace}
    {\it{}infinity}={\bf{}symbol}({\tt{}"infty"},{\tt{}"{\char92}{\char92}infty"});
    {\it{}real\_line}={\bf{}symbol}({\tt{}"R"},{\tt{}"{\char92}{\char92}mathbf{\char123}R{\char125}"});
    \LA{}add gar extension~{\nwtagstyle{}\subpageref{NWppJ6t-35j28o-1}}\RA{}
    {\it{}GiNaC}::{\it{}archive} {\it{}A};
    {\it{}std}::{\it{}ifstream} {\it{}ifs}({\it{}fn}.{\it{}c\_str}(),  {\it{}std}::{\it{}ifstream}::{\it{}in});

    {\it{}ifs} \begin{math}\gg\end{math} {\it{}A};
    \begin{math}\ast\end{math}{\it{}this}={\it{}ex\_to}\begin{math}<\end{math}{\bf{}figure}\begin{math}>\end{math}({\it{}A}.{\it{}unarchive\_ex}({\bf{}lst}{\nwlbrace}{\it{}infinity}, {\it{}real\_line}{\nwrbrace}, {\it{}fig\_name}));

    {\bf{}if} ({\it{}FIGURE\_DEBUG}) {\nwlbrace}
        {\it{}fn}={\tt{}"raw-read-"}+{\it{}fn};
        {\it{}ofstream} {\it{}out1}({\it{}fn}.{\it{}c\_str}());
        {\it{}A}.{\it{}printraw}({\it{}out1});
        {\it{}out1}.{\it{}close}();
        {\it{}out1}.{\it{}flush}();
    {\nwrbrace}
{\nwrbrace}

\nwused{\\{NWppJ6t-32jW6L-1}}\nwidentuses{\\{{\nwixident{archive}}{archive}}\\{{\nwixident{figure}}{figure}}\\{{\nwixident{FIGURE{\_}DEBUG}}{FIGURE:unDEBUG}}\\{{\nwixident{infinity}}{infinity}}\\{{\nwixident{k}}{k}}\\{{\nwixident{l}}{l}}\\{{\nwixident{m}}{m}}\\{{\nwixident{real{\_}line}}{real:unline}}\\{{\nwixident{realsymbol}}{realsymbol}}}\nwindexuse{\nwixident{archive}}{archive}{NWppJ6t-DKmU5-B}\nwindexuse{\nwixident{figure}}{figure}{NWppJ6t-DKmU5-B}\nwindexuse{\nwixident{FIGURE{\_}DEBUG}}{FIGURE:unDEBUG}{NWppJ6t-DKmU5-B}\nwindexuse{\nwixident{infinity}}{infinity}{NWppJ6t-DKmU5-B}\nwindexuse{\nwixident{k}}{k}{NWppJ6t-DKmU5-B}\nwindexuse{\nwixident{l}}{l}{NWppJ6t-DKmU5-B}\nwindexuse{\nwixident{m}}{m}{NWppJ6t-DKmU5-B}\nwindexuse{\nwixident{real{\_}line}}{real:unline}{NWppJ6t-DKmU5-B}\nwindexuse{\nwixident{realsymbol}}{realsymbol}{NWppJ6t-DKmU5-B}\nwendcode{}\nwbegindocs{868}\texttt{.gar} is the standard extension for \GiNaC\ archive files.
\nwenddocs{}\nwbegincode{869}\sublabel{NWppJ6t-35j28o-1}\nwmargintag{{\nwtagstyle{}\subpageref{NWppJ6t-35j28o-1}}}\moddef{add gar extension~{\nwtagstyle{}\subpageref{NWppJ6t-35j28o-1}}}\endmoddef\Rm{}\nwstartdeflinemarkup\nwusesondefline{\\{NWppJ6t-DKmU5-B}\\{NWppJ6t-DKmU5-C}}\nwenddeflinemarkup
    {\it{}string} {\it{}fn}={\it{}file\_name};
    {\it{}size\_t} {\it{}found} = {\it{}fn}.{\it{}find}({\tt{}".gar"});
    {\bf{}if} ({\it{}found} \begin{math}\equiv\end{math} {\it{}std}::{\it{}string}::{\it{}npos})
        {\it{}fn}={\it{}fn}+{\tt{}".gar"};

    {\it{}cout} \begin{math}\ll\end{math} {\tt{}"use filename: "} \begin{math}\ll\end{math} {\it{}fn} \begin{math}\ll\end{math} {\it{}endl};

\nwused{\\{NWppJ6t-DKmU5-B}\\{NWppJ6t-DKmU5-C}}\nwendcode{}\nwbegindocs{870}This method saves the figure to a file, which can be read by the
above constructor.
\nwenddocs{}\nwbegincode{871}\sublabel{NWppJ6t-DKmU5-C}\nwmargintag{{\nwtagstyle{}\subpageref{NWppJ6t-DKmU5-C}}}\moddef{figure class~{\nwtagstyle{}\subpageref{NWppJ6t-DKmU5-1}}}\plusendmoddef\Rm{}\nwstartdeflinemarkup\nwusesondefline{\\{NWppJ6t-32jW6L-1}}\nwprevnextdefs{NWppJ6t-DKmU5-B}{NWppJ6t-DKmU5-D}\nwenddeflinemarkup
{\bf{}void} {\bf{}figure}::{\it{}save}({\bf{}const} {\bf{}char}\begin{math}\ast\end{math} {\it{}file\_name}, {\bf{}const} {\bf{}char} \begin{math}\ast\end{math} {\it{}fig\_name}) {\bf{}const}\nwindexdefn{\nwixident{figure}}{figure}{NWppJ6t-DKmU5-C}
{\nwlbrace}
    \LA{}add gar extension~{\nwtagstyle{}\subpageref{NWppJ6t-35j28o-1}}\RA{}
    {\it{}GiNaC}::{\it{}archive} {\it{}A};
    {\it{}A}.{\it{}archive\_ex}(\begin{math}\ast\end{math}{\it{}this}, {\it{}fig\_name});
    {\it{}ofstream} {\it{}out}({\it{}fn}.{\it{}c\_str}());
    {\it{}out} \begin{math}\ll\end{math} {\it{}A};
    {\it{}out}.{\it{}flush}();
    {\it{}out}.{\it{}close}();
    {\bf{}if} ({\it{}FIGURE\_DEBUG}) {\nwlbrace}
        {\it{}fn}={\tt{}"raw-save-"}+{\it{}fn};
        {\it{}ofstream} {\it{}out1}({\it{}fn}.{\it{}c\_str}());
        {\it{}A}.{\it{}printraw}({\it{}out1});
        {\it{}out1}.{\it{}close}();
        {\it{}out1}.{\it{}flush}();
    {\nwrbrace}
{\nwrbrace}

\nwused{\\{NWppJ6t-32jW6L-1}}\nwidentdefs{\\{{\nwixident{figure}}{figure}}}\nwidentuses{\\{{\nwixident{archive}}{archive}}\\{{\nwixident{FIGURE{\_}DEBUG}}{FIGURE:unDEBUG}}\\{{\nwixident{save}}{save}}}\nwindexuse{\nwixident{archive}}{archive}{NWppJ6t-DKmU5-C}\nwindexuse{\nwixident{FIGURE{\_}DEBUG}}{FIGURE:unDEBUG}{NWppJ6t-DKmU5-C}\nwindexuse{\nwixident{save}}{save}{NWppJ6t-DKmU5-C}\nwendcode{}\nwbegindocs{872}\nwdocspar
\subsubsection{Addition of new cycles}
\label{sec:addition-new-cycles}

\nwenddocs{}\nwbegindocs{873}This method is merely a wrapper for the second form below.
\nwenddocs{}\nwbegincode{874}\sublabel{NWppJ6t-DKmU5-D}\nwmargintag{{\nwtagstyle{}\subpageref{NWppJ6t-DKmU5-D}}}\moddef{figure class~{\nwtagstyle{}\subpageref{NWppJ6t-DKmU5-1}}}\plusendmoddef\Rm{}\nwstartdeflinemarkup\nwusesondefline{\\{NWppJ6t-32jW6L-1}}\nwprevnextdefs{NWppJ6t-DKmU5-C}{NWppJ6t-DKmU5-E}\nwenddeflinemarkup
{\bf{}ex} {\bf{}figure}::{\it{}add\_point}({\bf{}const} {\bf{}ex} & {\it{}x}, {\it{}string} {\it{}name}, {\it{}string} {\it{}TeXname})
{\nwlbrace}
    \LA{}auto TeX name~{\nwtagstyle{}\subpageref{NWppJ6t-2tM1q2-1}}\RA{}
    {\bf{}symbol} {\it{}key}({\it{}name}, {\it{}TeXname\_new});
    {\bf{}return} {\it{}add\_point}({\it{}x}, {\it{}key});
{\nwrbrace}
\nwindexdefn{\nwixident{add{\_}point}}{add:unpoint}{NWppJ6t-DKmU5-D}\eatline
\nwused{\\{NWppJ6t-32jW6L-1}}\nwidentdefs{\\{{\nwixident{add{\_}point}}{add:unpoint}}}\nwidentuses{\\{{\nwixident{ex}}{ex}}\\{{\nwixident{figure}}{figure}}\\{{\nwixident{key}}{key}}\\{{\nwixident{name}}{name}}\\{{\nwixident{TeXname}}{TeXname}}}\nwindexuse{\nwixident{ex}}{ex}{NWppJ6t-DKmU5-D}\nwindexuse{\nwixident{figure}}{figure}{NWppJ6t-DKmU5-D}\nwindexuse{\nwixident{key}}{key}{NWppJ6t-DKmU5-D}\nwindexuse{\nwixident{name}}{name}{NWppJ6t-DKmU5-D}\nwindexuse{\nwixident{TeXname}}{TeXname}{NWppJ6t-DKmU5-D}\nwendcode{}\nwbegindocs{875}\nwdocspar
\nwenddocs{}\nwbegindocs{876}We start from check of parameters.
\nwenddocs{}\nwbegincode{877}\sublabel{NWppJ6t-DKmU5-E}\nwmargintag{{\nwtagstyle{}\subpageref{NWppJ6t-DKmU5-E}}}\moddef{figure class~{\nwtagstyle{}\subpageref{NWppJ6t-DKmU5-1}}}\plusendmoddef\Rm{}\nwstartdeflinemarkup\nwusesondefline{\\{NWppJ6t-32jW6L-1}}\nwprevnextdefs{NWppJ6t-DKmU5-D}{NWppJ6t-DKmU5-F}\nwenddeflinemarkup
{\bf{}ex} {\bf{}figure}::{\it{}add\_point}({\bf{}const} {\bf{}ex} & {\it{}x}, {\bf{}const} {\bf{}ex} & {\it{}key})
{\nwlbrace}
    {\bf{}if} ({\it{}not} ({\it{}is\_a}\begin{math}<\end{math}{\bf{}lst}\begin{math}>\end{math}({\it{}x}) {\it{}and} ({\it{}x}.{\it{}nops}() \begin{math}\equiv\end{math} {\it{}get\_dim}())))
        {\bf{}throw}({\it{}std}::{\it{}invalid\_argument}({\tt{}"figure::add\_point(const ex &, const ex &): "}
                                    {\tt{}"coordinates of a point shall be a lst of the right lenght"}));

    {\bf{}if} ({\it{}not} {\it{}is\_a}\begin{math}<\end{math}{\bf{}symbol}\begin{math}>\end{math}({\it{}key}))
        {\bf{}throw}({\it{}std}::{\it{}invalid\_argument}({\tt{}"figure::add\_point(const ex &, const ex &): the third"}
                                    {\tt{}" argument need to be a point"}));
\LA{}adding point with its parents~{\nwtagstyle{}\subpageref{NWppJ6t-1Ub2Xq-1}}\RA{}
\nwindexdefn{\nwixident{add{\_}point}}{add:unpoint}{NWppJ6t-DKmU5-E}\eatline
\nwused{\\{NWppJ6t-32jW6L-1}}\nwidentdefs{\\{{\nwixident{add{\_}point}}{add:unpoint}}}\nwidentuses{\\{{\nwixident{ex}}{ex}}\\{{\nwixident{figure}}{figure}}\\{{\nwixident{get{\_}dim()}}{get:undim()}}\\{{\nwixident{key}}{key}}\\{{\nwixident{nops}}{nops}}}\nwindexuse{\nwixident{ex}}{ex}{NWppJ6t-DKmU5-E}\nwindexuse{\nwixident{figure}}{figure}{NWppJ6t-DKmU5-E}\nwindexuse{\nwixident{get{\_}dim()}}{get:undim()}{NWppJ6t-DKmU5-E}\nwindexuse{\nwixident{key}}{key}{NWppJ6t-DKmU5-E}\nwindexuse{\nwixident{nops}}{nops}{NWppJ6t-DKmU5-E}\nwendcode{}\nwbegindocs{878}\nwdocspar
\nwenddocs{}\nwbegindocs{879}This part of the code is shared with {\Tt{}\Rm{}{\it{}move\_point}()\nwendquote}. We create two
ghost parents for a point, since the parameters of the cycle
representing depend from the metric, thus it shall not be hard-coded
into the node, see also Section~\ref{sec:figures-as-families}.
\nwenddocs{}\nwbegincode{880}\sublabel{NWppJ6t-1Ub2Xq-1}\nwmargintag{{\nwtagstyle{}\subpageref{NWppJ6t-1Ub2Xq-1}}}\moddef{adding point with its parents~{\nwtagstyle{}\subpageref{NWppJ6t-1Ub2Xq-1}}}\endmoddef\Rm{}\nwstartdeflinemarkup\nwusesondefline{\\{NWppJ6t-DKmU5-E}\\{NWppJ6t-DKmU5-W}}\nwprevnextdefs{\relax}{NWppJ6t-1Ub2Xq-2}\nwenddeflinemarkup
    {\bf{}int} {\it{}dim}={\it{}x}.{\it{}nops}();
    {\bf{}lst} {\it{}l0}, {\it{}rels};
    {\it{}rels}.{\it{}append}({\bf{}cycle\_relation}({\it{}key},{\it{}cycle\_orthogonal},{\bf{}false}));
    {\it{}rels}.{\it{}append}({\bf{}cycle\_relation}({\it{}infinity},{\it{}cycle\_different}));

    {\bf{}for}({\bf{}int} {\it{}i}=0; {\it{}i} \begin{math}<\end{math} {\it{}dim}; \protect\PP{\it{}i})
        {\it{}l0}.{\it{}append}({\bf{}numeric}(0));

    {\bf{}for}({\bf{}int} {\it{}i}=0; {\it{}i} \begin{math}<\end{math} {\it{}dim}; \protect\PP{\it{}i}) {\nwlbrace}
        {\it{}l0}[{\it{}i}]={\bf{}numeric}(1);
        {\bf{}char} {\it{}name}[8];
        {\it{}sprintf}({\it{}name}, {\tt{}"-(
        {\bf{}symbol} {\it{}mother}({\it{}ex\_to}\begin{math}<\end{math}{\bf{}symbol}\begin{math}>\end{math}({\it{}key}).{\it{}get\_name}()+{\it{}name});
        {\it{}nodes}[{\it{}mother}]={\bf{}cycle\_node}({\bf{}cycle\_data}({\bf{}numeric}(0),{\bf{}indexed}({\bf{}matrix}(1, {\it{}dim}, {\it{}l0}),
                                                               {\bf{}varidx}({\it{}mother}, {\it{}get\_dim}())),{\bf{}numeric}(2)\begin{math}\ast\end{math}{\it{}x}.{\it{}op}({\it{}i})),
                                 {\it{}GHOST\_GEN}, {\bf{}lst}{\nwlbrace}{\nwrbrace}, {\bf{}lst}{\nwlbrace}{\it{}key}{\nwrbrace});
        {\it{}l0}[{\it{}i}]={\bf{}numeric}(0);
        {\it{}rels}.{\it{}append}({\bf{}cycle\_relation}({\it{}mother},{\it{}cycle\_orthogonal}));
    {\nwrbrace}

\nwalsodefined{\\{NWppJ6t-1Ub2Xq-2}}\nwused{\\{NWppJ6t-DKmU5-E}\\{NWppJ6t-DKmU5-W}}\nwidentuses{\\{{\nwixident{cycle{\_}data}}{cycle:undata}}\\{{\nwixident{cycle{\_}different}}{cycle:undifferent}}\\{{\nwixident{cycle{\_}node}}{cycle:unnode}}\\{{\nwixident{cycle{\_}orthogonal}}{cycle:unorthogonal}}\\{{\nwixident{cycle{\_}relation}}{cycle:unrelation}}\\{{\nwixident{get{\_}dim()}}{get:undim()}}\\{{\nwixident{GHOST{\_}GEN}}{GHOST:unGEN}}\\{{\nwixident{infinity}}{infinity}}\\{{\nwixident{key}}{key}}\\{{\nwixident{name}}{name}}\\{{\nwixident{nodes}}{nodes}}\\{{\nwixident{nops}}{nops}}\\{{\nwixident{numeric}}{numeric}}\\{{\nwixident{op}}{op}}}\nwindexuse{\nwixident{cycle{\_}data}}{cycle:undata}{NWppJ6t-1Ub2Xq-1}\nwindexuse{\nwixident{cycle{\_}different}}{cycle:undifferent}{NWppJ6t-1Ub2Xq-1}\nwindexuse{\nwixident{cycle{\_}node}}{cycle:unnode}{NWppJ6t-1Ub2Xq-1}\nwindexuse{\nwixident{cycle{\_}orthogonal}}{cycle:unorthogonal}{NWppJ6t-1Ub2Xq-1}\nwindexuse{\nwixident{cycle{\_}relation}}{cycle:unrelation}{NWppJ6t-1Ub2Xq-1}\nwindexuse{\nwixident{get{\_}dim()}}{get:undim()}{NWppJ6t-1Ub2Xq-1}\nwindexuse{\nwixident{GHOST{\_}GEN}}{GHOST:unGEN}{NWppJ6t-1Ub2Xq-1}\nwindexuse{\nwixident{infinity}}{infinity}{NWppJ6t-1Ub2Xq-1}\nwindexuse{\nwixident{key}}{key}{NWppJ6t-1Ub2Xq-1}\nwindexuse{\nwixident{name}}{name}{NWppJ6t-1Ub2Xq-1}\nwindexuse{\nwixident{nodes}}{nodes}{NWppJ6t-1Ub2Xq-1}\nwindexuse{\nwixident{nops}}{nops}{NWppJ6t-1Ub2Xq-1}\nwindexuse{\nwixident{numeric}}{numeric}{NWppJ6t-1Ub2Xq-1}\nwindexuse{\nwixident{op}}{op}{NWppJ6t-1Ub2Xq-1}\nwendcode{}\nwbegindocs{881}We add relations to parents which define this point. All relations
are given in {\Tt{}\Rm{}{\it{}cycle\_metric}\nwendquote}, only self-orthogonality is given in
terms of {\Tt{}\Rm{}{\it{}point\_metric}\nwendquote}. This is done in sake of the parabolic point
space.
\nwenddocs{}\nwbegincode{882}\sublabel{NWppJ6t-1Ub2Xq-2}\nwmargintag{{\nwtagstyle{}\subpageref{NWppJ6t-1Ub2Xq-2}}}\moddef{adding point with its parents~{\nwtagstyle{}\subpageref{NWppJ6t-1Ub2Xq-1}}}\plusendmoddef\Rm{}\nwstartdeflinemarkup\nwusesondefline{\\{NWppJ6t-DKmU5-E}\\{NWppJ6t-DKmU5-W}}\nwprevnextdefs{NWppJ6t-1Ub2Xq-1}{\relax}\nwenddeflinemarkup
    {\it{}nodes}[{\it{}key}]={\bf{}cycle\_node}({\bf{}cycle\_data}(), 0, {\it{}rels});

\nwused{\\{NWppJ6t-DKmU5-E}\\{NWppJ6t-DKmU5-W}}\nwidentuses{\\{{\nwixident{cycle{\_}data}}{cycle:undata}}\\{{\nwixident{cycle{\_}node}}{cycle:unnode}}\\{{\nwixident{key}}{key}}\\{{\nwixident{nodes}}{nodes}}}\nwindexuse{\nwixident{cycle{\_}data}}{cycle:undata}{NWppJ6t-1Ub2Xq-2}\nwindexuse{\nwixident{cycle{\_}node}}{cycle:unnode}{NWppJ6t-1Ub2Xq-2}\nwindexuse{\nwixident{key}}{key}{NWppJ6t-1Ub2Xq-2}\nwindexuse{\nwixident{nodes}}{nodes}{NWppJ6t-1Ub2Xq-2}\nwendcode{}\nwbegindocs{883}Now, cycle date shall be generated.
\nwenddocs{}\nwbegincode{884}\sublabel{NWppJ6t-DKmU5-F}\nwmargintag{{\nwtagstyle{}\subpageref{NWppJ6t-DKmU5-F}}}\moddef{figure class~{\nwtagstyle{}\subpageref{NWppJ6t-DKmU5-1}}}\plusendmoddef\Rm{}\nwstartdeflinemarkup\nwusesondefline{\\{NWppJ6t-32jW6L-1}}\nwprevnextdefs{NWppJ6t-DKmU5-E}{NWppJ6t-DKmU5-G}\nwenddeflinemarkup
    {\bf{}if} (\begin{math}\neg\end{math} {\it{}info}({\it{}status\_flags}::{\it{}expanded}))
        {\it{}nodes}[{\it{}key}].{\it{}set\_cycles}({\it{}ex\_to}\begin{math}<\end{math}{\bf{}lst}\begin{math}>\end{math}({\it{}update\_cycle\_node}({\it{}key})));
    {\bf{}if} ({\it{}FIGURE\_DEBUG})
        {\it{}cerr} \begin{math}\ll\end{math} {\tt{}"Add the point: "} \begin{math}\ll\end{math} {\it{}x} \begin{math}\ll\end{math} {\tt{}" as the cycle: "} \begin{math}\ll\end{math} {\it{}nodes}[{\it{}key}] \begin{math}\ll\end{math} {\it{}endl};
    {\bf{}return} {\it{}key};
{\nwrbrace}

\nwused{\\{NWppJ6t-32jW6L-1}}\nwidentuses{\\{{\nwixident{FIGURE{\_}DEBUG}}{FIGURE:unDEBUG}}\\{{\nwixident{info}}{info}}\\{{\nwixident{key}}{key}}\\{{\nwixident{nodes}}{nodes}}\\{{\nwixident{update{\_}cycle{\_}node}}{update:uncycle:unnode}}}\nwindexuse{\nwixident{FIGURE{\_}DEBUG}}{FIGURE:unDEBUG}{NWppJ6t-DKmU5-F}\nwindexuse{\nwixident{info}}{info}{NWppJ6t-DKmU5-F}\nwindexuse{\nwixident{key}}{key}{NWppJ6t-DKmU5-F}\nwindexuse{\nwixident{nodes}}{nodes}{NWppJ6t-DKmU5-F}\nwindexuse{\nwixident{update{\_}cycle{\_}node}}{update:uncycle:unnode}{NWppJ6t-DKmU5-F}\nwendcode{}\nwbegindocs{885}Add a cycle at zero level with a prescribed data.
\nwenddocs{}\nwbegincode{886}\sublabel{NWppJ6t-DKmU5-G}\nwmargintag{{\nwtagstyle{}\subpageref{NWppJ6t-DKmU5-G}}}\moddef{figure class~{\nwtagstyle{}\subpageref{NWppJ6t-DKmU5-1}}}\plusendmoddef\Rm{}\nwstartdeflinemarkup\nwusesondefline{\\{NWppJ6t-32jW6L-1}}\nwprevnextdefs{NWppJ6t-DKmU5-F}{NWppJ6t-DKmU5-H}\nwenddeflinemarkup
{\bf{}ex} {\bf{}figure}::{\it{}add\_cycle}({\bf{}const} {\bf{}ex} & {\it{}C}, {\bf{}const} {\bf{}ex} & {\it{}key})
{\nwlbrace}
    {\bf{}ex} {\it{}lC}={\it{}ex\_to}\begin{math}<\end{math}{\bf{}cycle}\begin{math}>\end{math}({\it{}C}).{\it{}get\_l}();
    {\bf{}if} ({\it{}is\_a}\begin{math}<\end{math}{\bf{}indexed}\begin{math}>\end{math}({\it{}lC}))
        {\it{}nodes}[{\it{}key}]={\bf{}cycle\_node}({\it{}C}.{\it{}subs}({\it{}lC}.{\it{}op}(1)\begin{math}\equiv\end{math}{\it{}key}));
    {\bf{}else}
        {\it{}nodes}[{\it{}key}]={\bf{}cycle\_node}({\it{}C});
    {\bf{}if} ({\it{}FIGURE\_DEBUG})
        {\it{}cerr} \begin{math}\ll\end{math} {\tt{}"Add the cycle: "} \begin{math}\ll\end{math} {\it{}nodes}[{\it{}key}] \begin{math}\ll\end{math} {\it{}endl};
    {\bf{}return} {\it{}key};
{\nwrbrace}
\nwindexdefn{\nwixident{add{\_}cycle}}{add:uncycle}{NWppJ6t-DKmU5-G}\eatline
\nwused{\\{NWppJ6t-32jW6L-1}}\nwidentdefs{\\{{\nwixident{add{\_}cycle}}{add:uncycle}}}\nwidentuses{\\{{\nwixident{cycle{\_}node}}{cycle:unnode}}\\{{\nwixident{ex}}{ex}}\\{{\nwixident{figure}}{figure}}\\{{\nwixident{FIGURE{\_}DEBUG}}{FIGURE:unDEBUG}}\\{{\nwixident{key}}{key}}\\{{\nwixident{nodes}}{nodes}}\\{{\nwixident{op}}{op}}\\{{\nwixident{subs}}{subs}}}\nwindexuse{\nwixident{cycle{\_}node}}{cycle:unnode}{NWppJ6t-DKmU5-G}\nwindexuse{\nwixident{ex}}{ex}{NWppJ6t-DKmU5-G}\nwindexuse{\nwixident{figure}}{figure}{NWppJ6t-DKmU5-G}\nwindexuse{\nwixident{FIGURE{\_}DEBUG}}{FIGURE:unDEBUG}{NWppJ6t-DKmU5-G}\nwindexuse{\nwixident{key}}{key}{NWppJ6t-DKmU5-G}\nwindexuse{\nwixident{nodes}}{nodes}{NWppJ6t-DKmU5-G}\nwindexuse{\nwixident{op}}{op}{NWppJ6t-DKmU5-G}\nwindexuse{\nwixident{subs}}{subs}{NWppJ6t-DKmU5-G}\nwendcode{}\nwbegindocs{887}\nwdocspar
\nwenddocs{}\nwbegindocs{888}Add a cycle at zero level with a prescribed data.
\nwenddocs{}\nwbegincode{889}\sublabel{NWppJ6t-DKmU5-H}\nwmargintag{{\nwtagstyle{}\subpageref{NWppJ6t-DKmU5-H}}}\moddef{figure class~{\nwtagstyle{}\subpageref{NWppJ6t-DKmU5-1}}}\plusendmoddef\Rm{}\nwstartdeflinemarkup\nwusesondefline{\\{NWppJ6t-32jW6L-1}}\nwprevnextdefs{NWppJ6t-DKmU5-G}{NWppJ6t-DKmU5-I}\nwenddeflinemarkup
{\bf{}ex} {\bf{}figure}::{\it{}add\_cycle}({\bf{}const} {\bf{}ex} & {\it{}C}, {\it{}string} {\it{}name}, {\it{}string} {\it{}TeXname})
{\nwlbrace}
    \LA{}auto TeX name~{\nwtagstyle{}\subpageref{NWppJ6t-2tM1q2-1}}\RA{}
    {\bf{}symbol} {\it{}key}({\it{}name}, {\it{}TeXname\_new});
    {\bf{}return} {\it{}add\_cycle}({\it{}C}, {\it{}key});
 {\nwrbrace}

\nwused{\\{NWppJ6t-32jW6L-1}}\nwidentuses{\\{{\nwixident{add{\_}cycle}}{add:uncycle}}\\{{\nwixident{ex}}{ex}}\\{{\nwixident{figure}}{figure}}\\{{\nwixident{key}}{key}}\\{{\nwixident{name}}{name}}\\{{\nwixident{TeXname}}{TeXname}}}\nwindexuse{\nwixident{add{\_}cycle}}{add:uncycle}{NWppJ6t-DKmU5-H}\nwindexuse{\nwixident{ex}}{ex}{NWppJ6t-DKmU5-H}\nwindexuse{\nwixident{figure}}{figure}{NWppJ6t-DKmU5-H}\nwindexuse{\nwixident{key}}{key}{NWppJ6t-DKmU5-H}\nwindexuse{\nwixident{name}}{name}{NWppJ6t-DKmU5-H}\nwindexuse{\nwixident{TeXname}}{TeXname}{NWppJ6t-DKmU5-H}\nwendcode{}\nwbegindocs{890}\nwdocspar
\nwenddocs{}\nwbegincode{891}\sublabel{NWppJ6t-DKmU5-I}\nwmargintag{{\nwtagstyle{}\subpageref{NWppJ6t-DKmU5-I}}}\moddef{figure class~{\nwtagstyle{}\subpageref{NWppJ6t-DKmU5-1}}}\plusendmoddef\Rm{}\nwstartdeflinemarkup\nwusesondefline{\\{NWppJ6t-32jW6L-1}}\nwprevnextdefs{NWppJ6t-DKmU5-H}{NWppJ6t-DKmU5-J}\nwenddeflinemarkup
{\bf{}void} {\bf{}figure}::{\it{}set\_cycle}({\bf{}const} {\bf{}ex} & {\it{}key}, {\bf{}const} {\bf{}ex} & {\it{}C})\nwindexdefn{\nwixident{figure}}{figure}{NWppJ6t-DKmU5-I}
{\nwlbrace}
    {\bf{}if} ({\it{}nodes}.{\it{}find}({\it{}key}) \begin{math}\equiv\end{math} {\it{}nodes}.{\it{}end}())
        {\bf{}throw}({\it{}std}::{\it{}invalid\_argument}("{\bf{}figure}::{\it{}set\_cycle}(): {\it{}there} {\it{}is} {\it{}no} {\it{}node} {\it{}wi}\begin{math}\backslash\end{math}
{\it{}th} {\it{}the} {\it{}key} {\it{}given}"));

    {\bf{}if} ({\it{}nodes}[{\it{}key}].{\it{}get\_parents}().{\it{}nops}() \begin{math}>\end{math} 0)
        {\bf{}throw}({\it{}std}::{\it{}invalid\_argument}("{\bf{}figure}::{\it{}set\_cycle}(): {\it{}cannot} {\it{}modify} {\it{}data} \begin{math}\backslash\end{math}
{\it{}of} {\it{}a} {\bf{}cycle} {\it{}with} {\it{}parents}"));

    {\it{}nodes}[{\it{}key}].{\it{}set\_cycles}({\it{}C});

    {\bf{}if} ({\it{}FIGURE\_DEBUG})
        {\it{}cerr} \begin{math}\ll\end{math} {\tt{}"Replace the cycle: "} \begin{math}\ll\end{math} {\it{}nodes}[{\it{}key}] \begin{math}\ll\end{math} {\it{}endl};
{\nwrbrace}
\nwindexdefn{\nwixident{set{\_}cycle}}{set:uncycle}{NWppJ6t-DKmU5-I}\eatline
\nwused{\\{NWppJ6t-32jW6L-1}}\nwidentdefs{\\{{\nwixident{figure}}{figure}}\\{{\nwixident{set{\_}cycle}}{set:uncycle}}}\nwidentuses{\\{{\nwixident{ex}}{ex}}\\{{\nwixident{FIGURE{\_}DEBUG}}{FIGURE:unDEBUG}}\\{{\nwixident{key}}{key}}\\{{\nwixident{nodes}}{nodes}}\\{{\nwixident{nops}}{nops}}}\nwindexuse{\nwixident{ex}}{ex}{NWppJ6t-DKmU5-I}\nwindexuse{\nwixident{FIGURE{\_}DEBUG}}{FIGURE:unDEBUG}{NWppJ6t-DKmU5-I}\nwindexuse{\nwixident{key}}{key}{NWppJ6t-DKmU5-I}\nwindexuse{\nwixident{nodes}}{nodes}{NWppJ6t-DKmU5-I}\nwindexuse{\nwixident{nops}}{nops}{NWppJ6t-DKmU5-I}\nwendcode{}\nwbegindocs{892}\nwdocspar
\nwenddocs{}\nwbegindocs{893}\nwdocspar
\nwenddocs{}\nwbegincode{894}\sublabel{NWppJ6t-DKmU5-J}\nwmargintag{{\nwtagstyle{}\subpageref{NWppJ6t-DKmU5-J}}}\moddef{figure class~{\nwtagstyle{}\subpageref{NWppJ6t-DKmU5-1}}}\plusendmoddef\Rm{}\nwstartdeflinemarkup\nwusesondefline{\\{NWppJ6t-32jW6L-1}}\nwprevnextdefs{NWppJ6t-DKmU5-I}{NWppJ6t-DKmU5-K}\nwenddeflinemarkup
{\bf{}void} {\bf{}figure}::{\it{}move\_cycle}({\bf{}const} {\bf{}ex} & {\it{}key}, {\bf{}const} {\bf{}ex} & {\it{}C})\nwindexdefn{\nwixident{figure}}{figure}{NWppJ6t-DKmU5-J}
{\nwlbrace}
    {\bf{}if} ({\it{}nodes}.{\it{}find}({\it{}key}) \begin{math}\equiv\end{math} {\it{}nodes}.{\it{}end}())
        {\bf{}throw}({\it{}std}::{\it{}invalid\_argument}({\tt{}"figure::set\_cycle(): there is no node with the key given"}));

    {\bf{}if} ({\it{}nodes}[{\it{}key}].{\it{}get\_generation}() \begin{math}\neq\end{math} 0)
        {\bf{}throw}({\it{}std}::{\it{}invalid\_argument}({\tt{}"figure::set\_cycle(): cannot modify data of a cycle in"}
                                    {\tt{}" non-zero generation"}));
\nwindexdefn{\nwixident{move{\_}cycle}}{move:uncycle}{NWppJ6t-DKmU5-J}\eatline
\nwused{\\{NWppJ6t-32jW6L-1}}\nwidentdefs{\\{{\nwixident{figure}}{figure}}\\{{\nwixident{move{\_}cycle}}{move:uncycle}}}\nwidentuses{\\{{\nwixident{ex}}{ex}}\\{{\nwixident{get{\_}generation}}{get:ungeneration}}\\{{\nwixident{key}}{key}}\\{{\nwixident{nodes}}{nodes}}\\{{\nwixident{set{\_}cycle}}{set:uncycle}}}\nwindexuse{\nwixident{ex}}{ex}{NWppJ6t-DKmU5-J}\nwindexuse{\nwixident{get{\_}generation}}{get:ungeneration}{NWppJ6t-DKmU5-J}\nwindexuse{\nwixident{key}}{key}{NWppJ6t-DKmU5-J}\nwindexuse{\nwixident{nodes}}{nodes}{NWppJ6t-DKmU5-J}\nwindexuse{\nwixident{set{\_}cycle}}{set:uncycle}{NWppJ6t-DKmU5-J}\nwendcode{}\nwbegindocs{895}\nwdocspar
\nwenddocs{}\nwbegindocs{896}If we have at zero generation with parents, then they are ghost
parents of the point, so shall be deleted. We cannot do this by
{\Tt{}\Rm{}{\it{}remove\_cycle\_node}\nwendquote} since we do not want to remove all its grand childrens.
\nwenddocs{}\nwbegincode{897}\sublabel{NWppJ6t-DKmU5-K}\nwmargintag{{\nwtagstyle{}\subpageref{NWppJ6t-DKmU5-K}}}\moddef{figure class~{\nwtagstyle{}\subpageref{NWppJ6t-DKmU5-1}}}\plusendmoddef\Rm{}\nwstartdeflinemarkup\nwusesondefline{\\{NWppJ6t-32jW6L-1}}\nwprevnextdefs{NWppJ6t-DKmU5-J}{NWppJ6t-DKmU5-L}\nwenddeflinemarkup
    {\bf{}if} ({\it{}nodes}[{\it{}key}].{\it{}get\_parents}().{\it{}nops}() \begin{math}>\end{math} 0) {\nwlbrace}
        {\bf{}lst} {\it{}par}={\it{}nodes}[{\it{}key}].{\it{}get\_parent\_keys}();
        {\bf{}for}({\bf{}const} {\bf{}auto}& {\it{}it} : {\it{}par})
            {\bf{}if} ({\it{}nodes}[{\it{}it}].{\it{}get\_generation}() \begin{math}\equiv\end{math} {\it{}GHOST\_GEN})
                {\it{}nodes}.{\it{}erase}({\it{}it});
            {\bf{}else}
                {\it{}nodes}[{\it{}it}].{\it{}remove\_child}({\it{}key});
    {\nwrbrace}
    {\it{}nodes}[{\it{}key}].{\it{}parents}={\bf{}lst}{\nwlbrace}{\nwrbrace};

\nwused{\\{NWppJ6t-32jW6L-1}}\nwidentuses{\\{{\nwixident{get{\_}generation}}{get:ungeneration}}\\{{\nwixident{GHOST{\_}GEN}}{GHOST:unGEN}}\\{{\nwixident{key}}{key}}\\{{\nwixident{nodes}}{nodes}}\\{{\nwixident{nops}}{nops}}}\nwindexuse{\nwixident{get{\_}generation}}{get:ungeneration}{NWppJ6t-DKmU5-K}\nwindexuse{\nwixident{GHOST{\_}GEN}}{GHOST:unGEN}{NWppJ6t-DKmU5-K}\nwindexuse{\nwixident{key}}{key}{NWppJ6t-DKmU5-K}\nwindexuse{\nwixident{nodes}}{nodes}{NWppJ6t-DKmU5-K}\nwindexuse{\nwixident{nops}}{nops}{NWppJ6t-DKmU5-K}\nwendcode{}\nwbegindocs{898}Now, the cycle may be set.
\nwenddocs{}\nwbegincode{899}\sublabel{NWppJ6t-DKmU5-L}\nwmargintag{{\nwtagstyle{}\subpageref{NWppJ6t-DKmU5-L}}}\moddef{figure class~{\nwtagstyle{}\subpageref{NWppJ6t-DKmU5-1}}}\plusendmoddef\Rm{}\nwstartdeflinemarkup\nwusesondefline{\\{NWppJ6t-32jW6L-1}}\nwprevnextdefs{NWppJ6t-DKmU5-K}{NWppJ6t-DKmU5-M}\nwenddeflinemarkup
    {\it{}nodes}[{\it{}key}].{\it{}set\_cycles}({\it{}C});
    {\it{}update\_node\_lst}({\it{}nodes}[{\it{}key}].{\it{}get\_children}());

    {\bf{}if} ({\it{}FIGURE\_DEBUG})
        {\it{}cerr} \begin{math}\ll\end{math} {\tt{}"Replace the cycle: "} \begin{math}\ll\end{math} {\it{}nodes}[{\it{}key}] \begin{math}\ll\end{math} {\it{}endl};
{\nwrbrace}

\nwused{\\{NWppJ6t-32jW6L-1}}\nwidentuses{\\{{\nwixident{FIGURE{\_}DEBUG}}{FIGURE:unDEBUG}}\\{{\nwixident{key}}{key}}\\{{\nwixident{nodes}}{nodes}}\\{{\nwixident{update{\_}node{\_}lst}}{update:unnode:unlst}}}\nwindexuse{\nwixident{FIGURE{\_}DEBUG}}{FIGURE:unDEBUG}{NWppJ6t-DKmU5-L}\nwindexuse{\nwixident{key}}{key}{NWppJ6t-DKmU5-L}\nwindexuse{\nwixident{nodes}}{nodes}{NWppJ6t-DKmU5-L}\nwindexuse{\nwixident{update{\_}node{\_}lst}}{update:unnode:unlst}{NWppJ6t-DKmU5-L}\nwendcode{}\nwbegindocs{900}A cycle can be added by a single  {\Tt{}\Rm{}{\bf{}cycle\_relation}\nwendquote} or a {\Tt{}\Rm{}{\bf{}lst}\nwendquote} of
{\Tt{}\Rm{}{\bf{}cycle\_relation}\nwendquote}, but this is just a wrapper for a more general case
below.
\nwenddocs{}\nwbegincode{901}\sublabel{NWppJ6t-DKmU5-M}\nwmargintag{{\nwtagstyle{}\subpageref{NWppJ6t-DKmU5-M}}}\moddef{figure class~{\nwtagstyle{}\subpageref{NWppJ6t-DKmU5-1}}}\plusendmoddef\Rm{}\nwstartdeflinemarkup\nwusesondefline{\\{NWppJ6t-32jW6L-1}}\nwprevnextdefs{NWppJ6t-DKmU5-L}{NWppJ6t-DKmU5-N}\nwenddeflinemarkup
{\bf{}ex} {\bf{}figure}::{\it{}add\_cycle\_rel}({\bf{}const} {\bf{}ex} & {\it{}rel}, {\bf{}const} {\bf{}ex} & {\it{}key}) {\nwlbrace}
    {\bf{}if} ({\it{}is\_a}\begin{math}<\end{math}{\bf{}cycle\_relation}\begin{math}>\end{math}({\it{}rel}))
        {\bf{}return} {\it{}add\_cycle\_rel}({\bf{}lst}{\nwlbrace}{\it{}rel}{\nwrbrace}, {\it{}key});
    {\bf{}else}
        {\bf{}throw}({\it{}std}::{\it{}invalid\_argument}({\tt{}"figure::add\_cycle\_rel: a cycle shall be added "}
                                    {\tt{}"by a single expression, which is a cycle\_relation"}));
{\nwrbrace}
\nwindexdefn{\nwixident{add{\_}cycle{\_}rel}}{add:uncycle:unrel}{NWppJ6t-DKmU5-M}\eatline
\nwused{\\{NWppJ6t-32jW6L-1}}\nwidentdefs{\\{{\nwixident{add{\_}cycle{\_}rel}}{add:uncycle:unrel}}}\nwidentuses{\\{{\nwixident{cycle{\_}relation}}{cycle:unrelation}}\\{{\nwixident{ex}}{ex}}\\{{\nwixident{figure}}{figure}}\\{{\nwixident{key}}{key}}}\nwindexuse{\nwixident{cycle{\_}relation}}{cycle:unrelation}{NWppJ6t-DKmU5-M}\nwindexuse{\nwixident{ex}}{ex}{NWppJ6t-DKmU5-M}\nwindexuse{\nwixident{figure}}{figure}{NWppJ6t-DKmU5-M}\nwindexuse{\nwixident{key}}{key}{NWppJ6t-DKmU5-M}\nwendcode{}\nwbegindocs{902}\nwdocspar
\nwenddocs{}\nwbegindocs{903}And now we add a cycle defined the list of relations. The generation
of the new cycle is calculated by the rules described in
Sec.~\ref{sec:figures-as-families}.
\nwenddocs{}\nwbegincode{904}\sublabel{NWppJ6t-DKmU5-N}\nwmargintag{{\nwtagstyle{}\subpageref{NWppJ6t-DKmU5-N}}}\moddef{figure class~{\nwtagstyle{}\subpageref{NWppJ6t-DKmU5-1}}}\plusendmoddef\Rm{}\nwstartdeflinemarkup\nwusesondefline{\\{NWppJ6t-32jW6L-1}}\nwprevnextdefs{NWppJ6t-DKmU5-M}{NWppJ6t-DKmU5-O}\nwenddeflinemarkup
{\bf{}ex} {\bf{}figure}::{\it{}add\_cycle\_rel}({\bf{}const} {\bf{}lst} & {\it{}rel}, {\bf{}const} {\bf{}ex} & {\it{}key})
{\nwlbrace}
    {\bf{}lst} {\it{}cond};
    {\bf{}int} {\it{}gen}=0;

    {\bf{}for}({\bf{}const} {\bf{}auto}& {\it{}it} : {\it{}rel}) {\nwlbrace}
        {\bf{}if} ({\it{}ex\_to}\begin{math}<\end{math}{\bf{}cycle\_relation}\begin{math}>\end{math}({\it{}it}).{\it{}get\_parkey}() \begin{math}\neq\end{math} {\it{}key})
            {\it{}gen}={\it{}max}({\it{}gen}, {\it{}nodes}[{\it{}ex\_to}\begin{math}<\end{math}{\bf{}cycle\_relation}\begin{math}>\end{math}({\it{}it}).{\it{}get\_parkey}()].{\it{}get\_generation}());
        {\it{}nodes}[{\it{}ex\_to}\begin{math}<\end{math}{\bf{}cycle\_relation}\begin{math}>\end{math}({\it{}it}).{\it{}get\_parkey}()].{\it{}add\_child}({\it{}key});
    {\nwrbrace}

    {\it{}nodes}[{\it{}key}]={\bf{}cycle\_node}({\bf{}cycle\_data}(),{\it{}gen}+1,{\it{}rel});

\nwused{\\{NWppJ6t-32jW6L-1}}\nwidentuses{\\{{\nwixident{add{\_}cycle{\_}rel}}{add:uncycle:unrel}}\\{{\nwixident{cycle{\_}data}}{cycle:undata}}\\{{\nwixident{cycle{\_}node}}{cycle:unnode}}\\{{\nwixident{cycle{\_}relation}}{cycle:unrelation}}\\{{\nwixident{ex}}{ex}}\\{{\nwixident{figure}}{figure}}\\{{\nwixident{get{\_}generation}}{get:ungeneration}}\\{{\nwixident{key}}{key}}\\{{\nwixident{nodes}}{nodes}}}\nwindexuse{\nwixident{add{\_}cycle{\_}rel}}{add:uncycle:unrel}{NWppJ6t-DKmU5-N}\nwindexuse{\nwixident{cycle{\_}data}}{cycle:undata}{NWppJ6t-DKmU5-N}\nwindexuse{\nwixident{cycle{\_}node}}{cycle:unnode}{NWppJ6t-DKmU5-N}\nwindexuse{\nwixident{cycle{\_}relation}}{cycle:unrelation}{NWppJ6t-DKmU5-N}\nwindexuse{\nwixident{ex}}{ex}{NWppJ6t-DKmU5-N}\nwindexuse{\nwixident{figure}}{figure}{NWppJ6t-DKmU5-N}\nwindexuse{\nwixident{get{\_}generation}}{get:ungeneration}{NWppJ6t-DKmU5-N}\nwindexuse{\nwixident{key}}{key}{NWppJ6t-DKmU5-N}\nwindexuse{\nwixident{nodes}}{nodes}{NWppJ6t-DKmU5-N}\nwendcode{}\nwbegindocs{905}\nwdocspar
\nwenddocs{}\nwbegincode{906}\sublabel{NWppJ6t-DKmU5-O}\nwmargintag{{\nwtagstyle{}\subpageref{NWppJ6t-DKmU5-O}}}\moddef{figure class~{\nwtagstyle{}\subpageref{NWppJ6t-DKmU5-1}}}\plusendmoddef\Rm{}\nwstartdeflinemarkup\nwusesondefline{\\{NWppJ6t-32jW6L-1}}\nwprevnextdefs{NWppJ6t-DKmU5-N}{NWppJ6t-DKmU5-P}\nwenddeflinemarkup
    {\bf{}if} (\begin{math}\neg\end{math} {\it{}info}({\it{}status\_flags}::{\it{}expanded}))
        {\it{}nodes}[{\it{}key}].{\it{}set\_cycles}({\it{}ex\_to}\begin{math}<\end{math}{\bf{}lst}\begin{math}>\end{math}({\it{}update\_cycle\_node}({\it{}key})));

    {\bf{}if} ({\it{}FIGURE\_DEBUG})
        {\it{}cerr} \begin{math}\ll\end{math} {\tt{}"Add the cycle: "} \begin{math}\ll\end{math} {\it{}nodes}[{\it{}key}] \begin{math}\ll\end{math} {\it{}endl};

    {\bf{}return} {\it{}key};
{\nwrbrace}

\nwused{\\{NWppJ6t-32jW6L-1}}\nwidentuses{\\{{\nwixident{FIGURE{\_}DEBUG}}{FIGURE:unDEBUG}}\\{{\nwixident{info}}{info}}\\{{\nwixident{key}}{key}}\\{{\nwixident{nodes}}{nodes}}\\{{\nwixident{update{\_}cycle{\_}node}}{update:uncycle:unnode}}}\nwindexuse{\nwixident{FIGURE{\_}DEBUG}}{FIGURE:unDEBUG}{NWppJ6t-DKmU5-O}\nwindexuse{\nwixident{info}}{info}{NWppJ6t-DKmU5-O}\nwindexuse{\nwixident{key}}{key}{NWppJ6t-DKmU5-O}\nwindexuse{\nwixident{nodes}}{nodes}{NWppJ6t-DKmU5-O}\nwindexuse{\nwixident{update{\_}cycle{\_}node}}{update:uncycle:unnode}{NWppJ6t-DKmU5-O}\nwendcode{}\nwbegindocs{907}This version automatically supply \TeX\ label like \(c_{23}\) to
symbols with names {\Tt{}\Rm{}{\it{}c23}\nwendquote}.
\nwenddocs{}\nwbegincode{908}\sublabel{NWppJ6t-DKmU5-P}\nwmargintag{{\nwtagstyle{}\subpageref{NWppJ6t-DKmU5-P}}}\moddef{figure class~{\nwtagstyle{}\subpageref{NWppJ6t-DKmU5-1}}}\plusendmoddef\Rm{}\nwstartdeflinemarkup\nwusesondefline{\\{NWppJ6t-32jW6L-1}}\nwprevnextdefs{NWppJ6t-DKmU5-O}{NWppJ6t-DKmU5-Q}\nwenddeflinemarkup
{\bf{}ex} {\bf{}figure}::{\it{}add\_cycle\_rel}({\bf{}const} {\bf{}lst} & {\it{}rel}, {\it{}string} {\it{}name}, {\it{}string} {\it{}TeXname})
{\nwlbrace}
    \LA{}auto TeX name~{\nwtagstyle{}\subpageref{NWppJ6t-2tM1q2-1}}\RA{}
    {\bf{}return} {\it{}add\_cycle\_rel}({\it{}rel}, {\bf{}symbol}({\it{}name}, {\it{}TeXname\_new}));
{\nwrbrace}

\nwused{\\{NWppJ6t-32jW6L-1}}\nwidentuses{\\{{\nwixident{add{\_}cycle{\_}rel}}{add:uncycle:unrel}}\\{{\nwixident{ex}}{ex}}\\{{\nwixident{figure}}{figure}}\\{{\nwixident{name}}{name}}\\{{\nwixident{TeXname}}{TeXname}}}\nwindexuse{\nwixident{add{\_}cycle{\_}rel}}{add:uncycle:unrel}{NWppJ6t-DKmU5-P}\nwindexuse{\nwixident{ex}}{ex}{NWppJ6t-DKmU5-P}\nwindexuse{\nwixident{figure}}{figure}{NWppJ6t-DKmU5-P}\nwindexuse{\nwixident{name}}{name}{NWppJ6t-DKmU5-P}\nwindexuse{\nwixident{TeXname}}{TeXname}{NWppJ6t-DKmU5-P}\nwendcode{}\nwbegindocs{909}A similar method to add a cycle by a single relation.
\nwenddocs{}\nwbegincode{910}\sublabel{NWppJ6t-DKmU5-Q}\nwmargintag{{\nwtagstyle{}\subpageref{NWppJ6t-DKmU5-Q}}}\moddef{figure class~{\nwtagstyle{}\subpageref{NWppJ6t-DKmU5-1}}}\plusendmoddef\Rm{}\nwstartdeflinemarkup\nwusesondefline{\\{NWppJ6t-32jW6L-1}}\nwprevnextdefs{NWppJ6t-DKmU5-P}{NWppJ6t-DKmU5-R}\nwenddeflinemarkup
{\bf{}ex} {\bf{}figure}::{\it{}add\_cycle\_rel}({\bf{}const} {\bf{}ex} & {\it{}rel}, {\it{}string} {\it{}name}, {\it{}string} {\it{}TeXname})
{\nwlbrace}
    {\bf{}if} ({\it{}is\_a}\begin{math}<\end{math}{\bf{}cycle\_relation}\begin{math}>\end{math}({\it{}rel})) {\nwlbrace}
        \LA{}auto TeX name~{\nwtagstyle{}\subpageref{NWppJ6t-2tM1q2-1}}\RA{}
        {\bf{}return} {\it{}add\_cycle\_rel}({\bf{}lst}{\nwlbrace}{\it{}rel}{\nwrbrace}, {\bf{}symbol}({\it{}name}, {\it{}TeXname\_new}));
    {\nwrbrace} {\bf{}else}
        {\bf{}throw}({\it{}std}::{\it{}invalid\_argument}({\tt{}"figure::add\_cycle\_rel: a cycle shall be added "}
                                    {\tt{}"by a single expression, which is a cycle\_relation"}));
{\nwrbrace}

\nwused{\\{NWppJ6t-32jW6L-1}}\nwidentuses{\\{{\nwixident{add{\_}cycle{\_}rel}}{add:uncycle:unrel}}\\{{\nwixident{cycle{\_}relation}}{cycle:unrelation}}\\{{\nwixident{ex}}{ex}}\\{{\nwixident{figure}}{figure}}\\{{\nwixident{name}}{name}}\\{{\nwixident{TeXname}}{TeXname}}}\nwindexuse{\nwixident{add{\_}cycle{\_}rel}}{add:uncycle:unrel}{NWppJ6t-DKmU5-Q}\nwindexuse{\nwixident{cycle{\_}relation}}{cycle:unrelation}{NWppJ6t-DKmU5-Q}\nwindexuse{\nwixident{ex}}{ex}{NWppJ6t-DKmU5-Q}\nwindexuse{\nwixident{figure}}{figure}{NWppJ6t-DKmU5-Q}\nwindexuse{\nwixident{name}}{name}{NWppJ6t-DKmU5-Q}\nwindexuse{\nwixident{TeXname}}{TeXname}{NWppJ6t-DKmU5-Q}\nwendcode{}\nwbegindocs{911}This method adds a {\Tt{}\Rm{}{\bf{}subfigure} \nwendquote} as a single node. The generation
of the new node is again calculated by the rules described in
Sec.~\ref{sec:figures-as-families}.
\nwenddocs{}\nwbegincode{912}\sublabel{NWppJ6t-DKmU5-R}\nwmargintag{{\nwtagstyle{}\subpageref{NWppJ6t-DKmU5-R}}}\moddef{figure class~{\nwtagstyle{}\subpageref{NWppJ6t-DKmU5-1}}}\plusendmoddef\Rm{}\nwstartdeflinemarkup\nwusesondefline{\\{NWppJ6t-32jW6L-1}}\nwprevnextdefs{NWppJ6t-DKmU5-Q}{NWppJ6t-DKmU5-S}\nwenddeflinemarkup
{\bf{}ex} {\bf{}figure}::{\it{}add\_subfigure}({\bf{}const} {\bf{}ex} & {\it{}F}, {\bf{}const} {\bf{}lst} & {\it{}L}, {\bf{}const} {\bf{}ex} & {\it{}key})
{\nwlbrace}
    {\it{}GINAC\_ASSERT}({\it{}is\_a}\begin{math}<\end{math}{\bf{}figure}\begin{math}>\end{math}({\it{}F}));
    {\bf{}int} {\it{}gen}=0;

    {\bf{}for}({\bf{}const} {\bf{}auto}& {\it{}it} : {\it{}L}) {\nwlbrace}
        {\bf{}if} (\begin{math}\neg\end{math} {\it{}it}.{\it{}is\_equal}({\it{}key}))
            {\it{}gen}={\it{}max}({\it{}gen}, {\it{}nodes}[{\it{}it}].{\it{}get\_generation}());
        {\it{}nodes}[{\it{}it}].{\it{}add\_child}({\it{}key});
    {\nwrbrace}
    {\it{}nodes}[{\it{}key}]={\bf{}cycle\_node}({\bf{}cycle\_data}(),{\it{}gen}+1,{\bf{}lst}{\nwlbrace}{\bf{}subfigure}({\it{}F},{\it{}L}){\nwrbrace});
    {\bf{}if} (\begin{math}\neg\end{math} {\it{}info}({\it{}status\_flags}::{\it{}expanded}))
        {\it{}nodes}[{\it{}key}].{\it{}set\_cycles}({\it{}ex\_to}\begin{math}<\end{math}{\bf{}lst}\begin{math}>\end{math}({\it{}update\_cycle\_node}({\it{}key})));

    {\bf{}return} {\it{}key};
{\nwrbrace}
\nwindexdefn{\nwixident{add{\_}subfigure}}{add:unsubfigure}{NWppJ6t-DKmU5-R}\eatline
\nwused{\\{NWppJ6t-32jW6L-1}}\nwidentdefs{\\{{\nwixident{add{\_}subfigure}}{add:unsubfigure}}}\nwidentuses{\\{{\nwixident{cycle{\_}data}}{cycle:undata}}\\{{\nwixident{cycle{\_}node}}{cycle:unnode}}\\{{\nwixident{ex}}{ex}}\\{{\nwixident{figure}}{figure}}\\{{\nwixident{get{\_}generation}}{get:ungeneration}}\\{{\nwixident{info}}{info}}\\{{\nwixident{key}}{key}}\\{{\nwixident{nodes}}{nodes}}\\{{\nwixident{subfigure}}{subfigure}}\\{{\nwixident{update{\_}cycle{\_}node}}{update:uncycle:unnode}}}\nwindexuse{\nwixident{cycle{\_}data}}{cycle:undata}{NWppJ6t-DKmU5-R}\nwindexuse{\nwixident{cycle{\_}node}}{cycle:unnode}{NWppJ6t-DKmU5-R}\nwindexuse{\nwixident{ex}}{ex}{NWppJ6t-DKmU5-R}\nwindexuse{\nwixident{figure}}{figure}{NWppJ6t-DKmU5-R}\nwindexuse{\nwixident{get{\_}generation}}{get:ungeneration}{NWppJ6t-DKmU5-R}\nwindexuse{\nwixident{info}}{info}{NWppJ6t-DKmU5-R}\nwindexuse{\nwixident{key}}{key}{NWppJ6t-DKmU5-R}\nwindexuse{\nwixident{nodes}}{nodes}{NWppJ6t-DKmU5-R}\nwindexuse{\nwixident{subfigure}}{subfigure}{NWppJ6t-DKmU5-R}\nwindexuse{\nwixident{update{\_}cycle{\_}node}}{update:uncycle:unnode}{NWppJ6t-DKmU5-R}\nwendcode{}\nwbegindocs{913}\nwdocspar
\nwenddocs{}\nwbegindocs{914}This is again a wrapper for the previous method with the newly
defined symbol.
\nwenddocs{}\nwbegincode{915}\sublabel{NWppJ6t-DKmU5-S}\nwmargintag{{\nwtagstyle{}\subpageref{NWppJ6t-DKmU5-S}}}\moddef{figure class~{\nwtagstyle{}\subpageref{NWppJ6t-DKmU5-1}}}\plusendmoddef\Rm{}\nwstartdeflinemarkup\nwusesondefline{\\{NWppJ6t-32jW6L-1}}\nwprevnextdefs{NWppJ6t-DKmU5-R}{NWppJ6t-DKmU5-T}\nwenddeflinemarkup
{\bf{}ex} {\bf{}figure}::{\it{}add\_subfigure}({\bf{}const} {\bf{}ex} & {\it{}F}, {\bf{}const} {\bf{}lst} & {\it{}l}, {\it{}string} {\it{}name}, {\it{}string} {\it{}TeXname})
{\nwlbrace}
    \LA{}auto TeX name~{\nwtagstyle{}\subpageref{NWppJ6t-2tM1q2-1}}\RA{}
        {\bf{}return} {\it{}add\_subfigure}({\it{}F}, {\it{}l}, {\bf{}symbol}({\it{}name}, {\it{}TeXname\_new}));
{\nwrbrace}

\nwused{\\{NWppJ6t-32jW6L-1}}\nwidentuses{\\{{\nwixident{add{\_}subfigure}}{add:unsubfigure}}\\{{\nwixident{ex}}{ex}}\\{{\nwixident{figure}}{figure}}\\{{\nwixident{l}}{l}}\\{{\nwixident{name}}{name}}\\{{\nwixident{TeXname}}{TeXname}}}\nwindexuse{\nwixident{add{\_}subfigure}}{add:unsubfigure}{NWppJ6t-DKmU5-S}\nwindexuse{\nwixident{ex}}{ex}{NWppJ6t-DKmU5-S}\nwindexuse{\nwixident{figure}}{figure}{NWppJ6t-DKmU5-S}\nwindexuse{\nwixident{l}}{l}{NWppJ6t-DKmU5-S}\nwindexuse{\nwixident{name}}{name}{NWppJ6t-DKmU5-S}\nwindexuse{\nwixident{TeXname}}{TeXname}{NWppJ6t-DKmU5-S}\nwendcode{}\nwbegindocs{916}\nwdocspar
\nwenddocs{}\nwbegincode{917}\sublabel{NWppJ6t-2tM1q2-1}\nwmargintag{{\nwtagstyle{}\subpageref{NWppJ6t-2tM1q2-1}}}\moddef{auto TeX name~{\nwtagstyle{}\subpageref{NWppJ6t-2tM1q2-1}}}\endmoddef\Rm{}\nwstartdeflinemarkup\nwusesondefline{\\{NWppJ6t-DKmU5-D}\\{NWppJ6t-DKmU5-H}\\{NWppJ6t-DKmU5-P}\\{NWppJ6t-DKmU5-Q}\\{NWppJ6t-DKmU5-S}\\{NWppJ6t-DKmU5-2J}}\nwenddeflinemarkup
    {\it{}string} {\it{}TeXname\_new};
    {\it{}std}::{\it{}regex} {\it{}e} ({\tt{}"([[:alpha:]]+)([[:digit:]]+)"});
    {\it{}std}::{\it{}regex} {\it{}e1} ({\tt{}"([[:alnum:]]+)\_([[:alnum:]]+)"});
    {\bf{}if} ({\it{}TeXname} \begin{math}\equiv\end{math} {\tt{}""}) {\nwlbrace}
        {\bf{}if} ({\it{}std}::{\it{}regex\_match}({\it{}name}, {\it{}e}))
            {\it{}TeXname\_new}={\it{}std}::{\it{}regex\_replace} ({\it{}name},{\it{}e},{\tt{}"$1\_{\char123}$2{\char125}"});
        {\bf{}else} {\bf{}if} ({\it{}std}::{\it{}regex\_match}({\it{}name}, {\it{}e1}))
            {\it{}TeXname\_new}={\it{}std}::{\it{}regex\_replace} ({\it{}name},{\it{}e1},{\tt{}"$1\_{\char123}$2{\char125}"});
    {\nwrbrace} {\bf{}else}
        {\it{}TeXname\_new}={\it{}TeXname};

\nwused{\\{NWppJ6t-DKmU5-D}\\{NWppJ6t-DKmU5-H}\\{NWppJ6t-DKmU5-P}\\{NWppJ6t-DKmU5-Q}\\{NWppJ6t-DKmU5-S}\\{NWppJ6t-DKmU5-2J}}\nwidentuses{\\{{\nwixident{name}}{name}}\\{{\nwixident{TeXname}}{TeXname}}}\nwindexuse{\nwixident{name}}{name}{NWppJ6t-2tM1q2-1}\nwindexuse{\nwixident{TeXname}}{TeXname}{NWppJ6t-2tM1q2-1}\nwendcode{}\nwbegindocs{918}\nwdocspar
\subsubsection{Moving and removing cycles}
\label{sec:moving-remov-cycl}

\nwenddocs{}\nwbegindocs{919}The method to change a zero-generation cycle to a point with given coordinates.
\nwenddocs{}\nwbegincode{920}\sublabel{NWppJ6t-DKmU5-T}\nwmargintag{{\nwtagstyle{}\subpageref{NWppJ6t-DKmU5-T}}}\moddef{figure class~{\nwtagstyle{}\subpageref{NWppJ6t-DKmU5-1}}}\plusendmoddef\Rm{}\nwstartdeflinemarkup\nwusesondefline{\\{NWppJ6t-32jW6L-1}}\nwprevnextdefs{NWppJ6t-DKmU5-S}{NWppJ6t-DKmU5-U}\nwenddeflinemarkup
{\bf{}void} {\bf{}figure}::{\it{}move\_point}({\bf{}const} {\bf{}ex} & {\it{}key}, {\bf{}const} {\bf{}ex} & {\it{}x})\nwindexdefn{\nwixident{figure}}{figure}{NWppJ6t-DKmU5-T}
{\nwlbrace}
    {\bf{}if} ({\it{}not} ({\it{}is\_a}\begin{math}<\end{math}{\bf{}lst}\begin{math}>\end{math}({\it{}x}) {\it{}and} ({\it{}x}.{\it{}nops}() \begin{math}\equiv\end{math} {\it{}get\_dim}())))
        {\bf{}throw}({\it{}std}::{\it{}invalid\_argument}({\tt{}"figure::move\_point(const ex &, const ex &): "}
                                    {\tt{}"coordinates of a point shall be a lst of the right lenght"}));

    {\bf{}if} ({\it{}nodes}.{\it{}find}({\it{}key}) \begin{math}\equiv\end{math} {\it{}nodes}.{\it{}end}())
        {\bf{}throw}({\it{}std}::{\it{}invalid\_argument}({\tt{}"figure::move\_point(): there is no node with the key given"}));

    {\bf{}if} ({\it{}nodes}[{\it{}key}].{\it{}get\_generation}() \begin{math}\neq\end{math} 0)
        {\bf{}throw}({\it{}std}::{\it{}invalid\_argument}({\tt{}"figure::move\_point(): cannot modify data of a cycle in"}
                                    {\tt{}" non-zero generation"}));

    {\bf{}if} ({\it{}FIGURE\_DEBUG})
        {\it{}cerr} \begin{math}\ll\end{math} {\tt{}"A cycle is moved : "} \begin{math}\ll\end{math} {\it{}nodes}[{\it{}key}] \begin{math}\ll\end{math} {\it{}endl};
\nwindexdefn{\nwixident{move{\_}point}}{move:unpoint}{NWppJ6t-DKmU5-T}\eatline
\nwused{\\{NWppJ6t-32jW6L-1}}\nwidentdefs{\\{{\nwixident{figure}}{figure}}\\{{\nwixident{move{\_}point}}{move:unpoint}}}\nwidentuses{\\{{\nwixident{ex}}{ex}}\\{{\nwixident{FIGURE{\_}DEBUG}}{FIGURE:unDEBUG}}\\{{\nwixident{get{\_}dim()}}{get:undim()}}\\{{\nwixident{get{\_}generation}}{get:ungeneration}}\\{{\nwixident{key}}{key}}\\{{\nwixident{nodes}}{nodes}}\\{{\nwixident{nops}}{nops}}}\nwindexuse{\nwixident{ex}}{ex}{NWppJ6t-DKmU5-T}\nwindexuse{\nwixident{FIGURE{\_}DEBUG}}{FIGURE:unDEBUG}{NWppJ6t-DKmU5-T}\nwindexuse{\nwixident{get{\_}dim()}}{get:undim()}{NWppJ6t-DKmU5-T}\nwindexuse{\nwixident{get{\_}generation}}{get:ungeneration}{NWppJ6t-DKmU5-T}\nwindexuse{\nwixident{key}}{key}{NWppJ6t-DKmU5-T}\nwindexuse{\nwixident{nodes}}{nodes}{NWppJ6t-DKmU5-T}\nwindexuse{\nwixident{nops}}{nops}{NWppJ6t-DKmU5-T}\nwendcode{}\nwbegindocs{921}\nwdocspar
\nwenddocs{}\nwbegindocs{922}If number of parents was ``dimension plus 2'', so it was a proper point, we simply need to
replace the ghost parents.
\nwenddocs{}\nwbegincode{923}\sublabel{NWppJ6t-DKmU5-U}\nwmargintag{{\nwtagstyle{}\subpageref{NWppJ6t-DKmU5-U}}}\moddef{figure class~{\nwtagstyle{}\subpageref{NWppJ6t-DKmU5-1}}}\plusendmoddef\Rm{}\nwstartdeflinemarkup\nwusesondefline{\\{NWppJ6t-32jW6L-1}}\nwprevnextdefs{NWppJ6t-DKmU5-T}{NWppJ6t-DKmU5-V}\nwenddeflinemarkup
    {\bf{}lst} {\it{}par}={\it{}nodes}[{\it{}key}].{\it{}get\_parent\_keys}();
    {\bf{}unsigned} {\bf{}int} {\it{}dim}={\it{}x}.{\it{}nops}();
    {\bf{}lst} {\it{}l0};
    {\bf{}for}({\bf{}unsigned} {\bf{}int} {\it{}i}=0; {\it{}i}\begin{math}<\end{math}{\it{}dim}; \protect\PP{\it{}i})
        {\it{}l0}.{\it{}append}({\bf{}numeric}(0));

\nwused{\\{NWppJ6t-32jW6L-1}}\nwidentuses{\\{{\nwixident{key}}{key}}\\{{\nwixident{nodes}}{nodes}}\\{{\nwixident{nops}}{nops}}\\{{\nwixident{numeric}}{numeric}}}\nwindexuse{\nwixident{key}}{key}{NWppJ6t-DKmU5-U}\nwindexuse{\nwixident{nodes}}{nodes}{NWppJ6t-DKmU5-U}\nwindexuse{\nwixident{nops}}{nops}{NWppJ6t-DKmU5-U}\nwindexuse{\nwixident{numeric}}{numeric}{NWppJ6t-DKmU5-U}\nwendcode{}\nwbegindocs{924}We scan the name of parents to get number of components and
substitute their new values.
\nwenddocs{}\nwbegincode{925}\sublabel{NWppJ6t-DKmU5-V}\nwmargintag{{\nwtagstyle{}\subpageref{NWppJ6t-DKmU5-V}}}\moddef{figure class~{\nwtagstyle{}\subpageref{NWppJ6t-DKmU5-1}}}\plusendmoddef\Rm{}\nwstartdeflinemarkup\nwusesondefline{\\{NWppJ6t-32jW6L-1}}\nwprevnextdefs{NWppJ6t-DKmU5-U}{NWppJ6t-DKmU5-W}\nwenddeflinemarkup
    {\bf{}char} {\it{}label}[40];
    {\it{}sprintf}({\it{}label}, {\tt{}"
    {\bf{}if} ({\it{}par}.{\it{}nops}() \begin{math}\equiv\end{math} {\it{}dim}+2 ) {\nwlbrace}
        {\bf{}for}({\bf{}const} {\bf{}auto}& {\it{}it} : {\it{}par}) {\nwlbrace}
            {\bf{}unsigned} {\bf{}int} {\it{}i}={\it{}dim};
            {\bf{}int} {\it{}res}={\it{}sscanf}({\it{}ex\_to}\begin{math}<\end{math}{\bf{}symbol}\begin{math}>\end{math}({\it{}it}).{\it{}get\_name}().{\it{}c\_str}(), {\it{}label}, &{\it{}i});
            {\bf{}if} ({\it{}res}\begin{math}>\end{math}0 {\it{}and} {\it{}i}\begin{math}<\end{math}{\it{}dim}) {\nwlbrace}
                {\it{}l0}[{\it{}i}]={\bf{}numeric}(1);
                {\it{}nodes}[{\it{}it}].{\it{}set\_cycles}({\bf{}cycle\_data}({\bf{}numeric}(0),{\bf{}indexed}({\bf{}matrix}(1, {\it{}dim}, {\it{}l0}),
                                                                    {\bf{}varidx}({\it{}it}, {\it{}dim})), {\bf{}numeric}(2)\begin{math}\ast\end{math}{\it{}x}.{\it{}op}({\it{}i})));
                {\it{}l0}[{\it{}i}]={\bf{}numeric}(0);
            {\nwrbrace}
        {\nwrbrace}

\nwused{\\{NWppJ6t-32jW6L-1}}\nwidentuses{\\{{\nwixident{cycle{\_}data}}{cycle:undata}}\\{{\nwixident{key}}{key}}\\{{\nwixident{nodes}}{nodes}}\\{{\nwixident{nops}}{nops}}\\{{\nwixident{numeric}}{numeric}}\\{{\nwixident{op}}{op}}}\nwindexuse{\nwixident{cycle{\_}data}}{cycle:undata}{NWppJ6t-DKmU5-V}\nwindexuse{\nwixident{key}}{key}{NWppJ6t-DKmU5-V}\nwindexuse{\nwixident{nodes}}{nodes}{NWppJ6t-DKmU5-V}\nwindexuse{\nwixident{nops}}{nops}{NWppJ6t-DKmU5-V}\nwindexuse{\nwixident{numeric}}{numeric}{NWppJ6t-DKmU5-V}\nwindexuse{\nwixident{op}}{op}{NWppJ6t-DKmU5-V}\nwendcode{}\nwbegindocs{926}If the number of parents is zero, so it was a pre-defined cycle and
we need to create ghost parents for it.
\nwenddocs{}\nwbegincode{927}\sublabel{NWppJ6t-DKmU5-W}\nwmargintag{{\nwtagstyle{}\subpageref{NWppJ6t-DKmU5-W}}}\moddef{figure class~{\nwtagstyle{}\subpageref{NWppJ6t-DKmU5-1}}}\plusendmoddef\Rm{}\nwstartdeflinemarkup\nwusesondefline{\\{NWppJ6t-32jW6L-1}}\nwprevnextdefs{NWppJ6t-DKmU5-V}{NWppJ6t-DKmU5-X}\nwenddeflinemarkup
    {\nwrbrace} {\bf{}else} {\bf{}if}  ({\it{}par}.{\it{}nops}() \begin{math}\equiv\end{math} 0) {\nwlbrace}
        {\bf{}lst} {\it{}chil}={\it{}nodes}[{\it{}key}].{\it{}get\_children}();
        \LA{}adding point with its parents~{\nwtagstyle{}\subpageref{NWppJ6t-1Ub2Xq-1}}\RA{}
        {\it{}nodes}[{\it{}key}].{\it{}children}={\it{}chil};
    {\nwrbrace} {\bf{}else}
        {\bf{}throw}({\it{}std}::{\it{}invalid\_argument}({\tt{}"figure::move\_point(): strange number (neither 0 nor dim+2) of "}
                                    {\tt{}"parents, which zero-generation node shall have!"}));

    {\bf{}if} ({\it{}info}({\it{}status\_flags}::{\it{}expanded}))
        {\bf{}return};

    {\it{}nodes}[{\it{}key}].{\it{}set\_cycles}({\it{}ex\_to}\begin{math}<\end{math}{\bf{}lst}\begin{math}>\end{math}({\it{}update\_cycle\_node}({\it{}key})));
    {\it{}update\_node\_lst}({\it{}nodes}[{\it{}key}].{\it{}get\_children}());

\nwused{\\{NWppJ6t-32jW6L-1}}\nwidentuses{\\{{\nwixident{figure}}{figure}}\\{{\nwixident{info}}{info}}\\{{\nwixident{key}}{key}}\\{{\nwixident{move{\_}point}}{move:unpoint}}\\{{\nwixident{nodes}}{nodes}}\\{{\nwixident{nops}}{nops}}\\{{\nwixident{update{\_}cycle{\_}node}}{update:uncycle:unnode}}\\{{\nwixident{update{\_}node{\_}lst}}{update:unnode:unlst}}}\nwindexuse{\nwixident{figure}}{figure}{NWppJ6t-DKmU5-W}\nwindexuse{\nwixident{info}}{info}{NWppJ6t-DKmU5-W}\nwindexuse{\nwixident{key}}{key}{NWppJ6t-DKmU5-W}\nwindexuse{\nwixident{move{\_}point}}{move:unpoint}{NWppJ6t-DKmU5-W}\nwindexuse{\nwixident{nodes}}{nodes}{NWppJ6t-DKmU5-W}\nwindexuse{\nwixident{nops}}{nops}{NWppJ6t-DKmU5-W}\nwindexuse{\nwixident{update{\_}cycle{\_}node}}{update:uncycle:unnode}{NWppJ6t-DKmU5-W}\nwindexuse{\nwixident{update{\_}node{\_}lst}}{update:unnode:unlst}{NWppJ6t-DKmU5-W}\nwendcode{}\nwbegindocs{928}Then, to update all its children and grandchildren in all generations
excluding this node itself.
\nwenddocs{}\nwbegincode{929}\sublabel{NWppJ6t-DKmU5-X}\nwmargintag{{\nwtagstyle{}\subpageref{NWppJ6t-DKmU5-X}}}\moddef{figure class~{\nwtagstyle{}\subpageref{NWppJ6t-DKmU5-1}}}\plusendmoddef\Rm{}\nwstartdeflinemarkup\nwusesondefline{\\{NWppJ6t-32jW6L-1}}\nwprevnextdefs{NWppJ6t-DKmU5-W}{NWppJ6t-DKmU5-Y}\nwenddeflinemarkup
     {\it{}update\_node\_lst}({\it{}nodes}[{\it{}key}].{\it{}get\_children}());
     {\bf{}if} ({\it{}FIGURE\_DEBUG})
        {\it{}cerr} \begin{math}\ll\end{math} {\tt{}"Moved to: "} \begin{math}\ll\end{math} {\it{}x} \begin{math}\ll\end{math} {\it{}endl};
{\nwrbrace}

\nwused{\\{NWppJ6t-32jW6L-1}}\nwidentuses{\\{{\nwixident{FIGURE{\_}DEBUG}}{FIGURE:unDEBUG}}\\{{\nwixident{key}}{key}}\\{{\nwixident{nodes}}{nodes}}\\{{\nwixident{update{\_}node{\_}lst}}{update:unnode:unlst}}}\nwindexuse{\nwixident{FIGURE{\_}DEBUG}}{FIGURE:unDEBUG}{NWppJ6t-DKmU5-X}\nwindexuse{\nwixident{key}}{key}{NWppJ6t-DKmU5-X}\nwindexuse{\nwixident{nodes}}{nodes}{NWppJ6t-DKmU5-X}\nwindexuse{\nwixident{update{\_}node{\_}lst}}{update:unnode:unlst}{NWppJ6t-DKmU5-X}\nwendcode{}\nwbegindocs{930}Afterwards, to remove all children (includes grand children, grand grand
children\ldots) of the {\Tt{}\Rm{}{\bf{}cycle\_node}\nwendquote}.
\nwenddocs{}\nwbegincode{931}\sublabel{NWppJ6t-DKmU5-Y}\nwmargintag{{\nwtagstyle{}\subpageref{NWppJ6t-DKmU5-Y}}}\moddef{figure class~{\nwtagstyle{}\subpageref{NWppJ6t-DKmU5-1}}}\plusendmoddef\Rm{}\nwstartdeflinemarkup\nwusesondefline{\\{NWppJ6t-32jW6L-1}}\nwprevnextdefs{NWppJ6t-DKmU5-X}{NWppJ6t-DKmU5-Z}\nwenddeflinemarkup
{\bf{}void} {\bf{}figure}::{\it{}remove\_cycle\_node}({\bf{}const} {\bf{}ex} & {\it{}key})\nwindexdefn{\nwixident{figure}}{figure}{NWppJ6t-DKmU5-Y}
{\nwlbrace}
    {\bf{}lst}  {\it{}branches}={\it{}nodes}[{\it{}key}].{\it{}get\_children}();
    {\bf{}for} ({\bf{}const} {\bf{}auto}& {\it{}it} : {\it{}branches})
        {\it{}remove\_cycle\_node}({\it{}it});
\nwindexdefn{\nwixident{remove{\_}cycle{\_}node}}{remove:uncycle:unnode}{NWppJ6t-DKmU5-Y}\eatline
\nwused{\\{NWppJ6t-32jW6L-1}}\nwidentdefs{\\{{\nwixident{figure}}{figure}}\\{{\nwixident{remove{\_}cycle{\_}node}}{remove:uncycle:unnode}}}\nwidentuses{\\{{\nwixident{ex}}{ex}}\\{{\nwixident{key}}{key}}\\{{\nwixident{nodes}}{nodes}}}\nwindexuse{\nwixident{ex}}{ex}{NWppJ6t-DKmU5-Y}\nwindexuse{\nwixident{key}}{key}{NWppJ6t-DKmU5-Y}\nwindexuse{\nwixident{nodes}}{nodes}{NWppJ6t-DKmU5-Y}\nwendcode{}\nwbegindocs{932}\nwdocspar
\nwenddocs{}\nwbegindocs{933}Furthermore, to remove the {\Tt{}\Rm{}{\bf{}cycle\_node}\nwendquote} c from all its parents children lists.
\nwenddocs{}\nwbegincode{934}\sublabel{NWppJ6t-DKmU5-Z}\nwmargintag{{\nwtagstyle{}\subpageref{NWppJ6t-DKmU5-Z}}}\moddef{figure class~{\nwtagstyle{}\subpageref{NWppJ6t-DKmU5-1}}}\plusendmoddef\Rm{}\nwstartdeflinemarkup\nwusesondefline{\\{NWppJ6t-32jW6L-1}}\nwprevnextdefs{NWppJ6t-DKmU5-Y}{NWppJ6t-DKmU5-a}\nwenddeflinemarkup
    {\bf{}lst}  {\it{}par} = {\it{}nodes}[{\it{}key}].{\it{}get\_parent\_keys}();
    {\bf{}for} ({\bf{}const} {\bf{}auto}& {\it{}it} : {\it{}par}) {\nwlbrace}

\nwused{\\{NWppJ6t-32jW6L-1}}\nwidentuses{\\{{\nwixident{key}}{key}}\\{{\nwixident{nodes}}{nodes}}}\nwindexuse{\nwixident{key}}{key}{NWppJ6t-DKmU5-Z}\nwindexuse{\nwixident{nodes}}{nodes}{NWppJ6t-DKmU5-Z}\nwendcode{}\nwbegindocs{935}Parents of a point at gen-0 can be simply deleted as no other cycle
need them and they are not of interest. For other parents we modify
their {\Tt{}\Rm{}{\it{}cildren}\nwendquote} list.
\nwenddocs{}\nwbegincode{936}\sublabel{NWppJ6t-DKmU5-a}\nwmargintag{{\nwtagstyle{}\subpageref{NWppJ6t-DKmU5-a}}}\moddef{figure class~{\nwtagstyle{}\subpageref{NWppJ6t-DKmU5-1}}}\plusendmoddef\Rm{}\nwstartdeflinemarkup\nwusesondefline{\\{NWppJ6t-32jW6L-1}}\nwprevnextdefs{NWppJ6t-DKmU5-Z}{NWppJ6t-DKmU5-b}\nwenddeflinemarkup
        {\bf{}if} ({\it{}nodes}[{\it{}it}].{\it{}get\_generation}() \begin{math}\equiv\end{math} {\it{}GHOST\_GEN})
            {\it{}nodes}.{\it{}erase}({\it{}it});
        {\bf{}else}
            {\it{}nodes}[{\it{}it}].{\it{}remove\_child}({\it{}key});
    {\nwrbrace}

\nwused{\\{NWppJ6t-32jW6L-1}}\nwidentuses{\\{{\nwixident{get{\_}generation}}{get:ungeneration}}\\{{\nwixident{GHOST{\_}GEN}}{GHOST:unGEN}}\\{{\nwixident{key}}{key}}\\{{\nwixident{nodes}}{nodes}}}\nwindexuse{\nwixident{get{\_}generation}}{get:ungeneration}{NWppJ6t-DKmU5-a}\nwindexuse{\nwixident{GHOST{\_}GEN}}{GHOST:unGEN}{NWppJ6t-DKmU5-a}\nwindexuse{\nwixident{key}}{key}{NWppJ6t-DKmU5-a}\nwindexuse{\nwixident{nodes}}{nodes}{NWppJ6t-DKmU5-a}\nwendcode{}\nwbegindocs{937}Finally, remove the {\Tt{}\Rm{}{\bf{}cycle\_node}\nwendquote} from the figure.
\nwenddocs{}\nwbegincode{938}\sublabel{NWppJ6t-DKmU5-b}\nwmargintag{{\nwtagstyle{}\subpageref{NWppJ6t-DKmU5-b}}}\moddef{figure class~{\nwtagstyle{}\subpageref{NWppJ6t-DKmU5-1}}}\plusendmoddef\Rm{}\nwstartdeflinemarkup\nwusesondefline{\\{NWppJ6t-32jW6L-1}}\nwprevnextdefs{NWppJ6t-DKmU5-a}{NWppJ6t-DKmU5-c}\nwenddeflinemarkup
    {\it{}nodes}.{\it{}erase}({\it{}key});
    {\bf{}if} ({\it{}FIGURE\_DEBUG})
        {\it{}cerr} \begin{math}\ll\end{math} {\tt{}"The cycle is removed: "} \begin{math}\ll\end{math} {\it{}key} \begin{math}\ll\end{math} {\it{}endl} ;
{\nwrbrace}

\nwused{\\{NWppJ6t-32jW6L-1}}\nwidentuses{\\{{\nwixident{FIGURE{\_}DEBUG}}{FIGURE:unDEBUG}}\\{{\nwixident{key}}{key}}\\{{\nwixident{nodes}}{nodes}}}\nwindexuse{\nwixident{FIGURE{\_}DEBUG}}{FIGURE:unDEBUG}{NWppJ6t-DKmU5-b}\nwindexuse{\nwixident{key}}{key}{NWppJ6t-DKmU5-b}\nwindexuse{\nwixident{nodes}}{nodes}{NWppJ6t-DKmU5-b}\nwendcode{}\nwbegindocs{939}\nwdocspar
\subsubsection{Evaluattion of cycles and figure updates}
\label{sec:eval-cycl-from}

\nwenddocs{}\nwbegindocs{940}This procedure can solve a system of linear conditions or a system
with one quadratic equation. It was already observed
in~\citelist{\cite{FillmoreSpringer00a} \cite{Kisil12a}*{\S~5.5}}, see
Sec.~\ref{sec:conn-quadr-cycl}, that \(n\) tangency-type conditions
(each of them is quadratic) can be reduced to the single quadratic
condition \(\scalar{\cycle{}{}}{\cycle{}{}}=1\) and \(n\) linear
conditions like \(\scalar{\cycle{}{}}{\cycle{i}{}}=\lambda_i\).
\nwenddocs{}\nwbegincode{941}\sublabel{NWppJ6t-DKmU5-c}\nwmargintag{{\nwtagstyle{}\subpageref{NWppJ6t-DKmU5-c}}}\moddef{figure class~{\nwtagstyle{}\subpageref{NWppJ6t-DKmU5-1}}}\plusendmoddef\Rm{}\nwstartdeflinemarkup\nwusesondefline{\\{NWppJ6t-32jW6L-1}}\nwprevnextdefs{NWppJ6t-DKmU5-b}{NWppJ6t-DKmU5-d}\nwenddeflinemarkup
{\bf{}ex} {\bf{}figure}::{\it{}evaluate\_cycle}({\bf{}const} {\bf{}ex} & {\it{}symbolic}, {\bf{}const} {\bf{}lst} & {\it{}cond}) {\bf{}const}
{\nwlbrace}
    //cerr \begin{math}<\end{math}\begin{math}<\end{math} boolalpha \begin{math}<\end{math}\begin{math}<\end{math} "symbolic: "; symbolic.dbgprint();
    //cerr \begin{math}<\end{math}\begin{math}<\end{math} "condit: "; cond.dbgprint();
    {\bf{}bool} {\it{}first\_solution}={\bf{}true}, // whetehr the first solution is suitable
        {\it{}second\_solution}={\bf{}false}, // whetehr the second solution is suitable
        {\it{}is\_homogeneous}={\bf{}true}; // indicates whether all conditions are linear
\nwindexdefn{\nwixident{evaluate{\_}cycle}}{evaluate:uncycle}{NWppJ6t-DKmU5-c}\eatline
\nwused{\\{NWppJ6t-32jW6L-1}}\nwidentdefs{\\{{\nwixident{evaluate{\_}cycle}}{evaluate:uncycle}}}\nwidentuses{\\{{\nwixident{ex}}{ex}}\\{{\nwixident{figure}}{figure}}}\nwindexuse{\nwixident{ex}}{ex}{NWppJ6t-DKmU5-c}\nwindexuse{\nwixident{figure}}{figure}{NWppJ6t-DKmU5-c}\nwendcode{}\nwbegindocs{942}\nwdocspar
\nwenddocs{}\nwbegindocs{943}This method can be applied to cycles with numerical dimensions.
\nwenddocs{}\nwbegincode{944}\sublabel{NWppJ6t-DKmU5-d}\nwmargintag{{\nwtagstyle{}\subpageref{NWppJ6t-DKmU5-d}}}\moddef{figure class~{\nwtagstyle{}\subpageref{NWppJ6t-DKmU5-1}}}\plusendmoddef\Rm{}\nwstartdeflinemarkup\nwusesondefline{\\{NWppJ6t-32jW6L-1}}\nwprevnextdefs{NWppJ6t-DKmU5-c}{NWppJ6t-DKmU5-e}\nwenddeflinemarkup
    {\bf{}int} {\it{}D};
    {\bf{}if} ({\it{}is\_a}\begin{math}<\end{math}{\bf{}numeric}\begin{math}>\end{math}({\it{}get\_dim}()))
        {\it{}D}={\it{}ex\_to}\begin{math}<\end{math}{\bf{}numeric}\begin{math}>\end{math}({\it{}get\_dim}()).{\it{}to\_int}();
    {\bf{}else}
        {\bf{}throw} {\it{}logic\_error}({\tt{}"Could not resolve cycle relations if dimensionality is not numeric!"});

\nwused{\\{NWppJ6t-32jW6L-1}}\nwidentuses{\\{{\nwixident{get{\_}dim()}}{get:undim()}}\\{{\nwixident{numeric}}{numeric}}}\nwindexuse{\nwixident{get{\_}dim()}}{get:undim()}{NWppJ6t-DKmU5-d}\nwindexuse{\nwixident{numeric}}{numeric}{NWppJ6t-DKmU5-d}\nwendcode{}\nwbegindocs{945}Create the list of used symbols. The code is stolen from  \texttt{cycle.nw}
\nwenddocs{}\nwbegincode{946}\sublabel{NWppJ6t-DKmU5-e}\nwmargintag{{\nwtagstyle{}\subpageref{NWppJ6t-DKmU5-e}}}\moddef{figure class~{\nwtagstyle{}\subpageref{NWppJ6t-DKmU5-1}}}\plusendmoddef\Rm{}\nwstartdeflinemarkup\nwusesondefline{\\{NWppJ6t-32jW6L-1}}\nwprevnextdefs{NWppJ6t-DKmU5-d}{NWppJ6t-DKmU5-f}\nwenddeflinemarkup
    {\bf{}lst} {\it{}symbols}, {\it{}lin\_cond}, {\it{}nonlin\_cond};
    {\bf{}if} ({\it{}is\_a}\begin{math}<\end{math}{\bf{}symbol}\begin{math}>\end{math}({\it{}ex\_to}\begin{math}<\end{math}{\bf{}cycle\_data}\begin{math}>\end{math}({\it{}symbolic}).{\it{}get\_m}()))
        {\it{}symbols}.{\it{}append}({\it{}ex\_to}\begin{math}<\end{math}{\bf{}cycle\_data}\begin{math}>\end{math}({\it{}symbolic}).{\it{}get\_m}());
    {\bf{}for} ({\bf{}int} {\it{}i} = 0; {\it{}i} \begin{math}<\end{math} {\it{}D}; {\it{}i}\protect\PP)
        {\bf{}if} ({\it{}is\_a}\begin{math}<\end{math}{\bf{}symbol}\begin{math}>\end{math}({\it{}ex\_to}\begin{math}<\end{math}{\bf{}cycle\_data}\begin{math}>\end{math}({\it{}symbolic}).{\it{}get\_l}({\it{}i})))
            {\it{}symbols}.{\it{}append}({\it{}ex\_to}\begin{math}<\end{math}{\bf{}cycle\_data}\begin{math}>\end{math}({\it{}symbolic}).{\it{}get\_l}({\it{}i}));
    {\bf{}if} ({\it{}is\_a}\begin{math}<\end{math}{\bf{}symbol}\begin{math}>\end{math}({\it{}ex\_to}\begin{math}<\end{math}{\bf{}cycle\_data}\begin{math}>\end{math}({\it{}symbolic}).{\it{}get\_k}()))
        {\it{}symbols}.{\it{}append}({\it{}ex\_to}\begin{math}<\end{math}{\bf{}cycle\_data}\begin{math}>\end{math}({\it{}symbolic}).{\it{}get\_k}());

\nwused{\\{NWppJ6t-32jW6L-1}}\nwidentuses{\\{{\nwixident{cycle{\_}data}}{cycle:undata}}}\nwindexuse{\nwixident{cycle{\_}data}}{cycle:undata}{NWppJ6t-DKmU5-e}\nwendcode{}\nwbegindocs{947}If no symbols are found we assume that the cycle is uniquely defined
\nwenddocs{}\nwbegincode{948}\sublabel{NWppJ6t-DKmU5-f}\nwmargintag{{\nwtagstyle{}\subpageref{NWppJ6t-DKmU5-f}}}\moddef{figure class~{\nwtagstyle{}\subpageref{NWppJ6t-DKmU5-1}}}\plusendmoddef\Rm{}\nwstartdeflinemarkup\nwusesondefline{\\{NWppJ6t-32jW6L-1}}\nwprevnextdefs{NWppJ6t-DKmU5-e}{NWppJ6t-DKmU5-g}\nwenddeflinemarkup
    {\bf{}if} ({\it{}symbols}.{\it{}nops}() \begin{math}\equiv\end{math} 0)
        {\bf{}throw}({\it{}std}::{\it{}invalid\_argument}({\tt{}"figure::evaluate\_cycle(): could not construct the default list of "}
                                 {\tt{}"parameters"}));
    //cerr \begin{math}<\end{math}\begin{math}<\end{math} "symbols: "; symbols.dbgprint();

\nwused{\\{NWppJ6t-32jW6L-1}}\nwidentuses{\\{{\nwixident{evaluate{\_}cycle}}{evaluate:uncycle}}\\{{\nwixident{figure}}{figure}}\\{{\nwixident{nops}}{nops}}}\nwindexuse{\nwixident{evaluate{\_}cycle}}{evaluate:uncycle}{NWppJ6t-DKmU5-f}\nwindexuse{\nwixident{figure}}{figure}{NWppJ6t-DKmU5-f}\nwindexuse{\nwixident{nops}}{nops}{NWppJ6t-DKmU5-f}\nwendcode{}\nwbegindocs{949}Build matrix representation from equation system. The code is stolen
from  \texttt{ginac/inifcns.cpp}.
\nwenddocs{}\nwbegincode{950}\sublabel{NWppJ6t-DKmU5-g}\nwmargintag{{\nwtagstyle{}\subpageref{NWppJ6t-DKmU5-g}}}\moddef{figure class~{\nwtagstyle{}\subpageref{NWppJ6t-DKmU5-1}}}\plusendmoddef\Rm{}\nwstartdeflinemarkup\nwusesondefline{\\{NWppJ6t-32jW6L-1}}\nwprevnextdefs{NWppJ6t-DKmU5-f}{NWppJ6t-DKmU5-h}\nwenddeflinemarkup
    {\bf{}lst} {\it{}rhs};
    {\bf{}for} ({\it{}size\_t} {\it{}r}=0; {\it{}r}\begin{math}<\end{math}{\it{}cond}.{\it{}nops}(); {\it{}r}\protect\PP) {\nwlbrace}
        {\bf{}lst} {\it{}sys};
        {\bf{}ex} {\it{}eq} = ({\it{}cond}.{\it{}op}({\it{}r}).{\it{}op}(0)-{\it{}cond}.{\it{}op}({\it{}r}).{\it{}op}(1)).{\it{}expand}(); // lhs-rhs==0
        {\bf{}if} ({\it{}float\_evaluation})
            {\it{}eq}={\it{}eq}.{\it{}evalf}();
        //cerr \begin{math}<\end{math}\begin{math}<\end{math} "eq: "; eq.dbgprint();
        {\bf{}ex} {\it{}linpart} = {\it{}eq};
        {\bf{}for} ({\it{}size\_t} {\it{}c}=0; {\it{}c}\begin{math}<\end{math}{\it{}symbols}.{\it{}nops}(); {\it{}c}\protect\PP) {\nwlbrace}
            {\bf{}const} {\bf{}ex} {\it{}co} = {\it{}eq}.{\it{}coeff}({\it{}ex\_to}\begin{math}<\end{math}{\bf{}symbol}\begin{math}>\end{math}({\it{}symbols}.{\it{}op}({\it{}c})),1);
            {\it{}linpart} -= {\it{}co}\begin{math}\ast\end{math}{\it{}symbols}.{\it{}op}({\it{}c});
            {\it{}sys}.{\it{}append}({\it{}co});
        {\nwrbrace}
        {\it{}linpart} = {\it{}linpart}.{\it{}expand}();
        //cerr \begin{math}<\end{math}\begin{math}<\end{math} "sys: "; sys.dbgprint();
        //cerr \begin{math}<\end{math}\begin{math}<\end{math} "linpart: "; linpart.dbgprint();

\nwused{\\{NWppJ6t-32jW6L-1}}\nwidentuses{\\{{\nwixident{evalf}}{evalf}}\\{{\nwixident{ex}}{ex}}\\{{\nwixident{float{\_}evaluation}}{float:unevaluation}}\\{{\nwixident{nops}}{nops}}\\{{\nwixident{op}}{op}}}\nwindexuse{\nwixident{evalf}}{evalf}{NWppJ6t-DKmU5-g}\nwindexuse{\nwixident{ex}}{ex}{NWppJ6t-DKmU5-g}\nwindexuse{\nwixident{float{\_}evaluation}}{float:unevaluation}{NWppJ6t-DKmU5-g}\nwindexuse{\nwixident{nops}}{nops}{NWppJ6t-DKmU5-g}\nwindexuse{\nwixident{op}}{op}{NWppJ6t-DKmU5-g}\nwendcode{}\nwbegindocs{951}test if system is linear and fill vars matrix
\nwenddocs{}\nwbegincode{952}\sublabel{NWppJ6t-DKmU5-h}\nwmargintag{{\nwtagstyle{}\subpageref{NWppJ6t-DKmU5-h}}}\moddef{figure class~{\nwtagstyle{}\subpageref{NWppJ6t-DKmU5-1}}}\plusendmoddef\Rm{}\nwstartdeflinemarkup\nwusesondefline{\\{NWppJ6t-32jW6L-1}}\nwprevnextdefs{NWppJ6t-DKmU5-g}{NWppJ6t-DKmU5-i}\nwenddeflinemarkup
        {\bf{}bool} {\it{}is\_linear}={\bf{}true};
        {\bf{}for} ({\it{}size\_t} {\it{}i}=0; {\it{}i}\begin{math}<\end{math}{\it{}symbols}.{\it{}nops}(); {\it{}i}\protect\PP)
            {\bf{}if} ({\it{}sys}.{\it{}has}({\it{}symbols}.{\it{}op}({\it{}i})) \begin{math}\vee\end{math} {\it{}linpart}.{\it{}has}({\it{}symbols}.{\it{}op}({\it{}i})))
                {\it{}is\_linear} = {\bf{}false};
        //cerr \begin{math}<\end{math}\begin{math}<\end{math} "this equation linear? " \begin{math}<\end{math}\begin{math}<\end{math} is\_linear \begin{math}<\end{math}\begin{math}<\end{math} endl;

\nwused{\\{NWppJ6t-32jW6L-1}}\nwidentuses{\\{{\nwixident{nops}}{nops}}\\{{\nwixident{op}}{op}}}\nwindexuse{\nwixident{nops}}{nops}{NWppJ6t-DKmU5-h}\nwindexuse{\nwixident{op}}{op}{NWppJ6t-DKmU5-h}\nwendcode{}\nwbegindocs{953}To avoid an expensive expansion we use the previous calculations to
re-build the equation.
\nwenddocs{}\nwbegincode{954}\sublabel{NWppJ6t-DKmU5-i}\nwmargintag{{\nwtagstyle{}\subpageref{NWppJ6t-DKmU5-i}}}\moddef{figure class~{\nwtagstyle{}\subpageref{NWppJ6t-DKmU5-1}}}\plusendmoddef\Rm{}\nwstartdeflinemarkup\nwusesondefline{\\{NWppJ6t-32jW6L-1}}\nwprevnextdefs{NWppJ6t-DKmU5-h}{NWppJ6t-DKmU5-j}\nwenddeflinemarkup
        {\bf{}if} ({\it{}is\_linear}) {\nwlbrace}
            {\it{}lin\_cond}.{\it{}append}({\it{}sys});
            {\it{}rhs}.{\it{}append}({\it{}linpart});
            {\it{}is\_homogeneous} &= {\it{}linpart}.{\it{}normal}().{\it{}is\_zero}();
        {\nwrbrace} {\bf{}else}
            {\it{}nonlin\_cond}.{\it{}append}({\it{}cond}.{\it{}op}({\it{}r}));
    {\nwrbrace}
    //cerr \begin{math}<\end{math}\begin{math}<\end{math} "lin\_cond: "; lin\_cond.dbgprint();
    //cerr \begin{math}<\end{math}\begin{math}<\end{math} "nonlin\_cond: "; nonlin\_cond.dbgprint();

\nwused{\\{NWppJ6t-32jW6L-1}}\nwidentuses{\\{{\nwixident{op}}{op}}}\nwindexuse{\nwixident{op}}{op}{NWppJ6t-DKmU5-i}\nwendcode{}\nwbegindocs{955}Solving the linear part, the code is again stolen from  \texttt{ginac/inifcns.cpp}
\nwenddocs{}\nwbegincode{956}\sublabel{NWppJ6t-DKmU5-j}\nwmargintag{{\nwtagstyle{}\subpageref{NWppJ6t-DKmU5-j}}}\moddef{figure class~{\nwtagstyle{}\subpageref{NWppJ6t-DKmU5-1}}}\plusendmoddef\Rm{}\nwstartdeflinemarkup\nwusesondefline{\\{NWppJ6t-32jW6L-1}}\nwprevnextdefs{NWppJ6t-DKmU5-i}{NWppJ6t-DKmU5-k}\nwenddeflinemarkup
    {\bf{}lst} {\it{}subs\_lst1}, // The main list of substitutions of found solutions
        {\it{}subs\_lst2}, // The second solution lists for quadratic equations
        {\it{}free\_vars}; // List of free variables being parameters of the solution
    {\bf{}if} ({\it{}lin\_cond}.{\it{}nops}()\begin{math}>\end{math}0) {\nwlbrace}
        {\bf{}matrix} {\it{}solution};
        {\bf{}try} {\nwlbrace}
            {\it{}solution}={\it{}ex\_to}\begin{math}<\end{math}{\bf{}matrix}\begin{math}>\end{math}({\it{}lst\_to\_matrix}({\it{}lin\_cond})).{\it{}solve}({\bf{}matrix}({\it{}symbols}.{\it{}nops}(),1,{\it{}symbols}),
                                                                  {\bf{}matrix}({\it{}rhs}.{\it{}nops}(),1,{\it{}rhs}));

\nwused{\\{NWppJ6t-32jW6L-1}}\nwidentuses{\\{{\nwixident{main}}{main}}\\{{\nwixident{nops}}{nops}}}\nwindexuse{\nwixident{main}}{main}{NWppJ6t-DKmU5-j}\nwindexuse{\nwixident{nops}}{nops}{NWppJ6t-DKmU5-j}\nwendcode{}\nwbegindocs{957}If the system is incompatible no cycle data is returned (probably
singular matrix or otherwise overdetermined system, it is consistent to return an empty list)
\nwenddocs{}\nwbegincode{958}\sublabel{NWppJ6t-DKmU5-k}\nwmargintag{{\nwtagstyle{}\subpageref{NWppJ6t-DKmU5-k}}}\moddef{figure class~{\nwtagstyle{}\subpageref{NWppJ6t-DKmU5-1}}}\plusendmoddef\Rm{}\nwstartdeflinemarkup\nwusesondefline{\\{NWppJ6t-32jW6L-1}}\nwprevnextdefs{NWppJ6t-DKmU5-j}{NWppJ6t-DKmU5-l}\nwenddeflinemarkup
        {\nwrbrace} {\bf{}catch} ({\bf{}const} {\it{}std}::{\it{}runtime\_error} & {\it{}e}) {\nwlbrace}
            {\bf{}return} {\bf{}lst}{\nwlbrace}{\nwrbrace};
        {\nwrbrace}
        {\it{}GINAC\_ASSERT}({\it{}solution}.{\it{}cols}()\begin{math}\equiv\end{math}1);
        {\it{}GINAC\_ASSERT}({\it{}solution}.{\it{}rows}()\begin{math}\equiv\end{math}{\it{}symbols}.{\it{}nops}());

\nwused{\\{NWppJ6t-32jW6L-1}}\nwidentuses{\\{{\nwixident{nops}}{nops}}}\nwindexuse{\nwixident{nops}}{nops}{NWppJ6t-DKmU5-k}\nwendcode{}\nwbegindocs{959}Now we sort out the result: free variables will be used for
non-linear equation, resolved variables---for substitution.
\nwenddocs{}\nwbegincode{960}\sublabel{NWppJ6t-DKmU5-l}\nwmargintag{{\nwtagstyle{}\subpageref{NWppJ6t-DKmU5-l}}}\moddef{figure class~{\nwtagstyle{}\subpageref{NWppJ6t-DKmU5-1}}}\plusendmoddef\Rm{}\nwstartdeflinemarkup\nwusesondefline{\\{NWppJ6t-32jW6L-1}}\nwprevnextdefs{NWppJ6t-DKmU5-k}{NWppJ6t-DKmU5-m}\nwenddeflinemarkup
        {\bf{}for} ({\it{}size\_t} {\it{}i}=0; {\it{}i}\begin{math}<\end{math}{\it{}symbols}.{\it{}nops}(); {\it{}i}\protect\PP)
            {\bf{}if} ({\it{}symbols}.{\it{}op}({\it{}i})\begin{math}\equiv\end{math}{\it{}solution}({\it{}i},0))
                {\it{}free\_vars}.{\it{}append}({\it{}symbols}.{\it{}op}({\it{}i}));
            {\bf{}else}
                {\it{}subs\_lst1}.{\it{}append}({\it{}symbols}.{\it{}op}({\it{}i})\begin{math}\equiv\end{math}{\it{}solution}({\it{}i},0));
    {\nwrbrace}
    //cerr \begin{math}<\end{math}\begin{math}<\end{math} "Lin system is homogeneous: " \begin{math}<\end{math}\begin{math}<\end{math} is\_homogeneous \begin{math}<\end{math}\begin{math}<\end{math} endl;

\nwused{\\{NWppJ6t-32jW6L-1}}\nwidentuses{\\{{\nwixident{nops}}{nops}}\\{{\nwixident{op}}{op}}}\nwindexuse{\nwixident{nops}}{nops}{NWppJ6t-DKmU5-l}\nwindexuse{\nwixident{op}}{op}{NWppJ6t-DKmU5-l}\nwendcode{}\nwbegindocs{961}It is easy to solve a linear system, thus we immediate substitute
the result.
\nwenddocs{}\nwbegincode{962}\sublabel{NWppJ6t-DKmU5-m}\nwmargintag{{\nwtagstyle{}\subpageref{NWppJ6t-DKmU5-m}}}\moddef{figure class~{\nwtagstyle{}\subpageref{NWppJ6t-DKmU5-1}}}\plusendmoddef\Rm{}\nwstartdeflinemarkup\nwusesondefline{\\{NWppJ6t-32jW6L-1}}\nwprevnextdefs{NWppJ6t-DKmU5-l}{NWppJ6t-DKmU5-n}\nwenddeflinemarkup
    {\bf{}cycle\_data} {\it{}C\_new}, {\it{}C1\_new};
    {\bf{}if} ({\it{}nonlin\_cond}.{\it{}nops}() \begin{math}\equiv\end{math} 0) {\nwlbrace}
        {\it{}C\_new} = {\it{}ex\_to}\begin{math}<\end{math}{\bf{}cycle\_data}\begin{math}>\end{math}({\it{}symbolic}.{\it{}subs}({\it{}subs\_lst1})).{\it{}normalize}();
        //cerr \begin{math}<\end{math}\begin{math}<\end{math} "C\_new: "; C\_new.dbgprint();

\nwused{\\{NWppJ6t-32jW6L-1}}\nwidentuses{\\{{\nwixident{cycle{\_}data}}{cycle:undata}}\\{{\nwixident{nops}}{nops}}\\{{\nwixident{subs}}{subs}}}\nwindexuse{\nwixident{cycle{\_}data}}{cycle:undata}{NWppJ6t-DKmU5-m}\nwindexuse{\nwixident{nops}}{nops}{NWppJ6t-DKmU5-m}\nwindexuse{\nwixident{subs}}{subs}{NWppJ6t-DKmU5-m}\nwendcode{}\nwbegindocs{963}We check that the solution is not identical zero, which may happen
for homogeneous conditions, for example. For this we
prepare the respective norm of the cycle.
\nwenddocs{}\nwbegincode{964}\sublabel{NWppJ6t-DKmU5-n}\nwmargintag{{\nwtagstyle{}\subpageref{NWppJ6t-DKmU5-n}}}\moddef{figure class~{\nwtagstyle{}\subpageref{NWppJ6t-DKmU5-1}}}\plusendmoddef\Rm{}\nwstartdeflinemarkup\nwusesondefline{\\{NWppJ6t-32jW6L-1}}\nwprevnextdefs{NWppJ6t-DKmU5-m}{NWppJ6t-DKmU5-o}\nwenddeflinemarkup
    {\bf{}ex} {\it{}norm}={\it{}pow}({\it{}ex\_to}\begin{math}<\end{math}{\bf{}cycle\_data}\begin{math}>\end{math}({\it{}symbolic}).{\it{}get\_k}(),2)+{\it{}pow}({\it{}ex\_to}\begin{math}<\end{math}{\bf{}cycle\_data}\begin{math}>\end{math}({\it{}symbolic}).{\it{}get\_m}(),2);
    {\bf{}for} ({\bf{}int} {\it{}i} = 0; {\it{}i} \begin{math}<\end{math} {\it{}D}; {\it{}i}\protect\PP)
        {\it{}norm}+={\it{}pow}({\it{}ex\_to}\begin{math}<\end{math}{\bf{}cycle\_data}\begin{math}>\end{math}({\it{}symbolic}).{\it{}get\_l}({\it{}i}),2);
    {\it{}first\_solution} &= \begin{math}\neg\end{math} {\it{}is\_less\_than\_epsilon}({\it{}norm}.{\it{}subs}({\it{}subs\_lst1},
                                                       {\it{}subs\_options}::{\it{}algebraic} \begin{math}\mid\end{math} {\it{}subs\_options}::{\it{}no\_pattern}));

\nwused{\\{NWppJ6t-32jW6L-1}}\nwidentuses{\\{{\nwixident{cycle{\_}data}}{cycle:undata}}\\{{\nwixident{ex}}{ex}}\\{{\nwixident{is{\_}less{\_}than{\_}epsilon}}{is:unless:unthan:unepsilon}}\\{{\nwixident{subs}}{subs}}}\nwindexuse{\nwixident{cycle{\_}data}}{cycle:undata}{NWppJ6t-DKmU5-n}\nwindexuse{\nwixident{ex}}{ex}{NWppJ6t-DKmU5-n}\nwindexuse{\nwixident{is{\_}less{\_}than{\_}epsilon}}{is:unless:unthan:unepsilon}{NWppJ6t-DKmU5-n}\nwindexuse{\nwixident{subs}}{subs}{NWppJ6t-DKmU5-n}\nwendcode{}\nwbegindocs{965}If some non-linear equations present and there are free variables,
we sort out free and non-free variables.
\nwenddocs{}\nwbegincode{966}\sublabel{NWppJ6t-DKmU5-o}\nwmargintag{{\nwtagstyle{}\subpageref{NWppJ6t-DKmU5-o}}}\moddef{figure class~{\nwtagstyle{}\subpageref{NWppJ6t-DKmU5-1}}}\plusendmoddef\Rm{}\nwstartdeflinemarkup\nwusesondefline{\\{NWppJ6t-32jW6L-1}}\nwprevnextdefs{NWppJ6t-DKmU5-n}{NWppJ6t-DKmU5-p}\nwenddeflinemarkup
    {\nwrbrace} {\bf{}else} {\bf{}if} ({\it{}free\_vars}.{\it{}nops}() \begin{math}>\end{math} 0) {\nwlbrace}
        {\bf{}lst} {\it{}nonlin\_cond\_new};
        //cerr \begin{math}<\end{math}\begin{math}<\end{math} "free\_vars: "; free\_vars.dbgprint();
        //cerr \begin{math}<\end{math}\begin{math}<\end{math} "subs\_lst1: "; subs\_lst1.dbgprint();

\nwused{\\{NWppJ6t-32jW6L-1}}\nwidentuses{\\{{\nwixident{nops}}{nops}}}\nwindexuse{\nwixident{nops}}{nops}{NWppJ6t-DKmU5-o}\nwendcode{}\nwbegindocs{967}Only one non-linear (quadratic) equation can be treated by this method, so we
pick up the first from the list (hopefully other will be satisfied afterwards).
\nwenddocs{}\nwbegincode{968}\sublabel{NWppJ6t-DKmU5-p}\nwmargintag{{\nwtagstyle{}\subpageref{NWppJ6t-DKmU5-p}}}\moddef{figure class~{\nwtagstyle{}\subpageref{NWppJ6t-DKmU5-1}}}\plusendmoddef\Rm{}\nwstartdeflinemarkup\nwusesondefline{\\{NWppJ6t-32jW6L-1}}\nwprevnextdefs{NWppJ6t-DKmU5-o}{NWppJ6t-DKmU5-q}\nwenddeflinemarkup
        {\bf{}ex} {\it{}quadratic\_eq}={\it{}nonlin\_cond}.{\it{}op}(0).{\it{}subs}({\it{}subs\_lst1}, {\it{}subs\_options}::{\it{}algebraic}
                                               \begin{math}\mid\end{math} {\it{}subs\_options}::{\it{}no\_pattern});
        {\bf{}ex} {\it{}quadratic}=({\it{}quadratic\_eq}.{\it{}op}(0)-{\it{}quadratic\_eq}.{\it{}op}(1)).{\it{}expand}().{\it{}normal}()
            .{\it{}subs}({\it{}evaluation\_assist},{\it{}subs\_options}::{\it{}algebraic}).{\it{}normal}();
        {\bf{}if} ({\it{}float\_evaluation})
            {\it{}quadratic}={\it{}quadratic}.{\it{}evalf}();
        //cerr \begin{math}<\end{math}\begin{math}<\end{math} "quadratic: "; quadratic.dbgprint();

\nwused{\\{NWppJ6t-32jW6L-1}}\nwidentuses{\\{{\nwixident{evalf}}{evalf}}\\{{\nwixident{evaluation{\_}assist}}{evaluation:unassist}}\\{{\nwixident{ex}}{ex}}\\{{\nwixident{float{\_}evaluation}}{float:unevaluation}}\\{{\nwixident{op}}{op}}\\{{\nwixident{subs}}{subs}}}\nwindexuse{\nwixident{evalf}}{evalf}{NWppJ6t-DKmU5-p}\nwindexuse{\nwixident{evaluation{\_}assist}}{evaluation:unassist}{NWppJ6t-DKmU5-p}\nwindexuse{\nwixident{ex}}{ex}{NWppJ6t-DKmU5-p}\nwindexuse{\nwixident{float{\_}evaluation}}{float:unevaluation}{NWppJ6t-DKmU5-p}\nwindexuse{\nwixident{op}}{op}{NWppJ6t-DKmU5-p}\nwindexuse{\nwixident{subs}}{subs}{NWppJ6t-DKmU5-p}\nwendcode{}\nwbegindocs{969}We reduce the list of free variables to only present in the quadratic.
\nwenddocs{}\nwbegincode{970}\sublabel{NWppJ6t-DKmU5-q}\nwmargintag{{\nwtagstyle{}\subpageref{NWppJ6t-DKmU5-q}}}\moddef{figure class~{\nwtagstyle{}\subpageref{NWppJ6t-DKmU5-1}}}\plusendmoddef\Rm{}\nwstartdeflinemarkup\nwusesondefline{\\{NWppJ6t-32jW6L-1}}\nwprevnextdefs{NWppJ6t-DKmU5-p}{NWppJ6t-DKmU5-r}\nwenddeflinemarkup
        {\bf{}lst} {\it{}quadratic\_list};
        {\bf{}for} ({\it{}size\_t} {\it{}i}=0; {\it{}i} \begin{math}<\end{math} {\it{}free\_vars}.{\it{}nops}(); \protect\PP{\it{}i})
            {\bf{}if} ({\it{}quadratic}.{\it{}has}({\it{}free\_vars}.{\it{}op}({\it{}i})))
                {\it{}quadratic\_list}.{\it{}append}({\it{}free\_vars}.{\it{}op}({\it{}i}));
        {\it{}free\_vars}={\it{}ex\_to}\begin{math}<\end{math}{\bf{}lst}\begin{math}>\end{math}({\it{}quadratic\_list});
        //cerr \begin{math}<\end{math}\begin{math}<\end{math} "free\_vars which are present: "; free\_vars.dbgprint();

\nwused{\\{NWppJ6t-32jW6L-1}}\nwidentuses{\\{{\nwixident{nops}}{nops}}\\{{\nwixident{op}}{op}}}\nwindexuse{\nwixident{nops}}{nops}{NWppJ6t-DKmU5-q}\nwindexuse{\nwixident{op}}{op}{NWppJ6t-DKmU5-q}\nwendcode{}\nwbegindocs{971}We check homogeneity of the quadratic equation.
\nwenddocs{}\nwbegincode{972}\sublabel{NWppJ6t-DKmU5-r}\nwmargintag{{\nwtagstyle{}\subpageref{NWppJ6t-DKmU5-r}}}\moddef{figure class~{\nwtagstyle{}\subpageref{NWppJ6t-DKmU5-1}}}\plusendmoddef\Rm{}\nwstartdeflinemarkup\nwusesondefline{\\{NWppJ6t-32jW6L-1}}\nwprevnextdefs{NWppJ6t-DKmU5-q}{NWppJ6t-DKmU5-s}\nwenddeflinemarkup
        {\bf{}if} ({\it{}is\_homogeneous}) {\nwlbrace}
            {\bf{}ex} {\it{}Q}={\it{}quadratic};
            {\bf{}for} ({\it{}size\_t} {\it{}i}=1; {\it{}i} \begin{math}<\end{math} {\it{}free\_vars}.{\it{}nops}(); \protect\PP{\it{}i})
                {\it{}Q}={\it{}Q}.{\it{}subs}({\it{}free\_vars}.{\it{}op}({\it{}i})\begin{math}\equiv\end{math}{\it{}free\_vars}.{\it{}op}(0));
            {\it{}is\_homogeneous} &= ({\it{}Q}.{\it{}degree}({\it{}free\_vars}.{\it{}op}(0))\begin{math}\equiv\end{math}{\it{}Q}.{\it{}ldegree}({\it{}free\_vars}.{\it{}op}(0)));
        {\nwrbrace}
        //cerr \begin{math}<\end{math}\begin{math}<\end{math} "Quadratic part is homogeneous: " \begin{math}<\end{math}\begin{math}<\end{math} is\_homogeneous \begin{math}<\end{math}\begin{math}<\end{math} endl;

\nwused{\\{NWppJ6t-32jW6L-1}}\nwidentuses{\\{{\nwixident{ex}}{ex}}\\{{\nwixident{nops}}{nops}}\\{{\nwixident{op}}{op}}\\{{\nwixident{subs}}{subs}}}\nwindexuse{\nwixident{ex}}{ex}{NWppJ6t-DKmU5-r}\nwindexuse{\nwixident{nops}}{nops}{NWppJ6t-DKmU5-r}\nwindexuse{\nwixident{op}}{op}{NWppJ6t-DKmU5-r}\nwindexuse{\nwixident{subs}}{subs}{NWppJ6t-DKmU5-r}\nwendcode{}\nwbegindocs{973}The equation may be linear for a particular free variable, we will
search if it is.
\nwenddocs{}\nwbegincode{974}\sublabel{NWppJ6t-DKmU5-s}\nwmargintag{{\nwtagstyle{}\subpageref{NWppJ6t-DKmU5-s}}}\moddef{figure class~{\nwtagstyle{}\subpageref{NWppJ6t-DKmU5-1}}}\plusendmoddef\Rm{}\nwstartdeflinemarkup\nwusesondefline{\\{NWppJ6t-32jW6L-1}}\nwprevnextdefs{NWppJ6t-DKmU5-r}{NWppJ6t-DKmU5-t}\nwenddeflinemarkup
        {\bf{}bool} {\it{}is\_quadratic}={\bf{}true};
        {\it{}exmap} {\it{}flat\_var\_em}, {\it{}var1\_em}, {\it{}var2\_em};
        {\bf{}ex} {\it{}flat\_var}, {\it{}var1}, {\it{}var2};

\nwused{\\{NWppJ6t-32jW6L-1}}\nwidentuses{\\{{\nwixident{ex}}{ex}}}\nwindexuse{\nwixident{ex}}{ex}{NWppJ6t-DKmU5-s}\nwendcode{}\nwbegindocs{975}We now search if for some free variable the equation is linear
\nwenddocs{}\nwbegincode{976}\sublabel{NWppJ6t-DKmU5-t}\nwmargintag{{\nwtagstyle{}\subpageref{NWppJ6t-DKmU5-t}}}\moddef{figure class~{\nwtagstyle{}\subpageref{NWppJ6t-DKmU5-1}}}\plusendmoddef\Rm{}\nwstartdeflinemarkup\nwusesondefline{\\{NWppJ6t-32jW6L-1}}\nwprevnextdefs{NWppJ6t-DKmU5-s}{NWppJ6t-DKmU5-u}\nwenddeflinemarkup
        {\it{}size\_t} {\it{}i}=0;
        {\bf{}for} (; {\it{}i} \begin{math}<\end{math} {\it{}free\_vars}.{\it{}nops}(); \protect\PP{\it{}i}) {\nwlbrace}
            //cerr \begin{math}<\end{math}\begin{math}<\end{math} "degree: " \begin{math}<\end{math}\begin{math}<\end{math} quadratic.degree(free\_vars.op(i)) \begin{math}<\end{math}\begin{math}<\end{math} endl;
            {\bf{}if} ({\it{}quadratic}.{\it{}degree}({\it{}free\_vars}.{\it{}op}({\it{}i})) \begin{math}<\end{math} 2) {\nwlbrace}
                {\it{}is\_quadratic}={\bf{}false};
                //cerr \begin{math}<\end{math}\begin{math}<\end{math} "Equation is linear in "; free\_vars.op(i).dbgprint();
                {\bf{}break};
            {\nwrbrace}
        {\nwrbrace}

\nwused{\\{NWppJ6t-32jW6L-1}}\nwidentuses{\\{{\nwixident{nops}}{nops}}\\{{\nwixident{op}}{op}}}\nwindexuse{\nwixident{nops}}{nops}{NWppJ6t-DKmU5-t}\nwindexuse{\nwixident{op}}{op}{NWppJ6t-DKmU5-t}\nwendcode{}\nwbegindocs{977}If all equations are quadratic in any variable, we use homogenuity
to reduce the last free variable.
\nwenddocs{}\nwbegincode{978}\sublabel{NWppJ6t-DKmU5-u}\nwmargintag{{\nwtagstyle{}\subpageref{NWppJ6t-DKmU5-u}}}\moddef{figure class~{\nwtagstyle{}\subpageref{NWppJ6t-DKmU5-1}}}\plusendmoddef\Rm{}\nwstartdeflinemarkup\nwusesondefline{\\{NWppJ6t-32jW6L-1}}\nwprevnextdefs{NWppJ6t-DKmU5-t}{NWppJ6t-DKmU5-v}\nwenddeflinemarkup
        {\bf{}if} ({\it{}is\_quadratic}) {\nwlbrace}
            {\bf{}if} ({\it{}is\_homogeneous} \begin{math}\wedge\end{math} {\it{}free\_vars}.{\it{}nops}() \begin{math}>\end{math} 1) {\nwlbrace}
                {\it{}exmap} {\it{}erase\_var};
                {\it{}erase\_var}.{\it{}insert}({\it{}std}::{\it{}make\_pair}({\it{}free\_vars}.{\it{}op}({\it{}free\_vars}.{\it{}nops}()-1), {\bf{}numeric}(1)));
                {\it{}subs\_lst1}={\it{}ex\_to}\begin{math}<\end{math}{\bf{}lst}\begin{math}>\end{math}({\it{}subs\_lst1}.{\it{}subs}({\it{}erase\_var},
                                                    {\it{}subs\_options}::{\it{}algebraic} \begin{math}\mid\end{math} {\it{}subs\_options}::{\it{}no\_pattern}));
                {\it{}subs\_lst1}.{\it{}append}({\it{}free\_vars}.{\it{}op}({\it{}free\_vars}.{\it{}nops}()-1) \begin{math}\equiv\end{math} {\bf{}numeric}(1));
                {\it{}quadratic}={\it{}quadratic}.{\it{}subs}({\it{}free\_vars}.{\it{}op}({\it{}free\_vars}.{\it{}nops}()-1) \begin{math}\equiv\end{math} {\bf{}numeric}(1));
                {\it{}free\_vars}.{\it{}remove\_last}();
                //cerr \begin{math}<\end{math}\begin{math}<\end{math} "Quadratic reduced by homogenuity: "; quadratic.dbgprint();
            {\nwrbrace}

\nwused{\\{NWppJ6t-32jW6L-1}}\nwidentuses{\\{{\nwixident{nops}}{nops}}\\{{\nwixident{numeric}}{numeric}}\\{{\nwixident{op}}{op}}\\{{\nwixident{subs}}{subs}}}\nwindexuse{\nwixident{nops}}{nops}{NWppJ6t-DKmU5-u}\nwindexuse{\nwixident{numeric}}{numeric}{NWppJ6t-DKmU5-u}\nwindexuse{\nwixident{op}}{op}{NWppJ6t-DKmU5-u}\nwindexuse{\nwixident{subs}}{subs}{NWppJ6t-DKmU5-u}\nwendcode{}\nwbegindocs{979}and then proceed with solving of quadratic equation for each free
variable attempting to find root-free presentation.
\nwenddocs{}\nwbegincode{980}\sublabel{NWppJ6t-DKmU5-v}\nwmargintag{{\nwtagstyle{}\subpageref{NWppJ6t-DKmU5-v}}}\moddef{figure class~{\nwtagstyle{}\subpageref{NWppJ6t-DKmU5-1}}}\plusendmoddef\Rm{}\nwstartdeflinemarkup\nwusesondefline{\\{NWppJ6t-32jW6L-1}}\nwprevnextdefs{NWppJ6t-DKmU5-u}{NWppJ6t-DKmU5-w}\nwenddeflinemarkup
            {\bf{}ex} {\it{}A}, {\it{}B}, {\it{}C}, {\it{}D}, {\it{}sqrtD};
            {\bf{}for}({\it{}i}=0; {\it{}i} \begin{math}<\end{math} {\it{}free\_vars}.{\it{}nops}(); \protect\PP{\it{}i}) {\nwlbrace}
                {\it{}A}={\it{}quadratic}.{\it{}coeff}({\it{}free\_vars}.{\it{}op}({\it{}i}),2).{\it{}normal}();
                //cerr \begin{math}<\end{math}\begin{math}<\end{math} "A: "; A.dbgprint();
                {\it{}B}={\it{}quadratic}.{\it{}coeff}({\it{}free\_vars}.{\it{}op}({\it{}i}),1);
                {\it{}C}={\it{}quadratic}.{\it{}coeff}({\it{}free\_vars}.{\it{}op}({\it{}i}),0);
                {\it{}D}=({\it{}pow}({\it{}B},2)-{\bf{}numeric}(4)\begin{math}\ast\end{math}{\it{}A}\begin{math}\ast\end{math}{\it{}C}).{\it{}normal}();
                {\it{}sqrtD}={\it{}sqrt}({\it{}D});
                //cerr \begin{math}<\end{math}\begin{math}<\end{math} "D: "; D.dbgprint();

\nwused{\\{NWppJ6t-32jW6L-1}}\nwidentuses{\\{{\nwixident{ex}}{ex}}\\{{\nwixident{nops}}{nops}}\\{{\nwixident{numeric}}{numeric}}\\{{\nwixident{op}}{op}}}\nwindexuse{\nwixident{ex}}{ex}{NWppJ6t-DKmU5-v}\nwindexuse{\nwixident{nops}}{nops}{NWppJ6t-DKmU5-v}\nwindexuse{\nwixident{numeric}}{numeric}{NWppJ6t-DKmU5-v}\nwindexuse{\nwixident{op}}{op}{NWppJ6t-DKmU5-v}\nwendcode{}\nwbegindocs{981}For the condition of real coefficients, we are checking whether
another free variable survived in the discriminant of the quadratic equation.\\
TODO: this process need to be recursive for all free variables, not
just for one as it is now.
\nwenddocs{}\nwbegincode{982}\sublabel{NWppJ6t-DKmU5-w}\nwmargintag{{\nwtagstyle{}\subpageref{NWppJ6t-DKmU5-w}}}\moddef{figure class~{\nwtagstyle{}\subpageref{NWppJ6t-DKmU5-1}}}\plusendmoddef\Rm{}\nwstartdeflinemarkup\nwusesondefline{\\{NWppJ6t-32jW6L-1}}\nwprevnextdefs{NWppJ6t-DKmU5-v}{NWppJ6t-DKmU5-x}\nwenddeflinemarkup
                {\bf{}if} (//need\_reals &&
                    {\it{}free\_vars}.{\it{}nops}()\begin{math}>\end{math}1) {\nwlbrace}
                    {\bf{}int} {\it{}another}=0;
                    {\bf{}if} ({\it{}i}\begin{math}\equiv\end{math}0)
                        {\it{}another}=1;

\nwused{\\{NWppJ6t-32jW6L-1}}\nwidentuses{\\{{\nwixident{nops}}{nops}}}\nwindexuse{\nwixident{nops}}{nops}{NWppJ6t-DKmU5-w}\nwendcode{}\nwbegindocs{983}If another free variable, denoted \(x\) here, presents in the
discriminant \(D=A_1 x^2+B_1 x+C_1\), we try some hyperbolic or
trigonometric substitutions.
\nwenddocs{}\nwbegincode{984}\sublabel{NWppJ6t-DKmU5-x}\nwmargintag{{\nwtagstyle{}\subpageref{NWppJ6t-DKmU5-x}}}\moddef{figure class~{\nwtagstyle{}\subpageref{NWppJ6t-DKmU5-1}}}\plusendmoddef\Rm{}\nwstartdeflinemarkup\nwusesondefline{\\{NWppJ6t-32jW6L-1}}\nwprevnextdefs{NWppJ6t-DKmU5-w}{NWppJ6t-DKmU5-y}\nwenddeflinemarkup
                    {\bf{}if} ({\it{}not} {\it{}is\_less\_than\_epsilon}({\it{}D}) \begin{math}\wedge\end{math} {\it{}D}.{\it{}has}({\it{}free\_vars}.{\it{}op}({\it{}another}))) {\nwlbrace}
                        {\bf{}ex} {\it{}A1}={\it{}D}.{\it{}coeff}({\it{}free\_vars}.{\it{}op}({\it{}another}),2)
                        .{\it{}subs}({\it{}evaluation\_assist},{\it{}subs\_options}::{\it{}algebraic}).{\it{}normal}(),
                        {\it{}B1}={\it{}D}.{\it{}coeff}({\it{}free\_vars}.{\it{}op}({\it{}another}),1)
                        .{\it{}subs}({\it{}evaluation\_assist},{\it{}subs\_options}::{\it{}algebraic}).{\it{}normal}(),
                        {\it{}C1}={\it{}D}.{\it{}coeff}({\it{}free\_vars}.{\it{}op}({\it{}another}),0)
                        .{\it{}subs}({\it{}evaluation\_assist},{\it{}subs\_options}::{\it{}algebraic}).{\it{}normal}(),
                        {\it{}D1}=({\it{}pow}({\it{}B1},2)-4\begin{math}\ast\end{math}{\it{}A1}\begin{math}\ast\end{math}{\it{}C1}).{\it{}normal}();
                        //cerr \begin{math}<\end{math}\begin{math}<\end{math} "Atempt to resolve square root for A1=" \begin{math}<\end{math}\begin{math}<\end{math} A1;
                        //cerr \begin{math}<\end{math}\begin{math}<\end{math} ", B1=" \begin{math}<\end{math}\begin{math}<\end{math} B1 \begin{math}<\end{math}\begin{math}<\end{math} ", C1=" \begin{math}<\end{math}\begin{math}<\end{math} C1 \begin{math}<\end{math}\begin{math}<\end{math} ", D1=" \begin{math}<\end{math}\begin{math}<\end{math} D1 \begin{math}<\end{math}\begin{math}<\end{math} endl;

\nwused{\\{NWppJ6t-32jW6L-1}}\nwidentuses{\\{{\nwixident{evaluation{\_}assist}}{evaluation:unassist}}\\{{\nwixident{ex}}{ex}}\\{{\nwixident{is{\_}less{\_}than{\_}epsilon}}{is:unless:unthan:unepsilon}}\\{{\nwixident{op}}{op}}\\{{\nwixident{subs}}{subs}}}\nwindexuse{\nwixident{evaluation{\_}assist}}{evaluation:unassist}{NWppJ6t-DKmU5-x}\nwindexuse{\nwixident{ex}}{ex}{NWppJ6t-DKmU5-x}\nwindexuse{\nwixident{is{\_}less{\_}than{\_}epsilon}}{is:unless:unthan:unepsilon}{NWppJ6t-DKmU5-x}\nwindexuse{\nwixident{op}}{op}{NWppJ6t-DKmU5-x}\nwindexuse{\nwixident{subs}}{subs}{NWppJ6t-DKmU5-x}\nwendcode{}\nwbegindocs{985}If the expression is linear, we make a substitution \(D=B_1
x+C_1=y^2\), thus \(x=(y^2-C_1)/B_1\).
\nwenddocs{}\nwbegincode{986}\sublabel{NWppJ6t-DKmU5-y}\nwmargintag{{\nwtagstyle{}\subpageref{NWppJ6t-DKmU5-y}}}\moddef{figure class~{\nwtagstyle{}\subpageref{NWppJ6t-DKmU5-1}}}\plusendmoddef\Rm{}\nwstartdeflinemarkup\nwusesondefline{\\{NWppJ6t-32jW6L-1}}\nwprevnextdefs{NWppJ6t-DKmU5-x}{NWppJ6t-DKmU5-z}\nwenddeflinemarkup
                        {\bf{}if} ({\it{}is\_less\_than\_epsilon}({\it{}A1}) \begin{math}\wedge\end{math} {\it{}not} {\it{}is\_less\_than\_epsilon}({\it{}B1})) {\nwlbrace}
                            {\bf{}ex} {\it{}y}={\bf{}realsymbol}(),
                            {\it{}x}=({\it{}pow}({\it{}y},2)-{\it{}C1})\begin{math}\div\end{math}{\it{}B1};
                            {\it{}sqrtD}={\it{}y};
                            {\it{}flat\_var\_em}.{\it{}insert}({\it{}std}::{\it{}make\_pair}({\it{}free\_vars}.{\it{}op}({\it{}another}), {\it{}x}));
                            {\it{}flat\_var}=({\it{}free\_vars}.{\it{}op}({\it{}another})\begin{math}\equiv\end{math}{\it{}x});

\nwused{\\{NWppJ6t-32jW6L-1}}\nwidentuses{\\{{\nwixident{ex}}{ex}}\\{{\nwixident{is{\_}less{\_}than{\_}epsilon}}{is:unless:unthan:unepsilon}}\\{{\nwixident{op}}{op}}\\{{\nwixident{realsymbol}}{realsymbol}}}\nwindexuse{\nwixident{ex}}{ex}{NWppJ6t-DKmU5-y}\nwindexuse{\nwixident{is{\_}less{\_}than{\_}epsilon}}{is:unless:unthan:unepsilon}{NWppJ6t-DKmU5-y}\nwindexuse{\nwixident{op}}{op}{NWppJ6t-DKmU5-y}\nwindexuse{\nwixident{realsymbol}}{realsymbol}{NWppJ6t-DKmU5-y}\nwendcode{}\nwbegindocs{987}If \(A_1\) is positive, then the substitution depends on sign of the
second discriminant \(D_1=B_1^2-4A_1C_1\)
\nwenddocs{}\nwbegincode{988}\sublabel{NWppJ6t-DKmU5-z}\nwmargintag{{\nwtagstyle{}\subpageref{NWppJ6t-DKmU5-z}}}\moddef{figure class~{\nwtagstyle{}\subpageref{NWppJ6t-DKmU5-1}}}\plusendmoddef\Rm{}\nwstartdeflinemarkup\nwusesondefline{\\{NWppJ6t-32jW6L-1}}\nwprevnextdefs{NWppJ6t-DKmU5-y}{NWppJ6t-DKmU5-10}\nwenddeflinemarkup
                        {\nwrbrace} {\bf{}else} {\bf{}if} ({\it{}A1}.{\it{}evalf}().{\it{}info}({\it{}info\_flags}::{\it{}positive})) {\nwlbrace}

\nwused{\\{NWppJ6t-32jW6L-1}}\nwidentuses{\\{{\nwixident{evalf}}{evalf}}\\{{\nwixident{info}}{info}}}\nwindexuse{\nwixident{evalf}}{evalf}{NWppJ6t-DKmU5-z}\nwindexuse{\nwixident{info}}{info}{NWppJ6t-DKmU5-z}\nwendcode{}\nwbegindocs{989}Depending on the sign of \(D_1\) and thus \(C_1-B_1^2/(4A_1)\) we are using either
hyperbolic sine or cosine.
\nwenddocs{}\nwbegincode{990}\sublabel{NWppJ6t-DKmU5-10}\nwmargintag{{\nwtagstyle{}\subpageref{NWppJ6t-DKmU5-10}}}\moddef{figure class~{\nwtagstyle{}\subpageref{NWppJ6t-DKmU5-1}}}\plusendmoddef\Rm{}\nwstartdeflinemarkup\nwusesondefline{\\{NWppJ6t-32jW6L-1}}\nwprevnextdefs{NWppJ6t-DKmU5-z}{NWppJ6t-DKmU5-11}\nwenddeflinemarkup
                            {\bf{}if} ({\it{}D1}.{\it{}info}({\it{}info\_flags}::{\it{}negative})) {\nwlbrace}
                                {\bf{}ex} {\it{}y}={\bf{}realsymbol}(),
                                {\it{}x}=({\it{}sinh}({\it{}y})\begin{math}\ast\end{math}{\it{}sqrt}(-{\it{}D1})-{\it{}B1})\begin{math}\div\end{math}2\begin{math}\div\end{math}{\it{}A1};
                                {\it{}sqrtD}={\it{}sqrt}({\it{}C1}-{\it{}pow}({\it{}B1},2)\begin{math}\div\end{math}4\begin{math}\div\end{math}{\it{}A1})\begin{math}\ast\end{math}{\it{}cosh}({\it{}y});
                                {\it{}flat\_var\_em}.{\it{}insert}({\it{}std}::{\it{}make\_pair}({\it{}free\_vars}.{\it{}op}({\it{}another}), {\it{}x}));
                                {\it{}flat\_var}=({\it{}free\_vars}.{\it{}op}({\it{}another})\begin{math}\equiv\end{math}{\it{}x});
                            {\nwrbrace} {\bf{}else} {\bf{}if} ({\it{}D1}.{\it{}info}({\it{}info\_flags}::{\it{}positive})) {\nwlbrace}
                                {\bf{}ex} {\it{}y}={\bf{}realsymbol}(),
                                {\it{}x}=({\it{}cosh}({\it{}y})\begin{math}\ast\end{math}{\it{}sqrt}({\it{}D1})-{\it{}B1})\begin{math}\div\end{math}2\begin{math}\div\end{math}{\it{}A1};
                                {\it{}sqrtD}={\it{}sqrt}({\it{}pow}({\it{}B1},2)\begin{math}\div\end{math}4\begin{math}\div\end{math}{\it{}A1}-{\it{}C1})\begin{math}\ast\end{math}{\it{}sinh}({\it{}y});
                                {\it{}flat\_var\_em}.{\it{}insert}({\it{}std}::{\it{}make\_pair}({\it{}free\_vars}.{\it{}op}({\it{}another}), {\it{}x}));
                                {\it{}flat\_var}=({\it{}free\_vars}.{\it{}op}({\it{}another})\begin{math}\equiv\end{math}{\it{}x});
                            {\nwrbrace}

\nwused{\\{NWppJ6t-32jW6L-1}}\nwidentuses{\\{{\nwixident{ex}}{ex}}\\{{\nwixident{info}}{info}}\\{{\nwixident{op}}{op}}\\{{\nwixident{realsymbol}}{realsymbol}}}\nwindexuse{\nwixident{ex}}{ex}{NWppJ6t-DKmU5-10}\nwindexuse{\nwixident{info}}{info}{NWppJ6t-DKmU5-10}\nwindexuse{\nwixident{op}}{op}{NWppJ6t-DKmU5-10}\nwindexuse{\nwixident{realsymbol}}{realsymbol}{NWppJ6t-DKmU5-10}\nwendcode{}\nwbegindocs{991}If \(A_1\) is negative and \(C_1-B_1^2/(4A_1)>0\) we use the
trigonometric substitution \((2A_1 x+B_1)/\sqrt{4A_1C_1-B_1^2}=\cos y\).
\nwenddocs{}\nwbegincode{992}\sublabel{NWppJ6t-DKmU5-11}\nwmargintag{{\nwtagstyle{}\subpageref{NWppJ6t-DKmU5-11}}}\moddef{figure class~{\nwtagstyle{}\subpageref{NWppJ6t-DKmU5-1}}}\plusendmoddef\Rm{}\nwstartdeflinemarkup\nwusesondefline{\\{NWppJ6t-32jW6L-1}}\nwprevnextdefs{NWppJ6t-DKmU5-10}{NWppJ6t-DKmU5-12}\nwenddeflinemarkup
                        {\nwrbrace} {\bf{}else} {\bf{}if} ({\it{}A1}.{\it{}evalf}().{\it{}info}({\it{}info\_flags}::{\it{}negative})) {\nwlbrace}
                            {\bf{}if} ({\it{}D1}.{\it{}info}({\it{}info\_flags}::{\it{}negative})) {\nwlbrace}
                                {\bf{}ex} {\it{}y}={\bf{}realsymbol}(),
                                {\it{}x}=({\it{}sin}({\it{}y})\begin{math}\ast\end{math}{\it{}sqrt}(-{\it{}D1})-{\it{}B1})\begin{math}\div\end{math}2\begin{math}\div\end{math}{\it{}A1};
                                {\it{}sqrtD}={\it{}sqrt}(-{\it{}C1}+{\it{}pow}({\it{}B1},2)\begin{math}\div\end{math}4\begin{math}\div\end{math}{\it{}A1})\begin{math}\ast\end{math}{\it{}cos}({\it{}y});
                                {\it{}flat\_var\_em}.{\it{}insert}({\it{}std}::{\it{}make\_pair}({\it{}free\_vars}.{\it{}op}({\it{}another}), {\it{}x}));
                                {\it{}flat\_var}=({\it{}free\_vars}.{\it{}op}({\it{}another})\begin{math}\equiv\end{math}{\it{}x});

\nwused{\\{NWppJ6t-32jW6L-1}}\nwidentuses{\\{{\nwixident{evalf}}{evalf}}\\{{\nwixident{ex}}{ex}}\\{{\nwixident{info}}{info}}\\{{\nwixident{op}}{op}}\\{{\nwixident{realsymbol}}{realsymbol}}}\nwindexuse{\nwixident{evalf}}{evalf}{NWppJ6t-DKmU5-11}\nwindexuse{\nwixident{ex}}{ex}{NWppJ6t-DKmU5-11}\nwindexuse{\nwixident{info}}{info}{NWppJ6t-DKmU5-11}\nwindexuse{\nwixident{op}}{op}{NWppJ6t-DKmU5-11}\nwindexuse{\nwixident{realsymbol}}{realsymbol}{NWppJ6t-DKmU5-11}\nwendcode{}\nwbegindocs{993}If both are negative, we explicitly take out the imaginary part and
use the above hyperbolic substitution with \(\sinh\).
\nwenddocs{}\nwbegincode{994}\sublabel{NWppJ6t-DKmU5-12}\nwmargintag{{\nwtagstyle{}\subpageref{NWppJ6t-DKmU5-12}}}\moddef{figure class~{\nwtagstyle{}\subpageref{NWppJ6t-DKmU5-1}}}\plusendmoddef\Rm{}\nwstartdeflinemarkup\nwusesondefline{\\{NWppJ6t-32jW6L-1}}\nwprevnextdefs{NWppJ6t-DKmU5-11}{NWppJ6t-DKmU5-13}\nwenddeflinemarkup
                            {\nwrbrace} {\bf{}else} {\bf{}if} ({\it{}D1}.{\it{}info}({\it{}info\_flags}::{\it{}positive})) {\nwlbrace}
                                {\bf{}ex} {\it{}y}={\bf{}realsymbol}(),
                                {\it{}x}=({\it{}sinh}({\it{}y})\begin{math}\ast\end{math}{\it{}I}\begin{math}\ast\end{math}{\it{}sqrt}({\it{}D1})-{\it{}B1})\begin{math}\div\end{math}2\begin{math}\div\end{math}{\it{}A1};
                                {\it{}sqrtD}={\it{}I}\begin{math}\ast\end{math}{\it{}sqrt}({\it{}C1}-{\it{}pow}({\it{}B1},2)\begin{math}\div\end{math}4\begin{math}\div\end{math}{\it{}A1})\begin{math}\ast\end{math}{\it{}cosh}({\it{}y});
                                {\it{}flat\_var\_em}.{\it{}insert}({\it{}std}::{\it{}make\_pair}({\it{}free\_vars}.{\it{}op}({\it{}another}), {\it{}x}));
                                {\it{}flat\_var}=({\it{}free\_vars}.{\it{}op}({\it{}another})\begin{math}\equiv\end{math}{\it{}x});
                            {\nwrbrace}
                        {\nwrbrace}

\nwused{\\{NWppJ6t-32jW6L-1}}\nwidentuses{\\{{\nwixident{ex}}{ex}}\\{{\nwixident{info}}{info}}\\{{\nwixident{op}}{op}}\\{{\nwixident{realsymbol}}{realsymbol}}}\nwindexuse{\nwixident{ex}}{ex}{NWppJ6t-DKmU5-12}\nwindexuse{\nwixident{info}}{info}{NWppJ6t-DKmU5-12}\nwindexuse{\nwixident{op}}{op}{NWppJ6t-DKmU5-12}\nwindexuse{\nwixident{realsymbol}}{realsymbol}{NWppJ6t-DKmU5-12}\nwendcode{}\nwbegindocs{995}If a substitution was found we are staying with this solution.
\nwenddocs{}\nwbegincode{996}\sublabel{NWppJ6t-DKmU5-13}\nwmargintag{{\nwtagstyle{}\subpageref{NWppJ6t-DKmU5-13}}}\moddef{figure class~{\nwtagstyle{}\subpageref{NWppJ6t-DKmU5-1}}}\plusendmoddef\Rm{}\nwstartdeflinemarkup\nwusesondefline{\\{NWppJ6t-32jW6L-1}}\nwprevnextdefs{NWppJ6t-DKmU5-12}{NWppJ6t-DKmU5-14}\nwenddeflinemarkup
                    //cerr \begin{math}<\end{math}\begin{math}<\end{math} "real\_only sqrt(D): "; sqrtD.dbgprint();
                    {\bf{}if} ({\it{}not} ({\it{}sqrtD}-{\it{}sqrt}({\it{}D})).{\it{}is\_zero}())
                        {\bf{}break};
                {\nwrbrace}
            {\nwrbrace}
        {\nwrbrace}

\nwused{\\{NWppJ6t-32jW6L-1}}\nwendcode{}\nwbegindocs{997}Put index back to the range if needed.
\nwenddocs{}\nwbegincode{998}\sublabel{NWppJ6t-DKmU5-14}\nwmargintag{{\nwtagstyle{}\subpageref{NWppJ6t-DKmU5-14}}}\moddef{figure class~{\nwtagstyle{}\subpageref{NWppJ6t-DKmU5-1}}}\plusendmoddef\Rm{}\nwstartdeflinemarkup\nwusesondefline{\\{NWppJ6t-32jW6L-1}}\nwprevnextdefs{NWppJ6t-DKmU5-13}{NWppJ6t-DKmU5-15}\nwenddeflinemarkup
        {\bf{}if} ({\it{}i} \begin{math}\equiv\end{math} {\it{}free\_vars}.{\it{}nops}())
            \protect\MM{\it{}i};

\nwused{\\{NWppJ6t-32jW6L-1}}\nwidentuses{\\{{\nwixident{nops}}{nops}}}\nwindexuse{\nwixident{nops}}{nops}{NWppJ6t-DKmU5-14}\nwendcode{}\nwbegindocs{999}Small perturbations of the zero determinant can create the unwanted
imaginary entries, thus we treat  it as exactly zero. Also negligibly
small {\Tt{}\Rm{}{\it{}A}\nwendquote} corresponds to an effectively linear equation.
\nwenddocs{}\nwbegincode{1000}\sublabel{NWppJ6t-DKmU5-15}\nwmargintag{{\nwtagstyle{}\subpageref{NWppJ6t-DKmU5-15}}}\moddef{figure class~{\nwtagstyle{}\subpageref{NWppJ6t-DKmU5-1}}}\plusendmoddef\Rm{}\nwstartdeflinemarkup\nwusesondefline{\\{NWppJ6t-32jW6L-1}}\nwprevnextdefs{NWppJ6t-DKmU5-14}{NWppJ6t-DKmU5-16}\nwenddeflinemarkup
            {\bf{}if} ({\it{}is\_less\_than\_epsilon}({\it{}D}) \begin{math}\vee\end{math} ((\begin{math}\neg\end{math} {\it{}is\_less\_than\_epsilon}({\it{}B})) \begin{math}\wedge\end{math} {\it{}is\_less\_than\_epsilon}({\it{}A}\begin{math}\div\end{math}{\it{}B}))) {\nwlbrace}
                {\bf{}if} ({\it{}is\_less\_than\_epsilon}({\it{}D})) {\nwlbrace}
                    //cerr \begin{math}<\end{math}\begin{math}<\end{math} "zero determinant" \begin{math}<\end{math}\begin{math}<\end{math} endl;
                    {\it{}var1}=(-{\it{}B}\begin{math}\div\end{math}{\bf{}numeric}(2)\begin{math}\div\end{math}{\it{}A}).{\it{}subs}({\it{}flat\_var\_em},{\it{}subs\_options}::{\it{}algebraic}
                                                \begin{math}\mid\end{math} {\it{}subs\_options}::{\it{}no\_pattern}).{\it{}normal}();
                {\nwrbrace} {\bf{}else} {\nwlbrace}
                    //cerr \begin{math}<\end{math}\begin{math}<\end{math} "almost linear equation" \begin{math}<\end{math}\begin{math}<\end{math} endl;
                    {\it{}var1}=(-{\it{}C}\begin{math}\div\end{math}{\it{}B}).{\it{}subs}({\it{}flat\_var\_em},{\it{}subs\_options}::{\it{}algebraic}
                                     \begin{math}\mid\end{math} {\it{}subs\_options}::{\it{}no\_pattern}).{\it{}normal}();
                {\nwrbrace}
                {\it{}var1\_em}.{\it{}insert}({\it{}std}::{\it{}make\_pair}({\it{}free\_vars}.{\it{}op}({\it{}i}), {\it{}var1}));
                {\it{}subs\_lst1}={\it{}ex\_to}\begin{math}<\end{math}{\bf{}lst}\begin{math}>\end{math}({\it{}subs\_lst1}
                                     .{\it{}subs}({\it{}var1\_em},{\it{}subs\_options}::{\it{}algebraic} \begin{math}\mid\end{math} {\it{}subs\_options}::{\it{}no\_pattern}));
                {\it{}subs\_lst1}={\it{}ex\_to}\begin{math}<\end{math}{\bf{}lst}\begin{math}>\end{math}({\it{}subs\_lst1}.{\it{}append}({\it{}free\_vars}.{\it{}op}({\it{}i}) \begin{math}\equiv\end{math} {\it{}var1})
                                     .{\it{}subs}({\it{}flat\_var\_em},{\it{}subs\_options}::{\it{}algebraic} \begin{math}\mid\end{math} {\it{}subs\_options}::{\it{}no\_pattern}));
                {\bf{}if} ({\it{}flat\_var}.{\it{}nops}()\begin{math}>\end{math}0)
                    {\it{}subs\_lst1}.{\it{}append}({\it{}flat\_var});
                //cerr \begin{math}<\end{math}\begin{math}<\end{math} "subs\_lst1a: "; subs\_lst1.dbgprint();

\nwused{\\{NWppJ6t-32jW6L-1}}\nwidentuses{\\{{\nwixident{is{\_}less{\_}than{\_}epsilon}}{is:unless:unthan:unepsilon}}\\{{\nwixident{nops}}{nops}}\\{{\nwixident{numeric}}{numeric}}\\{{\nwixident{op}}{op}}\\{{\nwixident{subs}}{subs}}}\nwindexuse{\nwixident{is{\_}less{\_}than{\_}epsilon}}{is:unless:unthan:unepsilon}{NWppJ6t-DKmU5-15}\nwindexuse{\nwixident{nops}}{nops}{NWppJ6t-DKmU5-15}\nwindexuse{\nwixident{numeric}}{numeric}{NWppJ6t-DKmU5-15}\nwindexuse{\nwixident{op}}{op}{NWppJ6t-DKmU5-15}\nwindexuse{\nwixident{subs}}{subs}{NWppJ6t-DKmU5-15}\nwendcode{}\nwbegindocs{1001} For a non-zero discriminant we generate two solutions of the
quadratic equation.
\nwenddocs{}\nwbegincode{1002}\sublabel{NWppJ6t-DKmU5-16}\nwmargintag{{\nwtagstyle{}\subpageref{NWppJ6t-DKmU5-16}}}\moddef{figure class~{\nwtagstyle{}\subpageref{NWppJ6t-DKmU5-1}}}\plusendmoddef\Rm{}\nwstartdeflinemarkup\nwusesondefline{\\{NWppJ6t-32jW6L-1}}\nwprevnextdefs{NWppJ6t-DKmU5-15}{NWppJ6t-DKmU5-17}\nwenddeflinemarkup
            {\nwrbrace} {\bf{}else} {\nwlbrace}
                {\it{}second\_solution}={\bf{}true};
                {\it{}subs\_lst2}={\it{}subs\_lst1};
                {\it{}var1}=((-{\it{}B}+{\it{}sqrtD})\begin{math}\div\end{math}{\bf{}numeric}(2)\begin{math}\div\end{math}{\it{}A}).{\it{}subs}({\it{}flat\_var\_em},{\it{}subs\_options}::{\it{}algebraic}
                                                    \begin{math}\mid\end{math} {\it{}subs\_options}::{\it{}no\_pattern}).{\it{}normal}();
                {\it{}var1\_em}.{\it{}insert}({\it{}std}::{\it{}make\_pair}({\it{}free\_vars}.{\it{}op}({\it{}i}), {\it{}var1}));
                {\it{}var2}=((-{\it{}B}-{\it{}sqrtD})\begin{math}\div\end{math}{\bf{}numeric}(2)\begin{math}\div\end{math}{\it{}A}).{\it{}subs}({\it{}flat\_var\_em},{\it{}subs\_options}::{\it{}algebraic}
                                                    \begin{math}\mid\end{math} {\it{}subs\_options}::{\it{}no\_pattern}).{\it{}normal}();
                {\it{}var2\_em}.{\it{}insert}({\it{}std}::{\it{}make\_pair}({\it{}free\_vars}.{\it{}op}({\it{}i}), {\it{}var2}));
                {\it{}subs\_lst1}={\it{}ex\_to}\begin{math}<\end{math}{\bf{}lst}\begin{math}>\end{math}({\it{}subs\_lst1}
                                     .{\it{}subs}({\it{}var1\_em},{\it{}subs\_options}::{\it{}algebraic} \begin{math}\mid\end{math} {\it{}subs\_options}::{\it{}no\_pattern}));
                {\it{}subs\_lst1}={\it{}ex\_to}\begin{math}<\end{math}{\bf{}lst}\begin{math}>\end{math}({\it{}subs\_lst1}.{\it{}append}({\it{}free\_vars}.{\it{}op}({\it{}i}) \begin{math}\equiv\end{math} {\it{}var1})
                                     .{\it{}subs}({\it{}flat\_var\_em},{\it{}subs\_options}::{\it{}algebraic} \begin{math}\mid\end{math} {\it{}subs\_options}::{\it{}no\_pattern}));

\nwused{\\{NWppJ6t-32jW6L-1}}\nwidentuses{\\{{\nwixident{numeric}}{numeric}}\\{{\nwixident{op}}{op}}\\{{\nwixident{subs}}{subs}}}\nwindexuse{\nwixident{numeric}}{numeric}{NWppJ6t-DKmU5-16}\nwindexuse{\nwixident{op}}{op}{NWppJ6t-DKmU5-16}\nwindexuse{\nwixident{subs}}{subs}{NWppJ6t-DKmU5-16}\nwendcode{}\nwbegindocs{1003}Then we modify the second substitution list accordingly.
\nwenddocs{}\nwbegincode{1004}\sublabel{NWppJ6t-DKmU5-17}\nwmargintag{{\nwtagstyle{}\subpageref{NWppJ6t-DKmU5-17}}}\moddef{figure class~{\nwtagstyle{}\subpageref{NWppJ6t-DKmU5-1}}}\plusendmoddef\Rm{}\nwstartdeflinemarkup\nwusesondefline{\\{NWppJ6t-32jW6L-1}}\nwprevnextdefs{NWppJ6t-DKmU5-16}{NWppJ6t-DKmU5-18}\nwenddeflinemarkup
                {\it{}subs\_lst2}={\it{}ex\_to}\begin{math}<\end{math}{\bf{}lst}\begin{math}>\end{math}({\it{}subs\_lst2}
                                     .{\it{}subs}({\it{}var2\_em},{\it{}subs\_options}::{\it{}algebraic} \begin{math}\mid\end{math} {\it{}subs\_options}::{\it{}no\_pattern}));
                {\it{}subs\_lst2}={\it{}ex\_to}\begin{math}<\end{math}{\bf{}lst}\begin{math}>\end{math}({\it{}subs\_lst2}.{\it{}append}({\it{}free\_vars}.{\it{}op}({\it{}i}) \begin{math}\equiv\end{math} {\it{}var2})
                                     .{\it{}subs}({\it{}flat\_var\_em},{\it{}subs\_options}::{\it{}algebraic} \begin{math}\mid\end{math} {\it{}subs\_options}::{\it{}no\_pattern}));

\nwused{\\{NWppJ6t-32jW6L-1}}\nwidentuses{\\{{\nwixident{op}}{op}}\\{{\nwixident{subs}}{subs}}}\nwindexuse{\nwixident{op}}{op}{NWppJ6t-DKmU5-17}\nwindexuse{\nwixident{subs}}{subs}{NWppJ6t-DKmU5-17}\nwendcode{}\nwbegindocs{1005}We need to add the values of {\Tt{}\Rm{}{\it{}flat\_var}\nwendquote} which were assigned the
numeric value.
\nwenddocs{}\nwbegincode{1006}\sublabel{NWppJ6t-DKmU5-18}\nwmargintag{{\nwtagstyle{}\subpageref{NWppJ6t-DKmU5-18}}}\moddef{figure class~{\nwtagstyle{}\subpageref{NWppJ6t-DKmU5-1}}}\plusendmoddef\Rm{}\nwstartdeflinemarkup\nwusesondefline{\\{NWppJ6t-32jW6L-1}}\nwprevnextdefs{NWppJ6t-DKmU5-17}{NWppJ6t-DKmU5-19}\nwenddeflinemarkup
                {\bf{}if} ({\it{}flat\_var}.{\it{}nops}()\begin{math}>\end{math}0) {\nwlbrace}
                    {\it{}subs\_lst1}.{\it{}append}({\it{}flat\_var});
                    {\it{}subs\_lst2}.{\it{}append}({\it{}flat\_var});
                {\nwrbrace}
                //cerr \begin{math}<\end{math}\begin{math}<\end{math} "subs\_lst1b: "; subs\_lst1.dbgprint();
                //cerr \begin{math}<\end{math}\begin{math}<\end{math} "subs\_lst2b: "; subs\_lst2.dbgprint();
            {\nwrbrace}
            // end of the quadratic case

\nwused{\\{NWppJ6t-32jW6L-1}}\nwidentuses{\\{{\nwixident{nops}}{nops}}}\nwindexuse{\nwixident{nops}}{nops}{NWppJ6t-DKmU5-18}\nwendcode{}\nwbegindocs{1007}The non-linear equation is not quadratic in some variable,
e.g. is  \(mk+1=0\) then we are solving it as linear.
\nwenddocs{}\nwbegincode{1008}\sublabel{NWppJ6t-DKmU5-19}\nwmargintag{{\nwtagstyle{}\subpageref{NWppJ6t-DKmU5-19}}}\moddef{figure class~{\nwtagstyle{}\subpageref{NWppJ6t-DKmU5-1}}}\plusendmoddef\Rm{}\nwstartdeflinemarkup\nwusesondefline{\\{NWppJ6t-32jW6L-1}}\nwprevnextdefs{NWppJ6t-DKmU5-18}{NWppJ6t-DKmU5-1A}\nwenddeflinemarkup
        {\nwrbrace} {\bf{}else} {\nwlbrace}
            //cerr \begin{math}<\end{math}\begin{math}<\end{math} "The equation is not quadratic in a single variable"\begin{math}<\end{math}\begin{math}<\end{math}endl;
            //cerr \begin{math}<\end{math}\begin{math}<\end{math} "free\_vars: "; free\_vars.dbgprint();
            {\it{}var1}=-({\it{}quadratic}.{\it{}coeff}({\it{}free\_vars}.{\it{}op}({\it{}i}),0)\begin{math}\div\end{math}{\it{}quadratic}.{\it{}coeff}({\it{}free\_vars}.{\it{}op}({\it{}i}),1)).{\it{}normal}();
            {\it{}var1\_em}.{\it{}insert}({\it{}std}::{\it{}make\_pair}({\it{}free\_vars}.{\it{}op}({\it{}i}), {\it{}var1}));
            {\it{}subs\_lst1}={\it{}ex\_to}\begin{math}<\end{math}{\bf{}lst}\begin{math}>\end{math}({\it{}subs\_lst1}
                                 .{\it{}subs}({\it{}var1\_em},{\it{}subs\_options}::{\it{}algebraic} \begin{math}\mid\end{math} {\it{}subs\_options}::{\it{}no\_pattern}));
            {\it{}subs\_lst1}.{\it{}append}({\it{}free\_vars}.{\it{}op}({\it{}i}) \begin{math}\equiv\end{math} {\it{}var1});
            //cerr \begin{math}<\end{math}\begin{math}<\end{math} "non-quadratic subs\_lst1: "; subs\_lst1.dbgprint();
        {\nwrbrace}

\nwused{\\{NWppJ6t-32jW6L-1}}\nwidentuses{\\{{\nwixident{op}}{op}}\\{{\nwixident{subs}}{subs}}}\nwindexuse{\nwixident{op}}{op}{NWppJ6t-DKmU5-19}\nwindexuse{\nwixident{subs}}{subs}{NWppJ6t-DKmU5-19}\nwendcode{}\nwbegindocs{1009}Now we check that other non-linear conditions are satisfied by the
found solutions.
\nwenddocs{}\nwbegincode{1010}\sublabel{NWppJ6t-DKmU5-1A}\nwmargintag{{\nwtagstyle{}\subpageref{NWppJ6t-DKmU5-1A}}}\moddef{figure class~{\nwtagstyle{}\subpageref{NWppJ6t-DKmU5-1}}}\plusendmoddef\Rm{}\nwstartdeflinemarkup\nwusesondefline{\\{NWppJ6t-32jW6L-1}}\nwprevnextdefs{NWppJ6t-DKmU5-19}{NWppJ6t-DKmU5-1B}\nwenddeflinemarkup
            {\bf{}lst}::{\it{}const\_iterator} {\it{}it1}= {\it{}nonlin\_cond}.{\it{}begin}();
            \protect\PP{\it{}it1};
            //cerr \begin{math}<\end{math}\begin{math}<\end{math} "Subs list: "; subs\_lst1.dbgprint();
            {\bf{}lst} {\it{}subs\_f1}={\it{}ex\_to}\begin{math}<\end{math}{\bf{}lst}\begin{math}>\end{math}({\it{}subs\_lst1}.{\it{}evalf}()), {\it{}subs\_f2};
            //cerr \begin{math}<\end{math}\begin{math}<\end{math} "Subs list float: "; subs\_f1.dbgprint();
            {\bf{}if}({\it{}second\_solution})
                {\it{}subs\_f2}={\it{}ex\_to}\begin{math}<\end{math}{\bf{}lst}\begin{math}>\end{math}({\it{}subs\_lst2}.{\it{}evalf}());

\nwused{\\{NWppJ6t-32jW6L-1}}\nwidentuses{\\{{\nwixident{evalf}}{evalf}}}\nwindexuse{\nwixident{evalf}}{evalf}{NWppJ6t-DKmU5-1A}\nwendcode{}\nwbegindocs{1011}Since CAS is not as perfect as one may wish, we checked obtained
solutions in two ways: through float approximations and exact
calculations. If either works then the solution is accepted.
\nwenddocs{}\nwbegincode{1012}\sublabel{NWppJ6t-DKmU5-1B}\nwmargintag{{\nwtagstyle{}\subpageref{NWppJ6t-DKmU5-1B}}}\moddef{figure class~{\nwtagstyle{}\subpageref{NWppJ6t-DKmU5-1}}}\plusendmoddef\Rm{}\nwstartdeflinemarkup\nwusesondefline{\\{NWppJ6t-32jW6L-1}}\nwprevnextdefs{NWppJ6t-DKmU5-1A}{NWppJ6t-DKmU5-1C}\nwenddeflinemarkup
            {\bf{}for} (; {\it{}it1} \begin{math}\neq\end{math} {\it{}nonlin\_cond}.{\it{}end}(); \protect\PP{\it{}it1}) {\nwlbrace}
                {\it{}first\_solution} &= ({\it{}is\_less\_than\_epsilon}(({\it{}it1}\begin{math}\rightarrow\end{math}{\it{}op}(0)-{\it{}it1}\begin{math}\rightarrow\end{math}{\it{}op}(1)).{\it{}subs}({\it{}subs\_f1},
                                                                           {\it{}subs\_options}::{\it{}algebraic} \begin{math}\mid\end{math} {\it{}subs\_options}::{\it{}no\_pattern}))
                         \begin{math}\vee\end{math} (({\it{}it1}\begin{math}\rightarrow\end{math}{\it{}op}(0)-{\it{}it1}\begin{math}\rightarrow\end{math}{\it{}op}(1)).{\it{}subs}({\it{}subs\_lst1},
                                                          {\it{}subs\_options}::{\it{}algebraic} \begin{math}\mid\end{math} {\it{}subs\_options}::{\it{}no\_pattern})).{\it{}normal}().{\it{}is\_zero}());

\nwused{\\{NWppJ6t-32jW6L-1}}\nwidentuses{\\{{\nwixident{is{\_}less{\_}than{\_}epsilon}}{is:unless:unthan:unepsilon}}\\{{\nwixident{op}}{op}}\\{{\nwixident{subs}}{subs}}}\nwindexuse{\nwixident{is{\_}less{\_}than{\_}epsilon}}{is:unless:unthan:unepsilon}{NWppJ6t-DKmU5-1B}\nwindexuse{\nwixident{op}}{op}{NWppJ6t-DKmU5-1B}\nwindexuse{\nwixident{subs}}{subs}{NWppJ6t-DKmU5-1B}\nwendcode{}\nwbegindocs{1013}The same check for the second solution.
\nwenddocs{}\nwbegincode{1014}\sublabel{NWppJ6t-DKmU5-1C}\nwmargintag{{\nwtagstyle{}\subpageref{NWppJ6t-DKmU5-1C}}}\moddef{figure class~{\nwtagstyle{}\subpageref{NWppJ6t-DKmU5-1}}}\plusendmoddef\Rm{}\nwstartdeflinemarkup\nwusesondefline{\\{NWppJ6t-32jW6L-1}}\nwprevnextdefs{NWppJ6t-DKmU5-1B}{NWppJ6t-DKmU5-1D}\nwenddeflinemarkup
                {\bf{}if}({\it{}second\_solution})
                    {\it{}second\_solution} &= ({\it{}is\_less\_than\_epsilon}(({\it{}it1}\begin{math}\rightarrow\end{math}{\it{}op}(0)-{\it{}it1}\begin{math}\rightarrow\end{math}{\it{}op}(1)).{\it{}subs}({\it{}subs\_f2},
                                                                               {\it{}subs\_options}::{\it{}algebraic} \begin{math}\mid\end{math} {\it{}subs\_options}::{\it{}no\_pattern}))
                             \begin{math}\vee\end{math} (({\it{}it1}\begin{math}\rightarrow\end{math}{\it{}op}(0)-{\it{}it1}\begin{math}\rightarrow\end{math}{\it{}op}(1)).{\it{}subs}({\it{}subs\_lst2},
                                                              {\it{}subs\_options}::{\it{}algebraic} \begin{math}\mid\end{math} {\it{}subs\_options}::{\it{}no\_pattern})).{\it{}normal}().{\it{}is\_zero}());
            {\nwrbrace}

\nwused{\\{NWppJ6t-32jW6L-1}}\nwidentuses{\\{{\nwixident{is{\_}less{\_}than{\_}epsilon}}{is:unless:unthan:unepsilon}}\\{{\nwixident{op}}{op}}\\{{\nwixident{subs}}{subs}}}\nwindexuse{\nwixident{is{\_}less{\_}than{\_}epsilon}}{is:unless:unthan:unepsilon}{NWppJ6t-DKmU5-1C}\nwindexuse{\nwixident{op}}{op}{NWppJ6t-DKmU5-1C}\nwindexuse{\nwixident{subs}}{subs}{NWppJ6t-DKmU5-1C}\nwendcode{}\nwbegindocs{1015}If a solution is good, then we use it to generate the respective cycle.
\nwenddocs{}\nwbegincode{1016}\sublabel{NWppJ6t-DKmU5-1D}\nwmargintag{{\nwtagstyle{}\subpageref{NWppJ6t-DKmU5-1D}}}\moddef{figure class~{\nwtagstyle{}\subpageref{NWppJ6t-DKmU5-1}}}\plusendmoddef\Rm{}\nwstartdeflinemarkup\nwusesondefline{\\{NWppJ6t-32jW6L-1}}\nwprevnextdefs{NWppJ6t-DKmU5-1C}{NWppJ6t-DKmU5-1E}\nwenddeflinemarkup
            {\bf{}if} ({\it{}first\_solution})
                {\it{}C\_new}={\it{}symbolic}.{\it{}subs}({\it{}subs\_lst1}, {\it{}subs\_options}::{\it{}algebraic}
                                                    \begin{math}\mid\end{math} {\it{}subs\_options}::{\it{}no\_pattern});

            //cerr \begin{math}<\end{math}\begin{math}<\end{math} "C\_new: "; C\_new.dbgprint();
            {\bf{}if} ({\it{}second\_solution})
                {\it{}C1\_new}={\it{}symbolic}.{\it{}subs}({\it{}subs\_lst2}, {\it{}subs\_options}::{\it{}algebraic}
                                                     \begin{math}\mid\end{math} {\it{}subs\_options}::{\it{}no\_pattern});
            //cerr \begin{math}<\end{math}\begin{math}<\end{math} "C1\_new: "; C1\_new.dbgprint();
        {\nwrbrace}

\nwused{\\{NWppJ6t-32jW6L-1}}\nwidentuses{\\{{\nwixident{subs}}{subs}}}\nwindexuse{\nwixident{subs}}{subs}{NWppJ6t-DKmU5-1D}\nwendcode{}\nwbegindocs{1017}We check if any symbols survived after calculations\ldots
\nwenddocs{}\nwbegincode{1018}\sublabel{NWppJ6t-DKmU5-1E}\nwmargintag{{\nwtagstyle{}\subpageref{NWppJ6t-DKmU5-1E}}}\moddef{figure class~{\nwtagstyle{}\subpageref{NWppJ6t-DKmU5-1}}}\plusendmoddef\Rm{}\nwstartdeflinemarkup\nwusesondefline{\\{NWppJ6t-32jW6L-1}}\nwprevnextdefs{NWppJ6t-DKmU5-1D}{NWppJ6t-DKmU5-1F}\nwenddeflinemarkup
    {\bf{}lst} {\it{}repl};
    {\bf{}if} ({\it{}ex\_to}\begin{math}<\end{math}{\bf{}cycle\_data}\begin{math}>\end{math}({\it{}C\_new}).{\it{}has}({\it{}ex\_to}\begin{math}<\end{math}{\bf{}cycle\_data}\begin{math}>\end{math}({\it{}symbolic}).{\it{}get\_k}()))
        {\it{}repl}.{\it{}append}({\it{}ex\_to}\begin{math}<\end{math}{\bf{}cycle\_data}\begin{math}>\end{math}({\it{}symbolic}).{\it{}get\_k}()\begin{math}\equiv\end{math}{\bf{}realsymbol}());
    {\bf{}if} ({\it{}ex\_to}\begin{math}<\end{math}{\bf{}cycle\_data}\begin{math}>\end{math}({\it{}C\_new}).{\it{}has}({\it{}ex\_to}\begin{math}<\end{math}{\bf{}cycle\_data}\begin{math}>\end{math}({\it{}symbolic}).{\it{}get\_m}()))
        {\it{}repl}.{\it{}append}({\it{}ex\_to}\begin{math}<\end{math}{\bf{}cycle\_data}\begin{math}>\end{math}({\it{}symbolic}).{\it{}get\_m}()\begin{math}\equiv\end{math}{\bf{}realsymbol}());
    {\bf{}if} ({\it{}ex\_to}\begin{math}<\end{math}{\bf{}cycle\_data}\begin{math}>\end{math}({\it{}C\_new}).{\it{}has}({\it{}ex\_to}\begin{math}<\end{math}{\bf{}cycle\_data}\begin{math}>\end{math}({\it{}symbolic}).{\it{}get\_l}().{\it{}op}(0).{\it{}op}(0)))
        {\it{}repl}.{\it{}append}({\it{}ex\_to}\begin{math}<\end{math}{\bf{}cycle\_data}\begin{math}>\end{math}({\it{}symbolic}).{\it{}get\_l}().{\it{}op}(0).{\it{}op}(0)\begin{math}\equiv\end{math}{\bf{}realsymbol}());
    {\bf{}if} ({\it{}ex\_to}\begin{math}<\end{math}{\bf{}cycle\_data}\begin{math}>\end{math}({\it{}C\_new}).{\it{}has}({\it{}ex\_to}\begin{math}<\end{math}{\bf{}cycle\_data}\begin{math}>\end{math}({\it{}symbolic}).{\it{}get\_l}().{\it{}op}(0).{\it{}op}(1)))
        {\it{}repl}.{\it{}append}({\it{}ex\_to}\begin{math}<\end{math}{\bf{}cycle\_data}\begin{math}>\end{math}({\it{}symbolic}).{\it{}get\_l}().{\it{}op}(0).{\it{}op}(1)\begin{math}\equiv\end{math}{\bf{}realsymbol}());

\nwused{\\{NWppJ6t-32jW6L-1}}\nwidentuses{\\{{\nwixident{cycle{\_}data}}{cycle:undata}}\\{{\nwixident{op}}{op}}\\{{\nwixident{realsymbol}}{realsymbol}}}\nwindexuse{\nwixident{cycle{\_}data}}{cycle:undata}{NWppJ6t-DKmU5-1E}\nwindexuse{\nwixident{op}}{op}{NWppJ6t-DKmU5-1E}\nwindexuse{\nwixident{realsymbol}}{realsymbol}{NWppJ6t-DKmU5-1E}\nwendcode{}\nwbegindocs{1019}\ldots and if they are, then we replace them for new one
\nwenddocs{}\nwbegincode{1020}\sublabel{NWppJ6t-DKmU5-1F}\nwmargintag{{\nwtagstyle{}\subpageref{NWppJ6t-DKmU5-1F}}}\moddef{figure class~{\nwtagstyle{}\subpageref{NWppJ6t-DKmU5-1}}}\plusendmoddef\Rm{}\nwstartdeflinemarkup\nwusesondefline{\\{NWppJ6t-32jW6L-1}}\nwprevnextdefs{NWppJ6t-DKmU5-1E}{NWppJ6t-DKmU5-1G}\nwenddeflinemarkup
    {\bf{}if} ({\it{}repl}.{\it{}nops}()\begin{math}>\end{math}0) {\nwlbrace}
        {\bf{}if} ({\it{}first\_solution})
            {\it{}C\_new}={\it{}C\_new}.{\it{}subs}({\it{}repl});
        {\bf{}if} ({\it{}second\_solution})
            {\it{}C1\_new}={\it{}C1\_new}.{\it{}subs}({\it{}repl});
    {\nwrbrace}

    //cerr \begin{math}<\end{math}\begin{math}<\end{math} endl;

\nwused{\\{NWppJ6t-32jW6L-1}}\nwidentuses{\\{{\nwixident{nops}}{nops}}\\{{\nwixident{subs}}{subs}}}\nwindexuse{\nwixident{nops}}{nops}{NWppJ6t-DKmU5-1F}\nwindexuse{\nwixident{subs}}{subs}{NWppJ6t-DKmU5-1F}\nwendcode{}\nwbegindocs{1021}Finally, every constructed cycle is added to the result.
\nwenddocs{}\nwbegincode{1022}\sublabel{NWppJ6t-DKmU5-1G}\nwmargintag{{\nwtagstyle{}\subpageref{NWppJ6t-DKmU5-1G}}}\moddef{figure class~{\nwtagstyle{}\subpageref{NWppJ6t-DKmU5-1}}}\plusendmoddef\Rm{}\nwstartdeflinemarkup\nwusesondefline{\\{NWppJ6t-32jW6L-1}}\nwprevnextdefs{NWppJ6t-DKmU5-1F}{NWppJ6t-DKmU5-1H}\nwenddeflinemarkup
    {\bf{}lst} {\it{}res};
    {\bf{}if} ({\it{}first\_solution})
        {\it{}res}.{\it{}append}({\it{}float\_evaluation}?{\it{}C\_new}.{\it{}num\_normalize}().{\it{}evalf}():{\it{}C\_new}.{\it{}num\_normalize}());
    {\bf{}if} ({\it{}second\_solution})
        {\it{}res}.{\it{}append}({\it{}float\_evaluation}?{\it{}C1\_new}.{\it{}num\_normalize}().{\it{}evalf}():{\it{}C1\_new}.{\it{}num\_normalize}());

    {\bf{}return} {\it{}res};
{\nwrbrace}

\nwused{\\{NWppJ6t-32jW6L-1}}\nwidentuses{\\{{\nwixident{evalf}}{evalf}}\\{{\nwixident{float{\_}evaluation}}{float:unevaluation}}}\nwindexuse{\nwixident{evalf}}{evalf}{NWppJ6t-DKmU5-1G}\nwindexuse{\nwixident{float{\_}evaluation}}{float:unevaluation}{NWppJ6t-DKmU5-1G}\nwendcode{}\nwbegindocs{1023}This method runs recursively because we do not know in advance the
number of conditions glued by and/or. Also, some relations
(e.g. {\Tt{}\Rm{}{\it{}moebius\_trans}\nwendquote} or {\Tt{}\Rm{}{\bf{}subfigure}\nwendquote}) directly define the cycles,
and for others we need to solve some equations.
\nwenddocs{}\nwbegincode{1024}\sublabel{NWppJ6t-DKmU5-1H}\nwmargintag{{\nwtagstyle{}\subpageref{NWppJ6t-DKmU5-1H}}}\moddef{figure class~{\nwtagstyle{}\subpageref{NWppJ6t-DKmU5-1}}}\plusendmoddef\Rm{}\nwstartdeflinemarkup\nwusesondefline{\\{NWppJ6t-32jW6L-1}}\nwprevnextdefs{NWppJ6t-DKmU5-1G}{NWppJ6t-DKmU5-1I}\nwenddeflinemarkup
{\bf{}ex} {\bf{}figure}::{\it{}update\_cycle\_node}({\bf{}const} {\bf{}ex} & {\it{}key}, {\bf{}const} {\bf{}lst} & {\it{}eq\_cond}, {\bf{}const} {\bf{}lst} & {\it{}neq\_cond}, {\bf{}lst} {\it{}res}, {\it{}size\_t} {\it{}level})
{\nwlbrace}
    //cerr \begin{math}<\end{math}\begin{math}<\end{math} endl \begin{math}<\end{math}\begin{math}<\end{math} "level: " \begin{math}<\end{math}\begin{math}<\end{math} level \begin{math}<\end{math}\begin{math}<\end{math} "; cycle: "; nodes[key].dbgprint();
    {\bf{}if} ({\it{}level} \begin{math}\equiv\end{math} 0) {\nwlbrace}// set the iniail symbolic cycle for calculations
        \LA{}update node zero level~{\nwtagstyle{}\subpageref{NWppJ6t-3AODA9-1}}\RA{}
    {\nwrbrace}
\nwindexdefn{\nwixident{update{\_}cycle{\_}node}}{update:uncycle:unnode}{NWppJ6t-DKmU5-1H}\eatline
\nwused{\\{NWppJ6t-32jW6L-1}}\nwidentdefs{\\{{\nwixident{update{\_}cycle{\_}node}}{update:uncycle:unnode}}}\nwidentuses{\\{{\nwixident{ex}}{ex}}\\{{\nwixident{figure}}{figure}}\\{{\nwixident{key}}{key}}\\{{\nwixident{nodes}}{nodes}}}\nwindexuse{\nwixident{ex}}{ex}{NWppJ6t-DKmU5-1H}\nwindexuse{\nwixident{figure}}{figure}{NWppJ6t-DKmU5-1H}\nwindexuse{\nwixident{key}}{key}{NWppJ6t-DKmU5-1H}\nwindexuse{\nwixident{nodes}}{nodes}{NWppJ6t-DKmU5-1H}\nwendcode{}\nwbegindocs{1025}\nwdocspar
\nwenddocs{}\nwbegindocs{1026}If we get here, then some equations need to be solved. We advance
through the parents list to match the {\Tt{}\Rm{}{\it{}level}\nwendquote}.
\nwenddocs{}\nwbegincode{1027}\sublabel{NWppJ6t-DKmU5-1I}\nwmargintag{{\nwtagstyle{}\subpageref{NWppJ6t-DKmU5-1I}}}\moddef{figure class~{\nwtagstyle{}\subpageref{NWppJ6t-DKmU5-1}}}\plusendmoddef\Rm{}\nwstartdeflinemarkup\nwusesondefline{\\{NWppJ6t-32jW6L-1}}\nwprevnextdefs{NWppJ6t-DKmU5-1H}{NWppJ6t-DKmU5-1J}\nwenddeflinemarkup
    {\bf{}lst} {\it{}par} = {\it{}nodes}[{\it{}key}].{\it{}get\_parents}();
    {\bf{}lst}::{\it{}const\_iterator} {\it{}it} = {\it{}par}.{\it{}begin}();
    {\it{}std}::{\it{}advance}({\it{}it},{\it{}level});

    {\bf{}lst} {\it{}new\_cond}={\it{}ex\_to}\begin{math}<\end{math}{\bf{}lst}\begin{math}>\end{math}({\it{}ex\_to}\begin{math}<\end{math}{\bf{}cycle\_relation}\begin{math}>\end{math}(\begin{math}\ast\end{math}{\it{}it}).{\it{}rel\_to\_parent}({\it{}nodes}[{\it{}key}].{\it{}get\_cycles}().{\it{}op}(0),
                                                                     {\it{}point\_metric}, {\it{}cycle\_metric}, {\it{}nodes}));

\nwused{\\{NWppJ6t-32jW6L-1}}\nwidentuses{\\{{\nwixident{cycle{\_}metric}}{cycle:unmetric}}\\{{\nwixident{cycle{\_}relation}}{cycle:unrelation}}\\{{\nwixident{key}}{key}}\\{{\nwixident{nodes}}{nodes}}\\{{\nwixident{op}}{op}}\\{{\nwixident{point{\_}metric}}{point:unmetric}}}\nwindexuse{\nwixident{cycle{\_}metric}}{cycle:unmetric}{NWppJ6t-DKmU5-1I}\nwindexuse{\nwixident{cycle{\_}relation}}{cycle:unrelation}{NWppJ6t-DKmU5-1I}\nwindexuse{\nwixident{key}}{key}{NWppJ6t-DKmU5-1I}\nwindexuse{\nwixident{nodes}}{nodes}{NWppJ6t-DKmU5-1I}\nwindexuse{\nwixident{op}}{op}{NWppJ6t-DKmU5-1I}\nwindexuse{\nwixident{point{\_}metric}}{point:unmetric}{NWppJ6t-DKmU5-1I}\nwendcode{}\nwbegindocs{1028}We need to go through the cycle at least once at every {\Tt{}\Rm{}{\it{}level}\nwendquote} and
separate equations, which are used to calculate solutions, from
inequalities, which will be only checked on the obtained solution.
\nwenddocs{}\nwbegincode{1029}\sublabel{NWppJ6t-DKmU5-1J}\nwmargintag{{\nwtagstyle{}\subpageref{NWppJ6t-DKmU5-1J}}}\moddef{figure class~{\nwtagstyle{}\subpageref{NWppJ6t-DKmU5-1}}}\plusendmoddef\Rm{}\nwstartdeflinemarkup\nwusesondefline{\\{NWppJ6t-32jW6L-1}}\nwprevnextdefs{NWppJ6t-DKmU5-1I}{NWppJ6t-DKmU5-1K}\nwenddeflinemarkup
    {\bf{}for} ({\bf{}const} {\bf{}auto}& {\it{}it1} : {\it{}new\_cond}) {\nwlbrace}
        {\bf{}lst} {\it{}store\_cond}={\it{}neq\_cond};
        {\bf{}lst} {\it{}use\_cond}={\it{}eq\_cond};
        {\bf{}lst} {\it{}step\_cond}={\it{}ex\_to}\begin{math}<\end{math}{\bf{}lst}\begin{math}>\end{math}({\it{}it1});

\nwused{\\{NWppJ6t-32jW6L-1}}\nwendcode{}\nwbegindocs{1030}Iteration over the list of conditions
\nwenddocs{}\nwbegincode{1031}\sublabel{NWppJ6t-DKmU5-1K}\nwmargintag{{\nwtagstyle{}\subpageref{NWppJ6t-DKmU5-1K}}}\moddef{figure class~{\nwtagstyle{}\subpageref{NWppJ6t-DKmU5-1}}}\plusendmoddef\Rm{}\nwstartdeflinemarkup\nwusesondefline{\\{NWppJ6t-32jW6L-1}}\nwprevnextdefs{NWppJ6t-DKmU5-1J}{NWppJ6t-DKmU5-1L}\nwenddeflinemarkup
        {\bf{}for} ({\bf{}const} {\bf{}auto}& {\it{}it2} : {\it{}step\_cond})
            {\bf{}if} (({\it{}is\_a}\begin{math}<\end{math}{\bf{}relational}\begin{math}>\end{math}({\it{}it2}) \begin{math}\wedge\end{math} {\it{}ex\_to}\begin{math}<\end{math}{\bf{}relational}\begin{math}>\end{math}({\it{}it2}).{\it{}info}({\it{}info\_flags}::{\it{}relation\_equal})))
                {\it{}use\_cond}.{\it{}append}({\it{}it2});   // append the equation
            {\bf{}else} {\bf{}if} ({\it{}is\_a}\begin{math}<\end{math}{\bf{}cycle}\begin{math}>\end{math}({\it{}it2}))
                {\it{}res}.{\it{}append}({\it{}it2}); // append a solution
            {\bf{}else} {\nwlbrace}
                {\it{}store\_cond}.{\it{}append}(\begin{math}\ast\end{math}{\it{}it}); // store the pointer to parents producing inequality
            {\nwrbrace}
        //cerr \begin{math}<\end{math}\begin{math}<\end{math} "use\_cond: "; use\_cond.dbgprint();
        //cerr \begin{math}<\end{math}\begin{math}<\end{math} "store\_cond: "; store\_cond.dbgprint();

\nwused{\\{NWppJ6t-32jW6L-1}}\nwidentuses{\\{{\nwixident{info}}{info}}}\nwindexuse{\nwixident{info}}{info}{NWppJ6t-DKmU5-1K}\nwendcode{}\nwbegindocs{1032}When all conditions are unwrapped and there are equations to solve,
we call a solver. Solutions from {\Tt{}\Rm{}{\it{}res}\nwendquote} are copied there as well,
then {\Tt{}\Rm{}{\it{}res}\nwendquote} is cleared.
\nwenddocs{}\nwbegincode{1033}\sublabel{NWppJ6t-DKmU5-1L}\nwmargintag{{\nwtagstyle{}\subpageref{NWppJ6t-DKmU5-1L}}}\moddef{figure class~{\nwtagstyle{}\subpageref{NWppJ6t-DKmU5-1}}}\plusendmoddef\Rm{}\nwstartdeflinemarkup\nwusesondefline{\\{NWppJ6t-32jW6L-1}}\nwprevnextdefs{NWppJ6t-DKmU5-1K}{NWppJ6t-DKmU5-1M}\nwenddeflinemarkup
        {\bf{}if}({\it{}level} \begin{math}\equiv\end{math} {\it{}par}.{\it{}nops}()-1) {\nwlbrace} //if the last one in the parents list
            {\bf{}lst} {\it{}cnew};
            {\bf{}if} ({\it{}use\_cond}.{\it{}nops}()\begin{math}>\end{math}0)
                {\it{}cnew}={\it{}ex\_to}\begin{math}<\end{math}{\bf{}lst}\begin{math}>\end{math}({\it{}evaluate\_cycle}({\it{}nodes}[{\it{}key}].{\it{}get\_cycle\_data}(0), {\it{}use\_cond}));
            {\bf{}for} ({\bf{}const} {\bf{}auto}& {\it{}sol} : {\it{}res})
                {\it{}cnew}.{\it{}append}({\it{}sol});
            {\it{}res}={\bf{}lst}{\nwlbrace}{\nwrbrace};

\nwused{\\{NWppJ6t-32jW6L-1}}\nwidentuses{\\{{\nwixident{evaluate{\_}cycle}}{evaluate:uncycle}}\\{{\nwixident{key}}{key}}\\{{\nwixident{nodes}}{nodes}}\\{{\nwixident{nops}}{nops}}}\nwindexuse{\nwixident{evaluate{\_}cycle}}{evaluate:uncycle}{NWppJ6t-DKmU5-1L}\nwindexuse{\nwixident{key}}{key}{NWppJ6t-DKmU5-1L}\nwindexuse{\nwixident{nodes}}{nodes}{NWppJ6t-DKmU5-1L}\nwindexuse{\nwixident{nops}}{nops}{NWppJ6t-DKmU5-1L}\nwendcode{}\nwbegindocs{1034}Now we check which of the obtained solutions satisfy to the
restrictions in {\Tt{}\Rm{}{\it{}store\_cond}\nwendquote}
\nwenddocs{}\nwbegincode{1035}\sublabel{NWppJ6t-DKmU5-1M}\nwmargintag{{\nwtagstyle{}\subpageref{NWppJ6t-DKmU5-1M}}}\moddef{figure class~{\nwtagstyle{}\subpageref{NWppJ6t-DKmU5-1}}}\plusendmoddef\Rm{}\nwstartdeflinemarkup\nwusesondefline{\\{NWppJ6t-32jW6L-1}}\nwprevnextdefs{NWppJ6t-DKmU5-1L}{NWppJ6t-DKmU5-1N}\nwenddeflinemarkup
            //cerr\begin{math}<\end{math}\begin{math}<\end{math} "Store cond: "; store\_cond.dbgprint();
            //cerr\begin{math}<\end{math}\begin{math}<\end{math} "Use cond: "; use\_cond.dbgprint();
            {\bf{}for} ({\bf{}const} {\bf{}auto}& {\it{}inew}: {\it{}cnew}) {\nwlbrace}
                {\bf{}bool} {\it{}to\_add}={\bf{}true};
                {\bf{}for} ({\bf{}const} {\bf{}auto}& {\it{}icon}: {\it{}store\_cond}) {\nwlbrace}
                    {\bf{}lst} {\it{}suits}={\it{}ex\_to}\begin{math}<\end{math}{\bf{}lst}\begin{math}>\end{math}({\it{}ex\_to}\begin{math}<\end{math}{\bf{}cycle\_relation}\begin{math}>\end{math}({\it{}icon}).{\it{}rel\_to\_parent}({\it{}ex\_to}\begin{math}<\end{math}{\bf{}cycle\_data}\begin{math}>\end{math}({\it{}inew}),
                                                                                   {\it{}point\_metric}, {\it{}cycle\_metric}, {\it{}nodes}));
                    //cerr\begin{math}<\end{math}\begin{math}<\end{math} "Suit: "; suits.dbgprint();
                    {\bf{}for} ({\bf{}const} {\bf{}auto}& {\it{}is} : {\it{}suits})
                        {\bf{}for} ({\bf{}const} {\bf{}auto}& {\it{}ic} : {\it{}is}) {\nwlbrace}

\nwused{\\{NWppJ6t-32jW6L-1}}\nwidentuses{\\{{\nwixident{cycle{\_}data}}{cycle:undata}}\\{{\nwixident{cycle{\_}metric}}{cycle:unmetric}}\\{{\nwixident{cycle{\_}relation}}{cycle:unrelation}}\\{{\nwixident{nodes}}{nodes}}\\{{\nwixident{point{\_}metric}}{point:unmetric}}}\nwindexuse{\nwixident{cycle{\_}data}}{cycle:undata}{NWppJ6t-DKmU5-1M}\nwindexuse{\nwixident{cycle{\_}metric}}{cycle:unmetric}{NWppJ6t-DKmU5-1M}\nwindexuse{\nwixident{cycle{\_}relation}}{cycle:unrelation}{NWppJ6t-DKmU5-1M}\nwindexuse{\nwixident{nodes}}{nodes}{NWppJ6t-DKmU5-1M}\nwindexuse{\nwixident{point{\_}metric}}{point:unmetric}{NWppJ6t-DKmU5-1M}\nwendcode{}\nwbegindocs{1036}Two possibilities to check: either a {\Tt{}\Rm{}{\bf{}false}\nwendquote} relational or a number
close to zero.
\nwenddocs{}\nwbegincode{1037}\sublabel{NWppJ6t-DKmU5-1N}\nwmargintag{{\nwtagstyle{}\subpageref{NWppJ6t-DKmU5-1N}}}\moddef{figure class~{\nwtagstyle{}\subpageref{NWppJ6t-DKmU5-1}}}\plusendmoddef\Rm{}\nwstartdeflinemarkup\nwusesondefline{\\{NWppJ6t-32jW6L-1}}\nwprevnextdefs{NWppJ6t-DKmU5-1M}{NWppJ6t-DKmU5-1O}\nwenddeflinemarkup
                            {\bf{}if}  ({\it{}is\_a}\begin{math}<\end{math}{\bf{}relational}\begin{math}>\end{math}({\it{}ic})) {\nwlbrace}
                                {\bf{}if} (\begin{math}\neg\end{math}({\bf{}bool}){\it{}ex\_to}\begin{math}<\end{math}{\bf{}relational}\begin{math}>\end{math}({\it{}ic}))
                                    {\it{}to\_add}={\bf{}false};
                            {\nwrbrace} {\bf{}else} {\bf{}if} ({\it{}is\_less\_than\_epsilon}({\it{}ic}))
                                {\it{}to\_add}={\bf{}false};
                        {\nwrbrace}
                    {\bf{}if} (\begin{math}\neg\end{math} {\it{}to\_add})
                        {\bf{}break};
                {\nwrbrace}
                {\bf{}if} ({\it{}to\_add})
                    {\it{}res}.{\it{}append}({\it{}inew});
            {\nwrbrace}
        //cerr\begin{math}<\end{math}\begin{math}<\end{math} "Result: "; res.dbgprint();
        {\nwrbrace} {\bf{}else}
            {\it{}res}={\it{}ex\_to}\begin{math}<\end{math}{\bf{}lst}\begin{math}>\end{math}({\it{}update\_cycle\_node}({\it{}key}, {\it{}use\_cond}, {\it{}store\_cond}, {\it{}res}, {\it{}level}+1));
    {\nwrbrace}
    {\bf{}if} ({\it{}level} \begin{math}\equiv\end{math}0)
        {\bf{}return} {\it{}unique\_cycle}({\it{}res});
    {\bf{}else}
        {\bf{}return} {\it{}res};
{\nwrbrace}

\nwused{\\{NWppJ6t-32jW6L-1}}\nwidentuses{\\{{\nwixident{is{\_}less{\_}than{\_}epsilon}}{is:unless:unthan:unepsilon}}\\{{\nwixident{key}}{key}}\\{{\nwixident{unique{\_}cycle}}{unique:uncycle}}\\{{\nwixident{update{\_}cycle{\_}node}}{update:uncycle:unnode}}}\nwindexuse{\nwixident{is{\_}less{\_}than{\_}epsilon}}{is:unless:unthan:unepsilon}{NWppJ6t-DKmU5-1N}\nwindexuse{\nwixident{key}}{key}{NWppJ6t-DKmU5-1N}\nwindexuse{\nwixident{unique{\_}cycle}}{unique:uncycle}{NWppJ6t-DKmU5-1N}\nwindexuse{\nwixident{update{\_}cycle{\_}node}}{update:uncycle:unnode}{NWppJ6t-DKmU5-1N}\nwendcode{}\nwbegindocs{1038}If the cycle is defined by by a {\Tt{}\Rm{}{\bf{}subfigure}\nwendquote} all calculations are done
within it.
\nwenddocs{}\nwbegincode{1039}\sublabel{NWppJ6t-3AODA9-1}\nwmargintag{{\nwtagstyle{}\subpageref{NWppJ6t-3AODA9-1}}}\moddef{update node zero level~{\nwtagstyle{}\subpageref{NWppJ6t-3AODA9-1}}}\endmoddef\Rm{}\nwstartdeflinemarkup\nwusesondefline{\\{NWppJ6t-DKmU5-1H}}\nwprevnextdefs{\relax}{NWppJ6t-3AODA9-2}\nwenddeflinemarkup
    {\bf{}if} ( {\it{}nodes}[{\it{}key}].{\it{}get\_parents}().{\it{}nops}() \begin{math}\equiv\end{math} 1 \begin{math}\wedge\end{math} {\it{}is\_a}\begin{math}<\end{math}{\bf{}subfigure}\begin{math}>\end{math}({\it{}nodes}[{\it{}key}].{\it{}get\_parents}().{\it{}op}(0))) {\nwlbrace}
        {\bf{}figure} {\it{}F}={\it{}ex\_to}\begin{math}<\end{math}{\bf{}figure}\begin{math}>\end{math}({\it{}ex\_to}\begin{math}<\end{math}{\bf{}basic}\begin{math}>\end{math}({\it{}ex\_to}\begin{math}<\end{math}{\bf{}subfigure}\begin{math}>\end{math}({\it{}nodes}[{\it{}key}].{\it{}get\_parents}().{\it{}op}(0)).{\it{}get\_subf}())
                               .{\it{}clearflag}({\it{}status\_flags}::{\it{}expanded}));
        {\it{}F}={\it{}float\_evaluation}? {\it{}F}.{\it{}set\_float\_eval}(): {\it{}F}.{\it{}set\_exact\_eval}();

\nwalsodefined{\\{NWppJ6t-3AODA9-2}\\{NWppJ6t-3AODA9-3}}\nwused{\\{NWppJ6t-DKmU5-1H}}\nwidentuses{\\{{\nwixident{figure}}{figure}}\\{{\nwixident{float{\_}evaluation}}{float:unevaluation}}\\{{\nwixident{key}}{key}}\\{{\nwixident{nodes}}{nodes}}\\{{\nwixident{nops}}{nops}}\\{{\nwixident{op}}{op}}\\{{\nwixident{set{\_}exact{\_}eval}}{set:unexact:uneval}}\\{{\nwixident{set{\_}float{\_}eval}}{set:unfloat:uneval}}\\{{\nwixident{subfigure}}{subfigure}}}\nwindexuse{\nwixident{figure}}{figure}{NWppJ6t-3AODA9-1}\nwindexuse{\nwixident{float{\_}evaluation}}{float:unevaluation}{NWppJ6t-3AODA9-1}\nwindexuse{\nwixident{key}}{key}{NWppJ6t-3AODA9-1}\nwindexuse{\nwixident{nodes}}{nodes}{NWppJ6t-3AODA9-1}\nwindexuse{\nwixident{nops}}{nops}{NWppJ6t-3AODA9-1}\nwindexuse{\nwixident{op}}{op}{NWppJ6t-3AODA9-1}\nwindexuse{\nwixident{set{\_}exact{\_}eval}}{set:unexact:uneval}{NWppJ6t-3AODA9-1}\nwindexuse{\nwixident{set{\_}float{\_}eval}}{set:unfloat:uneval}{NWppJ6t-3AODA9-1}\nwindexuse{\nwixident{subfigure}}{subfigure}{NWppJ6t-3AODA9-1}\nwendcode{}\nwbegindocs{1040}We replace parameters of the {\Tt{}\Rm{}{\bf{}subfigure}\nwendquote} by current parents and
evaluate the result.
\nwenddocs{}\nwbegincode{1041}\sublabel{NWppJ6t-3AODA9-2}\nwmargintag{{\nwtagstyle{}\subpageref{NWppJ6t-3AODA9-2}}}\moddef{update node zero level~{\nwtagstyle{}\subpageref{NWppJ6t-3AODA9-1}}}\plusendmoddef\Rm{}\nwstartdeflinemarkup\nwusesondefline{\\{NWppJ6t-DKmU5-1H}}\nwprevnextdefs{NWppJ6t-3AODA9-1}{NWppJ6t-3AODA9-3}\nwenddeflinemarkup
        {\bf{}lst} {\it{}parkeys}={\it{}ex\_to}\begin{math}<\end{math}{\bf{}lst}\begin{math}>\end{math}({\it{}ex\_to}\begin{math}<\end{math}{\bf{}subfigure}\begin{math}>\end{math}({\it{}nodes}[{\it{}key}].{\it{}get\_parents}().{\it{}op}(0)).{\it{}get\_parlist}());
        {\bf{}unsigned} {\bf{}int} {\it{}var}=0;
        {\bf{}char} {\it{}name}[12];
        {\bf{}for} ({\bf{}const} {\bf{}auto}& {\it{}it} : {\it{}parkeys}) {\nwlbrace}
            {\it{}sprintf}({\it{}name}, {\tt{}"variable
            {\it{}F}.{\it{}set\_cycle}({\it{}F}.{\it{}get\_cycle\_label}({\it{}name}), {\it{}nodes}[{\it{}it}].{\it{}get\_cycles}());
            \protect\PP{\it{}var};
        {\nwrbrace}
        {\it{}F}.{\it{}set\_metric}({\it{}point\_metric},{\it{}cycle\_metric}); // this calls automatic figure re-calculation
        {\bf{}return} {\it{}F}.{\it{}get\_cycle}({\it{}F}.{\it{}get\_cycle\_label}({\tt{}"result"}));

\nwused{\\{NWppJ6t-DKmU5-1H}}\nwidentuses{\\{{\nwixident{cycle{\_}metric}}{cycle:unmetric}}\\{{\nwixident{figure}}{figure}}\\{{\nwixident{get{\_}cycle}}{get:uncycle}}\\{{\nwixident{get{\_}cycle{\_}label}}{get:uncycle:unlabel}}\\{{\nwixident{key}}{key}}\\{{\nwixident{name}}{name}}\\{{\nwixident{nodes}}{nodes}}\\{{\nwixident{op}}{op}}\\{{\nwixident{point{\_}metric}}{point:unmetric}}\\{{\nwixident{set{\_}cycle}}{set:uncycle}}\\{{\nwixident{set{\_}metric}}{set:unmetric}}\\{{\nwixident{subfigure}}{subfigure}}}\nwindexuse{\nwixident{cycle{\_}metric}}{cycle:unmetric}{NWppJ6t-3AODA9-2}\nwindexuse{\nwixident{figure}}{figure}{NWppJ6t-3AODA9-2}\nwindexuse{\nwixident{get{\_}cycle}}{get:uncycle}{NWppJ6t-3AODA9-2}\nwindexuse{\nwixident{get{\_}cycle{\_}label}}{get:uncycle:unlabel}{NWppJ6t-3AODA9-2}\nwindexuse{\nwixident{key}}{key}{NWppJ6t-3AODA9-2}\nwindexuse{\nwixident{name}}{name}{NWppJ6t-3AODA9-2}\nwindexuse{\nwixident{nodes}}{nodes}{NWppJ6t-3AODA9-2}\nwindexuse{\nwixident{op}}{op}{NWppJ6t-3AODA9-2}\nwindexuse{\nwixident{point{\_}metric}}{point:unmetric}{NWppJ6t-3AODA9-2}\nwindexuse{\nwixident{set{\_}cycle}}{set:uncycle}{NWppJ6t-3AODA9-2}\nwindexuse{\nwixident{set{\_}metric}}{set:unmetric}{NWppJ6t-3AODA9-2}\nwindexuse{\nwixident{subfigure}}{subfigure}{NWppJ6t-3AODA9-2}\nwendcode{}\nwbegindocs{1042}For a list of relations we simply set up a symbolic cycle and
proceed with calculations in recursion.
\nwenddocs{}\nwbegincode{1043}\sublabel{NWppJ6t-3AODA9-3}\nwmargintag{{\nwtagstyle{}\subpageref{NWppJ6t-3AODA9-3}}}\moddef{update node zero level~{\nwtagstyle{}\subpageref{NWppJ6t-3AODA9-1}}}\plusendmoddef\Rm{}\nwstartdeflinemarkup\nwusesondefline{\\{NWppJ6t-DKmU5-1H}}\nwprevnextdefs{NWppJ6t-3AODA9-2}{\relax}\nwenddeflinemarkup
    {\nwrbrace} {\bf{}else}
        {\it{}nodes}[{\it{}key}].{\it{}set\_cycles}({\bf{}cycle\_data}({\it{}k}, {\bf{}indexed}({\bf{}matrix}(1, {\it{}ex\_to}\begin{math}<\end{math}{\bf{}numeric}\begin{math}>\end{math}({\it{}get\_dim}()).{\it{}to\_int}(), {\it{}l}), {\bf{}varidx}({\it{}key}, {\it{}ex\_to}\begin{math}<\end{math}{\bf{}numeric}\begin{math}>\end{math}({\it{}get\_dim}()).{\it{}to\_int}())), {\it{}m}, {\bf{}false}));

\nwused{\\{NWppJ6t-DKmU5-1H}}\nwidentuses{\\{{\nwixident{cycle{\_}data}}{cycle:undata}}\\{{\nwixident{get{\_}dim()}}{get:undim()}}\\{{\nwixident{k}}{k}}\\{{\nwixident{key}}{key}}\\{{\nwixident{l}}{l}}\\{{\nwixident{m}}{m}}\\{{\nwixident{nodes}}{nodes}}\\{{\nwixident{numeric}}{numeric}}}\nwindexuse{\nwixident{cycle{\_}data}}{cycle:undata}{NWppJ6t-3AODA9-3}\nwindexuse{\nwixident{get{\_}dim()}}{get:undim()}{NWppJ6t-3AODA9-3}\nwindexuse{\nwixident{k}}{k}{NWppJ6t-3AODA9-3}\nwindexuse{\nwixident{key}}{key}{NWppJ6t-3AODA9-3}\nwindexuse{\nwixident{l}}{l}{NWppJ6t-3AODA9-3}\nwindexuse{\nwixident{m}}{m}{NWppJ6t-3AODA9-3}\nwindexuse{\nwixident{nodes}}{nodes}{NWppJ6t-3AODA9-3}\nwindexuse{\nwixident{numeric}}{numeric}{NWppJ6t-3AODA9-3}\nwendcode{}\nwbegindocs{1044}The figure is updated.
\nwenddocs{}\nwbegincode{1045}\sublabel{NWppJ6t-DKmU5-1O}\nwmargintag{{\nwtagstyle{}\subpageref{NWppJ6t-DKmU5-1O}}}\moddef{figure class~{\nwtagstyle{}\subpageref{NWppJ6t-DKmU5-1}}}\plusendmoddef\Rm{}\nwstartdeflinemarkup\nwusesondefline{\\{NWppJ6t-32jW6L-1}}\nwprevnextdefs{NWppJ6t-DKmU5-1N}{NWppJ6t-DKmU5-1P}\nwenddeflinemarkup
{\bf{}figure} {\bf{}figure}::{\it{}update\_cycles}()
{\nwlbrace}
    {\bf{}if} ({\it{}info}({\it{}status\_flags}::{\it{}expanded}))
        {\bf{}return} \begin{math}\ast\end{math}{\it{}this};
    {\bf{}lst} {\it{}all\_child};
    {\bf{}for} ({\bf{}auto}& {\it{}x}: {\it{}nodes})
        {\bf{}if} ({\it{}ex\_to}\begin{math}<\end{math}{\bf{}cycle\_node}\begin{math}>\end{math}({\it{}x}.{\it{}second}).{\it{}get\_generation}() \begin{math}\equiv\end{math} 0) {\nwlbrace}
            {\bf{}if} ({\it{}ex\_to}\begin{math}<\end{math}{\bf{}cycle\_node}\begin{math}>\end{math}({\it{}x}.{\it{}second}).{\it{}get\_parents}().{\it{}nops}() \begin{math}>\end{math} 0)
                {\it{}nodes}[{\it{}x}.{\it{}first}].{\it{}set\_cycles}({\it{}ex\_to}\begin{math}<\end{math}{\bf{}lst}\begin{math}>\end{math}({\it{}update\_cycle\_node}({\it{}x}.{\it{}first})));
\nwindexdefn{\nwixident{update{\_}cycles}}{update:uncycles}{NWppJ6t-DKmU5-1O}\eatline
\nwused{\\{NWppJ6t-32jW6L-1}}\nwidentdefs{\\{{\nwixident{update{\_}cycles}}{update:uncycles}}}\nwidentuses{\\{{\nwixident{cycle{\_}node}}{cycle:unnode}}\\{{\nwixident{figure}}{figure}}\\{{\nwixident{get{\_}generation}}{get:ungeneration}}\\{{\nwixident{info}}{info}}\\{{\nwixident{nodes}}{nodes}}\\{{\nwixident{nops}}{nops}}\\{{\nwixident{update{\_}cycle{\_}node}}{update:uncycle:unnode}}}\nwindexuse{\nwixident{cycle{\_}node}}{cycle:unnode}{NWppJ6t-DKmU5-1O}\nwindexuse{\nwixident{figure}}{figure}{NWppJ6t-DKmU5-1O}\nwindexuse{\nwixident{get{\_}generation}}{get:ungeneration}{NWppJ6t-DKmU5-1O}\nwindexuse{\nwixident{info}}{info}{NWppJ6t-DKmU5-1O}\nwindexuse{\nwixident{nodes}}{nodes}{NWppJ6t-DKmU5-1O}\nwindexuse{\nwixident{nops}}{nops}{NWppJ6t-DKmU5-1O}\nwindexuse{\nwixident{update{\_}cycle{\_}node}}{update:uncycle:unnode}{NWppJ6t-DKmU5-1O}\nwendcode{}\nwbegindocs{1046}\nwdocspar
\nwenddocs{}\nwbegindocs{1047}We collect all children of the zero-generation cycles for subsequent update.
\nwenddocs{}\nwbegincode{1048}\sublabel{NWppJ6t-DKmU5-1P}\nwmargintag{{\nwtagstyle{}\subpageref{NWppJ6t-DKmU5-1P}}}\moddef{figure class~{\nwtagstyle{}\subpageref{NWppJ6t-DKmU5-1}}}\plusendmoddef\Rm{}\nwstartdeflinemarkup\nwusesondefline{\\{NWppJ6t-32jW6L-1}}\nwprevnextdefs{NWppJ6t-DKmU5-1O}{NWppJ6t-DKmU5-1Q}\nwenddeflinemarkup
            {\bf{}lst} {\it{}ch}={\it{}ex\_to}\begin{math}<\end{math}{\bf{}cycle\_node}\begin{math}>\end{math}({\it{}x}.{\it{}second}).{\it{}get\_children}();
            {\bf{}for} ({\bf{}const} {\bf{}auto}& {\it{}it1} : {\it{}ch})
                {\it{}all\_child}.{\it{}append}({\it{}it1});
        {\nwrbrace}
    {\it{}all\_child}.{\it{}sort}();
    {\it{}all\_child}.{\it{}unique}();
    {\it{}update\_node\_lst}({\it{}all\_child});
    {\bf{}return} \begin{math}\ast\end{math}{\it{}this};
{\nwrbrace}

\nwused{\\{NWppJ6t-32jW6L-1}}\nwidentuses{\\{{\nwixident{cycle{\_}node}}{cycle:unnode}}\\{{\nwixident{update{\_}node{\_}lst}}{update:unnode:unlst}}}\nwindexuse{\nwixident{cycle{\_}node}}{cycle:unnode}{NWppJ6t-DKmU5-1P}\nwindexuse{\nwixident{update{\_}node{\_}lst}}{update:unnode:unlst}{NWppJ6t-DKmU5-1P}\nwendcode{}\nwbegindocs{1049}\nwdocspar
\subsubsection{Additional methods}
\label{sec:additional-methods}

\nwenddocs{}\nwbegindocs{1050}Set the new metric for the figure, repeating the previous code from
the constructor.
\nwenddocs{}\nwbegincode{1051}\sublabel{NWppJ6t-DKmU5-1Q}\nwmargintag{{\nwtagstyle{}\subpageref{NWppJ6t-DKmU5-1Q}}}\moddef{figure class~{\nwtagstyle{}\subpageref{NWppJ6t-DKmU5-1}}}\plusendmoddef\Rm{}\nwstartdeflinemarkup\nwusesondefline{\\{NWppJ6t-32jW6L-1}}\nwprevnextdefs{NWppJ6t-DKmU5-1P}{NWppJ6t-DKmU5-1R}\nwenddeflinemarkup
{\bf{}void} {\bf{}figure}::{\it{}set\_metric}({\bf{}const} {\bf{}ex} & {\it{}Mp}, {\bf{}const} {\bf{}ex} & {\it{}Mc})\nwindexdefn{\nwixident{figure}}{figure}{NWppJ6t-DKmU5-1Q}
{\nwlbrace}
    {\bf{}ex} {\it{}D}={\it{}get\_dim}();
    \LA{}set point metric in figure~{\nwtagstyle{}\subpageref{NWppJ6t-4ALcRV-1}}\RA{}
    \LA{}set cycle metric in figure~{\nwtagstyle{}\subpageref{NWppJ6t-1wPO4G-1}}\RA{}
    \LA{}check dimensionalities point and cycle metrics~{\nwtagstyle{}\subpageref{NWppJ6t-11clwc-1}}\RA{}
\nwindexdefn{\nwixident{set{\_}metric}}{set:unmetric}{NWppJ6t-DKmU5-1Q}\eatline
\nwused{\\{NWppJ6t-32jW6L-1}}\nwidentdefs{\\{{\nwixident{figure}}{figure}}\\{{\nwixident{set{\_}metric}}{set:unmetric}}}\nwidentuses{\\{{\nwixident{ex}}{ex}}\\{{\nwixident{get{\_}dim()}}{get:undim()}}}\nwindexuse{\nwixident{ex}}{ex}{NWppJ6t-DKmU5-1Q}\nwindexuse{\nwixident{get{\_}dim()}}{get:undim()}{NWppJ6t-DKmU5-1Q}\nwendcode{}\nwbegindocs{1052}\nwdocspar
\nwenddocs{}\nwbegindocs{1053}We check that the dimensionality of the new metric matches the old one.
\nwenddocs{}\nwbegincode{1054}\sublabel{NWppJ6t-DKmU5-1R}\nwmargintag{{\nwtagstyle{}\subpageref{NWppJ6t-DKmU5-1R}}}\moddef{figure class~{\nwtagstyle{}\subpageref{NWppJ6t-DKmU5-1}}}\plusendmoddef\Rm{}\nwstartdeflinemarkup\nwusesondefline{\\{NWppJ6t-32jW6L-1}}\nwprevnextdefs{NWppJ6t-DKmU5-1Q}{NWppJ6t-DKmU5-1S}\nwenddeflinemarkup
    {\bf{}if} (\begin{math}\neg\end{math} ({\it{}D}-{\it{}get\_dim}()).{\it{}is\_zero}())
        {\bf{}throw}({\it{}std}::{\it{}invalid\_argument}({\tt{}"New metric has a different dimensionality!"}));
    {\it{}update\_cycles}();
{\nwrbrace}

\nwused{\\{NWppJ6t-32jW6L-1}}\nwidentuses{\\{{\nwixident{get{\_}dim()}}{get:undim()}}\\{{\nwixident{update{\_}cycles}}{update:uncycles}}}\nwindexuse{\nwixident{get{\_}dim()}}{get:undim()}{NWppJ6t-DKmU5-1R}\nwindexuse{\nwixident{update{\_}cycles}}{update:uncycles}{NWppJ6t-DKmU5-1R}\nwendcode{}\nwbegindocs{1055}The method collects all key for nodes with generations in the range
{[}{\Tt{}\Rm{}{\it{}intgen}\nwendquote},{\Tt{}\Rm{}{\it{}maxgen}\nwendquote}{]} inclusively.
\nwenddocs{}\nwbegincode{1056}\sublabel{NWppJ6t-DKmU5-1S}\nwmargintag{{\nwtagstyle{}\subpageref{NWppJ6t-DKmU5-1S}}}\moddef{figure class~{\nwtagstyle{}\subpageref{NWppJ6t-DKmU5-1}}}\plusendmoddef\Rm{}\nwstartdeflinemarkup\nwusesondefline{\\{NWppJ6t-32jW6L-1}}\nwprevnextdefs{NWppJ6t-DKmU5-1R}{NWppJ6t-DKmU5-1T}\nwenddeflinemarkup
{\bf{}ex} {\bf{}figure}::{\it{}get\_all\_keys}({\bf{}const} {\bf{}int} {\it{}mingen}, {\bf{}const} {\bf{}int} {\it{}maxgen}) {\bf{}const} {\nwlbrace}
    {\bf{}lst} {\it{}keys};
    {\bf{}int} {\it{}mg}={\it{}get\_max\_generation}();
    {\bf{}if} ({\it{}maxgen} \begin{math}\neq\end{math} {\it{}GHOST\_GEN} \begin{math}\wedge\end{math} {\it{}maxgen} \begin{math}<\end{math} {\it{}mg})
            {\it{}mg} ={\it{}maxgen};
    {\bf{}for} ({\bf{}int} {\it{}i}={\it{}mingen}; {\it{}i} \begin{math}\leq\end{math} {\it{}mg}; \protect\PP{\it{}i})
        {\bf{}for} ({\bf{}const} {\bf{}auto}& {\it{}x}: {\it{}nodes}) {\nwlbrace}
            {\bf{}if} ({\it{}x}.{\it{}second}.{\it{}get\_generation}() \begin{math}\equiv\end{math} {\it{}i})
                {\it{}keys}.{\it{}append}({\it{}x}.{\it{}first});
    {\nwrbrace}
    {\bf{}return} {\it{}keys};
{\nwrbrace}
\nwindexdefn{\nwixident{get{\_}all{\_}keys}}{get:unall:unkeys}{NWppJ6t-DKmU5-1S}\eatline
\nwused{\\{NWppJ6t-32jW6L-1}}\nwidentdefs{\\{{\nwixident{get{\_}all{\_}keys}}{get:unall:unkeys}}}\nwidentuses{\\{{\nwixident{ex}}{ex}}\\{{\nwixident{figure}}{figure}}\\{{\nwixident{get{\_}generation}}{get:ungeneration}}\\{{\nwixident{get{\_}max{\_}generation}}{get:unmax:ungeneration}}\\{{\nwixident{GHOST{\_}GEN}}{GHOST:unGEN}}\\{{\nwixident{nodes}}{nodes}}}\nwindexuse{\nwixident{ex}}{ex}{NWppJ6t-DKmU5-1S}\nwindexuse{\nwixident{figure}}{figure}{NWppJ6t-DKmU5-1S}\nwindexuse{\nwixident{get{\_}generation}}{get:ungeneration}{NWppJ6t-DKmU5-1S}\nwindexuse{\nwixident{get{\_}max{\_}generation}}{get:unmax:ungeneration}{NWppJ6t-DKmU5-1S}\nwindexuse{\nwixident{GHOST{\_}GEN}}{GHOST:unGEN}{NWppJ6t-DKmU5-1S}\nwindexuse{\nwixident{nodes}}{nodes}{NWppJ6t-DKmU5-1S}\nwendcode{}\nwbegindocs{1057}\nwdocspar
\nwenddocs{}\nwbegindocs{1058}Scanning for the biggest number generation.
\nwenddocs{}\nwbegincode{1059}\sublabel{NWppJ6t-DKmU5-1T}\nwmargintag{{\nwtagstyle{}\subpageref{NWppJ6t-DKmU5-1T}}}\moddef{figure class~{\nwtagstyle{}\subpageref{NWppJ6t-DKmU5-1}}}\plusendmoddef\Rm{}\nwstartdeflinemarkup\nwusesondefline{\\{NWppJ6t-32jW6L-1}}\nwprevnextdefs{NWppJ6t-DKmU5-1S}{NWppJ6t-DKmU5-1U}\nwenddeflinemarkup
{\bf{}int} {\bf{}figure}::{\it{}get\_max\_generation}() {\bf{}const} {\nwlbrace}\nwindexdefn{\nwixident{figure}}{figure}{NWppJ6t-DKmU5-1T}
    {\bf{}int} {\it{}max\_gen} = {\it{}REAL\_LINE\_GEN};
    {\bf{}for} ({\bf{}const} {\bf{}auto}& {\it{}x}: {\it{}nodes})
        {\bf{}if} ({\it{}x}.{\it{}second}.{\it{}get\_generation}() \begin{math}>\end{math} {\it{}max\_gen})
            {\it{}max\_gen} = {\it{}x}.{\it{}second}.{\it{}get\_generation}();
    {\bf{}return} {\it{}max\_gen};
{\nwrbrace}
\nwindexdefn{\nwixident{get{\_}max{\_}generation}}{get:unmax:ungeneration}{NWppJ6t-DKmU5-1T}\eatline
\nwused{\\{NWppJ6t-32jW6L-1}}\nwidentdefs{\\{{\nwixident{figure}}{figure}}\\{{\nwixident{get{\_}max{\_}generation}}{get:unmax:ungeneration}}}\nwidentuses{\\{{\nwixident{get{\_}generation}}{get:ungeneration}}\\{{\nwixident{nodes}}{nodes}}\\{{\nwixident{REAL{\_}LINE{\_}GEN}}{REAL:unLINE:unGEN}}}\nwindexuse{\nwixident{get{\_}generation}}{get:ungeneration}{NWppJ6t-DKmU5-1T}\nwindexuse{\nwixident{nodes}}{nodes}{NWppJ6t-DKmU5-1T}\nwindexuse{\nwixident{REAL{\_}LINE{\_}GEN}}{REAL:unLINE:unGEN}{NWppJ6t-DKmU5-1T}\nwendcode{}\nwbegindocs{1060}\nwdocspar

\nwenddocs{}\nwbegindocs{1061}Return the list of cycles stored in the node with {\Tt{}\Rm{}{\it{}key}\nwendquote}.
\nwenddocs{}\nwbegincode{1062}\sublabel{NWppJ6t-DKmU5-1U}\nwmargintag{{\nwtagstyle{}\subpageref{NWppJ6t-DKmU5-1U}}}\moddef{figure class~{\nwtagstyle{}\subpageref{NWppJ6t-DKmU5-1}}}\plusendmoddef\Rm{}\nwstartdeflinemarkup\nwusesondefline{\\{NWppJ6t-32jW6L-1}}\nwprevnextdefs{NWppJ6t-DKmU5-1T}{NWppJ6t-DKmU5-1V}\nwenddeflinemarkup
{\bf{}ex} {\bf{}figure}::{\it{}get\_cycle}({\bf{}const} {\bf{}ex} & {\it{}key}, {\bf{}const} {\bf{}ex} & {\it{}metric}) {\bf{}const}
{\nwlbrace}
    {\it{}exhashmap}\begin{math}<\end{math}{\bf{}cycle\_node}\begin{math}>\end{math}::{\it{}const\_iterator}  {\it{}cnode}={\it{}nodes}.{\it{}find}({\it{}key});
    {\bf{}if} ({\it{}cnode} \begin{math}\equiv\end{math} {\it{}nodes}.{\it{}end}()) {\nwlbrace}
        {\bf{}if} ({\it{}FIGURE\_DEBUG})
            {\it{}cerr} \begin{math}\ll\end{math} {\tt{}"There is no key "} \begin{math}\ll\end{math} {\it{}key} \begin{math}\ll\end{math} {\tt{}" in the figure."} \begin{math}\ll\end{math} {\it{}endl};
        {\bf{}return} {\bf{}lst}{\nwlbrace}{\nwrbrace};
    {\nwrbrace} {\bf{}else}
        {\bf{}return} {\it{}cnode}\begin{math}\rightarrow\end{math}{\it{}second}.{\it{}get\_cycle}({\it{}metric});
{\nwrbrace}
\nwindexdefn{\nwixident{get{\_}cycle}}{get:uncycle}{NWppJ6t-DKmU5-1U}\eatline
\nwused{\\{NWppJ6t-32jW6L-1}}\nwidentdefs{\\{{\nwixident{get{\_}cycle}}{get:uncycle}}}\nwidentuses{\\{{\nwixident{cycle{\_}node}}{cycle:unnode}}\\{{\nwixident{ex}}{ex}}\\{{\nwixident{figure}}{figure}}\\{{\nwixident{FIGURE{\_}DEBUG}}{FIGURE:unDEBUG}}\\{{\nwixident{key}}{key}}\\{{\nwixident{nodes}}{nodes}}}\nwindexuse{\nwixident{cycle{\_}node}}{cycle:unnode}{NWppJ6t-DKmU5-1U}\nwindexuse{\nwixident{ex}}{ex}{NWppJ6t-DKmU5-1U}\nwindexuse{\nwixident{figure}}{figure}{NWppJ6t-DKmU5-1U}\nwindexuse{\nwixident{FIGURE{\_}DEBUG}}{FIGURE:unDEBUG}{NWppJ6t-DKmU5-1U}\nwindexuse{\nwixident{key}}{key}{NWppJ6t-DKmU5-1U}\nwindexuse{\nwixident{nodes}}{nodes}{NWppJ6t-DKmU5-1U}\nwendcode{}\nwbegindocs{1063}\nwdocspar
\nwenddocs{}\nwbegindocs{1064}Full reset of figure to the initial empty state.
\nwenddocs{}\nwbegincode{1065}\sublabel{NWppJ6t-DKmU5-1V}\nwmargintag{{\nwtagstyle{}\subpageref{NWppJ6t-DKmU5-1V}}}\moddef{figure class~{\nwtagstyle{}\subpageref{NWppJ6t-DKmU5-1}}}\plusendmoddef\Rm{}\nwstartdeflinemarkup\nwusesondefline{\\{NWppJ6t-32jW6L-1}}\nwprevnextdefs{NWppJ6t-DKmU5-1U}{NWppJ6t-DKmU5-1W}\nwenddeflinemarkup
{\bf{}void} {\bf{}figure}::{\it{}reset\_figure}()\nwindexdefn{\nwixident{figure}}{figure}{NWppJ6t-DKmU5-1V}
{\nwlbrace}
    {\it{}nodes}.{\it{}clear}();
    \LA{}set the infinity~{\nwtagstyle{}\subpageref{NWppJ6t-2j51zU-1}}\RA{}
    \LA{}set the real line~{\nwtagstyle{}\subpageref{NWppJ6t-2q1QAq-1}}\RA{}
{\nwrbrace}
\nwindexdefn{\nwixident{reset{\_}figure}}{reset:unfigure}{NWppJ6t-DKmU5-1V}\eatline
\nwused{\\{NWppJ6t-32jW6L-1}}\nwidentdefs{\\{{\nwixident{figure}}{figure}}\\{{\nwixident{reset{\_}figure}}{reset:unfigure}}}\nwidentuses{\\{{\nwixident{nodes}}{nodes}}}\nwindexuse{\nwixident{nodes}}{nodes}{NWppJ6t-DKmU5-1V}\nwendcode{}\nwbegindocs{1066}\nwdocspar
\nwenddocs{}\nwbegindocs{1067}Update nodes in the list and all their (grand)children subsequently.
\nwenddocs{}\nwbegincode{1068}\sublabel{NWppJ6t-DKmU5-1W}\nwmargintag{{\nwtagstyle{}\subpageref{NWppJ6t-DKmU5-1W}}}\moddef{figure class~{\nwtagstyle{}\subpageref{NWppJ6t-DKmU5-1}}}\plusendmoddef\Rm{}\nwstartdeflinemarkup\nwusesondefline{\\{NWppJ6t-32jW6L-1}}\nwprevnextdefs{NWppJ6t-DKmU5-1V}{NWppJ6t-DKmU5-1X}\nwenddeflinemarkup
{\bf{}void} {\bf{}figure}::{\it{}update\_node\_lst}({\bf{}const} {\bf{}ex} & {\it{}inlist})\nwindexdefn{\nwixident{figure}}{figure}{NWppJ6t-DKmU5-1W}
{\nwlbrace}
    {\bf{}if} ({\it{}info}({\it{}status\_flags}::{\it{}expanded}))
        {\bf{}return};

    {\bf{}lst} {\it{}intake}={\it{}ex\_to}\begin{math}<\end{math}{\bf{}lst}\begin{math}>\end{math}({\it{}inlist});
    {\bf{}while} ({\it{}intake}.{\it{}nops}() \begin{math}\neq\end{math}0) {\nwlbrace}
        {\bf{}int} {\it{}mingen}={\it{}nodes}[\begin{math}\ast\end{math}{\it{}intake}.{\it{}begin}()].{\it{}get\_generation}();
        {\bf{}for} ({\bf{}const} {\bf{}auto}& {\it{}it} : {\it{}intake})
            {\it{}mingen}={\it{}min}({\it{}mingen}, {\it{}nodes}[{\it{}it}].{\it{}get\_generation}());
        {\bf{}lst} {\it{}current}, {\it{}future};
        {\bf{}for} ({\bf{}const} {\bf{}auto}& {\it{}it} : {\it{}intake})
            {\bf{}if} ({\it{}nodes}[{\it{}it}].{\it{}get\_generation}() \begin{math}\equiv\end{math} {\it{}mingen})
                {\it{}current}.{\it{}append}({\it{}it});
            {\bf{}else}
                {\it{}future}.{\it{}append}({\it{}it});
\nwindexdefn{\nwixident{update{\_}node{\_}lst}}{update:unnode:unlst}{NWppJ6t-DKmU5-1W}\eatline
\nwused{\\{NWppJ6t-32jW6L-1}}\nwidentdefs{\\{{\nwixident{figure}}{figure}}\\{{\nwixident{update{\_}node{\_}lst}}{update:unnode:unlst}}}\nwidentuses{\\{{\nwixident{ex}}{ex}}\\{{\nwixident{get{\_}generation}}{get:ungeneration}}\\{{\nwixident{info}}{info}}\\{{\nwixident{nodes}}{nodes}}\\{{\nwixident{nops}}{nops}}}\nwindexuse{\nwixident{ex}}{ex}{NWppJ6t-DKmU5-1W}\nwindexuse{\nwixident{get{\_}generation}}{get:ungeneration}{NWppJ6t-DKmU5-1W}\nwindexuse{\nwixident{info}}{info}{NWppJ6t-DKmU5-1W}\nwindexuse{\nwixident{nodes}}{nodes}{NWppJ6t-DKmU5-1W}\nwindexuse{\nwixident{nops}}{nops}{NWppJ6t-DKmU5-1W}\nwendcode{}\nwbegindocs{1069}\nwdocspar
\nwenddocs{}\nwbegindocs{1070}All nodes at the current list are updated.
\nwenddocs{}\nwbegincode{1071}\sublabel{NWppJ6t-DKmU5-1X}\nwmargintag{{\nwtagstyle{}\subpageref{NWppJ6t-DKmU5-1X}}}\moddef{figure class~{\nwtagstyle{}\subpageref{NWppJ6t-DKmU5-1}}}\plusendmoddef\Rm{}\nwstartdeflinemarkup\nwusesondefline{\\{NWppJ6t-32jW6L-1}}\nwprevnextdefs{NWppJ6t-DKmU5-1W}{NWppJ6t-DKmU5-1Y}\nwenddeflinemarkup
    {\bf{}for} ({\bf{}const} {\bf{}auto}& {\it{}it} : {\it{}current}) {\nwlbrace}
        {\it{}nodes}[{\it{}it}].{\it{}set\_cycles}({\it{}ex\_to}\begin{math}<\end{math}{\bf{}lst}\begin{math}>\end{math}({\it{}update\_cycle\_node}({\it{}it})));
        {\bf{}lst} {\it{}nchild}={\it{}nodes}[{\it{}it}].{\it{}get\_children}();
        {\bf{}for} ({\bf{}const} {\bf{}auto}& {\it{}it1} : {\it{}nchild})
            {\it{}future}.{\it{}append}({\it{}it1});
    {\nwrbrace}

\nwused{\\{NWppJ6t-32jW6L-1}}\nwidentuses{\\{{\nwixident{nodes}}{nodes}}\\{{\nwixident{update{\_}cycle{\_}node}}{update:uncycle:unnode}}}\nwindexuse{\nwixident{nodes}}{nodes}{NWppJ6t-DKmU5-1X}\nwindexuse{\nwixident{update{\_}cycle{\_}node}}{update:uncycle:unnode}{NWppJ6t-DKmU5-1X}\nwendcode{}\nwbegindocs{1072}Future list becomes new intake.
\nwenddocs{}\nwbegincode{1073}\sublabel{NWppJ6t-DKmU5-1Y}\nwmargintag{{\nwtagstyle{}\subpageref{NWppJ6t-DKmU5-1Y}}}\moddef{figure class~{\nwtagstyle{}\subpageref{NWppJ6t-DKmU5-1}}}\plusendmoddef\Rm{}\nwstartdeflinemarkup\nwusesondefline{\\{NWppJ6t-32jW6L-1}}\nwprevnextdefs{NWppJ6t-DKmU5-1X}{NWppJ6t-DKmU5-1Z}\nwenddeflinemarkup
        {\it{}intake}={\it{}future};
        {\it{}intake}.{\it{}sort}();
        {\it{}intake}.{\it{}unique}();
    {\nwrbrace}
{\nwrbrace}

\nwused{\\{NWppJ6t-32jW6L-1}}\nwendcode{}\nwbegindocs{1074}Find a symbolic key for a cycle labelled by a {\Tt{}\Rm{}{\it{}name}\nwendquote}.
\nwenddocs{}\nwbegincode{1075}\sublabel{NWppJ6t-DKmU5-1Z}\nwmargintag{{\nwtagstyle{}\subpageref{NWppJ6t-DKmU5-1Z}}}\moddef{figure class~{\nwtagstyle{}\subpageref{NWppJ6t-DKmU5-1}}}\plusendmoddef\Rm{}\nwstartdeflinemarkup\nwusesondefline{\\{NWppJ6t-32jW6L-1}}\nwprevnextdefs{NWppJ6t-DKmU5-1Y}{NWppJ6t-DKmU5-1a}\nwenddeflinemarkup
{\bf{}ex} {\bf{}figure}::{\it{}get\_cycle\_label}({\it{}string} {\it{}name}) {\bf{}const}
{\nwlbrace}
    {\bf{}for} ({\bf{}const} {\bf{}auto}& {\it{}x}: {\it{}nodes})
        {\bf{}if} ({\it{}ex\_to}\begin{math}<\end{math}{\bf{}symbol}\begin{math}>\end{math}({\it{}x}.{\it{}first}).{\it{}get\_name}() \begin{math}\equiv\end{math} {\it{}name})
            {\bf{}return} {\it{}x}.{\it{}first};

    {\bf{}return} 0;
{\nwrbrace}
\nwindexdefn{\nwixident{get{\_}cycle{\_}label}}{get:uncycle:unlabel}{NWppJ6t-DKmU5-1Z}\eatline
\nwused{\\{NWppJ6t-32jW6L-1}}\nwidentdefs{\\{{\nwixident{get{\_}cycle{\_}label}}{get:uncycle:unlabel}}}\nwidentuses{\\{{\nwixident{ex}}{ex}}\\{{\nwixident{figure}}{figure}}\\{{\nwixident{name}}{name}}\\{{\nwixident{nodes}}{nodes}}}\nwindexuse{\nwixident{ex}}{ex}{NWppJ6t-DKmU5-1Z}\nwindexuse{\nwixident{figure}}{figure}{NWppJ6t-DKmU5-1Z}\nwindexuse{\nwixident{name}}{name}{NWppJ6t-DKmU5-1Z}\nwindexuse{\nwixident{nodes}}{nodes}{NWppJ6t-DKmU5-1Z}\nwendcode{}\nwbegindocs{1076}\nwdocspar
\nwenddocs{}\nwbegindocs{1077}\nwdocspar
\subsubsection{Drawing methods}
\label{sec:drawing-methods}

\nwenddocs{}\nwbegindocs{1078}Drawing the figure is possible only in two dimensions, thus we check
this at the start.
\nwenddocs{}\nwbegincode{1079}\sublabel{NWppJ6t-DKmU5-1a}\nwmargintag{{\nwtagstyle{}\subpageref{NWppJ6t-DKmU5-1a}}}\moddef{figure class~{\nwtagstyle{}\subpageref{NWppJ6t-DKmU5-1}}}\plusendmoddef\Rm{}\nwstartdeflinemarkup\nwusesondefline{\\{NWppJ6t-32jW6L-1}}\nwprevnextdefs{NWppJ6t-DKmU5-1Z}{NWppJ6t-DKmU5-1b}\nwenddeflinemarkup
{\bf{}void} {\bf{}figure}::{\it{}asy\_draw}({\it{}ostream} & {\it{}ost}, {\it{}ostream} & {\it{}err}, {\bf{}const} {\it{}string} {\it{}picture},\nwindexdefn{\nwixident{figure}}{figure}{NWppJ6t-DKmU5-1a}
                      {\bf{}const} {\bf{}ex} & {\it{}xmin}, {\bf{}const} {\bf{}ex} & {\it{}xmax}, {\bf{}const} {\bf{}ex} & {\it{}ymin}, {\bf{}const} {\bf{}ex} & {\it{}ymax},
                      {\it{}asy\_style} {\it{}style}, {\it{}label\_string} {\it{}lstring}, {\bf{}bool} {\it{}with\_realline},
                      {\bf{}bool} {\it{}with\_header}, {\bf{}int} {\it{}points\_per\_arc}, {\bf{}const} {\it{}string} {\it{}imaginary\_options},
                      {\bf{}bool} {\it{}with\_labels}) {\bf{}const}
{\nwlbrace}
    \LA{}check that dimensionality is 2~{\nwtagstyle{}\subpageref{NWppJ6t-A2asn-1}}\RA{}
\nwindexdefn{\nwixident{asy{\_}draw}}{asy:undraw}{NWppJ6t-DKmU5-1a}\eatline
\nwused{\\{NWppJ6t-32jW6L-1}}\nwidentdefs{\\{{\nwixident{asy{\_}draw}}{asy:undraw}}\\{{\nwixident{figure}}{figure}}}\nwidentuses{\\{{\nwixident{asy{\_}style}}{asy:unstyle}}\\{{\nwixident{ex}}{ex}}\\{{\nwixident{label{\_}string}}{label:unstring}}}\nwindexuse{\nwixident{asy{\_}style}}{asy:unstyle}{NWppJ6t-DKmU5-1a}\nwindexuse{\nwixident{ex}}{ex}{NWppJ6t-DKmU5-1a}\nwindexuse{\nwixident{label{\_}string}}{label:unstring}{NWppJ6t-DKmU5-1a}\nwendcode{}\nwbegindocs{1080}\nwdocspar
\nwenddocs{}\nwbegindocs{1081}\nwdocspar
\nwenddocs{}\nwbegincode{1082}\sublabel{NWppJ6t-A2asn-1}\nwmargintag{{\nwtagstyle{}\subpageref{NWppJ6t-A2asn-1}}}\moddef{check that dimensionality is 2~{\nwtagstyle{}\subpageref{NWppJ6t-A2asn-1}}}\endmoddef\Rm{}\nwstartdeflinemarkup\nwusesondefline{\\{NWppJ6t-DKmU5-1a}\\{NWppJ6t-DKmU5-1l}\\{NWppJ6t-DKmU5-1o}}\nwenddeflinemarkup
    {\bf{}if} (\begin{math}\neg\end{math} ({\it{}get\_dim}()-2).{\it{}is\_zero}())
        {\bf{}throw} {\it{}logic\_error}({\tt{}"Drawing is possible for two-dimensional figures only!"});

\nwused{\\{NWppJ6t-DKmU5-1a}\\{NWppJ6t-DKmU5-1l}\\{NWppJ6t-DKmU5-1o}}\nwidentuses{\\{{\nwixident{get{\_}dim()}}{get:undim()}}}\nwindexuse{\nwixident{get{\_}dim()}}{get:undim()}{NWppJ6t-A2asn-1}\nwendcode{}\nwbegindocs{1083}We will need to place different types of cycle into the different
places of the \Asymptote\ file.
\nwenddocs{}\nwbegincode{1084}\sublabel{NWppJ6t-DKmU5-1b}\nwmargintag{{\nwtagstyle{}\subpageref{NWppJ6t-DKmU5-1b}}}\moddef{figure class~{\nwtagstyle{}\subpageref{NWppJ6t-DKmU5-1}}}\plusendmoddef\Rm{}\nwstartdeflinemarkup\nwusesondefline{\\{NWppJ6t-32jW6L-1}}\nwprevnextdefs{NWppJ6t-DKmU5-1a}{NWppJ6t-DKmU5-1c}\nwenddeflinemarkup
    {\it{}stringstream} {\it{}preamble\_stream}, {\it{}main\_stream}, {\it{}labels\_stream};
    {\it{}string} {\it{}dots};
    {\it{}std}::{\it{}regex} {\it{}re}({\tt{}"dot{\char92}{\char92}("});

\nwused{\\{NWppJ6t-32jW6L-1}}\nwendcode{}\nwbegindocs{1085}Some bits will depend on the metric in the point space.
\nwenddocs{}\nwbegincode{1086}\sublabel{NWppJ6t-DKmU5-1c}\nwmargintag{{\nwtagstyle{}\subpageref{NWppJ6t-DKmU5-1c}}}\moddef{figure class~{\nwtagstyle{}\subpageref{NWppJ6t-DKmU5-1}}}\plusendmoddef\Rm{}\nwstartdeflinemarkup\nwusesondefline{\\{NWppJ6t-32jW6L-1}}\nwprevnextdefs{NWppJ6t-DKmU5-1b}{NWppJ6t-DKmU5-1d}\nwenddeflinemarkup
        {\bf{}int} {\it{}point\_metric\_signature}={\it{}ex\_to}\begin{math}<\end{math}{\bf{}numeric}\begin{math}>\end{math}({\it{}ex\_to}\begin{math}<\end{math}{\bf{}clifford}\begin{math}>\end{math}({\it{}point\_metric}).{\it{}get\_metric}({\bf{}idx}(0,2),{\bf{}idx}(0,2))
                                                  \begin{math}\ast\end{math}{\it{}ex\_to}\begin{math}<\end{math}{\bf{}clifford}\begin{math}>\end{math}({\it{}point\_metric}).{\it{}get\_metric}({\bf{}idx}(1,2),{\bf{}idx}(1,2)).{\it{}eval}()).{\it{}to\_int}();

    {\bf{}for} ({\bf{}const} {\bf{}auto}& {\it{}x}: {\it{}nodes}) {\nwlbrace}
        {\bf{}lst} {\it{}cycles}={\it{}ex\_to}\begin{math}<\end{math}{\bf{}lst}\begin{math}>\end{math}({\it{}x}.{\it{}second}.{\it{}get\_cycle}({\it{}point\_metric}));
        {\bf{}for} ({\bf{}const} {\bf{}auto}& {\it{}it1} : {\it{}cycles})
            {\bf{}try} {\nwlbrace}
                {\bf{}if} ( ({\it{}x}.{\it{}second}.{\it{}get\_generation}() \begin{math}>\end{math} {\it{}REAL\_LINE\_GEN}) \begin{math}\vee\end{math}
                     (({\it{}x}.{\it{}second}.{\it{}get\_generation}() \begin{math}\equiv\end{math} {\it{}REAL\_LINE\_GEN}) \begin{math}\wedge\end{math} {\it{}with\_realline})) {\nwlbrace}
                    {\it{}stringstream} {\it{}sstr};
                    {\bf{}if} ({\it{}with\_header})
                        {\it{}sstr} \begin{math}\ll\end{math} {\tt{}"// label: "} \begin{math}\ll\end{math} ({\it{}x}.{\it{}first}) \begin{math}\ll\end{math} {\it{}endl};

\nwused{\\{NWppJ6t-32jW6L-1}}\nwidentuses{\\{{\nwixident{get{\_}cycle}}{get:uncycle}}\\{{\nwixident{get{\_}generation}}{get:ungeneration}}\\{{\nwixident{nodes}}{nodes}}\\{{\nwixident{numeric}}{numeric}}\\{{\nwixident{point{\_}metric}}{point:unmetric}}\\{{\nwixident{REAL{\_}LINE{\_}GEN}}{REAL:unLINE:unGEN}}}\nwindexuse{\nwixident{get{\_}cycle}}{get:uncycle}{NWppJ6t-DKmU5-1c}\nwindexuse{\nwixident{get{\_}generation}}{get:ungeneration}{NWppJ6t-DKmU5-1c}\nwindexuse{\nwixident{nodes}}{nodes}{NWppJ6t-DKmU5-1c}\nwindexuse{\nwixident{numeric}}{numeric}{NWppJ6t-DKmU5-1c}\nwindexuse{\nwixident{point{\_}metric}}{point:unmetric}{NWppJ6t-DKmU5-1c}\nwindexuse{\nwixident{REAL{\_}LINE{\_}GEN}}{REAL:unLINE:unGEN}{NWppJ6t-DKmU5-1c}\nwendcode{}\nwbegindocs{1087}Produce the coulour and style for the cycle.
\nwenddocs{}\nwbegincode{1088}\sublabel{NWppJ6t-DKmU5-1d}\nwmargintag{{\nwtagstyle{}\subpageref{NWppJ6t-DKmU5-1d}}}\moddef{figure class~{\nwtagstyle{}\subpageref{NWppJ6t-DKmU5-1}}}\plusendmoddef\Rm{}\nwstartdeflinemarkup\nwusesondefline{\\{NWppJ6t-32jW6L-1}}\nwprevnextdefs{NWppJ6t-DKmU5-1c}{NWppJ6t-DKmU5-1e}\nwenddeflinemarkup
                {\bf{}lst} {\it{}colours};
                {\it{}string} {\it{}asy\_opt};
                {\bf{}if} ({\it{}x}.{\it{}second}.{\it{}custom\_asy}\begin{math}\equiv\end{math}{\tt{}""}) {\nwlbrace}
                    {\it{}asy\_opt}={\it{}style}({\it{}x}.{\it{}first}, ({\it{}it1}), {\it{}colours});
                {\nwrbrace} {\bf{}else}
                    {\it{}asy\_opt}={\it{}x}.{\it{}second}.{\it{}custom\_asy};

\nwused{\\{NWppJ6t-32jW6L-1}}\nwendcode{}\nwbegindocs{1089}Zero-radius cycles are treated specially, its centre become known to
Asymptote as a {\Tt{}\Rm{}{\it{}pair}\nwendquote}.
\nwenddocs{}\nwbegincode{1090}\sublabel{NWppJ6t-DKmU5-1e}\nwmargintag{{\nwtagstyle{}\subpageref{NWppJ6t-DKmU5-1e}}}\moddef{figure class~{\nwtagstyle{}\subpageref{NWppJ6t-DKmU5-1}}}\plusendmoddef\Rm{}\nwstartdeflinemarkup\nwusesondefline{\\{NWppJ6t-32jW6L-1}}\nwprevnextdefs{NWppJ6t-DKmU5-1d}{NWppJ6t-DKmU5-1f}\nwenddeflinemarkup
            {\bf{}if} ({\it{}is\_less\_than\_epsilon}({\it{}ex\_to}\begin{math}<\end{math}{\bf{}cycle}\begin{math}>\end{math}({\it{}it1}).{\it{}det}())) {\nwlbrace}
                {\bf{}double} {\it{}x1}={\it{}ex\_to}\begin{math}<\end{math}{\bf{}numeric}\begin{math}>\end{math}({\it{}ex\_to}\begin{math}<\end{math}{\bf{}cycle}\begin{math}>\end{math}({\it{}it1}).{\it{}center}({\it{}cycle\_metric}).{\it{}op}(0)
                                        .{\it{}evalf}()).{\it{}to\_double}(),
                    {\it{}y1}={\it{}ex\_to}\begin{math}<\end{math}{\bf{}numeric}\begin{math}>\end{math}({\it{}ex\_to}\begin{math}<\end{math}{\bf{}cycle}\begin{math}>\end{math}({\it{}it1}).{\it{}center}({\it{}cycle\_metric}).{\it{}op}(1)
                                        .{\it{}evalf}()).{\it{}to\_double}();
                {\it{}string} {\it{}var\_name}={\it{}regex\_replace}({\it{}ex\_to}\begin{math}<\end{math}{\bf{}symbol}\begin{math}>\end{math}({\it{}x}.{\it{}first}).{\it{}get\_name}(), {\it{}regex}({\tt{}"[[:space:]]+"}), {\tt{}"\_"});
                {\it{}preamble\_stream} \begin{math}\ll\end{math} {\tt{}"// label: "} \begin{math}\ll\end{math} ({\it{}x}.{\it{}first}) \begin{math}\ll\end{math} {\it{}endl}
                                \begin{math}\ll\end{math} {\tt{}"pair "} \begin{math}\ll\end{math} {\it{}var\_name} \begin{math}\ll\end{math} {\tt{}"=("}
                                \begin{math}\ll\end{math} {\it{}x1} \begin{math}\ll\end{math} {\tt{}","} \begin{math}\ll\end{math} {\it{}y1} \begin{math}\ll\end{math} {\tt{}");"} \begin{math}\ll\end{math} {\it{}endl};

\nwused{\\{NWppJ6t-32jW6L-1}}\nwidentuses{\\{{\nwixident{cycle{\_}metric}}{cycle:unmetric}}\\{{\nwixident{evalf}}{evalf}}\\{{\nwixident{is{\_}less{\_}than{\_}epsilon}}{is:unless:unthan:unepsilon}}\\{{\nwixident{numeric}}{numeric}}\\{{\nwixident{op}}{op}}}\nwindexuse{\nwixident{cycle{\_}metric}}{cycle:unmetric}{NWppJ6t-DKmU5-1e}\nwindexuse{\nwixident{evalf}}{evalf}{NWppJ6t-DKmU5-1e}\nwindexuse{\nwixident{is{\_}less{\_}than{\_}epsilon}}{is:unless:unthan:unepsilon}{NWppJ6t-DKmU5-1e}\nwindexuse{\nwixident{numeric}}{numeric}{NWppJ6t-DKmU5-1e}\nwindexuse{\nwixident{op}}{op}{NWppJ6t-DKmU5-1e}\nwendcode{}\nwbegindocs{1091}In the elliptic case we place the dot explicitly\ldots
\nwenddocs{}\nwbegincode{1092}\sublabel{NWppJ6t-DKmU5-1f}\nwmargintag{{\nwtagstyle{}\subpageref{NWppJ6t-DKmU5-1f}}}\moddef{figure class~{\nwtagstyle{}\subpageref{NWppJ6t-DKmU5-1}}}\plusendmoddef\Rm{}\nwstartdeflinemarkup\nwusesondefline{\\{NWppJ6t-32jW6L-1}}\nwprevnextdefs{NWppJ6t-DKmU5-1e}{NWppJ6t-DKmU5-1g}\nwenddeflinemarkup
                {\bf{}if} ({\it{}point\_metric\_signature} \begin{math}>\end{math} 0
                    \begin{math}\wedge\end{math} {\it{}xmin} \begin{math}\leq\end{math} {\it{}x1} \begin{math}\wedge\end{math} {\it{}x1}\begin{math}\leq\end{math}{\it{}xmax} \begin{math}\wedge\end{math} {\it{}ymin} \begin{math}\leq\end{math} {\it{}y1} \begin{math}\wedge\end{math} {\it{}y1}\begin{math}\leq\end{math}{\it{}ymax}) {\nwlbrace}
                        {\it{}sstr} \begin{math}\ll\end{math} {\tt{}"dot("} \begin{math}\ll\end{math} {\it{}var\_name} \begin{math}\ll\end{math} {\tt{}", rgb("}
                             \begin{math}\ll\end{math} {\it{}ex\_to}\begin{math}<\end{math}{\bf{}numeric}\begin{math}>\end{math}({\it{}colours}.{\it{}op}(0)).{\it{}to\_double}() \begin{math}\ll\end{math} {\tt{}","}
                             \begin{math}\ll\end{math} {\it{}ex\_to}\begin{math}<\end{math}{\bf{}numeric}\begin{math}>\end{math}({\it{}colours}.{\it{}op}(1)).{\it{}to\_double}() \begin{math}\ll\end{math}{\tt{}","}
                             \begin{math}\ll\end{math} {\it{}ex\_to}\begin{math}<\end{math}{\bf{}numeric}\begin{math}>\end{math}({\it{}colours}.{\it{}op}(2)).{\it{}to\_double}()
                             \begin{math}\ll\end{math} {\tt{}") "} \begin{math}\ll\end{math} {\tt{}");"} \begin{math}\ll\end{math} {\it{}endl};

\nwused{\\{NWppJ6t-32jW6L-1}}\nwidentuses{\\{{\nwixident{numeric}}{numeric}}\\{{\nwixident{op}}{op}}\\{{\nwixident{rgb}}{rgb}}}\nwindexuse{\nwixident{numeric}}{numeric}{NWppJ6t-DKmU5-1f}\nwindexuse{\nwixident{op}}{op}{NWppJ6t-DKmU5-1f}\nwindexuse{\nwixident{rgb}}{rgb}{NWppJ6t-DKmU5-1f}\nwendcode{}\nwbegindocs{1093}\ldots , otherwise output is handled by the {\Tt{}\Rm{}{\bf{}cycle2D}::{\it{}draw\_asy}\nwendquote} method
\nwenddocs{}\nwbegincode{1094}\sublabel{NWppJ6t-DKmU5-1g}\nwmargintag{{\nwtagstyle{}\subpageref{NWppJ6t-DKmU5-1g}}}\moddef{figure class~{\nwtagstyle{}\subpageref{NWppJ6t-DKmU5-1}}}\plusendmoddef\Rm{}\nwstartdeflinemarkup\nwusesondefline{\\{NWppJ6t-32jW6L-1}}\nwprevnextdefs{NWppJ6t-DKmU5-1f}{NWppJ6t-DKmU5-1h}\nwenddeflinemarkup
                {\nwrbrace} {\bf{}else} {\nwlbrace}
                    {\it{}ex\_to}\begin{math}<\end{math}{\bf{}cycle2D}\begin{math}>\end{math}({\it{}it1}).{\it{}asy\_draw}({\it{}sstr}, {\it{}picture}, {\it{}xmin}, {\it{}xmax},
                                             {\it{}ymin}, {\it{}ymax}, {\it{}colours}, {\it{}asy\_opt}, {\it{}with\_header}, {\it{}points\_per\_arc}, {\it{}imaginary\_options});

\nwused{\\{NWppJ6t-32jW6L-1}}\nwidentuses{\\{{\nwixident{asy{\_}draw}}{asy:undraw}}}\nwindexuse{\nwixident{asy{\_}draw}}{asy:undraw}{NWppJ6t-DKmU5-1g}\nwendcode{}\nwbegindocs{1095} Since in parabolic spaces zero-radius cycles are detached from the
their centres, which they denote we wish to have a hint on centres positions.
\nwenddocs{}\nwbegincode{1096}\sublabel{NWppJ6t-DKmU5-1h}\nwmargintag{{\nwtagstyle{}\subpageref{NWppJ6t-DKmU5-1h}}}\moddef{figure class~{\nwtagstyle{}\subpageref{NWppJ6t-DKmU5-1}}}\plusendmoddef\Rm{}\nwstartdeflinemarkup\nwusesondefline{\\{NWppJ6t-32jW6L-1}}\nwprevnextdefs{NWppJ6t-DKmU5-1g}{NWppJ6t-DKmU5-1i}\nwenddeflinemarkup
                    {\bf{}if} ({\it{}FIGURE\_DEBUG} \begin{math}\wedge\end{math} {\it{}point\_metric\_signature}\begin{math}\equiv\end{math}0
                        \begin{math}\wedge\end{math} {\it{}xmin} \begin{math}\leq\end{math} {\it{}x1} \begin{math}\wedge\end{math} {\it{}x1}\begin{math}\leq\end{math}{\it{}xmax} \begin{math}\wedge\end{math} {\it{}ymin} \begin{math}\leq\end{math} {\it{}y1} \begin{math}\wedge\end{math} {\it{}y1}\begin{math}\leq\end{math}{\it{}ymax})
                        {\it{}sstr} \begin{math}\ll\end{math} {\tt{}"dot("} \begin{math}\ll\end{math} {\it{}var\_name} \begin{math}\ll\end{math} {\tt{}", black+3pt);"} \begin{math}\ll\end{math} {\it{}endl};
                {\nwrbrace}

\nwused{\\{NWppJ6t-32jW6L-1}}\nwidentuses{\\{{\nwixident{FIGURE{\_}DEBUG}}{FIGURE:unDEBUG}}}\nwindexuse{\nwixident{FIGURE{\_}DEBUG}}{FIGURE:unDEBUG}{NWppJ6t-DKmU5-1h}\nwendcode{}\nwbegindocs{1097}Drawing a generic cycle through {\Tt{}\Rm{}{\bf{}cycle2D}::{\it{}draw\_asy}\nwendquote} method
\nwenddocs{}\nwbegincode{1098}\sublabel{NWppJ6t-DKmU5-1i}\nwmargintag{{\nwtagstyle{}\subpageref{NWppJ6t-DKmU5-1i}}}\moddef{figure class~{\nwtagstyle{}\subpageref{NWppJ6t-DKmU5-1}}}\plusendmoddef\Rm{}\nwstartdeflinemarkup\nwusesondefline{\\{NWppJ6t-32jW6L-1}}\nwprevnextdefs{NWppJ6t-DKmU5-1h}{NWppJ6t-DKmU5-1j}\nwenddeflinemarkup
            {\nwrbrace} {\bf{}else}
                {\it{}ex\_to}\begin{math}<\end{math}{\bf{}cycle2D}\begin{math}>\end{math}({\it{}it1}).{\it{}asy\_draw}({\it{}sstr}, {\it{}picture}, {\it{}xmin}, {\it{}xmax},
                                             {\it{}ymin}, {\it{}ymax}, {\it{}colours}, {\it{}asy\_opt}, {\it{}with\_header}, {\it{}points\_per\_arc}, {\it{}imaginary\_options});

\nwused{\\{NWppJ6t-32jW6L-1}}\nwidentuses{\\{{\nwixident{asy{\_}draw}}{asy:undraw}}}\nwindexuse{\nwixident{asy{\_}draw}}{asy:undraw}{NWppJ6t-DKmU5-1i}\nwendcode{}\nwbegindocs{1099} Dots and label will be drawn last to avoid over-painting.
\nwenddocs{}\nwbegincode{1100}\sublabel{NWppJ6t-DKmU5-1j}\nwmargintag{{\nwtagstyle{}\subpageref{NWppJ6t-DKmU5-1j}}}\moddef{figure class~{\nwtagstyle{}\subpageref{NWppJ6t-DKmU5-1}}}\plusendmoddef\Rm{}\nwstartdeflinemarkup\nwusesondefline{\\{NWppJ6t-32jW6L-1}}\nwprevnextdefs{NWppJ6t-DKmU5-1i}{NWppJ6t-DKmU5-1k}\nwenddeflinemarkup
                {\bf{}if} ({\it{}std}::{\it{}regex\_search}({\it{}sstr}.{\it{}str}(), {\it{}re}))
                    {\it{}dots}+={\it{}sstr}.{\it{}str}();
                {\bf{}else}
                    {\it{}main\_stream} \begin{math}\ll\end{math} {\it{}sstr}.{\it{}str}();

\nwused{\\{NWppJ6t-32jW6L-1}}\nwendcode{}\nwbegindocs{1101}Find the label position
\nwenddocs{}\nwbegincode{1102}\sublabel{NWppJ6t-DKmU5-1k}\nwmargintag{{\nwtagstyle{}\subpageref{NWppJ6t-DKmU5-1k}}}\moddef{figure class~{\nwtagstyle{}\subpageref{NWppJ6t-DKmU5-1}}}\plusendmoddef\Rm{}\nwstartdeflinemarkup\nwusesondefline{\\{NWppJ6t-32jW6L-1}}\nwprevnextdefs{NWppJ6t-DKmU5-1j}{NWppJ6t-DKmU5-1l}\nwenddeflinemarkup
                {\bf{}if} ({\it{}with\_labels})
                    {\it{}labels\_stream} \begin{math}\ll\end{math} {\it{}lstring}({\it{}x}.{\it{}first}, ({\it{}it1}), {\it{}sstr}.{\it{}str}());
                {\nwrbrace}
            {\nwrbrace} {\bf{}catch} ({\it{}exception} &{\it{}p}) {\nwlbrace}
                {\bf{}if} ({\it{}FIGURE\_DEBUG})
                    {\it{}err} \begin{math}\ll\end{math} {\tt{}"Failed to draw "} \begin{math}\ll\end{math} {\it{}x}.{\it{}first} \begin{math}\ll\end{math}{\tt{}": "} \begin{math}\ll\end{math} {\it{}x}.{\it{}second};
            {\nwrbrace}
    {\nwrbrace}
    //cerr \begin{math}<\end{math}\begin{math}<\end{math} "Dots: " \begin{math}<\end{math}\begin{math}<\end{math} dots;
    {\it{}ost} \begin{math}\ll\end{math} {\it{}preamble\_stream}.{\it{}str}()
        \begin{math}\ll\end{math} {\it{}main\_stream}.{\it{}str}()
        \begin{math}\ll\end{math} {\it{}dots}
        \begin{math}\ll\end{math} {\it{}labels\_stream}.{\it{}str}();
{\nwrbrace}

\nwused{\\{NWppJ6t-32jW6L-1}}\nwidentuses{\\{{\nwixident{FIGURE{\_}DEBUG}}{FIGURE:unDEBUG}}}\nwindexuse{\nwixident{FIGURE{\_}DEBUG}}{FIGURE:unDEBUG}{NWppJ6t-DKmU5-1k}\nwendcode{}\nwbegindocs{1103}\nwdocspar
\nwenddocs{}\nwbegincode{1104}\sublabel{NWppJ6t-DKmU5-1l}\nwmargintag{{\nwtagstyle{}\subpageref{NWppJ6t-DKmU5-1l}}}\moddef{figure class~{\nwtagstyle{}\subpageref{NWppJ6t-DKmU5-1}}}\plusendmoddef\Rm{}\nwstartdeflinemarkup\nwusesondefline{\\{NWppJ6t-32jW6L-1}}\nwprevnextdefs{NWppJ6t-DKmU5-1k}{NWppJ6t-DKmU5-1m}\nwenddeflinemarkup
{\bf{}void} {\bf{}figure}::{\it{}asy\_write}({\bf{}int} {\it{}size}, {\bf{}const} {\bf{}ex} & {\it{}xmin}, {\bf{}const} {\bf{}ex} & {\it{}xmax}, {\bf{}const} {\bf{}ex} & {\it{}ymin}, {\bf{}const} {\bf{}ex} & {\it{}ymax},\nwindexdefn{\nwixident{figure}}{figure}{NWppJ6t-DKmU5-1l}
                      {\it{}string} {\it{}name}, {\it{}string} {\it{}format},
                      {\it{}asy\_style} {\it{}style}, {\it{}label\_string} {\it{}lstring}, {\bf{}bool} {\it{}with\_realline},
                      {\bf{}bool} {\it{}with\_header}, {\bf{}int} {\it{}points\_per\_arc}, {\bf{}const} {\it{}string} {\it{}imaginary\_options},
                      {\bf{}bool} {\it{}rm\_asy\_file}, {\bf{}bool} {\it{}with\_labels}) {\bf{}const}
{\nwlbrace}
    \LA{}check that dimensionality is 2~{\nwtagstyle{}\subpageref{NWppJ6t-A2asn-1}}\RA{}
\nwindexdefn{\nwixident{asy{\_}write}}{asy:unwrite}{NWppJ6t-DKmU5-1l}\eatline
\nwused{\\{NWppJ6t-32jW6L-1}}\nwidentdefs{\\{{\nwixident{asy{\_}write}}{asy:unwrite}}\\{{\nwixident{figure}}{figure}}}\nwidentuses{\\{{\nwixident{asy{\_}style}}{asy:unstyle}}\\{{\nwixident{ex}}{ex}}\\{{\nwixident{label{\_}string}}{label:unstring}}\\{{\nwixident{name}}{name}}}\nwindexuse{\nwixident{asy{\_}style}}{asy:unstyle}{NWppJ6t-DKmU5-1l}\nwindexuse{\nwixident{ex}}{ex}{NWppJ6t-DKmU5-1l}\nwindexuse{\nwixident{label{\_}string}}{label:unstring}{NWppJ6t-DKmU5-1l}\nwindexuse{\nwixident{name}}{name}{NWppJ6t-DKmU5-1l}\nwendcode{}\nwbegindocs{1105}\nwdocspar
\nwenddocs{}\nwbegindocs{1106}Open the file.
\nwenddocs{}\nwbegincode{1107}\sublabel{NWppJ6t-DKmU5-1m}\nwmargintag{{\nwtagstyle{}\subpageref{NWppJ6t-DKmU5-1m}}}\moddef{figure class~{\nwtagstyle{}\subpageref{NWppJ6t-DKmU5-1}}}\plusendmoddef\Rm{}\nwstartdeflinemarkup\nwusesondefline{\\{NWppJ6t-32jW6L-1}}\nwprevnextdefs{NWppJ6t-DKmU5-1l}{NWppJ6t-DKmU5-1n}\nwenddeflinemarkup
    {\it{}string} {\it{}filename}={\it{}name}+{\tt{}".asy"};
    {\it{}ofstream} {\it{}out}({\it{}filename});
    {\it{}out} \begin{math}\ll\end{math} {\tt{}"size("} \begin{math}\ll\end{math} {\it{}size} \begin{math}\ll\end{math} {\tt{}");"} \begin{math}\ll\end{math} {\it{}endl};
    {\it{}asy\_draw}({\it{}out}, {\it{}cerr}, {\tt{}""}, {\it{}xmin}, {\it{}xmax}, {\it{}ymin}, {\it{}ymax},
             {\it{}style}, {\it{}lstring}, {\it{}with\_realline}, {\it{}with\_header}, {\it{}points\_per\_arc}, {\it{}imaginary\_options}, {\it{}with\_labels});
    {\bf{}if} ({\it{}name} \begin{math}\equiv\end{math} {\tt{}""})
        {\it{}out} \begin{math}\ll\end{math} {\tt{}"shipout();"} \begin{math}\ll\end{math} {\it{}endl};
    {\bf{}else}
        {\it{}out} \begin{math}\ll\end{math} {\tt{}"shipout({\char92}""} \begin{math}\ll\end{math} {\it{}name} \begin{math}\ll\end{math} {\tt{}"{\char92}");"} \begin{math}\ll\end{math} {\it{}endl};
    {\it{}out}.{\it{}flush}();
    {\it{}out}.{\it{}close}();

\nwused{\\{NWppJ6t-32jW6L-1}}\nwidentuses{\\{{\nwixident{asy{\_}draw}}{asy:undraw}}\\{{\nwixident{name}}{name}}}\nwindexuse{\nwixident{asy{\_}draw}}{asy:undraw}{NWppJ6t-DKmU5-1m}\nwindexuse{\nwixident{name}}{name}{NWppJ6t-DKmU5-1m}\nwendcode{}\nwbegindocs{1108}Preparation of \Asymptote\ call.
\nwenddocs{}\nwbegincode{1109}\sublabel{NWppJ6t-DKmU5-1n}\nwmargintag{{\nwtagstyle{}\subpageref{NWppJ6t-DKmU5-1n}}}\moddef{figure class~{\nwtagstyle{}\subpageref{NWppJ6t-DKmU5-1}}}\plusendmoddef\Rm{}\nwstartdeflinemarkup\nwusesondefline{\\{NWppJ6t-32jW6L-1}}\nwprevnextdefs{NWppJ6t-DKmU5-1m}{NWppJ6t-DKmU5-1o}\nwenddeflinemarkup
    {\bf{}char} {\it{}command}[256];
    {\it{}strcpy}({\it{}command}, {\it{}show\_asy\_graphics}? {\tt{}"asy -V"} : {\tt{}"asy"});
    {\bf{}if} ({\it{}format} \begin{math}\neq\end{math} {\tt{}""}) {\nwlbrace}
        {\it{}strcat}({\it{}command}, {\tt{}" -f "});
        {\it{}strcat}({\it{}command}, {\it{}format}.{\it{}c\_str}());
    {\nwrbrace}
    {\it{}strcat}({\it{}command}, {\tt{}" "});
    {\it{}strcat}({\it{}command}, {\it{}name}.{\it{}c\_str}());
    {\bf{}char} \begin{math}\ast\end{math} {\it{}pcommand}={\it{}command};
    {\it{}system}({\it{}pcommand});
    {\bf{}if} ({\it{}rm\_asy\_file})
        {\it{}remove}({\it{}filename}.{\it{}c\_str}());
{\nwrbrace}

\nwused{\\{NWppJ6t-32jW6L-1}}\nwidentuses{\\{{\nwixident{name}}{name}}\\{{\nwixident{show{\_}asy{\_}graphics}}{show:unasy:ungraphics}}}\nwindexuse{\nwixident{name}}{name}{NWppJ6t-DKmU5-1n}\nwindexuse{\nwixident{show{\_}asy{\_}graphics}}{show:unasy:ungraphics}{NWppJ6t-DKmU5-1n}\nwendcode{}\nwbegindocs{1110}This method animates figures with parameters.
\nwenddocs{}\nwbegincode{1111}\sublabel{NWppJ6t-DKmU5-1o}\nwmargintag{{\nwtagstyle{}\subpageref{NWppJ6t-DKmU5-1o}}}\moddef{figure class~{\nwtagstyle{}\subpageref{NWppJ6t-DKmU5-1}}}\plusendmoddef\Rm{}\nwstartdeflinemarkup\nwusesondefline{\\{NWppJ6t-32jW6L-1}}\nwprevnextdefs{NWppJ6t-DKmU5-1n}{NWppJ6t-DKmU5-1p}\nwenddeflinemarkup
{\bf{}void} {\bf{}figure}::{\it{}asy\_animate}({\bf{}const} {\bf{}ex} &{\it{}val},\nwindexdefn{\nwixident{figure}}{figure}{NWppJ6t-DKmU5-1o}
                         {\bf{}int} {\it{}size}, {\bf{}const} {\bf{}ex} & {\it{}xmin}, {\bf{}const} {\bf{}ex} & {\it{}xmax}, {\bf{}const} {\bf{}ex} & {\it{}ymin}, {\bf{}const} {\bf{}ex} & {\it{}ymax},
                         {\it{}string} {\it{}name}, {\it{}string} {\it{}format}, {\it{}asy\_style} {\it{}style}, {\it{}label\_string} {\it{}lstring}, {\bf{}bool} {\it{}with\_realline},
                         {\bf{}bool} {\it{}with\_header}, {\bf{}int} {\it{}points\_per\_arc}, {\bf{}const} {\it{}string} {\it{}imaginary\_options},
                         {\bf{}const} {\it{}string} {\it{}values\_position}, {\bf{}bool} {\it{}rm\_asy\_file}, {\bf{}bool} {\it{}with\_labels}) {\bf{}const}
{\nwlbrace}
    \LA{}check that dimensionality is 2~{\nwtagstyle{}\subpageref{NWppJ6t-A2asn-1}}\RA{}
    {\it{}string} {\it{}filename}={\it{}name}+{\tt{}".asy"};
    {\it{}ofstream} {\it{}out}({\it{}filename});
\nwindexdefn{\nwixident{asy{\_}animate}}{asy:unanimate}{NWppJ6t-DKmU5-1o}\eatline
\nwused{\\{NWppJ6t-32jW6L-1}}\nwidentdefs{\\{{\nwixident{asy{\_}animate}}{asy:unanimate}}\\{{\nwixident{figure}}{figure}}}\nwidentuses{\\{{\nwixident{asy{\_}style}}{asy:unstyle}}\\{{\nwixident{ex}}{ex}}\\{{\nwixident{label{\_}string}}{label:unstring}}\\{{\nwixident{name}}{name}}}\nwindexuse{\nwixident{asy{\_}style}}{asy:unstyle}{NWppJ6t-DKmU5-1o}\nwindexuse{\nwixident{ex}}{ex}{NWppJ6t-DKmU5-1o}\nwindexuse{\nwixident{label{\_}string}}{label:unstring}{NWppJ6t-DKmU5-1o}\nwindexuse{\nwixident{name}}{name}{NWppJ6t-DKmU5-1o}\nwendcode{}\nwbegindocs{1112}\nwdocspar
\nwenddocs{}\nwbegindocs{1113}Header of the file depends from format.
\nwenddocs{}\nwbegincode{1114}\sublabel{NWppJ6t-DKmU5-1p}\nwmargintag{{\nwtagstyle{}\subpageref{NWppJ6t-DKmU5-1p}}}\moddef{figure class~{\nwtagstyle{}\subpageref{NWppJ6t-DKmU5-1}}}\plusendmoddef\Rm{}\nwstartdeflinemarkup\nwusesondefline{\\{NWppJ6t-32jW6L-1}}\nwprevnextdefs{NWppJ6t-DKmU5-1o}{NWppJ6t-DKmU5-1q}\nwenddeflinemarkup
    {\bf{}if} ({\it{}format} \begin{math}\equiv\end{math} {\tt{}"pdf"})
        {\it{}out} \begin{math}\ll\end{math} {\tt{}"settings.tex={\char92}"pdflatex{\char92}";"} \begin{math}\ll\end{math} {\it{}endl}
            \begin{math}\ll\end{math} {\tt{}"settings.embed=true;"} \begin{math}\ll\end{math} {\it{}endl}
            \begin{math}\ll\end{math} {\tt{}"import animate;"} \begin{math}\ll\end{math} {\it{}endl}
            \begin{math}\ll\end{math} {\tt{}"size("} \begin{math}\ll\end{math} {\it{}size} \begin{math}\ll\end{math} {\tt{}");"} \begin{math}\ll\end{math} {\it{}endl}
            \begin{math}\ll\end{math} {\tt{}"animation a=animation({\char92}""} \begin{math}\ll\end{math} {\it{}name} \begin{math}\ll\end{math} {\tt{}"{\char92}");"} \begin{math}\ll\end{math} {\it{}endl};
    {\bf{}else}
        {\it{}out} \begin{math}\ll\end{math} {\tt{}"import animate;"} \begin{math}\ll\end{math} {\it{}endl}
            \begin{math}\ll\end{math} {\tt{}"size("} \begin{math}\ll\end{math} {\it{}size} \begin{math}\ll\end{math} {\tt{}");"} \begin{math}\ll\end{math} {\it{}endl}
            \begin{math}\ll\end{math} {\tt{}"animation a;"} \begin{math}\ll\end{math} {\it{}endl};

\nwused{\\{NWppJ6t-32jW6L-1}}\nwidentuses{\\{{\nwixident{name}}{name}}}\nwindexuse{\nwixident{name}}{name}{NWppJ6t-DKmU5-1p}\nwendcode{}\nwbegindocs{1115}For every element of {\Tt{}\Rm{}{\it{}val}\nwendquote} we perform the substitution and draw
the corresponding picture.
\nwenddocs{}\nwbegincode{1116}\sublabel{NWppJ6t-DKmU5-1q}\nwmargintag{{\nwtagstyle{}\subpageref{NWppJ6t-DKmU5-1q}}}\moddef{figure class~{\nwtagstyle{}\subpageref{NWppJ6t-DKmU5-1}}}\plusendmoddef\Rm{}\nwstartdeflinemarkup\nwusesondefline{\\{NWppJ6t-32jW6L-1}}\nwprevnextdefs{NWppJ6t-DKmU5-1p}{NWppJ6t-DKmU5-1r}\nwenddeflinemarkup
    {\bf{}for} ({\bf{}const} {\bf{}auto}& {\it{}it} : {\it{}ex\_to}\begin{math}<\end{math}{\bf{}lst}\begin{math}>\end{math}({\it{}val})) {\nwlbrace}
        {\it{}out} \begin{math}\ll\end{math} {\tt{}"save();"} \begin{math}\ll\end{math} {\it{}endl};
        {\it{}unfreeze}().{\it{}subs}({\it{}it}).{\it{}asy\_draw}({\it{}out}, {\it{}cerr}, {\tt{}""}, {\it{}xmin}, {\it{}xmax}, {\it{}ymin}, {\it{}ymax},
                                     {\it{}style}, {\it{}lstring}, {\it{}with\_realline}, {\it{}with\_header}, {\it{}points\_per\_arc}, {\it{}imaginary\_options}, {\it{}with\_labels});

\nwused{\\{NWppJ6t-32jW6L-1}}\nwidentuses{\\{{\nwixident{asy{\_}draw}}{asy:undraw}}\\{{\nwixident{save}}{save}}\\{{\nwixident{subs}}{subs}}\\{{\nwixident{unfreeze}}{unfreeze}}}\nwindexuse{\nwixident{asy{\_}draw}}{asy:undraw}{NWppJ6t-DKmU5-1q}\nwindexuse{\nwixident{save}}{save}{NWppJ6t-DKmU5-1q}\nwindexuse{\nwixident{subs}}{subs}{NWppJ6t-DKmU5-1q}\nwindexuse{\nwixident{unfreeze}}{unfreeze}{NWppJ6t-DKmU5-1q}\nwendcode{}\nwbegindocs{1117}We prepare the value string for output.
\nwenddocs{}\nwbegincode{1118}\sublabel{NWppJ6t-DKmU5-1r}\nwmargintag{{\nwtagstyle{}\subpageref{NWppJ6t-DKmU5-1r}}}\moddef{figure class~{\nwtagstyle{}\subpageref{NWppJ6t-DKmU5-1}}}\plusendmoddef\Rm{}\nwstartdeflinemarkup\nwusesondefline{\\{NWppJ6t-32jW6L-1}}\nwprevnextdefs{NWppJ6t-DKmU5-1q}{NWppJ6t-DKmU5-1s}\nwenddeflinemarkup
        {\it{}std}::{\it{}regex} {\it{}deq} ({\tt{}"=="});
        {\it{}stringstream} {\it{}sstr};
        {\it{}sstr} \begin{math}\ll\end{math}  ({\bf{}ex}){\it{}it};
        {\it{}string} {\it{}val\_str}={\it{}std}::{\it{}regex\_replace}({\it{}sstr}.{\it{}str}(),{\it{}deq},{\tt{}"="});

\nwused{\\{NWppJ6t-32jW6L-1}}\nwidentuses{\\{{\nwixident{ex}}{ex}}}\nwindexuse{\nwixident{ex}}{ex}{NWppJ6t-DKmU5-1r}\nwendcode{}\nwbegindocs{1119}We put the value of parameters to the figure in accordance with {\Tt{}\Rm{}{\it{}values\_position}\nwendquote}.
\nwenddocs{}\nwbegincode{1120}\sublabel{NWppJ6t-DKmU5-1s}\nwmargintag{{\nwtagstyle{}\subpageref{NWppJ6t-DKmU5-1s}}}\moddef{figure class~{\nwtagstyle{}\subpageref{NWppJ6t-DKmU5-1}}}\plusendmoddef\Rm{}\nwstartdeflinemarkup\nwusesondefline{\\{NWppJ6t-32jW6L-1}}\nwprevnextdefs{NWppJ6t-DKmU5-1r}{NWppJ6t-DKmU5-1t}\nwenddeflinemarkup
        {\bf{}if} ({\it{}values\_position}\begin{math}\equiv\end{math}{\tt{}"bl"})
            {\it{}out} \begin{math}\ll\end{math} {\tt{}"label({\char92}"{\char92}{\char92}texttt{\char123}"} \begin{math}\ll\end{math} {\it{}val\_str} \begin{math}\ll\end{math} {\tt{}"{\char125}{\char92}",("} \begin{math}\ll\end{math} {\it{}xmin} \begin{math}\ll\end{math} {\tt{}","} \begin{math}\ll\end{math} {\it{}ymin} \begin{math}\ll\end{math} {\tt{}"), SE);"};
        {\bf{}else} {\bf{}if} ({\it{}values\_position}\begin{math}\equiv\end{math}{\tt{}"br"})
            {\it{}out} \begin{math}\ll\end{math} {\tt{}"label({\char92}"{\char92}{\char92}texttt{\char123}"} \begin{math}\ll\end{math} {\it{}val\_str} \begin{math}\ll\end{math} {\tt{}"{\char125}{\char92}",("} \begin{math}\ll\end{math} {\it{}xmax} \begin{math}\ll\end{math} {\tt{}","} \begin{math}\ll\end{math} {\it{}ymin} \begin{math}\ll\end{math} {\tt{}"), SW);"};
        {\bf{}else} {\bf{}if} ({\it{}values\_position}\begin{math}\equiv\end{math}{\tt{}"tl"})
            {\it{}out} \begin{math}\ll\end{math} {\tt{}"label({\char92}"{\char92}{\char92}texttt{\char123}"} \begin{math}\ll\end{math} {\it{}val\_str} \begin{math}\ll\end{math} {\tt{}"{\char125}{\char92}",("} \begin{math}\ll\end{math} {\it{}xmin} \begin{math}\ll\end{math} {\tt{}","} \begin{math}\ll\end{math} {\it{}ymax} \begin{math}\ll\end{math} {\tt{}"), NE);"};
        {\bf{}else} {\bf{}if} ({\it{}values\_position}\begin{math}\equiv\end{math}{\tt{}"tr"})
            {\it{}out} \begin{math}\ll\end{math} {\tt{}"label({\char92}"{\char92}{\char92}texttt{\char123}"} \begin{math}\ll\end{math} {\it{}val\_str} \begin{math}\ll\end{math} {\tt{}"{\char125}{\char92}",("} \begin{math}\ll\end{math} {\it{}xmax} \begin{math}\ll\end{math} {\tt{}","} \begin{math}\ll\end{math} {\it{}ymax} \begin{math}\ll\end{math} {\tt{}"), NW);"};

        {\it{}out} \begin{math}\ll\end{math} {\tt{}"a.add();"} \begin{math}\ll\end{math} {\it{}endl}
            \begin{math}\ll\end{math} {\tt{}"restore();"} \begin{math}\ll\end{math} {\it{}endl};
    {\nwrbrace}

\nwused{\\{NWppJ6t-32jW6L-1}}\nwendcode{}\nwbegindocs{1121}For output in PDF, GIF, MNG or MP4 format we supply default
commands. User may do a custom command using {\Tt{}\Rm{}{\it{}format}\nwendquote} parameter.
\nwenddocs{}\nwbegincode{1122}\sublabel{NWppJ6t-DKmU5-1t}\nwmargintag{{\nwtagstyle{}\subpageref{NWppJ6t-DKmU5-1t}}}\moddef{figure class~{\nwtagstyle{}\subpageref{NWppJ6t-DKmU5-1}}}\plusendmoddef\Rm{}\nwstartdeflinemarkup\nwusesondefline{\\{NWppJ6t-32jW6L-1}}\nwprevnextdefs{NWppJ6t-DKmU5-1s}{NWppJ6t-DKmU5-1u}\nwenddeflinemarkup
    {\bf{}if} ({\it{}format} \begin{math}\equiv\end{math} {\tt{}"pdf"})
        {\it{}out} \begin{math}\ll\end{math} {\tt{}"label(a.pdf({\char92}"controls{\char92}",delay=250,keep=!settings.inlinetex));"} \begin{math}\ll\end{math} {\it{}endl};
    {\bf{}else} {\bf{}if} (({\it{}format} \begin{math}\equiv\end{math} {\tt{}"gif"}) \begin{math}\vee\end{math}  ({\it{}format} \begin{math}\equiv\end{math} {\tt{}"mp4"}) \begin{math}\vee\end{math} ({\it{}format} \begin{math}\equiv\end{math} {\tt{}"mng"}))
        {\it{}out} \begin{math}\ll\end{math} {\tt{}"a.movie(loops=10,delay=250);"} \begin{math}\ll\end{math} {\it{}endl};
    {\bf{}else}
        {\it{}out} \begin{math}\ll\end{math} {\it{}format} \begin{math}\ll\end{math} {\it{}endl};
    {\it{}out}.{\it{}flush}();
    {\it{}out}.{\it{}close}();

\nwused{\\{NWppJ6t-32jW6L-1}}\nwendcode{}\nwbegindocs{1123}Finally we run \Asymptote\ to produce an animation.
\nwenddocs{}\nwbegincode{1124}\sublabel{NWppJ6t-DKmU5-1u}\nwmargintag{{\nwtagstyle{}\subpageref{NWppJ6t-DKmU5-1u}}}\moddef{figure class~{\nwtagstyle{}\subpageref{NWppJ6t-DKmU5-1}}}\plusendmoddef\Rm{}\nwstartdeflinemarkup\nwusesondefline{\\{NWppJ6t-32jW6L-1}}\nwprevnextdefs{NWppJ6t-DKmU5-1t}{NWppJ6t-DKmU5-1v}\nwenddeflinemarkup
    {\bf{}char} {\it{}command}[256];
    {\it{}strcpy}({\it{}command}, {\it{}show\_asy\_graphics}? {\tt{}"asy -V "} : {\tt{}"asy "});
    {\bf{}if} (({\it{}format} \begin{math}\equiv\end{math} {\tt{}"gif"}) \begin{math}\vee\end{math} ({\it{}format} \begin{math}\equiv\end{math} {\tt{}"mp4"}) \begin{math}\vee\end{math} ({\it{}format} \begin{math}\equiv\end{math} {\tt{}"mng"})) {\nwlbrace}
        {\it{}strcat}({\it{}command}, {\tt{}" -f "});
        {\it{}strcat}({\it{}command}, {\it{}format}.{\it{}c\_str}());
        {\it{}strcat}({\it{}command}, {\tt{}" "});
    {\nwrbrace}
    {\it{}strcat}({\it{}command}, {\it{}name}.{\it{}c\_str}());
    {\bf{}char} \begin{math}\ast\end{math} {\it{}pcommand}={\it{}command};
    {\it{}system}({\it{}pcommand});
    {\bf{}if} ({\it{}rm\_asy\_file})
        {\it{}remove}({\it{}filename}.{\it{}c\_str}());
{\nwrbrace}

\nwused{\\{NWppJ6t-32jW6L-1}}\nwidentuses{\\{{\nwixident{name}}{name}}\\{{\nwixident{show{\_}asy{\_}graphics}}{show:unasy:ungraphics}}}\nwindexuse{\nwixident{name}}{name}{NWppJ6t-DKmU5-1u}\nwindexuse{\nwixident{show{\_}asy{\_}graphics}}{show:unasy:ungraphics}{NWppJ6t-DKmU5-1u}\nwendcode{}\nwbegindocs{1125}All cycles in generations starting from {\Tt{}\Rm{}{\it{}first\_gen}\nwendquote} (default value is
\(0\)) are dumped to a text file {\Tt{}\Rm{}{\it{}name}.{\it{}txt}\nwendquote}. Firstly, we check that the
figure is three dimensional and then open the file.
\nwenddocs{}\nwbegincode{1126}\sublabel{NWppJ6t-DKmU5-1v}\nwmargintag{{\nwtagstyle{}\subpageref{NWppJ6t-DKmU5-1v}}}\moddef{figure class~{\nwtagstyle{}\subpageref{NWppJ6t-DKmU5-1}}}\plusendmoddef\Rm{}\nwstartdeflinemarkup\nwusesondefline{\\{NWppJ6t-32jW6L-1}}\nwprevnextdefs{NWppJ6t-DKmU5-1u}{NWppJ6t-DKmU5-1w}\nwenddeflinemarkup
{\bf{}void} {\bf{}figure}::{\it{}arrangement\_write}({\it{}string} {\it{}name}, {\bf{}int} {\it{}first\_gen}) {\bf{}const}\nwindexdefn{\nwixident{figure}}{figure}{NWppJ6t-DKmU5-1v}
{\nwlbrace}
    {\bf{}if} (\begin{math}\neg\end{math} ({\it{}get\_dim}()-3).{\it{}is\_zero}())
        {\bf{}throw}({\it{}std}::{\it{}invalid\_argument}({\tt{}"figure::arrangement\_write(): the figure is not in 3D!"}));

    {\it{}string} {\it{}filename}={\it{}name}+{\tt{}".txt"};
    {\it{}ofstream} {\it{}out}({\it{}filename});
\nwindexdefn{\nwixident{arrangement{\_}write}}{arrangement:unwrite}{NWppJ6t-DKmU5-1v}\eatline
\nwused{\\{NWppJ6t-32jW6L-1}}\nwidentdefs{\\{{\nwixident{arrangement{\_}write}}{arrangement:unwrite}}\\{{\nwixident{figure}}{figure}}}\nwidentuses{\\{{\nwixident{get{\_}dim()}}{get:undim()}}\\{{\nwixident{name}}{name}}}\nwindexuse{\nwixident{get{\_}dim()}}{get:undim()}{NWppJ6t-DKmU5-1v}\nwindexuse{\nwixident{name}}{name}{NWppJ6t-DKmU5-1v}\nwendcode{}\nwbegindocs{1127}\nwdocspar
\nwenddocs{}\nwbegindocs{1128}We produce the iterator over all keys. This is a \GiNaC\ {\Tt{}\Rm{}{\bf{}lst}\nwendquote}
thus we need iterations through its components.
\nwenddocs{}\nwbegincode{1129}\sublabel{NWppJ6t-DKmU5-1w}\nwmargintag{{\nwtagstyle{}\subpageref{NWppJ6t-DKmU5-1w}}}\moddef{figure class~{\nwtagstyle{}\subpageref{NWppJ6t-DKmU5-1}}}\plusendmoddef\Rm{}\nwstartdeflinemarkup\nwusesondefline{\\{NWppJ6t-32jW6L-1}}\nwprevnextdefs{NWppJ6t-DKmU5-1v}{NWppJ6t-DKmU5-1x}\nwenddeflinemarkup
    {\bf{}lst} {\it{}keys}={\it{}ex\_to}\begin{math}<\end{math}{\bf{}lst}\begin{math}>\end{math}({\it{}get\_all\_keys}({\it{}first\_gen}));
    {\bf{}for} ({\bf{}const} {\bf{}auto}& {\it{}itk} : {\it{}keys}) {\nwlbrace}
        {\bf{}ex} {\it{}gen}={\it{}get\_generation}({\it{}itk});
        {\bf{}lst} {\it{}L}={\it{}ex\_to}\begin{math}<\end{math}{\bf{}lst}\begin{math}>\end{math}({\it{}get\_cycle}({\it{}itk}));

\nwused{\\{NWppJ6t-32jW6L-1}}\nwidentuses{\\{{\nwixident{ex}}{ex}}\\{{\nwixident{get{\_}all{\_}keys}}{get:unall:unkeys}}\\{{\nwixident{get{\_}cycle}}{get:uncycle}}\\{{\nwixident{get{\_}generation}}{get:ungeneration}}}\nwindexuse{\nwixident{ex}}{ex}{NWppJ6t-DKmU5-1w}\nwindexuse{\nwixident{get{\_}all{\_}keys}}{get:unall:unkeys}{NWppJ6t-DKmU5-1w}\nwindexuse{\nwixident{get{\_}cycle}}{get:uncycle}{NWppJ6t-DKmU5-1w}\nwindexuse{\nwixident{get{\_}generation}}{get:ungeneration}{NWppJ6t-DKmU5-1w}\nwendcode{}\nwbegindocs{1130} This is again a \GiNaC\ {\Tt{}\Rm{}{\bf{}lst}\nwendquote}, thus we need iterations through
its components again.
\nwenddocs{}\nwbegincode{1131}\sublabel{NWppJ6t-DKmU5-1x}\nwmargintag{{\nwtagstyle{}\subpageref{NWppJ6t-DKmU5-1x}}}\moddef{figure class~{\nwtagstyle{}\subpageref{NWppJ6t-DKmU5-1}}}\plusendmoddef\Rm{}\nwstartdeflinemarkup\nwusesondefline{\\{NWppJ6t-32jW6L-1}}\nwprevnextdefs{NWppJ6t-DKmU5-1w}{NWppJ6t-DKmU5-1y}\nwenddeflinemarkup
    {\bf{}for} ({\bf{}const} {\bf{}auto}& {\it{}it} : {\it{}L}) {\nwlbrace}
        {\bf{}cycle} {\it{}C}={\it{}ex\_to}\begin{math}<\end{math}{\bf{}cycle}\begin{math}>\end{math}({\it{}it});
        {\bf{}ex} {\it{}center} = {\it{}C}.{\it{}center}();

\nwused{\\{NWppJ6t-32jW6L-1}}\nwidentuses{\\{{\nwixident{ex}}{ex}}}\nwindexuse{\nwixident{ex}}{ex}{NWppJ6t-DKmU5-1x}\nwendcode{}\nwbegindocs{1132}A line of text represents a cycle by three coordinates of its
centre, radius, generation and label.
\nwenddocs{}\nwbegincode{1133}\sublabel{NWppJ6t-DKmU5-1y}\nwmargintag{{\nwtagstyle{}\subpageref{NWppJ6t-DKmU5-1y}}}\moddef{figure class~{\nwtagstyle{}\subpageref{NWppJ6t-DKmU5-1}}}\plusendmoddef\Rm{}\nwstartdeflinemarkup\nwusesondefline{\\{NWppJ6t-32jW6L-1}}\nwprevnextdefs{NWppJ6t-DKmU5-1x}{NWppJ6t-DKmU5-1z}\nwenddeflinemarkup
        {\it{}out} \begin{math}\ll\end{math}  {\it{}center}.{\it{}op}(0).{\it{}evalf}() \begin{math}\ll\end{math} {\tt{}" "} \begin{math}\ll\end{math} {\it{}center}.{\it{}op}(1).{\it{}evalf}() \begin{math}\ll\end{math} {\tt{}" "} \begin{math}\ll\end{math} {\it{}center}.{\it{}op}(2).{\it{}evalf}()
            \begin{math}\ll\end{math} {\tt{}" "} \begin{math}\ll\end{math} {\it{}sqrt}({\it{}C}.{\it{}radius\_sq}()).{\it{}evalf}()
            \begin{math}\ll\end{math} {\tt{}" "} \begin{math}\ll\end{math} {\it{}gen}
            \begin{math}\ll\end{math} {\tt{}" "} \begin{math}\ll\end{math} {\it{}itk}
            \begin{math}\ll\end{math} {\it{}endl};
        {\nwrbrace}
    {\nwrbrace}
    {\it{}out}.{\it{}flush}();
    {\it{}out}.{\it{}close}();
{\nwrbrace}

\nwused{\\{NWppJ6t-32jW6L-1}}\nwidentuses{\\{{\nwixident{evalf}}{evalf}}\\{{\nwixident{op}}{op}}}\nwindexuse{\nwixident{evalf}}{evalf}{NWppJ6t-DKmU5-1y}\nwindexuse{\nwixident{op}}{op}{NWppJ6t-DKmU5-1y}\nwendcode{}\nwbegindocs{1134}\nwdocspar
\subsubsection{Service utilities}
\label{sec:service-utilities}

\nwenddocs{}\nwbegindocs{1135}Here is the minimal set of service procedures which is reuired by
\GiNaC\ for derived classes.
\nwenddocs{}\nwbegincode{1136}\sublabel{NWppJ6t-DKmU5-1z}\nwmargintag{{\nwtagstyle{}\subpageref{NWppJ6t-DKmU5-1z}}}\moddef{figure class~{\nwtagstyle{}\subpageref{NWppJ6t-DKmU5-1}}}\plusendmoddef\Rm{}\nwstartdeflinemarkup\nwusesondefline{\\{NWppJ6t-32jW6L-1}}\nwprevnextdefs{NWppJ6t-DKmU5-1y}{NWppJ6t-DKmU5-20}\nwenddeflinemarkup
{\it{}return\_type\_t} {\bf{}figure}::{\it{}return\_type\_tinfo}() {\bf{}const}
{\nwlbrace}
    {\bf{}return} {\it{}make\_return\_type\_t}\begin{math}<\end{math}{\bf{}figure}\begin{math}>\end{math}();
{\nwrbrace}

\nwused{\\{NWppJ6t-32jW6L-1}}\nwidentuses{\\{{\nwixident{figure}}{figure}}}\nwindexuse{\nwixident{figure}}{figure}{NWppJ6t-DKmU5-1z}\nwendcode{}\nwbegindocs{1137}\nwdocspar
\nwenddocs{}\nwbegincode{1138}\sublabel{NWppJ6t-DKmU5-20}\nwmargintag{{\nwtagstyle{}\subpageref{NWppJ6t-DKmU5-20}}}\moddef{figure class~{\nwtagstyle{}\subpageref{NWppJ6t-DKmU5-1}}}\plusendmoddef\Rm{}\nwstartdeflinemarkup\nwusesondefline{\\{NWppJ6t-32jW6L-1}}\nwprevnextdefs{NWppJ6t-DKmU5-1z}{NWppJ6t-DKmU5-21}\nwenddeflinemarkup
{\bf{}int} {\bf{}figure}::{\it{}compare\_same\_type}({\bf{}const} {\bf{}basic} &{\it{}other}) {\bf{}const}\nwindexdefn{\nwixident{figure}}{figure}{NWppJ6t-DKmU5-20}
{\nwlbrace}
       {\it{}GINAC\_ASSERT}({\it{}is\_a}\begin{math}<\end{math}{\bf{}figure}\begin{math}>\end{math}({\it{}other}));
       {\bf{}return} {\it{}inherited}::{\it{}compare\_same\_type}({\it{}other});
{\nwrbrace}

\nwused{\\{NWppJ6t-32jW6L-1}}\nwidentdefs{\\{{\nwixident{figure}}{figure}}}\nwendcode{}\nwbegindocs{1139}To print the figure means to print all its nodes.
\nwenddocs{}\nwbegincode{1140}\sublabel{NWppJ6t-DKmU5-21}\nwmargintag{{\nwtagstyle{}\subpageref{NWppJ6t-DKmU5-21}}}\moddef{figure class~{\nwtagstyle{}\subpageref{NWppJ6t-DKmU5-1}}}\plusendmoddef\Rm{}\nwstartdeflinemarkup\nwusesondefline{\\{NWppJ6t-32jW6L-1}}\nwprevnextdefs{NWppJ6t-DKmU5-20}{NWppJ6t-DKmU5-22}\nwenddeflinemarkup
{\bf{}void} {\bf{}figure}::{\it{}do\_print}({\bf{}const} {\it{}print\_dflt} & {\it{}con}, {\bf{}unsigned} {\it{}level}) {\bf{}const} {\nwlbrace}\nwindexdefn{\nwixident{figure}}{figure}{NWppJ6t-DKmU5-21}
    {\bf{}lst} {\it{}keys}={\it{}ex\_to}\begin{math}<\end{math}{\bf{}lst}\begin{math}>\end{math}({\it{}get\_all\_keys}({\it{}FIGURE\_DEBUG}?{\it{}GHOST\_GEN}:{\it{}INFINITY\_GEN}));

    {\bf{}for} ({\bf{}const} {\bf{}auto}& {\it{}k}: {\it{}keys})
        {\it{}con}.{\it{}s} \begin{math}\ll\end{math} {\it{}k} \begin{math}\ll\end{math}{\tt{}": "} \begin{math}\ll\end{math} {\it{}get\_cycle\_node}({\it{}k});
    {\it{}con}.{\it{}s} \begin{math}\ll\end{math} {\tt{}"Altogether "} \begin{math}\ll\end{math} {\it{}keys}.{\it{}nops}() \begin{math}\ll\end{math} {\tt{}" cycle\_nodes."} \begin{math}\ll\end{math} {\it{}endl};
{\nwrbrace}

\nwused{\\{NWppJ6t-32jW6L-1}}\nwidentdefs{\\{{\nwixident{figure}}{figure}}}\nwidentuses{\\{{\nwixident{FIGURE{\_}DEBUG}}{FIGURE:unDEBUG}}\\{{\nwixident{get{\_}all{\_}keys}}{get:unall:unkeys}}\\{{\nwixident{get{\_}cycle{\_}node}}{get:uncycle:unnode}}\\{{\nwixident{GHOST{\_}GEN}}{GHOST:unGEN}}\\{{\nwixident{INFINITY{\_}GEN}}{INFINITY:unGEN}}\\{{\nwixident{k}}{k}}\\{{\nwixident{nops}}{nops}}}\nwindexuse{\nwixident{FIGURE{\_}DEBUG}}{FIGURE:unDEBUG}{NWppJ6t-DKmU5-21}\nwindexuse{\nwixident{get{\_}all{\_}keys}}{get:unall:unkeys}{NWppJ6t-DKmU5-21}\nwindexuse{\nwixident{get{\_}cycle{\_}node}}{get:uncycle:unnode}{NWppJ6t-DKmU5-21}\nwindexuse{\nwixident{GHOST{\_}GEN}}{GHOST:unGEN}{NWppJ6t-DKmU5-21}\nwindexuse{\nwixident{INFINITY{\_}GEN}}{INFINITY:unGEN}{NWppJ6t-DKmU5-21}\nwindexuse{\nwixident{k}}{k}{NWppJ6t-DKmU5-21}\nwindexuse{\nwixident{nops}}{nops}{NWppJ6t-DKmU5-21}\nwendcode{}\nwbegindocs{1141}This is a variation of printing in the float form.
\nwenddocs{}\nwbegincode{1142}\sublabel{NWppJ6t-DKmU5-22}\nwmargintag{{\nwtagstyle{}\subpageref{NWppJ6t-DKmU5-22}}}\moddef{figure class~{\nwtagstyle{}\subpageref{NWppJ6t-DKmU5-1}}}\plusendmoddef\Rm{}\nwstartdeflinemarkup\nwusesondefline{\\{NWppJ6t-32jW6L-1}}\nwprevnextdefs{NWppJ6t-DKmU5-21}{NWppJ6t-DKmU5-23}\nwenddeflinemarkup
{\bf{}void} {\bf{}figure}::{\it{}do\_print\_double}({\bf{}const} {\it{}print\_dflt} & {\it{}con}, {\bf{}unsigned} {\it{}level}) {\bf{}const} {\nwlbrace}\nwindexdefn{\nwixident{figure}}{figure}{NWppJ6t-DKmU5-22}
    {\bf{}for} ({\bf{}const} {\bf{}auto}& {\it{}x}: {\it{}nodes}) {\nwlbrace}
        {\bf{}if} ({\it{}x}.{\it{}second}.{\it{}get\_generation}() \begin{math}>\end{math} {\it{}GHOST\_GEN}  \begin{math}\vee\end{math} {\it{}FIGURE\_DEBUG}) {\nwlbrace}
            {\it{}con}.{\it{}s} \begin{math}\ll\end{math} {\it{}x}.{\it{}first} \begin{math}\ll\end{math}{\tt{}": "};
            {\it{}ex\_to}\begin{math}<\end{math}{\bf{}cycle\_node}\begin{math}>\end{math}({\it{}x}.{\it{}second}).{\it{}do\_print\_double}({\it{}con}, {\it{}level});
        {\nwrbrace}
    {\nwrbrace}
{\nwrbrace}

\nwused{\\{NWppJ6t-32jW6L-1}}\nwidentdefs{\\{{\nwixident{figure}}{figure}}}\nwidentuses{\\{{\nwixident{cycle{\_}node}}{cycle:unnode}}\\{{\nwixident{do{\_}print{\_}double}}{do:unprint:undouble}}\\{{\nwixident{FIGURE{\_}DEBUG}}{FIGURE:unDEBUG}}\\{{\nwixident{get{\_}generation}}{get:ungeneration}}\\{{\nwixident{GHOST{\_}GEN}}{GHOST:unGEN}}\\{{\nwixident{nodes}}{nodes}}}\nwindexuse{\nwixident{cycle{\_}node}}{cycle:unnode}{NWppJ6t-DKmU5-22}\nwindexuse{\nwixident{do{\_}print{\_}double}}{do:unprint:undouble}{NWppJ6t-DKmU5-22}\nwindexuse{\nwixident{FIGURE{\_}DEBUG}}{FIGURE:unDEBUG}{NWppJ6t-DKmU5-22}\nwindexuse{\nwixident{get{\_}generation}}{get:ungeneration}{NWppJ6t-DKmU5-22}\nwindexuse{\nwixident{GHOST{\_}GEN}}{GHOST:unGEN}{NWppJ6t-DKmU5-22}\nwindexuse{\nwixident{nodes}}{nodes}{NWppJ6t-DKmU5-22}\nwendcode{}\nwbegindocs{1143}\nwdocspar
\nwenddocs{}\nwbegincode{1144}\sublabel{NWppJ6t-DKmU5-23}\nwmargintag{{\nwtagstyle{}\subpageref{NWppJ6t-DKmU5-23}}}\moddef{figure class~{\nwtagstyle{}\subpageref{NWppJ6t-DKmU5-1}}}\plusendmoddef\Rm{}\nwstartdeflinemarkup\nwusesondefline{\\{NWppJ6t-32jW6L-1}}\nwprevnextdefs{NWppJ6t-DKmU5-22}{NWppJ6t-DKmU5-24}\nwenddeflinemarkup
{\bf{}ex} {\bf{}figure}::{\it{}op}({\it{}size\_t} {\it{}i}) {\bf{}const}
{\nwlbrace}
 {\it{}GINAC\_ASSERT}({\it{}i}\begin{math}<\end{math}{\it{}nops}());
    {\bf{}switch}({\it{}i}) {\nwlbrace}
    {\bf{}case} 0:
        {\bf{}return} {\it{}real\_line};
    {\bf{}case} 1:
        {\bf{}return} {\it{}infinity};
    {\bf{}case} 2:
        {\bf{}return} {\it{}point\_metric};
    {\bf{}case} 3:
        {\bf{}return} {\it{}cycle\_metric};
    {\bf{}default}:
        {\it{}exhashmap}\begin{math}<\end{math}{\bf{}cycle\_node}\begin{math}>\end{math}::{\it{}const\_iterator} {\it{}it}={\it{}nodes}.{\it{}begin}();
        {\bf{}for} ({\it{}size\_t} {\it{}n}=4; {\it{}n}\begin{math}<\end{math}{\it{}i};\protect\PP{\it{}n})
            \protect\PP{\it{}it};
        {\bf{}return} {\it{}it}\begin{math}\rightarrow\end{math}{\it{}second};
    {\nwrbrace}
{\nwrbrace}

\nwused{\\{NWppJ6t-32jW6L-1}}\nwidentuses{\\{{\nwixident{cycle{\_}metric}}{cycle:unmetric}}\\{{\nwixident{cycle{\_}node}}{cycle:unnode}}\\{{\nwixident{ex}}{ex}}\\{{\nwixident{figure}}{figure}}\\{{\nwixident{infinity}}{infinity}}\\{{\nwixident{nodes}}{nodes}}\\{{\nwixident{nops}}{nops}}\\{{\nwixident{op}}{op}}\\{{\nwixident{point{\_}metric}}{point:unmetric}}\\{{\nwixident{real{\_}line}}{real:unline}}}\nwindexuse{\nwixident{cycle{\_}metric}}{cycle:unmetric}{NWppJ6t-DKmU5-23}\nwindexuse{\nwixident{cycle{\_}node}}{cycle:unnode}{NWppJ6t-DKmU5-23}\nwindexuse{\nwixident{ex}}{ex}{NWppJ6t-DKmU5-23}\nwindexuse{\nwixident{figure}}{figure}{NWppJ6t-DKmU5-23}\nwindexuse{\nwixident{infinity}}{infinity}{NWppJ6t-DKmU5-23}\nwindexuse{\nwixident{nodes}}{nodes}{NWppJ6t-DKmU5-23}\nwindexuse{\nwixident{nops}}{nops}{NWppJ6t-DKmU5-23}\nwindexuse{\nwixident{op}}{op}{NWppJ6t-DKmU5-23}\nwindexuse{\nwixident{point{\_}metric}}{point:unmetric}{NWppJ6t-DKmU5-23}\nwindexuse{\nwixident{real{\_}line}}{real:unline}{NWppJ6t-DKmU5-23}\nwendcode{}\nwbegindocs{1145}\nwdocspar
\nwenddocs{}\nwbegincode{1146}\sublabel{NWppJ6t-DKmU5-24}\nwmargintag{{\nwtagstyle{}\subpageref{NWppJ6t-DKmU5-24}}}\moddef{figure class~{\nwtagstyle{}\subpageref{NWppJ6t-DKmU5-1}}}\plusendmoddef\Rm{}\nwstartdeflinemarkup\nwusesondefline{\\{NWppJ6t-32jW6L-1}}\nwprevnextdefs{NWppJ6t-DKmU5-23}{NWppJ6t-DKmU5-25}\nwenddeflinemarkup
\begin{math}\div\end{math}\begin{math}\ast\end{math}{\bf{}ex} & {\bf{}figure}::{\it{}let\_op}({\it{}size\_t} {\it{}i})
{\nwlbrace}
    {\it{}ensure\_if\_modifiable}();
    {\it{}GINAC\_ASSERT}({\it{}i}\begin{math}<\end{math}{\it{}nops}());
    {\bf{}switch}({\it{}i}) {\nwlbrace}
    {\bf{}case} 0:
        {\bf{}return} {\it{}real\_line};
    {\bf{}case} 1:
        {\bf{}return} {\it{}infinity};
    {\bf{}case} 2:
        {\bf{}return} {\it{}point\_metric};
    {\bf{}case} 3:
        {\bf{}return} {\it{}cycle\_metric};
    {\bf{}default}:
        {\it{}exhashmap}\begin{math}<\end{math}{\bf{}cycle\_node}\begin{math}>\end{math}::{\it{}iterator} {\it{}it}={\it{}nodes}.{\it{}begin}();
        {\bf{}for} ({\it{}size\_t} {\it{}n}=4; {\it{}n}\begin{math}<\end{math}{\it{}i};\protect\PP{\it{}n})
            \protect\PP{\it{}it};
        {\bf{}return} {\it{}nodes}[{\it{}it}\begin{math}\rightarrow\end{math}{\it{}first}];
    {\nwrbrace}
{\nwrbrace}\begin{math}\ast\end{math}\begin{math}\div\end{math}

\nwused{\\{NWppJ6t-32jW6L-1}}\nwidentuses{\\{{\nwixident{cycle{\_}metric}}{cycle:unmetric}}\\{{\nwixident{cycle{\_}node}}{cycle:unnode}}\\{{\nwixident{ex}}{ex}}\\{{\nwixident{figure}}{figure}}\\{{\nwixident{infinity}}{infinity}}\\{{\nwixident{nodes}}{nodes}}\\{{\nwixident{nops}}{nops}}\\{{\nwixident{point{\_}metric}}{point:unmetric}}\\{{\nwixident{real{\_}line}}{real:unline}}}\nwindexuse{\nwixident{cycle{\_}metric}}{cycle:unmetric}{NWppJ6t-DKmU5-24}\nwindexuse{\nwixident{cycle{\_}node}}{cycle:unnode}{NWppJ6t-DKmU5-24}\nwindexuse{\nwixident{ex}}{ex}{NWppJ6t-DKmU5-24}\nwindexuse{\nwixident{figure}}{figure}{NWppJ6t-DKmU5-24}\nwindexuse{\nwixident{infinity}}{infinity}{NWppJ6t-DKmU5-24}\nwindexuse{\nwixident{nodes}}{nodes}{NWppJ6t-DKmU5-24}\nwindexuse{\nwixident{nops}}{nops}{NWppJ6t-DKmU5-24}\nwindexuse{\nwixident{point{\_}metric}}{point:unmetric}{NWppJ6t-DKmU5-24}\nwindexuse{\nwixident{real{\_}line}}{real:unline}{NWppJ6t-DKmU5-24}\nwendcode{}\nwbegindocs{1147}We need to make substitution in the form of {\Tt{}\Rm{}{\it{}exmap}\nwendquote}.
\nwenddocs{}\nwbegincode{1148}\sublabel{NWppJ6t-DKmU5-25}\nwmargintag{{\nwtagstyle{}\subpageref{NWppJ6t-DKmU5-25}}}\moddef{figure class~{\nwtagstyle{}\subpageref{NWppJ6t-DKmU5-1}}}\plusendmoddef\Rm{}\nwstartdeflinemarkup\nwusesondefline{\\{NWppJ6t-32jW6L-1}}\nwprevnextdefs{NWppJ6t-DKmU5-24}{NWppJ6t-DKmU5-26}\nwenddeflinemarkup
{\bf{}figure} {\bf{}figure}::{\it{}subs}({\bf{}const} {\bf{}ex} & {\it{}e}, {\bf{}unsigned} {\it{}options}) {\bf{}const}
{\nwlbrace}
    {\it{}exmap} {\it{}m};
    {\bf{}if} ({\it{}e}.{\it{}info}({\it{}info\_flags}::{\it{}list})) {\nwlbrace}
        {\bf{}lst} {\it{}sl} = {\it{}ex\_to}\begin{math}<\end{math}{\bf{}lst}\begin{math}>\end{math}({\it{}e});
        {\bf{}for} ({\bf{}const} {\bf{}auto}& {\it{}i} : {\it{}sl})
            {\it{}m}.{\it{}insert}({\it{}std}::{\it{}make\_pair}({\it{}i}.{\it{}op}(0), {\it{}i}.{\it{}op}(1)));
    {\nwrbrace} {\bf{}else} {\bf{}if} ({\it{}is\_a}\begin{math}<\end{math}{\bf{}relational}\begin{math}>\end{math}({\it{}e})) {\nwlbrace}
        {\it{}m}.{\it{}insert}({\it{}std}::{\it{}make\_pair}({\it{}e}.{\it{}op}(0), {\it{}e}.{\it{}op}(1)));
    {\nwrbrace} {\bf{}else}
        {\bf{}throw}({\it{}std}::{\it{}invalid\_argument}({\tt{}"cycle::subs(): the parameter should be a relational or a lst"}));

    {\bf{}return} {\it{}ex\_to}\begin{math}<\end{math}{\bf{}figure}\begin{math}>\end{math}({\it{}subs}({\it{}m}, {\it{}options}));
{\nwrbrace}

\nwused{\\{NWppJ6t-32jW6L-1}}\nwidentuses{\\{{\nwixident{ex}}{ex}}\\{{\nwixident{figure}}{figure}}\\{{\nwixident{info}}{info}}\\{{\nwixident{m}}{m}}\\{{\nwixident{op}}{op}}\\{{\nwixident{subs}}{subs}}}\nwindexuse{\nwixident{ex}}{ex}{NWppJ6t-DKmU5-25}\nwindexuse{\nwixident{figure}}{figure}{NWppJ6t-DKmU5-25}\nwindexuse{\nwixident{info}}{info}{NWppJ6t-DKmU5-25}\nwindexuse{\nwixident{m}}{m}{NWppJ6t-DKmU5-25}\nwindexuse{\nwixident{op}}{op}{NWppJ6t-DKmU5-25}\nwindexuse{\nwixident{subs}}{subs}{NWppJ6t-DKmU5-25}\nwendcode{}\nwbegindocs{1149}\nwdocspar
\nwenddocs{}\nwbegincode{1150}\sublabel{NWppJ6t-DKmU5-26}\nwmargintag{{\nwtagstyle{}\subpageref{NWppJ6t-DKmU5-26}}}\moddef{figure class~{\nwtagstyle{}\subpageref{NWppJ6t-DKmU5-1}}}\plusendmoddef\Rm{}\nwstartdeflinemarkup\nwusesondefline{\\{NWppJ6t-32jW6L-1}}\nwprevnextdefs{NWppJ6t-DKmU5-25}{NWppJ6t-DKmU5-27}\nwenddeflinemarkup
{\bf{}ex} {\bf{}figure}::{\it{}subs}({\bf{}const} {\it{}exmap} & {\it{}m}, {\bf{}unsigned} {\it{}options}) {\bf{}const}
{\nwlbrace}
    {\it{}exhashmap}\begin{math}<\end{math}{\bf{}cycle\_node}\begin{math}>\end{math} {\it{}snodes};
    {\bf{}for} ({\bf{}const} {\bf{}auto}& {\it{}x}: {\it{}nodes})
        {\it{}snodes}[{\it{}x}.{\it{}first}]={\it{}ex\_to}\begin{math}<\end{math}{\bf{}cycle\_node}\begin{math}>\end{math}({\it{}x}.{\it{}second}.{\it{}subs}({\it{}m}, {\it{}options}));

    {\bf{}if} ({\it{}options} & {\it{}do\_not\_update\_subfigure})
        {\bf{}return} {\bf{}figure}({\it{}point\_metric}.{\it{}subs}({\it{}m}, {\it{}options}), {\it{}cycle\_metric}.{\it{}subs}({\it{}m}, {\it{}options}), {\it{}snodes});
    {\bf{}else}
        {\bf{}return} {\bf{}figure}({\it{}point\_metric}.{\it{}subs}({\it{}m}, {\it{}options}), {\it{}cycle\_metric}.{\it{}subs}({\it{}m}, {\it{}options}), {\it{}snodes}).{\it{}update\_cycles}();
{\nwrbrace}

\nwused{\\{NWppJ6t-32jW6L-1}}\nwidentuses{\\{{\nwixident{cycle{\_}metric}}{cycle:unmetric}}\\{{\nwixident{cycle{\_}node}}{cycle:unnode}}\\{{\nwixident{do{\_}not{\_}update{\_}subfigure}}{do:unnot:unupdate:unsubfigure}}\\{{\nwixident{ex}}{ex}}\\{{\nwixident{figure}}{figure}}\\{{\nwixident{m}}{m}}\\{{\nwixident{nodes}}{nodes}}\\{{\nwixident{point{\_}metric}}{point:unmetric}}\\{{\nwixident{subs}}{subs}}\\{{\nwixident{update{\_}cycles}}{update:uncycles}}}\nwindexuse{\nwixident{cycle{\_}metric}}{cycle:unmetric}{NWppJ6t-DKmU5-26}\nwindexuse{\nwixident{cycle{\_}node}}{cycle:unnode}{NWppJ6t-DKmU5-26}\nwindexuse{\nwixident{do{\_}not{\_}update{\_}subfigure}}{do:unnot:unupdate:unsubfigure}{NWppJ6t-DKmU5-26}\nwindexuse{\nwixident{ex}}{ex}{NWppJ6t-DKmU5-26}\nwindexuse{\nwixident{figure}}{figure}{NWppJ6t-DKmU5-26}\nwindexuse{\nwixident{m}}{m}{NWppJ6t-DKmU5-26}\nwindexuse{\nwixident{nodes}}{nodes}{NWppJ6t-DKmU5-26}\nwindexuse{\nwixident{point{\_}metric}}{point:unmetric}{NWppJ6t-DKmU5-26}\nwindexuse{\nwixident{subs}}{subs}{NWppJ6t-DKmU5-26}\nwindexuse{\nwixident{update{\_}cycles}}{update:uncycles}{NWppJ6t-DKmU5-26}\nwendcode{}\nwbegindocs{1151}\nwdocspar
\nwenddocs{}\nwbegincode{1152}\sublabel{NWppJ6t-DKmU5-27}\nwmargintag{{\nwtagstyle{}\subpageref{NWppJ6t-DKmU5-27}}}\moddef{figure class~{\nwtagstyle{}\subpageref{NWppJ6t-DKmU5-1}}}\plusendmoddef\Rm{}\nwstartdeflinemarkup\nwusesondefline{\\{NWppJ6t-32jW6L-1}}\nwprevnextdefs{NWppJ6t-DKmU5-26}{NWppJ6t-DKmU5-28}\nwenddeflinemarkup
{\bf{}ex} {\bf{}figure}::{\it{}evalf}({\bf{}int} {\it{}level}) {\bf{}const}
{\nwlbrace}
    {\it{}exhashmap}\begin{math}<\end{math}{\bf{}cycle\_node}\begin{math}>\end{math} {\it{}snodes};
    {\bf{}for} ({\bf{}const} {\bf{}auto}& {\it{}x}: {\it{}nodes})
{\bf{}\char35{}if}{\tt{} GINAC\_VERSION\_ATLEAST(1,7,0)}
        {\it{}snodes}[{\it{}x}.{\it{}first}]={\it{}ex\_to}\begin{math}<\end{math}{\bf{}cycle\_node}\begin{math}>\end{math}({\it{}x}.{\it{}second}.{\it{}evalf}());

    {\bf{}return} {\bf{}figure}({\it{}point\_metric}.{\it{}evalf}(), {\it{}cycle\_metric}.{\it{}evalf}(), {\it{}snodes});
{\bf{}\char35{}else}{\tt{}}
        {\it{}snodes}[{\it{}x}.{\it{}first}]={\it{}ex\_to}\begin{math}<\end{math}{\bf{}cycle\_node}\begin{math}>\end{math}({\it{}x}.{\it{}second}.{\it{}evalf}({\it{}level}));

    {\bf{}return} {\bf{}figure}({\it{}point\_metric}.{\it{}evalf}({\it{}level}), {\it{}cycle\_metric}.{\it{}evalf}({\it{}level}), {\it{}snodes});
{\bf{}\char35{}endif}{\tt{}}
{\nwrbrace}

\nwused{\\{NWppJ6t-32jW6L-1}}\nwidentuses{\\{{\nwixident{cycle{\_}metric}}{cycle:unmetric}}\\{{\nwixident{cycle{\_}node}}{cycle:unnode}}\\{{\nwixident{evalf}}{evalf}}\\{{\nwixident{ex}}{ex}}\\{{\nwixident{figure}}{figure}}\\{{\nwixident{nodes}}{nodes}}\\{{\nwixident{point{\_}metric}}{point:unmetric}}}\nwindexuse{\nwixident{cycle{\_}metric}}{cycle:unmetric}{NWppJ6t-DKmU5-27}\nwindexuse{\nwixident{cycle{\_}node}}{cycle:unnode}{NWppJ6t-DKmU5-27}\nwindexuse{\nwixident{evalf}}{evalf}{NWppJ6t-DKmU5-27}\nwindexuse{\nwixident{ex}}{ex}{NWppJ6t-DKmU5-27}\nwindexuse{\nwixident{figure}}{figure}{NWppJ6t-DKmU5-27}\nwindexuse{\nwixident{nodes}}{nodes}{NWppJ6t-DKmU5-27}\nwindexuse{\nwixident{point{\_}metric}}{point:unmetric}{NWppJ6t-DKmU5-27}\nwendcode{}\nwbegindocs{1153}\nwdocspar
\subsubsection{Archiving/Unarchiving utilities}
\label{sec:arch-util}

\nwenddocs{}\nwbegindocs{1154}\nwdocspar
\nwenddocs{}\nwbegincode{1155}\sublabel{NWppJ6t-DKmU5-28}\nwmargintag{{\nwtagstyle{}\subpageref{NWppJ6t-DKmU5-28}}}\moddef{figure class~{\nwtagstyle{}\subpageref{NWppJ6t-DKmU5-1}}}\plusendmoddef\Rm{}\nwstartdeflinemarkup\nwusesondefline{\\{NWppJ6t-32jW6L-1}}\nwprevnextdefs{NWppJ6t-DKmU5-27}{NWppJ6t-DKmU5-29}\nwenddeflinemarkup
{\bf{}void} {\bf{}figure}::{\it{}archive}({\it{}archive\_node} &{\it{}an}) {\bf{}const}\nwindexdefn{\nwixident{figure}}{figure}{NWppJ6t-DKmU5-28}
{\nwlbrace}
    {\it{}inherited}::{\it{}archive}({\it{}an});
    {\it{}an}.{\it{}add\_ex}({\tt{}"real\_line"}, {\it{}real\_line});
    {\it{}an}.{\it{}add\_ex}({\tt{}"infinity"}, {\it{}infinity});
    {\it{}an}.{\it{}add\_ex}({\tt{}"point\_metric"}, {\it{}ex\_to}\begin{math}<\end{math}{\bf{}clifford}\begin{math}>\end{math}({\it{}point\_metric}));
    {\it{}an}.{\it{}add\_ex}({\tt{}"cycle\_metric"}, {\it{}ex\_to}\begin{math}<\end{math}{\bf{}clifford}\begin{math}>\end{math}({\it{}cycle\_metric}));
    {\it{}an}.{\it{}add\_bool}({\tt{}"float\_evaluation"}, {\it{}float\_evaluation});

\nwused{\\{NWppJ6t-32jW6L-1}}\nwidentdefs{\\{{\nwixident{figure}}{figure}}}\nwidentuses{\\{{\nwixident{archive}}{archive}}\\{{\nwixident{cycle{\_}metric}}{cycle:unmetric}}\\{{\nwixident{float{\_}evaluation}}{float:unevaluation}}\\{{\nwixident{infinity}}{infinity}}\\{{\nwixident{point{\_}metric}}{point:unmetric}}\\{{\nwixident{real{\_}line}}{real:unline}}}\nwindexuse{\nwixident{archive}}{archive}{NWppJ6t-DKmU5-28}\nwindexuse{\nwixident{cycle{\_}metric}}{cycle:unmetric}{NWppJ6t-DKmU5-28}\nwindexuse{\nwixident{float{\_}evaluation}}{float:unevaluation}{NWppJ6t-DKmU5-28}\nwindexuse{\nwixident{infinity}}{infinity}{NWppJ6t-DKmU5-28}\nwindexuse{\nwixident{point{\_}metric}}{point:unmetric}{NWppJ6t-DKmU5-28}\nwindexuse{\nwixident{real{\_}line}}{real:unline}{NWppJ6t-DKmU5-28}\nwendcode{}\nwbegindocs{1156}{\Tt{}\Rm{}{\it{}exhashmap}\nwendquote} class does not have an archiving facility, thus we store it as two correponding lists.
\nwenddocs{}\nwbegincode{1157}\sublabel{NWppJ6t-DKmU5-29}\nwmargintag{{\nwtagstyle{}\subpageref{NWppJ6t-DKmU5-29}}}\moddef{figure class~{\nwtagstyle{}\subpageref{NWppJ6t-DKmU5-1}}}\plusendmoddef\Rm{}\nwstartdeflinemarkup\nwusesondefline{\\{NWppJ6t-32jW6L-1}}\nwprevnextdefs{NWppJ6t-DKmU5-28}{NWppJ6t-DKmU5-2A}\nwenddeflinemarkup
    {\bf{}lst} {\it{}keys}, {\it{}cnodes};
 {\bf{}for} ({\bf{}const} {\bf{}auto}& {\it{}x}: {\it{}nodes}) {\nwlbrace}
        {\it{}keys}.{\it{}append}({\it{}x}.{\it{}first});
        {\it{}cnodes}.{\it{}append}({\it{}x}.{\it{}second});
    {\nwrbrace}
    {\it{}an}.{\it{}add\_ex}({\tt{}"keys"}, {\it{}keys});
    {\it{}an}.{\it{}add\_ex}({\tt{}"cnodes"}, {\it{}cnodes});
{\nwrbrace}

\nwused{\\{NWppJ6t-32jW6L-1}}\nwidentuses{\\{{\nwixident{nodes}}{nodes}}}\nwindexuse{\nwixident{nodes}}{nodes}{NWppJ6t-DKmU5-29}\nwendcode{}\nwbegindocs{1158}\nwdocspar
\nwenddocs{}\nwbegincode{1159}\sublabel{NWppJ6t-DKmU5-2A}\nwmargintag{{\nwtagstyle{}\subpageref{NWppJ6t-DKmU5-2A}}}\moddef{figure class~{\nwtagstyle{}\subpageref{NWppJ6t-DKmU5-1}}}\plusendmoddef\Rm{}\nwstartdeflinemarkup\nwusesondefline{\\{NWppJ6t-32jW6L-1}}\nwprevnextdefs{NWppJ6t-DKmU5-29}{NWppJ6t-DKmU5-2B}\nwenddeflinemarkup
{\bf{}void} {\bf{}figure}::{\it{}read\_archive}({\bf{}const} {\it{}archive\_node} &{\it{}an}, {\bf{}lst} &{\it{}sym\_lst})\nwindexdefn{\nwixident{figure}}{figure}{NWppJ6t-DKmU5-2A}
{\nwlbrace}
    {\it{}inherited}::{\it{}read\_archive}({\it{}an}, {\it{}sym\_lst});
    {\bf{}ex} {\it{}e};
    {\it{}an}.{\it{}find\_ex}({\tt{}"point\_metric"}, {\it{}e}, {\it{}sym\_lst});
    {\it{}point\_metric}={\it{}ex\_to}\begin{math}<\end{math}{\bf{}clifford}\begin{math}>\end{math}({\it{}e});
    {\it{}an}.{\it{}find\_ex}({\tt{}"cycle\_metric"}, {\it{}e}, {\it{}sym\_lst});
    {\it{}cycle\_metric}={\it{}ex\_to}\begin{math}<\end{math}{\bf{}clifford}\begin{math}>\end{math}({\it{}e});
    {\bf{}lst} {\it{}all\_sym}={\it{}sym\_lst};
    {\bf{}ex} {\it{}keys}, {\it{}cnodes};
    {\it{}an}.{\it{}find\_ex}({\tt{}"real\_line"}, {\it{}real\_line}, {\it{}sym\_lst});
    {\it{}all\_sym}.{\it{}append}({\it{}real\_line});
    {\it{}an}.{\it{}find\_ex}({\tt{}"infinity"}, {\it{}infinity}, {\it{}sym\_lst});
    {\it{}all\_sym}.{\it{}append}({\it{}infinity});
    {\it{}an}.{\it{}find\_bool}({\tt{}"float\_evaluation"}, {\it{}float\_evaluation});

\nwused{\\{NWppJ6t-32jW6L-1}}\nwidentdefs{\\{{\nwixident{figure}}{figure}}}\nwidentuses{\\{{\nwixident{cycle{\_}metric}}{cycle:unmetric}}\\{{\nwixident{ex}}{ex}}\\{{\nwixident{float{\_}evaluation}}{float:unevaluation}}\\{{\nwixident{infinity}}{infinity}}\\{{\nwixident{point{\_}metric}}{point:unmetric}}\\{{\nwixident{read{\_}archive}}{read:unarchive}}\\{{\nwixident{real{\_}line}}{real:unline}}}\nwindexuse{\nwixident{cycle{\_}metric}}{cycle:unmetric}{NWppJ6t-DKmU5-2A}\nwindexuse{\nwixident{ex}}{ex}{NWppJ6t-DKmU5-2A}\nwindexuse{\nwixident{float{\_}evaluation}}{float:unevaluation}{NWppJ6t-DKmU5-2A}\nwindexuse{\nwixident{infinity}}{infinity}{NWppJ6t-DKmU5-2A}\nwindexuse{\nwixident{point{\_}metric}}{point:unmetric}{NWppJ6t-DKmU5-2A}\nwindexuse{\nwixident{read{\_}archive}}{read:unarchive}{NWppJ6t-DKmU5-2A}\nwindexuse{\nwixident{real{\_}line}}{real:unline}{NWppJ6t-DKmU5-2A}\nwendcode{}\nwbegindocs{1160}\nwdocspar
\nwenddocs{}\nwbegincode{1161}\sublabel{NWppJ6t-DKmU5-2B}\nwmargintag{{\nwtagstyle{}\subpageref{NWppJ6t-DKmU5-2B}}}\moddef{figure class~{\nwtagstyle{}\subpageref{NWppJ6t-DKmU5-1}}}\plusendmoddef\Rm{}\nwstartdeflinemarkup\nwusesondefline{\\{NWppJ6t-32jW6L-1}}\nwprevnextdefs{NWppJ6t-DKmU5-2A}{NWppJ6t-DKmU5-2C}\nwenddeflinemarkup
    //an.find\_ex("keys", keys, all\_sym);
    {\it{}an}.{\it{}find\_ex}({\tt{}"keys"}, {\it{}keys}, {\it{}sym\_lst});
    {\bf{}for} ({\bf{}const} {\bf{}auto}& {\it{}it} : {\it{}ex\_to}\begin{math}<\end{math}{\bf{}lst}\begin{math}>\end{math}({\it{}keys}))
        {\it{}all\_sym}.{\it{}append}({\it{}it});
    {\it{}all\_sym}.{\it{}sort}();
    {\it{}all\_sym}.{\it{}unique}();
    {\it{}an}.{\it{}find\_ex}({\tt{}"cnodes"}, {\it{}cnodes}, {\it{}all\_sym});
    {\bf{}lst}::{\it{}const\_iterator} {\it{}it1} = {\it{}ex\_to}\begin{math}<\end{math}{\bf{}lst}\begin{math}>\end{math}({\it{}cnodes}).{\it{}begin}();
    {\it{}nodes}.{\it{}clear}();
    {\bf{}for} ({\bf{}const} {\bf{}auto}& {\it{}it} : {\it{}ex\_to}\begin{math}<\end{math}{\bf{}lst}\begin{math}>\end{math}({\it{}keys})) {\nwlbrace}
        {\it{}nodes}[{\it{}it}]={\it{}ex\_to}\begin{math}<\end{math}{\bf{}cycle\_node}\begin{math}>\end{math}(\begin{math}\ast\end{math}{\it{}it1});
        \protect\PP{\it{}it1};
    {\nwrbrace}
{\nwrbrace}

\nwused{\\{NWppJ6t-32jW6L-1}}\nwidentuses{\\{{\nwixident{cycle{\_}node}}{cycle:unnode}}\\{{\nwixident{nodes}}{nodes}}}\nwindexuse{\nwixident{cycle{\_}node}}{cycle:unnode}{NWppJ6t-DKmU5-2B}\nwindexuse{\nwixident{nodes}}{nodes}{NWppJ6t-DKmU5-2B}\nwendcode{}\nwbegindocs{1162}\nwdocspar
\nwenddocs{}\nwbegincode{1163}\sublabel{NWppJ6t-DKmU5-2C}\nwmargintag{{\nwtagstyle{}\subpageref{NWppJ6t-DKmU5-2C}}}\moddef{figure class~{\nwtagstyle{}\subpageref{NWppJ6t-DKmU5-1}}}\plusendmoddef\Rm{}\nwstartdeflinemarkup\nwusesondefline{\\{NWppJ6t-32jW6L-1}}\nwprevnextdefs{NWppJ6t-DKmU5-2B}{NWppJ6t-DKmU5-2D}\nwenddeflinemarkup
{\it{}GINAC\_BIND\_UNARCHIVER}({\bf{}figure});\nwindexdefn{\nwixident{figure}}{figure}{NWppJ6t-DKmU5-2C}

\nwused{\\{NWppJ6t-32jW6L-1}}\nwidentdefs{\\{{\nwixident{figure}}{figure}}}\nwendcode{}\nwbegindocs{1164}\nwdocspar
\nwenddocs{}\nwbegincode{1165}\sublabel{NWppJ6t-DKmU5-2D}\nwmargintag{{\nwtagstyle{}\subpageref{NWppJ6t-DKmU5-2D}}}\moddef{figure class~{\nwtagstyle{}\subpageref{NWppJ6t-DKmU5-1}}}\plusendmoddef\Rm{}\nwstartdeflinemarkup\nwusesondefline{\\{NWppJ6t-32jW6L-1}}\nwprevnextdefs{NWppJ6t-DKmU5-2C}{NWppJ6t-DKmU5-2E}\nwenddeflinemarkup
{\bf{}bool} {\bf{}figure}::{\it{}info}({\bf{}unsigned} {\it{}inf}) {\bf{}const}
{\nwlbrace}
    {\bf{}switch} ({\it{}inf}) {\nwlbrace}
    {\bf{}case} {\it{}status\_flags}::{\it{}expanded}:
        {\bf{}return} ({\it{}inf} & {\it{}flags});
    {\nwrbrace}
    {\bf{}return} {\it{}inherited}::{\it{}info}({\it{}inf});
{\nwrbrace}

\nwused{\\{NWppJ6t-32jW6L-1}}\nwidentuses{\\{{\nwixident{figure}}{figure}}\\{{\nwixident{info}}{info}}}\nwindexuse{\nwixident{figure}}{figure}{NWppJ6t-DKmU5-2D}\nwindexuse{\nwixident{info}}{info}{NWppJ6t-DKmU5-2D}\nwendcode{}\nwbegindocs{1166}\nwdocspar
\subsubsection{Relations and measurements}
\label{sec:relat-meas}

\nwenddocs{}\nwbegindocs{1167}The method to check that two cycles are in a relation.
\nwenddocs{}\nwbegincode{1168}\sublabel{NWppJ6t-DKmU5-2E}\nwmargintag{{\nwtagstyle{}\subpageref{NWppJ6t-DKmU5-2E}}}\moddef{figure class~{\nwtagstyle{}\subpageref{NWppJ6t-DKmU5-1}}}\plusendmoddef\Rm{}\nwstartdeflinemarkup\nwusesondefline{\\{NWppJ6t-32jW6L-1}}\nwprevnextdefs{NWppJ6t-DKmU5-2D}{NWppJ6t-DKmU5-2F}\nwenddeflinemarkup
{\bf{}ex} {\bf{}figure}::{\it{}check\_rel}({\bf{}const} {\bf{}ex} & {\it{}key1}, {\bf{}const} {\bf{}ex} & {\it{}key2}, {\it{}PCR} {\it{}rel}, {\bf{}bool} {\it{}use\_cycle\_metric},
                     {\bf{}const} {\bf{}ex} & {\it{}parameter}, {\bf{}bool} {\it{}corresponds}) {\bf{}const}
{\nwlbrace}
    \LA{}run through all cycles in two nodes correspondingly~{\nwtagstyle{}\subpageref{NWppJ6t-1bkNr0-1}}\RA{}
    \LA{}add checked relation~{\nwtagstyle{}\subpageref{NWppJ6t-2SaBW8-1}}\RA{}
    \LA{}run through all cycles in two nodes async~{\nwtagstyle{}\subpageref{NWppJ6t-47PgDD-1}}\RA{}
    \LA{}add checked relation~{\nwtagstyle{}\subpageref{NWppJ6t-2SaBW8-1}}\RA{}
\nwindexdefn{\nwixident{check{\_}rel}}{check:unrel}{NWppJ6t-DKmU5-2E}\eatline
\nwused{\\{NWppJ6t-32jW6L-1}}\nwidentdefs{\\{{\nwixident{check{\_}rel}}{check:unrel}}}\nwidentuses{\\{{\nwixident{ex}}{ex}}\\{{\nwixident{figure}}{figure}}\\{{\nwixident{PCR}}{PCR}}}\nwindexuse{\nwixident{ex}}{ex}{NWppJ6t-DKmU5-2E}\nwindexuse{\nwixident{figure}}{figure}{NWppJ6t-DKmU5-2E}\nwindexuse{\nwixident{PCR}}{PCR}{NWppJ6t-DKmU5-2E}\nwendcode{}\nwbegindocs{1169}\nwdocspar
\nwenddocs{}\nwbegindocs{1170}This piece of code is common in {\Tt{}\Rm{}{\it{}check\_rel}\nwendquote} and {\Tt{}\Rm{}{\it{}measure}\nwendquote}.
\nwenddocs{}\nwbegincode{1171}\sublabel{NWppJ6t-1bkNr0-1}\nwmargintag{{\nwtagstyle{}\subpageref{NWppJ6t-1bkNr0-1}}}\moddef{run through all cycles in two nodes correspondingly~{\nwtagstyle{}\subpageref{NWppJ6t-1bkNr0-1}}}\endmoddef\Rm{}\nwstartdeflinemarkup\nwusesondefline{\\{NWppJ6t-DKmU5-2E}\\{NWppJ6t-DKmU5-2G}}\nwenddeflinemarkup
    {\bf{}lst} {\it{}res},
        {\it{}cycles1}={\it{}ex\_to}\begin{math}<\end{math}{\bf{}lst}\begin{math}>\end{math}({\it{}ex\_to}\begin{math}<\end{math}{\bf{}cycle\_node}\begin{math}>\end{math}({\it{}nodes}.{\it{}find}({\it{}key1})\begin{math}\rightarrow\end{math}{\it{}second})
                           .{\it{}get\_cycle}({\it{}use\_cycle\_metric}? {\it{}cycle\_metric} : {\it{}point\_metric})),
        {\it{}cycles2}={\it{}ex\_to}\begin{math}<\end{math}{\bf{}lst}\begin{math}>\end{math}({\it{}ex\_to}\begin{math}<\end{math}{\bf{}cycle\_node}\begin{math}>\end{math}({\it{}nodes}.{\it{}find}({\it{}key2})\begin{math}\rightarrow\end{math}{\it{}second})
                           .{\it{}get\_cycle}({\it{}use\_cycle\_metric}? {\it{}cycle\_metric} : {\it{}point\_metric}));

    {\bf{}if} ({\it{}corresponds} \begin{math}\wedge\end{math} {\it{}cycles1}.{\it{}nops}() \begin{math}\equiv\end{math} {\it{}cycles2}.{\it{}nops}()) {\nwlbrace}
        {\bf{}auto} {\it{}it2}={\it{}cycles2}.{\it{}begin}();
        {\bf{}for} ({\bf{}const} {\bf{}auto}& {\it{}it1} : {\it{}cycles1}) {\nwlbrace}
            {\bf{}lst} {\it{}calc}={\it{}ex\_to}\begin{math}<\end{math}{\bf{}lst}\begin{math}>\end{math}({\it{}rel}({\it{}it1},\begin{math}\ast\end{math}({\it{}it2}\protect\PP),{\it{}parameter}));
            {\bf{}for} ({\bf{}const} {\bf{}auto}& {\it{}itr} : {\it{}calc})

\nwused{\\{NWppJ6t-DKmU5-2E}\\{NWppJ6t-DKmU5-2G}}\nwidentuses{\\{{\nwixident{cycle{\_}metric}}{cycle:unmetric}}\\{{\nwixident{cycle{\_}node}}{cycle:unnode}}\\{{\nwixident{get{\_}cycle}}{get:uncycle}}\\{{\nwixident{nodes}}{nodes}}\\{{\nwixident{nops}}{nops}}\\{{\nwixident{point{\_}metric}}{point:unmetric}}}\nwindexuse{\nwixident{cycle{\_}metric}}{cycle:unmetric}{NWppJ6t-1bkNr0-1}\nwindexuse{\nwixident{cycle{\_}node}}{cycle:unnode}{NWppJ6t-1bkNr0-1}\nwindexuse{\nwixident{get{\_}cycle}}{get:uncycle}{NWppJ6t-1bkNr0-1}\nwindexuse{\nwixident{nodes}}{nodes}{NWppJ6t-1bkNr0-1}\nwindexuse{\nwixident{nops}}{nops}{NWppJ6t-1bkNr0-1}\nwindexuse{\nwixident{point{\_}metric}}{point:unmetric}{NWppJ6t-1bkNr0-1}\nwendcode{}\nwbegindocs{1172}We add corresponding relation. We wish to make output homogeneous
despite of the fact that {\Tt{}\Rm{}{\it{}rel}\nwendquote} can be of different type: either
returning {\Tt{}\Rm{}{\bf{}relational}\nwendquote} or not.
\nwenddocs{}\nwbegincode{1173}\sublabel{NWppJ6t-2SaBW8-1}\nwmargintag{{\nwtagstyle{}\subpageref{NWppJ6t-2SaBW8-1}}}\moddef{add checked relation~{\nwtagstyle{}\subpageref{NWppJ6t-2SaBW8-1}}}\endmoddef\Rm{}\nwstartdeflinemarkup\nwusesondefline{\\{NWppJ6t-DKmU5-2E}}\nwenddeflinemarkup
            {\nwlbrace}
                {\bf{}ex} {\it{}e}=({\it{}itr}.{\it{}op}(0)).{\it{}normal}();
                {\bf{}if} ({\it{}is\_a}\begin{math}<\end{math}{\bf{}relational}\begin{math}>\end{math}({\it{}e}))
                    {\it{}res}.{\it{}append}({\it{}e});
                {\bf{}else}
                    {\it{}res}.{\it{}append}({\it{}e}\begin{math}\equiv\end{math}0);
            {\nwrbrace}

\nwused{\\{NWppJ6t-DKmU5-2E}}\nwidentuses{\\{{\nwixident{ex}}{ex}}\\{{\nwixident{op}}{op}}}\nwindexuse{\nwixident{ex}}{ex}{NWppJ6t-2SaBW8-1}\nwindexuse{\nwixident{op}}{op}{NWppJ6t-2SaBW8-1}\nwendcode{}\nwbegindocs{1174}If cycles are treated asynchronously we run two independent loops.
\nwenddocs{}\nwbegincode{1175}\sublabel{NWppJ6t-47PgDD-1}\nwmargintag{{\nwtagstyle{}\subpageref{NWppJ6t-47PgDD-1}}}\moddef{run through all cycles in two nodes async~{\nwtagstyle{}\subpageref{NWppJ6t-47PgDD-1}}}\endmoddef\Rm{}\nwstartdeflinemarkup\nwusesondefline{\\{NWppJ6t-DKmU5-2E}\\{NWppJ6t-DKmU5-2G}}\nwenddeflinemarkup
      {\nwrbrace}
    {\nwrbrace} {\bf{}else} {\nwlbrace}
        {\bf{}for} ({\bf{}const} {\bf{}auto}& {\it{}it1} : {\it{}cycles1}) {\nwlbrace}
            {\bf{}for} ({\bf{}const} {\bf{}auto}& {\it{}it2} : {\it{}cycles2}) {\nwlbrace}
                {\bf{}lst} {\it{}calc}={\it{}ex\_to}\begin{math}<\end{math}{\bf{}lst}\begin{math}>\end{math}({\it{}rel}({\it{}it1},{\it{}it2},{\it{}parameter}));
                {\bf{}for} ({\bf{}const} {\bf{}auto}& {\it{}itr} : {\it{}calc})

\nwused{\\{NWppJ6t-DKmU5-2E}\\{NWppJ6t-DKmU5-2G}}\nwendcode{}\nwbegindocs{1176}Simply finish the routine with the right number of brackets.
\nwenddocs{}\nwbegincode{1177}\sublabel{NWppJ6t-DKmU5-2F}\nwmargintag{{\nwtagstyle{}\subpageref{NWppJ6t-DKmU5-2F}}}\moddef{figure class~{\nwtagstyle{}\subpageref{NWppJ6t-DKmU5-1}}}\plusendmoddef\Rm{}\nwstartdeflinemarkup\nwusesondefline{\\{NWppJ6t-32jW6L-1}}\nwprevnextdefs{NWppJ6t-DKmU5-2E}{NWppJ6t-DKmU5-2G}\nwenddeflinemarkup
            {\nwrbrace}
        {\nwrbrace}
    {\nwrbrace}
    {\bf{}return} {\it{}res};
{\nwrbrace}

\nwused{\\{NWppJ6t-32jW6L-1}}\nwendcode{}\nwbegindocs{1178}The method to measure certain quantity, it essentially copies code
from the previous method.
\nwenddocs{}\nwbegincode{1179}\sublabel{NWppJ6t-DKmU5-2G}\nwmargintag{{\nwtagstyle{}\subpageref{NWppJ6t-DKmU5-2G}}}\moddef{figure class~{\nwtagstyle{}\subpageref{NWppJ6t-DKmU5-1}}}\plusendmoddef\Rm{}\nwstartdeflinemarkup\nwusesondefline{\\{NWppJ6t-32jW6L-1}}\nwprevnextdefs{NWppJ6t-DKmU5-2F}{NWppJ6t-DKmU5-2H}\nwenddeflinemarkup
{\bf{}ex} {\bf{}figure}::{\it{}measure}({\bf{}const} {\bf{}ex} & {\it{}key1}, {\bf{}const} {\bf{}ex} & {\it{}key2}, {\it{}PCR} {\it{}rel}, {\bf{}bool} {\it{}use\_cycle\_metric},
                   {\bf{}const} {\bf{}ex} & {\it{}parameter}, {\bf{}bool} {\it{}corresponds}) {\bf{}const}
{\nwlbrace}
    \LA{}run through all cycles in two nodes correspondingly~{\nwtagstyle{}\subpageref{NWppJ6t-1bkNr0-1}}\RA{}
    {\it{}res}.{\it{}append}({\it{}itr}.{\it{}op}(0));
    \LA{}run through all cycles in two nodes async~{\nwtagstyle{}\subpageref{NWppJ6t-47PgDD-1}}\RA{}
    {\it{}res}.{\it{}append}({\it{}itr}.{\it{}op}(0));
                {\nwrbrace}
            {\nwrbrace}
        {\nwrbrace}
    {\bf{}return} {\it{}res};
{\nwrbrace}
\nwindexdefn{\nwixident{measure}}{measure}{NWppJ6t-DKmU5-2G}\eatline
\nwused{\\{NWppJ6t-32jW6L-1}}\nwidentdefs{\\{{\nwixident{measure}}{measure}}}\nwidentuses{\\{{\nwixident{ex}}{ex}}\\{{\nwixident{figure}}{figure}}\\{{\nwixident{op}}{op}}\\{{\nwixident{PCR}}{PCR}}}\nwindexuse{\nwixident{ex}}{ex}{NWppJ6t-DKmU5-2G}\nwindexuse{\nwixident{figure}}{figure}{NWppJ6t-DKmU5-2G}\nwindexuse{\nwixident{op}}{op}{NWppJ6t-DKmU5-2G}\nwindexuse{\nwixident{PCR}}{PCR}{NWppJ6t-DKmU5-2G}\nwendcode{}\nwbegindocs{1180}\nwdocspar
\nwenddocs{}\nwbegindocs{1181}We apply {\Tt{}\Rm{}{\it{}func}\nwendquote} to all cycles in the, figure one-by-one.
\nwenddocs{}\nwbegincode{1182}\sublabel{NWppJ6t-DKmU5-2H}\nwmargintag{{\nwtagstyle{}\subpageref{NWppJ6t-DKmU5-2H}}}\moddef{figure class~{\nwtagstyle{}\subpageref{NWppJ6t-DKmU5-1}}}\plusendmoddef\Rm{}\nwstartdeflinemarkup\nwusesondefline{\\{NWppJ6t-32jW6L-1}}\nwprevnextdefs{NWppJ6t-DKmU5-2G}{NWppJ6t-DKmU5-2I}\nwenddeflinemarkup
{\bf{}ex} {\bf{}figure}::{\it{}apply}({\it{}PEVAL} {\it{}func}, {\bf{}bool} {\it{}use\_cycle\_metric}, {\bf{}const} {\bf{}ex} & {\it{}param}) {\bf{}const}
{\nwlbrace}
    {\bf{}lst} {\it{}res};
    {\bf{}for} ({\bf{}const} {\bf{}auto}& {\it{}x}: {\it{}nodes}) {\nwlbrace}
        {\bf{}int} {\it{}i}=0;
        {\bf{}lst} {\it{}cycles}={\it{}ex\_to}\begin{math}<\end{math}{\bf{}lst}\begin{math}>\end{math}({\it{}x}.{\it{}second}.{\it{}get\_cycle}({\it{}use\_cycle\_metric}? {\it{}cycle\_metric} : {\it{}point\_metric}));
        {\bf{}for} ({\bf{}const} {\bf{}auto}& {\it{}itc} : {\it{}cycles}) {\nwlbrace}
            {\it{}res}.{\it{}append}({\bf{}lst}{\nwlbrace}{\it{}func}({\it{}itc}, {\it{}param}), {\it{}x}.{\it{}first}, {\it{}i}{\nwrbrace});
            \protect\PP{\it{}i};
        {\nwrbrace}
    {\nwrbrace}
    {\bf{}return} {\it{}res};
{\nwrbrace}
\nwindexdefn{\nwixident{apply}}{apply}{NWppJ6t-DKmU5-2H}\eatline
\nwused{\\{NWppJ6t-32jW6L-1}}\nwidentdefs{\\{{\nwixident{apply}}{apply}}}\nwidentuses{\\{{\nwixident{cycle{\_}metric}}{cycle:unmetric}}\\{{\nwixident{ex}}{ex}}\\{{\nwixident{figure}}{figure}}\\{{\nwixident{get{\_}cycle}}{get:uncycle}}\\{{\nwixident{nodes}}{nodes}}\\{{\nwixident{point{\_}metric}}{point:unmetric}}}\nwindexuse{\nwixident{cycle{\_}metric}}{cycle:unmetric}{NWppJ6t-DKmU5-2H}\nwindexuse{\nwixident{ex}}{ex}{NWppJ6t-DKmU5-2H}\nwindexuse{\nwixident{figure}}{figure}{NWppJ6t-DKmU5-2H}\nwindexuse{\nwixident{get{\_}cycle}}{get:uncycle}{NWppJ6t-DKmU5-2H}\nwindexuse{\nwixident{nodes}}{nodes}{NWppJ6t-DKmU5-2H}\nwindexuse{\nwixident{point{\_}metric}}{point:unmetric}{NWppJ6t-DKmU5-2H}\nwendcode{}\nwbegindocs{1183}\nwdocspar
\nwenddocs{}\nwbegindocs{1184}\nwdocspar
\subsubsection{Default Asymptote styles}
\label{sec:defa-asympt-styl}

\nwenddocs{}\nwbegindocs{1185}A simple \Asymptote\ style. We produce different colours for points,
lines and circles. No further options are specified.
\nwenddocs{}\nwbegincode{1186}\sublabel{NWppJ6t-DKmU5-2I}\nwmargintag{{\nwtagstyle{}\subpageref{NWppJ6t-DKmU5-2I}}}\moddef{figure class~{\nwtagstyle{}\subpageref{NWppJ6t-DKmU5-1}}}\plusendmoddef\Rm{}\nwstartdeflinemarkup\nwusesondefline{\\{NWppJ6t-32jW6L-1}}\nwprevnextdefs{NWppJ6t-DKmU5-2H}{NWppJ6t-DKmU5-2J}\nwenddeflinemarkup
{\it{}string} {\it{}asy\_cycle\_color}({\bf{}const} {\bf{}ex} & {\it{}label}, {\bf{}const} {\bf{}ex} & {\it{}C}, {\bf{}lst} & {\it{}color})
{\nwlbrace}
    {\it{}string} {\it{}asy\_options}={\tt{}""};
    {\bf{}if} ({\it{}is\_less\_than\_epsilon}({\it{}ex\_to}\begin{math}<\end{math}{\bf{}cycle}\begin{math}>\end{math}({\it{}C}).{\it{}det}())) {\nwlbrace}// point
        {\it{}color}={\bf{}lst}{\nwlbrace}0.5,0,0{\nwrbrace};
        {\it{}asy\_options}={\tt{}"dotted"};
    {\nwrbrace} {\bf{}else} {\bf{}if} ({\it{}is\_less\_than\_epsilon}({\it{}ex\_to}\begin{math}<\end{math}{\bf{}cycle}\begin{math}>\end{math}({\it{}C}).{\it{}get\_k}())) // straight line
        {\it{}color}={\bf{}lst}{\nwlbrace}0,0.5,0{\nwrbrace};
    {\bf{}else}  // a proper circle-hyperbola-parabola
        {\it{}color}={\bf{}lst}{\nwlbrace}0,0,0.5{\nwrbrace};

    {\bf{}return} {\it{}asy\_options};
{\nwrbrace}
\nwindexdefn{\nwixident{asy{\_}cycle{\_}color}}{asy:uncycle:uncolor}{NWppJ6t-DKmU5-2I}\eatline
\nwused{\\{NWppJ6t-32jW6L-1}}\nwidentdefs{\\{{\nwixident{asy{\_}cycle{\_}color}}{asy:uncycle:uncolor}}}\nwidentuses{\\{{\nwixident{ex}}{ex}}\\{{\nwixident{is{\_}less{\_}than{\_}epsilon}}{is:unless:unthan:unepsilon}}}\nwindexuse{\nwixident{ex}}{ex}{NWppJ6t-DKmU5-2I}\nwindexuse{\nwixident{is{\_}less{\_}than{\_}epsilon}}{is:unless:unthan:unepsilon}{NWppJ6t-DKmU5-2I}\nwendcode{}\nwbegindocs{1187}\nwdocspar
\nwenddocs{}\nwbegindocs{1188} A style to place labels.
\nwenddocs{}\nwbegincode{1189}\sublabel{NWppJ6t-DKmU5-2J}\nwmargintag{{\nwtagstyle{}\subpageref{NWppJ6t-DKmU5-2J}}}\moddef{figure class~{\nwtagstyle{}\subpageref{NWppJ6t-DKmU5-1}}}\plusendmoddef\Rm{}\nwstartdeflinemarkup\nwusesondefline{\\{NWppJ6t-32jW6L-1}}\nwprevnextdefs{NWppJ6t-DKmU5-2I}{NWppJ6t-DKmU5-2K}\nwenddeflinemarkup
{\it{}string} {\it{}label\_pos}({\bf{}const} {\bf{}ex} & {\it{}label}, {\bf{}const} {\bf{}ex} & {\it{}C}, {\bf{}const} {\it{}string} {\it{}draw\_str}) {\nwlbrace}
    {\it{}stringstream} {\it{}sstr};
    {\it{}sstr} \begin{math}\ll\end{math} {\it{}latex} \begin{math}\ll\end{math} {\it{}label};

    {\it{}string} {\it{}name}={\it{}ex\_to}\begin{math}<\end{math}{\bf{}symbol}\begin{math}>\end{math}({\it{}label}).{\it{}get\_name}(), {\it{}new\_TeXname};

    {\bf{}if} ({\it{}sstr}.{\it{}str}() \begin{math}\equiv\end{math} {\it{}name}) {\nwlbrace}
        {\it{}string} {\it{}TeXname};
        \LA{}auto TeX name~{\nwtagstyle{}\subpageref{NWppJ6t-2tM1q2-1}}\RA{}
        {\bf{}if} ({\it{}TeXname\_new} \begin{math}\equiv\end{math}{\tt{}""})
            {\it{}new\_TeXname}={\it{}name};
        {\bf{}else}
            {\it{}new\_TeXname}={\it{}TeXname\_new};
    {\nwrbrace} {\bf{}else}
        {\it{}new\_TeXname}={\it{}sstr}.{\it{}str}();
\nwindexdefn{\nwixident{label{\_}pos}}{label:unpos}{NWppJ6t-DKmU5-2J}\eatline
\nwused{\\{NWppJ6t-32jW6L-1}}\nwidentdefs{\\{{\nwixident{label{\_}pos}}{label:unpos}}}\nwidentuses{\\{{\nwixident{ex}}{ex}}\\{{\nwixident{name}}{name}}\\{{\nwixident{TeXname}}{TeXname}}}\nwindexuse{\nwixident{ex}}{ex}{NWppJ6t-DKmU5-2J}\nwindexuse{\nwixident{name}}{name}{NWppJ6t-DKmU5-2J}\nwindexuse{\nwixident{TeXname}}{TeXname}{NWppJ6t-DKmU5-2J}\nwendcode{}\nwbegindocs{1190}\nwdocspar
\nwenddocs{}\nwbegindocs{1191}We use {\Tt{}\Rm{}{\it{}regex}\nwendquote} to spot places for labels in the \Asymptote\ output.
\nwenddocs{}\nwbegincode{1192}\sublabel{NWppJ6t-DKmU5-2K}\nwmargintag{{\nwtagstyle{}\subpageref{NWppJ6t-DKmU5-2K}}}\moddef{figure class~{\nwtagstyle{}\subpageref{NWppJ6t-DKmU5-1}}}\plusendmoddef\Rm{}\nwstartdeflinemarkup\nwusesondefline{\\{NWppJ6t-32jW6L-1}}\nwprevnextdefs{NWppJ6t-DKmU5-2J}{\relax}\nwenddeflinemarkup
    {\it{}std}::{\it{}regex} {\it{}draw}({\tt{}"([.{\char92}{\char92}n{\char92}{\char92}r{\char92}{\char92}s]*)(draw){\char92}{\char92}(([{\char92}{\char92}w]+,)?((?:{\char92}{\char92}(.+?{\char92}{\char92})|{\char92}{\char92}{\char123}.+?{\char92}{\char92}{\char125}|[^-,0-9{\char92}{\char92}.])+),([.{\char92}{\char92}n{\char92}{\char92}r]*)"});
    {\it{}std}::{\it{}regex} {\it{}dot}({\tt{}"([.{\char92}{\char92}n{\char92}{\char92}r{\char92}{\char92}s]*)(dot){\char92}{\char92}(([{\char92}{\char92}w]*,)?((?:{\char92}{\char92}(.+?{\char92}{\char92})|{\char92}{\char92}{\char123}.+?{\char92}{\char92}{\char125}|[^-,0-9{\char92}{\char92}.])+|[{\char92}{\char92}w]+),([.{\char92}{\char92}n{\char92}{\char92}r]*)"});
    {\it{}std}::{\it{}regex} {\it{}e1}({\tt{}"symbolLaTeXname"});

    {\bf{}if} ({\it{}std}::{\it{}regex\_search}({\it{}draw\_str}, {\it{}draw})) {\nwlbrace}
            {\it{}string} {\it{}labelstr}={\it{}std}::{\it{}regex\_replace} ({\it{}draw\_str}, {\it{}draw},
                                {\tt{}"label($3{\char92}"$symbolLaTeXname${\char92}", point($4,0.1), SE);{\char92}n"},
                                {\it{}std}::{\it{}regex\_constants}::{\it{}format\_no\_copy} \begin{math}\mid\end{math} {\it{}std}::{\it{}regex\_constants}::{\it{}format\_first\_only});
            {\bf{}return} {\it{}std}::{\it{}regex\_replace} ({\it{}labelstr}, {\it{}e1}, {\it{}new\_TeXname});
    {\nwrbrace} {\bf{}else} {\bf{}if} ({\it{}std}::{\it{}regex\_search}({\it{}draw\_str}, {\it{}dot})) {\nwlbrace}
            {\it{}string} {\it{}labelstr}={\it{}std}::{\it{}regex\_replace} ({\it{}draw\_str}, {\it{}dot},
                                {\tt{}"label($3{\char92}"$symbolLaTeXname${\char92}", $4, SE);{\char92}n"},
                                {\it{}std}::{\it{}regex\_constants}::{\it{}format\_no\_copy});
            {\bf{}return} {\it{}std}::{\it{}regex\_replace} ({\it{}labelstr}, {\it{}e1}, {\it{}new\_TeXname});
    {\nwrbrace} {\bf{}else}
        {\bf{}return} {\tt{}""};
{\nwrbrace}

\nwused{\\{NWppJ6t-32jW6L-1}}\nwendcode{}\nwbegindocs{1193}\nwdocspar
\subsection{Functions defining cycle relations}
\label{sec:cycles-relations}

\nwenddocs{}\nwbegindocs{1194}This is collection of linear cycle relations which do not require a parameter.
\nwenddocs{}\nwbegincode{1195}\sublabel{NWppJ6t-3fVAGh-1}\nwmargintag{{\nwtagstyle{}\subpageref{NWppJ6t-3fVAGh-1}}}\moddef{add cycle relations~{\nwtagstyle{}\subpageref{NWppJ6t-3fVAGh-1}}}\endmoddef\Rm{}\nwstartdeflinemarkup\nwusesondefline{\\{NWppJ6t-32jW6L-1}}\nwprevnextdefs{\relax}{NWppJ6t-3fVAGh-2}\nwenddeflinemarkup
{\bf{}ex} {\it{}cycle\_orthogonal}({\bf{}const} {\bf{}ex} & {\it{}C1}, {\bf{}const} {\bf{}ex} & {\it{}C2}, {\bf{}const} {\bf{}ex} & {\it{}pr})
{\nwlbrace}
    {\bf{}return} {\bf{}lst}{\nwlbrace}({\bf{}ex}){\bf{}lst}{\nwlbrace}{\it{}ex\_to}\begin{math}<\end{math}{\bf{}cycle}\begin{math}>\end{math}({\it{}C1}).{\it{}is\_orthogonal}({\it{}ex\_to}\begin{math}<\end{math}{\bf{}cycle}\begin{math}>\end{math}({\it{}C2})){\nwrbrace}{\nwrbrace};
{\nwrbrace}
\nwindexdefn{\nwixident{cycle{\_}orthogonal}}{cycle:unorthogonal}{NWppJ6t-3fVAGh-1}\eatline
\nwalsodefined{\\{NWppJ6t-3fVAGh-2}\\{NWppJ6t-3fVAGh-3}\\{NWppJ6t-3fVAGh-4}\\{NWppJ6t-3fVAGh-5}\\{NWppJ6t-3fVAGh-6}\\{NWppJ6t-3fVAGh-7}\\{NWppJ6t-3fVAGh-8}\\{NWppJ6t-3fVAGh-9}\\{NWppJ6t-3fVAGh-A}\\{NWppJ6t-3fVAGh-B}\\{NWppJ6t-3fVAGh-C}\\{NWppJ6t-3fVAGh-D}\\{NWppJ6t-3fVAGh-E}\\{NWppJ6t-3fVAGh-F}\\{NWppJ6t-3fVAGh-G}\\{NWppJ6t-3fVAGh-H}\\{NWppJ6t-3fVAGh-I}}\nwused{\\{NWppJ6t-32jW6L-1}}\nwidentdefs{\\{{\nwixident{cycle{\_}orthogonal}}{cycle:unorthogonal}}}\nwidentuses{\\{{\nwixident{ex}}{ex}}\\{{\nwixident{is{\_}orthogonal}}{is:unorthogonal}}}\nwindexuse{\nwixident{ex}}{ex}{NWppJ6t-3fVAGh-1}\nwindexuse{\nwixident{is{\_}orthogonal}}{is:unorthogonal}{NWppJ6t-3fVAGh-1}\nwendcode{}\nwbegindocs{1196}\nwdocspar
\nwenddocs{}\nwbegindocs{1197}\nwdocspar
\nwenddocs{}\nwbegincode{1198}\sublabel{NWppJ6t-3fVAGh-2}\nwmargintag{{\nwtagstyle{}\subpageref{NWppJ6t-3fVAGh-2}}}\moddef{add cycle relations~{\nwtagstyle{}\subpageref{NWppJ6t-3fVAGh-1}}}\plusendmoddef\Rm{}\nwstartdeflinemarkup\nwusesondefline{\\{NWppJ6t-32jW6L-1}}\nwprevnextdefs{NWppJ6t-3fVAGh-1}{NWppJ6t-3fVAGh-3}\nwenddeflinemarkup
{\bf{}ex} {\it{}cycle\_f\_orthogonal}({\bf{}const} {\bf{}ex} & {\it{}C1}, {\bf{}const} {\bf{}ex} & {\it{}C2}, {\bf{}const} {\bf{}ex} & {\it{}pr})
{\nwlbrace}
    {\bf{}return} {\bf{}lst}{\nwlbrace}({\bf{}ex}){\bf{}lst}{\nwlbrace}{\it{}ex\_to}\begin{math}<\end{math}{\bf{}cycle}\begin{math}>\end{math}({\it{}C1}).{\it{}is\_f\_orthogonal}({\it{}ex\_to}\begin{math}<\end{math}{\bf{}cycle}\begin{math}>\end{math}({\it{}C2})){\nwrbrace}{\nwrbrace};
{\nwrbrace}
\nwindexdefn{\nwixident{cycle{\_}f{\_}orthogonal}}{cycle:unf:unorthogonal}{NWppJ6t-3fVAGh-2}\eatline
\nwused{\\{NWppJ6t-32jW6L-1}}\nwidentdefs{\\{{\nwixident{cycle{\_}f{\_}orthogonal}}{cycle:unf:unorthogonal}}}\nwidentuses{\\{{\nwixident{ex}}{ex}}\\{{\nwixident{is{\_}f{\_}orthogonal}}{is:unf:unorthogonal}}}\nwindexuse{\nwixident{ex}}{ex}{NWppJ6t-3fVAGh-2}\nwindexuse{\nwixident{is{\_}f{\_}orthogonal}}{is:unf:unorthogonal}{NWppJ6t-3fVAGh-2}\nwendcode{}\nwbegindocs{1199}\nwdocspar
\nwenddocs{}\nwbegindocs{1200}\nwdocspar
\nwenddocs{}\nwbegincode{1201}\sublabel{NWppJ6t-3fVAGh-3}\nwmargintag{{\nwtagstyle{}\subpageref{NWppJ6t-3fVAGh-3}}}\moddef{add cycle relations~{\nwtagstyle{}\subpageref{NWppJ6t-3fVAGh-1}}}\plusendmoddef\Rm{}\nwstartdeflinemarkup\nwusesondefline{\\{NWppJ6t-32jW6L-1}}\nwprevnextdefs{NWppJ6t-3fVAGh-2}{NWppJ6t-3fVAGh-4}\nwenddeflinemarkup
{\bf{}ex} {\it{}cycle\_adifferent}({\bf{}const} {\bf{}ex} & {\it{}C1}, {\bf{}const} {\bf{}ex} & {\it{}C2}, {\bf{}const} {\bf{}ex} & {\it{}pr})
{\nwlbrace}
    {\bf{}return} {\bf{}lst}{\nwlbrace}({\bf{}ex}){\bf{}lst}{\nwlbrace}{\bf{}cycle\_data}({\it{}C1}).{\it{}is\_almost\_equal}({\it{}ex\_to}\begin{math}<\end{math}{\bf{}basic}\begin{math}>\end{math}({\bf{}cycle\_data}({\it{}C2})),{\bf{}true})? 0: 1{\nwrbrace}{\nwrbrace};
{\nwrbrace}
\nwindexdefn{\nwixident{cycle{\_}adifferent}}{cycle:unadifferent}{NWppJ6t-3fVAGh-3}\eatline
\nwused{\\{NWppJ6t-32jW6L-1}}\nwidentdefs{\\{{\nwixident{cycle{\_}adifferent}}{cycle:unadifferent}}}\nwidentuses{\\{{\nwixident{cycle{\_}data}}{cycle:undata}}\\{{\nwixident{ex}}{ex}}\\{{\nwixident{is{\_}almost{\_}equal}}{is:unalmost:unequal}}}\nwindexuse{\nwixident{cycle{\_}data}}{cycle:undata}{NWppJ6t-3fVAGh-3}\nwindexuse{\nwixident{ex}}{ex}{NWppJ6t-3fVAGh-3}\nwindexuse{\nwixident{is{\_}almost{\_}equal}}{is:unalmost:unequal}{NWppJ6t-3fVAGh-3}\nwendcode{}\nwbegindocs{1202}\nwdocspar
\nwenddocs{}\nwbegindocs{1203}To \emph{check} the tangential property we use the condition from
\cite{Kisil12a}*{Ex.~5.26(i)}
\begin{equation}
  \label{eq:tangent-condition}
  (\scalar{C_1}{C_2})^2-\scalar{C_1}{C_1}\scalar{C_2}{C_2}=0.
\end{equation}
\nwenddocs{}\nwbegincode{1204}\sublabel{NWppJ6t-3fVAGh-4}\nwmargintag{{\nwtagstyle{}\subpageref{NWppJ6t-3fVAGh-4}}}\moddef{add cycle relations~{\nwtagstyle{}\subpageref{NWppJ6t-3fVAGh-1}}}\plusendmoddef\Rm{}\nwstartdeflinemarkup\nwusesondefline{\\{NWppJ6t-32jW6L-1}}\nwprevnextdefs{NWppJ6t-3fVAGh-3}{NWppJ6t-3fVAGh-5}\nwenddeflinemarkup
{\bf{}ex} {\it{}check\_tangent}({\bf{}const} {\bf{}ex} & {\it{}C1}, {\bf{}const} {\bf{}ex} & {\it{}C2}, {\bf{}const} {\bf{}ex} & {\it{}pr})
{\nwlbrace}
    {\bf{}return} {\bf{}lst}{\nwlbrace}({\bf{}ex}){\bf{}lst}{\nwlbrace}{\it{}pow}({\it{}ex\_to}\begin{math}<\end{math}{\bf{}cycle}\begin{math}>\end{math}({\it{}C1}).{\it{}cycle\_product}({\it{}ex\_to}\begin{math}<\end{math}{\bf{}cycle}\begin{math}>\end{math}({\it{}C2})),2)
                -{\it{}ex\_to}\begin{math}<\end{math}{\bf{}cycle}\begin{math}>\end{math}({\it{}C1}).{\it{}cycle\_product}({\it{}ex\_to}\begin{math}<\end{math}{\bf{}cycle}\begin{math}>\end{math}({\it{}C1}))
                \begin{math}\ast\end{math}{\it{}ex\_to}\begin{math}<\end{math}{\bf{}cycle}\begin{math}>\end{math}({\it{}C2}).{\it{}cycle\_product}({\it{}ex\_to}\begin{math}<\end{math}{\bf{}cycle}\begin{math}>\end{math}({\it{}C2})) \begin{math}\equiv\end{math} 0{\nwrbrace}{\nwrbrace};
{\nwrbrace}
\nwindexdefn{\nwixident{check{\_}tangent}}{check:untangent}{NWppJ6t-3fVAGh-4}\eatline
\nwused{\\{NWppJ6t-32jW6L-1}}\nwidentdefs{\\{{\nwixident{check{\_}tangent}}{check:untangent}}}\nwidentuses{\\{{\nwixident{ex}}{ex}}}\nwindexuse{\nwixident{ex}}{ex}{NWppJ6t-3fVAGh-4}\nwendcode{}\nwbegindocs{1205}\nwdocspar
\nwenddocs{}\nwbegindocs{1206}To \emph{define} tangential property, theoretically we can
use~\eqref{eq:tangent-condition} as well.
However, a system of several such quadratic conditions will be difficult to
resolve. Thus, we use a single quadratic relations
\(\scalar{C_1}{C_1}=-1\) which allows to linearise the tangential
property to a pair of identities: \(\scalar{C_1}{C_2}\pm\sqrt{\scalar{C_2}{C_2}}=0\).
\nwenddocs{}\nwbegincode{1207}\sublabel{NWppJ6t-3fVAGh-5}\nwmargintag{{\nwtagstyle{}\subpageref{NWppJ6t-3fVAGh-5}}}\moddef{add cycle relations~{\nwtagstyle{}\subpageref{NWppJ6t-3fVAGh-1}}}\plusendmoddef\Rm{}\nwstartdeflinemarkup\nwusesondefline{\\{NWppJ6t-32jW6L-1}}\nwprevnextdefs{NWppJ6t-3fVAGh-4}{NWppJ6t-3fVAGh-6}\nwenddeflinemarkup
{\bf{}ex} {\it{}cycle\_tangent}({\bf{}const} {\bf{}ex} & {\it{}C1}, {\bf{}const} {\bf{}ex} & {\it{}C2}, {\bf{}const} {\bf{}ex} & {\it{}pr})
{\nwlbrace}
    {\bf{}return} {\bf{}lst}{\nwlbrace}{\bf{}lst}{\nwlbrace}{\it{}ex\_to}\begin{math}<\end{math}{\bf{}cycle}\begin{math}>\end{math}({\it{}C1}).{\it{}cycle\_product}({\it{}ex\_to}\begin{math}<\end{math}{\bf{}cycle}\begin{math}>\end{math}({\it{}C1}))+{\bf{}numeric}(1)\begin{math}\equiv\end{math}0,
                    {\it{}ex\_to}\begin{math}<\end{math}{\bf{}cycle}\begin{math}>\end{math}({\it{}C1}).{\it{}cycle\_product}({\it{}ex\_to}\begin{math}<\end{math}{\bf{}cycle}\begin{math}>\end{math}({\it{}C2}))
                   -{\it{}sqrt}({\it{}abs}({\it{}ex\_to}\begin{math}<\end{math}{\bf{}cycle}\begin{math}>\end{math}({\it{}C2}).{\it{}cycle\_product}({\it{}ex\_to}\begin{math}<\end{math}{\bf{}cycle}\begin{math}>\end{math}({\it{}C2}))))\begin{math}\equiv\end{math}0{\nwrbrace},
               {\bf{}lst}{\nwlbrace}{\it{}ex\_to}\begin{math}<\end{math}{\bf{}cycle}\begin{math}>\end{math}({\it{}C1}).{\it{}cycle\_product}({\it{}ex\_to}\begin{math}<\end{math}{\bf{}cycle}\begin{math}>\end{math}({\it{}C1}))-{\bf{}numeric}(1)\begin{math}\equiv\end{math}0,
                    {\it{}ex\_to}\begin{math}<\end{math}{\bf{}cycle}\begin{math}>\end{math}({\it{}C1}).{\it{}cycle\_product}({\it{}ex\_to}\begin{math}<\end{math}{\bf{}cycle}\begin{math}>\end{math}({\it{}C2}))
                   -{\it{}sqrt}({\it{}abs}({\it{}ex\_to}\begin{math}<\end{math}{\bf{}cycle}\begin{math}>\end{math}({\it{}C2}).{\it{}cycle\_product}({\it{}ex\_to}\begin{math}<\end{math}{\bf{}cycle}\begin{math}>\end{math}({\it{}C2}))))\begin{math}\equiv\end{math}0{\nwrbrace},
               {\bf{}lst}{\nwlbrace}{\it{}ex\_to}\begin{math}<\end{math}{\bf{}cycle}\begin{math}>\end{math}({\it{}C1}).{\it{}cycle\_product}({\it{}ex\_to}\begin{math}<\end{math}{\bf{}cycle}\begin{math}>\end{math}({\it{}C1}))+{\bf{}numeric}(1)\begin{math}\equiv\end{math}0,
                    {\it{}ex\_to}\begin{math}<\end{math}{\bf{}cycle}\begin{math}>\end{math}({\it{}C1}).{\it{}cycle\_product}({\it{}ex\_to}\begin{math}<\end{math}{\bf{}cycle}\begin{math}>\end{math}({\it{}C2}))
                   +{\it{}sqrt}({\it{}abs}({\it{}ex\_to}\begin{math}<\end{math}{\bf{}cycle}\begin{math}>\end{math}({\it{}C2}).{\it{}cycle\_product}({\it{}ex\_to}\begin{math}<\end{math}{\bf{}cycle}\begin{math}>\end{math}({\it{}C2}))))\begin{math}\equiv\end{math}0{\nwrbrace},
               {\bf{}lst}{\nwlbrace}{\it{}ex\_to}\begin{math}<\end{math}{\bf{}cycle}\begin{math}>\end{math}({\it{}C1}).{\it{}cycle\_product}({\it{}ex\_to}\begin{math}<\end{math}{\bf{}cycle}\begin{math}>\end{math}({\it{}C1}))-{\bf{}numeric}(1)\begin{math}\equiv\end{math}0,
                    {\it{}ex\_to}\begin{math}<\end{math}{\bf{}cycle}\begin{math}>\end{math}({\it{}C1}).{\it{}cycle\_product}({\it{}ex\_to}\begin{math}<\end{math}{\bf{}cycle}\begin{math}>\end{math}({\it{}C2}))
                   +{\it{}sqrt}({\it{}abs}({\it{}ex\_to}\begin{math}<\end{math}{\bf{}cycle}\begin{math}>\end{math}({\it{}C2}).{\it{}cycle\_product}({\it{}ex\_to}\begin{math}<\end{math}{\bf{}cycle}\begin{math}>\end{math}({\it{}C2}))))\begin{math}\equiv\end{math}0{\nwrbrace}{\nwrbrace};
{\nwrbrace}
\nwindexdefn{\nwixident{cycle{\_}tangent}}{cycle:untangent}{NWppJ6t-3fVAGh-5}\eatline
\nwused{\\{NWppJ6t-32jW6L-1}}\nwidentdefs{\\{{\nwixident{cycle{\_}tangent}}{cycle:untangent}}}\nwidentuses{\\{{\nwixident{ex}}{ex}}\\{{\nwixident{numeric}}{numeric}}}\nwindexuse{\nwixident{ex}}{ex}{NWppJ6t-3fVAGh-5}\nwindexuse{\nwixident{numeric}}{numeric}{NWppJ6t-3fVAGh-5}\nwendcode{}\nwbegindocs{1208}\nwdocspar
\nwenddocs{}\nwbegindocs{1209}\nwdocspar
\nwenddocs{}\nwbegincode{1210}\sublabel{NWppJ6t-3fVAGh-6}\nwmargintag{{\nwtagstyle{}\subpageref{NWppJ6t-3fVAGh-6}}}\moddef{add cycle relations~{\nwtagstyle{}\subpageref{NWppJ6t-3fVAGh-1}}}\plusendmoddef\Rm{}\nwstartdeflinemarkup\nwusesondefline{\\{NWppJ6t-32jW6L-1}}\nwprevnextdefs{NWppJ6t-3fVAGh-5}{NWppJ6t-3fVAGh-7}\nwenddeflinemarkup
{\bf{}ex} {\it{}cycle\_tangent\_o}({\bf{}const} {\bf{}ex} & {\it{}C1}, {\bf{}const} {\bf{}ex} & {\it{}C2}, {\bf{}const} {\bf{}ex} & {\it{}pr})
{\nwlbrace}
    {\bf{}return} {\bf{}lst}{\nwlbrace}{\bf{}lst}{\nwlbrace}{\it{}ex\_to}\begin{math}<\end{math}{\bf{}cycle}\begin{math}>\end{math}({\it{}C1}).{\it{}cycle\_product}({\it{}ex\_to}\begin{math}<\end{math}{\bf{}cycle}\begin{math}>\end{math}({\it{}C1}))+{\bf{}numeric}(1)\begin{math}\equiv\end{math}0,
                {\it{}ex\_to}\begin{math}<\end{math}{\bf{}cycle}\begin{math}>\end{math}({\it{}C1}).{\it{}cycle\_product}({\it{}ex\_to}\begin{math}<\end{math}{\bf{}cycle}\begin{math}>\end{math}({\it{}C2}))
                -{\it{}sqrt}({\it{}abs}({\it{}ex\_to}\begin{math}<\end{math}{\bf{}cycle}\begin{math}>\end{math}({\it{}C2}).{\it{}cycle\_product}({\it{}ex\_to}\begin{math}<\end{math}{\bf{}cycle}\begin{math}>\end{math}({\it{}C2}))))\begin{math}\equiv\end{math}0{\nwrbrace},
            {\bf{}lst}{\nwlbrace}{\it{}ex\_to}\begin{math}<\end{math}{\bf{}cycle}\begin{math}>\end{math}({\it{}C1}).{\it{}cycle\_product}({\it{}ex\_to}\begin{math}<\end{math}{\bf{}cycle}\begin{math}>\end{math}({\it{}C1}))-{\bf{}numeric}(1)\begin{math}\equiv\end{math}0,
                    {\it{}ex\_to}\begin{math}<\end{math}{\bf{}cycle}\begin{math}>\end{math}({\it{}C1}).{\it{}cycle\_product}({\it{}ex\_to}\begin{math}<\end{math}{\bf{}cycle}\begin{math}>\end{math}({\it{}C2}))
                    -{\it{}sqrt}({\it{}abs}({\it{}ex\_to}\begin{math}<\end{math}{\bf{}cycle}\begin{math}>\end{math}({\it{}C2}).{\it{}cycle\_product}({\it{}ex\_to}\begin{math}<\end{math}{\bf{}cycle}\begin{math}>\end{math}({\it{}C2}))))\begin{math}\equiv\end{math}0{\nwrbrace}{\nwrbrace};
{\nwrbrace}
\nwindexdefn{\nwixident{cycle{\_}tangent{\_}o}}{cycle:untangent:uno}{NWppJ6t-3fVAGh-6}\eatline
\nwused{\\{NWppJ6t-32jW6L-1}}\nwidentdefs{\\{{\nwixident{cycle{\_}tangent{\_}o}}{cycle:untangent:uno}}}\nwidentuses{\\{{\nwixident{ex}}{ex}}\\{{\nwixident{numeric}}{numeric}}}\nwindexuse{\nwixident{ex}}{ex}{NWppJ6t-3fVAGh-6}\nwindexuse{\nwixident{numeric}}{numeric}{NWppJ6t-3fVAGh-6}\nwendcode{}\nwbegindocs{1211}\nwdocspar
\nwenddocs{}\nwbegindocs{1212}\nwdocspar
\nwenddocs{}\nwbegincode{1213}\sublabel{NWppJ6t-3fVAGh-7}\nwmargintag{{\nwtagstyle{}\subpageref{NWppJ6t-3fVAGh-7}}}\moddef{add cycle relations~{\nwtagstyle{}\subpageref{NWppJ6t-3fVAGh-1}}}\plusendmoddef\Rm{}\nwstartdeflinemarkup\nwusesondefline{\\{NWppJ6t-32jW6L-1}}\nwprevnextdefs{NWppJ6t-3fVAGh-6}{NWppJ6t-3fVAGh-8}\nwenddeflinemarkup
{\bf{}ex} {\it{}cycle\_tangent\_i}({\bf{}const} {\bf{}ex} & {\it{}C1}, {\bf{}const} {\bf{}ex} & {\it{}C2}, {\bf{}const} {\bf{}ex} & {\it{}pr})
{\nwlbrace}
    {\bf{}return} {\bf{}lst}{\nwlbrace}{\bf{}lst}{\nwlbrace}{\it{}ex\_to}\begin{math}<\end{math}{\bf{}cycle}\begin{math}>\end{math}({\it{}C1}).{\it{}cycle\_product}({\it{}ex\_to}\begin{math}<\end{math}{\bf{}cycle}\begin{math}>\end{math}({\it{}C1}))+{\bf{}numeric}(1)\begin{math}\equiv\end{math}0,
                {\it{}ex\_to}\begin{math}<\end{math}{\bf{}cycle}\begin{math}>\end{math}({\it{}C1}).{\it{}cycle\_product}({\it{}ex\_to}\begin{math}<\end{math}{\bf{}cycle}\begin{math}>\end{math}({\it{}C2}))
                +{\it{}sqrt}({\it{}abs}({\it{}ex\_to}\begin{math}<\end{math}{\bf{}cycle}\begin{math}>\end{math}({\it{}C2}).{\it{}cycle\_product}({\it{}ex\_to}\begin{math}<\end{math}{\bf{}cycle}\begin{math}>\end{math}({\it{}C2}))))\begin{math}\equiv\end{math}0{\nwrbrace},
            {\bf{}lst}{\nwlbrace}{\it{}ex\_to}\begin{math}<\end{math}{\bf{}cycle}\begin{math}>\end{math}({\it{}C1}).{\it{}cycle\_product}({\it{}ex\_to}\begin{math}<\end{math}{\bf{}cycle}\begin{math}>\end{math}({\it{}C1}))-{\bf{}numeric}(1)\begin{math}\equiv\end{math}0,
                    {\it{}ex\_to}\begin{math}<\end{math}{\bf{}cycle}\begin{math}>\end{math}({\it{}C1}).{\it{}cycle\_product}({\it{}ex\_to}\begin{math}<\end{math}{\bf{}cycle}\begin{math}>\end{math}({\it{}C2}))
                    +{\it{}sqrt}({\it{}abs}({\it{}ex\_to}\begin{math}<\end{math}{\bf{}cycle}\begin{math}>\end{math}({\it{}C2}).{\it{}cycle\_product}({\it{}ex\_to}\begin{math}<\end{math}{\bf{}cycle}\begin{math}>\end{math}({\it{}C2}))))\begin{math}\equiv\end{math}0{\nwrbrace}{\nwrbrace};
{\nwrbrace}
\nwindexdefn{\nwixident{cycle{\_}tangent{\_}i}}{cycle:untangent:uni}{NWppJ6t-3fVAGh-7}\eatline
\nwused{\\{NWppJ6t-32jW6L-1}}\nwidentdefs{\\{{\nwixident{cycle{\_}tangent{\_}i}}{cycle:untangent:uni}}}\nwidentuses{\\{{\nwixident{ex}}{ex}}\\{{\nwixident{numeric}}{numeric}}}\nwindexuse{\nwixident{ex}}{ex}{NWppJ6t-3fVAGh-7}\nwindexuse{\nwixident{numeric}}{numeric}{NWppJ6t-3fVAGh-7}\nwendcode{}\nwbegindocs{1214}\nwdocspar
\nwenddocs{}\nwbegindocs{1215}\nwdocspar
\nwenddocs{}\nwbegincode{1216}\sublabel{NWppJ6t-3fVAGh-8}\nwmargintag{{\nwtagstyle{}\subpageref{NWppJ6t-3fVAGh-8}}}\moddef{add cycle relations~{\nwtagstyle{}\subpageref{NWppJ6t-3fVAGh-1}}}\plusendmoddef\Rm{}\nwstartdeflinemarkup\nwusesondefline{\\{NWppJ6t-32jW6L-1}}\nwprevnextdefs{NWppJ6t-3fVAGh-7}{NWppJ6t-3fVAGh-9}\nwenddeflinemarkup
{\bf{}ex} {\it{}cycle\_different}({\bf{}const} {\bf{}ex} & {\it{}C1}, {\bf{}const} {\bf{}ex} & {\it{}C2}, {\bf{}const} {\bf{}ex} & {\it{}pr})
{\nwlbrace}
    {\bf{}return} {\bf{}lst}{\nwlbrace}({\bf{}ex}){\bf{}lst}{\nwlbrace}{\it{}ex\_to}\begin{math}<\end{math}{\bf{}cycle}\begin{math}>\end{math}({\it{}C1}).{\it{}is\_equal}({\it{}ex\_to}\begin{math}<\end{math}{\bf{}basic}\begin{math}>\end{math}({\it{}C2}), {\bf{}true})? 0: 1{\nwrbrace}{\nwrbrace};
{\nwrbrace}
\nwindexdefn{\nwixident{cycle{\_}different}}{cycle:undifferent}{NWppJ6t-3fVAGh-8}\eatline
\nwused{\\{NWppJ6t-32jW6L-1}}\nwidentdefs{\\{{\nwixident{cycle{\_}different}}{cycle:undifferent}}}\nwidentuses{\\{{\nwixident{ex}}{ex}}}\nwindexuse{\nwixident{ex}}{ex}{NWppJ6t-3fVAGh-8}\nwendcode{}\nwbegindocs{1217}\nwdocspar
\nwenddocs{}\nwbegindocs{1218}If the cycle product has imaginary part we return the false
statement. For a real cycle product we check its sign.
\nwenddocs{}\nwbegincode{1219}\sublabel{NWppJ6t-3fVAGh-9}\nwmargintag{{\nwtagstyle{}\subpageref{NWppJ6t-3fVAGh-9}}}\moddef{add cycle relations~{\nwtagstyle{}\subpageref{NWppJ6t-3fVAGh-1}}}\plusendmoddef\Rm{}\nwstartdeflinemarkup\nwusesondefline{\\{NWppJ6t-32jW6L-1}}\nwprevnextdefs{NWppJ6t-3fVAGh-8}{NWppJ6t-3fVAGh-A}\nwenddeflinemarkup
{\bf{}ex} {\it{}product\_sign}({\bf{}const} {\bf{}ex} & {\it{}C1}, {\bf{}const} {\bf{}ex} & {\it{}C2}, {\bf{}const} {\bf{}ex} & {\it{}pr})
{\nwlbrace}
    {\bf{}if} ({\it{}is\_less\_than\_epsilon}({\it{}ex\_to}\begin{math}<\end{math}{\bf{}cycle}\begin{math}>\end{math}({\it{}C1}).{\it{}cycle\_product}({\it{}ex\_to}\begin{math}<\end{math}{\bf{}cycle}\begin{math}>\end{math}({\it{}C1})).{\it{}evalf}().{\it{}imag\_part}()))
        {\bf{}return} {\bf{}lst}{\nwlbrace}({\bf{}ex}){\bf{}lst}{\nwlbrace}{\it{}pr}\begin{math}\ast\end{math}({\it{}ex\_to}\begin{math}<\end{math}{\bf{}cycle}\begin{math}>\end{math}({\it{}C1}).{\it{}cycle\_product}({\it{}ex\_to}\begin{math}<\end{math}{\bf{}cycle}\begin{math}>\end{math}({\it{}C1})).{\it{}evalf}().{\it{}real\_part}() - {\it{}epsilon}) \begin{math}<\end{math}0{\nwrbrace}{\nwrbrace};
    {\bf{}else}
        {\bf{}return} {\bf{}lst}{\nwlbrace}({\bf{}ex}){\bf{}lst}{\nwlbrace}{\bf{}numeric}(1) \begin{math}<\end{math}0{\nwrbrace}{\nwrbrace};
{\nwrbrace}
\nwindexdefn{\nwixident{product{\_}sign}}{product:unsign}{NWppJ6t-3fVAGh-9}\eatline
\nwused{\\{NWppJ6t-32jW6L-1}}\nwidentdefs{\\{{\nwixident{product{\_}sign}}{product:unsign}}}\nwidentuses{\\{{\nwixident{epsilon}}{epsilon}}\\{{\nwixident{evalf}}{evalf}}\\{{\nwixident{ex}}{ex}}\\{{\nwixident{is{\_}less{\_}than{\_}epsilon}}{is:unless:unthan:unepsilon}}\\{{\nwixident{numeric}}{numeric}}}\nwindexuse{\nwixident{epsilon}}{epsilon}{NWppJ6t-3fVAGh-9}\nwindexuse{\nwixident{evalf}}{evalf}{NWppJ6t-3fVAGh-9}\nwindexuse{\nwixident{ex}}{ex}{NWppJ6t-3fVAGh-9}\nwindexuse{\nwixident{is{\_}less{\_}than{\_}epsilon}}{is:unless:unthan:unepsilon}{NWppJ6t-3fVAGh-9}\nwindexuse{\nwixident{numeric}}{numeric}{NWppJ6t-3fVAGh-9}\nwendcode{}\nwbegindocs{1220}\nwdocspar
\nwenddocs{}\nwbegindocs{1221}Now we define the relation between cycles to ``intersect with
certain angle'' (but the ``intersection'' may be imaginary). If cycles
are intersecting indeed then the value of {\Tt{}\Rm{}{\it{}pr}\nwendquote} is the cosine of the
angle.
\nwenddocs{}\nwbegincode{1222}\sublabel{NWppJ6t-3fVAGh-A}\nwmargintag{{\nwtagstyle{}\subpageref{NWppJ6t-3fVAGh-A}}}\moddef{add cycle relations~{\nwtagstyle{}\subpageref{NWppJ6t-3fVAGh-1}}}\plusendmoddef\Rm{}\nwstartdeflinemarkup\nwusesondefline{\\{NWppJ6t-32jW6L-1}}\nwprevnextdefs{NWppJ6t-3fVAGh-9}{NWppJ6t-3fVAGh-B}\nwenddeflinemarkup
{\bf{}ex} {\it{}cycle\_angle}({\bf{}const} {\bf{}ex} & {\it{}C1}, {\bf{}const} {\bf{}ex} & {\it{}C2}, {\bf{}const} {\bf{}ex} & {\it{}pr})
{\nwlbrace}
    {\bf{}return} {\bf{}lst}{\nwlbrace}{\bf{}lst}{\nwlbrace}{\it{}ex\_to}\begin{math}<\end{math}{\bf{}cycle}\begin{math}>\end{math}({\it{}C1}).{\it{}cycle\_product}({\it{}ex\_to}\begin{math}<\end{math}{\bf{}cycle}\begin{math}>\end{math}({\it{}C2}).{\it{}normalize\_norm}())-{\it{}pr}\begin{math}\equiv\end{math}0,
                {\it{}ex\_to}\begin{math}<\end{math}{\bf{}cycle}\begin{math}>\end{math}({\it{}C1}).{\it{}cycle\_product}({\it{}ex\_to}\begin{math}<\end{math}{\bf{}cycle}\begin{math}>\end{math}({\it{}C1}))+{\bf{}numeric}(1)\begin{math}\equiv\end{math}0{\nwrbrace},
            {\bf{}lst}{\nwlbrace}{\it{}ex\_to}\begin{math}<\end{math}{\bf{}cycle}\begin{math}>\end{math}({\it{}C1}).{\it{}cycle\_product}({\it{}ex\_to}\begin{math}<\end{math}{\bf{}cycle}\begin{math}>\end{math}({\it{}C2}).{\it{}normalize\_norm}())-{\it{}pr}\begin{math}\equiv\end{math}0,
                    {\it{}ex\_to}\begin{math}<\end{math}{\bf{}cycle}\begin{math}>\end{math}({\it{}C1}).{\it{}cycle\_product}({\it{}ex\_to}\begin{math}<\end{math}{\bf{}cycle}\begin{math}>\end{math}({\it{}C1}))-{\bf{}numeric}(1)\begin{math}\equiv\end{math}0{\nwrbrace}{\nwrbrace};
{\nwrbrace}
\nwindexdefn{\nwixident{cycle{\_}angle}}{cycle:unangle}{NWppJ6t-3fVAGh-A}\eatline
\nwused{\\{NWppJ6t-32jW6L-1}}\nwidentdefs{\\{{\nwixident{cycle{\_}angle}}{cycle:unangle}}}\nwidentuses{\\{{\nwixident{ex}}{ex}}\\{{\nwixident{numeric}}{numeric}}}\nwindexuse{\nwixident{ex}}{ex}{NWppJ6t-3fVAGh-A}\nwindexuse{\nwixident{numeric}}{numeric}{NWppJ6t-3fVAGh-A}\nwendcode{}\nwbegindocs{1223}\nwdocspar
\nwenddocs{}\nwbegindocs{1224}The next relation defines tangential distance between cycles.
\nwenddocs{}\nwbegincode{1225}\sublabel{NWppJ6t-3fVAGh-B}\nwmargintag{{\nwtagstyle{}\subpageref{NWppJ6t-3fVAGh-B}}}\moddef{add cycle relations~{\nwtagstyle{}\subpageref{NWppJ6t-3fVAGh-1}}}\plusendmoddef\Rm{}\nwstartdeflinemarkup\nwusesondefline{\\{NWppJ6t-32jW6L-1}}\nwprevnextdefs{NWppJ6t-3fVAGh-A}{NWppJ6t-3fVAGh-C}\nwenddeflinemarkup
{\bf{}ex} {\it{}steiner\_power}({\bf{}const} {\bf{}ex} & {\it{}C1}, {\bf{}const} {\bf{}ex} & {\it{}C2}, {\bf{}const} {\bf{}ex} & {\it{}pr})
{\nwlbrace}
    {\bf{}cycle} {\it{}C}={\it{}ex\_to}\begin{math}<\end{math}{\bf{}cycle}\begin{math}>\end{math}({\it{}C2}).{\it{}normalize}();
    {\bf{}return} {\bf{}lst}{\nwlbrace}{\bf{}lst}{\nwlbrace}{\it{}ex\_to}\begin{math}<\end{math}{\bf{}cycle}\begin{math}>\end{math}({\it{}C1}).{\it{}cycle\_product}({\it{}C})+{\it{}sqrt}({\it{}abs}({\it{}C}.{\it{}cycle\_product}({\it{}C})))
                -{\it{}pr}\begin{math}\ast\end{math}{\it{}ex\_to}\begin{math}<\end{math}{\bf{}cycle}\begin{math}>\end{math}({\it{}C1}).{\it{}get\_k}()\begin{math}\equiv\end{math}0,
                {\it{}ex\_to}\begin{math}<\end{math}{\bf{}cycle}\begin{math}>\end{math}({\it{}C1}).{\it{}cycle\_product}({\it{}ex\_to}\begin{math}<\end{math}{\bf{}cycle}\begin{math}>\end{math}({\it{}C1}))+{\bf{}numeric}(1)\begin{math}\equiv\end{math}0{\nwrbrace},
            {\bf{}lst}{\nwlbrace}{\it{}ex\_to}\begin{math}<\end{math}{\bf{}cycle}\begin{math}>\end{math}({\it{}C1}).{\it{}cycle\_product}({\it{}C})+{\it{}sqrt}({\it{}abs}({\it{}C}.{\it{}cycle\_product}({\it{}C})))
                        -{\it{}pr}\begin{math}\ast\end{math}{\it{}ex\_to}\begin{math}<\end{math}{\bf{}cycle}\begin{math}>\end{math}({\it{}C1}).{\it{}get\_k}()\begin{math}\equiv\end{math}0,
                    {\it{}ex\_to}\begin{math}<\end{math}{\bf{}cycle}\begin{math}>\end{math}({\it{}C1}).{\it{}cycle\_product}({\it{}ex\_to}\begin{math}<\end{math}{\bf{}cycle}\begin{math}>\end{math}({\it{}C1}))-{\bf{}numeric}(1)\begin{math}\equiv\end{math}0{\nwrbrace}{\nwrbrace};
{\nwrbrace}
\nwindexdefn{\nwixident{steiner{\_}power}}{steiner:unpower}{NWppJ6t-3fVAGh-B}\eatline
\nwused{\\{NWppJ6t-32jW6L-1}}\nwidentdefs{\\{{\nwixident{steiner{\_}power}}{steiner:unpower}}}\nwidentuses{\\{{\nwixident{ex}}{ex}}\\{{\nwixident{numeric}}{numeric}}}\nwindexuse{\nwixident{ex}}{ex}{NWppJ6t-3fVAGh-B}\nwindexuse{\nwixident{numeric}}{numeric}{NWppJ6t-3fVAGh-B}\nwendcode{}\nwbegindocs{1226}\nwdocspar
\nwenddocs{}\nwbegindocs{1227}Cross tangential distance is different by a sign of one term.
\nwenddocs{}\nwbegincode{1228}\sublabel{NWppJ6t-3fVAGh-C}\nwmargintag{{\nwtagstyle{}\subpageref{NWppJ6t-3fVAGh-C}}}\moddef{add cycle relations~{\nwtagstyle{}\subpageref{NWppJ6t-3fVAGh-1}}}\plusendmoddef\Rm{}\nwstartdeflinemarkup\nwusesondefline{\\{NWppJ6t-32jW6L-1}}\nwprevnextdefs{NWppJ6t-3fVAGh-B}{NWppJ6t-3fVAGh-D}\nwenddeflinemarkup
{\bf{}ex} {\it{}cycle\_cross\_t\_distance}({\bf{}const} {\bf{}ex} & {\it{}C1}, {\bf{}const} {\bf{}ex} & {\it{}C2}, {\bf{}const} {\bf{}ex} & {\it{}pr})
{\nwlbrace}
    {\bf{}cycle} {\it{}C}={\it{}ex\_to}\begin{math}<\end{math}{\bf{}cycle}\begin{math}>\end{math}({\it{}C2}).{\it{}normalize}();
    {\bf{}return} {\bf{}lst}{\nwlbrace}{\bf{}lst}{\nwlbrace}{\it{}ex\_to}\begin{math}<\end{math}{\bf{}cycle}\begin{math}>\end{math}({\it{}C1}).{\it{}cycle\_product}({\it{}C})-{\it{}sqrt}({\it{}abs}({\it{}C}.{\it{}cycle\_product}({\it{}C})))
                -{\it{}pow}({\it{}pr},2)\begin{math}\ast\end{math}{\it{}ex\_to}\begin{math}<\end{math}{\bf{}cycle}\begin{math}>\end{math}({\it{}C1}).{\it{}get\_k}()\begin{math}\equiv\end{math}0,
                {\it{}ex\_to}\begin{math}<\end{math}{\bf{}cycle}\begin{math}>\end{math}({\it{}C1}).{\it{}cycle\_product}({\it{}ex\_to}\begin{math}<\end{math}{\bf{}cycle}\begin{math}>\end{math}({\it{}C1}))+{\bf{}numeric}(1)\begin{math}\equiv\end{math}0{\nwrbrace},
            {\bf{}lst}{\nwlbrace}{\it{}ex\_to}\begin{math}<\end{math}{\bf{}cycle}\begin{math}>\end{math}({\it{}C1}).{\it{}cycle\_product}({\it{}C})-{\it{}sqrt}({\it{}abs}({\it{}C}.{\it{}cycle\_product}({\it{}C})))
                -{\it{}pow}({\it{}pr},2)\begin{math}\ast\end{math}{\it{}ex\_to}\begin{math}<\end{math}{\bf{}cycle}\begin{math}>\end{math}({\it{}C1}).{\it{}get\_k}()\begin{math}\equiv\end{math}0,
                {\it{}ex\_to}\begin{math}<\end{math}{\bf{}cycle}\begin{math}>\end{math}({\it{}C1}).{\it{}cycle\_product}({\it{}ex\_to}\begin{math}<\end{math}{\bf{}cycle}\begin{math}>\end{math}({\it{}C1}))-{\bf{}numeric}(1)\begin{math}\equiv\end{math}0{\nwrbrace}{\nwrbrace};
{\nwrbrace}
\nwindexdefn{\nwixident{cycle{\_}cross{\_}t{\_}distance}}{cycle:uncross:unt:undistance}{NWppJ6t-3fVAGh-C}\eatline
\nwused{\\{NWppJ6t-32jW6L-1}}\nwidentdefs{\\{{\nwixident{cycle{\_}cross{\_}t{\_}distance}}{cycle:uncross:unt:undistance}}}\nwidentuses{\\{{\nwixident{ex}}{ex}}\\{{\nwixident{numeric}}{numeric}}}\nwindexuse{\nwixident{ex}}{ex}{NWppJ6t-3fVAGh-C}\nwindexuse{\nwixident{numeric}}{numeric}{NWppJ6t-3fVAGh-C}\nwendcode{}\nwbegindocs{1229}\nwdocspar
\nwenddocs{}\nwbegindocs{1230}Check that all coefficients of the first cycle are real.
\nwenddocs{}\nwbegincode{1231}\sublabel{NWppJ6t-3fVAGh-D}\nwmargintag{{\nwtagstyle{}\subpageref{NWppJ6t-3fVAGh-D}}}\moddef{add cycle relations~{\nwtagstyle{}\subpageref{NWppJ6t-3fVAGh-1}}}\plusendmoddef\Rm{}\nwstartdeflinemarkup\nwusesondefline{\\{NWppJ6t-32jW6L-1}}\nwprevnextdefs{NWppJ6t-3fVAGh-C}{NWppJ6t-3fVAGh-E}\nwenddeflinemarkup
{\bf{}ex} {\it{}coefficients\_are\_real}({\bf{}const} {\bf{}ex} & {\it{}C1}, {\bf{}const} {\bf{}ex} & {\it{}C2}, {\bf{}const} {\bf{}ex} & {\it{}pr})
{\nwlbrace}
    {\bf{}cycle} {\it{}C}={\it{}ex\_to}\begin{math}<\end{math}{\bf{}cycle}\begin{math}>\end{math}({\it{}ex\_to}\begin{math}<\end{math}{\bf{}cycle}\begin{math}>\end{math}({\it{}C1}.{\it{}evalf}()).{\it{}imag\_part}());
    {\bf{}if} (\begin{math}\neg\end{math} ({\it{}is\_less\_than\_epsilon}({\it{}C}.{\it{}get\_k}()) \begin{math}\wedge\end{math} {\it{}is\_less\_than\_epsilon}({\it{}C}.{\it{}get\_m}())))
        {\bf{}return} {\bf{}lst}{\nwlbrace}({\bf{}ex}){\bf{}lst}{\nwlbrace}0{\nwrbrace}{\nwrbrace};
    {\bf{}for} ({\bf{}int} {\it{}i}=0; {\it{}i} \begin{math}<\end{math} {\it{}ex\_to}\begin{math}<\end{math}{\bf{}cycle}\begin{math}>\end{math}({\it{}C1}).{\it{}get\_dim}(); \protect\PP{\it{}i})
        {\bf{}if} (\begin{math}\neg\end{math} {\it{}is\_less\_than\_epsilon}({\it{}C}.{\it{}get\_l}({\it{}i})))
            {\bf{}return} {\bf{}lst}{\nwlbrace}({\bf{}ex}){\bf{}lst}{\nwlbrace}0{\nwrbrace}{\nwrbrace};

    {\bf{}return} {\bf{}lst}{\nwlbrace}({\bf{}ex}){\bf{}lst}{\nwlbrace}1{\nwrbrace}{\nwrbrace};
{\nwrbrace}
\nwindexdefn{\nwixident{coefficients{\_}are{\_}real}}{coefficients:unare:unreal}{NWppJ6t-3fVAGh-D}\eatline
\nwused{\\{NWppJ6t-32jW6L-1}}\nwidentdefs{\\{{\nwixident{coefficients{\_}are{\_}real}}{coefficients:unare:unreal}}}\nwidentuses{\\{{\nwixident{evalf}}{evalf}}\\{{\nwixident{ex}}{ex}}\\{{\nwixident{get{\_}dim()}}{get:undim()}}\\{{\nwixident{is{\_}less{\_}than{\_}epsilon}}{is:unless:unthan:unepsilon}}}\nwindexuse{\nwixident{evalf}}{evalf}{NWppJ6t-3fVAGh-D}\nwindexuse{\nwixident{ex}}{ex}{NWppJ6t-3fVAGh-D}\nwindexuse{\nwixident{get{\_}dim()}}{get:undim()}{NWppJ6t-3fVAGh-D}\nwindexuse{\nwixident{is{\_}less{\_}than{\_}epsilon}}{is:unless:unthan:unepsilon}{NWppJ6t-3fVAGh-D}\nwendcode{}\nwbegindocs{1232}\nwdocspar
\nwenddocs{}\nwbegindocs{1233}\nwdocspar
\subsubsection{Measured quantities}
\label{sec:measured-quantities}

\nwenddocs{}\nwbegindocs{1234}This function measures relative powers of two cycles, which turn to
be their cycle product for norm-normalised vectors.
\nwenddocs{}\nwbegincode{1235}\sublabel{NWppJ6t-3fVAGh-E}\nwmargintag{{\nwtagstyle{}\subpageref{NWppJ6t-3fVAGh-E}}}\moddef{add cycle relations~{\nwtagstyle{}\subpageref{NWppJ6t-3fVAGh-1}}}\plusendmoddef\Rm{}\nwstartdeflinemarkup\nwusesondefline{\\{NWppJ6t-32jW6L-1}}\nwprevnextdefs{NWppJ6t-3fVAGh-D}{NWppJ6t-3fVAGh-F}\nwenddeflinemarkup
{\bf{}ex} {\it{}angle\_is}({\bf{}const} {\bf{}ex} & {\it{}C1}, {\bf{}const} {\bf{}ex} & {\it{}C2}, {\bf{}const} {\bf{}ex} & {\it{}pr})
{\nwlbrace}
    {\bf{}return} {\bf{}lst}{\nwlbrace}({\bf{}ex}){\bf{}lst}{\nwlbrace}{\it{}ex\_to}\begin{math}<\end{math}{\bf{}cycle}\begin{math}>\end{math}({\it{}C1}).{\it{}normalize\_norm}().{\it{}cycle\_product}({\it{}ex\_to}\begin{math}<\end{math}{\bf{}cycle}\begin{math}>\end{math}({\it{}C2}).{\it{}normalize\_norm}()){\nwrbrace}{\nwrbrace};
{\nwrbrace}
\nwindexdefn{\nwixident{angle{\_}is}}{angle:unis}{NWppJ6t-3fVAGh-E}\eatline
\nwused{\\{NWppJ6t-32jW6L-1}}\nwidentdefs{\\{{\nwixident{angle{\_}is}}{angle:unis}}}\nwidentuses{\\{{\nwixident{ex}}{ex}}}\nwindexuse{\nwixident{ex}}{ex}{NWppJ6t-3fVAGh-E}\nwendcode{}\nwbegindocs{1236}\nwdocspar
\nwenddocs{}\nwbegindocs{1237}This function measures relative powers of two cycles, which turn to
be their cycle product for \(k\)-normalised vectors.
\nwenddocs{}\nwbegincode{1238}\sublabel{NWppJ6t-3fVAGh-F}\nwmargintag{{\nwtagstyle{}\subpageref{NWppJ6t-3fVAGh-F}}}\moddef{add cycle relations~{\nwtagstyle{}\subpageref{NWppJ6t-3fVAGh-1}}}\plusendmoddef\Rm{}\nwstartdeflinemarkup\nwusesondefline{\\{NWppJ6t-32jW6L-1}}\nwprevnextdefs{NWppJ6t-3fVAGh-E}{NWppJ6t-3fVAGh-G}\nwenddeflinemarkup
{\bf{}ex} {\it{}power\_is}({\bf{}const} {\bf{}ex} & {\it{}C1}, {\bf{}const} {\bf{}ex} & {\it{}C2}, {\bf{}const} {\bf{}ex} & {\it{}pr})
{\nwlbrace}
    {\bf{}cycle} {\it{}Ca}={\it{}ex\_to}\begin{math}<\end{math}{\bf{}cycle}\begin{math}>\end{math}({\it{}C1}).{\it{}normalize}(), {\it{}Cb}={\it{}ex\_to}\begin{math}<\end{math}{\bf{}cycle}\begin{math}>\end{math}({\it{}C2}).{\it{}normalize}();

    {\bf{}return} {\bf{}lst}{\nwlbrace}({\bf{}ex}){\bf{}lst}{\nwlbrace}{\it{}Ca}.{\it{}cycle\_product}({\it{}Cb})+{\it{}pr}\begin{math}\ast\end{math}{\it{}sqrt}({\it{}abs}({\it{}Ca}.{\it{}cycle\_product}({\it{}Ca})\begin{math}\ast\end{math}{\it{}Cb}.{\it{}cycle\_product}({\it{}Cb}))){\nwrbrace}{\nwrbrace};
{\nwrbrace}
\nwindexdefn{\nwixident{power{\_}is}}{power:unis}{NWppJ6t-3fVAGh-F}\eatline
\nwused{\\{NWppJ6t-32jW6L-1}}\nwidentdefs{\\{{\nwixident{power{\_}is}}{power:unis}}}\nwidentuses{\\{{\nwixident{ex}}{ex}}}\nwindexuse{\nwixident{ex}}{ex}{NWppJ6t-3fVAGh-F}\nwendcode{}\nwbegindocs{1239}\nwdocspar
\nwenddocs{}\nwbegindocs{1240}\nwdocspar
\nwenddocs{}\nwbegincode{1241}\sublabel{NWppJ6t-3fVAGh-G}\nwmargintag{{\nwtagstyle{}\subpageref{NWppJ6t-3fVAGh-G}}}\moddef{add cycle relations~{\nwtagstyle{}\subpageref{NWppJ6t-3fVAGh-1}}}\plusendmoddef\Rm{}\nwstartdeflinemarkup\nwusesondefline{\\{NWppJ6t-32jW6L-1}}\nwprevnextdefs{NWppJ6t-3fVAGh-F}{NWppJ6t-3fVAGh-H}\nwenddeflinemarkup
{\bf{}ex} {\it{}cycle\_moebius}({\bf{}const} {\bf{}ex} & {\it{}C1}, {\bf{}const} {\bf{}ex} & {\it{}C2}, {\bf{}const} {\bf{}ex} & {\it{}pr})
{\nwlbrace}
    {\bf{}return} {\bf{}lst}{\nwlbrace}({\bf{}ex}){\bf{}lst}{\nwlbrace}{\it{}ex\_to}\begin{math}<\end{math}{\bf{}cycle}\begin{math}>\end{math}({\it{}C2}).{\it{}matrix\_similarity}({\it{}pr}.{\it{}op}(0),{\it{}pr}.{\it{}op}(1),{\it{}pr}.{\it{}op}(2),{\it{}pr}.{\it{}op}(3)){\nwrbrace}{\nwrbrace};
{\nwrbrace}
\nwindexdefn{\nwixident{cycle{\_}moebius}}{cycle:unmoebius}{NWppJ6t-3fVAGh-G}\eatline
\nwused{\\{NWppJ6t-32jW6L-1}}\nwidentdefs{\\{{\nwixident{cycle{\_}moebius}}{cycle:unmoebius}}}\nwidentuses{\\{{\nwixident{ex}}{ex}}\\{{\nwixident{op}}{op}}}\nwindexuse{\nwixident{ex}}{ex}{NWppJ6t-3fVAGh-G}\nwindexuse{\nwixident{op}}{op}{NWppJ6t-3fVAGh-G}\nwendcode{}\nwbegindocs{1242}\nwdocspar
\nwenddocs{}\nwbegindocs{1243}That relations works only for real matrices, thus we start from the relevant checks.
\nwenddocs{}\nwbegincode{1244}\sublabel{NWppJ6t-3fVAGh-H}\nwmargintag{{\nwtagstyle{}\subpageref{NWppJ6t-3fVAGh-H}}}\moddef{add cycle relations~{\nwtagstyle{}\subpageref{NWppJ6t-3fVAGh-1}}}\plusendmoddef\Rm{}\nwstartdeflinemarkup\nwusesondefline{\\{NWppJ6t-32jW6L-1}}\nwprevnextdefs{NWppJ6t-3fVAGh-G}{NWppJ6t-3fVAGh-I}\nwenddeflinemarkup
{\bf{}cycle\_relation} {\it{}sl2\_transform}({\bf{}const} {\bf{}ex} & {\it{}key}, {\bf{}bool} {\it{}cm}, {\bf{}const} {\bf{}ex} & {\bf{}matrix}) {\nwlbrace}
    {\bf{}if} ({\it{}is\_a}\begin{math}<\end{math}{\bf{}lst}\begin{math}>\end{math}({\bf{}matrix}) \begin{math}\wedge\end{math} {\bf{}matrix}.{\it{}op}(0).{\it{}info}({\it{}info\_flags}::{\it{}real}) \begin{math}\wedge\end{math} {\bf{}matrix}.{\it{}op}(1).{\it{}info}({\it{}info\_flags}::{\it{}real})
        \begin{math}\wedge\end{math} {\bf{}matrix}.{\it{}op}(2).{\it{}info}({\it{}info\_flags}::{\it{}real}) \begin{math}\wedge\end{math} {\bf{}matrix}.{\it{}op}(3).{\it{}info}({\it{}info\_flags}::{\it{}real}))
            {\bf{}return} {\bf{}cycle\_relation}({\it{}key}, {\it{}cycle\_sl2}, {\it{}cm}, {\bf{}matrix});
    {\bf{}else}
        {\bf{}throw}({\it{}std}::{\it{}invalid\_argument}({\tt{}"sl2\_transform(): shall be applied only with a matrix having"}
                                    {\tt{}" real entries"}));
{\nwrbrace}
\nwindexdefn{\nwixident{sl2{\_}transform}}{sl2:untransform}{NWppJ6t-3fVAGh-H}\eatline
\nwused{\\{NWppJ6t-32jW6L-1}}\nwidentdefs{\\{{\nwixident{sl2{\_}transform}}{sl2:untransform}}}\nwidentuses{\\{{\nwixident{cycle{\_}relation}}{cycle:unrelation}}\\{{\nwixident{cycle{\_}sl2}}{cycle:unsl2}}\\{{\nwixident{ex}}{ex}}\\{{\nwixident{info}}{info}}\\{{\nwixident{key}}{key}}\\{{\nwixident{op}}{op}}}\nwindexuse{\nwixident{cycle{\_}relation}}{cycle:unrelation}{NWppJ6t-3fVAGh-H}\nwindexuse{\nwixident{cycle{\_}sl2}}{cycle:unsl2}{NWppJ6t-3fVAGh-H}\nwindexuse{\nwixident{ex}}{ex}{NWppJ6t-3fVAGh-H}\nwindexuse{\nwixident{info}}{info}{NWppJ6t-3fVAGh-H}\nwindexuse{\nwixident{key}}{key}{NWppJ6t-3fVAGh-H}\nwindexuse{\nwixident{op}}{op}{NWppJ6t-3fVAGh-H}\nwendcode{}\nwbegindocs{1245}\nwdocspar
\nwenddocs{}\nwbegindocs{1246}That relations works only in two dimensions, thus we start from the relevant checks.
\nwenddocs{}\nwbegincode{1247}\sublabel{NWppJ6t-3fVAGh-I}\nwmargintag{{\nwtagstyle{}\subpageref{NWppJ6t-3fVAGh-I}}}\moddef{add cycle relations~{\nwtagstyle{}\subpageref{NWppJ6t-3fVAGh-1}}}\plusendmoddef\Rm{}\nwstartdeflinemarkup\nwusesondefline{\\{NWppJ6t-32jW6L-1}}\nwprevnextdefs{NWppJ6t-3fVAGh-H}{\relax}\nwenddeflinemarkup
{\bf{}ex} {\it{}cycle\_sl2}({\bf{}const} {\bf{}ex} & {\it{}C1}, {\bf{}const} {\bf{}ex} & {\it{}C2}, {\bf{}const} {\bf{}ex} & {\it{}pr})
{\nwlbrace}
    {\bf{}if} ({\it{}ex\_to}\begin{math}<\end{math}{\bf{}cycle}\begin{math}>\end{math}({\it{}C2}).{\it{}get\_dim}() \begin{math}\equiv\end{math} 2)
        {\bf{}return} {\bf{}lst}{\nwlbrace}({\bf{}ex}){\bf{}lst}{\nwlbrace}{\it{}ex\_to}\begin{math}<\end{math}{\bf{}cycle}\begin{math}>\end{math}({\it{}C2}).{\it{}sl2\_similarity}({\it{}pr}.{\it{}op}(0),{\it{}pr}.{\it{}op}(1),{\it{}pr}.{\it{}op}(2),{\it{}pr}.{\it{}op}(3),
                                                       {\it{}ex\_to}\begin{math}<\end{math}{\bf{}cycle}\begin{math}>\end{math}({\it{}C2}).{\it{}get\_unit}()){\nwrbrace}{\nwrbrace};
    {\bf{}else}
        {\bf{}throw}({\it{}std}::{\it{}invalid\_argument}({\tt{}"cycle\_sl2(): shall be applied only in two dimensions"}));
{\nwrbrace}
\nwindexdefn{\nwixident{cycle{\_}sl2}}{cycle:unsl2}{NWppJ6t-3fVAGh-I}\eatline
\nwused{\\{NWppJ6t-32jW6L-1}}\nwidentdefs{\\{{\nwixident{cycle{\_}sl2}}{cycle:unsl2}}}\nwidentuses{\\{{\nwixident{ex}}{ex}}\\{{\nwixident{get{\_}dim()}}{get:undim()}}\\{{\nwixident{op}}{op}}}\nwindexuse{\nwixident{ex}}{ex}{NWppJ6t-3fVAGh-I}\nwindexuse{\nwixident{get{\_}dim()}}{get:undim()}{NWppJ6t-3fVAGh-I}\nwindexuse{\nwixident{op}}{op}{NWppJ6t-3fVAGh-I}\nwendcode{}\nwbegindocs{1248}\nwdocspar
\nwenddocs{}\nwbegindocs{1249}\nwdocspar
\subsection{Additional functions}
\label{sec:additional-functions}

\nwenddocs{}\nwbegindocs{1250}Equality of {\Tt{}\Rm{}{\bf{}cycle}\nwendquote}s.
\nwenddocs{}\nwbegincode{1251}\sublabel{NWppJ6t-3LImg1-1}\nwmargintag{{\nwtagstyle{}\subpageref{NWppJ6t-3LImg1-1}}}\moddef{addional functions~{\nwtagstyle{}\subpageref{NWppJ6t-3LImg1-1}}}\endmoddef\Rm{}\nwstartdeflinemarkup\nwusesondefline{\\{NWppJ6t-32jW6L-1}}\nwprevnextdefs{\relax}{NWppJ6t-3LImg1-2}\nwenddeflinemarkup
{\bf{}bool} {\it{}is\_almost\_equal}({\bf{}const} {\bf{}ex} & {\it{}A}, {\bf{}const} {\bf{}ex} & {\it{}B})
{\nwlbrace}
    {\bf{}if} (({\it{}not} {\it{}is\_a}\begin{math}<\end{math}{\bf{}cycle}\begin{math}>\end{math}({\it{}A})) \begin{math}\vee\end{math} ({\it{}not} {\it{}is\_a}\begin{math}<\end{math}{\bf{}cycle}\begin{math}>\end{math}({\it{}B})))
        {\bf{}return} {\bf{}false};

    {\bf{}const} {\bf{}cycle} {\it{}C1} = {\it{}ex\_to}\begin{math}<\end{math}{\bf{}cycle}\begin{math}>\end{math}({\it{}A}),
        {\it{}C2} = {\it{}ex\_to}\begin{math}<\end{math}{\bf{}cycle}\begin{math}>\end{math}({\it{}B});
    {\bf{}ex} {\it{}factor}=0, {\it{}ofactor}=0;

    // Check that coefficients are scalar multiples of C2
    {\bf{}if} ({\it{}not} {\it{}is\_less\_than\_epsilon}(({\it{}C1}.{\it{}get\_m}()\begin{math}\ast\end{math}{\it{}C2}.{\it{}get\_k}()-{\it{}C2}.{\it{}get\_m}()\begin{math}\ast\end{math}{\it{}C1}.{\it{}get\_k}()).{\it{}normal}()))
        {\bf{}return} {\bf{}false};
    // Set up coefficients for proportionality
    {\bf{}if} ({\it{}C1}.{\it{}get\_k}().{\it{}normal}().{\it{}is\_zero}()) {\nwlbrace}
        {\it{}factor}={\it{}C1}.{\it{}get\_m}();
        {\it{}ofactor}={\it{}C2}.{\it{}get\_m}();
    {\nwrbrace} {\bf{}else} {\nwlbrace}
        {\it{}factor}={\it{}C1}.{\it{}get\_k}();
        {\it{}ofactor}={\it{}C2}.{\it{}get\_k}();
    {\nwrbrace}
\nwindexdefn{\nwixident{is{\_}almost{\_}equal}}{is:unalmost:unequal}{NWppJ6t-3LImg1-1}\eatline
\nwalsodefined{\\{NWppJ6t-3LImg1-2}\\{NWppJ6t-3LImg1-3}\\{NWppJ6t-3LImg1-4}\\{NWppJ6t-3LImg1-5}\\{NWppJ6t-3LImg1-6}\\{NWppJ6t-3LImg1-7}\\{NWppJ6t-3LImg1-8}\\{NWppJ6t-3LImg1-9}\\{NWppJ6t-3LImg1-A}}\nwused{\\{NWppJ6t-32jW6L-1}}\nwidentdefs{\\{{\nwixident{is{\_}almost{\_}equal}}{is:unalmost:unequal}}}\nwidentuses{\\{{\nwixident{ex}}{ex}}\\{{\nwixident{is{\_}less{\_}than{\_}epsilon}}{is:unless:unthan:unepsilon}}}\nwindexuse{\nwixident{ex}}{ex}{NWppJ6t-3LImg1-1}\nwindexuse{\nwixident{is{\_}less{\_}than{\_}epsilon}}{is:unless:unthan:unepsilon}{NWppJ6t-3LImg1-1}\nwendcode{}\nwbegindocs{1252}\nwdocspar
\nwenddocs{}\nwbegindocs{1253}Now we iterate through the coefficients of {\Tt{}\Rm{}{\it{}l}\nwendquote}.
\nwenddocs{}\nwbegincode{1254}\sublabel{NWppJ6t-3LImg1-2}\nwmargintag{{\nwtagstyle{}\subpageref{NWppJ6t-3LImg1-2}}}\moddef{addional functions~{\nwtagstyle{}\subpageref{NWppJ6t-3LImg1-1}}}\plusendmoddef\Rm{}\nwstartdeflinemarkup\nwusesondefline{\\{NWppJ6t-32jW6L-1}}\nwprevnextdefs{NWppJ6t-3LImg1-1}{NWppJ6t-3LImg1-3}\nwenddeflinemarkup
       {\bf{}for} ({\bf{}unsigned} {\bf{}int} {\it{}i}=0; {\it{}i}\begin{math}<\end{math}{\it{}C1}.{\it{}get\_l}().{\it{}nops}(); {\it{}i}\protect\PP)
        // search the the first non-zero coefficient
        {\bf{}if} ({\it{}factor}.{\it{}is\_zero}()) {\nwlbrace}
            {\it{}factor}={\it{}C1}.{\it{}get\_l}({\it{}i});
            {\it{}ofactor}={\it{}C2}.{\it{}get\_l}({\it{}i});
        {\nwrbrace} {\bf{}else}
            {\bf{}if} (\begin{math}\neg\end{math} {\it{}is\_less\_than\_epsilon}(({\it{}C1}.{\it{}get\_l}({\it{}i})\begin{math}\ast\end{math}{\it{}ofactor}-{\it{}C2}.{\it{}get\_l}({\it{}i})\begin{math}\ast\end{math}{\it{}factor}).{\it{}normal}()))
                {\bf{}return} {\bf{}false};
    {\bf{}return} {\bf{}true};
{\nwrbrace}

\nwused{\\{NWppJ6t-32jW6L-1}}\nwidentuses{\\{{\nwixident{is{\_}less{\_}than{\_}epsilon}}{is:unless:unthan:unepsilon}}\\{{\nwixident{nops}}{nops}}}\nwindexuse{\nwixident{is{\_}less{\_}than{\_}epsilon}}{is:unless:unthan:unepsilon}{NWppJ6t-3LImg1-2}\nwindexuse{\nwixident{nops}}{nops}{NWppJ6t-3LImg1-2}\nwendcode{}\nwbegindocs{1255}\nwdocspar
\nwenddocs{}\nwbegincode{1256}\sublabel{NWppJ6t-3LImg1-3}\nwmargintag{{\nwtagstyle{}\subpageref{NWppJ6t-3LImg1-3}}}\moddef{addional functions~{\nwtagstyle{}\subpageref{NWppJ6t-3LImg1-1}}}\plusendmoddef\Rm{}\nwstartdeflinemarkup\nwusesondefline{\\{NWppJ6t-32jW6L-1}}\nwprevnextdefs{NWppJ6t-3LImg1-2}{NWppJ6t-3LImg1-4}\nwenddeflinemarkup
{\bf{}ex} {\it{}midpoint\_constructor}()
{\nwlbrace}
    {\bf{}figure} {\it{}SF}={\it{}ex\_to}\begin{math}<\end{math}{\bf{}figure}\begin{math}>\end{math}(({\bf{}new} {\bf{}figure})\begin{math}\rightarrow\end{math}{\it{}setflag}({\it{}status\_flags}::{\it{}expanded}));

    {\bf{}ex} {\it{}v1}={\it{}SF}.{\it{}add\_cycle}({\bf{}cycle\_data}(),{\tt{}"variable000"});
    {\bf{}ex} {\it{}v2}={\it{}SF}.{\it{}add\_cycle}({\bf{}cycle\_data}(),{\tt{}"variable001"});
    {\bf{}ex} {\it{}v3}={\it{}SF}.{\it{}add\_cycle}({\bf{}cycle\_data}(),{\tt{}"variable002"});
\nwindexdefn{\nwixident{midpoint{\_}constructor}}{midpoint:unconstructor}{NWppJ6t-3LImg1-3}\eatline
\nwused{\\{NWppJ6t-32jW6L-1}}\nwidentdefs{\\{{\nwixident{midpoint{\_}constructor}}{midpoint:unconstructor}}}\nwidentuses{\\{{\nwixident{add{\_}cycle}}{add:uncycle}}\\{{\nwixident{cycle{\_}data}}{cycle:undata}}\\{{\nwixident{ex}}{ex}}\\{{\nwixident{figure}}{figure}}}\nwindexuse{\nwixident{add{\_}cycle}}{add:uncycle}{NWppJ6t-3LImg1-3}\nwindexuse{\nwixident{cycle{\_}data}}{cycle:undata}{NWppJ6t-3LImg1-3}\nwindexuse{\nwixident{ex}}{ex}{NWppJ6t-3LImg1-3}\nwindexuse{\nwixident{figure}}{figure}{NWppJ6t-3LImg1-3}\nwendcode{}\nwbegindocs{1257}\nwdocspar
\nwenddocs{}\nwbegindocs{1258}Join three point by an "interval" cycle.
\nwenddocs{}\nwbegincode{1259}\sublabel{NWppJ6t-3LImg1-4}\nwmargintag{{\nwtagstyle{}\subpageref{NWppJ6t-3LImg1-4}}}\moddef{addional functions~{\nwtagstyle{}\subpageref{NWppJ6t-3LImg1-1}}}\plusendmoddef\Rm{}\nwstartdeflinemarkup\nwusesondefline{\\{NWppJ6t-32jW6L-1}}\nwprevnextdefs{NWppJ6t-3LImg1-3}{NWppJ6t-3LImg1-5}\nwenddeflinemarkup
    {\bf{}ex} {\it{}v4}={\it{}SF}.{\it{}add\_cycle\_rel}({\bf{}lst}{\nwlbrace}{\bf{}cycle\_relation}({\it{}v1},{\it{}cycle\_orthogonal}),
                {\bf{}cycle\_relation}({\it{}v2},{\it{}cycle\_orthogonal}),
                {\bf{}cycle\_relation}({\it{}v3},{\it{}cycle\_orthogonal}){\nwrbrace},
        {\tt{}"v4"});

\nwused{\\{NWppJ6t-32jW6L-1}}\nwidentuses{\\{{\nwixident{add{\_}cycle{\_}rel}}{add:uncycle:unrel}}\\{{\nwixident{cycle{\_}orthogonal}}{cycle:unorthogonal}}\\{{\nwixident{cycle{\_}relation}}{cycle:unrelation}}\\{{\nwixident{ex}}{ex}}}\nwindexuse{\nwixident{add{\_}cycle{\_}rel}}{add:uncycle:unrel}{NWppJ6t-3LImg1-4}\nwindexuse{\nwixident{cycle{\_}orthogonal}}{cycle:unorthogonal}{NWppJ6t-3LImg1-4}\nwindexuse{\nwixident{cycle{\_}relation}}{cycle:unrelation}{NWppJ6t-3LImg1-4}\nwindexuse{\nwixident{ex}}{ex}{NWppJ6t-3LImg1-4}\nwendcode{}\nwbegindocs{1260} A cycle ortogonal to the above interval.
\nwenddocs{}\nwbegincode{1261}\sublabel{NWppJ6t-3LImg1-5}\nwmargintag{{\nwtagstyle{}\subpageref{NWppJ6t-3LImg1-5}}}\moddef{addional functions~{\nwtagstyle{}\subpageref{NWppJ6t-3LImg1-1}}}\plusendmoddef\Rm{}\nwstartdeflinemarkup\nwusesondefline{\\{NWppJ6t-32jW6L-1}}\nwprevnextdefs{NWppJ6t-3LImg1-4}{NWppJ6t-3LImg1-6}\nwenddeflinemarkup
    {\bf{}ex} {\it{}v5}={\it{}SF}.{\it{}add\_cycle\_rel}({\bf{}lst}{\nwlbrace}{\bf{}cycle\_relation}({\it{}v1},{\it{}cycle\_orthogonal}),
                {\bf{}cycle\_relation}({\it{}v2},{\it{}cycle\_orthogonal}),
                {\bf{}cycle\_relation}({\it{}v4},{\it{}cycle\_orthogonal}){\nwrbrace},
        {\tt{}"v5"});

\nwused{\\{NWppJ6t-32jW6L-1}}\nwidentuses{\\{{\nwixident{add{\_}cycle{\_}rel}}{add:uncycle:unrel}}\\{{\nwixident{cycle{\_}orthogonal}}{cycle:unorthogonal}}\\{{\nwixident{cycle{\_}relation}}{cycle:unrelation}}\\{{\nwixident{ex}}{ex}}}\nwindexuse{\nwixident{add{\_}cycle{\_}rel}}{add:uncycle:unrel}{NWppJ6t-3LImg1-5}\nwindexuse{\nwixident{cycle{\_}orthogonal}}{cycle:unorthogonal}{NWppJ6t-3LImg1-5}\nwindexuse{\nwixident{cycle{\_}relation}}{cycle:unrelation}{NWppJ6t-3LImg1-5}\nwindexuse{\nwixident{ex}}{ex}{NWppJ6t-3LImg1-5}\nwendcode{}\nwbegindocs{1262} The perpendicular to the interval and the cycle passing the midpoint.
\nwenddocs{}\nwbegincode{1263}\sublabel{NWppJ6t-3LImg1-6}\nwmargintag{{\nwtagstyle{}\subpageref{NWppJ6t-3LImg1-6}}}\moddef{addional functions~{\nwtagstyle{}\subpageref{NWppJ6t-3LImg1-1}}}\plusendmoddef\Rm{}\nwstartdeflinemarkup\nwusesondefline{\\{NWppJ6t-32jW6L-1}}\nwprevnextdefs{NWppJ6t-3LImg1-5}{NWppJ6t-3LImg1-7}\nwenddeflinemarkup
    {\bf{}ex} {\it{}v6}={\it{}SF}.{\it{}add\_cycle\_rel}({\bf{}lst}{\nwlbrace}{\bf{}cycle\_relation}({\it{}v3},{\it{}cycle\_orthogonal}),
                {\bf{}cycle\_relation}({\it{}v4},{\it{}cycle\_orthogonal}),
                {\bf{}cycle\_relation}({\it{}v5},{\it{}cycle\_orthogonal}){\nwrbrace},
        {\tt{}"v6"});

\nwused{\\{NWppJ6t-32jW6L-1}}\nwidentuses{\\{{\nwixident{add{\_}cycle{\_}rel}}{add:uncycle:unrel}}\\{{\nwixident{cycle{\_}orthogonal}}{cycle:unorthogonal}}\\{{\nwixident{cycle{\_}relation}}{cycle:unrelation}}\\{{\nwixident{ex}}{ex}}}\nwindexuse{\nwixident{add{\_}cycle{\_}rel}}{add:uncycle:unrel}{NWppJ6t-3LImg1-6}\nwindexuse{\nwixident{cycle{\_}orthogonal}}{cycle:unorthogonal}{NWppJ6t-3LImg1-6}\nwindexuse{\nwixident{cycle{\_}relation}}{cycle:unrelation}{NWppJ6t-3LImg1-6}\nwindexuse{\nwixident{ex}}{ex}{NWppJ6t-3LImg1-6}\nwendcode{}\nwbegindocs{1264}The mid point as the intersection point.
\nwenddocs{}\nwbegincode{1265}\sublabel{NWppJ6t-3LImg1-7}\nwmargintag{{\nwtagstyle{}\subpageref{NWppJ6t-3LImg1-7}}}\moddef{addional functions~{\nwtagstyle{}\subpageref{NWppJ6t-3LImg1-1}}}\plusendmoddef\Rm{}\nwstartdeflinemarkup\nwusesondefline{\\{NWppJ6t-32jW6L-1}}\nwprevnextdefs{NWppJ6t-3LImg1-6}{NWppJ6t-3LImg1-8}\nwenddeflinemarkup
    {\bf{}ex} {\it{}r}={\bf{}symbol}({\tt{}"result"});
 {\it{}SF}.{\it{}add\_cycle\_rel}({\bf{}lst}{\nwlbrace}{\bf{}cycle\_relation}({\it{}v4},{\it{}cycle\_orthogonal}),
             {\bf{}cycle\_relation}({\it{}v6},{\it{}cycle\_orthogonal}),
             {\bf{}cycle\_relation}({\it{}r},{\it{}cycle\_orthogonal},{\bf{}false}),
             {\bf{}cycle\_relation}({\it{}v3},{\it{}cycle\_adifferent}){\nwrbrace},
     {\it{}r});

    {\bf{}return} {\it{}SF};
{\nwrbrace}

\nwused{\\{NWppJ6t-32jW6L-1}}\nwidentuses{\\{{\nwixident{add{\_}cycle{\_}rel}}{add:uncycle:unrel}}\\{{\nwixident{cycle{\_}adifferent}}{cycle:unadifferent}}\\{{\nwixident{cycle{\_}orthogonal}}{cycle:unorthogonal}}\\{{\nwixident{cycle{\_}relation}}{cycle:unrelation}}\\{{\nwixident{ex}}{ex}}}\nwindexuse{\nwixident{add{\_}cycle{\_}rel}}{add:uncycle:unrel}{NWppJ6t-3LImg1-7}\nwindexuse{\nwixident{cycle{\_}adifferent}}{cycle:unadifferent}{NWppJ6t-3LImg1-7}\nwindexuse{\nwixident{cycle{\_}orthogonal}}{cycle:unorthogonal}{NWppJ6t-3LImg1-7}\nwindexuse{\nwixident{cycle{\_}relation}}{cycle:unrelation}{NWppJ6t-3LImg1-7}\nwindexuse{\nwixident{ex}}{ex}{NWppJ6t-3LImg1-7}\nwendcode{}\nwbegindocs{1266}This is an auxiliary function which removes duplicated cycles from a
list {\Tt{}\Rm{}{\it{}L}\nwendquote}.
\nwenddocs{}\nwbegincode{1267}\sublabel{NWppJ6t-3LImg1-8}\nwmargintag{{\nwtagstyle{}\subpageref{NWppJ6t-3LImg1-8}}}\moddef{addional functions~{\nwtagstyle{}\subpageref{NWppJ6t-3LImg1-1}}}\plusendmoddef\Rm{}\nwstartdeflinemarkup\nwusesondefline{\\{NWppJ6t-32jW6L-1}}\nwprevnextdefs{NWppJ6t-3LImg1-7}{NWppJ6t-3LImg1-9}\nwenddeflinemarkup
{\bf{}ex} {\it{}unique\_cycle}({\bf{}const} {\bf{}ex} & {\it{}L})
{\nwlbrace}
    {\bf{}if}({\it{}is\_a}\begin{math}<\end{math}{\bf{}lst}\begin{math}>\end{math}({\it{}L}) \begin{math}\wedge\end{math} ({\it{}L}.{\it{}nops}() \begin{math}>\end{math} 1) ) {\nwlbrace}
        {\bf{}lst} {\it{}res};
        {\bf{}lst}::{\it{}const\_iterator} {\it{}it} = {\it{}ex\_to}\begin{math}<\end{math}{\bf{}lst}\begin{math}>\end{math}({\it{}L}).{\it{}begin}();
        {\bf{}if} ({\it{}is\_a}\begin{math}<\end{math}{\bf{}cycle\_data}\begin{math}>\end{math}(\begin{math}\ast\end{math}{\it{}it})) {\nwlbrace}
            {\it{}res}.{\it{}append}(\begin{math}\ast\end{math}{\it{}it});
            \protect\PP{\it{}it};
            {\bf{}for} (; {\it{}it} \begin{math}\neq\end{math} {\it{}ex\_to}\begin{math}<\end{math}{\bf{}lst}\begin{math}>\end{math}({\it{}L}).{\it{}end}(); \protect\PP{\it{}it}) {\nwlbrace}
                {\bf{}bool} {\it{}is\_new}={\bf{}true};
                {\bf{}if} (\begin{math}\neg\end{math} {\it{}is\_a}\begin{math}<\end{math}{\bf{}cycle\_data}\begin{math}>\end{math}(\begin{math}\ast\end{math}{\it{}it}))
                    {\bf{}break}; // a non-cycle detected, get out

                {\bf{}for} ({\bf{}const} {\bf{}auto}& {\it{}it1} : {\it{}res})
                    {\bf{}if} ({\it{}ex\_to}\begin{math}<\end{math}{\bf{}cycle\_data}\begin{math}>\end{math}(\begin{math}\ast\end{math}{\it{}it}).{\it{}is\_almost\_equal}({\it{}ex\_to}\begin{math}<\end{math}{\bf{}basic}\begin{math}>\end{math}({\it{}it1}),{\bf{}true})
                        \begin{math}\vee\end{math} {\it{}ex\_to}\begin{math}<\end{math}{\bf{}cycle\_data}\begin{math}>\end{math}(\begin{math}\ast\end{math}{\it{}it}).{\it{}is\_equal}({\it{}ex\_to}\begin{math}<\end{math}{\bf{}basic}\begin{math}>\end{math}({\it{}it1}),{\bf{}true})) {\nwlbrace}
                        {\it{}is\_new}={\bf{}false}; // is a duplicate
                        {\bf{}break};
                    {\nwrbrace}
                {\bf{}if} ({\it{}is\_new})
                    {\it{}res}.{\it{}append}(\begin{math}\ast\end{math}{\it{}it});
            {\nwrbrace}
            {\bf{}if} ({\it{}it} \begin{math}\equiv\end{math} {\it{}ex\_to}\begin{math}<\end{math}{\bf{}lst}\begin{math}>\end{math}({\it{}L}).{\it{}end}()) // all are processed, no non-cycle is detected
                {\bf{}return} {\it{}res};
        {\nwrbrace}
    {\nwrbrace}
    {\bf{}return} {\it{}L};
{\nwrbrace}
\nwindexdefn{\nwixident{unique{\_}cycle}}{unique:uncycle}{NWppJ6t-3LImg1-8}\eatline
\nwused{\\{NWppJ6t-32jW6L-1}}\nwidentdefs{\\{{\nwixident{unique{\_}cycle}}{unique:uncycle}}}\nwidentuses{\\{{\nwixident{cycle{\_}data}}{cycle:undata}}\\{{\nwixident{ex}}{ex}}\\{{\nwixident{is{\_}almost{\_}equal}}{is:unalmost:unequal}}\\{{\nwixident{nops}}{nops}}}\nwindexuse{\nwixident{cycle{\_}data}}{cycle:undata}{NWppJ6t-3LImg1-8}\nwindexuse{\nwixident{ex}}{ex}{NWppJ6t-3LImg1-8}\nwindexuse{\nwixident{is{\_}almost{\_}equal}}{is:unalmost:unequal}{NWppJ6t-3LImg1-8}\nwindexuse{\nwixident{nops}}{nops}{NWppJ6t-3LImg1-8}\nwendcode{}\nwbegindocs{1268}\nwdocspar
\nwenddocs{}\nwbegindocs{1269}The debug output may be switched on and switched off by the
following methods.
\nwenddocs{}\nwbegincode{1270}\sublabel{NWppJ6t-3LImg1-9}\nwmargintag{{\nwtagstyle{}\subpageref{NWppJ6t-3LImg1-9}}}\moddef{addional functions~{\nwtagstyle{}\subpageref{NWppJ6t-3LImg1-1}}}\plusendmoddef\Rm{}\nwstartdeflinemarkup\nwusesondefline{\\{NWppJ6t-32jW6L-1}}\nwprevnextdefs{NWppJ6t-3LImg1-8}{NWppJ6t-3LImg1-A}\nwenddeflinemarkup
{\bf{}void} {\it{}figure\_debug\_on}() {\nwlbrace} {\it{}FIGURE\_DEBUG} = {\bf{}true}; {\nwrbrace}\nwindexdefn{\nwixident{figure{\_}debug{\_}on}}{figure:undebug:unon}{NWppJ6t-3LImg1-9}
{\bf{}void} {\it{}figure\_debug\_off}() {\nwlbrace} {\it{}FIGURE\_DEBUG} = {\bf{}false}; {\nwrbrace}\nwindexdefn{\nwixident{figure{\_}debug{\_}off}}{figure:undebug:unoff}{NWppJ6t-3LImg1-9}
{\bf{}bool} {\it{}figure\_ask\_debug\_status}() {\nwlbrace} {\bf{}return} {\it{}FIGURE\_DEBUG}; {\nwrbrace}
\nwindexdefn{\nwixident{figure{\_}debug{\_}on}}{figure:undebug:unon}{NWppJ6t-3LImg1-9}\nwindexdefn{\nwixident{figure{\_}debug{\_}off}}{figure:undebug:unoff}{NWppJ6t-3LImg1-9}\nwindexdefn{\nwixident{figure{\_}ask{\_}debug{\_}status}}{figure:unask:undebug:unstatus}{NWppJ6t-3LImg1-9}\eatline
\nwused{\\{NWppJ6t-32jW6L-1}}\nwidentdefs{\\{{\nwixident{figure{\_}ask{\_}debug{\_}status}}{figure:unask:undebug:unstatus}}\\{{\nwixident{figure{\_}debug{\_}off}}{figure:undebug:unoff}}\\{{\nwixident{figure{\_}debug{\_}on}}{figure:undebug:unon}}}\nwidentuses{\\{{\nwixident{FIGURE{\_}DEBUG}}{FIGURE:unDEBUG}}}\nwindexuse{\nwixident{FIGURE{\_}DEBUG}}{FIGURE:unDEBUG}{NWppJ6t-3LImg1-9}\nwendcode{}\nwbegindocs{1271}\nwdocspar
\nwenddocs{}\nwbegindocs{1272}Setting variable {\Tt{}\Rm{}{\it{}show\_asy\_graphics}\nwendquote} to switch \Asymptote\ display
on and off.
\nwenddocs{}\nwbegincode{1273}\sublabel{NWppJ6t-3LImg1-A}\nwmargintag{{\nwtagstyle{}\subpageref{NWppJ6t-3LImg1-A}}}\moddef{addional functions~{\nwtagstyle{}\subpageref{NWppJ6t-3LImg1-1}}}\plusendmoddef\Rm{}\nwstartdeflinemarkup\nwusesondefline{\\{NWppJ6t-32jW6L-1}}\nwprevnextdefs{NWppJ6t-3LImg1-9}{\relax}\nwenddeflinemarkup
{\bf{}void} {\it{}show\_asy\_on}() {\nwlbrace} {\it{}show\_asy\_graphics}={\bf{}true}; {\nwrbrace}\nwindexdefn{\nwixident{show{\_}asy{\_}on}}{show:unasy:unon}{NWppJ6t-3LImg1-A}
{\bf{}void} {\it{}show\_asy\_off}() {\nwlbrace} {\it{}show\_asy\_graphics}={\bf{}false}; {\nwrbrace}\nwindexdefn{\nwixident{show{\_}asy{\_}off}}{show:unasy:unoff}{NWppJ6t-3LImg1-A}
\nwindexdefn{\nwixident{show{\_}asy{\_}on}}{show:unasy:unon}{NWppJ6t-3LImg1-A}\nwindexdefn{\nwixident{show{\_}asy{\_}off}}{show:unasy:unoff}{NWppJ6t-3LImg1-A}\eatline
\nwused{\\{NWppJ6t-32jW6L-1}}\nwidentdefs{\\{{\nwixident{show{\_}asy{\_}off}}{show:unasy:unoff}}\\{{\nwixident{show{\_}asy{\_}on}}{show:unasy:unon}}}\nwidentuses{\\{{\nwixident{show{\_}asy{\_}graphics}}{show:unasy:ungraphics}}}\nwindexuse{\nwixident{show{\_}asy{\_}graphics}}{show:unasy:ungraphics}{NWppJ6t-3LImg1-A}\nwendcode{}\nwbegindocs{1274}\nwdocspar
\nwenddocs{}\nwbegindocs{1275}\nwdocspar
\section{Change Log}
\label{sec:changes-log}

\begin{description}
\item[3.2] The following changes are committed:
  \begin{itemize}
  \item Methods {\Tt{}\Rm{}{\it{}get\_all\_keys}()\nwendquote} and {\Tt{}\Rm{}{\bf{}figure}::{\it{}do\_print}()\nwendquote} sort output
    from lower to higher generations.
  \item Better structure of the \Asymptote\ output.
  \item Add {\Tt{}\Rm{}{\bf{}figure}::{\it{}get\_max\_generation}()\nwendquote} method.
  \item Fix archiving/unarchiving of figure.
  \item {\Tt{}\Rm{}{\bf{}cycle\_node}\nwendquote} is archiving its custom \Asymptote\ style.
  \item Minor improvements of code and documentation.
  \item Introduce {\Tt{}\Rm{}{\it{}do\_print\_double}()\nwendquote} for a more compact output of figures.
  \end{itemize}
\item[3.1] The following changes are committed:
  \begin{itemize}
  \item  Updated cycle solver to handle homogeneous equations properly
    and produce root-free parametrisation in some cases.
  \item Theoretical aspects are revised in documentation.
  \item In cycles with numerous instances only corresponding cycles
    may be checked for a relation.
  \item Numerous other small improvements.
 \end{itemize}
\item[3.0]  The following changes are committed:
  \begin{itemize}
  \item Functions {\Tt{}\Rm{}{\it{}sl2\_clifford}()\nwendquote} and {\Tt{}\Rm{}{\it{}sl2\_similarity}()\nwendquote}
    work for hypercomplex matrices as well.
  \item Cycle library is able to work both in vector and paravector
    formalisms.
  \item Add flag {\Tt{}\Rm{}{\it{}ignore\_unit}\nwendquote} to {\Tt{}\Rm{}{\bf{}cycle}::{\it{}is\_equal}()\nwendquote}.
  \item Add {\Tt{}\Rm{}{\it{}with\_label}\nwendquote} parameter to {\Tt{}\Rm{}{\bf{}figure}::{\it{}asy\_write}()\nwendquote}.
  \item Improved the example with modular group action.
  \item Numerous small improvements to code and documentations.
  \end{itemize}
\item[2.7]  The following changes are committed:
  \begin{itemize}
  \item Container ([lst]) assignments are using curly brackets
    now.
  \item Some fixes for upcoming \GiNaC\ 1.7.0.
  \end{itemize}
\item[2.6]  The following changes are committed:
  \begin{itemize}
  \item Installation instructions are updated and tested.
  \item \texttt{PyGiNaC} (refreshed) is added as a subproject.
  \end{itemize}
\item[2.5] The following changes are committed:
  \begin{itemize}
  \item Documentation is updated.
  \item 3D visualiser is added as a subproject.
  \item Minor fixes and adjustments.
  \end{itemize}
\item[2.4] The following minor changes are committed:
  \begin{itemize}
  \item Embedded PDF animation can be produced.
  \item Numerous improvements to documentation.
  \end{itemize}
\item[2.3] The following minor changes are committed:
  \begin{itemize}
  \item The stereometric example is done with the symbolic parameter.
  \item A concise mathematical introduction is written.
  \item Re-shape code of figure library.
  \item Use both symbolic and float checks to analyse newly evaluated cycles.
  \item Some minor code improvements.
  \end{itemize}
\item[2.2] The following minor changes are committed:
  \begin{itemize}
  \item New cycle relations {\Tt{}\Rm{}{\it{}moebius\_transform}\nwendquote} and
    {\Tt{}\Rm{}{\it{}sl2\_transform}\nwendquote} are added.
  \item Example programme with modular group action is added.
  \item Method {\Tt{}\Rm{}{\it{}add\_cycle\_rel}\nwendquote} may take a single relation now.
  \item Numerous internal fixes.
  \end{itemize}
\item[2.1] The following minor changes are committed:
  \begin{itemize}
  \item The method {\Tt{}\Rm{}{\bf{}figure}::{\it{}get\_all\_keys}()\nwendquote} is added
  \item Debug output may be switched on/off from the code.
  \item Improvements to documentation.
  \item Initialisation of cycles in {\Python} wrapper are corrected.
  \end{itemize}
\item[2.0] The two-dimension restriction is removed from the {\Tt{}\Rm{}{\bf{}figure}\nwendquote}
  library.  This breaks APIs, thus the major version number is increased.
\item[1.0] First official stable release with all essential functionality.
\end{description}
\nwenddocs{}\nwbegindocs{1276}\nwdocspar
\nwenddocs{}\nwbegincode{1277}\sublabel{NWppJ6t-1fWRR8-9}\nwmargintag{{\nwtagstyle{}\subpageref{NWppJ6t-1fWRR8-9}}}\moddef{separating chunk~{\nwtagstyle{}\subpageref{NWppJ6t-1fWRR8-1}}}\plusendmoddef\Rm{}\nwstartdeflinemarkup\nwprevnextdefs{NWppJ6t-1fWRR8-8}{\relax}\nwenddeflinemarkup

\nwendcode{}\nwbegindocs{1278}\nwdocspar
\section{License}
\label{sec:license}
This programme is distributed under GNU GPLv3~\cite{GNUGPL}.
\nwenddocs{}\nwbegincode{1279}\sublabel{NWppJ6t-ZXuKx-1}\nwmargintag{{\nwtagstyle{}\subpageref{NWppJ6t-ZXuKx-1}}}\moddef{license~{\nwtagstyle{}\subpageref{NWppJ6t-ZXuKx-1}}}\endmoddef\Rm{}\nwstartdeflinemarkup\nwusesondefline{\\{NWppJ6t-3rkk58-1}\\{NWppJ6t-1kKRnF-1}\\{NWppJ6t-2SYluR-1}\\{NWppJ6t-2t2vfS-1}\\{NWppJ6t-30Tc7D-1}\\{NWppJ6t-2zWvnC-1}\\{NWppJ6t-RDDRz-1}\\{NWppJ6t-2pgnFW-1}\\{NWppJ6t-3NCnUp-1}\\{NWppJ6t-32jW6L-1}}\nwenddeflinemarkup
// The library for ensembles of interrelated cycles in non-Euclidean geometry
//
//  Copyright (C) 2014-2018 Vladimir V. Kisil \begin{math}<\end{math}kisilv@maths.leeds.ac.uk\begin{math}>\end{math}
//
//  This program is free software: you can redistribute it and/or modify
//  it under the terms of the GNU General Public License as published by
//  the Free Software Foundation, either version 3 of the License, or
//  (at your option) any later version.
//
//  This program is distributed in the hope that it will be useful,
//  but WITHOUT ANY WARRANTY; without even the implied warranty of
//  MERCHANTABILITY or FITNESS FOR A PARTICULAR PURPOSE.  See the
//  GNU General Public License for more details.
//
//  You should have received a copy of the GNU General Public License
//  along with this program.  If not, see \begin{math}<\end{math}http://www.gnu.org/licenses/\begin{math}>\end{math}.

\nwused{\\{NWppJ6t-3rkk58-1}\\{NWppJ6t-1kKRnF-1}\\{NWppJ6t-2SYluR-1}\\{NWppJ6t-2t2vfS-1}\\{NWppJ6t-30Tc7D-1}\\{NWppJ6t-2zWvnC-1}\\{NWppJ6t-RDDRz-1}\\{NWppJ6t-2pgnFW-1}\\{NWppJ6t-3NCnUp-1}\\{NWppJ6t-32jW6L-1}}\nwendcode{}\nwbegindocs{1280}\nwdocspar
\section{Index of Identifiers}
\label{sec:index-identifiers}
\small
\nowebindex

\nwenddocs{}

\nwixlogsorted{c}{{*}{NWppJ6t-1p0Y9w-1}{\nwixd{NWppJ6t-1p0Y9w-1}}}%
\nwixlogsorted{c}{{3D-figure-example-common}{NWppJ6t-2pgnFW-1}{\nwixu{NWppJ6t-4Yd8hY-1}\nwixd{NWppJ6t-2pgnFW-1}\nwixd{NWppJ6t-2pgnFW-2}\nwixu{NWppJ6t-u6tXb-1}}}%
\nwixlogsorted{c}{{3D-figure-example.cpp}{NWppJ6t-4Yd8hY-1}{\nwixd{NWppJ6t-4Yd8hY-1}\nwixd{NWppJ6t-4Yd8hY-2}\nwixd{NWppJ6t-4Yd8hY-3}}}%
\nwixlogsorted{c}{{3D-figure-visualise.cpp}{NWppJ6t-u6tXb-1}{\nwixd{NWppJ6t-u6tXb-1}\nwixd{NWppJ6t-u6tXb-2}\nwixd{NWppJ6t-u6tXb-3}}}%
\nwixlogsorted{c}{{add checked relation}{NWppJ6t-2SaBW8-1}{\nwixu{NWppJ6t-DKmU5-2E}\nwixd{NWppJ6t-2SaBW8-1}}}%
\nwixlogsorted{c}{{add cycle relations}{NWppJ6t-3fVAGh-1}{\nwixu{NWppJ6t-32jW6L-1}\nwixd{NWppJ6t-3fVAGh-1}\nwixd{NWppJ6t-3fVAGh-2}\nwixd{NWppJ6t-3fVAGh-3}\nwixd{NWppJ6t-3fVAGh-4}\nwixd{NWppJ6t-3fVAGh-5}\nwixd{NWppJ6t-3fVAGh-6}\nwixd{NWppJ6t-3fVAGh-7}\nwixd{NWppJ6t-3fVAGh-8}\nwixd{NWppJ6t-3fVAGh-9}\nwixd{NWppJ6t-3fVAGh-A}\nwixd{NWppJ6t-3fVAGh-B}\nwixd{NWppJ6t-3fVAGh-C}\nwixd{NWppJ6t-3fVAGh-D}\nwixd{NWppJ6t-3fVAGh-E}\nwixd{NWppJ6t-3fVAGh-F}\nwixd{NWppJ6t-3fVAGh-G}\nwixd{NWppJ6t-3fVAGh-H}\nwixd{NWppJ6t-3fVAGh-I}}}%
\nwixlogsorted{c}{{add gar extension}{NWppJ6t-35j28o-1}{\nwixu{NWppJ6t-DKmU5-B}\nwixd{NWppJ6t-35j28o-1}\nwixu{NWppJ6t-DKmU5-C}}}%
\nwixlogsorted{c}{{add symbols to match dimensionality}{NWppJ6t-3QR99o-1}{\nwixu{NWppJ6t-DKmU5-7}\nwixd{NWppJ6t-3QR99o-1}\nwixu{NWppJ6t-DKmU5-A}}}%
\nwixlogsorted{c}{{adding point with its parents}{NWppJ6t-1Ub2Xq-1}{\nwixu{NWppJ6t-DKmU5-E}\nwixd{NWppJ6t-1Ub2Xq-1}\nwixd{NWppJ6t-1Ub2Xq-2}\nwixu{NWppJ6t-DKmU5-W}}}%
\nwixlogsorted{c}{{addional functions}{NWppJ6t-3LImg1-1}{\nwixu{NWppJ6t-32jW6L-1}\nwixd{NWppJ6t-3LImg1-1}\nwixd{NWppJ6t-3LImg1-2}\nwixd{NWppJ6t-3LImg1-3}\nwixd{NWppJ6t-3LImg1-4}\nwixd{NWppJ6t-3LImg1-5}\nwixd{NWppJ6t-3LImg1-6}\nwixd{NWppJ6t-3LImg1-7}\nwixd{NWppJ6t-3LImg1-8}\nwixd{NWppJ6t-3LImg1-9}\nwixd{NWppJ6t-3LImg1-A}}}%
\nwixlogsorted{c}{{additional functions header}{NWppJ6t-1lh2X2-1}{\nwixd{NWppJ6t-1lh2X2-1}\nwixd{NWppJ6t-1lh2X2-2}\nwixd{NWppJ6t-1lh2X2-3}\nwixd{NWppJ6t-1lh2X2-4}\nwixd{NWppJ6t-1lh2X2-5}\nwixu{NWppJ6t-3NCnUp-3}}}%
\nwixlogsorted{c}{{asy styles}{NWppJ6t-43RSeU-1}{\nwixu{NWppJ6t-3NCnUp-3}\nwixd{NWppJ6t-43RSeU-1}\nwixd{NWppJ6t-43RSeU-2}}}%
\nwixlogsorted{c}{{auto TeX name}{NWppJ6t-2tM1q2-1}{\nwixu{NWppJ6t-DKmU5-D}\nwixu{NWppJ6t-DKmU5-H}\nwixu{NWppJ6t-DKmU5-P}\nwixu{NWppJ6t-DKmU5-Q}\nwixu{NWppJ6t-DKmU5-S}\nwixd{NWppJ6t-2tM1q2-1}\nwixu{NWppJ6t-DKmU5-2J}}}%
\nwixlogsorted{c}{{auxillary function}{NWppJ6t-1KYtgD-1}{\nwixu{NWppJ6t-32jW6L-1}\nwixd{NWppJ6t-1KYtgD-1}\nwixd{NWppJ6t-1KYtgD-2}}}%
\nwixlogsorted{c}{{build medioscribed cycle}{NWppJ6t-2LTMm-1}{\nwixu{NWppJ6t-30Tc7D-1}\nwixd{NWppJ6t-2LTMm-1}\nwixd{NWppJ6t-2LTMm-2}\nwixd{NWppJ6t-2LTMm-3}\nwixd{NWppJ6t-2LTMm-4}\nwixd{NWppJ6t-2LTMm-5}\nwixd{NWppJ6t-2LTMm-6}\nwixd{NWppJ6t-2LTMm-7}\nwixd{NWppJ6t-2LTMm-8}\nwixd{NWppJ6t-2LTMm-9}\nwixd{NWppJ6t-2LTMm-A}\nwixd{NWppJ6t-2LTMm-B}\nwixd{NWppJ6t-2LTMm-C}\nwixd{NWppJ6t-2LTMm-D}\nwixd{NWppJ6t-2LTMm-E}\nwixd{NWppJ6t-2LTMm-F}\nwixu{NWppJ6t-2zWvnC-1}}}%
\nwixlogsorted{c}{{check cycles are valid}{NWppJ6t-1qiQ4a-1}{\nwixu{NWppJ6t-26cGQh-1}\nwixu{NWppJ6t-1HqLYY-3}\nwixd{NWppJ6t-1qiQ4a-1}\nwixu{NWppJ6t-1HqLYY-4}\nwixu{NWppJ6t-1HqLYY-5}\nwixu{NWppJ6t-1HqLYY-B}}}%
\nwixlogsorted{c}{{check dimensionalities point and cycle metrics}{NWppJ6t-11clwc-1}{\nwixu{NWppJ6t-DKmU5-7}\nwixd{NWppJ6t-11clwc-1}\nwixu{NWppJ6t-DKmU5-1Q}}}%
\nwixlogsorted{c}{{check parents are valid}{NWppJ6t-2RJcfD-1}{\nwixu{NWppJ6t-1HqLYY-3}\nwixd{NWppJ6t-2RJcfD-1}\nwixu{NWppJ6t-1HqLYY-4}\nwixu{NWppJ6t-1HqLYY-5}}}%
\nwixlogsorted{c}{{check that cycles are tangent}{NWppJ6t-2XKMIQ-1}{\nwixu{NWppJ6t-30Tc7D-3}\nwixd{NWppJ6t-2XKMIQ-1}\nwixu{NWppJ6t-30Tc7D-5}\nwixu{NWppJ6t-30Tc7D-6}\nwixu{NWppJ6t-30Tc7D-7}\nwixu{NWppJ6t-30Tc7D-8}}}%
\nwixlogsorted{c}{{check that dimensionality is 2}{NWppJ6t-A2asn-1}{\nwixu{NWppJ6t-DKmU5-1a}\nwixd{NWppJ6t-A2asn-1}\nwixu{NWppJ6t-DKmU5-1l}\nwixu{NWppJ6t-DKmU5-1o}}}%
\nwixlogsorted{c}{{check the theorem}{NWppJ6t-4d7QoM-1}{\nwixu{NWppJ6t-2LTMm-F}\nwixd{NWppJ6t-4d7QoM-1}\nwixu{NWppJ6t-30Tc7D-5}\nwixu{NWppJ6t-30Tc7D-6}\nwixu{NWppJ6t-30Tc7D-7}\nwixu{NWppJ6t-30Tc7D-8}}}%
\nwixlogsorted{c}{{common part of float output}{NWppJ6t-tXDC3-1}{\nwixu{NWppJ6t-2oFbmT-8}\nwixd{NWppJ6t-tXDC3-1}\nwixu{NWppJ6t-2oFbmT-A}\nwixu{NWppJ6t-2oFbmT-B}}}%
\nwixlogsorted{c}{{cycle data class}{NWppJ6t-2oFbmT-1}{\nwixu{NWppJ6t-32jW6L-1}\nwixd{NWppJ6t-2oFbmT-1}\nwixd{NWppJ6t-2oFbmT-2}\nwixd{NWppJ6t-2oFbmT-3}\nwixd{NWppJ6t-2oFbmT-4}\nwixd{NWppJ6t-2oFbmT-5}\nwixd{NWppJ6t-2oFbmT-6}\nwixd{NWppJ6t-2oFbmT-7}\nwixd{NWppJ6t-2oFbmT-8}\nwixd{NWppJ6t-2oFbmT-9}\nwixd{NWppJ6t-2oFbmT-A}\nwixd{NWppJ6t-2oFbmT-B}\nwixd{NWppJ6t-2oFbmT-C}\nwixd{NWppJ6t-2oFbmT-D}\nwixd{NWppJ6t-2oFbmT-E}\nwixd{NWppJ6t-2oFbmT-F}\nwixd{NWppJ6t-2oFbmT-G}\nwixd{NWppJ6t-2oFbmT-H}\nwixd{NWppJ6t-2oFbmT-I}\nwixd{NWppJ6t-2oFbmT-J}\nwixd{NWppJ6t-2oFbmT-K}\nwixd{NWppJ6t-2oFbmT-L}\nwixd{NWppJ6t-2oFbmT-M}\nwixd{NWppJ6t-2oFbmT-N}\nwixd{NWppJ6t-2oFbmT-O}}}%
\nwixlogsorted{c}{{cycle data header}{NWppJ6t-8XFRe-1}{\nwixu{NWppJ6t-3NCnUp-3}\nwixd{NWppJ6t-8XFRe-1}\nwixd{NWppJ6t-8XFRe-2}\nwixd{NWppJ6t-8XFRe-3}}}%
\nwixlogsorted{c}{{cycle node class}{NWppJ6t-1HqLYY-1}{\nwixu{NWppJ6t-32jW6L-1}\nwixd{NWppJ6t-1HqLYY-1}\nwixd{NWppJ6t-1HqLYY-2}\nwixd{NWppJ6t-1HqLYY-3}\nwixd{NWppJ6t-1HqLYY-4}\nwixd{NWppJ6t-1HqLYY-5}\nwixd{NWppJ6t-1HqLYY-6}\nwixd{NWppJ6t-1HqLYY-7}\nwixd{NWppJ6t-1HqLYY-8}\nwixd{NWppJ6t-1HqLYY-9}\nwixd{NWppJ6t-1HqLYY-A}\nwixd{NWppJ6t-1HqLYY-B}\nwixd{NWppJ6t-1HqLYY-C}\nwixd{NWppJ6t-1HqLYY-D}\nwixd{NWppJ6t-1HqLYY-E}\nwixd{NWppJ6t-1HqLYY-F}\nwixd{NWppJ6t-1HqLYY-G}\nwixd{NWppJ6t-1HqLYY-H}\nwixd{NWppJ6t-1HqLYY-I}\nwixd{NWppJ6t-1HqLYY-J}\nwixd{NWppJ6t-1HqLYY-K}\nwixd{NWppJ6t-1HqLYY-L}\nwixd{NWppJ6t-1HqLYY-M}\nwixd{NWppJ6t-1HqLYY-N}\nwixd{NWppJ6t-1HqLYY-O}\nwixd{NWppJ6t-1HqLYY-P}\nwixd{NWppJ6t-1HqLYY-Q}\nwixd{NWppJ6t-1HqLYY-R}\nwixd{NWppJ6t-1HqLYY-S}}}%
\nwixlogsorted{c}{{cycle node header}{NWppJ6t-2uztPH-1}{\nwixu{NWppJ6t-3NCnUp-3}\nwixd{NWppJ6t-2uztPH-1}\nwixd{NWppJ6t-2uztPH-2}\nwixd{NWppJ6t-2uztPH-3}\nwixd{NWppJ6t-2uztPH-4}\nwixd{NWppJ6t-2uztPH-5}\nwixd{NWppJ6t-2uztPH-6}\nwixd{NWppJ6t-2uztPH-7}\nwixd{NWppJ6t-2uztPH-8}\nwixd{NWppJ6t-2uztPH-9}\nwixd{NWppJ6t-2uztPH-A}\nwixd{NWppJ6t-2uztPH-B}\nwixd{NWppJ6t-2uztPH-C}\nwixd{NWppJ6t-2uztPH-D}\nwixd{NWppJ6t-2uztPH-E}\nwixd{NWppJ6t-2uztPH-F}\nwixd{NWppJ6t-2uztPH-G}}}%
\nwixlogsorted{c}{{cycle relation class}{NWppJ6t-3bfNK9-1}{\nwixu{NWppJ6t-32jW6L-1}\nwixd{NWppJ6t-3bfNK9-1}\nwixd{NWppJ6t-3bfNK9-2}\nwixd{NWppJ6t-3bfNK9-3}\nwixd{NWppJ6t-3bfNK9-4}\nwixd{NWppJ6t-3bfNK9-5}\nwixd{NWppJ6t-3bfNK9-6}\nwixd{NWppJ6t-3bfNK9-7}\nwixd{NWppJ6t-3bfNK9-8}\nwixd{NWppJ6t-3bfNK9-9}\nwixd{NWppJ6t-3bfNK9-A}\nwixd{NWppJ6t-3bfNK9-B}\nwixd{NWppJ6t-3bfNK9-C}\nwixd{NWppJ6t-3bfNK9-D}\nwixd{NWppJ6t-3bfNK9-E}}}%
\nwixlogsorted{c}{{cycle relations}{NWppJ6t-1d2GZW-1}{\nwixu{NWppJ6t-3NCnUp-3}\nwixd{NWppJ6t-1d2GZW-1}\nwixd{NWppJ6t-1d2GZW-2}\nwixd{NWppJ6t-1d2GZW-3}\nwixd{NWppJ6t-1d2GZW-4}\nwixd{NWppJ6t-1d2GZW-5}\nwixd{NWppJ6t-1d2GZW-6}\nwixd{NWppJ6t-1d2GZW-7}\nwixd{NWppJ6t-1d2GZW-8}\nwixd{NWppJ6t-1d2GZW-9}\nwixd{NWppJ6t-1d2GZW-A}\nwixd{NWppJ6t-1d2GZW-B}}}%
\nwixlogsorted{c}{{defining types}{NWppJ6t-1s4LDF-1}{\nwixd{NWppJ6t-1s4LDF-1}\nwixu{NWppJ6t-3NCnUp-3}\nwixd{NWppJ6t-1s4LDF-2}}}%
\nwixlogsorted{c}{{end to print cycle node}{NWppJ6t-1tclWh-1}{\nwixu{NWppJ6t-1HqLYY-G}\nwixu{NWppJ6t-1HqLYY-H}\nwixd{NWppJ6t-1tclWh-1}\nwixd{NWppJ6t-1tclWh-2}\nwixd{NWppJ6t-1tclWh-3}}}%
\nwixlogsorted{c}{{figure class}{NWppJ6t-DKmU5-1}{\nwixu{NWppJ6t-32jW6L-1}\nwixd{NWppJ6t-DKmU5-1}\nwixd{NWppJ6t-DKmU5-2}\nwixd{NWppJ6t-DKmU5-3}\nwixd{NWppJ6t-DKmU5-4}\nwixd{NWppJ6t-DKmU5-5}\nwixd{NWppJ6t-DKmU5-6}\nwixd{NWppJ6t-DKmU5-7}\nwixd{NWppJ6t-DKmU5-8}\nwixd{NWppJ6t-DKmU5-9}\nwixd{NWppJ6t-DKmU5-A}\nwixd{NWppJ6t-DKmU5-B}\nwixd{NWppJ6t-DKmU5-C}\nwixd{NWppJ6t-DKmU5-D}\nwixd{NWppJ6t-DKmU5-E}\nwixd{NWppJ6t-DKmU5-F}\nwixd{NWppJ6t-DKmU5-G}\nwixd{NWppJ6t-DKmU5-H}\nwixd{NWppJ6t-DKmU5-I}\nwixd{NWppJ6t-DKmU5-J}\nwixd{NWppJ6t-DKmU5-K}\nwixd{NWppJ6t-DKmU5-L}\nwixd{NWppJ6t-DKmU5-M}\nwixd{NWppJ6t-DKmU5-N}\nwixd{NWppJ6t-DKmU5-O}\nwixd{NWppJ6t-DKmU5-P}\nwixd{NWppJ6t-DKmU5-Q}\nwixd{NWppJ6t-DKmU5-R}\nwixd{NWppJ6t-DKmU5-S}\nwixd{NWppJ6t-DKmU5-T}\nwixd{NWppJ6t-DKmU5-U}\nwixd{NWppJ6t-DKmU5-V}\nwixd{NWppJ6t-DKmU5-W}\nwixd{NWppJ6t-DKmU5-X}\nwixd{NWppJ6t-DKmU5-Y}\nwixd{NWppJ6t-DKmU5-Z}\nwixd{NWppJ6t-DKmU5-a}\nwixd{NWppJ6t-DKmU5-b}\nwixd{NWppJ6t-DKmU5-c}\nwixd{NWppJ6t-DKmU5-d}\nwixd{NWppJ6t-DKmU5-e}\nwixd{NWppJ6t-DKmU5-f}\nwixd{NWppJ6t-DKmU5-g}\nwixd{NWppJ6t-DKmU5-h}\nwixd{NWppJ6t-DKmU5-i}\nwixd{NWppJ6t-DKmU5-j}\nwixd{NWppJ6t-DKmU5-k}\nwixd{NWppJ6t-DKmU5-l}\nwixd{NWppJ6t-DKmU5-m}\nwixd{NWppJ6t-DKmU5-n}\nwixd{NWppJ6t-DKmU5-o}\nwixd{NWppJ6t-DKmU5-p}\nwixd{NWppJ6t-DKmU5-q}\nwixd{NWppJ6t-DKmU5-r}\nwixd{NWppJ6t-DKmU5-s}\nwixd{NWppJ6t-DKmU5-t}\nwixd{NWppJ6t-DKmU5-u}\nwixd{NWppJ6t-DKmU5-v}\nwixd{NWppJ6t-DKmU5-w}\nwixd{NWppJ6t-DKmU5-x}\nwixd{NWppJ6t-DKmU5-y}\nwixd{NWppJ6t-DKmU5-z}\nwixd{NWppJ6t-DKmU5-10}\nwixd{NWppJ6t-DKmU5-11}\nwixd{NWppJ6t-DKmU5-12}\nwixd{NWppJ6t-DKmU5-13}\nwixd{NWppJ6t-DKmU5-14}\nwixd{NWppJ6t-DKmU5-15}\nwixd{NWppJ6t-DKmU5-16}\nwixd{NWppJ6t-DKmU5-17}\nwixd{NWppJ6t-DKmU5-18}\nwixd{NWppJ6t-DKmU5-19}\nwixd{NWppJ6t-DKmU5-1A}\nwixd{NWppJ6t-DKmU5-1B}\nwixd{NWppJ6t-DKmU5-1C}\nwixd{NWppJ6t-DKmU5-1D}\nwixd{NWppJ6t-DKmU5-1E}\nwixd{NWppJ6t-DKmU5-1F}\nwixd{NWppJ6t-DKmU5-1G}\nwixd{NWppJ6t-DKmU5-1H}\nwixd{NWppJ6t-DKmU5-1I}\nwixd{NWppJ6t-DKmU5-1J}\nwixd{NWppJ6t-DKmU5-1K}\nwixd{NWppJ6t-DKmU5-1L}\nwixd{NWppJ6t-DKmU5-1M}\nwixd{NWppJ6t-DKmU5-1N}\nwixd{NWppJ6t-DKmU5-1O}\nwixd{NWppJ6t-DKmU5-1P}\nwixd{NWppJ6t-DKmU5-1Q}\nwixd{NWppJ6t-DKmU5-1R}\nwixd{NWppJ6t-DKmU5-1S}\nwixd{NWppJ6t-DKmU5-1T}\nwixd{NWppJ6t-DKmU5-1U}\nwixd{NWppJ6t-DKmU5-1V}\nwixd{NWppJ6t-DKmU5-1W}\nwixd{NWppJ6t-DKmU5-1X}\nwixd{NWppJ6t-DKmU5-1Y}\nwixd{NWppJ6t-DKmU5-1Z}\nwixd{NWppJ6t-DKmU5-1a}\nwixd{NWppJ6t-DKmU5-1b}\nwixd{NWppJ6t-DKmU5-1c}\nwixd{NWppJ6t-DKmU5-1d}\nwixd{NWppJ6t-DKmU5-1e}\nwixd{NWppJ6t-DKmU5-1f}\nwixd{NWppJ6t-DKmU5-1g}\nwixd{NWppJ6t-DKmU5-1h}\nwixd{NWppJ6t-DKmU5-1i}\nwixd{NWppJ6t-DKmU5-1j}\nwixd{NWppJ6t-DKmU5-1k}\nwixd{NWppJ6t-DKmU5-1l}\nwixd{NWppJ6t-DKmU5-1m}\nwixd{NWppJ6t-DKmU5-1n}\nwixd{NWppJ6t-DKmU5-1o}\nwixd{NWppJ6t-DKmU5-1p}\nwixd{NWppJ6t-DKmU5-1q}\nwixd{NWppJ6t-DKmU5-1r}\nwixd{NWppJ6t-DKmU5-1s}\nwixd{NWppJ6t-DKmU5-1t}\nwixd{NWppJ6t-DKmU5-1u}\nwixd{NWppJ6t-DKmU5-1v}\nwixd{NWppJ6t-DKmU5-1w}\nwixd{NWppJ6t-DKmU5-1x}\nwixd{NWppJ6t-DKmU5-1y}\nwixd{NWppJ6t-DKmU5-1z}\nwixd{NWppJ6t-DKmU5-20}\nwixd{NWppJ6t-DKmU5-21}\nwixd{NWppJ6t-DKmU5-22}\nwixd{NWppJ6t-DKmU5-23}\nwixd{NWppJ6t-DKmU5-24}\nwixd{NWppJ6t-DKmU5-25}\nwixd{NWppJ6t-DKmU5-26}\nwixd{NWppJ6t-DKmU5-27}\nwixd{NWppJ6t-DKmU5-28}\nwixd{NWppJ6t-DKmU5-29}\nwixd{NWppJ6t-DKmU5-2A}\nwixd{NWppJ6t-DKmU5-2B}\nwixd{NWppJ6t-DKmU5-2C}\nwixd{NWppJ6t-DKmU5-2D}\nwixd{NWppJ6t-DKmU5-2E}\nwixd{NWppJ6t-DKmU5-2F}\nwixd{NWppJ6t-DKmU5-2G}\nwixd{NWppJ6t-DKmU5-2H}\nwixd{NWppJ6t-DKmU5-2I}\nwixd{NWppJ6t-DKmU5-2J}\nwixd{NWppJ6t-DKmU5-2K}}}%
\nwixlogsorted{c}{{figure define}{NWppJ6t-3ONz84-1}{\nwixu{NWppJ6t-3NCnUp-3}\nwixd{NWppJ6t-3ONz84-1}}}%
\nwixlogsorted{c}{{figure header}{NWppJ6t-7vmg0-1}{\nwixu{NWppJ6t-3NCnUp-3}\nwixd{NWppJ6t-7vmg0-1}\nwixd{NWppJ6t-7vmg0-2}\nwixd{NWppJ6t-7vmg0-3}\nwixd{NWppJ6t-7vmg0-4}\nwixd{NWppJ6t-7vmg0-5}\nwixd{NWppJ6t-7vmg0-6}\nwixd{NWppJ6t-7vmg0-7}\nwixd{NWppJ6t-7vmg0-8}}}%
\nwixlogsorted{c}{{figure library variables and constants}{NWppJ6t-A2bag-1}{\nwixd{NWppJ6t-A2bag-1}\nwixd{NWppJ6t-A2bag-2}\nwixu{NWppJ6t-32jW6L-1}\nwixd{NWppJ6t-A2bag-3}\nwixd{NWppJ6t-A2bag-4}\nwixd{NWppJ6t-A2bag-5}}}%
\nwixlogsorted{c}{{figure-ortho-anlytic-proof.cpp}{NWppJ6t-2t2vfS-1}{\nwixd{NWppJ6t-2t2vfS-1}\nwixd{NWppJ6t-2t2vfS-2}\nwixd{NWppJ6t-2t2vfS-3}\nwixd{NWppJ6t-2t2vfS-4}\nwixd{NWppJ6t-2t2vfS-5}\nwixd{NWppJ6t-2t2vfS-6}\nwixd{NWppJ6t-2t2vfS-7}\nwixd{NWppJ6t-2t2vfS-8}\nwixd{NWppJ6t-2t2vfS-9}\nwixd{NWppJ6t-2t2vfS-A}\nwixd{NWppJ6t-2t2vfS-B}\nwixd{NWppJ6t-2t2vfS-C}\nwixd{NWppJ6t-2t2vfS-D}}}%
\nwixlogsorted{c}{{figure.cpp}{NWppJ6t-32jW6L-1}{\nwixd{NWppJ6t-32jW6L-1}}}%
\nwixlogsorted{c}{{figure.h}{NWppJ6t-3NCnUp-1}{\nwixd{NWppJ6t-3NCnUp-1}\nwixd{NWppJ6t-3NCnUp-2}\nwixd{NWppJ6t-3NCnUp-3}}}%
\nwixlogsorted{c}{{fillmore-springer-example.cpp}{NWppJ6t-RDDRz-1}{\nwixd{NWppJ6t-RDDRz-1}\nwixd{NWppJ6t-RDDRz-2}\nwixd{NWppJ6t-RDDRz-3}\nwixd{NWppJ6t-RDDRz-4}\nwixd{NWppJ6t-RDDRz-5}\nwixd{NWppJ6t-RDDRz-6}\nwixd{NWppJ6t-RDDRz-7}\nwixd{NWppJ6t-RDDRz-8}}}%
\nwixlogsorted{c}{{GiNaC declarations}{NWppJ6t-3vIsfh-1}{\nwixu{NWppJ6t-32jW6L-1}\nwixd{NWppJ6t-3vIsfh-1}}}%
\nwixlogsorted{c}{{hello-cycle-anim.cpp}{NWppJ6t-1kKRnF-1}{\nwixd{NWppJ6t-1kKRnF-1}\nwixd{NWppJ6t-1kKRnF-2}\nwixd{NWppJ6t-1kKRnF-3}\nwixd{NWppJ6t-1kKRnF-4}\nwixd{NWppJ6t-1kKRnF-5}\nwixd{NWppJ6t-1kKRnF-6}\nwixd{NWppJ6t-1kKRnF-7}}}%
\nwixlogsorted{c}{{hello-cycle.cpp}{NWppJ6t-3rkk58-1}{\nwixd{NWppJ6t-3rkk58-1}\nwixd{NWppJ6t-3rkk58-2}\nwixd{NWppJ6t-3rkk58-3}\nwixd{NWppJ6t-3rkk58-4}\nwixd{NWppJ6t-3rkk58-5}}}%
\nwixlogsorted{c}{{identify infinity and real line}{NWppJ6t-1pIH3B-1}{\nwixu{NWppJ6t-DKmU5-3}\nwixd{NWppJ6t-1pIH3B-1}\nwixu{NWppJ6t-DKmU5-A}}}%
\nwixlogsorted{c}{{initial data for drawing}{NWppJ6t-17vXT1-1}{\nwixu{NWppJ6t-30Tc7D-1}\nwixd{NWppJ6t-17vXT1-1}}}%
\nwixlogsorted{c}{{initial data for proof}{NWppJ6t-2y20ZP-1}{\nwixu{NWppJ6t-2zWvnC-1}\nwixd{NWppJ6t-2y20ZP-1}\nwixd{NWppJ6t-2y20ZP-2}}}%
\nwixlogsorted{c}{{initialise the dimension and vector}{NWppJ6t-2V7UoL-1}{\nwixd{NWppJ6t-2V7UoL-1}\nwixu{NWppJ6t-2j51zU-1}\nwixu{NWppJ6t-1NAwBA-1}}}%
\nwixlogsorted{c}{{license}{NWppJ6t-ZXuKx-1}{\nwixu{NWppJ6t-3rkk58-1}\nwixu{NWppJ6t-1kKRnF-1}\nwixu{NWppJ6t-2SYluR-1}\nwixu{NWppJ6t-2t2vfS-1}\nwixu{NWppJ6t-30Tc7D-1}\nwixu{NWppJ6t-2zWvnC-1}\nwixu{NWppJ6t-RDDRz-1}\nwixu{NWppJ6t-2pgnFW-1}\nwixu{NWppJ6t-3NCnUp-1}\nwixu{NWppJ6t-32jW6L-1}\nwixd{NWppJ6t-ZXuKx-1}}}%
\nwixlogsorted{c}{{member of figure class}{NWppJ6t-4OZeJw-1}{\nwixu{NWppJ6t-7vmg0-1}\nwixd{NWppJ6t-4OZeJw-1}\nwixd{NWppJ6t-4OZeJw-2}\nwixd{NWppJ6t-4OZeJw-3}\nwixd{NWppJ6t-4OZeJw-4}\nwixd{NWppJ6t-4OZeJw-5}}}%
\nwixlogsorted{c}{{modular-group.cpp}{NWppJ6t-2SYluR-1}{\nwixd{NWppJ6t-2SYluR-1}\nwixd{NWppJ6t-2SYluR-2}\nwixd{NWppJ6t-2SYluR-3}\nwixd{NWppJ6t-2SYluR-4}\nwixd{NWppJ6t-2SYluR-5}\nwixd{NWppJ6t-2SYluR-6}\nwixd{NWppJ6t-2SYluR-7}\nwixd{NWppJ6t-2SYluR-8}\nwixd{NWppJ6t-2SYluR-9}\nwixd{NWppJ6t-2SYluR-A}\nwixd{NWppJ6t-2SYluR-B}}}%
\nwixlogsorted{c}{{nine-points-thm-symb.cpp}{NWppJ6t-2zWvnC-1}{\nwixd{NWppJ6t-2zWvnC-1}\nwixd{NWppJ6t-2zWvnC-2}}}%
\nwixlogsorted{c}{{nine-points-thm.cpp}{NWppJ6t-30Tc7D-1}{\nwixd{NWppJ6t-30Tc7D-1}\nwixd{NWppJ6t-30Tc7D-2}\nwixd{NWppJ6t-30Tc7D-3}\nwixd{NWppJ6t-30Tc7D-4}\nwixd{NWppJ6t-30Tc7D-5}\nwixd{NWppJ6t-30Tc7D-6}\nwixd{NWppJ6t-30Tc7D-7}\nwixd{NWppJ6t-30Tc7D-8}\nwixd{NWppJ6t-30Tc7D-9}\nwixd{NWppJ6t-30Tc7D-A}\nwixd{NWppJ6t-30Tc7D-B}\nwixd{NWppJ6t-30Tc7D-C}\nwixu{NWppJ6t-1p0Y9w-1}}}%
\nwixlogsorted{c}{{predefined cycle relations}{NWppJ6t-2iyHmp-1}{\nwixd{NWppJ6t-2iyHmp-1}\nwixd{NWppJ6t-2iyHmp-2}\nwixd{NWppJ6t-2iyHmp-3}\nwixd{NWppJ6t-2iyHmp-4}\nwixd{NWppJ6t-2iyHmp-5}\nwixd{NWppJ6t-2iyHmp-6}\nwixd{NWppJ6t-2iyHmp-7}\nwixd{NWppJ6t-2iyHmp-8}\nwixd{NWppJ6t-2iyHmp-9}\nwixd{NWppJ6t-2iyHmp-A}\nwixd{NWppJ6t-2iyHmp-B}\nwixd{NWppJ6t-2iyHmp-C}\nwixd{NWppJ6t-2iyHmp-D}\nwixd{NWppJ6t-2iyHmp-E}\nwixd{NWppJ6t-2iyHmp-F}\nwixu{NWppJ6t-1d2GZW-9}}}%
\nwixlogsorted{c}{{public methods for cycle relation}{NWppJ6t-1HOJwC-1}{\nwixd{NWppJ6t-1HOJwC-1}\nwixu{NWppJ6t-1d2GZW-2}}}%
\nwixlogsorted{c}{{public methods for subfigure}{NWppJ6t-25h4gT-1}{\nwixd{NWppJ6t-25h4gT-1}\nwixu{NWppJ6t-4UgdbB-1}}}%
\nwixlogsorted{c}{{public methods in figure class}{NWppJ6t-AWRj3-1}{\nwixd{NWppJ6t-AWRj3-1}\nwixd{NWppJ6t-AWRj3-2}\nwixd{NWppJ6t-AWRj3-3}\nwixd{NWppJ6t-AWRj3-4}\nwixd{NWppJ6t-AWRj3-5}\nwixd{NWppJ6t-AWRj3-6}\nwixd{NWppJ6t-AWRj3-7}\nwixd{NWppJ6t-AWRj3-8}\nwixd{NWppJ6t-AWRj3-9}\nwixd{NWppJ6t-AWRj3-A}\nwixd{NWppJ6t-AWRj3-B}\nwixd{NWppJ6t-AWRj3-C}\nwixd{NWppJ6t-AWRj3-D}\nwixd{NWppJ6t-AWRj3-E}\nwixd{NWppJ6t-AWRj3-F}\nwixd{NWppJ6t-AWRj3-G}\nwixd{NWppJ6t-AWRj3-H}\nwixd{NWppJ6t-AWRj3-I}\nwixd{NWppJ6t-AWRj3-J}\nwixd{NWppJ6t-AWRj3-K}\nwixd{NWppJ6t-AWRj3-L}\nwixd{NWppJ6t-AWRj3-M}\nwixd{NWppJ6t-AWRj3-N}\nwixd{NWppJ6t-AWRj3-O}\nwixd{NWppJ6t-AWRj3-P}\nwixd{NWppJ6t-AWRj3-Q}\nwixd{NWppJ6t-AWRj3-R}\nwixd{NWppJ6t-AWRj3-S}\nwixd{NWppJ6t-AWRj3-T}\nwixd{NWppJ6t-AWRj3-U}\nwixu{NWppJ6t-7vmg0-4}\nwixd{NWppJ6t-AWRj3-V}\nwixd{NWppJ6t-AWRj3-W}\nwixd{NWppJ6t-AWRj3-X}\nwixd{NWppJ6t-AWRj3-Y}}}%
\nwixlogsorted{c}{{relations to check}{NWppJ6t-iB8QG-1}{\nwixd{NWppJ6t-iB8QG-1}\nwixd{NWppJ6t-iB8QG-2}\nwixd{NWppJ6t-iB8QG-3}\nwixd{NWppJ6t-iB8QG-4}\nwixd{NWppJ6t-iB8QG-5}\nwixd{NWppJ6t-iB8QG-6}\nwixd{NWppJ6t-iB8QG-7}\nwixu{NWppJ6t-1d2GZW-5}}}%
\nwixlogsorted{c}{{run through all cycles in two nodes async}{NWppJ6t-47PgDD-1}{\nwixu{NWppJ6t-DKmU5-2E}\nwixd{NWppJ6t-47PgDD-1}\nwixu{NWppJ6t-DKmU5-2G}}}%
\nwixlogsorted{c}{{run through all cycles in two nodes correspondingly}{NWppJ6t-1bkNr0-1}{\nwixu{NWppJ6t-DKmU5-2E}\nwixd{NWppJ6t-1bkNr0-1}\nwixu{NWppJ6t-DKmU5-2G}}}%
\nwixlogsorted{c}{{separating chunk}{NWppJ6t-1fWRR8-1}{\nwixd{NWppJ6t-1fWRR8-1}\nwixd{NWppJ6t-1fWRR8-2}\nwixd{NWppJ6t-1fWRR8-3}\nwixd{NWppJ6t-1fWRR8-4}\nwixd{NWppJ6t-1fWRR8-5}\nwixd{NWppJ6t-1fWRR8-6}\nwixd{NWppJ6t-1fWRR8-7}\nwixd{NWppJ6t-1fWRR8-8}\nwixd{NWppJ6t-1fWRR8-9}}}%
\nwixlogsorted{c}{{set cycle data to the node}{NWppJ6t-26cGQh-1}{\nwixu{NWppJ6t-1HqLYY-2}\nwixd{NWppJ6t-26cGQh-1}}}%
\nwixlogsorted{c}{{set cycle metric in figure}{NWppJ6t-1wPO4G-1}{\nwixu{NWppJ6t-DKmU5-4}\nwixd{NWppJ6t-1wPO4G-1}\nwixd{NWppJ6t-1wPO4G-2}\nwixd{NWppJ6t-1wPO4G-3}\nwixd{NWppJ6t-1wPO4G-4}\nwixu{NWppJ6t-DKmU5-1Q}}}%
\nwixlogsorted{c}{{set point metric in figure}{NWppJ6t-4ALcRV-1}{\nwixu{NWppJ6t-DKmU5-2}\nwixd{NWppJ6t-4ALcRV-1}\nwixd{NWppJ6t-4ALcRV-2}\nwixu{NWppJ6t-DKmU5-1Q}}}%
\nwixlogsorted{c}{{set the infinity}{NWppJ6t-2j51zU-1}{\nwixu{NWppJ6t-DKmU5-1}\nwixd{NWppJ6t-2j51zU-1}\nwixu{NWppJ6t-1NAwBA-1}\nwixu{NWppJ6t-DKmU5-1V}}}%
\nwixlogsorted{c}{{set the real line}{NWppJ6t-2q1QAq-1}{\nwixu{NWppJ6t-DKmU5-1}\nwixd{NWppJ6t-2q1QAq-1}\nwixu{NWppJ6t-1NAwBA-1}\nwixu{NWppJ6t-DKmU5-1V}}}%
\nwixlogsorted{c}{{setup real line and infinity}{NWppJ6t-1NAwBA-1}{\nwixu{NWppJ6t-DKmU5-8}\nwixd{NWppJ6t-1NAwBA-1}\nwixu{NWppJ6t-DKmU5-A}}}%
\nwixlogsorted{c}{{start to print cycle node}{NWppJ6t-1o3xH5-1}{\nwixu{NWppJ6t-1HqLYY-G}\nwixu{NWppJ6t-1HqLYY-H}\nwixd{NWppJ6t-1o3xH5-1}}}%
\nwixlogsorted{c}{{subfigure class}{NWppJ6t-3dX0u0-1}{\nwixu{NWppJ6t-32jW6L-1}\nwixd{NWppJ6t-3dX0u0-1}\nwixd{NWppJ6t-3dX0u0-2}\nwixd{NWppJ6t-3dX0u0-3}\nwixd{NWppJ6t-3dX0u0-4}\nwixd{NWppJ6t-3dX0u0-5}\nwixd{NWppJ6t-3dX0u0-6}\nwixd{NWppJ6t-3dX0u0-7}\nwixd{NWppJ6t-3dX0u0-8}\nwixd{NWppJ6t-3dX0u0-9}}}%
\nwixlogsorted{c}{{subfigure header}{NWppJ6t-4UgdbB-1}{\nwixu{NWppJ6t-3NCnUp-3}\nwixd{NWppJ6t-4UgdbB-1}\nwixd{NWppJ6t-4UgdbB-2}\nwixd{NWppJ6t-4UgdbB-3}}}%
\nwixlogsorted{c}{{update node zero level}{NWppJ6t-3AODA9-1}{\nwixu{NWppJ6t-DKmU5-1H}\nwixd{NWppJ6t-3AODA9-1}\nwixd{NWppJ6t-3AODA9-2}\nwixd{NWppJ6t-3AODA9-3}}}%
\nwixlogsorted{c}{{using all namespaces}{NWppJ6t-3uHj2c-1}{\nwixu{NWppJ6t-3rkk58-1}\nwixd{NWppJ6t-3uHj2c-1}\nwixu{NWppJ6t-1kKRnF-1}\nwixu{NWppJ6t-2SYluR-1}\nwixu{NWppJ6t-2t2vfS-1}\nwixu{NWppJ6t-30Tc7D-1}\nwixu{NWppJ6t-2zWvnC-1}\nwixu{NWppJ6t-RDDRz-1}\nwixu{NWppJ6t-2pgnFW-1}}}%
\nwixlogsorted{i}{{\nwixident{{\_}{\_}{\_}{\_}figure{\_}{\_}}}{:un:un:un:unfigure:un:un}}%
\nwixlogsorted{i}{{\nwixident{add{\_}cycle}}{add:uncycle}}%
\nwixlogsorted{i}{{\nwixident{add{\_}cycle{\_}rel}}{add:uncycle:unrel}}%
\nwixlogsorted{i}{{\nwixident{add{\_}point}}{add:unpoint}}%
\nwixlogsorted{i}{{\nwixident{add{\_}subfigure}}{add:unsubfigure}}%
\nwixlogsorted{i}{{\nwixident{angle{\_}is}}{angle:unis}}%
\nwixlogsorted{i}{{\nwixident{apply}}{apply}}%
\nwixlogsorted{i}{{\nwixident{archive}}{archive}}%
\nwixlogsorted{i}{{\nwixident{arrangement{\_}write}}{arrangement:unwrite}}%
\nwixlogsorted{i}{{\nwixident{asy{\_}animate}}{asy:unanimate}}%
\nwixlogsorted{i}{{\nwixident{asy{\_}cycle{\_}color}}{asy:uncycle:uncolor}}%
\nwixlogsorted{i}{{\nwixident{asy{\_}draw}}{asy:undraw}}%
\nwixlogsorted{i}{{\nwixident{asy{\_}style}}{asy:unstyle}}%
\nwixlogsorted{i}{{\nwixident{asy{\_}write}}{asy:unwrite}}%
\nwixlogsorted{i}{{\nwixident{check{\_}rel}}{check:unrel}}%
\nwixlogsorted{i}{{\nwixident{check{\_}tangent}}{check:untangent}}%
\nwixlogsorted{i}{{\nwixident{coefficients{\_}are{\_}real}}{coefficients:unare:unreal}}%
\nwixlogsorted{i}{{\nwixident{cross{\_}t{\_}distance}}{cross:unt:undistance}}%
\nwixlogsorted{i}{{\nwixident{cycle{\_}adifferent}}{cycle:unadifferent}}%
\nwixlogsorted{i}{{\nwixident{cycle{\_}angle}}{cycle:unangle}}%
\nwixlogsorted{i}{{\nwixident{cycle{\_}cross{\_}t{\_}distance}}{cycle:uncross:unt:undistance}}%
\nwixlogsorted{i}{{\nwixident{cycle{\_}data}}{cycle:undata}}%
\nwixlogsorted{i}{{\nwixident{cycle{\_}different}}{cycle:undifferent}}%
\nwixlogsorted{i}{{\nwixident{cycle{\_}f{\_}orthogonal}}{cycle:unf:unorthogonal}}%
\nwixlogsorted{i}{{\nwixident{cycle{\_}metric}}{cycle:unmetric}}%
\nwixlogsorted{i}{{\nwixident{cycle{\_}moebius}}{cycle:unmoebius}}%
\nwixlogsorted{i}{{\nwixident{cycle{\_}node}}{cycle:unnode}}%
\nwixlogsorted{i}{{\nwixident{cycle{\_}orthogonal}}{cycle:unorthogonal}}%
\nwixlogsorted{i}{{\nwixident{cycle{\_}orthogonal{\_}e}}{cycle:unorthogonal:une}}%
\nwixlogsorted{i}{{\nwixident{cycle{\_}orthogonal{\_}h}}{cycle:unorthogonal:unh}}%
\nwixlogsorted{i}{{\nwixident{cycle{\_}orthogonal{\_}p}}{cycle:unorthogonal:unp}}%
\nwixlogsorted{i}{{\nwixident{cycle{\_}power}}{cycle:unpower}}%
\nwixlogsorted{i}{{\nwixident{cycle{\_}relation}}{cycle:unrelation}}%
\nwixlogsorted{i}{{\nwixident{cycle{\_}sl2}}{cycle:unsl2}}%
\nwixlogsorted{i}{{\nwixident{cycle{\_}tangent}}{cycle:untangent}}%
\nwixlogsorted{i}{{\nwixident{cycle{\_}tangent{\_}i}}{cycle:untangent:uni}}%
\nwixlogsorted{i}{{\nwixident{cycle{\_}tangent{\_}o}}{cycle:untangent:uno}}%
\nwixlogsorted{i}{{\nwixident{do{\_}not{\_}update{\_}subfigure}}{do:unnot:unupdate:unsubfigure}}%
\nwixlogsorted{i}{{\nwixident{do{\_}print{\_}double}}{do:unprint:undouble}}%
\nwixlogsorted{i}{{\nwixident{epsilon}}{epsilon}}%
\nwixlogsorted{i}{{\nwixident{evalf}}{evalf}}%
\nwixlogsorted{i}{{\nwixident{evaluate{\_}cycle}}{evaluate:uncycle}}%
\nwixlogsorted{i}{{\nwixident{evaluation{\_}assist}}{evaluation:unassist}}%
\nwixlogsorted{i}{{\nwixident{ex}}{ex}}%
\nwixlogsorted{i}{{\nwixident{figure}}{figure}}%
\nwixlogsorted{i}{{\nwixident{figure{\_}ask{\_}debug{\_}status}}{figure:unask:undebug:unstatus}}%
\nwixlogsorted{i}{{\nwixident{FIGURE{\_}DEBUG}}{FIGURE:unDEBUG}}%
\nwixlogsorted{i}{{\nwixident{figure{\_}debug{\_}off}}{figure:undebug:unoff}}%
\nwixlogsorted{i}{{\nwixident{figure{\_}debug{\_}on}}{figure:undebug:unon}}%
\nwixlogsorted{i}{{\nwixident{float{\_}evaluation}}{float:unevaluation}}%
\nwixlogsorted{i}{{\nwixident{freeze}}{freeze}}%
\nwixlogsorted{i}{{\nwixident{get{\_}all{\_}keys}}{get:unall:unkeys}}%
\nwixlogsorted{i}{{\nwixident{get{\_}asy{\_}style}}{get:unasy:unstyle}}%
\nwixlogsorted{i}{{\nwixident{get{\_}cycle}}{get:uncycle}}%
\nwixlogsorted{i}{{\nwixident{get{\_}cycle{\_}label}}{get:uncycle:unlabel}}%
\nwixlogsorted{i}{{\nwixident{get{\_}cycle{\_}metric}}{get:uncycle:unmetric}}%
\nwixlogsorted{i}{{\nwixident{get{\_}cycle{\_}node}}{get:uncycle:unnode}}%
\nwixlogsorted{i}{{\nwixident{get{\_}dim()}}{get:undim()}}%
\nwixlogsorted{i}{{\nwixident{get{\_}generation}}{get:ungeneration}}%
\nwixlogsorted{i}{{\nwixident{get{\_}infinity}}{get:uninfinity}}%
\nwixlogsorted{i}{{\nwixident{get{\_}max{\_}generation}}{get:unmax:ungeneration}}%
\nwixlogsorted{i}{{\nwixident{get{\_}point{\_}metric}}{get:unpoint:unmetric}}%
\nwixlogsorted{i}{{\nwixident{get{\_}real{\_}line}}{get:unreal:unline}}%
\nwixlogsorted{i}{{\nwixident{GHOST{\_}GEN}}{GHOST:unGEN}}%
\nwixlogsorted{i}{{\nwixident{infinity}}{infinity}}%
\nwixlogsorted{i}{{\nwixident{INFINITY{\_}GEN}}{INFINITY:unGEN}}%
\nwixlogsorted{i}{{\nwixident{info}}{info}}%
\nwixlogsorted{i}{{\nwixident{is{\_}adifferent}}{is:unadifferent}}%
\nwixlogsorted{i}{{\nwixident{is{\_}almost{\_}equal}}{is:unalmost:unequal}}%
\nwixlogsorted{i}{{\nwixident{is{\_}different}}{is:undifferent}}%
\nwixlogsorted{i}{{\nwixident{is{\_}f{\_}orthogonal}}{is:unf:unorthogonal}}%
\nwixlogsorted{i}{{\nwixident{is{\_}less{\_}than{\_}epsilon}}{is:unless:unthan:unepsilon}}%
\nwixlogsorted{i}{{\nwixident{is{\_}orthogonal}}{is:unorthogonal}}%
\nwixlogsorted{i}{{\nwixident{is{\_}real{\_}cycle}}{is:unreal:uncycle}}%
\nwixlogsorted{i}{{\nwixident{is{\_}tangent}}{is:untangent}}%
\nwixlogsorted{i}{{\nwixident{is{\_}tangent{\_}i}}{is:untangent:uni}}%
\nwixlogsorted{i}{{\nwixident{is{\_}tangent{\_}o}}{is:untangent:uno}}%
\nwixlogsorted{i}{{\nwixident{k}}{k}}%
\nwixlogsorted{i}{{\nwixident{key}}{key}}%
\nwixlogsorted{i}{{\nwixident{l}}{l}}%
\nwixlogsorted{i}{{\nwixident{label{\_}pos}}{label:unpos}}%
\nwixlogsorted{i}{{\nwixident{label{\_}string}}{label:unstring}}%
\nwixlogsorted{i}{{\nwixident{m}}{m}}%
\nwixlogsorted{i}{{\nwixident{main}}{main}}%
\nwixlogsorted{i}{{\nwixident{make{\_}angle}}{make:unangle}}%
\nwixlogsorted{i}{{\nwixident{measure}}{measure}}%
\nwixlogsorted{i}{{\nwixident{metric{\_}e}}{metric:une}}%
\nwixlogsorted{i}{{\nwixident{metric{\_}h}}{metric:unh}}%
\nwixlogsorted{i}{{\nwixident{metric{\_}p}}{metric:unp}}%
\nwixlogsorted{i}{{\nwixident{midpoint{\_}constructor}}{midpoint:unconstructor}}%
\nwixlogsorted{i}{{\nwixident{MoebInv}}{MoebInv}}%
\nwixlogsorted{i}{{\nwixident{moebius{\_}transform}}{moebius:untransform}}%
\nwixlogsorted{i}{{\nwixident{move{\_}cycle}}{move:uncycle}}%
\nwixlogsorted{i}{{\nwixident{move{\_}point}}{move:unpoint}}%
\nwixlogsorted{i}{{\nwixident{name}}{name}}%
\nwixlogsorted{i}{{\nwixident{nodes}}{nodes}}%
\nwixlogsorted{i}{{\nwixident{nops}}{nops}}%
\nwixlogsorted{i}{{\nwixident{numeric}}{numeric}}%
\nwixlogsorted{i}{{\nwixident{only{\_}reals}}{only:unreals}}%
\nwixlogsorted{i}{{\nwixident{op}}{op}}%
\nwixlogsorted{i}{{\nwixident{PCR}}{PCR}}%
\nwixlogsorted{i}{{\nwixident{point{\_}metric}}{point:unmetric}}%
\nwixlogsorted{i}{{\nwixident{power{\_}is}}{power:unis}}%
\nwixlogsorted{i}{{\nwixident{product{\_}nonpositive}}{product:unnonpositive}}%
\nwixlogsorted{i}{{\nwixident{product{\_}sign}}{product:unsign}}%
\nwixlogsorted{i}{{\nwixident{read{\_}archive}}{read:unarchive}}%
\nwixlogsorted{i}{{\nwixident{real{\_}line}}{real:unline}}%
\nwixlogsorted{i}{{\nwixident{REAL{\_}LINE{\_}GEN}}{REAL:unLINE:unGEN}}%
\nwixlogsorted{i}{{\nwixident{realsymbol}}{realsymbol}}%
\nwixlogsorted{i}{{\nwixident{remove{\_}cycle{\_}node}}{remove:uncycle:unnode}}%
\nwixlogsorted{i}{{\nwixident{reset{\_}figure}}{reset:unfigure}}%
\nwixlogsorted{i}{{\nwixident{rgb}}{rgb}}%
\nwixlogsorted{i}{{\nwixident{save}}{save}}%
\nwixlogsorted{i}{{\nwixident{set{\_}asy{\_}style}}{set:unasy:unstyle}}%
\nwixlogsorted{i}{{\nwixident{set{\_}cycle}}{set:uncycle}}%
\nwixlogsorted{i}{{\nwixident{set{\_}exact{\_}eval}}{set:unexact:uneval}}%
\nwixlogsorted{i}{{\nwixident{set{\_}float{\_}eval}}{set:unfloat:uneval}}%
\nwixlogsorted{i}{{\nwixident{set{\_}metric}}{set:unmetric}}%
\nwixlogsorted{i}{{\nwixident{show{\_}asy{\_}graphics}}{show:unasy:ungraphics}}%
\nwixlogsorted{i}{{\nwixident{show{\_}asy{\_}off}}{show:unasy:unoff}}%
\nwixlogsorted{i}{{\nwixident{show{\_}asy{\_}on}}{show:unasy:unon}}%
\nwixlogsorted{i}{{\nwixident{sl2{\_}transform}}{sl2:untransform}}%
\nwixlogsorted{i}{{\nwixident{sq{\_}cross{\_}t{\_}distance{\_}is}}{sq:uncross:unt:undistance:unis}}%
\nwixlogsorted{i}{{\nwixident{sq{\_}t{\_}distance{\_}is}}{sq:unt:undistance:unis}}%
\nwixlogsorted{i}{{\nwixident{steiner{\_}power}}{steiner:unpower}}%
\nwixlogsorted{i}{{\nwixident{subfigure}}{subfigure}}%
\nwixlogsorted{i}{{\nwixident{subs}}{subs}}%
\nwixlogsorted{i}{{\nwixident{tangential{\_}distance}}{tangential:undistance}}%
\nwixlogsorted{i}{{\nwixident{TeXname}}{TeXname}}%
\nwixlogsorted{i}{{\nwixident{unfreeze}}{unfreeze}}%
\nwixlogsorted{i}{{\nwixident{unique{\_}cycle}}{unique:uncycle}}%
\nwixlogsorted{i}{{\nwixident{update{\_}cycle{\_}node}}{update:uncycle:unnode}}%
\nwixlogsorted{i}{{\nwixident{update{\_}cycles}}{update:uncycles}}%
\nwixlogsorted{i}{{\nwixident{update{\_}node{\_}lst}}{update:unnode:unlst}}%
\nwbegindocs{1281}\nwdocspar
\fi
\end{document}